\documentclass[11pt]{article}
\usepackage{amsmath}
\usepackage{graphicx}
\usepackage{amsfonts}
\usepackage{amssymb}
\usepackage{epsf}
\usepackage{latexsym}

\def\80{\hspace{0.8in}}
\def\ni{\noindent}
\def\be{\begin{equation}}
\def\ee{\end{equation}}
\def\bea{\begin{eqnarray}}
\def\eea{\end{eqnarray}}
\def\pa{\partial}

\def\fn{\footnote}
\def\vc{V^{\frac{2}{3}}}

\def\case#1/#2{\textstyle\frac{#1}{#2}}

\begin{document}
\begin{titlepage}

\vspace{.7in}

\begin{center}
\Huge
{\bf GEOMETRODYNAMICS:}

\vspace{.2in}

{\bf SPACETIME OR SPACE?}

\normalsize 

\vspace{.4in}

\Huge{\bf Edward Anderson}

\vspace{.2in}

\Large{\bf Submitted for Ph.D. 03 / 2004}\normalsize

\vspace{.4in}

\Large{\em  Astronomy Unit, School of Mathematical Sciences, 

\vspace{.2in}

Queen Mary, University of London, E1 4NS, U.K. }

\normalsize

\vspace{.2in}
\vspace{.3in}
\baselineskip=24pt

\vspace{.3in}


\vspace{5.65in}

1

\end{center}

\end{titlepage}

\vspace{10in}

%

\noindent\Large\bf Acknowledgments \normalfont\normalsize

\mbox{ }

\noindent To the memory of my Father, who taught me mathematics and much more. 
To Claire for her caring support, and to my Mother.  

\mbox{ }

\noindent To all my friends, with thanks.  Yves, Becca, Suzy, Lynnette and Mark helped me 
survive my Cambridge years.  Yves, Becca and Suzy have stayed in touch, while I often enjoyed 
Lynnette and Chris's company though my PhD years, and Mark was a familiar and welcome figure at QMUL.    
I also thank the good friends I made toward the end of my Cambridge years: Ed, Matt, Bjoern, Alex, 
Angela and Alison.  And Bryony, for being a good friend during the difficult last year.  
Thanks also to all my other friends, office mates, and kind and entertaining people
I have crossed paths with.

\mbox{ }

\noindent To Professors Malcolm MacCallum and Reza Tavakol, with thanks for agreeing to supervise me, 
and for their encouragement and wisdom.  I also thank Dr James Lidsey for supervision, and both 
Dr Lidsey and Professor Tavakol for active collaborations.  I also thank Dr. Julian Barbour 
and Professor Niall \'{O} Murchadha for teaching me many things and for collaboration, 
and the Barbour family for much hospitality.  I also thank Brendan Foster and Dr. Bryan Kelleher 
for many discussions and for collaboration, and Dr. Harvey Brown for many discussions.   
I thank Professor MacCallum, Dr Barbour and Professor Tavakol for 
reading, criticizing and commenting on earlier drafts of this thesis.  

\mbox{ }

\noindent I thank the following people for one or more of: 
discussion, collaboration, reading final drafts of my papers, 
providing useful references, providing encouragement.  
Dr. Martin Bojowald, 
Professor Bernard Carr, 
Dr. David Coule, 
Professor Naresh Dadhich, 
Dr. Fabio Dahia, 
Dr. Steven Davis, 
Dr. Carl Dolby, 
Dr. Fay Dowker, 
Dr Henk van Elst, 
Richard Frewin,
Dr. Cristiano Germani, 
Dr. Domenico Giulini, 
Dr. Anne Green, 
Dr Tomohiro Harada, 
Rachael Hawkins, 
Dr Gregory Huey, 
Gian Paolo Imponente, 
Dr Deborah Konkowski, 
Dr Janna Levin,   
Professor Roy Maartens, 
Dr Joao Magueijo, 
Professor Shahn Majid, 
Dr. Marc Mars, 
Dr. Nikolaos Mavromatos, 
Dr. Filipe Mena, 
Simone Mercuri, 
Professor John Moffat, 
David Mulryne, 
Professor James Nester, 
Professor Don Page, 
Dr. Alexander Polnarev, 
Dr. Oliver Pooley, 
Dr. Istvan Racz,
Dr. Carlos Romero, 
Mike Roper, 
Professor Misao Sasaki, 
Dr. Chopin Soo,
Professor Raphael Sorkin, 
Professor Marek Szydlowski, 
Professor John Stachel, 
John Taylor, 
Professor William Unruh, 
Dr. Raul Vera, 
Professor James York and  
Professor Jorge Zanelli.  
\normalfont

\mbox{ }

\noindent I would like to thank PPARC for funding me, the 
Astronomy Unit for being generous with my allotted conference money, 
and the subsidy of the Heraeus Foundation for the Bad Honnef Quantum Gravity Workshop.

\vspace{5.3 in}

\noindent\Large{\bf Abstract}\normalsize

\mbox{ }

\noindent The work in this thesis concerns the split of Einstein field equations (EFE's) 
with respect to nowhere-null hypersurfaces, the  GR Cauchy and Initial Value problems (CP and IVP), 
the Canonical formulation of GR and its interpretation, and the 
Foundations of Relativity.  

\mbox{ }

\noindent I address Wheeler's question about the why 
of the form of the GR Hamiltonian constraint ``from plausible first principles".   
I consider Hojman--Kuchar--Teitelboim's spacetime-based first principles, and especially 
the new 3-space approach (TSA) first principles studied by Barbour, 
Foster, \'{O} Murchadha and myself.  The latter are relational, and assume 
less structure, but from these Dirac's procedure picks out GR 
as one of a few consistent possibilities.  The alternative possibilities are 
Strong gravity theories and some 
new Conformal theories.  The latter have privileged slicings similar to 
the maximal and constant mean curvature slicings of the Conformal 
IVP method.  
  
\mbox{ }

\noindent The plausibility of the TSA first principles are 
tested by coupling to fundamental matter.  Yang--Mills theory works.  
I criticize the original form of the TSA since I find that tacit assumptions remain 
and Dirac fields are not permitted.  However, comparison with Kucha\v{r}'s hypersurface 
formalism allows me to argue that all the known fundamental 
matter fields can be incorporated into the TSA.  The spacetime 
picture appears to possess more kinematics than  strictly necessary for 
building Lagrangians for physically-realized fundamental matter fields.  I debate whether 
space may be regarded as primary rather than spacetime.    The 
emergence (or not) of the Special Relativity Principles and 4-d 
General Covariance in the various TSA alternatives is investigated, as 
is the Equivalence Principle, and the Problem of Time in 
Quantum Gravity.  

\mbox{ }

\noindent Further results concern Elimination versus Conformal 
IVP methods, the badness of the timelike split of the EFE's, and 
reinterpreting Embeddings and Braneworlds guided by CP and IVP knowledge.  

\vspace{6in}

\noindent\Large\bf{Overview}\normalfont\normalsize  

\mbox{ }

\noindent Here I explain what is in this thesis and on which of my articles each topic is based.  

\mbox{ }

\noindent My Introduction serves the following purposes.  In I.0-1 I summarize the standard 
foundations of classical physics for later comparison (mostly in Part A); these sections also 
serve to establish much notation for the thesis.  In I.2 I provide the split formulation of GR, 
to be used both in Part A and Part B.  Much of this material was included in early sections of my 
publications.  This section also includes some original material filling up some gaps in the older 
literature.  I.3 briefly provides the quantum physics and quantum gravity needed to understand what 
is being attempted in Part A, and how Part A and Part B fit into current physical thought.  
 
\mbox{ }

\noindent{\bf Part A:} II considers work by Barbour, Bertotti, \'{O} Murchadha, Foster and myself 
({\sl The 3-space Approach}), on the foundations of particle dynamics, gauge theory and especially 
Relativity, starting from relational first principles.   I have carefully reworked this material and 
have added to it both from my papers \cite{Sanderson, Vanderson} and as an original contribution to this Thesis.  
III considers {\sl strong} and {\sl conformal} alternative theories of gravity that arise alongside 
GR in this 3-space approach, and is based on the pure gravity halves of my paper \cite{Sanderson} and 
of my collaboration with Barbour, Foster and \'{O} Murchadha \cite{CG}.  IV furthermore includes 
bosonic matter and is based on the matter halves of the papers \cite{Sanderson, CG} and on my 
paper with Barbour \cite{AB}.  VI is based on my critical article on the 3-space approach 
\cite{Vanderson}.  This comes to grips with what the 3-space approach's spatial principles really are, 
and how they may be related to Kucha\v{r}'s principles that presuppose spacetime.  My modified 
3-space principles are capable of accommodating spin-$\frac{1}{2}$ fermions and more.  
V.1 and VII are original contributions of mine to this Thesis which update the 3-space approach to 
General Relativity and attempt to relate it to the standard approach's Relativity Principles and 
Principle of Equivalence.  I hope to publish the new material in II, V.I, and VII as a new article 
\cite{LAnderson}.  V.2 provides more material on the conformal alternatives; this or related material 
in preparation may appear in a solo article as well as in a collaboration with 
Barbour, Foster, Kelleher and \'{O} Murchadha \cite{ABFKO}.  VIII is an incipient account of 
quantum gravitational issues which underly and motivate motivate much of Part A.  I briefly mentioned 
this in the conference proceedings \cite{Rio1}, but it remains work in progress.  

\mbox{ }

\noindent{\bf Part B:} This is a separate, largely critical application of the split formulation of 
GR.  I explain in B.1 how the {\sl sideways} analogue of the usual split, cases of which have 
recently been suggested in various guises, is causally and mathematically undesirable.  This work 
began when Lidsey and I came across and used a sideways method \cite{AL}, followed by my criticism of 
it in my contribution to the article with Dahia, Lidsey and Romero \cite{ADLR}, further criticism in 
my preprint with Tavakol \cite{ATpap}, and yet further related work filling in some gaps in the early 
Initial Value Problem literature, which were most conveniently presented in I.2.  
\cite{ATpap} also considers a multitude of other treatments of proper and sideways splits for 
GR-like braneworlds.  Some of this material (in B2) was published in my letter with Tavakol 
\cite{ATlett}, some of it will appear in my conference proceedings \cite{Rio2} (a summary of B.3), and 
some of it will form the basis for further articles that Tavakol and I have in mind.

\mbox{ } 

\noindent Appendix C on elliptic equations contains further technical material useful in both Parts.

\vspace{4in}

\noindent\Huge\bf{Table of Contents}\normalfont\normalsize
\bf{

\begin{tabbing}

\ni\bf{Acknowledgments} \= \hspace{4.18in} \= \mbox{ }\mbox{ }\mbox{ }2 \\

\mbox{ } \\

\ni\bf{Abstract} \> \> \mbox{ }\mbox{ }\mbox{ }3 \\

\mbox{ } \\

\ni\bf{Overview} \> \> \mbox{ }\mbox{ }\mbox{ }4 \\

\mbox{ } \\

\ni\bf{Table of Contents} \>  \> \mbox{ }\mbox{ }\mbox{ }5 \\

\mbox{ } \\

\ni\bf{\Large{I Introduction}} \= \bf{\mbox{ }} \hspace{4.15in} \= \mbox{ }10 \\

\mbox{ } \> \mbox{ } \>  \\


\ni\bf{0} \80 \= \bf{Geometry} \> \mbox{ }10 \\

\bf{0.1} \> \bf{Geometry and physics} \> \mbox{ }10  \\

\noindent 0.2 \> Intrinsic geometry of the 2-sphere \> \mbox{ }12 \\

\noindent 0.3 \> Differential geometry \> \mbox{ }12 \\

\noindent App 0.A \> Densities and integration \> \mbox{ }16 \\

\noindent 1 \> Classical physics \> \mbox{ }16 \\

\noindent 1.1 \> Newtonian physics \> \mbox{ }16 \\

\noindent 1.2 \> Principles of dynamics \> \mbox{ }17 \\

\noindent 1.2.1 \> Configuration space and the Euler--Lagrange equations \> \mbox{ }17 \\

\noindent 1.2.2 \> Legendre transformations, Jacobi's principle and  \> \\ 
\> Hamiltonian dynamics \> \mbox{ }18 \\

\noindent 1.2.3 \> Dirac's generalized Hamiltonian dynamics \> \mbox{ }19 \\

\noindent 1.2.4 \> Motivation for use of principles of dynamics \> \mbox{ }20 \\

\ni 1.3 \> Electromagnetism \> \mbox{ }21 \\

\ni 1.4 \> Special relativity \> \mbox{ }22 \\

\ni 1.5 \> General relativity \> \mbox{ }24 \\

\ni 1.6 \> Many routes to relativity \> \mbox{ }26 \\

\ni 1.7 \> Other classical matter fields \>  \mbox{ }28 \\

\ni 1.7.1 \> Scalars and spin-$\frac{1}{2}$ fermions \> \mbox{ }28 \\

\ni 1.7.2 \> Electromagnetism, U(1) scalar gauge theory and QED \> \mbox{ }29 \\

\ni 1.7.3 \> Yang--Mills theory, QCD and Weinberg--Salam theory  \> \mbox{ }30 \\

\ni 2 \> On the split formulation of Einstein's field equations \> \mbox{ }33 \\

\ni 2.1 \> Geometry of hypersurfaces \> \mbox{ }33 \\ 

\ni 2.1.1 \> Extrinsic curvature \> \mbox{ }33 \\

\ni 2.1.2 \> Gauss' outstanding theorem \> \mbox{ }33 \\

\ni 2.1.3 \> Hypersurface geometry  \> \mbox{ }33 \\

\ni 2.2 \> Split of the EFE's with respect to nowhere-null \> \\
\ni \> hypersurfaces \> \mbox{ }35 \\

\ni 2.3 \> Analytic approach to the GR ``CP and IVP" \> \mbox{ }37 \\

\ni 2.3.1 \> Cauchy problems \> \mbox{ }37 \\

\ni 2.3.2 \> Simple signature-independent features of the GR CP \> \mbox{ }38 \\

\ni 2.3.3 \> Signature-dependent GR CP results \> \mbox{ }40 \\

\ni 2.3.4 \> CP's for GR including fundamental matter \> \mbox{ }42 \\

\ni 2.3.5 \> Global results \> \mbox{ }43 \\

\ni 2.4 \> Variational principles for GR \> \mbox{ }43 \\

\ni 2.5 \> Inclusion of fundamental matter fields \> \mbox{ }44 \\

\ni 2.6 \> The Dirac--ADM Hamiltonian formulation of GR \>  \mbox{ }44 \\

\ni 2.7 \> Superspace and geometrodynamics \> \mbox{ }47 \\

\ni 2.7.1 \> Geometry on Riem and Superspace \> \mbox{ }47 \\

\ni 2.7.2 \> Geometrodynamics: old RMW interpretation \> \mbox{ }49 \\

\ni 2.7.3 \> Geometrodynamics: modern interpretation \> \mbox{ }50 \\

\ni 2.8 \> The Baierlein--Sharp--Wheeler action for GR \> \mbox{ }50 \\

\ni 2.9 \> The GR IVP   \>   \mbox{ }51 \\

\ni 2.9.1 \> The thin sandwich method and conjecture   \>   \mbox{ }51 \\

\ni 2.9.2 \> Componentwise methods, including traditional elimination \> \\ 
\ni       \>        methods \> \mbox{ }52 \\

\ni 2.9.2.1 \> A systematic treatment of componentwise methods \> \mbox{ }52 \\

\ni 2.9.2.2 \> The result in Hawking and Ellis \> \mbox{ }53 \\

\ni 2.9.2.3 \> Bruhat's criticisms and Magaard's argument \> \mbox{ }54 \\

\ni 2.9.3 \> The conformal method of Lichnerowicz and York \> \mbox{ }56 \\

\ni 2.9.4 \> Further approaches and related mathematics used in \> \\ 
\ni       \> (3, 0; -1) case \> \mbox{ }58 \\

\ni 2.9.4.1 \> The conformal thin sandwich method \> \mbox{ }58 \\

\ni 2.9.4.2 \> Conformal superspace \> \mbox{ }58 \\

\ni 2.9.4.3 \> 3-d Conformal tensors and the York 1971 formulation \> \mbox{ }59 \\

\ni 2.9.5 \> A new elimination method in terms of irreducibles \> \mbox{ }59 \\

\ni 2.9.6 \> The IVP with fundamental matter and for alternative  \> \\
\ni       \> theories of gravity \> \mbox{ }60 \\

\ni 2.10 \> Maximal and CMC slicings \> \mbox{ }60 \\

\ni 2.11 \> Numerical applications \> \mbox{ }62 \\

\ni 2.12 \> Further formulations of the EFE's \> \mbox{ }62 \\

\ni 2.12.1 \> Metric formulations \> \mbox{ }62 \\

\ni 2.12.2 \> `Bein' formulations \> \mbox{ }63 \\

\ni 2.12.3 \> Ashtekar variables  \> \mbox{ }63 \\

\ni 2.12.4 \> Formulations using different splits \> \mbox{ }63  \\

\ni 3 \> Quantum physics \> \mbox{ }64 \\

\ni 3.1 \>  Finite systems \> \mbox{ }64 \\

\ni 3.2 \> Infinite systems: quantum field theory  \> \mbox{ }64 \\

\ni 3.3 \> Approaches to quatum gravity \> \mbox{ }65 \\

\ni 3.3.1 \> Outline of need for and approaches to quantum gravity \> \mbox{ }65 \\

\ni 3.3.2 \> The canonical approach and the Problem of time \> \mbox{ }66 \\

\ni 3.3.3 \> Minisuperspace quantum cosmology and other toys \> \mbox{ }68 \\

\ni 3.3.4 \> The covariant approach and unification  \> \mbox{ }68 \\


\mbox{ } \> \mbox{ } \\

\ni{\Large{\bf A The 3-space approach}} \> \> \mbox{ }70 \\

\mbox{ } \> \mbox{ } \\

\ni{\Large{\bf II Answering Wheeler's question}} \> \> \mbox{ }70 \\ 

\mbox{ } \> \mbox{ } \\

\ni\bf{1} \> \bf{Hojman, Kucha\v{r} and Teitelboim's answer}    \> \mbox{ }70             \\

\bf{2}             \> \bf{Barbour, Foster and O Murchadha's answer: the 3-space}  \\ 
\ni \mbox{}        \>  approach (TSA) \> \mbox{ }72 \\

\ni 2.1 \> The Barbour--Bertotti 1982 theory and best matching \> \mbox{ }72 \\

\ni 2.1.1 \> Machian point-particle theories \> \mbox{ }72 \\

\ni 2.1.2 \> General best-matched reparameterization invariant schemes \> \mbox{ }77 \\

\ni 2.2 \> Relativity without relativity  \> \mbox{ }78 \\

\ni 2.2.1 \> Local square roots \> \mbox{ }78 \\

\ni 2.2.2 \> Uniqueness of consistent BSW-type actions \> \mbox{ }81 \\

\ni 2.2.3 \> Interpretation of the consistency condition \> \mbox{ }82 \\

\ni 2.2.4 \> Deducing BM from BSW alone \> \mbox{ }83 \\

\ni 2.2.5 \> Higher derivative potentials \> \mbox{ }84 \\

\ni 2.2.6 \> The inclusion of matter \> \mbox{ }84 \\

\ni 3 \> Spacetime or space? \> \mbox{ }85 \\

\ni App II.A \> Deriving the relativity without relativity result \> \mbox{ }87 \\

\ni App II.B \> 3-space approach and gauge theory \> \mbox{ }87 \\

\mbox{ } \> \mbox{ } \> \\

\ni\bf{\Large{III Alternative TSA theories}} \> \> \mbox{ }90  \\

\mbox{ } \> \mbox{ } \\

\ni\bf{1} \> \bf{Strong gravity alternatives} \> \mbox{ }90 \\

\ni\bf{1.1} \> \bf{Introduction} \> \mbox{ }90 \\

\ni 1.2 \> Strong gravity and the TSA: X is arbitrary \> \mbox{ }91 \\

\ni 1.3 \> On the meaning of the theories: application to scalar-tensor \\
\ni     \> theories  \> \mbox{ }92 \\

\ni 1.4 \> Application to conformal gravity \> \mbox{ }94 \\

\ni 1.5 \> Difficulty with implementation of Ashtekar variables \> \mbox{ }94 \\

\ni 1.6 \> PDE problems for strong gravity theories \> \mbox{ }95 \\

\ni 2 \> Conformal alternatives \> \mbox{ }96 \\

\ni 2.1 \> Introduction \> \mbox{ }96 \\

\ni 2.2 \> Scale-invariant particle dynamics model \> \mbox{ }97 \\

\ni 2.3 \> Two auxiliary coordinate formulation of conformal gravity \> \mbox{ }99 \\

\ni 2.4 \> Single auxiliary Lagrangian formulation of conformal gravity \> 101 \\ 

\ni 2.5 \> Integral conditions and the cosmological constant \> 106 \\

\ni 2.6 \> Hamiltonian formulation and alternative theories \> 108 \\

\ni App III.2.A \>  Supporting material on conformal IVP method \> 111 \\

\ni App III.2.B \>  Other scale-invariant theories \> 112 \\

\mbox{ } \> \mbox{ } \\ 

\ni\bf{\Large{IV TSA: coupling of bosonic matter}} \>  \> 113 \\

\mbox{ } \> \mbox{ } \\

\ni\bf{1} \> \bf{Matter in TSA formulation of GR} \> 114 \\

\ni\bf{1.1} \> \bf{Original BFO work} \>  114 \\

\ni 1.1.1 \> Scalar fields \> 114 \\

\ni 1.1.2 \> A single 1-form field \> 115 \\

\ni 1.2  \> Matter workings from the perspective of a general theorem \> 117 \\

\ni 1.3 \> Examples \> 119 \\

\ni 1.3.1 \> Minimally-coupled scalar fields \> 119 \\

\ni 1.3.2 \> K interacting 1-form fields \> 119 \\

\ni 1.3.3 \> Discussion \> 124 \\

\ni 1.4 \>  Coupling matter to strong gravity \> 125 \\

\ni 1.4.1 \>  Scalar fields \> 125 \\

\ni 1.4.2 \> K interacting 1-form fields \> 126 \\

\ni 1.4.3 \> Discussion of strong gravity matter-coupling results \> 127 \\

\ni 2 \> Coupling of matter to conformal gravity \> 127 \\

\ni 2.1 \> General theorems \> 128 \\ 

\ni 2.2 \> Examples \> 130 \\

\ni 2.2.1 \>  Scalar fields \> 130 \\

\ni 2.2.2 \> 1-form fields \> 131 \\

\ni App IV.A \>  Na\"{\i}ve renormalizability \> 134 \\

\ni App IV.B \>  Gell-Mann--Glashow theorem \> 135 \\

\ni App IV.C \> Teitelboim's inclusion of matter into the HKT route \> 136 \\

\mbox{ } \> \mbox{ } \\

\ni{\bf{\Large V TSA: Discussion and interpretation}} \>  \> 137  \\

\mbox{ } \> \mbox{ } \\

\ni\bf{1} \> \bf{TSA versus the principles of relativity} \> 137 \\

\ni\bf{2} \> \bf{Discussion of conformal theories} \> 139 \\ 

\ni 2.1 \> Discussion from ABFO paper \>  139 \\

\ni 2.1.1 \> On the weak field limit \> 139 \\

\ni 2.1.2 \> Cosmology \> 140 \\

\ni 2.1.3 \> Brief quantum outline \> 141 \\

\ni 2.2 \> Further conformal alternatives \> 142 \\

\ni 2.2.1 \> First-principles formulation of CS + V theory \> 142 \\

\ni 2.2.2 \> Recovery of conformal gravity \> 143 \\

\ni 2.2.3 \> Further alternative conformal theories \> 144 \\

\ni 2.3 \>  Discussion and interpretation \> 145 \\

\ni 2.3.1 \> Conformal gravity and CS + V theory as PDE systems \> 145 \\

\ni 2.3.1.1 \> Traditional thin sandwiches \> 145 \\

\ni 2.3.1.2 \> Conformal thin sandwiches \> 146 \\ 

\ni 2.3.1.3 \> IVP-CP formulation \> 146 \\

\ni 2.3.2 \> Cosmology \> 147 \\

\ni 2.3.3 \> CS + V theory: interpretation and possible tests \> 148 \\

\mbox{ } \> \mbox{ } \\

\ni{\bf{\Large VI TSA: criticism}} \> \> 150 \\

\mbox{ } \> \mbox{ } \\

\ni\bf{1} \> \bf{Problems with the use of BSW actions} \> 150 \\

\ni\bf{1.1} \>  \bf{Insights from mechanics} \> 150 \\

\ni 1.2 \> Criticism of BSW--Jacobi analogy \>  151 \\

\ni 1.3 \> Lack of validity of the BSW form  \>  152 \\    
 
\ni 1.3.1 \> Conformal Problem of zeros \>  152 \\

\ni 1.3.2 \> The BSW form is an unknown notion of distance \>  152 \\
 
\ni 1.4 \> The fermionic contribution to the action is linear \> 153 \\ 

\ni 1.5 \> Higher derivative theories \> 156 \\  

\ni 1.6 \> Lapse-uneliminated actions \> 157 \\ 

\ni 2 \> The 3-space approach and the split spacetime framework \> 158 \\

\ni 2.1 \> Kucha\v{r}'s hypersurface or split spacetime framework \> 158 \\

\ni 2.2 \> SSF as a TSA tool \> 159 \\

\ni 2.3 \> The TSA allows more than the fields of nature \> 163 \\

\ni 2.4 \> Derivative coupling and the 3-space 1-form ansatz \> 164 \\

\ni 3 \> Connection between space and spacetime viewpoints \> 166 \\
 
\ni 4 \> Spin-$\frac{1}{2}$ fermions and the TSA \> 167 \\
 
\mbox{ } \> \mbox{ } \\

\ni{\bf{\Large VII More on TSA matter schemes}} \> \> 171 \\

\mbox{ } \> \mbox{ } \\

\ni\bf{1} \> \bf{Update of TSA with a single 1-form} \> 171 \\

\ni 1.1 \> A means of including Proca theory \> 171 \\

\ni 1.2 \> Further nonuniqueness examples \> 172 \\

\ni 2 \> TSA and the principle of equivalence \> 173 \\

\ni 2.1 \> On the origin of gauge theory \> 174 \\

\mbox{ } \> \mbox{ } \\

\ni {\bf{\Large VIII Toward a quantum TSA?}} \>  \> 175 \\

\mbox{ } \> \mbox{ } \\ 

\ni 1 \> Barbour's suggestion for quantization \> 175 \\

\ni 2 \>  Wheeler--DeWitt approach \> 176 \\

\ni 2.1 \> Strong gravities \> 176 \\

\ni 2.2 \> Conformal gravity \> 176 \\

\ni 3 \> Other approaches \> 178 \\

\ni 3.1 \> Internal time approach \> 178 \\

\mbox{ } \> \mbox{ } \\

\ni {\bf{\Large B Higher-dimensional spacetime?}} \> \> 179 \\

\mbox{ } \> \mbox{ } \\

\ni 1 \> Sideways problem in absence of thin matter sheets  \> 180 \\

\ni 1.1  \> Bad or unexplored behaviour of sideways analogue of GR CP \> 181 \\

\ni 1.2 \> Extra difficulty with sideways Campbell--Magaard \> 182 \\

\ni 1.3 \> Arguments against Campbell's theorem: conclusion \> 182 \\

\ni 1.4 \> What parts of conformal IVP method survive? \> 184 \\

\ni 1.5 \> Other methods and formulations \> 184 \\

\ni 1.6  \> Applying (r, 1; 1) methods to remove singularities \> 185 \\

\ni 1.6.1 \> Secondary objects in the study of singularities \> 185 \\

\ni 1.6.2  \> Relating the embedded and embedding secondary objects \> 186 \\

\ni 1.6.3 \>  Embeddings and geodesic incompleteness \> 187 \\

\ni 1.6.4 \>  Embedding flat FLRW in Minkowski \> 188 \\

\ni 1.6.5 \> Nonrigorousness of singularity removal by embedding \> 189 \\

\ni 2 \> (r, 1; 1) methods with thin matter sheets \> 189 \\

\ni 2.1 \> Origin of the junction conditions \> 190 \\

\ni 2.2 \> SMS's braneworld EFE's \> 191 \\

\ni 2.3  \> Ambiguity in the formulation of the BEFE's \> 192 \\

\ni 2.4 \> Examples of formulations of BEFE's with no quadratic terms  \> 196 \\

\ni 2.5 \> Two further comments about building SMS-type braneworlds \> 198 \\

\ni 2.6 \> Sideways York and thin-sandwich schemes with thin \\
\ni     \> matter sheets \> 199 \\

\ni 2.7 \>(n, 1; 1) singularity removal and thin matter sheets \> 200 \\

\ni 3 \> (n, 0; -1) methods with thin matter sheets \> 201 \\

\ni 3.1 \> Hierarchy of problems and their difficulties \> 201 \\

\ni 3.2 \> Thin matter sheet IVP \> 203 \\

\ni 3.3 \> Thick matter sheet IVP \> 208 \\

\ni 3.4 \> Discussion: modelling assumptions and stringy features \> 209 \\

\mbox{ } \> \mbox{ } \\

\ni {\bf{\Large C Appendix on elliptic equations}} \> \> 211 \\

\mbox{ } \> \mbox{ } \\

\ni 1 \> Linear elliptic equations \> 211 \\ 

\ni 2 \> Lichnerowicz--York equation \> 212 \\

\ni 2.1 \> Linearization \> 212 \\

\ni 2.2 \> The full equation and its simplifications \> 213 \\

\ni 2.3 \> Boundary conditions \> 213 \\

\ni 3 \> Techniques for solving the momentum constraint \> 214 \\

\mbox{ } \> \mbox{ } \\

\ni{\bf{References}} \> \> 215 \\

\mbox{ } \> \mbox{ } \\

\ni{\bf{Useful list of acronyms} } \> \> 226 \\

\end{tabbing}

\vspace{6in}

\noindent\Huge\bf{I Introduction}\normalfont\normalsize

\mbox{ }

\noindent\Large{\bf 0 Geometry}\normalsize

\mbox{ }

\noindent \it ``He who attempts natural philosophy without geometry is lost'' \normalfont Galileo Galilei 
\cite{dialogo}

\mbox{ }

\noindent \large{\bf 0.1 Geometry and physics}\normalsize 

\mbox{}

\noindent During the development of classical Newtonian physics, it was assumed that there was 
an absolute time.  This defined a sequence of simultaneities representing nature at each of 
its instants.  Each simultaneity was a copy of the apparent Euclidean 3-geometry of space, 
containing a collection of particles (possibly constituting extended objects).  This setting 
was also assumed for the study of further emergent branches of physics, both those with firm 
foundations in terms of particles (such as fluid mechanics and sound) and those in which the 
role of particles became increasingly less clear (such as gravitation, optics, electricity and 
magnetism).  Scientists began to favour rather the description of the latter in terms of mysterious 
fields pervading space.  Field theory led to a significant advance: Maxwell 
successfully showed electricity and magnetism to be two aspects of a unified physical 
phenomenon, electromagnetism, and as a consequence that light is none other than 
electromagnetic waves.  Contemporary experience strongly suggested that waves arise as 
excitations of some material medium, so space was presumably pervaded with some  
electromagnetic medium for which there was as yet no direct observational evidence.  However, 
explanations in terms of this medium eventually turned out to be inconsistent with experimental 
optics, motivating Einstein to begin his quest to rid physics of all additional unverifiable 
structures and make it exclusively the province of the physical laws themselves.  

This first led him to special relativity (SR) \cite{OEOMB} in which the simultaneity concept 
was destroyed.  Minkowski \cite{Mink} argued that this made it appropriate to not treat space 
and time separately as 3-d and 1-d entities respectively, but rather as a single 4-d entity: 
{\it spacetime}.  Now already the geometry of Minkowski spacetime differs from the usual 
Euclidean geometry (see below).  Furthermore, Einstein was not satisfied with SR (see I.1.5), 
partly because, although his characteristic wish for it to incorporate not just 
electromagnetism but \sl all \normalfont the known laws of physics was almost completely 
successful, the gravitational field remained elusive.  Einstein eventually realized that its 
incorporation required the geometry of spacetime to be not that of Minkowski, but rather a 
geometry curved by the presence of gravitating matter \cite{poe}.  This led him 
\cite{EinGR} toward general relativity (GR).  

That curved spacetime is usually perceived as the arena for physics including gravitation is 
the first reason why the development of modern physics relies heavily on the 19th century 
developments in geometry.  Originally Euclidean geometry was thought to be the only geometry.  
However, Gauss discovered a means of having others.  Simple examples of these have 
decidedly non-Euclidean properties which are all interpretable as manifestations of a property 
called \it curvature \normalfont.  Absence of curvature is \it flatness\normalfont: Euclidean 
geometry is flat.  

\mbox{ }

I consider first a modern approach to 
(without loss of generality 3-d) Euclidean geometry $\mbox{{\sl E}}^3$, which is well-adapted to the passage 
to curved geometry.  Lengths and angles may be characterized in terms of the inner products 
(i.p's) between vectors \b{x} with components $x_{\bar{a}}$, relative to some 
Cartesian basis:
$$
(\mbox{ length of } \mbox{ } \mbox{\b{x}}) = \sqrt{\mbox{\b{x}}\cdot\mbox{\b{x}}} \equiv |x| \mbox{ } , \mbox{ } 
(\mbox{ angle between } \mbox{\b{x}} \mbox{ and } \mbox{\b{y}}) = \mbox{arccos}\left(\frac{\mbox{\b{x}}\cdot\mbox{\b{y}}}{|x||y|}\right).
$$ 
\fn{In this thesis, I adopt the convention of barred indices for flat space(time) and unbarred 
indices for general space(time)s. In simultaneously treating higher- and lower-d space(time)s  
I use capital and lower-case indices respectively.  Repeated indices are to be summed over, 
unless this is explicitly overruled.}  The matrix associated with this i.p is the  
{\it Euclidean metric} which can always be put into the form 
$$
\delta_{\bar{a}\bar{c}} = \left(   \begin{array}{lll} 
                                 1  & 0  & 0 \\
                                 0  & 1  & 0 \\
                                 0  & 0  & 1 \\
                                    \end{array}   \right)_{\bar{a}\bar{c}}
$$
with respect to some basis.  The rigid motions in $\mbox{{\sl E}}^3$ (rotations, translations and 
reflections) are those which preserve lengths and angles, and thus the i.p, and thus the 
Euclidean metric, and are thus termed {\it isometries}.  The rotations form the {\it special 
orthogonal group} SO(3).  Together with the translations, they form the (special) 
{\it Euclidean group} Eucl.\fn{These are all 
continuous transformations. Reflections are discrete.  Whereas continuous transformations can 
be built up from infinitesimal transformations, discrete transformations are {\it large} 
(they have no such constituent pieces); the word `special' here means that such are excluded.}  
The action of the generators of these on a 3-vector are
\be
T_{\mbox{\scriptsize k\normalsize}}: \mbox{\b{x}} \longrightarrow \mbox{\b{x}} + \mbox{\b{k}} \mbox{ } ,
\label{trac}
\ee
\be
R_{\Omega}: \mbox{\b{x}} \longrightarrow \mbox{\b{x}} - \mbox{\b{$\Omega$}} \mbox{ \scriptsize $\times$ \normalsize}\mbox{\b{x}} \mbox{ } .
\label{rotac}
\ee

One can similarly characterize lengths and angles in 4-d Minkowski spacetime in terms of another 
i.p that is associated with the Minkowski metric $\eta_{\bar{A}\bar{C}}$ which can always be 
put into the form
$$
\eta_{\bar{A}\bar{C}} =  \left(   \begin{array}{llll} 
                      -1  &  0  &  0  & 0  \\
              \mbox{ } 0  &  1  &  0  & 0  \\
              \mbox{ } 0  &  0  &  1  & 0  \\
              \mbox{ } 0  &  0  &  0  & 1  \\
                       \end{array}   \right)_{\bar{A}\bar{C}}
$$
with respect to some basis.  Note crucially that whereas $\delta_{\bar{a}\bar{c}}$ 
is positive-definite, $\eta_{\bar{A}\bar{C}}$ is indefinite.  Thus whereas nonzero vectors 
in $\mbox{{\sl E}}^3$ space always have positive norms $|x|^2$, there are three types of 
nonzero vector in Minkowski spacetime:  
those with negative norm termed \it timelike\normalfont, those with zero norm termed 
\it null \normalfont and those with positive norm 
termed \it spacelike\normalfont.  The  existence of these three types is central to the physical 
interpretation of SR (see I.1.4).  Furthermore these notions and their interpretation locally 
extend to GR (see I.1.5).  

The rigid motions of 4-d Minkowski spacetime are those motions which leave invariant the 
Minkowski metric.   They are standard rotations, \it boosts \normalfont (rotations 
involving the time direction), space- and time-translations and space- and time-rotations.  
The rotations and boosts form the {\it special Lorentz group} SO(3, 1).  Together  with 
the space- and time-translations, these form the {\it Poincar\'{e} group}.\fn{Time-reflections are 
usually undesirable.  Space reflections are usually included to form the {\it orthochronous} Lorentz and Poincar\'{e} groups.}  The action of the generators 
on Minkowskian 4-vectors \v{x} = [t, \b{x}] is analogous to (\ref{trac}, \ref{rotac}).

These notions of metric and isometry readily generalize to the study of simple 
curved geometries, such as those that Einstein very successfully chose to model 
the gravitational field.  The (spacetime) metric itself replaces the single Newtonian scalar as the 
gravitational field in GR.  

\mbox{ }

Before discussing curved geometry, I give further reasons for its importance in 
modern physics.  The dynamical study of the laws of physics themselves, regardless of 
whichever arena they are set in, leads to associated abstract curved spaces, which often help 
conceptualize physical issues and solve specific physical problems.  This thesis uses 
such abstract spaces (configuration spaces, phase spaces and relative configuration spaces) for a 
wide range of accepted physical theories.  Both the arena aspect and the dynamical aspect play 
an important role in attempting to provide competing physical theories, often by extending or 
generalizing the geometrical structure of accepted theories.  III.2 and Part B provide two very different 
examples of this.  Geometry may both be used to provide single all-encompassing formulations 
of the accepted laws of nature, and to attempt to supplant 
current formulations by more unified schemes.

\mbox{ }

\noindent \large\bf{0.2 Intrinsic geometry of the 2-sphere}\normalfont\normalsize  

\mbox{ }

\begin{figure}[h]
\centerline{\def\epsfsize#1#2{0.5#1}\epsffile{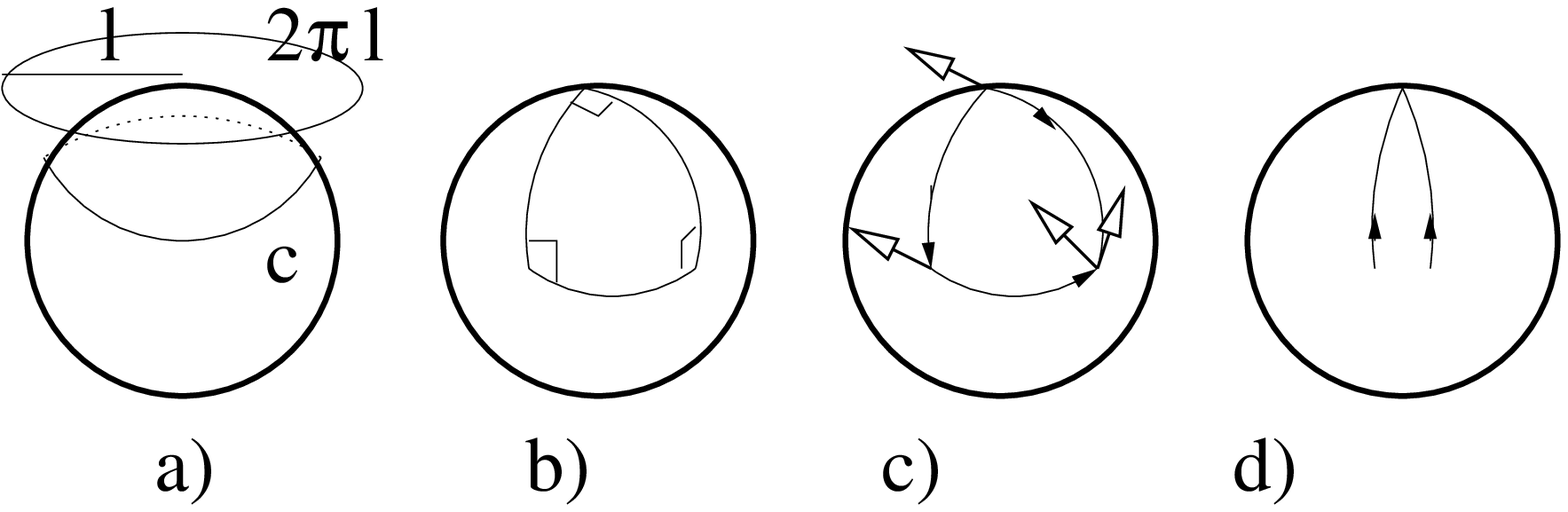}}
\caption[]{\label{fig1.ps}}
\end{figure}

\noindent Consider the surface of a 2-sphere.  
Some of its properties happen to be independent of the conventional visualization's 
surrounding $\mbox{{\sl E}}^3$.  These are the {\it intrinsic} properties of the sphere itself, 
which would be noted by beings whose motions and perceptions are restricted to the surface of 
the sphere only.  Geometers among such beings could in principle notice that although 
their world is locally $\mbox{{\sl E}}^2$, it has curiously non-Euclidean properties on 
sufficiently large scales.  For example (Fig 1) a) if they fixed one end of a taut thread of sufficient 
length $l$ and traced out the locus of the other end, they would find that its circumference 
$c < 2\pi l$.  b) Angles made between three of the straightest possible lines might 
not add up to $\pi$ e.g octants contain 3 right angles. c) Gyroscopes moved between two fixed 
points may have a path-dependent final orientation of the precession axis.  
d) Initially-parallel straightest possible lines may converge.  It turns out that all these 
symptoms can be characterized in terms of a single quantity: the {\sl intrinsic curvature}.  
I next investigate this more formally, as required for general higher-d surfaces. 

\mbox{ }
 
\noindent\large\bf{0.3 Differential geometry}\normalfont\normalsize  

\mbox{ }

\noindent Begin with as little structure as possible.  Consider a \it real topological 
manifold \normalfont {\sl M} i.e a 
particular kind of 
topological space that is everywhere locally $\Re^n$.  
Since one wishes to study tensorial \sl differential equations \normalfont that represent 
physical law, one would normally require the manifold to be sufficiently smooth.   
Tricks used are \cite{Stewart, Wald, Stephani}: 1) to work with manifolds that are locally 
$\Re^n$ and consider a collection of \it charts \normalfont (mappings from an open set in {\sl M} 
$\longrightarrow \Re^n$) that cover the whole 
manifold, moreover for which the notion of nearby points in all the overlaps between charts is 
consistent.  Then one can establish a differential structure on the manifold.   
2) To introduce vectors on the manifold as the tangents to \it curves \normalfont 
(mappings {\sl I} $\longrightarrow$ {\sl M} for {\sl I} an interval of $\Re$).  
3) To then compose these maps to make use of the ordinary $C^k$ calculus 
of $\Re^p$ to $\Re^q$, and moreover show that all the notions involved are chart-independent.  

4) To use the machinery of linear algebra to form the dual vector and all the 
higher-rank tensors on the manifold.  From the perspective here, a \it dual vector \normalfont 
at a point p on the manifold is a linear map 
$T_{\mbox{\scriptsize p\normalsize}}(\mbox{{\sl M}}) \longrightarrow \Re$ where 
$T_{\mbox{\scriptsize p\normalsize}}(\mbox{{\sl M}})$ is the tangent space at p, a 
\it vector \normalfont at p is a linear map 
$T_{\mbox{\scriptsize p\normalsize}}^*(\mbox{{\sl M}}) \longrightarrow \Re$ where 
$T^*_{\mbox{\scriptsize p\normalsize}}(\mbox{{\sl M}})$ is the cotangent space at p (the 
dual of the tangent space), and a \it rank ($k$, $l$) tensor \normalfont at p is a 
multilinear map from the product of $k$ copies of 
$T_{\mbox{\scriptsize p\normalsize}}^*(\mbox{{\sl M}})$ and $l$ copies of 
$T_{\mbox{\scriptsize p\normalsize}}(\mbox{{\sl M}})$ to $\Re$.
A collection of vectors, one at each p $\in \mbox{{\sl M}}$, constitutes a \it vector field 
\normalfont over {\sl M}, and \it tensor fields \normalfont are similarly defined.  The more 
old-fashioned equivalent definition of a ($k$, $l$) tensor is in terms of components: that these 
\be
\mbox{transform according to }
\mbox{\hspace{0.7in}}
{T^{i_1^{\prime}...i_k^{\prime}}}_{j_1^{\prime}...j_l^{\prime}} = 
{L^{i_1^{\prime}}}_{i_1}...{L^{i_k^{\prime}}}_{i_k}{L^{j_1}}_{j_1^{\prime}}...{L^{j_l}}_{j_l^{\prime}}
{T^{i_1...i_k}}_{j_1...j_l}
\label{tentranslaw}
\mbox{\hspace{1.3in}}
\ee
in passing between unprimed and primed coordinate systems, where 
${L^i}_{j^{\prime}} = \frac{\pa x^i}{\pa x^{j^{\prime}}}$ is the general curvilinear 
coordinate transformation.

5) It is not straightforward to introduce derivatives acting on vector fields  
fields because derivatives involve taking the limit of the difference between vectors 
at different points.  So then these vectors belong to different tangent spaces.  Whereas in 
$\Re$ one can just move the vectors to the same point, this is not a straightforward procedure 
on a curved surface [c.f fig 1c)].   The usual partial derivation is undesirable since it does 
not map tensors to tensors.  However, it suffices to construct such a derivation acting on 
vectors and acting trivially on scalars, since 
then one can easily obtain its action on all the other tensors by the Leibniz rule.  
  
\it Lie derivation \normalfont does map tensors to tensors, but is directional because it is 
with respect to some additional vector field $\xi^i$ along which the tensors are dragged.  I denote this 
derivative by $\pounds_{\xi}$.  It is established from dragging first principles \cite{Stewart} 
to act on scalars and vectors as
\be
\pounds_{\xi}S = \xi^i\pa_i S
\mbox{ } ,
\label{slie}
\ee
\be
\pounds_{\xi}V^a = \xi^i\pa_iV^{a} - \pa_i{\xi^a}V^i  
\mbox{ } . 
\ee

To have a non-directional derivative that maps tensors to tensors, one corrects the partial 
derivative 
by introducing an extra structure: the \it affine connection\normalfont, which is a rank 
(1, 2) non-tensorial object with components denoted by ${\Gamma^a}_{bc}$, transforming not like 
(\ref{tentranslaw}) but rather 
\be
\mbox{according to }
\mbox{\hspace{0.7in}}
{\Gamma^{i^{\prime}}}_{j^{\prime}k^{\prime}} = 
{L^{i^{\prime}}_{i}}{L^{j}}_{j^{\prime}}{L^{k}}_{k^{\prime}}
{\Gamma^{i}}_{jk} + {L^{i^{\prime}}}_{i_1}{L^{j}}_{j^{\prime}}\pa_j{L^{i}}_{k^{\prime}} 
\mbox{ } . 
\mbox{\hspace{1.7in}}
\label{Christrans}
\ee
The non-tensoriality of the connection compensates for that 
of the partial derivative so that tensors are mapped to tensors.  
The derivative obtained thus is the \it covariant derivative\normalfont.  
I denote it by $D$ or $\nabla$.  It is just the partial derivative when acting on scalars, but is
\be
D_av^b = \pa_av^b + {\Gamma^a}_{bc}v^c								
\ee
when acting on vectors.  For later use, note that Lie derivatives can be re-expressed in terms 
\be
\mbox{of covariant derivatives} 
\mbox{\hspace{0.6in}}
\pounds_{\xi}V^a = \xi^i D_iV^{a} - D_i{\xi^a}V^i 
\mbox{\hspace{2.2in}}
\ee
when the manifold happens to have affine structure.  

\begin{figure}[h]
\centerline{\def\epsfsize#1#2{0.4#1}\epsffile{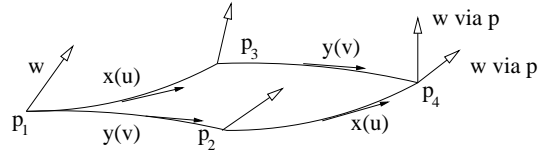}}
\caption[]{\label{TO1.ps}
\footnotesize Two different ways of transporting a vector $w^a$.  
$u$, $v$ parameterize 2 arbitrary curves with tangents $x^a$ and $y^a$ respectively.\normalsize}
\end{figure}
A consequence of the non-tensorial transformation law of the affine connection is that 
coordinates may be found in which the affine connection is zero at any particular point,  
which is crucial in the development of GR.    
The affine connection may be interpreted as giving a notion of straightest possible transport of 
vectors along curves.  Now, one finds that such \it parallel transport \normalfont along two 
paths generally  depends on the order in which the two paths are traversed (Fig 2).  
One finds that a combination of derivatives and squares of the affine connection, the 
\it Riemann curvature tensor\normalfont:
\be
{R^a}_{bcd} \equiv \pa_c {\Gamma^a}_{bd} - \pa_d {\Gamma^a}_{bc} + {\Gamma^e}_{bd}{\Gamma^a}_{ec} - {\Gamma^e}_{bc}{\Gamma^a}_{ed} \mbox{ },
\ee 
can be associated with this property of the transport of a vector $w^a$:
\be
{R^a}_{bcd}w^bx^cy^d = 
\begin{array}{c}
\mbox{lim} \\ 
\mbox{\scriptsize $\Delta$\normalsize} u,\mbox{ } \mbox{\scriptsize $\Delta$\normalsize} v \longrightarrow 0 
\end{array} 
\left( 
\frac{\mbox{\scriptsize $\Delta$\normalsize} w^a}
{\mbox{\scriptsize $\Delta$\normalsize} u\mbox{\scriptsize $\Delta$\normalsize} v}
\right)
\mbox{ }.
\label{transportdef}
\ee 
Instead one can define the Riemann curvature tensor from the `Ricci lemma'
\be
2\nabla_{[a}\nabla_{b]}w^c = {R^c}_{dab}w^d 
\mbox{ } . 
\label{Riccilemma}
\ee

In affine geometry the `locally straightest paths' or \it affine geodesics \normalfont may be 
parameterized 
\be
\mbox{in terms of some $\nu$ so as to have the form } 
\mbox{\hspace{0.4in}}
\frac{D\dot{x}^i}{D\nu} \equiv \dot{x}^aD_a\dot{x}^i = 
\ddot{x}^i + {\Gamma^i}_{jk}\dot{x}^j\dot{x}^k = 0 
\mbox{ } ,
\mbox{\hspace{1.5in}}
\label{afgeoeq}
\ee
where a dot denotes $\frac{\pa}{\pa \nu}$.  If one considers two neighbouring geodesics 
with initially-parallel tangent vectors $\dot{x}^a$, then one can arrive at the Riemann 
curvature tensor by considering the relative acceleration 
$\frac{{D}^2Z^a}{{D}\nu^2}$ of two 
neighbouring geodesics with connecting vector $z^a$ (fig 3) 
\be
\frac{D^2Z^a}{D\nu^2} = - {R^a}_{bcd}\dot{x}^b\dot{x}^cz^d
\label{geodeveq} 
\mbox{ } .
\ee
\begin{figure}[h]
\centerline{\def\epsfsize#1#2{0.4#1}\epsffile{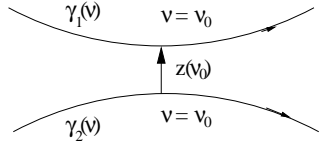}}
\caption[]{\label{TO2.ps}
\noindent \footnotesize For 2 nearby geodesics $\gamma_1$, $\gamma_2$ in a congruence, 
each parameterized by $\nu$, the {\it connecting vector} $z$ is the tangent to the curve 
connecting equal-$\nu$ points.  \normalsize}
\end{figure}
Note that this is a nonlocal effect.  
\be
\mbox{ }\mbox{ It is necessary to exclude \it torsion \normalfont }  
\mbox{\hspace{1in}}
{T^a}_{bc} \equiv {\Gamma^a}_{[bc]} = 0 \mbox{ }  
\mbox{\hspace{1.6in}}
\ee
for formulae (\ref{transportdef}),(\ref{Riccilemma}) and (\ref{geodeveq}) to hold and for it to 
be true that it is necessarily curvature that underlies these symptoms (since the torsion 
corrections to the formulae by themselves produce similar symptoms). 
\be
\mbox{ }
\mbox{The Riemann tensor obeys } 
\mbox{\hspace{1.35in}}
{R^a}_{bcd} = - {R^a}_{bdc} 
\mbox{\hspace{1.6in}}
\ee
\be
\mbox{and the first Bianchi identity } 
\mbox{\hspace{1.3in}}
{R^a}_{[bcd]} = 0 
\mbox{ } ,
\mbox{\hspace{4in}}
\ee
\be
\mbox{whereas its derivatives are related by the \it (second) Bianchi identity\normalfont: } 
\mbox{ } \mbox{ }
\nabla_{[e|}{R^a}_{b|cd]} = 0 
\mbox{ } .
\mbox{\hspace{4in}}
\label{Bianchi2}
\ee
\be
\mbox{The \it Ricci tensor \normalfont is }
\mbox{\hspace{2.0in}}
R_{bd} \equiv {R^a}_{bad} 
\mbox{ } , 
\mbox{\hspace{4in}}
\ee
\be
\mbox{which is symmetric: } 
\mbox{\hspace{1.9in}}
R_{ab} = R_{ba} 
\mbox{ } .
\mbox{\hspace{4in}}
\ee

Another structure is required in order to have a notion of distance (in the spacetime or space 
application this is desirable to model how observers perform measurements of length-duration).  
A \it metric tensor \normalfont $g_{ab}$ is brought in for this purpose. It is usually taken to 
be symmetric, nondegenerate and a function of the coordinates alone.  
Then $(\mbox{{\sl M}}, g_{ab})$ is a 
(semi-)Riemannian manifold.\fn{Genuine \it Riemannian \normalfont metrics are those which are 
positive-definite; those which are indefinite are termed \it semi-Riemannian\normalfont.  
I also refer to these as {\sl Euclidean-} and {\sl Lorentzian-signature} metrics.  I use $g_{AB}$ and 
$h_{ab}$ for metrics when simultaneously treating higher- and lower-d manifolds, 
with corresponding covariant derivatives $\nabla_A$ and $D_a$.  Higher-d objects are 
generally distinguished with checks.  All the formulae of this section should be transcribed 
accordingly prior to their application.  Round brackets surrounding more than one index of any 
type denote symmetrization and square brackets denote antisymmetrization; indices which are not 
part of this (anti)symmetrization are set between vertical lines.}  I denote the 
determinant of the metric by $g$ and its inverse 
$$
\mbox{by $g^{ab}$.  The metric defines 
\it length \normalfont along a path with tangent $\dot{x}^A$ by  } 
\mbox{ }
s = \int_{\nu_1}^{\nu_2}\sqrt{g_{ab}\dot{x}^a\dot{x}^b}\textrm{d}\nu
\mbox{\hspace{4in}}
$$
$$
\mbox{and \it (hyper)volume \normalfont by } 
\mbox{\hspace{1.8in}}
s = \int \sqrt{g}\textrm{d}^nx 
\mbox{ } .  
\mbox{\hspace{2.2in}}
$$

(Semi-)Riemannian geometry has a \it metric connection (Christoffel symbol)
\normalfont 
\be
\left\{
\begin{array}{l}
\mbox{\scriptsize $a$\normalsize} \\ 
\mbox{\scriptsize $b$ \normalsize} \mbox{\scriptsize $c$\normalsize}
\end{array}
\right\}
\equiv \frac{1}{2}g^{ad}(\pa_c g_{bd} + \pa_b g_{cd} - \pa_d g_{bc})
\label{Chris}
\ee
with corresponding metric geodesics (paths locally of extremal length) parameterizable as
\be
\ddot{x}^a + \left\{
\begin{array}{l}
\mbox{\scriptsize $a$\normalsize} \\ 
\mbox{\scriptsize $b$ \normalsize} \mbox{\scriptsize $c$\normalsize}
\end{array}
\right\}
\dot{x}^b\dot{x}^c = 0 \mbox{ } .
\ee
In (semi-)Riemannian geometry all intrinsic properties follow from the metric: one 
assumes 
\be
\mbox{that the affine connection is the metric connection: }
\mbox{\hspace{0.7in}}
\left\{
\begin{array}{l}
\mbox{\scriptsize $a$\normalsize} \\ 
\mbox{\scriptsize $b$ \normalsize} \mbox{\scriptsize $c$\normalsize}
\end{array}
\right\}
= {\Gamma^a}_{bc} 
\mbox{ } ,
\mbox{\hspace{0.8in}}
\label{metcon}
\ee
so then geodesics and the Riemann tensor are expressible in terms of the metric.  Einstein assumed 
this simple geometry for GR, which has so far been consistent with observation.  If however 
one does not assume the above equality, then the difference of the two connections would constitute an 
additional tensor generally composed of torsion and nonmetricity parts.  Apart from allowing these, 
other ways of having more complicated geometry are for the metric to be non-symmetric, or 
degenerate, or to depend on the velocities as well as the coordinates (as in \it Finslerian geometry\normalfont).  
These will not be considered as options for the geometrization of space(time), but the last two 
occur in my dynamical study. 

The metric and its inverse may be used to lower and raise indices on other tensors.  
Thus one  can define the Riemann tensor with all its indices downstairs by
\be
R_{ebcd} \equiv g_{ea}{R^a}_{bcd} 
\ee
\be
\mbox{which has the additional symmetry property }
\mbox{\hspace{0.9in}}
R_{abcd} =  R_{cdab} 
\mbox{ } .
\mbox{\hspace{4in}}
\ee
\be
\mbox{One can also now obtain the \it Ricci scalar \normalfont }
\mbox{\hspace{1.1in}}
R \equiv g^{bd}R_{bd} 
\mbox{ } .
\mbox{\hspace{4in}}
\ee
The curvature tensors contain the following numbers of independent pieces of information:
\be
\#R_{abcd} = \frac{n^2(n^2 - 1)}{12} \mbox{ } , \mbox{ } \mbox{ } \#R_{ab} = \frac{n(n + 1)}{2} \mbox{ } , \mbox{ } \mbox{ } \#R = 1 
\mbox{ } .
\ee
This establishes which tensor suffices to describe intrinsic curvature 
in each dimension: $R$ suffices in 2-d, $R_{ab}$ in 3-d and all of ${R^a}_{bcd}$ is required 
in all higher-d.  It also establishes whether the numbers of equations and unknowns make sense in different 
possible physical theories based on such geometrical objects.  
The (irreducible) part of ${R^a}_{bcd}$ not in $R_{ab}$ is 
\be
\mbox{the \it Weyl tensor \normalfont }
\mbox{\hspace{0.2in}}
{W^a}_{bcd} \equiv {R^a}_{bcd} - \frac{2}{n - 2}({\delta^a}_{[c}R_{d]b} - g_{b[c}{R_{d]}}^a) 
- \frac{2}{(n - 1)(n - 2)}{\delta^a}_{[d}g_{c]b}R 
\mbox{ } , 
\label{Weyldef}  
\ee
which inherits all the symmetry properties of the Riemann tensor.

The \it Einstein tensor \normalfont (used in the construction of the Einstein field equations), 
\be
G_{ab} = R_{ab} - \frac{R}{2}g_{ab}
\ee
has the important properties of being divergenceless by the contracted Bianchi identity  
\be
\nabla^aG_{ab} = 0 
\mbox{ } , 
\label{contbi}
\ee
and a symmetric (0, 2) tensor like the metric. 

\mbox{ }

\noindent \bf{App 0.A Densities and Integration}\normalfont

\mbox{ }

\noindent
A $(n, m)$ \it tensor density \normalfont of weight $w$ is an object which transforms like
\be
{T^{\hat{a}_1...\hat{a}_n}}_{\hat{b}_1...\hat{b}_m} = 
(\sqrt{det L})^w{L^{\hat{a}_1}}_{a_1}...{L^{\hat{a}_n}}_{a_n}{L^{b_1}}_{\hat{b}_1}...{L^{b_1}}_{\hat{b}_m}{T^{a_1...a_n}}_{b_1...b_m}
\ee
I denote $w = 1$ and $w = -1$ objects by overlines and underlines respectively.  
In particular, the square root of the metric determinant is a scalar density of weight 1.  
I need the following 
\be
\mbox{ingredients to build up derivatives of densities: }
\mbox{\hspace{0.45in}}
\pounds_{\xi}\sqrt{h} = \sqrt{h}D^a{\xi_a} 
\mbox{\hspace{2in}}
\ee
\be
\mbox{\hspace{2in}}
D_a\sqrt{h} = 0 
\mbox{ } .  
\label{0covderiv}
\ee

For integration to make sense on a manifold, integrands must be weight-1 scalars 
(since it does not make sense to add higher-rank tensors at distinct points).
Furthermore, (\ref{0covderiv}) 
$$
\mbox{allows one to have a divergence theorem }
\mbox{\hspace{0.9in}}
\int_{\partial V}\overline{A}^a\textrm{d}\Sigma_a = \int_VD_a\overline{A}^a\textrm{d}\Omega
\mbox{\hspace{2.8in}}
$$
for $\overline{A}^a$ a (1, 0) vector density of weight 1 and where $\textrm{d}\Sigma_a$  and $\textrm{d}\Omega$ 
are the obvious (hyper)surface and (hyper)volume elements.

\section{Classical physics}

\subsection{Newtonian physics}

Newton \cite{Principia13} gave us three fundamental laws of motion.

\noindent{\bf N1} Every body continues in its state of rest, or of uniform motion in a right line unless 
it is compelled to change that state by forces impressed upon it.   

\noindent{\bf N2} The change of motion is proportional to the motive force impressed; and is made in the 

\noindent direction of the right line in which that force is impressed.  

\noindent{\bf N3} To every action there is always opposed an equal reaction.  

In the modern notation adopted, the position of a Newtonian particle is 
\b{x}, and its velocity is $\dot{\mbox{\b{x}}} \equiv \frac{  \textrm{d}\mbox{\b{x}}  }{  \textrm{d}t  }$ where $t$ is 
Newton's notion of time (see below).  Then {\bf N2} reads 
\be
\mbox{Impressed force } \mbox{\b{F}} = \frac{  \textrm{d}\mbox{\b{p}}  }{  \textrm{d}t } \mbox{ } \mbox{ , for } \mbox{\b{p}} \mbox{ the momentum . }   
\label{NII}
\ee
Newton's momentum is \b{p} = $m\dot{\mbox{\b{x}}}$; in most applications the mass $m$ of the 
particle is taken to be constant.  From (\ref{NII}) it follows that in the absence of 
the action of external forces the momentum \b{p} is conserved.  

Newton explained \cite{Scholium} that his laws are to be interpreted as occurring in his  
all-pervading, similar and immovable \it absolute space \normalfont while his external 
\it absolute time \normalfont flows.  

There are a number of frames of reference in which {\bf N1} holds, called 
\it inertial frames\normalfont, which are at rest in absolute space or moving uniformly through it along 
a right line.  These inertial frames are inter-related by the \it transformations of 
Galilean relativity \normalfont :    
\be
\mbox{\b{x}} \longrightarrow \mbox{\b{x}}^{\prime} = \mbox{\b{x}} + \mbox{\b{v}}t + \mbox{\b{k}} \mbox{ } ,
\label{Gal1}
\ee
\be
t \longrightarrow t + l \mbox{ } ,
\label{Gal2}
\ee
for constant \b{v}, \b{k} and $l$.  Non-inertial observers perceive additional 
\it {fictitious forces}\normalfont.  Note that (\ref{Gal2}) is a statement of existence 
of absolute time up to choice of time origin.  Newton's laws are not 
necessarily Galileo-invariant.  This would further require \cite{DOD} 

\noindent `{\bf N4}' The masses, and the strengths of the forces are independent of the motion 
of the centre of mass of the system of particles relative to absolute space.  

The immovable external character of absolute space and time is abhorred by 
{\it relationalists}, who maintain that the laws of physics should rather be based on relative notions 
alone (see I.1.5 and II--VIII).  

Newton also provided an explicit form for a particular universal force: 

\noindent {\bf Newton's law of gravitation}: The gravitational force on a particle of mass 
$m_{(1)}$ at position $\mbox{\b{x}}_{(1)}$ 
due to a particle of mass $m_{(2)}$ at position $\mbox{\b{x}}_{(2)}$ is given by\fn{Units: 
in I.1.1--5 I keep $G$ (Newton's constant), $c$ (the speed of light) and $\hbar$ 
(Planck's constant) explicit.  I then set $c = \hbar = 1$, $8\pi G = 1$. 

I use $(i)$-indices to run over particle labels, and 
$\mbox{\b{r}}_{(i)(j)} \equiv \mbox{\b{x}}_{(i)} - \mbox{\b{x}}_{(j)}$.} 
\be
\mbox{\b{F}}_{(1)(2)}^{\mbox{\scriptsize g\normalfont}} = 
- \frac{Gm_{(1)}m_{(2)}}{|\mbox{\b{r}}_{(1)(2)}|^3}\mbox{\b{r}}_{(1)(2)} \mbox{ } .
\label{NLOG} 
\ee
$\mbox{\b{F}}_{(1)(2)}^{\mbox{\scriptsize g\normalfont}}$ may be written as  
$\mbox{\b{F}}_{(1)(2)}^{\mbox{\scriptsize g\normalfont}} 
= - m_{(1)}\mbox{\b{$\pa $}}\phi_{(1)(2)}$  
for $\phi_{(1)(2)} = \frac{m_{(2)}}{|\mbox{\b{r}}_{(1)(2)}|}$ the gravitational potential at $\mbox{\b{x}}_{(1)}$ 
due to the particle of mass $m_{(2)}$ at $\mbox{\b{x}}_{(2)}$.  Newtonian gravity is linear, so one can build up by superposition 
the total gravitational force at \b{x} due to all the particles in the universe, 
$\mbox{\b{F}}^{\mbox{\scriptsize g\normalfont}}(\mbox{\b{x}}) = 
- [m\mbox{\b{$\pa $}}\phi](\mbox{\b{x}})$.   Combining this with (\ref{NII}), one obtains the 
equation of motion for a (constant-mass) particle (in the absence of any other type of forces):
\be
\ddot{\mbox{\b{x}}} = - [\mbox{\b{$\pa $}}\phi](\mbox{\b{x}}) \mbox{ } .
\label{NII + NLOG}
\ee
Consider this for two neighbouring particles at positions \b{x} and \b{x} + 
$\mbox{\scriptsize $\Delta$\normalsize}$\b{x}.  Then subtracting and by the definition of 
derivative one obtains the \it tidal equation \normalfont
\be
\mbox{\scriptsize $\Delta$\normalsize}\ddot{\mbox{\b{x}}} = 
- \mbox{\b{$\pa $}}(  \mbox{\b{$\pa $}} \cdot \mbox{\scriptsize $\Delta$\normalsize} \mbox{\b{x}}  )
\label{tidal}
\ee
for the relative acceleration of the two particles.  

Finally, one requires a field equation describing how sources produce gravitation.  
In 
\be
\mbox{differential form, for $\rho_{\mbox{\scriptsize m\normalfont}}$ the mass density, 
it is Poisson's law }
\mbox{\hspace{0.4in}} 
D^2\phi = 4\pi G \rho_{\mbox{\scriptsize m\normalfont}} 
\mbox{ } . 
\mbox{\hspace{0.4in}}
\label{GPoisson}
\ee  

Newtonian gravity is very successful at describing the solar system: Kepler's laws and 
perturbations due to planet-planet interactions follow from it.  
However, a small anomalous residual perihelion precession of Mercury was eventually observed.

\subsection{Principles of dynamics}

{\noindent\bf 1.2.1 Configuration space and the Euler--Lagrange equations}

\mbox{ }

\noindent Theorists can choose to describe the position of a particle in (absolute) Euclidean 3-space 
$\mbox{{\sl E}}^3$ by 3 curvilinear coordinates.  Convenience and not any physical reality underlies 
which particular choice.  
Mechanical systems are usually taken to be second order, so that the initial position of the 
particle does not suffice to determine the motion.  One requires also e.g the initial 
velocity or the initial momentum.  For $n$ particles in $\mbox{\sl{E}}^3$, na\"{\i}vely one 
requires the prescription of $3n$ coordinates to describe their positions.  But, the particles 
may not be free to move in all possible ways, e.g some of them could be attached via strings, 
springs or rods.  Such constitute {\it constrained} mechanical systems, describable in terms of less 
than $3n$ independent coordinates, 
$q_{\mbox{\sffamily\scriptsize A\normalsize\normalfont}}$.\fn{My capital sans-serif indices 
are general indices, and my $<i>$-indices run over indexing sets.  I use capital sans-serif 
letters for objects used in the principles of dynamics.}  

The space of coordinates that describe a system is its \it configuration space \normalfont  
\sffamily Q\normalfont.  The motion of the whole system is represented by a curve in this 
configuration space.  
Sometimes one may attempt to work with a \sffamily Q \normalfont that is larger than necessary 
e.g a $3n$-d \sffamily Q \normalfont for a $n$-particle system with $p$ constraints.  
However, some aspects of the system would be better understood if one were able to take into 
account the constraints and pass to a ($3n$ -- $p$)-d 
$\mbox{\sffamily Q\normalfont}_{\mbox{\scriptsize reduced\normalsize}}$.

Whereas one may attempt to study such particle systems directly using Newton's laws, this may be 
cumbersome and requires knowledge of all the forces acting at each point in the system.  A 
method which is often more practical, and which 
extends to field theory, is that based on energy considerations, as 
formalized by Lagrange, Euler and others \cite{Lanczos}.  One computes the kinetic energy 
\sffamily T \normalfont of the system  and the potential energy \sffamily V\normalfont.  
Then one forms the \it Lagrangian \normalfont 
\sffamily L \normalfont = \sffamily T \normalfont -- \sffamily V\normalfont, knowledge of which 
permits one to write down a set of equations of motion equivalent to Newton's.  
These equations arise by declaring the standard prescription of the calculus of variations:  
the equations of motion are such that the action  
\sffamily I \normalfont$ = \int \textrm{d}$t\sffamily L \normalfont is stationary with respect to  
$q_{\mbox{\scriptsize \sffamily A\normalfont\normalsize}}$ when one considers the true motion 
between two particular fixed endpoints $e_1$ and $e_2$ together with the set of varied 
paths (subject to the same fixed endpoints) about this motion.  Then for a second-order 
mechanical system, 
\be
0= \delta \mbox{\sffamily I\normalfont} = \int_{e_1}^{e_2} \textrm{d}t
\left(
\frac{\pa \mbox{\sffamily L\normalfont}}{\pa q_{\mbox{\sffamily\scriptsize A\normalsize\normalfont}}}\delta q_{\mbox{\sffamily\scriptsize A\normalsize\normalfont}}  +  
                                                           \frac{\pa \mbox{\sffamily L\normalfont}}{\pa \dot{q}_{\mbox{\sffamily\scriptsize A\normalsize\normalfont}}}\delta \dot{q}_{\mbox{\sffamily\scriptsize A\normalsize\normalfont}}
\right)  
= \int_{e_1}^{e_2} \textrm{d}t 
\left[
\frac{\pa \mbox{\sffamily L\normalfont}}{\pa q_{\mbox{\sffamily\scriptsize A\normalsize\normalfont}}} - \frac{\pa}{\pa t}
\left(
\frac{\pa \mbox{\sffamily L\normalfont}}{\pa \dot{q}_{\mbox{\sffamily\scriptsize A\normalsize\normalfont}}}
\right)
\right]
\delta q_{\mbox{\sffamily\scriptsize A\normalsize\normalfont}} + 
\left[
\frac{    \pa\mbox{\sffamily L\normalfont}    }{    \pa \dot{q}_{\mbox{\sffamily\scriptsize A\normalsize\normalfont}}    }
\delta q_{\mbox{\sffamily\scriptsize A\normalsize\normalfont}}
\right]_{e_1}^{e_2}
\label{ELDERIV}
\ee
by parts.  Now note that the last term vanishes because the endpoints have been taken to be 
fixed.  As $\delta q_{\mbox{\sffamily\scriptsize A\normalsize\normalfont}}$ is an arbitrary 
variation, the laws of motion in their differential form 
\be
\mbox{follow: the {\it Euler--Lagrange equations} (ELE's) }
\mbox{\hspace{0.7in}} 
\frac{\pa}{\pa t}\left( \frac{\pa \mbox{\sffamily L\normalfont}}{\pa \dot{q}_{\mbox{\sffamily\scriptsize A\normalsize\normalfont}}}\right) - \frac{\pa \mbox{\sffamily L\normalfont}}{\pa q_{\mbox{\sffamily\scriptsize A\normalsize\normalfont}}} = 0 
\mbox{ }. 
\mbox{\hspace{0.7in}}
\label{ELEQ}
\ee

These equations are easier to integrate for two special cases of coordinate.  Thus one main 
theme in the principles of dynamics is to judiciously choose a coordinate system with as many 
special case coordinates as possible.  This extends the well-known exploitation of 
conservation of momentum and angular momentum.  The special cases are 

\noindent 1) \it Lagrange multipliers\normalfont: if \sffamily L \normalfont is independent of $\dot{q}_n$, the corresponding 
ELE is  
\be
\frac{\pa \mbox{\sffamily L\normalfont}}{\pa q_n} = 0 \mbox{ } .
\label{lmel}
\ee

\noindent 2) \it Cyclic coordinates\normalfont: if \sffamily L \normalfont is independent of $q_n$, the corresponding ELE is  
\be
\frac{\pa \mbox{\sffamily L\normalfont}}{\pa \dot{q}_n} = \mbox{ constant } \mbox{ }.
\label{cyclicel}
\ee

\noindent{\bf 1.2.2 Legendre transformations, Jacobi's principle and Hamiltonian dynamics}

\mbox{ }

\noindent Suppose one has a function 
$\mbox{\sffamily F\normalfont}(x_{\mbox{\sffamily\scriptsize A\normalsize\normalfont}}: 
{\mbox{\sffamily A\normalfont}} = 1 \mbox{ to } m; 
y_{\mbox{\sffamily\scriptsize W\normalsize\normalfont}})$  and one wishes to use  
$z_{\mbox{\sffamily\scriptsize A\normalsize\normalfont}} 
= \frac{\pa \mbox{\sffamily\scriptsize F\normalsize\normalfont}}
{\pa x_{\mbox{\sffamily\tiny A\normalsize\normalfont}}}$
as variables in place of the 
$x_{\mbox{\sffamily\scriptsize A\normalsize\normalfont}}$.  To obtain equivalent 
equations of motion, one has to pass to a function  
$\mbox{\sffamily G\normalfont} = x_{\mbox{\sffamily\scriptsize A\normalsize\normalfont}}
z_{\mbox{\sffamily\scriptsize A\normalsize\normalfont}} - 
\mbox{\sffamily F\normalfont}(x_{\mbox{\sffamily\scriptsize A\normalsize\normalfont}}, 
y_{\mbox{\sffamily\scriptsize W\normalsize\normalfont}})$
which may always be written as 
$\mbox{\sffamily G\normalfont}(z_{\mbox{\sffamily\scriptsize A\normalsize\normalfont}}, 
y_{\mbox{\sffamily\scriptsize W\normalsize\normalfont}})$.  This is the \it Legendre 
transformation\normalfont.  It is symmetric between  
$x_{\mbox{\sffamily\scriptsize A\normalsize\normalfont}}$ and  
$z_{\mbox{\sffamily\scriptsize A\normalsize\normalfont}}$: if one defines     
$x_{\mbox{\sffamily\scriptsize A\normalsize\normalfont}} = 
\frac{\pa \mbox{\sffamily\scriptsize G\normalsize\normalfont}}
{\pa z_{\mbox{\sffamily\tiny A\normalsize\normalfont}}}$, the passage is to 
$\mbox{\sffamily F\normalfont}(x_{\mbox{\sffamily\scriptsize A\normalsize\normalfont}}, 
y_{\mbox{\sffamily\scriptsize W\normalsize\normalfont}}) = 
x_{\mbox{\sffamily\scriptsize A\normalsize\normalfont}}
z_{\mbox{\sffamily\scriptsize A\normalsize\normalfont}} - 
\mbox{\sffamily G\normalfont}(z_{\mbox{\sffamily\scriptsize A\normalsize\normalfont}}, 
y_{\mbox{\sffamily\scriptsize W\normalsize\normalfont}})$.  
In particular, if the function one has is a Lagrangian $\mbox{\sffamily L\normalfont}(q_{\mbox{\sffamily\scriptsize A\normalsize\normalfont}}, \dot{q}_{\mbox{\sffamily\scriptsize A\normalsize\normalfont}})$, one may wish to use some 
of the  \it (generalized) conjugate momenta \normalfont 
\be
p^{\mbox{\sffamily\scriptsize A\normalsize\normalfont}} 
\equiv \frac{\pa \mbox{\sffamily L\normalfont}}{\pa\dot{q}_{\mbox{\sffamily\scriptsize A\normalsize\normalfont}}}
\label{MOMENTA}
\ee
as variables in place of the corresponding velocities.  

A first example of Legendre transformation occurs in {\it Routhian reduction}: given a 
Lagrangian with cyclic coordinate  
$Q, \mbox{ } \mbox{\sffamily L\normalfont}(
q_{\mbox{\sffamily\scriptsize B\normalsize\normalfont}}; 
\dot{q}_{\mbox{\sffamily\scriptsize B\normalsize\normalfont}}, \dot{Q})$, then  
$P \equiv \frac{\pa \mbox{\sffamily\scriptsize L\normalsize\normalfont}}{\pa\dot{Q}} = \mbox{ constant }$,  
so that one may pass from \sffamily L \normalfont to the {\it Routhian }
${\mbox{\sffamily R\normalfont}}(
q_{\mbox{\sffamily\scriptsize B\normalsize\normalfont}}; 
\dot{q}_{\mbox{\sffamily\scriptsize B\normalsize\normalfont}}) = 
\mbox{\sffamily L\normalfont}(q_{\mbox{\sffamily\scriptsize B\normalsize\normalfont}}; 
\dot{q}_{\mbox{\sffamily\scriptsize B\normalsize\normalfont}}, \dot{Q}) - P\dot{Q}$.  

A second example, is the formulation of Jacobi's principle \cite{Lanczos}.  
If one has a $\mbox{\sffamily L\normalfont}(q_{\mbox{\sffamily\scriptsize A\normalsize\normalfont}}, \dot{q}_{\mbox{\sffamily\scriptsize A\normalsize\normalfont}})$ which 
does not depend explicitly on time, $t$.  Then one can express $\mbox{\sffamily I\normalfont} = \int \textrm{d}t\mbox{\sffamily L\normalfont}$ as 
\be
\mbox{\sffamily I\normalfont} = \int\textrm{d}\tau[t^{\prime}\mbox{\sffamily L\normalfont}(q_{\mbox{\sffamily\scriptsize A\normalsize\normalfont}}; q_{\mbox{\sffamily\scriptsize A\normalsize\normalfont}}^{\prime}, t^{\prime})]
\label{Ltadjac}
\ee
i.e $\tau$-parameterize the action by adjoining time to the positions, where $^{\prime} \equiv \frac{\pa}{\pa\tau}$.  Now as a consequence of the explicit $t$-independence of the original Lagrangian, $t$ is a cyclic coordinate in
(\ref{Ltadjac}) so (as a special case of Routhian reduction) one may pass to 
\be
\mbox{$\mbox{\sffamily L\normalfont}_{\mbox{\scriptsize J\normalsize}}(q_{\mbox{\sffamily\scriptsize A\normalsize\normalfont}}; q_{\mbox{\sffamily\scriptsize A\normalsize\normalfont}}^{\prime}) = \mbox{\sffamily L\normalfont}(q_{\mbox{\sffamily\scriptsize A\normalsize\normalfont}}; q_{\mbox{\sffamily\scriptsize A\normalsize\normalfont}}^{\prime})t^{\prime} - p^tt^{\prime}$ 
\hspace{0.5in} for }
\mbox{\hspace{0.5in}}
\frac{\pa \mbox{\sffamily L\normalfont}}{\pa t^{\prime}} = p^t = - \mbox{\sffamily E\normalfont} \mbox{ } , \mbox{ } \mbox{ constant }.  
\mbox{\hspace{4in}}
\label{tprimeeq}
\ee
Minimization of $\mbox{\sffamily I\normalfont}_{\mbox{\scriptsize J\normalsize}} = 
\int \textrm{d}\tau \mbox{\sffamily L\normalfont}_{\mbox{\scriptsize J\normalsize}}$ is 
\it Jacobi's principle\normalfont.  It is a {\it geodesic principle} 
since its use reduces the problem of motion to the problem of finding the geodesics associated 
with some geometry.  I particularly use the special case of a mechanical system which is 
homogeneous 
\be
\mbox{quadratic in its velocities: } 
\mbox{\hspace{1.1in}}
\mbox{\sffamily L\normalfont} = \mbox{\sffamily T\normalfont} - \mbox{\sffamily V\normalfont} = \frac{1}{2}M^{\mbox{\sffamily\scriptsize AB\normalsize\normalfont}}\dot{q}_{\mbox{\sffamily\scriptsize A\normalsize\normalfont}}\dot{q}_{\mbox{\sffamily\scriptsize B\normalsize\normalfont}} - \mbox{\sffamily V\normalfont}(q_{\mbox{\sffamily\scriptsize A\normalsize\normalfont}}) 
\mbox{ } ,
\mbox{\hspace{1.5in}}
\label{QUADRATIC}
\ee
$$
\mbox{where $M^{\mbox{\sffamily\scriptsize AB\normalsize\normalfont}}$ is the `kinetic matrix'.  Then, }
\mbox{ } 
\mbox{\sffamily I\normalfont} = \int \textrm{d}\tau t^{\prime}
\left(
\frac{1}{2t^{\prime 2}}M^{\mbox{\sffamily\scriptsize AB\normalsize\normalfont}}(q_{\mbox{\sffamily\scriptsize A\normalsize\normalfont}})
{q}^{\prime}_{\mbox{\sffamily\scriptsize A\normalsize\normalfont}}
{q}^{\prime}_{\mbox{\sffamily\scriptsize B\normalsize\normalfont}} - \mbox{\sffamily V\normalfont}(q_{\mbox{\sffamily\scriptsize A\normalsize\normalfont}})
\right) 
\mbox{ } .
\mbox{\hspace{2in}}
$$
$$
\mbox{Using (\ref{tprimeeq}) as an equation to eliminate $t^{\prime}$, }
\mbox{\sffamily E\normalfont}  
=  \frac{1}{2t^{\prime 2}}
M^{\mbox{\sffamily\scriptsize AB\normalsize\normalfont}}
{q}^{\prime}_{\mbox{\sffamily\scriptsize A\normalsize\normalfont}}
{q}^{\prime}_{\mbox{\sffamily\scriptsize B\normalsize\normalfont}} 
+ \mbox{\sffamily V\normalfont}(q_{\mbox{\sffamily\scriptsize A\normalsize\normalfont}}) 
\mbox{ } \Rightarrow \mbox{ } t^{\prime} = \sqrt{\frac{\mbox{\sffamily T\normalfont}}{\mbox{\sffamily E\normalfont}  
-  \mbox{\sffamily V\normalfont}}} 
\mbox{ } .
\mbox{\hspace{0.5in}}
$$
\be
\mbox{Thus }
\mbox{\hspace{1.6in}}
\mbox{\sffamily I\normalfont}_{\mbox{\scriptsize J\normalsize}} 
= \int \textrm{d} \tau 2\sqrt{\mbox{\sffamily E\normalfont} - \mbox{\sffamily V\normalfont}    }
\sqrt{\mbox{\sffamily T\normalfont}} 
= \int \sqrt{2(\mbox{\sffamily E\normalfont}  - \mbox{\sffamily V\normalfont})}\textrm{d}\sigma
\label{Quadjacac}
\mbox{\hspace{2in}} 
\ee
for $\textrm{d}\sigma^2 = M_{\mbox{\sffamily\scriptsize AB\normalsize\normalfont}}\textrm{d}q^{\mbox{\sffamily\scriptsize A\normalsize\normalfont}}
\textrm{d}q^{\mbox{\sffamily\scriptsize B\normalsize\normalfont}}$ the line element of 
the configuration space geometry.  Thus in this case use of Jacobi's principle translates 
the problem of finding the motion of the mechanical system into the well-defined, 
well-studied problem of finding the geodesics of a {\sl Riemannian} geometry with line 
element ${\textrm{d}}\tilde{\sigma}^2$ related to $\textrm{d}\sigma^2$ by the conformal 
transformation 

\noindent${\textrm{d}}\tilde{\sigma}^2 = 2(\mbox{\sffamily E\normalfont} - \mbox{\sffamily V\normalfont})\textrm{d}\sigma^2$.  

Further examples of Legendre transformation include the passage to first-order form in VI, in 
addition to well-known moves in thermodynamics and QM.  There is also the passage 
$$
\mbox{from the Lagrangian to the {\it Hamiltonian}}
\mbox{\hspace{0.6in}}
\mbox{\sffamily H\normalfont}(q_{\mbox{\sffamily\scriptsize A\normalsize\normalfont}}; p^{\mbox{\sffamily\scriptsize A\normalsize\normalfont}}; t) = p^{\mbox{\sffamily\scriptsize A\normalsize\normalfont}}\dot{q}_{\mbox{\sffamily\scriptsize A\normalsize\normalfont}} - \mbox{\sffamily L\normalfont}(q_{\mbox{\sffamily\scriptsize A\normalsize\normalfont}}; \dot{q}_{\mbox{\sffamily\scriptsize A\normalsize\normalfont}}; t)
\mbox{\hspace{2in}} \mbox{ } ,
$$
which makes use of \sl all \normalfont the conjugate momenta.    
The equations of motion are now 
\be
\mbox{\it Hamilton's equations \normalfont }
\mbox{\hspace{1.3in}}
\dot{q}_{\mbox{\sffamily\scriptsize A\normalsize\normalfont}} = \frac{\pa \mbox{\sffamily H\normalfont}}{\pa p^{\mbox{\sffamily\scriptsize A\normalsize\normalfont}}} \mbox{ } , \mbox{ } \mbox{ } \dot{p}^{\mbox{\sffamily\scriptsize A\normalsize\normalfont}} = - \frac{\pa \mbox{\sffamily H\normalfont}}{\pa q_{\mbox{\sffamily\scriptsize A\normalsize\normalfont}}} 
\mbox{ } .
\mbox{\hspace{2in}}
\ee
Instead of working in configuration space \sffamily Q\normalfont, one now works in the (na\"{\i}vely $6n$-d) phase space 
({\sffamily Q\normalfont}, {\sffamily P\normalfont}) where {\sffamily P\normalfont} is the set of the momenta $\mbox{\b{p}}^{(i)}$.  
There are canonical transformations on phase space which generally mix up what are the positions and what are the momenta: 
\be
q_{\mbox{\sffamily\scriptsize A\normalsize\normalfont}}, p^{\mbox{\sffamily\scriptsize A\normalsize\normalfont}} \longrightarrow Q_{\mbox{\sffamily\scriptsize A\normalsize\normalfont}}, P^{\mbox{\sffamily\scriptsize A\normalsize\normalfont}} \mbox{ } . 
\ee
Whereas configuration space has a (commonly Riemannian) geometry on it with a symmetric metric $M_{\mbox{\sffamily\scriptsize AB\normalsize\normalfont}}$ as its preserved object, phase space has a preserved object 
$p^{\mbox{\sffamily\scriptsize A\normalsize\normalfont}}\textrm{d}q_{\mbox{\sffamily\scriptsize A\normalsize\normalfont}}$ 
up to a complete differential.  This gives rise to an antisymmetric \it symplectic 
structure\normalfont.  This is most usefully phrased 
via the introduction of \it Poisson brackets \normalfont
\be
\{f(q_{\mbox{\sffamily\scriptsize A\normalsize\normalfont}}, p^{\mbox{\sffamily\scriptsize A\normalsize\normalfont}})
, g(q_{\mbox{\sffamily\scriptsize A\normalsize\normalfont}}, p^{\mbox{\sffamily\scriptsize A\normalsize\normalfont}}) \} 
\equiv \frac{\pa f}{\pa q_{\mbox{\sffamily\scriptsize A\normalsize\normalfont}}}\frac{\pa g}{\pa p^{\mbox{\sffamily\scriptsize A\normalsize\normalfont}}}     -   \frac{\pa g}{\pa q_{\mbox{\sffamily\scriptsize A\normalsize\normalfont}}}\frac{\pa f}{\pa p^{\mbox{\sffamily\scriptsize A\normalsize\normalfont}}} 
\mbox{ } ,
\ee
$$
\mbox{whereupon the equations of motion take the form }
\mbox{\hspace{0.7in}}
\{q_{\mbox{\sffamily\scriptsize A\normalsize\normalfont}}, \mbox{\sffamily H\normalfont}\} = \dot{q}_{\mbox{\sffamily\scriptsize A\normalsize\normalfont}} \mbox{ } , \mbox{ } \mbox{ } \{p^{\mbox{\sffamily\scriptsize A\normalsize\normalfont}} , \mbox{\sffamily H\normalfont}     \} = \dot{p}^{\mbox{\sffamily\scriptsize A\normalsize\normalfont}} 
\mbox{ } . 
\mbox{\hspace{0.7in}}
$$

\noindent{\bf 1.2.3 Dirac's generalized Hamiltonian dynamics}

\mbox{ }

\noindent 
This subsection provides a systematic means of studying constrained systems.  The 
field-theoretical version of this is of particular importance in this thesis.  
I use \#\sffamily A \normalfont to denote the number of degrees of freedom (d.o.f's) in the 
set indexed by \sffamily A\normalfont. If not all the conjugate momenta 
$p^{\mbox{\scriptsize\sffamily A\normalfont\normalsize}}  = \frac{\partial \mbox{\sffamily\scriptsize L\normalsize\normalfont}}  
{\partial\dot{q}_{\mbox{\scriptsize\sffamily A\normalfont\normalsize}}  }$  
can be inverted to give the $\dot{q}_{\mbox{\scriptsize\sffamily A\normalfont\normalsize}}$ 
in terms of the $p^{\mbox{\scriptsize\sffamily A\normalfont\normalsize}}$, then the theory has 
\it primary constraints \normalfont ${\cal C }_{\mbox{\scriptsize\sffamily P\normalfont\normalsize}}(q_{\mbox{\scriptsize\sffamily A\normalfont\normalsize}},
p^{\mbox{\scriptsize\sffamily A\normalfont\normalsize}}  ) = 0$ solely by virtue of the form of \mbox{\sffamily L\normalfont}.  As Dirac noted \cite{Dirac}, in such a 
case a theory described by a Hamiltonian $\mbox{\sffamily H\normalfont}(q_{\mbox{\scriptsize\sffamily A\normalfont\normalsize}} , 
p^{\mbox{\scriptsize\sffamily A\normalfont\normalsize}}  )$ could just as well be
\be 
\mbox{described by a Hamiltonian } 
\mbox{\hspace{1.5in}}
\mbox{\sffamily H\normalfont}^{\mbox{\scriptsize Total\normalsize}} = \mbox{\sffamily H\normalfont} + N_{\mbox{\scriptsize\sffamily P\normalfont\normalsize}}
{\cal C}^{\mbox{\scriptsize\sffamily P\normalfont\normalsize}},  
\mbox{\hspace{3in}}
\label{hamtot} 
\ee 
for arbitrary functions $N_{\mbox{\scriptsize\sffamily P\normalfont\normalsize}}$.  Moreover, 
one needs to check that the primary constraints and any further \it secondary constraints 
\normalfont ${\cal C}_{\mbox{\scriptsize\sffamily G\normalfont\normalsize}}(q_{\mbox{\scriptsize\sffamily A\normalfont\normalsize}}, 
p^{\mbox{\scriptsize\sffamily A\normalfont\normalsize}})$ (obtained as true variational equations ${\cal C}_{\mbox{\scriptsize\sffamily G\normalfont\normalsize}} = 0$) 
are propagated by the evolution equations.  If they are, then the constraint algebra indexed by 
$\mbox{\sffamily I\normalfont}_{<1>} = {\mbox{\sffamily P\normalfont}}  \bigcup  {\mbox{\sffamily G\normalfont}}$ closes, and a classically-consistent theory is obtained.  
This happens when $\dot{{\cal C }}_{  \mbox{\scriptsize\sffamily I\normalfont\normalsize}_{<1>}  }$ vanishes either by virtue of the 
ELE's alone or 
additionally by virtue of the vanishing of ${\cal C}_{  \mbox{\scriptsize\sffamily I\normalfont\normalsize}_{<1>}  }$. Following Dirac, I denote this 
\it weak vanishing \normalfont
by $\dot{{\cal C}}_{  \mbox{\scriptsize\sffamily I\normalfont\normalsize}_{<1>}  } \approx 0$.   

If ${\cal C }_{  \mbox{\scriptsize\sffamily I\normalfont\normalsize}_{<1>}  }$ does not vanish 
weakly, then it must contain further independently-vanishing expressions 
${\cal C}_{  \mbox{\scriptsize\sffamily S\normalfont\normalsize}_{<1>}  }
(q_{\mbox{\scriptsize\sffamily A\normalfont\normalsize}}, 
p^{\mbox{\scriptsize\sffamily A\normalfont\normalsize}})$ in order for the theory to be 
consistent.  One must then enlarge the indexing set to
${\mbox{\sffamily I\normalfont}}_{<2>} =  {\mbox{\sffamily I\normalfont}}_{<1>} \bigcup  {\mbox{\sffamily S\normalfont}}_{<1>}$ 
and see if $\dot{{\cal C }}_{ {\mbox{\scriptsize\sffamily I\normalfont\normalsize}}_{<2>}} \approx 0$.  
In principle, this becomes an iterative process by which one may build toward   
a full \it constraint algebra \normalfont indexed by ${\mbox{\sffamily I\normalfont}}_{<\mbox{\scriptsize final\normalsize}>}$ by successive enlargements 
$ {\mbox{\sffamily I\normalfont}}_{<i+1>} =  {\mbox{\sffamily I\normalfont}}_{<i>} \bigcup  {\mbox{\sffamily S\normalfont}}_{<i>}$. 

In practice, however, the process cannot continue for many steps since $\# {\mbox{\sffamily I\normalfont}}_{<i + 1>} > 
\# {\mbox{\sffamily I\normalfont}}_{<i>}$, $\# {\mbox{\sffamily A\normalfont}}$ is a small number (for field theories, a small number per space point), 
and it is important for true number of d.o.f's to satisfy $\# {\mbox{\sffamily A\normalfont}} - \# {\mbox{\sffamily I\normalfont}}_{<\mbox{\scriptsize final\normalsize}>} 
> 0$ if one is seeking nontrivial theories.  It should be emphasized that there is no 
guarantee that a given Lagrangian corresponds to
any consistent system.  The perception of the first point post-dates Dirac (see e.g 
\cite{Thiemann, AB}, and also my comment in V), whereas Dirac's book contains a famously 
simple example of the second point: $ {\mbox{\sffamily L\normalfont}} = q$ gives 
the ELE $1 = 0$.  


Finally, constraints are \it first-class \normalfont if they have vanishing Poisson 
brackets with all the other constraints.  Otherwise, they are of \it second-class\normalfont.  
If there are second-class constraints, 
\be
\mbox{one may define the \it Dirac bracket\normalfont }
\mbox{\hspace{0.35in}} 
\{f ,g\}^* \equiv \{f , g\} - \{f  , {\cal C}_{\mbox{\sffamily\scriptsize D\normalsize\normalfont}} \}    
(     \{ {\cal C}_{\mbox{\sffamily\scriptsize D\normalsize\normalfont}} , 
{\cal C}_{{\mbox{\sffamily\scriptsize D\normalsize\normalfont}}^{\prime}} \}    )^{-1}
\{{\cal C}_{{\mbox{\sffamily\scriptsize D\normalsize\normalfont}}^{\prime}}, g\}
\mbox{\hspace{0.35in}}
\label{Dibra}
\ee
where \sffamily D \normalfont and $\mbox{\sffamily D\normalfont}^{\prime}$ index the smallest 
possible set of second-class constraints with respect to the original Poisson bracket.   The Dirac 
bracket is built so that if one uses it in place of the Poisson bracket, there are no 
second-class constraints.  Thus one may always choose to work with first-class constraints 
alone.  

\mbox{ }

\noindent \bf{1.2.4 Motivation for use of principles of dynamics}\normalfont

\mbox{ }

\noindent 
The principles of dynamics developed for classical mechanics have turned out to be 
applicable and very useful in the study of relativistic field theory to which the rest 
of this chapter is dedicated.  In particular, the ease of derivation of Einstein's field 
equations of GR using the principles of dynamics (See I.1.6 and I.2.4-5) made physicists 
appreciate the value of these methods.  II--VII contain many further uses 
of the principles of dynamics in the study of GR.

In the development of QM (see I.3) the Hamiltonian plays a central role.  Quantization 
itself may be na\"{\i}vely perceived as the passage from classical Poisson brackets to 
{\it commutators} of the corresponding quantum-mechanical operators:
\be
\{f, g\} \longrightarrow \frac{\hbar}{i}|[\hat{f}, \hat{g}]| 
\equiv \frac{\hbar}{i}(\hat{f}\hat{g} - \hat{g}\hat{f}) 
\mbox{ } .
\label{cpbtoqc}
\ee

Finally, if one replaces $p^{\mbox{\sffamily\scriptsize A\normalsize\normalfont}}$ by 
$\frac{\pa \mbox{\sffamily\scriptsize S\normalsize\normalfont}} 
{\pa q_{\mbox{\sffamily\tiny A\normalsize\normalfont}}}$ in \sffamily H \normalfont,  
one obtains the \it Hamilton--Jacobi equation \normalfont 
\be
\frac{\pa \mbox{\sffamily S\normalfont}}{\pa t} + \mbox{\sffamily H\normalfont}
\left(
q_{\mbox{\sffamily\scriptsize A\normalsize\normalfont}}; \frac{\pa \mbox{\sffamily S\normalfont}}{\pa q_{\mbox{\sffamily\scriptsize A\normalsize\normalfont}}}; t
\right) = 0 \mbox{ } , 
\label{HJEQ}
\ee
to be solved as a p.d.e for the as-yet undetermined \it Jacobi's principal function \normalfont 
{\sffamily S\normalfont}.  If the system has constraints (see e.g \cite{HTbook}) 
${\cal C}_{\mbox{\sffamily\scriptsize X\normalsize\normalfont}}
(q_{\mbox{\sffamily\scriptsize A\normalsize\normalfont}}; 
p^{\mbox{\sffamily\scriptsize A\normalsize\normalfont}} )$, (\ref{HJEQ}) would be supplemented by 
\be
{\cal C}_{\mbox{\sffamily\scriptsize X\normalsize\normalfont}}
\left(
q_{\mbox{\sffamily\scriptsize A\normalsize\normalfont}}; 
\frac{      \pa \mbox{\sffamily S}      }
     {      \pa q_{\mbox{\sffamily\scriptsize A\normalsize\normalfont}}      } 
\right) 
= 0 
\mbox{ }
\label{HJSUPPL} 
\ee
The Hamilton--Jacobi formulation is close to the semiclassical approximation to QM.

\subsection{Electromagnetism}

Many sorts of forces appear in Newtonian physics.  Physicists have tried to subdivide these 
into a few fundamental forces and consequent forces. 
The gravitational force (\ref{NLOG}) is regarded as a fundamental force; the theories of the 
electric and magnetic forces discussed below are also regarded as fundamental forces, in terms of 
which many other forces such as friction and chemical bonding are to be understood.  

The electric and magnetic force laws are 
\be
\mbox{\b{F}}_{(1)(2)}^{\mbox{\scriptsize e\normalfont}} = 
\frac{1}{4\pi\epsilon_0}\frac{q_{(1)}q_{(2)}}{|\mbox{\b{r}}_{(1)(2)}|^3}\mbox{\b{r}}_{(1)(2)} 
\mbox{ } , 
\mbox{\hspace{0.4in}}
\label{coulomb} 
\ee
\be
\mbox{\b{F}}_{(1)(2)}^{\mbox{\scriptsize m\normalfont}} = 
\frac{\mu_0}{4\pi}\frac{I_{(1)}I_{(2)}}{|\mbox{\b{r}}_{(1)(2)}|^3}\textrm{d}\mbox{\b{x}}_{(1)} 
\mbox{ \scriptsize $\times$ \normalsize} \textrm{d}\mbox{\b{x}}_{(2)} 
\mbox{ \scriptsize $\times$ \normalsize} \mbox{\b{r}}_{(1)(2)} 
\mbox{ } .
\label{magforce} 
\ee
(\ref{coulomb}) is for the force on a charge $q_{(1)}$ at position $\mbox{\b{x}}_{(1)}$ due to a 
charge $q_{(2)}$ at position $\mbox{\b{x}}_{(2)}$. 
(\ref{magforce}) is for the force on a wire element at position $\mbox{\b{x}}_{(1)}$ with current $I_{(1)}$ 
due to a wire element at position $\mbox{\b{x}}_{(2)}$ with current $I_{(2)}$.  $\epsilon_0$ and $\mu_0$ 
are respectively the permettivity and permeability in vacuo.  

Just as in Newtonian gravity, it is useful to think in terms of fields.  One often uses 
$\mbox{\b{F}}^{\mbox{\scriptsize e\normalfont}} = q\mbox{\b{E}}$ for \b{E} the electric field, and 
$\mbox{\b{F}}^{\mbox{\scriptsize m\normalfont}} = I \textrm{d}\mbox{\b{x}} \mbox{ \scriptsize $\times$ \normalsize} \mbox{\b{B}}$ for \b{B} the 
magnetic field.  Some of the field equations are then, in traditional integral form: 
\be
\epsilon_0\int_{\pa\Omega} \mbox{\b{E}}\cdot\textrm{d}\mbox{\b{S}} = 
\int_{\Omega}\rho_{\mbox{\scriptsize e\normalfont}}\textrm{d}^3x \mbox{ } 
\mbox{\hspace{1.65in}}
\mbox{ Gauss's Law }, 
\label{intemgauss}
\ee
\be
\oint_{\mbox{\scriptsize wire loop\normalfont}}\mbox{\b{B}}\cdot\textrm{d}\mbox{\b{x}} = \mu_0I_{\mbox{\scriptsize enclosed by wire loop\normalfont}} \mbox{ } 
\mbox{\hspace{0.8in}}
\mbox{ Amp\`{e}re's Law }, 
\label{ampere}
\ee
where $\rho_{\mbox{\scriptsize e\normalfont}}$ is electric charge density.
In the presence of changing magnetic fields, (\ref{coulomb}) was found no longer to hold; rather 
one has 
\be
\oint_{\mbox{\scriptsize wire loop\normalfont}}\mbox{\b{E}}\cdot\textrm{d}\mbox{\b{x}} = - \frac{d}{dt}\int_{\pa\Omega}\mbox{\b{B}}\cdot\textrm{d}\mbox{\b{S}}  \mbox{ } 
\mbox{\hspace{1in}}
\mbox{ Faraday's Law , }
\label{faraday}
\ee 
which suggests that electricity and magnetism are not independent concepts.  

Provided that Amp\`{e}re's law in fact also includes a \it displacement current \normalfont correction, 
Maxwell found that he could consistently explain electricity and magnetism as aspects of a single 
phenomenon: \it electromagnetism \normalfont.  The differential form of the equations 
now reads 
\be
\mbox{\hspace{1in}}
\mbox{\b{$\pa $}}\cdot\mbox{\b{E}} = \frac{\rho_{\mbox{\scriptsize e\normalfont}}}{\epsilon_0}  
\mbox{\hspace{2.5in}}
\mbox{ } \mbox{ Gauss' Law }, 
\mbox{\hspace{0.1in}}
\label{M1} 
\ee
\be 
\mbox{\hspace{1in}}
\mbox{\b{$\pa $}} \mbox{ \scriptsize $\times$ \normalsize} \mbox{\b{E}} = - 
\dot{\mbox{\b{B}}}
\mbox{\hspace{2.2in}} 
\mbox{ } \mbox{ Faraday's Law }, 
\mbox{\hspace{0.1in}}
\label{M2} 
\ee
\be 
\mbox{\hspace{1in}}
\mbox{\b{$\pa $}} \cdot \mbox{\b{B}} = 0 \mbox{ } 
\mbox{\hspace{0.9in}}
\mbox{ non-existence of magnetic monopoles },
\mbox{\hspace{0.15in}}
\label{M3}
\ee
\be
\mbox{\hspace{1in}}
\mbox{\b{$\pa $}} \mbox{ \scriptsize $\times$ \normalsize} \mbox{\b{B}} = 
\mu_0(\mbox{\b{j}}_{\mbox{\scriptsize e\normalfont}} + \epsilon_0\dot{\mbox{\b{E}}}) 
\mbox{\hspace{0.05in}}
\mbox{Amp\`{e}re's law with displacement current },   
\mbox{\hspace{0.05in}}
\label{M4}
\ee
where $\mbox{\b{j}}_{\mbox{\scriptsize e\normalfont}}$ is the electric current.  
The displacement current was then experimentally verified.  N.B it is often convenient to 
rewrite Maxwell's equations in terms of a 
\it magnetic potential \normalfont \b{A} and an \it electric potential \normalfont $\Phi$, s.t. 
\b{B} = $\mbox{\b{$\pa$}} \mbox{ \scriptsize $\times$ \normalsize} \mbox{\b{A}}$ and \b{E} = 
$ - \mbox{\b{$\pa$}}\Phi - \dot{\mbox{\b{A}}}$.
 
Here are some useful comments for later.  
First, from this unification of electricity and magnetism into electromagnetism, 
Maxwell unexpectedly deduced that light is nothing but electromagnetic radiation [by combining the 
above equations to form two wave equations whose propagation speed is the speed of light 
$c = (\epsilon_0\mu_0)^{-\frac{1}{2}}]$.  

Second, whereas (\ref{M2}) [and (\ref{M4})] are the usual sort of evolutionary laws, 
(\ref{M1}) [and (\ref{M3})] are constraints i.e instantaneous laws about the permissible 
configurations.  

Third, Maxwell's equations are field equations, determining how sources produce 
electromagnetic fields.  To have a full grasp of electromagnetism, one also requires a law to 
compare the motion of (constant mass) charged and uncharged particles in the presence of 
\be
\mbox{an electromagnetic field: the \it Lorentz force law \normalfont  }
\mbox{\hspace{0.7in}}
\ddot{\mbox{\b{x}}} = \frac{e}{m}(\mbox{\b{E }} + \mbox{\b{v}} \mbox{ \scriptsize $\times$ \normalsize} \mbox{\b{B}}) 
\mbox{ } .  
\mbox{\hspace{2in}} 
\label{LFL}
\ee

Fourth, if Maxwell's equations provide an account for light waves, then the contemporary 
experience with other types of waves suggested that there should be an 
associated medium, the \it luminiferous Aether\normalfont, of which light is an excitation.  
Fifth, Maxwell's equations are not invariant under the Galilean transformations 
(\ref{Gal1}, \ref{Gal2}).  Instead, one may define a new group of transformations which leave 
Maxwell's equations invariant: the Lorentz group, most conveniently described as follows.  
It consists of rotations and {\it boosts}.  
\be
t \longrightarrow t^{\prime} = \gamma[t - vx/c^2] \mbox{ } , \mbox{ } \gamma = 1/\sqrt{1 - (v/c)^2}
\ee
\be
x \longrightarrow x^{\prime} = \gamma(x - vt)
\ee
\be
y \longrightarrow y^{\prime} = y
\ee
\be
z \longrightarrow z^{\prime} = z 
\ee
is the boost for passing from a frame at rest to a frame moving 
with constant velocity $v$ in the $x$ direction.  For other directions of motion, 
rotate the axes, then apply the above and then rotate back.

\subsection{Special relativity}

The last two issues above have profound consequences.   Given the existence of the Aether, 
its rest frame would be expected to be be privileged by Maxwell's equations, so the lack of 
Galilean invariance was not perceived as an immediate impasse.  This led to the proposal that 
as electromagnetism is not Galileo-invariant, experiments involving electromagnetism could 
then be used to determine motion with respect to the Aether rest frame.  There were furthermore 
speculation that this Aether rest-frame might coincide with absolute space.  

However, in the Michelson--Morley experiment a null result\fn{Today this is known to be 
null to 1 part in $10^{15}$ \cite{exptinrin}.}  was obtained for the velocity of the Earth 
relative to the Aether.  Furthermore, within the framework of Aether theory, this was in 
contradiction with Bradley's observation of stellar aberration.  Although Fitzgerald and 
Lorentz \cite{Lorentz} attempted to explain these observations  \it constructively \normalfont 
in terms of somehow the inter-particle distances of particles travelling parallel to the 
aether flow being contracted, Einstein had a different, \it axiomatic \normalfont strategy 
akin \cite{CGconstructive, CGspacetime} to how thermodynamics is based on the non-existence of perpetual 
motion machines.  He elevated the outcome of the Michelson--Morley experiment from a null 
result about motion and electromagnetism to a \it universal \normalfont postulate.  Rather than 
there being Galilean invariance for mechanics, Lorentz invariance for electromagnetism 
and goodness knows what invariance for other branches of physics, he postulated that

\noindent{\bf RP1} (relativity principle) 
all inertial frames are equivalent for the formulation of all physical laws.

From this it follows that the laws of nature share a universal transformation group under 
which they are invariant.  There is then the issue of which transformation group this should be.  
{\bf RP1} narrows this down to two obvious physical possibilities, distinguished by whether the laws of 
nature contain a finite or infinite propagation speed 
$v_{\mbox{\scriptsize prop\normalsize}}$.\fn{If one wishes furthermore to avoid causality paradoxes 
one chooses this to ba a maximum speed (rather than a minimun one for tachyons).}  
If one adopts absolute time as a second postulate (\bf Galilean RP2\normalfont), the infinite is 
selected, and one has universally Galileo-invariant physics.  
The finite is selected if one adopts instead a constant velocity postulate such as 

\noindent{\bf Lorentzian RP2}: light signals in vacuo are propagated rectilinearly with the same velocity 
at all times, in all directions, in all inertial frames.

The chosen velocity serves universally [and so is unique, so taking 
$v_{\mbox{\scriptsize prop\normalsize}}$ =  (the speed of light $c$) is without loss of generality].  One has then a 
universally Lorentz-invariant physics.  In the former case, which amounts to upholding `{\bf N4}', electromagnetism must be 
corrected, whereas in the latter case Newtonian mechanics must be corrected.  
Einstein chose the latter.  Notice that this is the option given by a law of nature and not 
some postulated absolute structure; also whereas there was ample experimental evidence 
for Maxwellian electromagnetism, existing experimental evidence for Newtonian mechanics 
was confined to the low velocity ($v{\ll}c$) regime for which Galilean transformations are an excellent 
approximation to Lorentz transformations.  Indeed the investigation of the high velocity 
regime promptly verified Einstein's corrections to Newtonian mechanics.  This example of the 
great predictive power of special relativity is compounded by the universality:  for each 
branch of physics, one obtains specific corrections by requiring the corresponding laws 
to be Lorentz-invariant.  The concept of non-materially substantiated media and the 
above proposal were thus destroyed, and physics was rebuilt on the premise that there 
was no room in any of its branches for analogous concepts and proposals.  

Minkowski pointed out that whilst Newton's notions of absolute space and time are also destroyed 
because there are no longer privileged surfaces of simultaneity, one could geometrize space and 
time together as spacetime, in which the null cones are privileged.  These correspond to the 
surfaces on which the free motion of light occurs (and of all other massless particles, by 
Einstein's postulates: one has a \it universal null cone structure \normalfont in classical 
physics).  And massive particles are permitted only to travel from an event 
(spacetime point) into the interior of the future null cone of that event.  Of particular significance, in free `inertial' 
motion all massive particles follow timelike straight lines whereas all massless particles 
follow null straight lines.  Following from such a geometrization, it makes sense to implement 
the laws of physics in terms of the 4-tensors corresponding to Minkowski's 4-d spacetime.  

First, reconsider electromagnetism.  Whereas in this special case the main laws 
(Maxwell's equations) are already Lorentz-invariant and thus require no corrections, 
they can be cast in an elegant spacetime notation, and the new conceptual framework 
greatly facilitates the study of electrodynamics \cite{OEOMB}.  
Introducing the {\it electromagnetic field tensor } $\mbox{\v{F}}_{\bar{A}\bar{P}}$ 
\be
\mbox{(s.t $F_{\bar{a}\bar{0}} = E_{\bar{a}}$,  $F_{\bar{a}\bar{p}} = \epsilon_{\bar{a}\bar{p}\bar{c}}B^{\bar{c}}$) the Maxwell equations are }
\mbox{\hspace{0.6in}}
\pa_{\bar{A}}\mbox{\v{F}}^{\bar{A}\bar{B}} = -\mu_0\mbox{\v{j}}^{\bar{B}}_{\mbox{\scriptsize e\normalsize}}
\mbox{\hspace{0.7in}}
\label{inhommax}
\ee
\be
\mbox{\hspace{3.5in}}
\pa_{[\bar{A}}\mbox{\v{F}}_{\bar{B}\bar{C}]} = 0
\label{hommax}
\ee
\be
\mbox{If one uses an \it electromagnetic 4-potential \normalfont $\mbox{\v{A}} = [-\Phi, \mbox{\b{A}}]$ s.t  }
\mbox{\hspace{0.5in}}
\mbox{\v{F}}_{\bar{A}\bar{B}} = 2\pa_{[\bar{A}}\mbox{\v{A}}_{\bar{B}]} 
\mbox{ } , 
\mbox{\hspace{1in}}
\label{Adef}
\ee
\be
\mbox{then (\ref{hommax}) holds trivially and one is left with }
\mbox{\hspace{0.5in}}
\Box \mbox{\v{A}}^{\bar{A}} - \pa^{\bar{A}}\pa_{\bar{B}}\mbox{\v{A}}^{\bar{B}} = 
- \mu_0\mbox{\v{\j}}^{\bar{A}}_{\mbox{\scriptsize e\normalsize}}
\mbox{\hspace{1.5in}}
\label{Amaxfull}
\ee
where $\mbox{\v{\j}}^{\bar{A}}_{\mbox{\scriptsize e\normalfont}} = 
[\rho_{\mbox{\scriptsize e\normalfont}}, \mbox{\b{j}}_{\mbox{\scriptsize e\normalfont}}]$.  Na\"{\i}vely $\mbox{\v{A}}_{\bar{A}}$ 
would have 4 d.o.f's.  But there is the constraint (\ref{M1}); 
also (not unrelatedly, it turns out) (\ref{Adef}) is clearly invariant under the 
\it gauge transformations \normalfont 
\be
\mbox{\v{A}}_{\bar{A}} \longrightarrow \mbox{\v{A}}_{\bar{A}} + \pa_{\bar{A}}\Lambda 
\label{Abelgau}
\ee
for any function $\Lambda$.  Thus electromagnetism has 2 d.o.f's per space point, 
corresponding to 

\noindent light having 2 polarizations.  
For later use, the energy-momentum tensor of electromagnetism is\fn{I use $A\circ B$ to denote 
the trace of the matrix product of $A$ and $B$.}
\be
\mbox{\v{T}}^{\bar{A}\bar{B}}{\mbox{\scriptsize em\normalfont}} = 
\frac{1}{\mu_0}(\mbox{\v{F}}^{\bar{A}\bar{C}}{\mbox{\v{F}}^{\bar{B}}}_{\bar{C}} 
- \frac{1}{4}\eta^{\bar{A}\bar{B}}\mbox{\v{F}} \circ \mbox{\v{F}})
\ee
which is symmetric and is conserved.  
The Lorentz force law (\ref{LFL}) becomes
\be
\frac{    d^2x^{\bar{A}}    }{    d\tau^2    } = 
\frac{e}{m}{\mbox{\v{F}}^{\bar{A}}}_{\bar{B}}
\frac{    dx^{\bar{B}}    }{    d\tau    }
\label{4LFL}
\ee
where $\tau$ is the time measured in the particle's rest-frame (the \it proper time\normalfont). 

Next it was required to change the forms of all the laws of nature.\fn{Since this occurred in 
1905, we mean all the other classical laws of nature known at that time.}  For Newtonian 
Mechanics, {\bf N2} and the definition of momentum are still correct, provided that proper time 
is employed.  The relativistic laws of nature are a great success.  
Indeed in many applications an important step toward proposing new laws of physics 
is to consider only the Lorentz-invariant possibilities.  
However, Einstein found that attempting to accommodate gravity in this scheme 
presented significant difficulties.

\subsection{General relativity}

Nearby freely-falling particles in a (non-uniform) gravitational field experience a relative 
acceleration.  This leads to the need to replace the inertial frames of 
Newtonian mechanics (which are supposedly of infinite extent) by local inertial frames.  In  
order to be able to define these it is crucial that inertial mass be identically proportional 
to gravitational mass for all materials, for else each material would require its own 
definition of local inertial frame.  This is the \it principle of equivalence \normalfont 
({\bf POE}).   Einstein \cite{poe} then adopted the somewhat stronger supposition\fn{Tests 
distinguishing between the various forms of the principle of equivalence have been devised 
and used to experimentally confirm each of these forms to high accuracy \cite{Will}.} that 
gravitation is not locally distinguishable from acceleration by physical experiments anywhere 
in the universe, and can thus be transformed away by passing to the suitable local inertial 
frame.  He then guessed\fn{This guess is now experimentally supported \cite{Stewart} by 
Pound--Rebka type experiments \cite{Will} to 2 parts in $10^{-4}$.} that the inertial frames of SR were to be 
identified with the local inertial frames of freely-falling particles.  To Einstein the 
{\bf POE} strongly suggested \cite{EG13} that gravitation could be included within 
relativity by boldly postulating that spacetime with gravitation would not be flat Minkowski 
spacetime but rather a spacetime curved by the sources of gravitation so that the straight 
timelike lines followed by free particles in Minkowski spacetime are bent into the curves 
followed by relatively-accelerated freely-falling particles.  The straight null lines 
constituting the lightcones of Minkowski spacetime would then likewise be bent by the sources 
of gravitation.

The mathematics of the connection permits the incorporation of the above features of the 
gravitational field.  The coordinates in which the connection may be set to zero at each 
particular point are to correspond to the freely-falling frame at that point.  
The privileged curves followed by freely falling particles and by light rays are to be the 
timelike and null affine geodesics of the geometry; at any point in the freely-falling frame 
these reduce to the straight lines of Minkowski spacetime.\fn{Thus this implementation of the 
{\bf POE} tacitly includes the signature assumption contained in {\bf Lorentzian RP2}.}  The geodesic 
equation (\ref{afgeoeq}) is of the form of the combination (\ref{NII + NLOG}) of \bf N2 \normalfont 
and Newton's law of gravitation for a connection whose only nonzero components are  
${\check{\Gamma}^i}_{00} = \pa_i\phi$; from this it follows that the only nonzero Riemann 
tensor components are 
\be
{\mbox{\v{R}}^i}_{0j0} = \pa_{i}\pa_{j}\phi
\label{NewtR}
\ee
so that one obtains agreement between the Newtonian tidal equation (\ref{tidal}) and the corresponding 
case of the geodesic deviation equation (\ref{geodeveq}).  
In GR the geodesic deviation equation plays an analogous role to that of the Lorentz force law (\ref{4LFL}) in electrodynamics 
\cite{WheelerGRT}.  
  
Furthermore, Einstein introduced a semi-Riemannian metric $g_{AB}$ on spacetime, both to 
account for observers in spacetime having the ability to measure lengths and times if equipped 
with standard rods and clocks (paralleling the development of SR),
and furthermore to geometrize the gravitational field.  For simplicity, he assumed a 
symmetric metric and it turned out that the aforementioned connection was the metric one \cite{EinGR}.  As $g_{AB}$ reduces 
locally to SR's $\eta_{\bar{A}\bar{B}}$ everywhere locally the laws of physics take their SR form.  


This is not yet a gravitational theory: field equations remain to 
be found.  
Einstein \cite{VEinstein} `derived' his field equations (EFE's)\footnote{$\check{T}_{AB}$ is 
the curved spacetime energy-momentum tensor.}
\be
\mbox{\v{G}}_{AB} = \mbox{\v{R}}_{ AB } - \frac{1}{2} g_{AB}\mbox{\v{R}} = 
\left(
\frac{8\pi G}{c^4}
\right)
\mbox{\v{T}}_{AB}^{\mbox{\scriptsize Matter\normalsize}} 
\label{Vefes}
\ee
by demanding 

\noindent  {\bf GRP} (the General Relativity Principle) that all frames are equivalent 
embodied in spacetime general covariance (the field equations are to be a 4-tensor equation).  

\noindent {\bf GR Newtonian Limit} that the correct Newtonian limit be recovered in 
situations with low velocities $v \ll c$ and 
weak gravitational fields $\phi \ll c^2$.  Note that by (\ref{NewtR}) 
Poisson's 
\be
\mbox{equation of Newtonian gravity (\ref{GPoisson}) may now be written as }
\mbox{\hspace{0.7in}} 
\mbox{\v{R}}_{00} = 4\pi G\rho 
\mbox{\hspace{1in}} 
\ee
which is suggestive that some curvature term should be equated to the energy-momentum causing 
the gravitation.

\noindent {\bf GR divergencelessness} since $\mbox{\v{T}}_{AB}$ 
is conserved (divergenceless: $\nabla_A\mbox{\v{T}}^{AB} = 0$) and symmetric, this curvature 
term should also have these properties.  

\noindent Thus, from the end of 0.3, $\mbox{\v{G}}_{AB}$ is a good choice of curvature 
term. 

It is also helpful that both the EFEs' Einstein tensor and the metric which solves them 
are (0, 2) symmetric tensors, so that the EFE's are neither over- nor under-determined.  
In principle if one attempted other geometrizations of the 
gravitational field, one would usually face mismatches rather than the above coincidences 
\cite{Schrodinger50}.  

I leave the disputed r\^{o}le of {\it Mach's principle} \cite{Mach} (about how to abolish absolute space) 
both in the above conception of GR and within GR itself to I.2.7.3 and II.2.  

The above considerations are all physical.  But in fact the following 
mathematical {\bf GR Cartan simplicities} \cite{VCartan} are also required to axiomatize GR: 
that $\mbox{\v{G}}_{AB}^{\mbox{\scriptsize trial\normalsize}}$ contains at most second-order derivatives and is linear in these.  
The {\bf GR Lovelock simplicities} \cite{Lovelocktensor} eliminated the linearity assumption in dimension $n \leq 4$.    
One should note that throughout $\Lambda g_{AB}$ is an acceptable 
second term on the left hand side by all these considerations.  Such a $\Lambda$ is a 
\it cosmological constant \normalfont which is thus a theoretically-optional feature, the need 
for which is rather an issue of fitting cosmological observations.  

\mbox{ }

The credibility of GR was rapidly established by its precise explanation of the perihelion 
shift of Mercury and experimental verification of the bending of lightrays by the sun.  Today 
such solar system tests (also including the time-delay effect) show no significant deviation 
from GR to 1 part in $10^4$ \cite{Will}.  Binary pulsar data are also in very good agreement 
with GR.  Note however that all these tests are of the weak-field regime of GR.  There 
may soon be tests of a more strong-field regime from gravity wave experiments (see I.2.11).  

GR however may be interpreted as having theoretical impasses.  It predicts its own 
inapplicability in extreme circumstances (in black holes, and during a tiny 
time interval after a cosmological Big Bang) that are moreover likely to occur in our universe 
(by the Hawking--Penrose singularity theorems \cite{HE, Wald}). Perhaps relatedly, one does 
not know how to combine GR and QM to form a theory of quantum gravity necessary for the study 
of these extreme regimes (see I.3).

\subsection{Many routes to relativity}

There are a number of other routes to obtaining GR.  It is important to have as many routes 
as possible to a physical theory since a number of them will allow new insights into how the 
theory works or be particularly adapted to the solution of otherwise intractable problems.  
Furthermore, different routes may suggest different alternative or more general theories for 
the current theory to be tested against.  

Wheeler listed 6 such routes to GR in 1973 \cite{MTW}. The first is Einstein's above.  The second is Hilbert's, from variation of the Einstein--Hilbert action \cite{VWeyl} 
\be
\mbox{\sffamily I\normalfont}_{\mbox{\scriptsize EH\normalsize}} = \int \textrm{d}^4x \sqrt{|g|}
(\mbox{\v{R}} + \mbox{\sffamily L\normalfont}_{\mbox{\scriptsize Matter\normalsize}}) \mbox{ } .
\label{VEinsteinHilbert}
\ee 
A simplicity proof equivalent to Cartan's but for actions was given by Weyl \cite{VWeyl}.  This derivation of the EFE's is very straightforward (see I.2.4), which helped to 
bring variational principles to the attention of the physics community.

The third and fourth routes are the two-way workings between (\ref{VEinsteinHilbert}) and the Arnowitt--Deser--Misner (ADM) action, which (in vacuo for simplicity) takes the 
form \cite{ADM}
\bea
\mbox{\sffamily I\normalfont}_{\mbox{\scriptsize ADM\normalsize}} = \int \textrm{d}t \int \textrm{d}^3x ( p \circ \dot{h} - \alpha{\cal H} - \beta^i{\cal H}_{i})
\label{VADM} \mbox{ } , \\
{\cal H} \equiv \frac{1}{\sqrt{h}}\left(p\circ p - \frac{p^2}{2}\right) - \sqrt{h}R  = 0 \mbox{ } ,
\label{Vham} \\
{\cal H}_i \equiv -2D_j{p_i}^j = 0 \mbox{ } ,
\label{Vmom}
\eea
(up to a divergence term), which follow from the `ADM' split of the spacetime metric with respect to a sequence of spatial slices, according to 
\be
\begin{array}{ll}
g_{AB} = \left(\begin{array}{ll} \beta_k\beta^k - \alpha^2   &    \beta_j    
\\
                                           \beta_i            &    h_{ij}
                   \end{array}\right) \mbox{ , so that }
&
g^{AB} = \left(\begin{array}{ll} -\frac{1}{\alpha^2}  & 
\frac{\beta^j}{\alpha^2}  \\
                                          \frac{\beta^i}{\alpha^2} & h^{ij} - 
\frac{\beta^i\beta^j}{\alpha^2}
                    \end{array}\right) 

\end{array} \mbox{ } .
\label{VADMsplit}
\ee
Here, $h_{ij}$ is the metric induced on the spatial slice (see II.2.1 and fig 4) 
with determinant $h$, the \it lapse \normalfont $\alpha$ is the change in proper time as one 
moves off the spatial surface and the \it shift \normalfont $\beta_i$ is is the displacement 
in identification of the spatial coordinates between two adjacent slices.  The dot is the 
derivative in the (time) direction perpendicular to the slice, $\frac{\pa}{\pa t}$.  
Conventionally, $h_{ij}$, $\beta_i$ and $\alpha$ are regarded as canonical coordinates, 
$p^{ij}$ is the momentum conjugate to $h_{ij}$ and there is no momentum associated 
with $\beta_i$ nor $\alpha$: these are Lagrange multipliers.  Thus the true gravitational 
d.o.f's in GR are contained in 
\be
\mbox{Riem} = \{\mbox{space of Riemannian 3-metrics}\} 
\ee
on a fixed topology, which I usually take to be compact without boundary (CWB).   
But the true d.o.f's are furthermore subjected to the Hamiltonian 
and momentum constraints ${\cal H}$ and ${\cal H}_i$ respectively.   
\begin{figure}[h] 
\centerline{\def\epsfsize#1#2{0.4#1}\epsffile{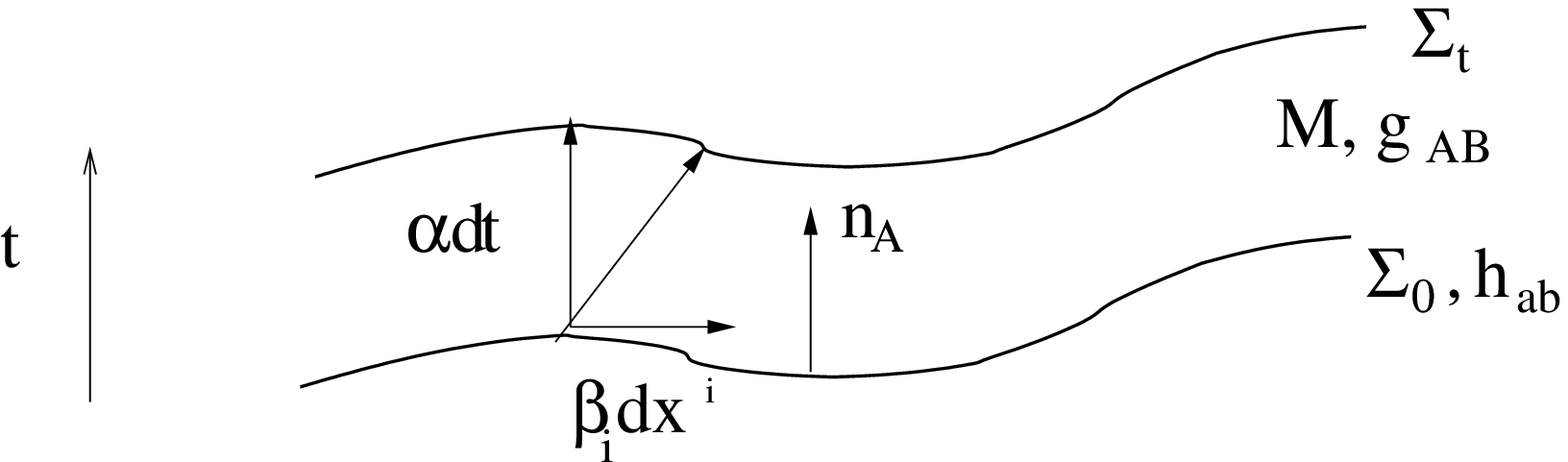}}
\caption[]{\label{TO3.ps}}
\end{figure}

The 3-space approach (TSA) of this thesis is mainly concerned with the third and fourth routes, 
particularly the extension of the derivation of ADM's system of field equations without 
starting from any spacetime formulation of GR.  For completeness, the fifth and sixth routes 
mentioned are the Fierz--Pauli spin-2 field in an unobservable flat background \cite{VFP} and 
Sakharov's idea that gravitation is an effective elasticity of space that arises from particle 
physics \cite{VSak}.   

\mbox{ }

One could add a number of (mostly) more recent routes to Wheeler's list.  
Three elements are common to many new routes: changes of variables, splits and 
discretizations.  Changes of variables include the use of first-order formulations such 
as those of Cartan \cite{Cartan25} (credited by Wheeler as a route in his earlier 
account \cite{Wheeler}) or Palatini \cite{MTW}, as well as the use of various sorts of spinors, 
twistors and the Newman--Penrose formalism \cite{Stewart, PenRind}.  Among these the Ashtekar 
variable formulations (see I.2.12.3) stand out for allowing new quantization possibilities. 
At least some
of these formulations can be cast in terms of spacetime or in terms of a 
foliation by spatial hypersurfaces.  There are also other interpretations of the standard 
split such as the thin sandwich (I.2.9), and indeed other splits (see I.2.12.4).  
Discretizations can also be spacetime-based (e.g Regge calculus \cite{MTW}) or space-based 
(dynamical triangulation \cite{Carlipbook}).  These might be taken as approximations for 
numerics, but that logic can be reversed to suggest that these are the true formulations and 
that GR is but a macroscopic limit.

Other elements to new routes are more bizarre -- first principles or apparently--unrelated 
mathematics from which GR somehow drops out.  Examples include obtaining GR from the algebra 
of deformations of a spatial hypersurface (II.1, VI), obtaining GR from relational 
3-space principles in the TSA, and routes obtaining GR as an 
effective theory such as Sakharov's route above, or Jacobson's \cite{Jacobbh} (in which 
GR is deduced from black hole thermodynamics).  When QM considerations are 
included, even more colourful routes emerge: in particular the EFE's arise from the closed 
string spectrum (see I.3.3.4). There are also as--yet incomplete attempts to recover spacetime 
(see the reviews \cite{Carliprev, qgravrevI00}). 

Among all these routes, I distinguish those to GR alone,\fn{Furthermore, some routes do not 
quite lead to standard GR (e.g one might get Euclidean GR or 10-d GR, or a dilatonic rather 
than minimal coupling which could be thought of as interfering with the geodesic postulate 
part of GR).} those to GR with matter fields  `added on', and those which attempt to be more: 
unifications of GR with other branches of physics.  I emphasize the need to be comprehensive 
in including a full enough set of fields to satisfactorily describe all of nature to lend good 
credibility to the principles behind the route.    
Among the unified theories, I distinguish between total and partial attempts at unification  
and between the classical and those which furthermore {\sl include} QM.  For 
example, already-unified Rainich--Misner--Wheeler theory (I.2.7.2), 
Weyl's theory (App III.A) and Kaluza--Klein theory (I.3.3.4) are attempted classical 
unifications of electromagnetism and gravity.  
The five superstring theories [or an as-yet undetermined, all-embracing M-theory] (I.3.3) are 
Theories of Everything -- attempts at total unification of all classical and quantum physics.  
Credible unification should unexpectedly include extra physics (like light in electromagnetism) 
and (or) lead to verified predictions.  In the absence of this one should apply Occam's razor.  

\mbox{ }

Finally, some attempted routes or the dropping of some simplicity postulates suggest 
modifications to GR.  This is extra structure for its own sake rather than geared to wards the 
inclusion of other branches of physics.  Relevant well--known alternative theories include 
higher derivative gravities, Lovelock gravity (see 
p 65) and dilatonically-coupled 
scalar--tensor-type theories such as Brans--Dicke theory (see III.1.3).  One can think of 
these theories as providing alternatives against which to test GR in certain regimes. 
Brans--Dicke theory provided a theory to test GR against in the solar system.  I worry 
higher-derivative theories might alter both the physics and the numerical modelling of compact 
binaries.

\subsection{Other classical matter fields}  

There is much treatment of matter in the thesis so as to upgrade the TSA route to 
relativity alone to being a route with matter `added on'.  So I wish to couple a full 
set of matter fields capable of describing nature.  I consider the fundamental fields,\fn{The 
fundamental fields are those it makes sense to 
quantize and in terms of which all other matter fields may in principle be described.  
Of course this is to some extent a matter of taste: e.g particle physics may change in what 
is regarded as fundamental or GR might turn out not to be fundamental.}
which are important since their flat spacetime versions account for all accepted physics bar 
gravity.  Consequently these are valued as theoretical schemes, so I also hope to learn from 
the classical and quantum theory of these flat-spacetime theories so as to better understand 
the structure of GR (see below, I.2.3.1, I.3.2, App II.A).  For the moment I present the usual 
fundamental fields in flat spacetime below, establishing much notation.  
The curved versions coupled to GR are introduced later 
(in I.2.5-6 for standard formulations, and in IV and VI for the TSA).  Other 
(in some cases decidedly nastier) fields come into play to wards the end of 
the thesis (VI-VII) as theoretical possibilities to help sharpen the TSA,  
and to see to which extent its workings were genuinely selective rather than 
selective by unwarranted simplicity assumptions.  Here the possibilities that the TSA 
``hints at partial unification'' or might have some control over the equivalence principle 
are investigated.  The below treatment includes not just the usual Lagrangian formulation but 
also the Hamiltonian formulation because this thesis is a 3+1 space--time split study, 
geared toward canonical quantization.  

\mbox{ }

\noindent \bf{1.7.1 Scalars and spin-$\frac{1}{2}$ fermions}\normalfont

\mbox{ }

\noindent
The Klein--Gordon theory of a scalar field $\varsigma$ with mass $m_{\varsigma}$ has the Lagrangian 
\be  
\bar{\mbox{\sffamily L\normalfont}}^{\varsigma}_{\mbox{\scriptsize KG\normalsize}}  
= - \frac{1}{2}(\pa_{\bar{A}}\varsigma\pa^{\bar{A}}\varsigma + m_{\varsigma}^2\varsigma^2) 
= \frac{1}{2}(\dot{\varsigma}^2 - \mbox{\b{$\pa$}}\varsigma\cdot\mbox{\b{$\pa$}}\varsigma 
- m_{\varsigma}^2\varsigma^2) 
\mbox{ } .
\label{KG}
\ee
\be 
\mbox{The conjugate momentum is }
\mbox{\hspace{1.7in}}
\pi_{\varsigma} \equiv \frac{\pa \bar{\mbox{\sffamily L\normalfont}}}{\pa\varsigma} = \dot{\varsigma} 
\mbox{ } , 
\label{KGmom}
\mbox{\hspace{3in}} 
\ee
\be
\mbox{and the Hamiltonian is }
\mbox{\hspace{1.6in}}
\bar{\mbox{\sffamily H\normalfont}}^{\varsigma}_{\mbox{\scriptsize KG\normalsize}}  = \frac{1}{2}(\pi_{\varsigma}^2 
+ \mbox{\b{$\pa$}}\varsigma\cdot\mbox{\b{$\pa$}}\varsigma + m_{\varsigma}^2\varsigma^2) 
\mbox{ } .
\mbox{\hspace{3in}} 
\label{KGham}
\ee
\be
\mbox{The corresponding ELE is the Klein--Gordon equation }
\Box\varsigma + m_{\varsigma}^2\varsigma = 0 
\mbox{ } , 
\mbox{\hspace{2in}} 
\label{KGeq}
\ee 
which is a flat-space relativistic (i.e Lorentz-invariant) wave equation.  
For $m_{\varsigma} = 0$, it \sl is \normalfont the flat-space wave equation.  

For use below, the Klein--Gordon theory for two scalar fields 
$\varsigma_1$ and $\varsigma_2$ may be re-expressed as the Klein--Gordon theory for a 
single complex scalar field, with Lagrangian 
\be
\bar{\mbox{\sffamily L\normalfont}}_{\mbox{\scriptsize KG\normalsize}}^{\varsigma, \varsigma^*} 
= - \pa_{\bar A}\varsigma\pa^{\bar A}\varsigma^* + m_{\varsigma}^2\varsigma^*\varsigma
\label{CKG}
\ee
for $\varsigma = \frac{1}{\sqrt{2}}(\varsigma_1 + \varsigma_2)$.  
There is no reason to believe that Klein--Gordon theory is realized as a fundamental theory 
in nature.   I use it as a toy (in I.2.3.1 and I.3), and note that its classical canonical 
structure is unaffected if one uses a potential of form $V(\varsigma)$ in place of 
$m_{\varsigma}^2\varsigma^2$.    To date, fundamental scalars play only hypothetical roles in 
physics.  Higgs scalars are postulated to give masses to certain fields, and inflaton(s) 
are postulated to explain certain features of the early universe.  More complicated scalars 
also occur in scalar-tensor theory (III.1.3).   

\mbox{ }

Dirac investigated whether the Klein--Gordon operator might be the square of some linear 
operator.  Indeed, he thus arrived at the linear Dirac equation\fn{I use lower-case greek indices running over 1 to 4 
for 4-component spinor indices, and typewriter capital indices running over 1 to 2 for 2-component 
spinor indices. I often suppress these spinorial indices.}   
\be
({{i\gamma^{\bar{A}}}_{\rho}}^{\sigma}\pa_{\bar{A}} - m_{\psi}{\delta_{\rho}}^{\sigma})\psi^{\rho} = 0
\label{Diraceq}
\ee
\be
\mbox{for $\gamma^{\bar{A}}$ the Dirac matrices obeying the \it Dirac algebra \normalfont} 
\mbox{\hspace{0.5in}}
\gamma^{\bar{A}}\gamma^{\bar{B}} + \gamma^{\bar{A}}\gamma^{\bar{B}} 
= 2\eta^{\bar{A}\bar{B}} 
\mbox{ } . 
\mbox{\hspace{0.5in}} 
\label{Diracalgebra}
\ee 
\be 
\mbox{I work in the chiral representation in which the 4-spinor }
\mbox{\hspace{0.6in}}
\psi^{\rho} = 
\left[
\begin{array}{l}
\psi_{\mbox{\scriptsize D\normalsize}}^{\mbox{\tt A\normalfont}} \\
\psi_{\mbox{\scriptsize L\normalsize}}^{\mbox{\tt A\normalfont}} \\
\end{array}
\right] 
\mbox{ } , 
\mbox{\hspace{1.2in}} 
\label{chiralspinor}
\ee 
\be
\mbox{where D and L stand for right- and left-handed, and }
\gamma^{\bar{0}} = 
\left(
\begin{array}{ll}
0 & 1 \\ 1 & 0
\end{array}
\right)
\mbox{ } , 
\mbox{ } 
\gamma^{\bar{a}} = \left(
\begin{array}{ll}
0 & \sigma^{\bar{a}} \\ -\sigma^{\bar{a}} & 0
\end{array}
\right)
\label{chiralgamma}
\ee
are the chiral representation Dirac matrices, and
\be
\sigma^{\bar{1}} =
\left(
\begin{array}{ll}
0 & 1 \\ 1 & 0
\end{array}
\right)
\mbox{ } \mbox{ } , 
\mbox{ } \mbox{ }
\sigma^{\bar{2}} =
\left(
\begin{array}{ll}
0 & -i \\ i & 0
\end{array}
\right)
\mbox{ } \mbox{ } , 
\mbox{ } \mbox{ }
\sigma^{\bar{3}} =
\left(
\begin{array}{ll}
1 & 0 \\ 0 & -1
\end{array}
\right)
\ee 
are the Pauli matrices.  
\be
\mbox{ }
\mbox{ The Dirac equation is then } 
\mbox{ }
i(-\pa_{\bar{0}}      +    \sigma^{\bar{a}}\pa_{\bar{a}})\psi_{\mbox{\scriptsize D\normalsize}} 
= m_{\psi}\psi_{\mbox{\scriptsize L\normalsize}}
\mbox{ } \mbox{ } , 
\mbox{ } \mbox{ }
i(-\pa_{\bar{0}}      -    \sigma^{\bar{a}}\pa_{\bar{a}})\psi_{\mbox{\scriptsize L\normalsize}} 
= m_{\psi}\psi_{\mbox{\scriptsize D\normalsize}} 
\mbox{ } .
\mbox{\hspace{0.1in}} 
\ee 
Clearly if $m_{\psi} = 0$, this decouples, into two 2-component \it Weyl equations\normalfont:
\be
i(-\pa_{\bar{0}}      +    \sigma^{\bar{a}}\pa_{\bar{a}})\psi_{\mbox{\scriptsize D\normalsize}} 
= 0
\mbox{ } \mbox{ } , 
\mbox{ } \mbox{ }
i(-\pa_{\bar{0}}      -    \sigma^{\bar{a}}\pa_{\bar{a}})\psi_{\mbox{\scriptsize L\normalsize}} 
= 0 
\mbox{ } .
\ee
The Dirac equation gives a highly successful theory of the electron (and of the predicted and 
successfully observed positron), whereas the L-Weyl equation gives a theory of the neutrino.  
Note that I am treating these things classically here.  
They are usually treated quantum-mechanically.   

I next provide 2 forms of Dirac theory Lagrangian.    
Defining $\bar{\psi} = \psi^{\dagger}\gamma^{\bar{0}}$,   
\be
\bar{\mbox{\sffamily L\normalfont}}^{\psi,\bar{\psi}}_{\mbox{\scriptsize Dirac\normalsize}} 
= i\bar{\psi}\gamma^{\bar{A}}\pa_{\bar{A}}\psi - m_{\psi}\bar{\psi}\psi  = 
\frac{1}{2}
\left(
\bar{\psi}\gamma^{\bar{A}}\pa_{\bar{A}}\psi
- \pa_{\bar{A}}\bar{\psi}\gamma^{\bar{A}}\psi 
\right)
- m_{\psi}\bar{\psi}\psi
\label{Diraclag2} 
\mbox{ } .
\ee
\be 
\mbox{The conjugate momentum of Dirac theory is curious: }
\mbox{\hspace{0.6in}} 
\pi_{\psi} \equiv \frac{\pa\mbox{\sffamily L\normalfont}}{\pa\dot{\psi}} 
= -i\psi^{\dagger} 
\mbox{ } . 
\mbox{\hspace{1in}} 
\label{Diracmom}
\ee
\be
\mbox{The Dirac Hamiltonian is then }
\mbox{\hspace{0.8in}}  
\bar{\mbox{\sffamily H\normalfont}}_{\mbox{\scriptsize Dirac\normalsize}} = 
- i\psi^{\dagger}\mbox{\b{$\gamma$}}\cdot\mbox{\b{$\pa$}}\psi + m\bar{\psi}\psi \mbox{ } ( = i\psi^{\dagger}\dot{\psi}) 
\mbox{ } .
\label{Diracham}
\mbox{\hspace{1in}} 
\ee

Finally, {\it Yukawa theory} (which may be regarded as a toy, a failed theory of the nuclear forces 
or how the electron may have gained its mass from a Higgs scalar) follows from   
\be
\bar{\mbox{\sffamily L\normalfont}}^{\psi, \bar{\psi}, \varsigma}_{\mbox{\scriptsize Yukawa\normalsize}} 
= \bar{\mbox{\sffamily L\normalfont}}_{\mbox{\scriptsize KG\normalsize}}^{\varsigma}
+ \bar{\mbox{\sffamily L\normalfont}}_{\mbox{\scriptsize D\normalsize}}^{\psi, \bar{\psi}} 
- \mbox{\tt g\normalfont}_{\mbox{\scriptsize Y\normalsize}}\bar{\psi}\psi\varsigma \mbox{ } .
\label{Yukawa}
\ee
Note that the interaction term does not disrupt the classical canonical structure.  

\mbox{ }

\noindent \bf{1.7.2 Electromagnetism, U(1) scalar gauge theory and QED}\normalfont

\mbox{ }

\noindent
Pure electromagnetism (no sources) follows from the Lagrangian 
\be
\bar{\mbox{\sffamily L\normalfont}}_{\mbox{\scriptsize em\normalfont}}^{\mbox{\scriptsize A\normalsize}} = 
- \frac{1}{4}\mbox{\v{F}}\circ \mbox{\v{F}} = - \frac{1}{2}[(\dot{\mbox{\b{A}}} + \mbox{\b{$\pa$}}\Phi)^2  - B^2] \mbox{ } .
\label{emlag}
\ee
\be 
\mbox{The conjugate momentum is } 
\mbox{\hspace{1.3in}} 
\pi^{\bar{a}} \equiv \frac{\pa\bar{\mbox{\sffamily L\normalfont}}}{\pa\dot{A}_{\bar{a}}} 
= - \dot{A}^{\bar{a}} - \pa^{\bar{a}}\Phi = E^{\bar{a}} 
\mbox{ } . 
\mbox{\hspace{2in}} 
\label{emmom}
\ee
\be
\mbox{The Hamiltonian is then } 
\mbox{\hspace{1.2in}}
\mbox{\sffamily H\normalfont}_{\mbox{\scriptsize em\normalfont}} 
= -\frac{1}{2}(\pi^2 + B^2) + \Phi\mbox{\b{$\pa$}}\cdot\mbox{\b{$\pi$}}
\mbox{\hspace{2in}} 
\label{emham} 
\ee
up to a total divergence.  The last term has the form of an appended constraint.  

\be
\mbox{ } \mbox{ A common gauge choice is 
the Lorenz gauge} \mbox{ } \pa_{\bar{A}}\check{A}^{\bar{A}} = 0 \mbox{ so } (\ref{Amaxfull}) 
\mbox{ in vacuo} \mbox{ } \Rightarrow \mbox{ } \Box \check{A}^{\bar{A}} = 0,
\label{Lorgau}
\ee

I now consider coupling the theories of the previous section to the electromagnetic field.  
The usual treatment is to consider a complex scalar field, note that its Lagrangian is 
\be
\mbox{globally U(1) symmetric, i.e invariant under } 
\mbox{\hspace{0.7in}} 
\varsigma \longrightarrow e^{ie\Lambda}\varsigma \mbox{ } , \mbox{ }  \Lambda = \mbox{const } , 
\mbox{\hspace{2in}} 
\label{U1gauxform}
\ee
and demand that the symmetry also holds locally i.e for $\Lambda = \Lambda(x)$.  
Then one finds that $\pa_A\varsigma$ would contain an extra unwanted term.  
Just as on 
p 8, one introduces a new term whose transformation properties compensate 
for the unwanted term.  This is \it gauging\normalfont, a procedure originally due to Weyl 
\cite{CGWeylGauge}.  Again the introduced term is a connection,
whereby one 
\be 
\mbox{forms a covariant derivative } 
\mbox{\hspace{0.5in}} 
\bar{\nabla}^{\mbox{\scriptsize e\normalsize}}_{\bar{A}}\varsigma 
= \pa_{\bar{A}}\varsigma + ie\check{A}_{\bar{A}}\varsigma \mbox{ } \mbox{ } , \mbox{ } \mbox{ } 
\bar{\nabla}^{\mbox{\scriptsize e\normalsize}}_{\bar{A}}\varsigma^* 
= \pa_{\bar{A}}\varsigma^* - ie\check{A}_{\bar{A}}\varsigma^* \mbox{ } , 
\mbox{\hspace{0.5in}}
\label{U1covderiv}
\ee
Then in place of the Lagrangian of complex Klein--Gordon theory, one arrives instead at 
\be
\bar{\mbox{\sffamily L\normalfont}}^{\varsigma, \varsigma^*}_{\mbox{\scriptsize U(1) gauged\normalsize}}  
= -\bar{\nabla}^{\mbox{\scriptsize e\normalsize}}_{\bar{A}}\varsigma
\bar{\nabla}^{\mbox{\scriptsize e\normalsize}\bar{A}}\varsigma^* + m_{\varsigma}\varsigma\varsigma^* \mbox{ } .        
\label{U1gauKG}
\ee
One then requires $A_{\bar{A}}$ to be a dynamical field.  Standardly, this is taken to be 
the field uniquely selected by Lorentz-, gauge- and parity-invariance plus the flat-space QFT restriction 
on higher-order interaction terms (explained in I.3).  
This is of the same form as (and is furthermore identified with) the electromagnetic potential.  
So the resulting U(1) scalar 
\be
\mbox{gauge theory Lagrangian is } 
\mbox{\hspace{1.2in}}
\bar{\mbox{\sffamily L\normalfont}}^{\varsigma, \varsigma^*, \mbox{\scriptsize A\normalsize}}_{\mbox{\scriptsize U(1)\normalsize}} 
= \bar{\mbox{\sffamily L\normalfont}}^{\varsigma, \varsigma^*}_{\mbox{\scriptsize U(1) gauged\normalsize}}  
+ \bar{\mbox{\sffamily L\normalfont}}^{\mbox{\scriptsize A\normalsize}}_{\mbox{\scriptsize em\normalsize}} 
\mbox{ } .
\label{U1gautheo} 
\mbox{\hspace{1.2in}} 
\ee
\be
\mbox{ }\mbox{ For spin-$\frac{1}{2}$ fermions, there is a global U(1) invariance under }
\psi \longrightarrow e^{ie\Lambda} \psi 
\mbox{ } \mbox{ } , \mbox{ } \mbox{ } \Lambda = \mbox{const } . 
\mbox{\hspace{0.3in}}
\ee 
Gauging this, the Dirac Lagrangian may be replaced by 
\be
\bar{\mbox{\sffamily L\normalfont}}^{\psi, \bar{\psi}}_{\mbox{\scriptsize U(1) gauged\normalsize}} = 
i\bar{\psi}\gamma^{\bar{A}}\bar{\nabla}^{\mbox{\scriptsize e\normalsize}}_{\bar{A}}
\psi - m_{\psi}\bar{\psi}\psi  = 
\frac{1}{2}
\left(
\bar{\psi}\gamma^{\bar{A}}\bar{\nabla}^{\mbox{\scriptsize e\normalsize}}_{\bar{A}}
\psi - \bar{\nabla}^{\mbox{\scriptsize e\normalsize}}_{\bar{A}}\bar{\psi}\gamma^{\bar{A}}\psi 
\right)
- m_{\psi}\bar{\psi}\psi \mbox{ } .
\label{U1gauDirac}
\ee
Adjoining this to the electromagnetic Lagrangian to make $A_{\bar{A}}$ into a dynamical field, 
one has the classical theory whose QM counterpart is 
\it quantum electrodynamics (QED)\normalfont:  
\be
\bar{\mbox{\sffamily L\normalfont}}^{\psi, \bar{\psi}, \mbox{\scriptsize A\normalsize}}_{\mbox{\scriptsize QED\normalsize}} =
\bar{\mbox{\sffamily L\normalfont}}^{\psi, \bar{\psi}}_{\mbox{\scriptsize U(1) gauge\normalsize}} + 
\bar{\mbox{\sffamily L\normalfont}}^{\mbox{\scriptsize A\normalsize}}_{\mbox{\scriptsize em\normalsize}} \mbox{ } .
\label{QEDlag}
\ee

Note that the classical canonical structure of the free theories does not significantly differ 
from that of the coupled, interacting theories.  Also, in the above theories the 
electromagnetic vacuum laws are replaced by non-vacuum laws  
with fundamental source terms 
\be
\mbox{ } \mbox{ } \mbox{ } j_{\varsigma}^{\bar{A}} \equiv ie
\varsigma 
\stackrel{\longleftrightarrow}{    {\bar{\nabla}}^{\mbox{\scriptsize G\normalsize}\bar{A}}    }
\varsigma^*
\mbox{ } \mbox{ } , 
\mbox{ } \mbox{ }
j_{\psi}^{\bar{A}} \equiv e\bar{\psi}\gamma^{\bar{A}}\psi 
\mbox{ } ,
\ee
where the big double arrow indicates the derivative acting to the right minus the derivative 
acting to the left.  

QED is regarded as a highly successful theory: it predicted the currently-observed 
anomalous values of the magnetic moment and charge radius of the electron, 
as well as the small Lamb and Uehling shifts in spectral lines \cite{Weinberg}.  

\mbox{ }

\noindent \bf{1.7.3 Yang--Mills theory, QCD and Weinberg--Salam theory}\normalfont

\mbox{ } 

\noindent
There remains the issue of the powerful but apparently short-range nuclear forces.  
The Yang--Mills idea was to consider gauging with respect to a larger gauge group G 
(they considered 

\noindent SU(2) \cite{YM}, 
but it was soon generalized \cite{Utiyama} to a large class\fn{This class was precisely 
determined by Gell-Mann and Glashow, see App IV.B.} of gauge groups\fn{The gauge groups 
considered are {\it internal}, as opposed to the space(time) groups of I.0.1.  All these 
groups are {\it Lie groups}, i.e simultaneously groups, topological spaces and analytic 
manifolds.  Near the identity they may be studied via the corresponding {\it Lie algebra} \sc g\normalfont.  
This is an algebra with a product $|[\mbox{ },\mbox{ }]| : {\mbox{\sc g} } \times {\mbox{\sc g} } 
\longrightarrow {\mbox{\sc g}} $
that is bilinear, antisymmetric and obeys the {\it Jacobi identity}
\be
|[g_1 , |[g_2 , g_3]| ]| + |[g_2 , |[g_3 , g_1]| ]|  + |[g_3 , |[g_1 , g_2]| ]| = 0 
\ee
$\forall \mbox{ }$ $g_1$, $g_2$, $g_3$ $\in$ {\sc g}.
The $\tau$ are generators.  I usually use [i]-indices over these, but in the particular case of internal gauge groups, I use bold Capital indices.  
The $Q$'s are auxiliaries parameterizing the amount of each generator.  
$|[ \tau_{\mbox{\tiny\bf A\normalfont\normalsize}} , 
    \tau_{\mbox{\tiny\bf B\normalfont\normalsize}} ]| = 
i{f^{\mbox{\tiny\bf C\normalfont\normalsize}}}_{\mbox{\tiny\bf AB\normalfont\normalsize}}$, 
where ${f^{\mbox{\bf\tiny C\normalsize\normalfont}}}_{\mbox{\bf\tiny AB\normalsize\normalfont}}$ 
are the \it{structure constants}\normalfont.  From the properties of 
$|[\mbox{ },\mbox{ }]|$, these obey 
$$
f_{{\mbox{\bf\tiny IJK\normalsize\normalfont}}} = f_{{\mbox{\bf\tiny I\normalsize\normalfont}}[{\mbox{\bf\tiny JK\normalsize\normalfont}}]} 
\mbox{ (antisymmetry) } , 
$$
\be
{f^{\mbox{\bf\tiny I\normalsize\normalfont}}}_{\mbox{\bf\tiny JK\normalsize\normalfont}}
f_{\mbox{\bf\tiny ILM\normalsize\normalfont}} + 
{f^{\mbox{\bf\tiny I\normalsize\normalfont}}}_{\mbox{\bf\tiny JM\normalsize\normalfont}}
f_{{\mbox{\bf\tiny IKL\normalsize\normalfont}}} + 
{f^{\mbox{\bf\tiny I\normalsize\normalfont}}}_{\mbox{\bf\tiny JL\normalsize\normalfont}}
f_{\mbox{\bf\tiny IMK\normalsize\normalfont}} = 0 \mbox{ }                       \mbox{ (Jacobi identity) }.
\label{firstJac}
\ee  }).   

Consider a multiplet of scalar fields $\varsigma$ with arbitrary potential $V(\varsigma)$ 
with a Lagrangian 
\be
\mbox{invariant under the global symmetry }
\mbox{\hspace{1.0in}} 
\varsigma \longrightarrow e^{iQ_{\mbox{\tiny\bf A\normalfont\normalsize}}\tau^{\mbox{\tiny\bf A\normalfont\normalsize}}}\varsigma 
\mbox{ } , \mbox{ } Q_{\mbox{\scriptsize\bf A\normalfont\normalsize}} \mbox{ const}
\mbox{\hspace{2.5in}}
\ee
and demand that the symmetry also holds locally i.e for 
$Q_{\mbox{\scriptsize\bf A\normalfont\normalsize}} = Q_{\mbox{\scriptsize\bf A\normalfont\normalsize}}(x)$.  
Then one finds that $\pa_A\varsigma$ would contain an extra unwanted term.  
Thus one requires to include a compensatory 
\be
\mbox{connection obeying }
\mbox{\hspace{1.2in}} 
\check{A}_{\bar{A}}^{\mbox{\scriptsize\bf A\normalfont\normalsize}}
\tau_{\mbox{\scriptsize\bf A\normalfont\normalsize}} 
\longrightarrow g\check{A}_{\bar{A}}^{\mbox{\scriptsize\bf A\normalfont\normalsize}}
\tau_{\mbox{\scriptsize\bf A\normalfont\normalsize}} g^{-1} 
- (\pa_{\bar{A}}g)g^{-1} \mbox{ } ,    
\label{YMgauxform}
\mbox{\hspace{2.5in}}
\ee
\be
\mbox{thus leading to the formation of a covariant derivative }
\mbox{\hspace{0.4in}} 
\bar{\nabla}^{\mbox{\scriptsize G\normalsize}}_{\bar{A}}\varsigma  
\equiv \pa_{\bar{A}}\varsigma + i\mbox{\sffamily g\normalfont}_{\mbox{\scriptsize c\normalsize}}
\check{A}_{\bar{A}}^{\mbox{\scriptsize\bf A\normalfont\normalsize}}
\tau_{\mbox{\scriptsize\bf A\normalfont\normalsize}}\varsigma  
\mbox{ } . 
\mbox{\hspace{2.5in}}
\label{Gcovderiv}
\ee
Thus one has passed from the $\varsigma$-Lagrangian to a 
G-gauged $\varsigma$-Lagrangian.  
\be
\bar{\mbox{\sffamily L\normalfont}}^{\varsigma, \varsigma^*}_{\mbox{\scriptsize G gauged\normalsize}}  
= -\sum_{\varsigma \mbox{\scriptsize -multiplet\normalsize}}\bar{\nabla}^{\mbox{\scriptsize G\normalsize}}_{\bar{A}}\varsigma
\bar{\nabla}^{\mbox{\scriptsize G\normalsize}\bar{A}}\varsigma + V(\varsigma) \mbox{ } .
\label{Ggauscalar}
\ee
One may similarly pass from the Dirac Lagrangian to  
\be
\bar{\mbox{\sffamily L\normalfont}}^{\bar{\psi}, \psi}_{\mbox{\scriptsize G gauged\normalsize}} = 
\bar{\psi}\gamma^{\bar{A}}
\bar{\nabla}^{\mbox{\scriptsize G\normalsize}}_{\bar{A}}\psi 
 - m_{\psi}\bar{\psi}\psi 
  = 
\frac{1}{2}
\left(
\bar{\psi}\gamma^{\bar{A}}
\bar{\nabla}^{\mbox{\scriptsize G\normalsize}}_{\bar{A}}\psi 
- (\nabla^{\mbox{\scriptsize G\normalsize}}_{\bar{A}}\bar{\psi})\gamma^{\bar{A}}\psi
\right) 
- m_{\psi}\bar{\psi}\psi  
\mbox{ } . 
\label{flatYMDlag}
\ee

One then requires $A^{\mbox{\bf\scriptsize J\normalsize\normalfont}}_{\bar{A}}$ to be a dynamical field.  Again, by   
Lorentz-, gauge-, parity invariance and the QM argument which cuts down on higher-order interactions, 
one arrives at the pure Yang--Mills theory Lagrangian\fn{This is self interacting theory of spin 1 bosons alone rather 
than Yang--Mills gauge theory, taken to mean 
Yang--Mills--`scalar and (or) Dirac' theory.} 
\be 
\bar{\mbox{\sffamily L\normalfont}}^{A_{\mbox{\bf\tiny I\normalsize\normalfont}}}_{\mbox{\scriptsize YM(G)\normalsize}} = 
- \frac{1}{4}\mbox{\v{F}}_{\mbox{\scriptsize\bf I\normalfont\normalsize}}\circ \mbox{\v{F}}^{\mbox{\scriptsize\bf I\normalfont\normalsize}} 
\label{flatYMlag}
\ee
where the Yang--Mills field tensor is $\mbox{\v{F}}_{\mbox{\scriptsize\bf I\normalfont\normalsize}\bar{A}\bar{B}} = 
\pa_{\bar{B}}\check{A}_{\mbox{\scriptsize\bf I\normalfont\normalsize}\bar{A}} - \pa_{\bar{A}}\check{A}_{\mbox{\scriptsize\bf I\normalfont\normalsize}\bar{B}} 
+ \mbox{\sffamily g\normalfont}_{\mbox{\scriptsize c\normalsize}}|[ \check{A}_{\bar{A}}, \check{A}_{\bar{B}} ]|_{\mbox{\scriptsize\bf I\normalfont\normalsize}}$. 
$\mbox{\sffamily g\normalfont}_{\mbox{\scriptsize c\normalfont}}$ is the 
{\it coupling constant } of 
the theory. 

The pure Yang--Mills field equations are 
\be
\bar{\nabla}^{\mbox{\scriptsize G\normalsize}}_{\bar{A}} 
\bar{\nabla}^{\mbox{\scriptsize G\normalsize}
[\bar{A}}\check{A}^{\bar{B}]}_{\mbox{\bf\scriptsize I\normalsize\normalfont}} = 0  
\mbox{ } \mbox{ i.e } \mbox{ } 
\pa_{\bar{A}}\pa^{[\bar{A}} \check{A}^{\bar{B}]}_{\mbox{\bf\scriptsize I\normalsize\normalfont}}
= -\mbox{\sffamily g\normalfont}_{\mbox{\scriptsize c\normalfont}}
f_{{\mbox{\bf\scriptsize IJK\normalsize\normalfont}}}
\check{A}^{\mbox{\bf\scriptsize J\normalsize\normalfont}}_{\bar{A}}
\check{F}^{{\mbox{\bf\scriptsize K\normalsize\normalfont}}\bar{A}\bar{B}} 
\mbox{ } ,
\label{flatYM4}
\ee
so one can see this as a bigger, self-sourcing and hence nonlinear version of 
electromagnetism, corresponding to messenger particles which, unlike the photon, 
are themselves carriers of the theory's charge.  
The 3 + 1 split is
\be
\bar{\mbox{\sffamily L\normalfont}}^{\mbox{\scriptsize A\normalfont}
_{\mbox{\tiny\bf I\normalfont\normalsize}}}_{3+1(\mbox{\scriptsize YM(G)\normalsize})} 
= - \frac{1}{4}F_{\mbox{\bf\scriptsize I\normalsize\normalfont}}
\circ 
F^{\mbox{\bf\scriptsize I\normalsize\normalfont}} + \frac{1}{2}(\pa_{\bar{0}}
A_{\mbox{\bf\scriptsize I\normalsize\normalfont}\bar{a}} - \pa_{\bar{a}}
A_{\mbox{\bf\scriptsize I\normalsize\normalfont}\bar{0}} 
+ |[A_{\bar{a}}, A_{\bar{0}} ]|_{\mbox{\bf\scriptsize I\normalsize\normalfont}})(\pa^{\bar{0}}
A^{\mbox{\bf\scriptsize I\normalsize\normalfont}\bar{a}} - \pa^{\bar{a}}
A^{\mbox{\bf\scriptsize I\normalsize\normalfont}\bar{0}} 
+ |[A^{\bar{a}}, A^{\bar{0}} ]|^{\mbox{\bf\scriptsize I\normalsize\normalfont}}) \mbox{ } ,
\ee
\be
{\cal G}_{\mbox{\bf\scriptsize J\normalsize\normalfont}} \equiv D_{\bar{a}}A^{\bar{a}}_{\mbox{\bf\scriptsize J\normalsize\normalfont}} = 
\pa_{\bar{a}}A^{\bar{a}}_{\mbox{\bf\scriptsize J\normalsize\normalfont}} 
- \mbox{\sffamily g\normalfont}_{\mbox{\scriptsize c\normalfont}} f_{\mbox{\bf\scriptsize IJK\normalsize\normalfont}}
A^{\mbox{\bf\scriptsize K\normalsize\normalfont}}_{\bar{a}}\pi^{\bar{a}}_{\mbox{\bf\scriptsize I\normalsize\normalfont}}    = 0 \mbox{ } ,
\label{flatYMgauss}
\ee
\be
\bar{D}_{\bar{a}}^{\mbox{\scriptsize G\normalsize}}\bar{D}^{\mbox{\scriptsize G\normalsize}[\bar{a}}
A^{\bar{b}]}_{\mbox{\bf\scriptsize I\normalsize\normalfont}} 
= \frac{\pa}{\pa t}(\dot{A}^{\bar{b}}_{\mbox{\bf\scriptsize I\normalsize\normalfont}} - 
\pa^{\bar{b}}\Phi_{\mbox{\bf\scriptsize I\normalsize\normalfont}})
\mbox{ } . 
\label{flatYMevol}
\ee
\be 
\mbox{ }\mbox{ The conjugate momenta are }  
\mbox{\hspace{1.6in}} 
\pi_{\mbox{\scriptsize\bf I\normalfont\normalsize}}^{\bar{a}}  \equiv 
\frac{\pa\bar{\mbox{\sffamily L\normalfont}}}{\pa \dot{A}_{\bar{a}}^{\mbox{\scriptsize\bf I\normalfont\normalsize}} } \mbox{ } ,
\label{flatYMmom}
\mbox{\hspace{3in}} 
\ee
\be 
\mbox{and the Yang--Mills Hamiltonian is } 
\mbox{\hspace{1in}} 
\bar{\mbox{\sffamily H\normalfont}}^{A_{\mbox{\bf\scriptsize I\normalsize\normalfont}}}_{\mbox{\scriptsize YM\normalsize}} 
= \pi_{\mbox{\bf\scriptsize I\normalsize\normalfont}}^{\bar{a}}\pi^{\mbox{\bf\scriptsize I\normalsize\normalfont}}_{\bar{a}} 
+ \frac{1}{4} F_{\mbox{\bf\scriptsize I\normalsize\normalfont}} \circ F^{\mbox{\bf\scriptsize I\normalsize\normalfont}} 
+ A_{\bar{0}}^{\mbox{\bf\scriptsize J\normalsize\normalfont}}{\cal G}_{\mbox{\bf\scriptsize J\normalsize\normalfont}} \mbox{ } .
\label{flatYMham}
\mbox{\hspace{2in}} 
\ee
\be
\mbox{ } 
\mbox{ One then has }
\mbox{\hspace{1in}}
\bar{\mbox{\sffamily L\normalfont}}^{\varsigma\mbox{\scriptsize -multiplet\normalsize}, \mbox{\scriptsize A\normalsize}_{\mbox{\tiny\bf I\normalfont\normalsize}}}_{\mbox{\scriptsize G\normalsize}} =
\bar{\mbox{\sffamily L\normalfont}}^{\varsigma\mbox{\scriptsize -multiplet\normalsize}}_{\mbox{\scriptsize G gauged\normalsize}} + 
\bar{\mbox{\sffamily L\normalfont}}_{\mbox{\scriptsize YM(G)\normalsize}}^{\mbox{\scriptsize A\normalsize}_{\mbox{\tiny\bf I\normalfont\normalsize}}} \mbox{ } ,
\mbox{\hspace{2in}} 
\label{Varda}
\ee
\be
\bar{\mbox{\sffamily L\normalfont}}^{\bar{\psi}, \psi, \mbox{\scriptsize A\normalsize}_{\mbox{\tiny\bf I\normalfont\normalsize}}}_{\mbox{\scriptsize G \normalsize}} = 
\bar{\mbox{\sffamily L\normalfont}}^{\bar{\psi}, \psi}_{\mbox{\scriptsize G-gauged\normalsize}} + 
\bar{\mbox{\sffamily L\normalfont}}_{\mbox{\scriptsize YM(G)\normalsize}}^{\mbox{\scriptsize A\normalsize}_{\mbox{\tiny\bf I\normalfont\normalsize}}} 
\mbox{ } .  
\label{Elbereth}
\ee
In these, additionally to the self-sourcing in (\ref{flatYM4}), there are source currents similar to those of the previous subsection.    

\mbox{ }

Yang--Mills gauge theory was originally ridiculed as a theory for the nuclear forces, for surely 
such short range forces would be associated with massive messengers.   However, by a distinct 
argument in each case, this difficulty has been overcome to form successful theories for the weak 
and strong nuclear forces.  

For the weak force, the idea is Higgs' \it local spontaneous symmetry breaking \normalfont \cite{Higgs, Weinberg}. 
The messenger bosons may indeed have mass, since they may acquire 
it from as-yet experimentally hypothetical Higgs scalar fields.  
This study requires the Yang--Mills--scalar gauge theory.

The weak nuclear force accounts for observations of parity violation.  The accepted theory for 
this force is contained within the {\it Weinberg--Salam unified theory of electroweak 
interactions}, which is based on the spontaneously-broken SU(2) $\times$ U(1) Yang--Mills  
theory.  The theory's Lagrangian is a complicated combination of Yang--Mills and appropriately 
gauge-corrected Dirac and Weyl pieces together with multiplets of Higgs scalars to break the 
symmetry, in which the left and right multiplets are treated differently.  This is the theory 
of 3 messenger bosons ($W^+$, $W^-$ and $Z^0$)
$^{21}$ in addition to the photon, and of electrons, 
positrons and neutrinos.  Weinberg--Salam theory explained a number of nuclear reactions, 
and correctly predicted the existence of (and masses for) $W^+$, $W^-$ and $Z^0$ bosons and  
effects due to the neutral currents associated with $Z^0$.

The strong force is not accounted for by spontaneous symmetry breaking, but rather by the idea 
of {\it confinement}: that the coupling constant rises over long distances.  There is evidence for the strong force 
having 3 charge types from scattering cross-sections.  This led to postulating an SU(3) theory,$^{21}$ 
\it quantum chromodynamics (QCD)\normalfont, which in turn led to the eightfold way\fn{SU(n) 
has dimension $n$ and $n^2$ - 1 independent generators.} `periodic table' of nuclear physics,  a pattern partly explaining the known spectrum of 
hadrons and partly correctly predicting hadrons not yet known at the time.  
The charge is called {\it colour}, the messengers are the 8 coloured {\it gluons}, and the hadrons are composed of coloured, 
fractionally-electromagnetically-charged particles called {\it quarks}.  
Confinement, which follows from a theoretical property of QCD called \it asymptotic 
freedom\normalfont, accounts for quarks and gluons not being observed.   But jets in 
high-energy collisions are regarded as indirect evidence for quarks.  QCD also explains some 
low-energy physics of pions and nucleons [but this has the U(1) Problem, see IV.1.3.3].  

The {\it Standard Model} (SM) of particle physics is the SU(3) $\times$ SU(2) $\times$ U(1) collection 
of the above, repeated 3 times over [due to the known particles of nature curiously and 
unexplainedly apppearing to belong to 3 increasingly more massive copies (generations) 
of an otherwise identical set of particles].

\section{On the split formulation of Einstein's field equations}

\subsection{Geometry of Hypersurfaces}

\noindent \bf{2.1.1 Extrinsic curvature} \normalfont

\mbox{ }

\noindent Consider 2 nearby points $p_1$ and $p_2$ lying on a curve in $\Re^2$.  
The arclength between these is $\mbox{\scriptsize$\Delta$\normalsize}s 
= \sqrt{\mbox{\scriptsize $\Delta$\normalsize}x^2 + \mbox{\scriptsize $\Delta$\normalsize} y^2}$.  
Let the tangents at $p_1$  and $p_2$ be $t_1$ and $t_2$ and the angle between them 
$$
\mbox{be $\mbox{\scriptsize $\Delta$\normalsize} \psi$.  
Then another type of curvature at $p_1$ is given by }
\mbox{\hspace{0.1in}}
\kappa = 
\left. 
\left.
\begin{array}{c} 
\mbox{lim} \\ 
\mbox{\scriptsize$\Delta$\normalsize} s \longrightarrow 0
\end{array}
 \frac{\mbox{\scriptsize$\Delta$\normalsize} \psi}{\mbox{\scriptsize$\Delta$\normalsize} s}
\right|
_{p_1} = \frac{d \psi}{ds}
\right|
_{p_1}.
$$
Note that this can also be interpreted as the rate of change of the normal, as used instead 
in the modern generalized definition below.  
Next, if the curve is in $\Re^3$, then near each $p_1$ the curve lies within a plane.  
Then use the above notion of curvature.  
This notion of curvature is \it extrinsic\normalfont, i.e referring to 
how the curve is bent relative to its surrounding $\Re^3$ exterior.  
To generalize this notion of extrinsic curvature from curves to 2-surfaces within Euclidean 
3-space $\Re^3$,  draw each plane through the point $p$ in turn.  
In each plane the 2-surface traces a curve.  Apply the 2-d notion above to determine the 
curvature.  The maximum and minimum values thus found are the 
\it principal curvatures \normalfont $\kappa_1$ and $\kappa_2$.  
Then one has the following measures of how the surface is bent within $\Re^3$:\fn{The overall 
numerical factors in these definitions are unimportant.  The mean curvature is often defined 
as a true mean i.e one half of the above. The forms given are the sum and product of the 
eigenvalues i.e the trace and determinant of the extrinsic curvature matrix.}  
\be
\mbox{ (mean curvature) } = \kappa_1 + \kappa_2 \mbox{ } ,
\label{meanK}
\ee
\be
\mbox{ (Gauss curvature) } = \kappa_1\kappa_2 \mbox{ } .  
\label{GaussK}
\ee

\noindent \bf{2.1.2 Gauss' outstanding theorem}\normalfont

\mbox{ }

\noindent Despite being defined in very different ways, it turns out that the intrinsic and extrinsic notions 
of curvature of a surface are related.  For a 2-surface embedded in $\Re^3$, the intrinsic curvature is in fact equal to 
(in the above convention twice) the Gauss curvature:
\be
R = 2\kappa_1\kappa_2,
\ee
which is \it Gauss' outstanding theorem\normalfont .  

\mbox{ }

\noindent \bf{2.1.3 Hypersurface geometry}\normalfont

\mbox{ }

\noindent I now generalize the above result.  First, allow the embedding space itself to also 
be curved.  Then the result is in fact about a projection of 
$0 = \check{R}_{ABCD}$, the flat-space Riemann tensor.  
Thus one is to set up a theory of projections.  Although for the above one can get away with 
less (the projections of $0 = \check{R}_{AB}$ suffice for 3-d), the second generalization is 
to arbitrary-d with the role of the surfaces being played by hypersurfaces of codimension 1, 
which generally requires the projections of the full Riemann tensor.    
The third generalization is to permit the embedded and embedding spaces to be of arbitrary 
signature.  I denote a manifold with $r$ independent spatial directions and $s$ independent 
timelike directions as an ($r$, $s$) space (of 
dimension $n = r + s$ and `signature' $s$) and denote the embedding itself by 
($r$, $s$; $\epsilon$) where the `added dimension' $\mu$ is spacelike for $\epsilon = 1$ and 
timelike for $\epsilon = -1$.

To set up the theory of projections,  
one takes the $(n + 1)$-d metric $g_{CD}$ and expands it in terms of a basis consisting 
of the normal to the hypersurface $n_A$ and the $n$ \it projectors \normalfont ${e^a}_A$: 
\be
g_{CD} = h_{CD} + \epsilon n_Cn_D 
\mbox{ } ,  
\ee
where the first term is the metric \it induced \normalfont by $g_{CD}$ on the hypersurface, 
$h_{CD} = g_{cd}{e^c}_C{e^d}_D$.  
I am interested in the study of projections onto the hypersurface and onto the normal.  
I use $\check{O}_{...a...} = \check{O}_{...A...}{e^A}_a$ to denote projections onto the ($r$, $s$) hypersurface $\upsilon$, and 

\noindent $\check{O}_{...\perp...} = \check{O}_{...A...}n^A$ to denote projections onto the normal.  
Then for the metric, 
\be
g_{\perp\perp} = \epsilon \mbox{ } , \mbox{ } \mbox{ } g_{a\perp} = 0 
\mbox{ } .
\ee

In general, the higher-d metric is split according to  
\be
\begin{array}{ll}
g_{CD} = \left( \begin{array}{ll} \mbox{ } \mbox{ } \mbox{ } \mbox{ } \beta_i\beta^i + \epsilon\alpha^2  & 
\beta_d \\ \mbox{ } \mbox{ } \mbox{ } \mbox{ } \beta_c &  h_{cd} \mbox{ } \mbox{ } \mbox{ } \mbox{ } \mbox{ } \end{array}\right)
\mbox{ }  , \mbox{  }  \mbox{ so that } \mbox{ }  
g^{CD} =  \left( \begin{array}{ll}  \epsilon\frac{1}{\alpha^2} & 
-\epsilon\frac{\beta^d}{\alpha^2}
\\ -\epsilon\frac{\beta^{c}}{\alpha^2} & h^{cd} + \epsilon 
\frac{\beta^c\beta^d}{\alpha^2}\\ \end{array}\right)
\end{array}.
\label{ADMs}
\ee
\begin{figure}[h] 
\centerline{\def\epsfsize#1#2{0.4#1}\epsffile{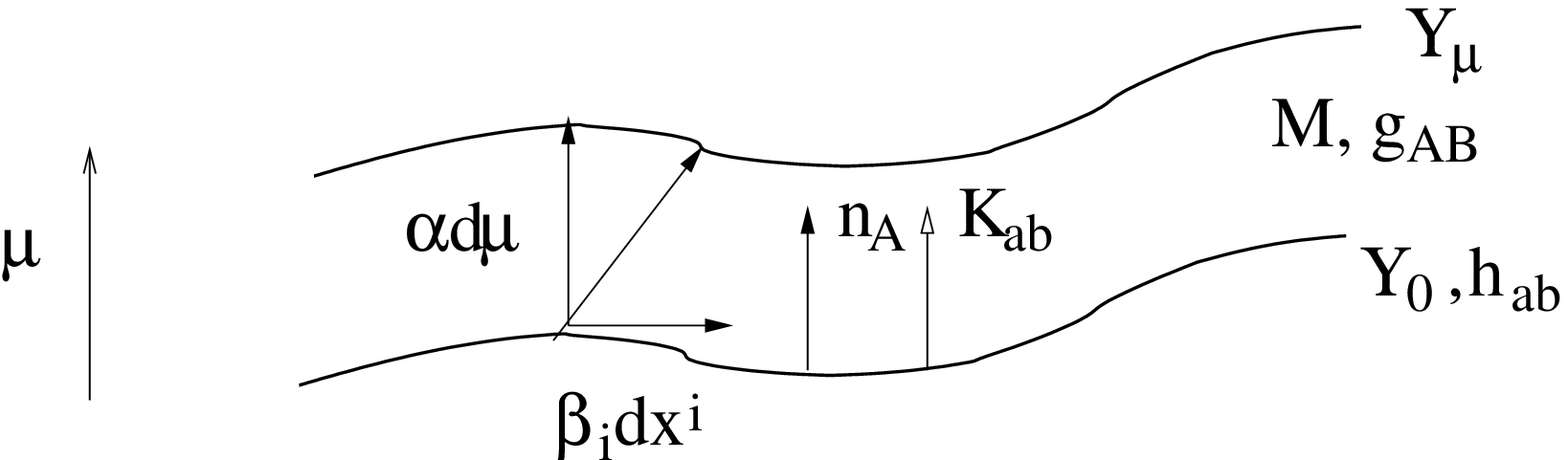}}
\caption[]{\label{TO4.ps}}
\end{figure}

\be
\mbox{The \it extrinsic curvature \normalfont is } 
\mbox{\hspace{1.2in}}
K_{ab} = -\nabla_a n_b  = -\frac{1}{2\alpha}\delta_{\check{\beta}}{h}_{ab} 
\mbox{ } ,  
\mbox{\hspace{1.5in}}
\label{ecd}
\ee
which is a measure in terms of the rate of change of the normals of how much the hypersurface is bent in the 
surrounding spacetime.\fn{The {\it hypersurface derivative} is 
$\delta_{\check{\beta}} \equiv \frac{\pa}{\pa\mu} - \pounds_{\beta}$.  Note that the correction 
to $\frac{\pa h_{ab}}{\pa\mu}$ is 
$\pounds_{\beta}h_{ab} = 2D_{(a}\beta_{b)} \equiv (|\mbox{K}\beta)_{ab}$ for  
$|\mbox{K}$ the {\it Killing form}.} The second form of (\ref{ecd}) shows that the extrinsic curvature is manifestly symmetric.  Its other notable 
property is that it is both a $n$- and ($n$+1)-d tensor since 
$K_{A\perp} = 0$.

The three projections of the Riemann tensor (see e.g \cite{Eisenhart26, KucharII}) 
are respectively the \it Gauss equation\normalfont , the \it Codazzi equation \normalfont 
and the \it Ricci equation\normalfont :  
\be
\check{R}_{abcd} = R_{abcd} - 2\epsilon K_{a[c}K_{d]b} 
\mbox{ } ,
\label{Gaussful}
\ee
\be
\check{R}_{\perp abc} =  -2\epsilon D_{[c}K_{b]a} 
\mbox{ } ,
\label{codful}
\ee
\be
\check{R}_{\perp a\perp b} =  \frac{1}{\alpha}(\delta_{\check{\beta}} K_{ab} - \epsilon D_bD_a\alpha) + {K_a}^cK_{cb} 
\mbox{ } .
\label{thirdproj}
\ee
From these it follows by one contraction that 
\be
\check{R}_{bd} - \epsilon \check{R}_{\perp b\perp d} =  R_{bd} - \epsilon(KK_{bd} - {K_{b}}^{c}K_{cd}) 
\mbox{ } ,
\label{contG}
\ee
\be
\check{G}_{a\perp} = \check{R}_{a\perp} = -{\epsilon}(D_b{K^b}_a - D_aK) 
\mbox{ } ,
\label{Gpa}
\ee
\be
\check{R}_{\perp\perp} =  \frac{\delta_{\check{\beta}} K - \epsilon D^2\alpha}{\alpha} - 
K\circ K 
\mbox{ } ,
\label{tpcont}
\ee 
and the Gauss equation can be contracted a second time to give  
\be
2\check{G}_{\perp\perp} = 2\check{R}_{\perp\perp} - \epsilon\check{R} = - \epsilon R + K^2 - K\circ K 
\mbox{ } ,
\label{Gpp}
\ee 
where I exhibit the useful relations to the projections of the Einstein 
tensor $\check{G}_{AB}$.

(\ref{Gpp}) is indeed Gauss' outstanding theorem for a 2-surface embedded in flat 3-space:
$$
R = K^2 - K\circ K = \left(\sum_{\Delta = 1, 2} \kappa_{\Delta}\right)^2 
- \sum_{\Delta = 1, 2}\kappa_{\Delta}^2  
= \kappa_1^2 + 2\kappa_1\kappa_2 + \kappa_2^2 - (\kappa_1^2 + \kappa_2^2) = 2\kappa_1\kappa_2 
\mbox{ } .
$$

The following geometrical identities are useful for GR.  
(\ref{thirdproj}) and (\ref{contG}) $\Rightarrow$ 
\be
\frac{1}{\alpha}(\delta_{\check{\beta}} {K}_{ab} - \epsilon D_bD_a\alpha) - KK_{ab} +2{K_{a}}^cK_{bc} + \epsilon R_{ab} = \epsilon\check{R}_{ab} 
\mbox{ } , 
\label{protoev}
\ee
which has the trace [a linear combination of (\ref{Gpp}) and (\ref{tpcont})]  
\be
\frac{1}{\alpha}(\delta_{\check{\beta}} {K} - \epsilon D^2\alpha) - K^2 + \epsilon R = \epsilon\check{R}  - \check{R}_{\perp\perp}
\label{prototrev} 
\mbox{ } .
\ee
$$
\mbox{Another linear combination of (\ref{Gpp}) and (\ref{tpcont}) is } 
\mbox{\hspace{0.2in}}
\frac{\delta_{\check{\beta}}K - \epsilon D^2\alpha}{\alpha} - \frac{K\circ K + K^2}{2} 
= \epsilon\frac{\check{R}}{2} 
\mbox{ } .
\mbox{\hspace{0.2in}}
$$
$$
\mbox{From this and (\ref{protoev}) it follows that }
\frac{1}{\alpha}
\left[
\epsilon(\delta_{\check{\beta}}K_{ab} - h_{ab}\delta_{\check{\beta}}K) - D_bD_a\alpha + h_{ab}D^2\alpha
\right]  
\mbox{\hspace{1in}} 
$$
\be
\mbox{\hspace{.5in}}
+ \epsilon
\left(
2{K_a}^cK_{bc} - KK_{ab} + \frac{K\circ K + K^2}{2}h_{ab}
\right)
+ G_{ab} = \check{G}_{ab}
\label{gid}
\ee
which forms the last projection of $\check{G}_{AB}$. 

The above equations (\ref{Gaussful}--\ref{Gpp}) are the starting point for embedding theorems with 
codimension 1.  If one sets all the higher-d curvature tensors to be zero in 
(\ref{Gaussful}--\ref{thirdproj}) one has the equations determining embeddability into flat manifolds.  
If the Ricci curvature is set to zero in (\ref{contG}--\ref{tpcont}) one has the equations determining 
embeddability into Ricci-flat (vacuum) manifolds.  Locally, whereas the former is not guaranteed, the 
latter is (assuming analyticity: the Campbell--Magaard result, see I.2.3 and B.1). 
The higher codimension generalization permits all spaces to be locally embeddable in higher-d flat 
manifolds.  Global results involve much harder mathematics and much larger lower bounds on the required 
codimension \cite{Greene}.

\subsection{Split of the EFE's with respect to nowhere-null hypersurfaces}

Consider slicing some region of spacetime into a sequence of hypersurfaces of codimension 1.  
The working is for general ($r$, $s$; $\epsilon$).  The hypersurfaces are held to be everywhere of fixed signature and are not allowed to be null.  
Whereas my main interest is in the 

\noindent(3, 0; --1) case i.e in foliating the usual 4-d spacetime with a 
sequence of 3-d spatial slices, I will also investigate the crucial role of these particular values of 
$s$ and $\epsilon$ (the overall dimension itself is of little consequence).  Also, the application to 
thin matter sheets (Part B) has been conventionally studied as the (2, 1; 1) case in the ordinary GR or as the 

\noindent (3, 1; 1) 
case in GR-based braneworld scenarios.  The framework of this section is broad enough to be of use for all these 
applications.  One should also note that whereas one can perform an infinity of different 
slicings\fn{By a slicing I mean a sequence of non-intersecting hypersurfaces that cover some 
spacetime region.  This is also known as a foliation.  GR happens to be a foliation-invariant theory 
as explained in (I.2.6). } in GR, particular slices and slicings are useful in numerical relativity 
(see I.2.9--11) and privileged slices and slicings occur 
both in some alternative theories in II.2 and as the thin matter sheets themselves in Part B.

I use the arbitrary ($r$, $s$; $\epsilon$) ``ADM--type'' split (\ref{ADMs})\fn{I use unnecessarily 
doubled quotation marks to cover either general-signature or specifically 
nonstandard-signature analogues of concepts usually associated with a standard signature.} 
of the higher-d metric $g_{CD}$ (\ref{ADMs}) with respect to the ($r$, $s$) `initial' 
hypersurface $\Upsilon_0$.  The following restrictions are associated with such splits.  
One requires the foliating hypersurfaces to be of a fixed topology.  One may require further 
restrictions on the spacetime to ensure good causal behaviour.  In the ($n$, 0; --1) case, note that the 
first of these prevents consideration of topology change, while the second takes the form that 
the spacetime is to be \it globally hyperbolic \normalfont \cite{HE, Wald}), although sometimes 
one can get away with working locally instead.      

In the ($n$, 0; --1) case, the foliation is to be interpreted in terms of a choice of time $t$ and 
associated timeflow $t^i$.  There are an  infinity of choices of such a $t$.  One usually 
demands the spacetime to be time-orientable so that it is always possible to consistently 
allocate notions of past and future.  For this ($n$, 0, --1) case $t$ is truly temporal.  
In the other cases, to start off with one can think of the sequence of hypersurfaces 
being parameterized by an ``independent dynamical variable" (IDV) $\mu$.     

In the ($n$, 0; --1) case, the lapse may be interpreted as the change in proper time 
$d\tau = \alpha(t, x^i)dt$, $\alpha = n^at_a$.  The interpretation of the shift is the 
displacement 
in identification of the spatial coordinates between 2 adjacent slices, $\beta^i = h^{ij}t_j$.   
Analogous notions can be defined for the general case.  

\mbox{ }

Applying an ``ADM--type"  
split of the metric splits the EFE's into two 
systems of equations to be studied as two separate steps.  In the ($n$, 0; --1) case these 
steps are the GR Initial Value problem (IVP) and the GR Cauchy problem (CP).  By {\sl problem} I 
mean a p.d.e system that holds in some region $\Omega$ together with prescribed data on (some 
portion of) the boundary $\pa \Omega$.

Here I consider the general ($r$, $s$; $\epsilon$) case with phenomenological matter.  One 
could instead have fundamental matter (see I.2.3.4).  In sufficiently complicated cases, the 
matter is governed by separate field equations which should then be adjoined to the EFE's to 
complete a coupled Einstein--Matter system.  The ``IVP'' system consists of $n$ + 1 
constraints (4 for the usual GR): the \it Gauss and Codazzi ``constraints"\normalfont 
\be
K^2 - K\circ K  - \epsilon R = 2\check{G}_{\perp\perp} = 2\check{T}_{\perp\perp} 
\equiv 2\rho,
\label{gauss}
\ee
\be
-\epsilon(D_b{K^b}_a - D_aK) = \check{G}_{a\perp} = \check{T}_{a\perp} \equiv j_a.
\label{cod}
\ee
These are obtained by use of the EFE's in (\ref{Gpp}) and (\ref{Gpa}).  They are ``constraints" because they contain none of the 
highest derivatives with respect to the IDV $\mu$.  Their solution involves only the ``initial hypersurface".  

The remaining $\frac{n(n + 1)}{2}$ equations (6 for the usual GR) are ``evolution" equations with respect to  
$\mu$ 
\be
\frac{1}{\alpha}(\delta_{\check{\beta}} {K}_{ab} - \epsilon D_bD_a\alpha) - KK_{ab} +2{K_{a}}^cK_{bc} + \epsilon R_{ab} = 
\epsilon\check{R}_{ab} = \epsilon
\left(
S_{ab} - \frac{S}{n - 1}h_{ab}
\right) - \frac{{\rho}}{n - 1}h_{ab} 
\mbox{ } ,
\label{evK}
\ee
obtained from the EFE's and (\ref{protoev}).  Here $\rho$, $j_a$ and $S_{ab} \equiv \check{T}_{ab}$ 
(N.B $S \neq \check{T}$) are general matter terms which are usually prescribed as 
functions of matter fields that are governed by usually-separate field equations.  
A useful equation is the trace of (\ref{evK})
\be
\delta_{\check{\beta}} K = \alpha
\left(
K^2 - \epsilon R - \frac{n\rho + \epsilon S}{n - 1}
\right) 
+ \epsilon D^2\alpha =
\alpha
\left[
K\circ K + \frac{(n-2)\rho - \epsilon S}{n - 1}
\right] 
+ \epsilon D^2\alpha 
\mbox{ } ,
\label{trevK}
\ee
the second equality of which follows from the Gauss constraint (\ref{gauss}).

An alternative `$G_{ab}$' rather than `$R_{ab}$' form for the evolution equations follows from 
the 
$$
\mbox{identity (\ref{gid}): } 
\mbox{\hspace{0.6in}}  
\frac{1}{\alpha}
\left[
\epsilon(\delta_{\check{\beta}}K_{ab} - h_{ab}\delta_{\check{\beta}}K) - D_bD_a\alpha + h_{ab}D^2\alpha
\right] 
\mbox{\hspace{1.4in}}
$$
\be
\mbox{\hspace{0.4in}}
+ \epsilon
\left(
2{K_a}^cK_{bc} - KK_{ab} + \frac{K\circ K + K^2}{2}h_{ab} 
\right)
+ G_{ab} = S_{ab} 
\mbox{ } .
\label{GevK}
\ee
Note how the form (\ref{evK}) has simpler gravity terms but more complicated matter terms whilst the 
opposite is true for (\ref{GevK}).
 
I often use the decomposition of symmetric second-rank tensors $\Sigma_{ab}$ into 
their \it trace part \normalfont
\be
\Sigma \equiv \Sigma_{ab}h^{ab}
\ee 
\be
\mbox{and their  \it tracefree part \normalfont  }
\mbox{\hspace{1.04in}}
\Sigma^{\mbox{\scriptsize T\normalsize}}_{ab} \equiv \Sigma_{ab} - \frac{\Sigma}{n}h_{ab}.
\mbox{\hspace{2.2in}}
\ee  
In particular if this split is applied to $K_{ab}$,  (\ref{gauss}) and (\ref{cod}) take the forms 
\be
K^{\mbox{\scriptsize T\normalsize}}\circ K^{\mbox{\scriptsize T\normalsize}} - \frac{n - 1}{n}K^2 + \epsilon R + 2\rho = 0 
\mbox{ } ,
\label{Agauss}
\ee
\be
D_b {K^{\mbox{\scriptsize T\normalsize}b}}_a - \frac{n - 1}{n}D_aK + \epsilon j_a = 0 
\mbox{ } ,
\label{Acod}
\ee
whilst the second form of (\ref{trevK}) may be rewritten as 
\be
\frac{1}{\alpha}(\delta_{\check{\beta}} K - \epsilon D^2\alpha) - \frac{K^2}{n} = 
K^{\mbox{\scriptsize T\normalsize}}\circ K^{\mbox{\scriptsize T\normalsize}} + \frac{(n-2)\rho - \epsilon S}{n - 1} 
\mbox{ } .
\label{protoRay}
\ee

\subsection{Analytic approach to the GR ``CP and IVP"}

\mbox{ }

\noindent{\bf 2.3.1 Cauchy problems \normalfont}

\mbox{ }

\noindent For a given d.e system of order $p$ to be solved in a spacetime region $\Omega$ 
for functions $\psi_{A}$, the {\it Cauchy problem} (CP) consists of the d.e system together with 
the $\psi_A$ and their time derivatives up to $p - 1$ th order on some initial hypersurface 
$\Sigma_0$.  It is required that the initial hypersurface be nowhere-characteristic i.e a 
genuine spacelike surface.  One could alternatively work with characteristic hypersurfaces 
(I.2.12.4).  
 
I consider CP's for a number of physically well-motivated systems, 
in order to isolate in simpler examples a number of features of the GR CP.  
These second-order systems recurringly provide a test-bed for GR ideas and 
all the field-theoretical examples will furthermore be coupled to GR.  I work from the 
viewpoint of the {\it analytic method}, stressing the 
importance of what input to prescribe and what output to expect from the 
equations of motion.  In other words, which p.d.e problems are meaningful for the equations 
that describe nature?   

In Newtonian mechanics, if one is prescribed the positions and velocities of all the particles 
at a given instant of time (the Cauchy data)  
then one can in principle predict\fn{Or retrodict -- all the accepted dynamical laws 
of physics are time-reversal invariant.} these at all future times by solving 
the second-order o.d.e that is Newton's second law (\ref{NII}).  Existence, uniqueness and other important properties are well-understood for o.d.e's, 
although in practice limited knowledge of real-world initial data may result in the onset of chaotic behaviour.  

Most of the laws of physics are however represented by p.d.e's, which are harder to treat 
than o.d.e's.  For example if one is prescribed a flat spacetime Klein--Gordon scalar 
$\varsigma$ and its velocity everywhere (the Cauchy data) within a compact region at a given time, one 
can in principle predict the wave and its velocity everywhere within certain future and past 
regions\fn{I will be precise about what is meant by these regions in the next subsection.  
This is to do with the Klein--Gordon equation respecting causality, a feature shared by all the 
laws of physics below.} by solving the Klein--Gordon equation (\ref{KGeq}).  As a first simple 
case, if the Cauchy data are analytic functions, one can assert the local existence and uniqueness of an 
analytic solution to the wave equation from the following theorem.  

\mbox{ }

\noindent\bf Cauchy--Kovalevskaya theorem: \normalfont Suppose one has \sffamily A \normalfont 
unknowns $\Psi_{\mbox{\sffamily\scriptsize A\normalsize\normalfont}}$ which are functions of 
the IDV $\mu$ and of \sffamily Z \normalfont other independent variables 
$x_{\mbox{\sffamily\scriptsize Z\normalsize\normalfont}}$.  Suppose one is prescribed on some 
domain $\Omega$ the {\sffamily A } p.d.e's of order $k$ of form 
$\frac{  \pa^{(k)}\Psi_{\mbox{\sffamily\tiny A\normalsize\normalfont}}  }
{  \pa \mu^{(k)}  } = F_{\mbox{\sffamily\scriptsize A\normalsize\normalfont}}$  for the 
$F_{\mbox{\sffamily\scriptsize A\normalsize\normalfont}}$  functions of $\mu$, 
$x_{\mbox{\sffamily\scriptsize Z\normalsize\normalfont}}$, 
$\Psi_{\mbox{\sffamily\scriptsize A\normalsize\normalfont}}$  and of derivatives of 
$\Psi_{\mbox{\sffamily\scriptsize A\normalsize\normalfont}}$ up to 
($k$ -- 1)th order with respect to $\mu$, where furthermore in $\Omega$ the 
$F_{\mbox{\sffamily\scriptsize A\normalsize\normalfont}}$  
are \sl analytic functions \normalfont of their arguments.  Suppose one is further prescribed 
{\sl analytic data} $\Psi_{\mbox{\sffamily\scriptsize A\normalsize\normalfont}}(0, x_{\mbox{\sffamily\scriptsize Z\normalsize\normalfont}}) 
= {}^{(0)}d_{\mbox{\sffamily\scriptsize A\normalsize\normalfont}}(x_{\mbox{\sffamily\scriptsize Z\normalsize\normalfont}} )$, ... , 
$\frac{  \pa^{(k-1)}\Psi_{\mbox{\sffamily\tiny A\normalsize\normalfont}}  }
{  \pa \mu^{(k-1)}  }(0, x_{\mbox{\sffamily\scriptsize Z\normalsize\normalfont}}) = 
{ }^{(k -1)}d_{\mbox{\sffamily\scriptsize A\normalsize\normalfont}}
(x_{\mbox{\sffamily\scriptsize Z\normalsize\normalfont}})$ on some piece $U$ of a \sl 
nowhere-characteristic surface\normalfont.  Then this problem \sl locally \normalfont 
possesses precisely one \sl analytic \normalfont solution.  

\mbox{ }

\noindent The meaning and applicability of this theorem are elaborated in I.2.3.2 and B.1.

Complications arise in attempting to repeat the above simple approach for 

\noindent electromagnetism.  
Upon using $\mbox{\b{E}} = - \mbox{\b{$\pa$}}A_{\bar{0}} - \dot{\mbox{\b{A}}}$ in the  
vacuum version of the inhomogeneous Maxwell equations (\ref{M1}, \ref{M2}), 
one finds that one of the equations contains no time derivatives of \b{E} whereas the other three do.  
This corresponds to electromagnetism being constrained.  
The presence of the linear vacuum Gauss constraint $\mbox{\b{$\pa$}}\cdot \mbox{\b{E}} = 0$ 
is associated with the gauge freedom (\ref{Abelgau}).    
Because of the constraint, electromagnetism is strictly everywhere-characteristic, 
but one usually forms the problem for the three evolution equations (\ref{M2}) 
subject to already-constrained initial data and refers to this as the Cauchy problem for 
electromagnetism.  To set this up one requires 

\noindent 1) the electromagnetic field and its velocity at a given instant 
throughout some region. 

\noindent 2) To fix the gauge freedom.  In Lorenz gauge (\ref{Lorgau}) the evolution equations 
are of the correct form to invoke the Cauchy--Kovalevskaya theorem.  

\noindent Note that for the theory to make sense nothing is special about the initial time.  
So one would expect the constraint 
to hold for all times.  Fortunately, the evolution equations 
$$
\mbox{guarantee this happens: }
\mbox{\hspace{1.1in}}
\frac{\pa}{\pa t} (\mbox{\b{$\pa$}}\cdot\mbox{\b{E}}) = \mbox{\b{$\pa$}}\cdot\frac{\pa\mbox{\b{E}}}{\pa t} = \mbox{\b{$\pa$}}\cdot\mbox{\b{$\pa$}}\mbox{ \scriptsize$\times$ \normalsize} \mbox{\b{B}} = 0
\mbox{\hspace{1.1in}}
$$
i.e the evolution equations {\sl propagate the constraint.}

The above analysis can be repeated for the 4{\bf K} vacuum flat-space Yang--Mills equations, 
giving rise to {\bf K} constraints (\ref{flatYMgauss}) corresponding to {\bf K} gauge freedoms 
(\ref{YMgauxform}), and 3{\bf K} evolution equations (\ref{flatYMevol}).   
The {\bf K} constraints are similarly propagated by the evolution equations.  
A Lorenz-type gauge fixing permits use of the Cauchy--Kovalevskaya theorem as before.  The novel feature of these as compared to 
electromagnetism is that they are nonlinear equations which vastly complicates their behaviour 
and solution (eg superposition is no longer of use).  

The EFE's manifest all the above complications, being a constraint--gauge system and nonlinear (in fact the EFE's are quasilinear).  They also have further complications of their 
own: one is dealing with a theory of spacetime itself rather than working on a fixed background, which is considerably more difficult particularly from a global perspective (see 2.3.6).  
I now consider the simpler results of the (3 ,0; --1) GR CP, most of which hold irrespective of both $\epsilon$ and $s$.   

\mbox{ }

\noindent{\bf 2.3.2 Simple signature-independent features of the GR CP \normalfont}

\mbox{ }
 
\noindent First, it was soon realized (e.g by Hilbert \cite{Stachel} and 
Birkhoff--Langer \cite{BL22}) that there are $n$ + 1 restrictions on the system 
(4 for the usual GR).  Thus there are $n$ + 1 constraints (\ref{gauss}--\ref{cod}).  
These were explicitly provided by Darmois \cite{Darmois23, Darmois27}, although it 
has escaped attention that Campbell also provided them.\fn{Darmois' EFE split 
used normal coordinates $\alpha = 1$, \b{$\beta$} = 0 while Campbell used $\alpha$ arbitrary, 
\b{$\beta$} = 0.} The solution of the EFE's is then to be a 2-step procedure.   

\noindent The \bf ``initial value" or data construction step \normalfont is 
to construct data $(h_{ab}, K_{ab})$ 
together with coordinate functions $\rho$ and $j_a$ all on some portion $U_0$ of $\Upsilon_0$ 
by solving the constraints (see I.2.9).  

\noindent The \bf ``Evolution" or embedding step \normalfont is, given such data, 
to solve the evolution equations (\ref{ecd}) and (\ref{evK}) as a prescription for 
evaluating  $(h_{ab}, K_{ab})$ on some portion $U_{\mu}$ of a nearby $\Upsilon_{\mu}$ 
i.e for a small increment of the dynamical variable $\mu$.

\noindent This is the logical order of the steps, but historically the second was tackled before the first.  

\noindent The issue of what input is to be prescribed and what output is to be expected has 
played an important role in developing the above procedure, through the works of 
Lichnerowicz \cite{CGLich} and (Choquet)-Bruhat \cite{B52, B56, Cauchylit}, and through Wheeler's 
questions \cite{WheelerGRT, Wheeler} and those works answering these questions (see I.2.9).    

Second, provided that the constraints hold on some ``initial" $U_0$, they are guaranteed to 
hold throughout the embedding spacetime region.  This is most quickly seen from the contracted 
Bianchi identity (\ref{contbi}).  Understanding of this is implicit in \cite{BL22} and 
it is made explicit in \cite{Darmois27}, and is standard knowledge today: together with 
conservation of energy--momentum, and then making use of the Einstein tensor projection form 
of the constraints 
${\cal H}_A \equiv {\check{G}_A}^{\perp} - {\check{T}_A}^{\perp}$ and the EFE's: 
$$
0 = \nabla_B{\check{G}^B}_A - \nabla_B{\check{T}^B}_A 
\mbox{\hspace{1.2in}}
\mbox{\hspace{5in}}
$$
\be
\mbox{\hspace{0.02in}}
= \frac{\pa}{\pa t}({\check{G}_A}{}^{\perp} - {\check{T}_A}{}^{\perp}) 
+ \pa_i({\check{G}_A}{}^i - {\check{T}_A}{}^i ) - ({\check{G}_A}{}^B - {\check{T}_A}{}^B)
{\check{\Gamma}^{\perp}}{}_{B\perp} + ({\check{G}_B}{}^{\perp} -{\check{T}_B}{}^{\perp}){\check{\Gamma}^B}{}_{A\perp} 
\approx \dot{{\cal H}}_A 
\mbox{ } . 
\ee

Third, constraints come hand-in-hand with gauge freedoms.  Thus $n$ + 1 of the metric 
components are freely-specifiable: the coordinate functions painted onto the geometry have 
nothing to do with the geometry itself.  This is the content of general covariance.  
I will explain how many applications are tied to particular, careful gauge choices.  
For the moment I just consider either the \it normal coordinate gauge \normalfont 
($\alpha = 1$, $\beta_i = 0$) or the \it harmonic gauge \normalfont \cite{Cauchylit, Wald}: coordinates 
$x^{A}$ such that $\nabla^2x^A = 0$ \cite{deDonder} 
or equivalently $\check{\Gamma}^A \equiv g^{BC}{\check{\Gamma}^A}_{BC}$ \cite{Lanczos22}.  

Fourth, by a suitable choice of gauge one can ascertain by the Cauchy--Kovalevskaya theorem 
for admissible analytic data that there locally exists a unique solution to the evolution 
equations (\ref{evK}).  For this and further applications in the GR CP, the harmonic gauge is 
usually chosen.   However, I make the following observations about the use of this theorem. 

\noindent 1) It is the most basic theorem for p.d.e's. 

\noindent 2) It is a very general theorem in the sense that it holds for all sorts of p.d.e's.  
This is reflected by the signature (principal symbol)-independence of the applications and by 
the irrelevance of what analytic function 
$F_{\mbox{\sffamily\scriptsize A\normalsize\normalfont}}$ is, in, e.g, extending 
vacuum proofs to include fundamental as well as phenomenological sources.  

\noindent 3) Moreover, it is a very restrictive theorem in that it relies entirely on the 
functions being analytic, which seriously limits its applicability.  Without the analytic 
functions, no extremally general theorems in the sense of 2) are known or plausible.  

Furthermore, the analytic functions are inappropriate for any 
causal study (i.e in any sort of relativistic physics) since they are {\it rigid } (a change 
in a small region of an analytic data set induces a causally-unwanted change of the entire 
data set).  

Thus the application of the Cauchy--Kovalevskaya theorem to GR was never a serious 
consideration in the mainstream literature.  It would have been regarded as trivial 
because of observation 1) and not of serious interest because of observations 2) and 3).  
The point of the pre-1938 literature is that \sl no better results than this 
were available\normalfont, a serious lack.  

Fifth, whereas it appears to simplify the EFE's, the use of the normal coordinate gauge in the 
GR CP is discouraged because in practice it typically breaks down quickly.  This follows from the normal gauge 
Raychaudhuri equation that follows from (\ref{protoRay}): 
\be
\mbox{\hspace{0.3in}}
\dot{K} - \frac{K^2}{n} = K^{\mbox{\scriptsize T\normalsize}} \circ K^{\mbox{\scriptsize T\normalsize}} + \check{R}_{\perp\perp} \geq 0,  
\mbox{\hspace{0.3in}}
\label{normRay}
\ee 
where the inequality follows from the definition of the strong energy condition: 
\be
\mbox{the general Raychaudhuri equation term } 
\mbox{\hspace{0.5in}}
\check{R}_{AB}u^Au^B \geq 0 \mbox{ } \mbox{ } \forall \mbox{ unit timelike } 
u^A 
\mbox{ } , 
\mbox{\hspace{0.5in}}
\label{protoSEC}
\ee
by picking $u^A = n^A = t^A$.  To make (\ref{protoSEC}) manifestly an energy condition, 
one uses the 
\be
\mbox{EFE's to obtain }
\mbox{\hspace{1.1in}}
\check{T}_{AB}u^Au^B \geq -\frac{\check{T}}{n - 1} \mbox{ } \mbox{ } \forall \mbox{ unit timelike } u^A 
\mbox{ } .
\mbox{\hspace{1.1in}}
\label{SEC}
\ee
Integrating the differential inequality in (\ref{normRay}) shows that if $K_0 \geq 0$, then 
$|K| \longrightarrow \infty$ within a finite amount of parameter 
$\pi = \frac{n}{|K_0|}$ along $t^A$.  This blowup is by definition a \it caustic \normalfont and signifies a breakdown of the normal coordinates.  However, for the 
($r$, 1; 1) embedding, $n^A = z^A$ is spacelike.  There is then no good reason for $\check{R}_{\perp\perp}$ to be always positive, so caustics may be less likely to 
form.  

Elsewhere in the literature there is an old and supposedly little-known embedding result 
\cite{Campbell, Magaard, RTZ}.  However, I identified it \cite{ADLR, ATpap} 
as partly following from the above results, and consequently argue that a number of 
its supposed modern applications are highly questionable (see B.1.2--3).   

\mbox{ }

\noindent\bf Campbell--Magaard embedding theorem: \normalfont Any $n$-space can be `locally 
surrounded'\fn{By B `being surrounded' by A I mean that there is an embedding of B into A.  
What is meant by `locally' here is not particularly clear in the literature, and will be developed 
in 2.9.2.} by ($n$ + 1)-d vacuum space (where both spaces in question are non-null 
but otherwise of fixed arbitrary signature, and are both analytic).  

\mbox{ }

\noindent This followed from Campbell splitting the vacuum ($r$, $s$, $\epsilon$) EFE's 
independently of Darmois' work that blossomed into the GR CP and IVP.  As above, the given 
proof of the result is subdivided into an embedding or ``evolutionary'' step considered here 
and a data construction step, ``Magaard's method'', considered in I.2.9.2.3.  
What Campbell 
offered as `proof' for the former was incomplete and indeed spurious by the Bianchi identity.
All of the actual proof is contained in Magaard's use of the Cauchy--Kovalevskaya theorem, 
which is thus just a version of the fourth result above.  Thus it is well-known for 
the GR CP, and follows for the other ($r$, $s$; $\epsilon$) embedding cases from the 
well-known insensitivity of the Cauchy--Kovalevskaya theorem to ($r$, $s$; $\epsilon$).  
Thus this step is in fact obvious and well-known.  Moreover, as argued above, it is trivial 
and inappropriate.   Strictly, the 
Campbell--Magaard result is the vacuum case i.e $\check{\mbox{T}}_{AB} = 0$, a fact which has been 
much vaunted, but its proof may trivially be extended to any (analytic) functional form of 
$\check{\mbox{T}}_{AB}$ \cite{ADLR} 
(since the use throughout of the Cauchy--Kovalevskaya theorem is totally insensitive to 
changes in such subleading order terms, which merely contribute to  
$F_{\mbox{\sffamily\scriptsize A\normalsize\normalfont}}$).  Thus there is a `generalized 
Campbell--Magaard theorem'.  Moreover, the above results for the ($n$, 0; 1) GR CP case 
are protected and extended by a host of further theorems which, as argued in the next 
subsection, do not carry over to other signatures.  There are also difficulties with 
``Magaard's method" given in I.2.9.2.3 for $s = 0$, $\epsilon = -1$ and especially in B.1 
for $s = 1$, $\epsilon = 1$, where I conclude about this issue.      

\mbox{ }

\noindent{\bf 2.3.3 Signature-dependent GR CP results\normalfont}

\mbox{ }

\noindent The vast majority of serious results \cite{B52, Leray, B56, Cauchylit, HE, HKM, CBY, RF, Klainerman} 
for the GR CP turn out to be entirely dependent on the lower-d signature $s = 0$.  
In other words, the choice of methods which properly respect the difference between space and 
time is absolutely crucial.  In this subsection I restrict to documenting the 
standard results.  I explain why these techniques quite simply do not 
have any direct counterparts for other signatures in Part B.  

To have a sensible problem in mathematical physics, one requires not just local existence and 
uniqueness of solutions but rather \it well-posedness\normalfont.  
Whereas well-posedness always includes local existence and uniqueness, 
it also always includes continuous dependence on the data, without which an arbitrarily small change in the data could 
cause an arbitrarily large immediate\fn{See e.g page 229 of \cite{CH}: I do mean 
{\sl immediate},  
not an issue of chaos or unwanted growing modes (though well-posedness often {\sl also} bounds 
the growth of such modes \cite{RF}).} change in the evolution i.e there is no guarantee of physical predictability from such a problem.  
For a hyperbolic-type system one further requires a domain of dependence (DOD) property to enforce a 
sensible notion of causality.  If the data are known only on a closed achronal set $S$ (that is, 
a piece of a spacelike hypersurface), then the evolution can only be predicted within a region 
${\cal D}^+[S_0]$ (the future DOD), defined as the set of points such that all past-inextendible 
causal curves through each point [represented by the $\gamma$'s leading to the point $p$ in 
fig 5] intersect $S_0$.  In my opinion (substantiated in B.1) causality can effectively be studied 
only in settings where the IDV is time.    
\begin{figure}[h]
\centerline{\def\epsfsize#1#2{0.6#1}\epsffile{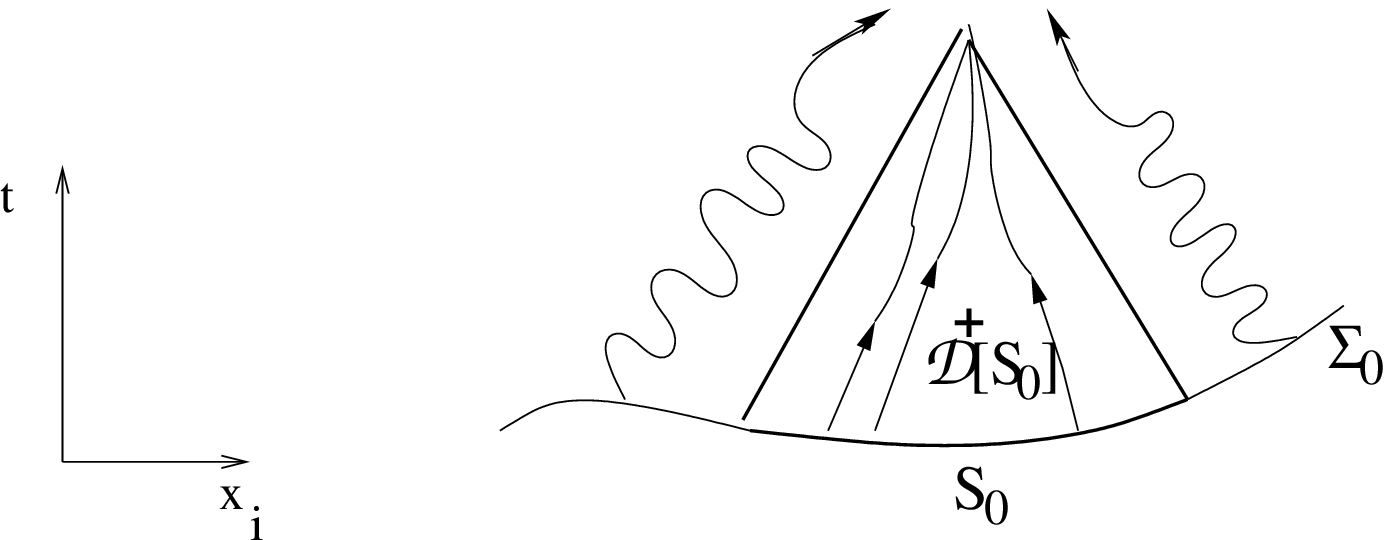}}
\caption[]{\label{TO5.ps}\footnotesize Domain of dependence property. \normalsize    

\noindent \footnotesize One should not claim to be able to predict too great a portion of 
the future.  Given data for a hyperbolic system on a piece of a spacelike surface, 
one can predict the future only in the forward wedge (domain of dependence) within which all causal curves, 
i.e allowed paths of information flow, emanate from the data.  Outside this wedge, for all we know, large 
disturbances could influence the future arbitrarily soon.  \normalsize}
\end{figure}

Note that further grounds of insufficiency for the use of the 
Cauchy--Kovalevskaya theorem are that it gives no control whatsoever over these last two 
properties.  The analytic functions are also undesirable because they are too restrictive 
a function space\fn{Whereas Hawking and Ellis \cite{HE} argue that the choice of function space used 
to model nature does not matter since it is not experimentally-determinable and 
in any case is only an approximation due to QM,  I would refine this argument to say that 
the difference between the analytic functions and other function spaces is important because of 
the rigidity inappropriateness.  Beyond that, I do not know if the particular function spaces  
used to prove rigorous theorems about the EFE's and about the extendibility of spacetime \cite{Clarke}  
may be substituted in these applications by approximations based on other function spaces.  
In this case, it may be a mere matter of convenience: one wants to use whichever sufficiently 
general function spaces accessibly give rise to the strongest possible theorems.    }
to cover a number of interesting studies (e.g involving boundaries, discontinuities  
or low differentiability upon approaching some spacetime singularities).  Piecewise, rougher 
function spaces are required.  

Serious theorems for GR were obtained in 1938 by Stellmacher \cite{Stellmacher38}, and 
more extensively by  Choquet-Bruhat in 1952 (\cite{B52}) and 1956 (\cite{B56}).  These 
advances were tied to progress in the (signature-specific!) general theory of 
nonlinear hyperbolic p.d.e's.  For example, Bruhat made use of the ($n$, 0; --1) 
EFE system (\ref{gauss}--\ref{evK}) 
being of 
\be
\mbox{the correct quasilinear hyperbolic form }
\mbox{\hspace{0.1in}}
{\cal L}^{\mbox{\sffamily\scriptsize CD\normalsize\normalfont}}
(x,\Psi_{\mbox{\sffamily\scriptsize A\normalsize\normalfont}},
\nabla_{\mbox{\sffamily\scriptsize A\normalsize\normalfont}}
\Psi_{\mbox{\sffamily\scriptsize B\normalsize\normalfont}}) 
\nabla_{\mbox{\sffamily\scriptsize C\normalsize\normalfont}}
\nabla_{\mbox{\sffamily\scriptsize D\normalsize\normalfont}}
\Psi_{\mbox{\sffamily\scriptsize E\normalsize\normalfont}} 
= {\cal F}_{\mbox{\sffamily\scriptsize E\normalsize\normalfont}}
(x,\Psi_{\mbox{\sffamily\scriptsize A\normalsize\normalfont}},
\nabla_{\mbox{\sffamily\scriptsize A\normalsize\normalfont}}
\Psi_{\mbox{\sffamily\scriptsize B\normalsize\normalfont}} )
\mbox{\hspace{0.3in}}
\ee 
when cast in harmonic coordinates 
(where ${\cal L}^{\mbox{\sffamily\scriptsize CD\normalsize\normalfont}}$ 
is a Lorentzian metric; both this and the function ${\cal F}_{\mbox{\sffamily\scriptsize E\normalsize\normalfont}}$ are smooth) to enable use of Leray's theorem \cite{Leray, Wald}, which 
guarantees local existence and uniqueness, and furthermore continuous dependence on the 
initial data and the DOD property.  
\begin{figure}[h]
\centerline{\def\epsfsize#1#2{0.6#1}\epsffile{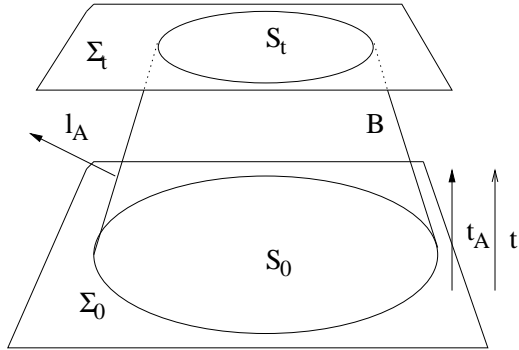}}
\caption[]{\label{TO6.ps}
\footnotesize The bucket-shaped construction that motivates the use of Sobolev spaces for the Cauchy problem for the 
Klein--Gordon equation. \normalsize 
}
\end{figure}

The above proofs of the four well-posedness criteria for the GR CP are now usually done 
using \it Sobolev spaces \normalfont \cite{Wald, HE, HKM, CBY, Clarke, Klainerman}. 
At first these lend themselves to less involved proofs than Bruhat's, although 
if one seeks yet stronger results the functional analysis becomes extremely unpleasant.  
That Sobolev spaces are appropriate follows from simple consideration \cite{Wald} of 
the flat spacetime Klein--Gordon equation.        
Given data on a bounded region $S_0$ of a spacelike hypersurface $\Sigma_0$, one can draw the 
future DOD ${\cal D}^+[S_0]$ [fig 6)] which is the region affected solely by this data due 
to the finite propagation speed of light.  One can then consider the values of $\varsigma$ and its 
first derivatives on $S_t = {\cal D}^+[S_0] \cap \Sigma_t, t > 0$.  Then using the 
construction in fig 7, Gauss's (divergence) 
theorem and energy-momentum

\noindent 
\be
\mbox{conservation, }
\mbox{\hspace{1.3in}}
\int_{S_0}\check{T}_{AC}t^{A}t^{C} + 
\int_{B}\check{T}_{AC}l^{A}t^{C} =
\int_{S_t}\check{T}_{AC}t^{A}t^{C} 
\mbox{ } ,
\mbox{\hspace{1.3in}}
\ee
and the second term $\geq 0$ provided that the matter obeys the dominant energy condition (DEC):  
\be
-{\check{T}^A}_Cu^C \mbox{ is a future-directed timelike or null vector } \mbox{ }\forall \mbox{ } \mbox{future-directed timelike } u^A
\ee
and that $t^{A}$ is timelike.  Then the definition of the energy-momentum tensor gives
\be
\int_{S_t}\left[(\dot{\varsigma})^2 + (\nabla\varsigma)^2 + m^2\varsigma^2\right] \leq
\int_{S_0}\left[(\dot{\varsigma})^2 + (\nabla\varsigma)^2 + m^2\varsigma^2\right]
\label{sums}
\ee
Because each integrand is the sum of squares (which are necessarily positive), this means that control over the data on $S_0$  
gives control of the solution on $S_t$.  
The idea of a Sobolev norm is a generalization of these last two `energy' integrals \cite{Clarke}.  
The Sobolev class $H^w$ has a bounded norm of this type including up to $w$th order derivatives.
There is then the following theorem \cite{HKM, Clarke}. 

\mbox{ }

\noindent \bf Hughes--Kato--Marsden Theorem: \normalfont 
Existence is guaranteed for the $n$-d EFE's in harmonic coordinates if the Sobolev class of the 
induced metric is no rougher than $H^{n + 1}$ and that of the extrinsic curvature is no rougher than 
$H^n$.

\mbox{ }

\noindent This is the roughest mathematics fully worked out to date.
The limitation on roughness comes from the definition of products, which is a necessary complication 
because the EFE's are nonlinear.   

\mbox{ }

\noindent{\bf 2.3.4 CP's for GR including fundamental matter}   

\mbox{ }

\noindent Matter theories on a fixed curved 
spacetime are built according to $\eta_{IJ} \longrightarrow g_{IJ}$ and the 
`$\pa \longrightarrow \nabla$' rule.  This undeniably produces the simplest field theories, but it 
is not certain that these are the ones chosen by nature, since the particle accelerators 
by which we judge our field theories are located in a rather flat region.  Thus the study of these theories is subject to our ignorance of nature's unexplored high-curvature regime.  
In particular, the `$\pa \longrightarrow \nabla$' rule could in principle be ambiguous 
\cite{MTW} (but for electromagnetism the order is dictated by current conservation) 
or not realized in nature due to putative further symmetry reasons (e.g instead of 
minimal coupling for scalars, one might argue for conformal coupling \cite{VKiefer}).  
We note that some of these theories would violate some form or other of the {\bf POE}.  
By the nature of many such theories (e.g Brans--Dicke theory), this possibility \sl cannot \normalfont 
be overruled by direct tests of the {\bf POE}.

The curved spacetime version of the equations of Klein--Gordon, Maxwell and Yang--Mills 
theory (see I.2.5) remain of the correct form to invoke Leray's theorem.  The coupled version 
of these fields with GR may be built by pairing the corresponding equations with the EFE's 
(\ref{gauss}, \ref{cod}, \ref{evK}) sourced by the corresponding energy--momentum tensor pieces 
(\ref{EMTen}).  In each of these cases, the total Leray form is blockwise the GR and matter 
Leray forms.  The GR Leray form {\sl is} disrupted by some of the nastier matter fields.  This 
goes hand in hand with some of the unpleaseantries of VI and VIII.  
Dirac theory is sufficiently different that I defer 
its treatment to VI.4.

\mbox{ }

\noindent{\bf 2.3.5 Global results \normalfont}

\mbox{ }

\noindent Global results for the Einstein evolution equations are considerably harder to obtain than local ones.  The need for global results 
stems from the generic onset of singularities from the evolution of smooth data.  Were naked singularities to arise thus, 
all the above nice notions of predictability would break down.  Penrose \cite{PenroseCCP69} conjectured that naked singularities do not arise generically 
(the cosmic censorship conjecture).  Whereas this has been studied for some simple cases, it is an open question.    
Among these simple cases are those with high-symmetry, particular asymptotics, or a `small' departure from exact data \cite{Rendall02}.

\subsection{Variational principles for GR}

I next consider the ADM split from the point of view of the principles of dynamics.  The applications in this thesis 
requires this just for the usual (3, 0; --1) case.  
Here are some useful results for variation with respect to the (arbitrary-d) metric $g_{ab}$: 
\be
\delta\sqrt{|g|} = \frac{1}{2}\sqrt{|g|}g^{ab}\delta g_{ab} 
\mbox{ } ,
\ee
\be
\delta g^{ab} = - g ^{ai}g^{bj} \delta g_{ij} 
\mbox{ } ,
\ee
\be
\delta {\Gamma^a}_{bc} = \frac{1}{2}g^{ad} \left[ 
\nabla_c(\delta g_{db}) + \nabla_b(\delta g_{dc}) - \nabla_d(\delta g_{bc}) 
\right] 
\mbox{ } ,
\label{deltaGamma}
\ee
\be
\delta {\Gamma^c}_{bc} = \frac{1}{2}g^{cd}\nabla_b(\delta g_{cd}) 
\mbox{ } ,
\ee
\be
\delta {R^a}_{bcd} = 2\nabla_{[d|}\left(\delta {\Gamma^a}_{b|c]}\right) 
\mbox{ } ,
\ee
\be
\delta R_{bd} = 2\nabla_{[d|}\left(\delta {\Gamma^c}_{b|c]}\right) 
\mbox{ } ,
\ee
\be
\delta R =  - R^{ij}\delta g_{ij} + 2\nabla^b\left(\nabla_{[b}\delta{g_{c]}}^c\right) 
\mbox{ } .
\ee
\be
\mbox{ } \mbox{ Then variation of the Einstein--Hilbert action }
\mbox{\hspace{0.3in}}
\mbox{\sffamily I\normalfont}_{\mbox{\scriptsize EH\normalfont}} = \int \textrm{d}^4x \sqrt{|g|}(\check{R}
+ \overline{\mbox{\sffamily L\normalfont}}_{\mbox{\scriptsize Matter\normalsize}}) 
\mbox{\hspace{0.3in}}
\ee
$$
\mbox{easily gives }
\mbox{\hspace{1.2in}}
0 = \delta \mbox{\sffamily I\normalfont} = \int \textrm{d}^{n + 1}x 
\left[
\delta 
\left(
\sqrt{|g|}\check{R}
\right) 
+ \delta\overline{\mbox{\sffamily L\normalfont}}_{\mbox{\scriptsize Matter\normalsize}}
\right]
\mbox{\hspace{1.2in}}
$$
$$ 
= \int d^4x\sqrt{|g|}
\left(
-
\left[
\left(
\check{R}^{AB} - \frac{\check{R}}{2}g^{AB}
\right) 
+ \frac{\delta \mbox{\sffamily L\normalfont}_{\mbox{\scriptsize Matter\normalsize}}}{\delta g_{AB}})\delta g_{AB} 
\right]
+  2\nabla^B
\left[
\nabla_{[B}\delta{  g_{C]}  }^C
\right]
\right)
\mbox{ } , 
$$
\be
\mbox{which upon discarding 
the divergence term and adopting }
\mbox{\hspace{0.4in}}
\check{T}_{AB} = \frac{1}{\sqrt{|g|}}\frac{\delta \overline{\mbox{\sffamily L\normalfont}}_{\mbox{\scriptsize Matter\normalsize}}}{\delta g_{AB}}
\mbox{\hspace{0.4in}}
\ee 
as the curved spacetime definition of the energy--momentum tensor yields the EFE's.  

Note that the pure gravity Einstein--Hilbert action may be rewritten in split form as
\be
\int \textrm{d}t \int \textrm{d}^3x \alpha\sqrt{h}
[  R - \epsilon(K\circ K - K^2)  ] \equiv \mbox{\sffamily I\normalfont}_{\mbox{\scriptsize 3+1 (GR)\normalsize}} 
\mbox{ } .
\ee
upon discarding a divergence.  
Variation of 
this yields the split form of the EFE's of I.2.2. 
It 
is also used below in further GR principles of dynamics work.

\subsection{Inclusion of fundamental matter fields}

There is no difficulty in including simple phenomenological matter in the above.  Likewise, 
there is no difficulty in including fundamental matter fields at least to start off with.  
I use 

\noindent the below in IV, whereas VI--VII discusses difficulties with more complicated fields.    
\be
\mbox{\sffamily I\normalfont}_{\mbox{\scriptsize KG\normalsize}}^{\varsigma} = \int\int \alpha\sqrt{h}
\left[
\frac{(\delta_{\check{\beta}}\varsigma)^2}{\alpha^2} - 
(h^{ab}\pa_a\varsigma\pa_b\varsigma + m_{\varsigma}\varsigma^2)
\right]
\ee
\be
\mbox{\sffamily I\normalfont}_{\mbox{\scriptsize em\normalsize}}^{\mbox{\scriptsize A\normalsize}}  = \int\int \alpha\sqrt{h}
\left[
\frac{        (\delta_{\check{\beta}} A_a - \pa_a A_0)(\delta_{\check{\beta}} A^a - \pa^a A_0)        }{       2\alpha^2        } 
- \frac{1}{4}F \circ F
\right]
\label{Lemcssplit}
\ee
\be
\mbox{\sffamily I\normalfont}_{\mbox{\scriptsize YM\normalsize}}^{\mbox{\scriptsize A\normalsize}_{\mbox{\tiny\bf I\normalfont\normalsize}}}  
= \int\int \alpha\sqrt{h}
\left[
\frac{          (\delta_{\check{\beta}}A^{\mbox{\bf\scriptsize I\normalsize\normalfont}}_a\mbox{--}\pa_a A^{\mbox{\bf\scriptsize I\normalsize\normalfont}}_0\mbox{+}
                 \mbox{\sffamily g\normalfont}_{\mbox{\scriptsize c\normalfont}}{f^{\mbox{\bf\scriptsize I\normalsize\normalfont}}}_{\mbox{\bf\scriptsize LM\normalsize\normalfont}}A_{{\mbox{\bf\scriptsize L\normalsize\normalfont}}a}A_{0{\mbox{\bf\scriptsize M\normalsize\normalfont}}})        
                (\delta_{\check{\beta}}A^a_{\mbox{\bf\scriptsize I\normalsize\normalfont}}\mbox{--}\pa^a_{\mbox{\bf\scriptsize I\normalsize\normalfont}} A_0\mbox{+} 
                 \mbox{\sffamily g\normalfont}_{\mbox{\scriptsize c\normalfont}}f_{\mbox{\bf\scriptsize IJK\normalsize\normalfont}}A^{\mbox{\bf\scriptsize J\normalsize\normalfont}a}A^{{\mbox{\bf\scriptsize K\normalsize\normalfont}0}})             }{          2\alpha^2           } 
\mbox{--}\frac{1}{4}F_{\mbox{\bf\scriptsize I\normalsize\normalfont}} \circ F^{\mbox{\bf\scriptsize I\normalsize\normalfont}}
\right]
\label{LYMcssplit}
\ee
The corresponding split field equations are 
\be
\delta_{\check{\beta}}
\left(
\frac{\sqrt{h}}{N}
\delta_{\check{\beta}}\varsigma
\right)
= \sqrt{h}D^a(N\pa_a\varsigma) + m_{\varsigma}\varsigma 
\mbox{ } ,
\ee
\be
\left\{
\begin{array}{l}
{\cal G} \equiv D_i
\left[
\frac{\sqrt{h}}{2\alpha}
\left(
\delta_{\check{\beta}}A^j - \pa^j\Phi
\right)
\right] 
 = 0  \\
\label{Gaussics}
D_iD^{[i}(NA^{j]}) = \delta_{\check{\beta}}
\left[
\frac{\sqrt{h}}{2\alpha}
\left(
\delta_{\check{\beta}}A^j - \pa^j\Phi
\right)
\right]
\end{array}
\right. 
\mbox{ } , 
\ee
\be
\left\{
\begin{array}{l}
{\cal G}_{\mbox{\bf\scriptsize J\normalsize\normalfont}} \equiv 
D^{\mbox{\scriptsize G\normalsize}}_i
\left[
\frac{\sqrt{h}}{2\alpha}
\left(
\delta_{\check{\beta}}A^j_{\mbox{\bf\scriptsize I\normalsize\normalfont}} 
- \pa^j\Phi_{\mbox{\bf\scriptsize I\normalsize\normalfont}}
\right)
\right] 
= 0  \\
\label{YMGaussics}
D^{\mbox{\scriptsize G\normalsize}}_iD^{{\mbox{\scriptsize G\normalsize}}[i}
(NA^{j]}_{\mbox{\bf\scriptsize I\normalsize\normalfont}}) 
= \delta_{\check{\beta}}
\left[
\frac{\sqrt{h}}{2\alpha}
\left(
\delta_{\check{\beta}}A^j_{\mbox{\bf\scriptsize I\normalsize\normalfont}} - \pa^j\Phi_{\mbox{\bf\scriptsize I\normalsize\normalfont}}
\right)
\right]
\end{array}
\right. 
\mbox{ } . 
\ee
To make coupled Einstein--matter systems, use these together with the split
\be
\begin{array}{cccc}
\mbox{Theory}            &            \rho            &            j_a           &        S_{ab}
\\
\mbox{KG}                & 
-\frac{1}{2}
\left[ 
\frac{(\delta_{\check{\beta}}\varsigma)^2}{\alpha^2}  + |\pa\varsigma|^2 + m_{\varsigma}^2\varsigma^2
\right]         
&     
-\pa_a\varsigma\frac{\delta_{\check{\beta}}\varsigma}{\alpha} 
&  
\begin{array}{c}
-\pa_a\varsigma\pa_b\varsigma \\ 
+ \frac{1}{2}h_{ab}
\left[
\pa^c\varsigma\pa_c\varsigma - (\delta_{\check{\beta}}\varsigma)^2 + m_{\varsigma}^2\varsigma^2 
\right]
\end{array}
\\
\mbox{e/m}                
& 
\frac{1}{2}(E^2 + B^2) 
&  
\frac{1}{2}(E \mbox{ \scriptsize $\times$ \normalfont} B )_a
& 
-\frac{1}{2}[B_aB_b + E_aE_b - h_{ab}(E^2 + B^2)]
\\
\end{array}
\label{EMTen}
\ee
of the energy--momentum tensor\fn{ Note that 
the 00 components are field energy expressions, the 0$a$ components are field momentum flux 
expressions (such as the Poynting vector of electromagnetism) and the $ab$  
components are field stresses.  The Yang--Mills energy-momentum tensor looks just like the 
electromagnetic one, with 
$B^i_{\mbox{\tiny\bf I\normalfont\normalsize}}$, 
$E^i_{\mbox{\tiny\bf I\normalfont\normalsize}}$ in place of $B^i$, $E^i$.  } in the split 
EFE's.

\subsection{The Dirac--ADM Hamiltonian formulation of GR}

In the case of (3, 0; --1) GR where the foliating spatial hypersurfaces are assumed to be 
compact without boundary (CWB),
$\overline{  \mbox{\sffamily L\normalfont}  } = \alpha\sqrt{h}(R + K\circ K - K^2)$ so 
\be
p^{ab} = \frac{  \pa\bar{\mbox{\sffamily L\normalfont}}  }{\pa\dot{h}_{ab}} = \frac{ \sqrt{h} }{ 2\alpha }
(h^{ac}h^{bd}  - h^{ab}h^{cd})\delta_{\check{\beta}}h_{bd} 
\mbox{ } .
\ee
\be
\mbox{This can be rearranged to }
\dot{h}_{cd} = \frac{2\alpha}{\sqrt{h}}(h_{ac}h_{bd}  - \frac{1}{2}h_{ab}h_{cd} )p^{ab} + 2D_{(c}\beta_{d)} 
\mbox{ } .
\mbox{\hspace{1.3in}}
\label{smll}
\ee
\be
\mbox{ It is also worth noting that for GR } 
\mbox{\hspace{0.9in}}
p^{ab} = -\sqrt{h}(K^{ab} -  h^{ab}K) 
\mbox{ } , 
\mbox{\hspace{1.9in}}
\label{pk}
\ee
\be
\mbox{\hspace{1.55in}} 
p \mbox{ } = \mbox{ } 2\sqrt{h}K 
\mbox{ } . 
\label{trptrk}
\ee
Then by Legendre transformation 
$$
\overline{    \mbox{\sffamily H\normalfont}    }(h_{ij}, p^{kl}; \alpha, p_{\alpha}; \beta_m, p^n_{\beta}) 
=  p\circ \dot{h} - 
\overline{    \mbox{\sffamily L\normalfont}    }(h_{ij}, \dot{h}_{kl}; \alpha, \dot{\alpha}; \beta_m, \dot{\beta}_n)   
$$
$$
= \frac{2\alpha}{\sqrt{h}}
\left(
p\circ p - \frac{p^2}{2}
\right)
+ p\circ (D\beta) -\alpha
\left[
\sqrt{h}R + \frac{1}{\sqrt{h}}
\left(
p\circ p - \frac{p^2}{2}
\right)
\right]
$$
$$
\Rightarrow \mbox{\sffamily H\normalfont}_{\mbox{\scriptsize ADM\normalsize}} = 
\int\textrm{d}^3x\overline{\mbox{\sffamily H\normalfont}} = \int
\left[
\alpha\left(\frac{1}{\sqrt{h}}\left(p\circ p - \frac{p^2}{2}
\right)
- \sqrt{h}R
\right) 
+ \beta^i
\left(
-2D_j{p^j}_i
\right)
\right]
$$
$$
\equiv \int\textrm{d}^3x(\alpha{\cal H} + \beta^i{\cal H}_i) 
\mbox{ } .  
$$
According to the standard interpretation, $\alpha$ and $\beta_i$ are Lagrange multipliers.  
Variation with respect to these multipliers gives the constraints ${\cal H} = 0$ and ${\cal H}_i = 0$.  
Thus the GR Hamiltonian is zero (for the CWB case; else there would be 
surface terms).  But this does not mean GR is trivial, since its Hamiltonian is only weakly zero:
\be
\mbox{\sffamily H\normalfont}_{\mbox{\scriptsize GR\normalsize}} = 0 \mbox{ } , \mbox{ but }
\mbox{\sffamily H\normalfont}_{\mbox{\scriptsize GR\normalsize}}^{\mbox{\scriptsize Total\normalsize}} = 
\mbox{\sffamily H\normalfont}_{\mbox{\scriptsize GR\normalsize}} + \int\textrm{d}^3x(\alpha{\cal H} + \beta^i{\cal H}_i) = \int\textrm{d}^3x(\alpha{\cal H} + \beta^i{\cal H}_i) = 
\mbox{\sffamily H\normalfont}_{\mbox{\scriptsize ADM\normalsize}}
\ee 
\be 
\mbox{Hamilton's equations are then the evolution equations } 
\mbox{\hspace{0.4in}}
\delta_{\check{\beta}}h_{ab} =  \frac{2\alpha}{\sqrt{h}}
\left(
p_{ab} - \frac{p}{2}h_{ab}
\right) 
\mbox{ } , 
\mbox{\hspace{0.4in}}
\label{CGsmallevol}  
\ee
$$
\delta_{\check{\beta}}p^{ab} = \sqrt{h}\alpha
\left(
\frac{R}{2}h^{ab} - R^{ab}
\right)
- {\sqrt{h}}(  h^{ab}D^2\alpha - D^aD^b\alpha  )  
+ \frac{\alpha}{2\sqrt{h}}h^{ab}
\left(
p\circ p - \frac{p^2}{2}
\right) 
$$
\be
- \frac{2\alpha}{\sqrt{h}}
\left(
p^{ac}{p^b}_c - \frac{p}{2}p^{ab}
\right).
\label{CGBSWEL}
\ee
The first of these is just a rewrite of (\ref{smll}).    

The constraints indeed propagate.  In Poisson bracket language, using the basic 
\be 
\mbox{gravitational bracket } 
\mbox{\hspace{1.1in}} 
\{h_{ij}(x), p^{kl}(x^{\prime})\} = {\delta^k}_{(i} {\delta^l}_{j)}{\delta}^{(3)}(x , x^{\prime}) 
\mbox{\hspace{1.1in}}
\label{gpb}
\ee
and the smearing out of the constraints by arbitrary functions $\check{\beta}_I = [\alpha, \beta_i]$ so as 
to obtain undensitized scalar functions on the gravitational phase space 
\be
{\cal H}(\alpha) \equiv \int_{\Sigma}\textrm{d}^3x\alpha{\cal H} \mbox{ } , \mbox{ } \mbox{ } 
\mbox{\b{${\cal H}$}}(\alpha) \equiv \int_{\Sigma}\textrm{d}^3x\beta^i{\cal H}_i 
\mbox{ } ,  
\ee
\be 
\mbox{then }
\mbox{\hspace{1.9in}}
\{f(h_{ij}, p^{kl}), \mbox{\b{${\cal H}$}}(\beta)\} = \pounds_{\beta}f 
\mbox{ } .
\mbox{\hspace{1.9in}}
\label{momaction}
\ee
Thus the momentum constraint's action is associated with dragging around within each hypersurface i.e with the 3-diffeomorphisms.  
From (\ref{momaction}) it follows that 
\be
\{\mbox{\b{${\cal H}$}}(\beta^{\prime}), \mbox{\b{${\cal H}$}}(\beta)\} = \pounds_{\beta}\mbox{\b{${\cal H}$}}(\beta^{\prime}) \mbox{ } ,
\label{mommom}
\ee
\be
\{{\cal H}(\alpha^{\prime}), \mbox{\b{${\cal H}$}}(\beta)\} = \pounds_{\beta}{\cal H}(\alpha^{\prime}) \mbox{ }.
\label{hammom}
\ee
A little extra effort is required to obtain the Poisson bracket of 
two Hamiltonian constraints: 
\be
\{{\cal H}(\alpha), {\cal H}(\alpha^{\prime})\} = \mbox{\b{${\cal H}$}}(\gamma) \mbox{ } , \mbox{ }\mbox{ } 
\gamma^i = hh^{ij}(\alpha\pa_j\alpha^{\prime} - \alpha^{\prime}\pa_j\alpha)  
\mbox{ } . 
\label{hamham}
\ee

Because this last bracket involves the metric, this is no true Lie algebra (for which only 
structure {\sl constants} would appear, see 
footnote 19).  So although the 4-diffeomorphisms 
form a true Lie algebra and the 3-diffeomorphisms also, splitting up 
the 4-diffeomorphisms into 3-diffeomorphisms and `other diffeomorphisms' becomes messy.  
Unlike the 

\noindent 3-diffeomorphism Lie-dragging invariance associated with the momentum constraint, 
there is no manifest symmetry associated with the Hamiltonian constraint.  There is instead  
a remarkable \sl hidden \normalfont symmetry from the 
perspective of the split formulation of GR: invariance under {\it refoliation} 
(or of choices of time function).  
The symmetry is hidden because in the split formulation one is working within a particular foliation, so that what happens under 
refoliation becomes obscured.  It is not known if or how the time function and the 
true d.o.f's of GR may be disentangled.  This is part of the {\it Problem of Time} (see I.3.3).  

\mbox{ } 

I next consider the passage to the Hamiltonian in the presence of matter fields 
\cite{ADM, KucharII, KucharIII} $\Psi$, denoting the Hamiltonian and momentum constraints obtained in this 
case by $^{\Psi}{\cal H}$ and $^{\Psi}{\cal H }_i$, and  the contributions to these from the 
matter fields by $_{\Psi}{\cal H}$ and $_{\Psi}{\cal H }_i$.  For example, for 
Einstein--Klein--Gordon theory, Einstein--Maxwell--theory and Einstein--Yang--Mills theory, 
the field momenta are 
\be
\pi \equiv  \frac{\pa \mbox{\sffamily L\normalfont}}{\pa \dot{\varsigma}} =  \frac{2\sqrt{h}}{\alpha}(\dot{\varsigma} - \pounds_{\beta}\varsigma) \mbox{ } ,
\ee
\be
\pi^i \equiv \frac{\pa \mbox{\sffamily L\normalfont}}{\pa \dot{A}_i}=  \frac{\sqrt{h}}{\alpha}(\dot{A}^i - \pounds_{\beta}A^i - \pa^iA_0) \mbox{ } ,
\ee
\be
\pi^i_{\mbox{\scriptsize\bf I\normalsize\normalfont}} \equiv 
\frac{\pa \mbox{\sffamily L\normalfont}}{\pa \dot{A}_i^{\mbox{\scriptsize\bf I\normalsize\normalfont}}} 
= \frac{\sqrt{h}}{\alpha}(\dot{A}^{i{\mbox{\scriptsize\bf I\normalsize\normalfont}}} 
- \pounds_{\beta}A^{i{\mbox{\scriptsize\bf I\normalsize\normalfont}}} 
- \pa^iA_0^{\mbox{\scriptsize\bf I\normalsize\normalfont}} + 
\mbox{\sffamily g\normalfont}_{\mbox{\scriptsize c\normalfont}}
{f^{\mbox{\scriptsize\bf I\normalsize\normalfont}}}_{\mbox{\scriptsize\bf JK\normalsize\normalfont}}
A_a^{\mbox{\scriptsize\bf J\normalsize\normalfont}}A_0^{\mbox{\scriptsize\bf K\normalsize\normalfont}}) 
\mbox{ } ,
\ee 
the Hamiltonians are 
\be
\mbox{\sffamily H\normalfont}_{\mbox{\scriptsize KG\normalsize}} = \int\textrm{d}^3x(\alpha{\cal H} + \beta^i{\cal H}_i) \mbox{ } ,
\ee
\be
_{\varsigma}{\cal H} \equiv \frac{\pi^2}{4\sqrt{h}} + \sqrt{h}(|\pa\varsigma|^2 + m_{\varsigma}\varsigma^2) \mbox{ } ,
\ee
\be
_{\varsigma}{\cal H}_i \equiv \pi\pa_i\varsigma \mbox{ } ,
\ee
\be
\mbox{\sffamily H\normalfont}_{\mbox{\scriptsize em\normalsize}} = \int\textrm{d}^3x(\alpha{\cal H} + \beta^i{\cal H}_i + A_0{\cal G}) \mbox{ } ,
\ee
\be
_{\mbox{\scriptsize A\normalsize}}{\cal H} \equiv \frac{1}{2\sqrt{h}}\pi_i\pi^i + \frac{\sqrt{h}}{4} F \circ F \mbox{ } ,
\ee
\be
_{\mbox{\scriptsize A\normalsize}}{\cal H}_i = \pi^c(D_iA_c - D_cA_i) - A_iD_c\pi^c \mbox{ } ,
\ee
\be
{\cal G} \equiv \pa_a\pi^a = 0 \mbox{ } ,
\label{curemgau}
\ee
\be
\mbox{\sffamily H\normalfont}_{\mbox{\scriptsize YM(G)\normalsize}} = \int\textrm{d}^3x(\alpha{\cal H} + \beta^i{\cal H}_i 
+ A_0^{\mbox{\bf\scriptsize J\normalsize\normalfont}}{\cal G}_{\mbox{\bf\scriptsize J\normalsize\normalfont}}) \mbox{ } ,
\ee
\be
_{\mbox{\scriptsize A\normalsize}_{\mbox{\bf\tiny I\normalsize\normalfont}}}{\cal H} \equiv 
\frac{1}{2\sqrt{h}}\pi^{\mbox{\bf\scriptsize I\normalsize\normalfont}}_i\pi_{\mbox{\bf\scriptsize I\normalsize\normalfont}}^i + \frac{\sqrt{h}}{4} 
F_{\mbox{\bf\scriptsize I\normalsize\normalfont}} \circ F^{\mbox{\bf\scriptsize I\normalsize\normalfont}} 
\mbox{ } ,
\ee
\be
_{\mbox{\scriptsize A\normalsize}_{\mbox{\bf\tiny I\normalsize\normalfont}}}{\cal H}_i = \pi^c_{\mbox{\bf\scriptsize I\normalsize\normalfont}}(D_iA^{\mbox{\bf\scriptsize I\normalsize\normalfont}}_c 
- D_cA^{\mbox{\bf\scriptsize I\normalsize\normalfont}}_i) - A_i^{\mbox{\bf\scriptsize I\normalsize\normalfont}}D_c\pi^c_{\mbox{\bf\scriptsize I\normalsize\normalfont}} 
\mbox{ } ,
\ee
\be
{\cal G}_{\mbox{\bf\scriptsize J\normalsize\normalfont}} =  D^{\mbox{\scriptsize G\normalsize}}_a\pi^a_{\mbox{\bf\scriptsize J\normalsize\normalfont}} = 0 
\mbox{ } ,
\label{curYMgau}
\ee
and the evolution equations are
\be
\dot{\pi} = \sqrt{h}D^a(N\pa_a\varsigma) + m_{\varsigma}\varsigma + \pounds_{\beta}\pi 
\mbox{ } ,
\ee
\be
\dot{\pi}^i = D_jD^{[j}(NA^{i]}) + \pounds_{\beta}\pi^i 
\mbox{ } ,
\ee
\be
\dot{\pi}^i_{\mbox{\bf\scriptsize I\normalsize\normalfont}} =  D^{\mbox{\scriptsize G\normalsize}}_jD^{{\mbox{\scriptsize G\normalsize}}[j}
(NA^{i]}_{\mbox{\bf\scriptsize I\normalsize\normalfont}}) + \pounds_{\beta}\pi^i_{\mbox{\bf\scriptsize I\normalsize\normalfont}}        
\mbox{ } .
\ee

\subsection{Superspace and geometrodynamics}

The evolution of a mechanical system may be viewed as a curve (parameterized by some label $\lambda$) traced in the configuration space {\sffamily Q\normalfont}. 
For GR, na\"{\i}vely the configuration space is the space Riem of 3-metrics on a manifold of fixed topology, taken here to be CWB.  
The evolution may then be viewed as a curve in Riem, i.e as a curve of 3-metrics.    
However, GR has constraints, so one would like to pass to a reduced configuration space in which these are satisfied.  This would correspond to the space of true 
d.o.f's of the gravitational field.  It is relatively straightforward to take into account the momentum constraint since it generates the infinitesimal spatial coordinate 
transformations, so that the true d.o.f's of 
\be
\mbox{GR lie within }
\mbox{\hspace{1.8in}}
\mbox{Superspace} = \frac {\mbox{Riem}}
{\mbox{Diff}} 
\mbox{ } .
\mbox{\hspace{1.8in}}
\ee
This is the space of 3-geometries, not 3-metrics:  how one paints the coordinates onto the geometries does not affect the physics.  Thus GR may be interpreted as a 
theory of evolving 3-geometries, or `geometrodynamics', as Wheeler put it \cite{MW57, Wheeler}.  It is however not known 
how to take the Hamiltonian constraint into consideration in order to pass to a fully reduced 
configuration space.

\mbox{ } 

\noindent\bf{2.7.1 Geometry on Riem and Superspace}\normalfont 

\mbox{ }

\noindent One can introduce on Riem the geometry associated with  
\be
F = \sqrt{\int\int f^{ijk^{\prime}l^{\prime}}\dot{h}_{ij}\dot{h}^{\prime}_{k^{\prime}l^{\prime}}
\sqrt{h}\textrm{d}^3x\sqrt{h^{\prime}}\textrm{d}^3x^{\prime}}
\ee 
\cite{VDeWitt70} where $f^{ijkl}$ is an invertible tempered distribution.   
One usually talks of the geometry in terms of the general supermetric $G_{ijkl}^{\mbox{\scriptsize general\normalsize}}$ which is related to $f^{abcd}$ by 
$
G_{ijkl}^{\mbox{\scriptsize general\normalsize}} = \frac{1}{\sqrt{h}}e_{ijkl}
$
$$
\mbox{where $e_{ijkl}$ is the inverse of $f^{ijkl}$ in that } 
\mbox{\hspace{0.05in}}
\int\textrm{d}^3x^{\prime\prime}\sqrt{h^{\prime\prime}}f^{abc^{\prime\prime}d^{\prime\prime}}e_{c^{\prime\prime}d^{\prime\prime}e^{\prime}f^{\prime}} = 
{\delta^a}_{(e^{\prime}}{\delta^b}_{f^{\prime})} \delta^{(3)}(x, x^{\prime})
\mbox{\hspace{0.05in}}.
$$
I concentrate on the \it ultralocal supermetrics \normalfont \cite{DeWitt}.\fn{A function is 
ultralocal in some variable (here the metric) if it contains no spatial derivatives of that 
variable.}  These generally consist of three terms  $h_{ik}h_{jl}$, $h_{il}h_{jk}$ and $h_{ij}h_{kl}$.  Thus there is a 2-parameter family of ultralocal supermetrics up to overall scale.  However, as far 
as the usual applications are concerned, there is just a 1-parameter family of these, 
\be
G^{\mbox{\scriptsize X\normalsize}}_{ijlk} = \frac{1}{\sqrt{h}}\left(h_{il}h_{jk} - \frac{X}{2}h_{ij}h_{kl} \right)
\label{ULSUPER}
\ee
because the supermetric appears in these applications in products of the form $G^{ijkl}s_{ij}s_{kl}$ for $s^{ij}$ symmetric, 
rendering equivalent the $h_{ik}h_{jl}$ and $h_{il}h_{jk}$ contributions.  
The inverse of 
\be 
\mbox{this supermetric is }
\mbox{\hspace{1in}}
G^{ijkl}_{\mbox{\scriptsize W\normalsize}} = \sqrt{h}\left(h^{il}h^{jk} - Wh^{ij}h^{kl}\right)
\mbox{ } ,
\mbox{\hspace{3in}}
\ee
\be
\mbox{where $X$ and $W$ are related by }
\mbox{\hspace{1.4in}}
X = \frac{2W}{3W - 1} 
\mbox{ } .
\mbox{\hspace{2in}}
\ee
The inverse supermetric does not exist for $X = \frac{2}{3}$ while the supermetric itself does not exist for  $W = \frac{1}{3}$.
By use of DeWitt's 2-index to 1-index map \cite{DeWitt},\fn{The new indices thus obtained run from 
1 to 6 and are written as hatted capitals in this thesis.}
\be
h_{aa} = h^{\hat{a}} \mbox{ (no sum) } \mbox{ } , \mbox{ } \mbox{ } 
h_{23} = \frac{1}{\sqrt{2}}h^{\hat{4}} \mbox{ } , \mbox{ } \mbox{ } 
h_{31} = \frac{1}{\sqrt{2}}h^{\hat{5}} \mbox{ } , \mbox{ } \mbox{ }
h_{12} = \frac{1}{\sqrt{2}}h^{\hat{6}} 
\mbox{ } ,
\label{2TO1}
\ee
one can see that for $W < \frac{1}{3}$ the ultralocal supermetric is positive-definite and for 
$W > \frac{1}{3}$ it is indefinite, having one `timelike' direction.  For $W = \frac{1}{3}$ it is degenerate. 
See III.1 and VIII for implications of these different cases.    

Whereas all the above ultralocal supermetrics play a role in this thesis, the supermetric  
\be
\mbox{in GR is the DeWitt supermetric \cite{DeWitt} }
\mbox{\hspace{0.35in}}
G_{ijkl} = \frac{1}{\sqrt{h}}
\left(
h_{i(l|}h_{j|k)} - \frac{1}{2}h_{ij}h_{kl} 
\right)
\mbox{\hspace{0.15in}}
\label{dwsm}
\ee
\be
\mbox{which occurs in ${\cal H}$.  Its inverse is } 
\mbox{\hspace{0.9in}}
G^{ijkl} = \sqrt{h}\left(h^{i(l|}h^{j|k)} - h^{ij}h^{kl}\right)
\mbox{\hspace{0.9in}}
\ee
(the symmetrizations in these expressions may be dropped in the applications mentioned above).

All the ultralocal supermetrics have the following symmetries
\be
G^{\mbox{\scriptsize X\normalsize}}_{abcd} = G^{\mbox{\scriptsize X\normalsize}}_{badc} \mbox{ } , \mbox{ } \mbox{ } 
s^{bd}G^{\mbox{\scriptsize X\normalsize}}_{bedi} = s^{bd}G^{\mbox{\scriptsize X\normalsize}}_{bide} 
\mbox{ } .
\ee
Additionally, the DeWitt supermetric has the further independent symmetry $G_{ijkl} = G_{klji}$
\be
\mbox{ whereas its inverse has the symmetries }
\mbox{\hspace{1in}}
G^{ijkl} = - G^{ikjl} = G^{klij} 
\mbox{ } .
\mbox{\hspace{2in}}
\label{invsym}
\ee

As explained below, the above geometries give just pointwise geometries on Superspace.    
DeWitt showed that the natural geometry inherited on Superspace has barriers causing 
it to be geodesically-incomplete, although the decomposition 
\be
G_{\hat{A}\hat{B}}= 
\left(\begin{array}{ll}
-1 & 0 \\
 0 & \frac{3}{32}\zeta^2\bar{G}_{\hat{a}\hat{b}}
\end{array}\right)
\label{55BLOCK}
\ee
of (\ref{dwsm}) by use of $\zeta = \sqrt{\frac{32}{3}}h^{\frac{1}{4}}$ and any other 
five coordinates orthogonal to it has a geodesically-complete geometry associated with its  
$5\times 5$ block.  

Further study by Fischer \cite{VFischer70} (see also DeWitt's accompanying commentary 
\cite{VDeWitt70}) demonstrated that Superspace is no manifold.  It is rather a collection 
of manifolds called a \it stratified manifold\normalfont, in which the individual strata 
are manifolds.  One must realize that the quotienting out of 3-diffeomorphisms depends 
on the particular properties of each 3-metric.  In particular it is easily demonstrated that 
different 3-metrics $h_{ij}$ have isometry groups Isom($h_{ij}$) of different dimension: 
$\mbox{dim}(\mbox{Isom}(\delta_{ij})) = \mbox{dim}(Eucl) = 6$, 
$\mbox{dim}(\mbox{Isom}(\mbox{generic } h_{ij})) = 0$.  Thus, in a manner clearly related 
to the underlying symmetries, at different `points' (i.e 3-geometries), Superspace differs 
in dimension.  Thus it is not a manifold, but rather a collection of manifolds each 
labelled by the conjugacy classes in the 3-diffeomorphisms of Isom.  This type of 
mathematical structure is to play an important role in II--VIII.

Two schemes have been proposed to overcome the incompleteness or non-manifold structure of 
Superspace.  First, DeWitt suggested \cite{VDeWitt70} that when one reaches the edge 
of one of the constituent manifolds (where the next stratum starts), one could reflect the 
path in Superspace that represents the evolution of the 3-geometry.  
Second, Fischer later alternatively proposed \cite{Fischer86} to replace Superspace with a 
nonsingular extended space built using the theory of fibre bundles. 

DeWitt had yet another Problem with interpreting paths on Superspace \cite{VDeWitt70}.  
Because the lapse is undetermined one is in fact dealing with whole sheaves of paths 
rather than with individual paths.  



\mbox{ }

\noindent{\bf Minisuperspace}

\mbox{ }

\noindent It is substantially simpler to consider the finite truncation of Superspace to the homogeneous 
geometries: {\it Minisuperspace} \cite{M69}.  Then the above geometry is the full geometry 
rather than just the pointwise geometry, since the geometry is the same at each point.  The 
restriction of the DeWitt supermetric to this case \cite{VMisner} is then just a $3 \times 3$ 
array rather than a $6 \times 6$ array at each space point.  Thus there are 3 -- 1 = 2 d.o.f's 
rather than 6 -- 3 -- 1 = 2 per space point.  For later use, Minisuperspace includes nine 
Bianchi classes labelled with slanted Roman numerals there are also 
even more symmetric 1 d.o.f solutions (e.g the Taub universes) which then simultaneously 
belong to several Bianchi types. Minisuperspace is potentially an important testing-ground 
for classical and quantum cosmology/gravity, since it is conjectured that the behaviour of the 
general GR solution upon approaching the initial singularity is an independent Kasner universe 
at each spatial point.  Belinskii, Khalatnikov and Lifshitz (BKL) conjectured mixmaster 
behaviour (a sequence of Kasner epochs as occurs in Bianchi {\sl XI} spacetimes) at each spatial 
point (see e.g \cite{SGBKL, M69, MTW}) whilst straightforward Kasner (Bianchi {\sl I}) behaviour at each spatial 
point can sometimes occur \cite{SGVDTandAVDT1, SGVDTandAVDT2}.  There is growing numerical 
evidence for these behaviours \cite{SGBKLmod}.  I mostly consider Minisuperspace as a toy model 
to test ideas against (see I.3.3.3, V.1, VI.1.4).  Although one might try to 
extrapolate ideas about and techniques developed for Minisuperspace, this is often fruitless 
because these break down in the full Superspace.  

\mbox{ }

\noindent\bf{2.7.2 Geometrodynamics: old RMW interpretation}\normalfont

\mbox{ }

\noindent How is the two-layered body of mathematics in I.2.6 and I.2.7.1 to be interpreted?  
Wheeler made a first attempt at this \cite{RMW2}.  He perceived that GR could be viewed as a 
theory of evolving 3-geometries: geometrodynamics.   The central object of this 
scheme is the still-remaining vacuum Hamiltonian constraint ${\cal H}$.  
Whereas the momentum constraint is conceptually (if not technically) easy to deal with,
${\cal H}$ leads to the Problem of Time which plagues geometrodynamics.  
${\cal H}$ is also central in the sense that since ${\cal H} \equiv \check{G}_{\perp\perp} 
= 0$, if this is known in all frames of reference i.e choices of projection operator, 
then the EFE's $\check{G}_{AB} = 0$ are known (assuming general covariance):  
`field equations without field equations'.  ${\cal H}$ is also central in Gerlach's formulation 
\cite{Gerlach}.  I will also discuss how ${\cal H}_i$ may be 
regarded as an integrability of ${\cal H}$ rather than on an equal footing to it (II.2.2.4).  

Wheeler's early hope was that vacuum geometrodynamics might turn out to be a Theory of Everything.  
This was based on \cite{MW57}, in which a number of other physical features were argued to be 
incorporable into this vacuum geometrodynamics.  The properties of a distant isolated mass 
were argued to be recoverable from \it geons \normalfont i.e gravity waves almost completely 
trapped in some region (`mass without mass').  Aspects of charge could be recovered from the 
mere topology of empty space (`charge without charge').  The electromagnetic field 
tensor\fn{Another difficulty is that only the non-null case of this was included.} was shown to be 
expressible in terms of the energy-momentum tensor, from which the Einstein--Maxwell system 
could be interpreted as reducible to a theory of empty space alone.  The first part of the 
latter result was a rediscovery of a result by Rainich \cite{Rainich}.  The idea was then to 
investigate whether all other known forms of fundamental matter could be similarly 
incorporated into this Rainich--Misner--Wheeler (RMW) already-unified theory of gravity and 
electromagnetism.  In those days the Yang--Mills-type theories of the nuclear forces were not 
yet known, so the RMW theory was a unification of all the understood fundamental forces.  But 
no way to incorporate spin-$\frac{1}{2}$ fermion fields was found, the massless neutrino field 
presenting much difficulty for this conceptual framework.  As a result of this 
already in 1959 \cite{W59} Wheeler thought it unlikely that vacuum geometrodynamics would be 
sufficient to describe nature.  RMW theory also turned out to have a somewhat defective IVP 
\cite{Penrose60} and as far as I know, nobody ever succeeded in writing down an action for it 
\cite{MW57} or in recovering the superposition principle for electromagnetism.  

This interpretation was not a waste of time since it prompted the first exercises in the 
construction of 3-geometries as initial data sets, which underlies much 
of modern numerical relativity.


\mbox{ }

\noindent\bf{2.7.3 Geometrodynamics: modern interpretation}\normalfont

\mbox{ }

\noindent Wheeler asked some important questions about the Superspace interpretation of 
geometrodynamics \cite{Wheeler}.\fn{The quantum `conceptual difficulties' in I.3.3.3 are 
also Wheeler's.}  The Hamiltonian constraint remains central but is to be understood to be 
with matter `added on' \cite{Wheeler}.  The first question (quoted at the start of) is why 
the closely-related Einstein--Hamilton--Jacobi equation takes the form it does, to be answered 
without assuming GR.  This thesis chiefly (II--VIII) consists of an answer to this question 
along the lines of the recent proposal by Barbour, Foster and \'{O} Murchadha.  
There is also an older rather different attempt to answer this question by Hojman, 
Kucha\v{r} and Teitelboim (see II.1.1).   One of the issues at stake with these answers is 
that they should past the test of natural inclusion of all the known fundamental matter fields.  
As in 
the RMW interpretation of geometrodynamics, it is the natural inclusion of spin-$\frac{1}{2}$ 
fermions that is troublesome.    

Second, what kind of structure does Superspace have?  It was reported \cite{Haus} that Stern 
proved that Superspace is Hausdorff-separated.  There is also the indefinite DeWitt supermetric 
naturally defined on Superspace.  There is than the issue of whether spatial topology change 
is permitted in classical geometrodynamics.  It is not, as shown by Geroch \cite{Geroch67}.  
So classically one must consider one separate Superspace per choice of spatial topology.  
Finally, is there any classical explanation for the 3-d 
Riemannian character of space and its ability to accommodate the known matter fields of nature.  

Third, what is two-thirds of Superspace? I.e how can the true d.o.f's in GR be represented?  
This question was almost answered by York (see I.2.9.4.2) and is also further discussed in 
III.2.  

Note that Wheeler had in mind the case of evolving CWB geometries, which have the 
following merits and plausibilities.  Einstein's arguments for a 
closed universe are based on its simplicity and on attempting to realize Mach's principle 
\cite{Ein34, Ein50} (it is displeasing for `absolutist' boundary conditions at infinity to 
interfere with local physics).  Wheeler used the thin sandwich formulation (see below) to 
attempt to conceptualize classical and quantum geometrodynamics 
\cite{W59, WheelerGRT, Wheeler}, and to implement Mach's 
principle \cite{WheelerGRT}.  The conformal IVP (also see below) was found to provide more 
adequate protection in the form 
of rigorous mathematical theorems than the thin sandwich, causing Wheeler later 
\cite{WI79, W88} to shift the interpretation of Mach to be instead in terms of constant mean 
curvature and conformal 3-geometry.  This led to the idea of `Wheeler--Einstein--Mach universes' 
\cite{Isenberg81}.  A further simplicity argument for CWB geometries is than that the GR 
conformal IVP is simplest for these.  

Whether the favoured cosmological model for the universe is open or closed has shifted 
around during the history of observational cosmology.  
One should note however that observationally-open universes may close on a larger-than-observed 
scale e.g along the lines of Lindquist's gluing together of many Schwarzchild solution pieces \cite{Lindquist}, or 
of topological identification \cite{Luminet}.  Conversely, closed-looking could actually be 
open via hitherto unprobeably small holes leading to open regions.  So it looks to be down to 
a matter of taste.      

Finally, the inclusion of simple forms of matter such as Klein-Gordon scalars, 
electromagnetism and Yang--Mills theory does not disrupt the above interpretation of vacuum 
geometrodynamics (see IV, VI.2).  However, nastier matter fields \sl do \normalfont disrupt 
it (VI.2.4).

\subsection{The Baierlein--Sharp--Wheeler action for GR}

I will often use the manifestly reparameterization-invariant Baierlein--Sharp--Wheeler (BSW) \cite{BSW} 
formulation of GR, which arose by analogy with QM transition amplitudes during Wheeler's 
attempts to interpret geometrodynamics.  This formulation arises from rewriting the Lagrangian 
formulation of GR 
\be
\mbox{\sffamily I\normalfont}_{{\mbox{\scriptsize 3+1(GR)\normalsize}}}  = 
\int \textrm{d}t \int \textrm{d}^3x\sqrt{h}\alpha
\left(
R + \frac{\mbox{\sffamily T\normalfont}^{\mbox{\scriptsize g\normalsize}}(\delta_{\check{\beta}} h_{ij})}{4\alpha^2}
\right) 
\mbox{ } , 
\mbox{\sffamily T\normalfont}^{\mbox{\scriptsize g\normalsize}} = (h^{ik}h^{jl} - h^{ij}h^{kl})\delta_{\check{\beta}} h_{ij}\delta_{\check{\beta}}h_{kl} 
\mbox{ } , 
\label{VBSWmethod}
\ee
\be
\mbox{by using the $\alpha$-multiplier equation }  
\mbox{\hspace{1in}}
\alpha = \pm \frac{1}{2}\sqrt{ \frac{\mbox{\sffamily T\normalfont}^{\mbox{\scriptsize g\normalsize}}}{R}} 
\mbox{ } ,
\mbox{\hspace{2in}}
\ee
to \sl algebraically \normalfont eliminate $\alpha$ from(\ref{VBSWmethod}) (notice the analogy 
with setting up the homogeneous Jacobi principle).  Thus (assuming $R \neq 0$ everywhere in the 
region of interest) one arrives at the BSW action 
\be
\mbox{\sffamily I\normalfont}_{\mbox{\scriptsize BSW\normalsize}} = \int \textrm{d}t \int \textrm{d}^3x \sqrt{h} 
\sqrt{R\mbox{\sffamily T\normalfont}^{\mbox{\scriptsize g\normalsize}}} 
\mbox{ } .
\label{VBashwe}
\ee
This formulation is the start of the first method we consider below for the GR IVP.

\subsection{The GR IVP}

GR is constrained.  
I now explore how the solution of these constraints has been approached.  
I include arbitrary dimension and signature to start off with, although I often then 
specialize to the usual (3, 0; --1) case, deferring treatment of the other cases to 
Part B where they are required.  
I include phenomenological matter.    
Several methods have been proposed, of which I only consider those which are {\sl general} 
(i.e independent of symmetry restrictions).  BSW proposed the {\it thin sandwich approach} 
(see I.2.9.1).  Although largely unexplored, there is evidence that this approach is of 
limited success.  Two older types of method are the {\it elimination} and {\it conformal} 
methods (in I.2.9.2 and I.2.9.3 respectively).  
Elimination methods are intuitive in that the prescribed quantities are all physical, but rely 
on brute force and have a number of undesirable mathematical features. In contrast, the 
conformal method is counterintuitive in that the prescribed quantities are unphysical, but 
this method exploits well the mathematical properties of the constraint system in order to 
decouple it.  Lichnerowicz proposed and argued for the conformal method in 1944 \cite{CGLich}.  
Bruhat compared conformal and elimination methods in 1956 \cite{B56}, accumulating good reasons to favour 
the conformal method.  However, I found that her arguments by themselves do not suffice to 
dismiss Magaard's 1962 method \cite{Magaard} and others.  I fill in this gap below by 
extending her arguments.  York and others substantially developed the conformal method in the 
1970's \cite{York71, York72, York73, MTW, cwb, CGSYkin, Y79book}; this is commonly used as a 
cornerstone of numerical relativity (see I.2.11).  Further methods for the GR IVP are outlined 
in I.2.9.4--5 including a small work of mine.  
Fundamental matter is included in I.2.9.6.

\mbox{ }


\noindent\bf{2.9.1 The thin sandwich method and conjecture}\normalfont

\mbox{ }

\noindent The thin sandwich method is a Lagrangian method: freely specify the metric $h_{ij}$ and its 
\be
\mbox{velocity $\dot{h}_{ij}$. Eliminate $\alpha$ from its multiplier equation }
\mbox{\hspace{0.6in}}
\alpha = \frac{1}{2}\sqrt{\frac{ \mbox{\sffamily T\normalfont}^{\mbox{\scriptsize g\normalsize}}        }{ \sigma R - 2\rho }} 
\mbox{ } , 
\mbox{\hspace{1.6in}}
\label{29aeq}
\ee
\be 
\mbox{thus forming }
\mbox{\hspace{1in}}  
\mbox{\sffamily I\normalfont}_{\mbox{\scriptsize TS\normalsize}} 
= \int \textrm{d}t \int \textrm{d}^3x \sqrt{h}
\left( 
\sqrt{(\sigma R - 2\rho) \mbox{\sffamily T\normalfont}^{\mbox{\scriptsize g\normalsize}}} + \beta_ij^i 
\right)
\mbox{ } ,
\mbox{\hspace{2in}}
\label{tsac}
\ee  
where $\sigma = - \epsilon$ and $\rho$ is the energy-momentum tensor component 
$\mbox{T}_{00}$.  
Then treat the $\beta_i$ multiplier equation  (which replaces the Codazzi constraint) 
as a p.d.e for 
$\beta_i$ itself: 
\be
D^a
\left[
\sqrt{    \frac{    \sigma R - 2\rho   }{   (     h^{ik}h^{jl} - h^{ij}h^{kl}    )\delta_{\check{\beta}} h_{ij}\delta_{\check{\beta}}h_{kl}   }    }
\left(
\delta_{\check{\beta}}h_{ab} - h_{ab}h^{cd}\delta_{\check{\beta}}h_{cd}
\right)
\right] 
= j_b  
\mbox{ } ,
\label{thinsaneq}
\ee
obtained using the explicit expressions for $\alpha$ and 
$\mbox{\sffamily T\normalfont}^{\mbox{\scriptsize g\normalsize}}$.  
The conjecture is that (\ref{thinsaneq}) has a 
unique solution under a 
suitably broad range of circumstances.  If this is true, then it is trivial to find $\alpha$ 
using  (\ref{29aeq}) and then to find $K_{ij}$ from its definition (\ref{ecd}).  This last 
step is `filling in the sandwich'.

There are in fact two sandwich methods.  The `thick sandwich' involves prescribing the metric 
on two nearby slices and constructing the spacetime between them, whereas the other 
discussed above is the limit of this as the two nearby slices become arbitrarily close.  

As regards the conjecture, although progress has been made, a regular method to solve this has 
not been found and counterexamples exist.  This is all for the $s = 0$, $\sigma = 1$ case with 
phenomenological matter.  

To avoid the `Problem of zeros'$,^{37}$ 
one assumes $2\rho - \sigma R > 0$ in the region studied.  
Then solutions to the thin sandwich equation with Dirichlet boundary conditions are unique 
\cite{thin sandwich1, thin sandwich2}.  
But one has no guarantee of existence, as shown by counterexamples in 
\cite{thin sandwich1, York83}.  

This is however not the end of the thin sandwich idea.  See C.3 for the effect 
of considering an {\sl Einstein--Maxwell} thin sandwich, and I.2.9.4.1 for the modern 
{\sl conformal} thin sandwich formulation.  

\mbox{ }

\noindent{\bf 2.9.2 Componentwise methods, including traditional elimination methods}

\mbox{ }

\noindent{\bf 2.9.2.1 A systematic treatment of componentwise methods}

\mbox{ }

\noindent The Gauss--Codazzi constraint system consists of $n + 1$ equations relating 
$n(n + 1)$ functions  
($h_{ij}$, $K_{ij}$).  A simple idea is to treat this system \it componentwise\normalfont: 
consider $(n - 1)(n + 1)$ of the ($h_{ij}$, $K_{ij}$) as knowns and attempt to solve the 
system for the remaining $n + 1$ components.  Hawking and Ellis \cite{HE} state that in the 
usual $n = 3$ case it is possible to solve this system for any such choice of knowns and 
unknowns.   I investigate this below, but find it profitable to consider first what the different 
sorts of componentwise methods are.  I attempt to cast the Gauss--Codazzi 
constraint system as some ``evolution" system with respect to some auxiliary IDV 
(without loss of generality $x_1$).  In this subsection, I simply attempt to cast the system in 
Cauchy--Kovalevskaya form, which would suffice to establish existence and uniqueness 
when the coefficients in the system and the `data for the data' at the ``initial" value of 
$x_1$ are all analytic.    

Some of the componentwise methods are \it algebraic elimination methods\normalfont.  
For example when one considers an embedding of a given lower-d metric, then the Gauss 
equation is a mere algebraic expression in the unknowns (which are by default some selection 
of extrinsic curvature components).  There are two such componentwise procedures. 

\noindent {\bf Method 1.} If a diagonal component, without loss of generality $K_{11}$, is available among the unknowns, 
one can choose to interpret the Gauss constraint as an equation for this which is linear by 
the antisymmetry (\ref{invsym}) of the DeWitt supermetric:
\be
2K_{11}G^{11uw}K_{uw} + G^{1u1w}K_{1u}K_{1w} + G^{uwxy}K_{uw}K_{xy} = - (\epsilon R + 2\rho)
\ee
where $u$, $w$, $x$, $y$ $\neq 1$).  Thus if $Z \equiv G^{11uw}K_{uw}$ 
has no zeros in the region of interest, $K_{11}$ can be straightforwardly eliminated in the 
Codazzi constraint, which one may then attempt to treat as a p.d.e system for the other unknowns. 
This method was criticized by Bruhat \cite{B56} in the usual-dimension-and-signature 
case.  These criticisms are discussed and extended in I.2.9.2.3.  

\noindent {\bf Method 2.} Since there are $n + 1$ unknowns and only $n$ diagonal components, it is 
always true that there will be a nondiagonal component, without loss of generality $K_{12}$, among the unknowns.  
One may then interpret the Gauss constraint as a quadratic equation in this: 
\be
G^{1212}K_{12}K_{12} + G^{12cd}K_{12}K_{cd} + G^{abcd}K_{ab}K_{cd} = - (\epsilon R + 2\rho)
\ee
for $ab$, $cd$ $\neq 12$ or $21$.  So long as 
$\bar{Z} = G^{1212} = h^{11}h^{22} - h^{12}h^{12}$
has no zeros in the region of interest, the solutions of the Gauss constraint may be 
substituted into the Codazzi constraint, and one may then attempt to treat this as a p.d.e 
system for the other unknowns.   

\noindent {\bf Method 3.} If in contrast with the above two methods, there are metric 
components among the unknowns, the Gauss constraint may be treated as a p.d.e. for one of 
these.  The resulting componentwise methods do not then 
involve any algebraic elimination.  
These are \it componentwise p.d.e methods\normalfont.  
Note that p.d.e's rather than a mixture of p.d.e's and algebraic equations generally lend 
themselves to making stronger existence and uniqueness proofs.  

\mbox{ }

\noindent{\bf 2.9.2.2 The result in Hawking and Ellis}

\mbox{ }

\noindent{\bf Claim} (In 3-d) one can show that one can prescribe 8 of the 12 independent components 
of ($h_{ij}$, $K_{ij}$) and solve the constraint equations to find the other 4.

\mbox{ }

\noindent Although Hawking and Ellis \cite{HE} support this merely by citing \cite{Cauchylit} where certain cases 
for the split into known and unknown components are considered, I prove this here for all cases 
in certain small regions\fn{The domain of applicability of each subcase is subject to one of the 
following kinds of `Problem of zeros' of varying severity.  
First kind: in setting up the p.d.e system, if division is required by a function of unknowns 
(over which one has therefore no control) 
then the emergence of zeros of this function invalidates the procedure in regions to be discovered.  
Second kind: if the function is rather 
of knowns, one at least knows a priori in which regions the system will and will not be valid in.   
Third kind: the system is always valid, but the casting of it into Cauchy--Kovalevskaya form may 
involve division by a function which potentially has zeros, or more generally inversion of a matrix 
whose determinant may possess zeros.} for the analytic functions.  Because I proceed via the 
Cauchy--Kovalevskaya theorem, my proof is clearly signature-independent.  Moreover, there are 
additional complications if the dimension is increased.  

\mbox{ }

\noindent{\bf Proof}  It is important first to notice which $\pa_1$ terms are 
present in the uneliminated 
Gauss--Codazzi system since I am treating this as an ``evolution" system with respect to $x_1$.  
The Gauss equation is linear in $\pa_1^2h_{uw}$, quadratic in $\pa_1 h_{u1}$ and linear 
in $\pa_1h_{11}$.  The 1-component of the Codazzi constraint contains the $\pa_1$ derivatives 
of all the variables, whereas the $u$-component (for any $u \neq 1$) contains only the $\pa_1$ 
derivatives of $K_{ua}$ 
and of all the metric components.  I next consider the different possibilities for the 
unknowns case by case.  

Suppose the 4 unknowns are all components of $K_{ab}$.  Then one may attempt to use method 1 
or method 2, that is one eliminates some component $E$ from the Codazzi constraint by use of 
the Gauss constraint.   
The 2-component of the Codazzi equation may be solved for a $K_{2b}$ component, 
the 3-component of the Codazzi equation may be solved for a $K_{3b}$ component, and
the 1-component of the Codazzi equation may be solved for any other $K_{ab}$ component, $C$.  
This never causes any trouble since there must always be such components among the 4 unknowns.  
The $E$-eliminated Codazzi system may then be cast into a 3-equation first-order 
Cauchy--Kovalevskaya  
form, and the final unknown $E$ is then to be read off the Gauss constraint.  

If just one of the unknowns is a metric component $m$ and the other unknowns do not share an 
absent index (without loss of generality are not $K_{11}$, $K_{12}$ and $K_{22}$ which share an absent  
$3$-index), then one can set up a 4- or 5-equation first-order 
Cauchy--Kovalevskaya form.  The 5-equation form corresponds to when $m$ is some $h_{uw}$ so 
that the Gauss constraint is second-order and can then be decoupled via $\pa_1 m = y$; the 
4-equation form corresponds to the other possible 
choices of $m$.  The Codazzi constraint may be solved for the 3 extrinsic curvature components.  
If there is a shared absent coordinate without loss of generality $x_3$, then the corresponding 
3-component of the Codazzi constraint contains no partial derivatives with respect to the dynamical 
variable of the extrinsic curvature unknowns and so the above scheme cannot be used.  
Then for the Gauss constraint, use method 1 or 2, giving a 3-component Codazzi system.

If precisely two of the unknowns are metric components, then without loss of generality one of the extrinsic curvature 
unknowns carries a 2-index.  Then one may solve the 2-component of the Codazzi constraint for this 
component and the 1-component for whichever other extrinsic curvature unknown one has declared.  
Then the 3-component of the Codazzi constraint and the Gauss constraint are to be solved for the two 
metric unknowns.   Thus one may obtain a 4 or 5-d first-order Cauchy--Kovalevskaya form.  

If there is just one extrinsic curvature component unknown, then the 1-component of the Codazzi system 
may be solved for it and all the remaining equations be solved for the unknown metric components.   
One then obtains a 5-equation first-order Cauchy--Kovalevskaya form.    

If all the unknowns are metric components there are no difficulties in writing down a 
4- or 5-equation first-order Cauchy--Kovalevskaya form.  For example without loss of generality a metric component 
$h_{uw}$ must be among the unknowns (since there are $4$ unknown metric 
components and only $3$ can contain a 1-index).  
Then the Gauss equation is of the form
\be
(h^{uw}h^{11} - h^{u1}h^{w1})\pa_1\pa_1h_{uw} = 
F(h_{ij}, \pa_kh_{ij}, \pa_k\pa_lh_{ij} \mbox{ with }  k,l \mbox{ not both } 1, K_{ij}, \rho)
\label{oversizeCK}
\ee
which for $h^{uw}h^{11} - h^{u1}h^{w1} \neq 0$ can be combined with the Codazzi equations 
arranged as equations for the $d$-derivatives of any $n$ other unknown metric components to form 
a system to which the Cauchy--Kovalevskaya theorem is applicable [e.g decouple 
(\ref{oversizeCK}) to form a system of $5$ first-order equations].

\mbox{ }

Finally, note that the above simple scheme, in which the Gauss--Codazzi constraint 
system is cast as an evolution system with respect to some auxiliary IDV, does not 
by itself suffice to investigate whether Hawking and Ellis' result holds for higher-d.    
For, consider the unknowns to be $K_{11}$, $K_{12}$, $K_{13}$, $K_{22}$ and $K_{23}$.  Then the 4-component of the Codazzi constraint 
does not contain any 1-derivatives and so it is a constraint rather than an ``evolution" 
equation with respect to the auxiliary IDV $x_1$.  

\mbox{ }

\noindent{\bf 2.9.2.3 Bruhat's criticisms and Magaard's argument}

\mbox{ }

\noindent Bruhat's first criticism of elimination method 1 \cite{B56} is that it is a non-covariant 
procedure.  This contributes to it being highly ambiguous, since the nature of the 
prescription depends on the choice of coordinates.  It is also ambiguous because there 
is no unique clear-cut way of choosing which components are to be regarded as the knowns and 
unknowns.  This criticism holds for all the componentwise methods above.   

Bruhat's second criticism of elimination method 1 is that the eliminated system is not valid if the region 
of interest contains zeros of $Z$.  Some form of this criticism holds for all three methods, 
although different kinds of the `Problem of zeros' may occur.  In all these cases, the 
occurrence of zeros indicates that the proof is not necessarily valid everywhere within a 
{\sl given} region of interest.

Whenever the first kind of Problem of zeros occurs,  nevertheless data construction for small 
regions is permitted by {\it Magaard's argument} (originally stated for the particular case below).  
First, one is entitled to declare `data for the data' on some 
$(n - 1)$-d (partial) boundary set 
${\cal X}_1 = \{x_1 = \mbox{ some constant}\}$.  
For, although $Z$ is a function of unknowns in the region of interest away from this set, 
these unknowns are declared to be known on the set itself, so one can choose to 
prescribe them there so that the resulting $Z|_{x_1 = c}$ is bounded away from zero.    
Then by continuity, there are no zeros of $Z$ near the set on which the `data for the data' is
prescribed.   

Magaard's particular method treats all the lower-d metric $h_{ab}$'s components as knowns because his  
aim was to prove an embedding theorem (the Campbell--Magaard theorem).      
Magaard's method is the special case of componentwise algebraic elimination method 1, in 
which  after eliminating $K_{11}$ by use of the Gauss constraint, the Codazzi constraint is 
treated as a p.d.e system for unknowns ${\cal U} \equiv $ \{$K_{1w}$, some other $K_{uv}$ 
component denoted $E$\}.  This satisfies the criteria for the Cauchy--Kovalevskaya theorem if 
one treats ${\cal U}^{\prime} \equiv$ \{all the components of $K_{ij}$ bar $K_{11}$, $K_{1u}$ 
and $E$\} as known functions on $x_1$ and provided that the p.d.e's coefficients and the data 
are analytic. So a unique solution exists.  One can then group \cite{Magaard, ADLR} this method 
and the local existence of a unique evolution to form the Campbell--Magaard statement that a 
($r$, $s$) spacetime with prescribed analytic metric $h_{ab}$ may be embedded with an extra 
space or time dimension for any analytic functional form of the energy-momentum tensor.  
This statement suggests very many embeddings exist, which as argued below and in Part B, is 
a disaster for supposed applicability of these embeddings to build meaningful higher-d worlds in 
brane cosmology or `noncompact KK theory'.     

Magaard's method leads to two prices to pay later on.  Although the `data for the data' (values of ${\cal U}$ 
on ${\cal X}_1 = \{x_1 = \mbox{ some constant }\}$) can be validly picked  so that $Z \neq 0$ on ${\cal X}_1$, 
I note that the protecting continuity argument is only guaranteed to produce a thin strip $0 \leq x_1 < \eta$ 
before zeros of $Z$ develop.  Now, the first price to pay is that one cannot expect to be able to patch 
such strips together to make extended patches of data.  
For, since the strip construction ends where $Z$ picks up a zero for some $x_1 = \eta$, while 
restarting the  procedure with $x_1 = \eta$ in place of $x_1 = 0$ is valid, the two data strips 
thus produced will have a discontinuity across $x_1 = \eta$.  So what one produces is a 
collection of strips belonging to different possible global data sets.  The evolution of each 
of these strips would produce pieces of different higher-d manifolds.  
So the statement that an empty ($n$ + 1)-d manifold `locally surrounds' any $n$-d manifold has a 
more complicated meaning than might be na\"{\i}vely expected.  Concretely, what has been proven 
is that any $n$-d manifold can be cut up in an infinite number of ways 
(choices of the $x_1$ coordinate) into many pieces (which are a priori undetermined), 
each of which can be separately bent in an infinite number of ways 
(corresponding to the freedom in choosing the components of $K_{ab}$ in ${\cal U}^{\prime}$ 
on each set of `data for the data'), and for each of these bent regions one thicken the region 
with respect to the extra dimension $\mu$ to find a piece of an ($n$ + 1)-d manifold.  
Furthermore, all of this can be done for every possible analytic function form of 
$\check{T}_{AB}$ (corresponding to the generalized Campbell--Magaard result) \cite{ADLR}.  
This excess richness compromises attaching any physical significance to any particular such 
construction.  

I found three other indications that Magaard's method is not well thought-out.  
First it is not considered how far the `data for the data' extends along $x_1 = 0$.  Clearly 
the topology of the $n$-d manifold is an important input, for if it is not CWB, 
there is a missing boundary or asymptotics prescription required to make 
Magaard's method rigourous.  Also the topology of the $x_1 = 0$ set itself has not been 
brought into consideration.  For example, the method does not guarantee continuity if 
$x_1 = 0$ contains loops.  It may also 
be that the coordinate condition $x_1 = 0$ breaks down within the region of interest.  In the 
original Campbell--Magaard theorem statement the word `local' is left to cover these and yet 
other aspects!  This makes the theorem less powerful in truth than in superficial appearance.  
 
My second point applies only to the $s = 1$, $\epsilon = 1$ case in Part B, for which 
the existence of an evolutionary region after the data construction step may easily be ruined 
by an `information leak Problem'.  The usual signature case 
$s = 0$, $\epsilon = -1$ does not suffer from this thanks to the DOD property.  
Again this is an implicit limitation, in this case either the 
signature-independence of the theorem or of its physical applicability is found to be untrue 
(depending on how exactly the theorem is phrased).  

Third, Bruhat's second criticism still holds: Magaard's method lacks any 
$n$-d general covariance since it involves the choice of a coordinate $x_1$ and the 
elimination of a ${11}$-component of a tensor.  

The small strip Problem and the second point above are badly convoluted for the Magaard method 
since the zeros of $Z$ are not known until one has solved very cumbersome p.d.e's arising 
from the Codazzi constraint.  This substantially inhibits the construction of specific examples of the 
pathologies and the study of how widespread they might be.  Because of this, my proof 
based on method 2 supercedes Magaard's.  Now, the zeros of $\bar{Z}$ 
are known from the start, making this a cleaner procedure, and also one for which counterexamples 
can be read off: any $n$-metric for which $h^{11}h^{22} - h^{12}h^{12}$ is 0 within the region of 
interest will do.  However, for this proof Magaard's idea of guaranteeing no zeros on the 
`data for the data' itself also fails because the zeros are now entirely controlled by knowns.  
What one has is a statement about existence in certain regions known beforehand.  
A yet more satisfactory method is provided in I.2.9.5.  



\mbox{ }

\noindent\bf{2.9.3 The conformal method of Lichnerowicz and York}\normalfont

\mbox{ }

\noindent The arguments of Lichnerowicz \cite{CGLich} and Bruhat \cite{B56} led to the GR IVP 
taking a very different route from the above sort of brute-force elimination methods.  
Lichnerowicz' method, which was much developed and extended by York 
\cite{York71, York72, York73, MTW}, is instead preferred.  Its treatment below is 
as far as possible for general ($r$, $s$; $\epsilon$), and then specializes to the 

\noindent($n$, 0; --1) and particularly to the (3, 0; --1) case.

\mbox{ } 

In the conformal method one chooses to treat $h_{ij}$ as a known metric which is moreover 
{\sl not} the physical metric but rather only conformally-related to it by 
\be
\tilde{h}_{ij} = \psi^{\eta}h_{ij} 
\mbox{ } ,  
\label{metscal}
\ee
where the \it conformal factor \normalfont $\psi$ is a positive suitably smooth function.  
Then  
\be
\tilde{h}^{ab} = \psi^{-\eta}h^{ab} \mbox{ } ,
\label{confinv}
\ee
\be
\tilde{h} = \psi^{n\eta}h \mbox{ } ,
\label{confdet}
\ee
\be
{\tilde{\Gamma}^a}{}_{bc} = {{\Gamma}^a}_{bc} + \frac{\eta}{2\psi}
\left(
{\delta^a}_b\pa_c\psi + {\delta^a}_c\pa_b\psi - h_{bc}\pa^a\psi
\right) \mbox{ } ,
\label{confChris}
\ee
\be
{\tilde{\Gamma}^a}{}_{ba} ={{\Gamma}^a}_{ba} + \frac{n\eta\pa_b\psi}{2\psi} \mbox{ } , 
\label{confChriscont}
\ee
\be
\tilde{R} = \psi^{-\eta}
\left[
R + \eta(n - 1)
\left(
1 - \eta\frac{n - 2}{4}
\right)
\frac{|D\psi|^2}{\psi^2}   -   \frac{D^2\psi}{\psi} 
\right] \mbox{ } . 
\label{confR}
\ee
A metric is {\it conformally-flat } if there exists a conformal transformation from it to the 
corresponding flat metric.
Furthermore, when it exists ($n \geq 4$), the Weyl tensor serves as a 
{\it conformal curvature tensor} because it has the following properties:
\be
\mbox{1) it is conformally invariant } \mbox{\hspace{1.5in}} 
\widetilde{W}^a{}_{bcd} = {W^a}_{bcd}  \mbox{ }, 
\mbox{\hspace{2in}} 
\ee
\be
\mbox{2) }
\mbox{\hspace{1.35in}}
{W^{a}}_{bcd} = 0 \Leftrightarrow \mbox{ the metric is conformally flat .}
\mbox{\hspace{2.35in}}
\ee

\mbox{ } 

Work in terms of the $K^{\mbox{\scriptsize T\normalsize}}_{ij}$ and $K$ split,\fn{One 
can instead (and in close parallel with the above) choose to work with momenta in 
place of extrinsic curvature.  Most of the applications in thesis are formulated in terms of momenta.} 
and permit $K^{\mbox{\scriptsize T\normalsize}ij}$, $j^i$ and $\rho$ to conformally 
\be 
\mbox{transform according to }
\mbox{\hspace{0.7in}}
\tilde{K}^{\mbox{\scriptsize T\normalsize}ij} = \psi^{\zeta - 2\eta}K^{\mbox{\scriptsize T\normalsize}ij} \mbox{ } , \mbox{ } \mbox{ } \tilde{j}^i = \psi^{\xi} 
\mbox{ } , \mbox{ } \mbox{ } \tilde{\rho} = \psi^{\omega}\rho 
\mbox{ } , 
\mbox{\hspace{2in}}
\label{otherscal}
\ee
whilst crucially demanding the {\it constant mean curvature}\fn{This is the same notion as in I.2.1.} 
(CMC) condition   
\be
K = \mbox{hypersurface constant, } C(\mu)
\ee
holds and is conformally-invariant.  This includes as a subcase Lichnerowicz's earlier use of 
\be 
\mbox{the {\it maximal condition} } 
\mbox{\hspace{2.0in}}
K = 0 
\mbox{ } . 
\mbox{\hspace{2.0in}}
\ee

One then demands that the (raised) Codazzi constraint (\ref{Acod}) is to be 
conformally-
$$
\mbox{invariant.  Since }
\mbox{\hspace{0.4in}} 
\tilde{D}_a\tilde{K}^{\mbox{\scriptsize T\normalsize}ab} = \psi^{\zeta - 2\eta}
\left[
D_aK^{\mbox{\scriptsize T\normalsize}ab} + 
\left(
\zeta - 2\eta + \eta\frac{n + 2}{2}
\right)K^{\mbox{\scriptsize T\normalsize}ab}\frac{D_a\psi}{\psi}
\right]
\mbox{\hspace{0.8in}}
$$
\be 
\mbox{then one requires by (\ref{confChris}) that }
\mbox{\hspace{1.2in}}
-\eta\frac{n + 2}{2} \mbox{ } = \zeta - 2\eta = \xi 
\mbox{ } .
\mbox{\hspace{1.2in}}
\label{twiddle}
\ee
Furthermore it is desired that the conformally-transformed Gauss equation (\ref{Agauss}), i.e 
\be
R - \eta(n - 1)\frac{D^2\psi}{\psi} + \eta(n - 1)\left( 1 - \eta\frac{n - 2}{4}\right)\frac{|D\psi|^2}{\psi^2} + \epsilon(M\psi^{2\zeta - \eta} - 
m^2\psi^{\eta} + 2\rho\psi^{\omega + \eta}) = 0
\label{prolich}
\ee
by (\ref{confR}), where $M = K^{\mbox{\scriptsize T\normalsize}} \circ K^{\mbox{\scriptsize T\normalsize}}$ 
and $m$ is proportional to $K$, should contain no $|D\psi|^2$ term.  Thus,  
$\eta = \frac{4}{n - 2}$,  $\zeta = - 2$ and $\xi = -2\frac{n + 2}{n - 2}$  
[making use of (\ref{twiddle})].  Now, regardless of ($r$, $s$; $\epsilon$), provided that the ($n$ + 1)-d DEC is to be preserved by the
conformal transformation,  $\rho^2$ must conformally-transform like $j^aj_a$, implying that 
$\omega = -2\frac{n + 1}{n - 2}$.  
Then (\ref{prolich}) becomes the ($r$, $s$; $\epsilon$) version of the 
Lichnerowicz--York equation\fn{The main feature due to $s$ is hidden in $D^2$.  I use rather 
$\triangle$ (the Laplacian, which is an elliptic operator) for $s = 0$ and 
$\Box$ (the wave operator, which is a hyperbolic operator) for $s = 1$ to bring out this important difference in applications.} 
\be
D^2\psi = -\frac{\epsilon}{4}\frac{n - 2}{n - 1}\psi
\left( 
-\epsilon R - M\psi^{-4\frac{n - 1}{n - 2} } + m^2\psi^{\frac{4}{n - 2}} - 2\rho\psi^{-2\frac{n - 1}{n - 2}}
\right)
\label{ndlich}
\ee
for the conformal factor $\psi$.  Lichnerowicz's original equation had no $m$ term in it and 
no phenomenological matter term $\rho$; the study of these additional features was carried out 
in the usual (3, 0; --1) case by York and \'{O} Murchadha \cite{cwb}.

Now the solution of the Codazzi constraint is decoupled from the solution of the Gauss constraint.  
The former proceeds by a traceless-transverse (TT)--traceless-longitudinal (TL) splitting \cite{York73, York74}
\be
K^{\mbox{\scriptsize T\normalsize}ij} = K^{\mbox{\scriptsize TT\normalsize}ij} + K^{\mbox{\scriptsize TL\normalsize}ij} \mbox{ } , \mbox{ } \mbox{ } 
D_iK^{\mbox{\scriptsize TT\normalsize}ij} \equiv 0 \mbox{ } , \mbox{ } \mbox{ } 
K^{\mbox{\scriptsize TL\normalsize}ij} = 2\left(D^{(i}W^{j)} - \frac{1}{n}h^{ij}D_cW^c\right) \equiv (|\mbox{L}W)^{ij} 
\label{bigsplit}
\ee
(for some vector potential $W^c$ and for $|\mbox{L}W$ the {\it conformal Killing form}), which 
along with the trace-tracefree split is conformally-invariant \cite{York74} and thus unaffected 
by the solution of the latter, which has become the p.d.e (\ref{ndlich}) for the conformal factor.  
The simplest case is $K^{\mbox{\scriptsize T\normalsize}ij} = K^{\mbox{\scriptsize TT\normalsize}ij}$ 
which is possible for $j^a  = 0$.  Then all one needs is any $K^{\mbox{\scriptsize TT\normalsize}}$.  
The standard case is that $K^{\mbox{\scriptsize TT\normalsize}}$ is known and that one is to solve the Codazzi 
constraint as a second-order p.d.e for the potential $W^i$ out of which $K^{\mbox{\scriptsize TL\normalsize}}$ 
is built.  See C.2--3 for such procedures, along with tricks and simplifications for solving the 
Lichnerowicz--York equation.    

Once the Codazzi constraint has been solved, $K^{\mbox{\scriptsize T\normalsize}}_{ij}$ is known 
so $M$ is known.  Then one can attempt to solve the Lichnerowicz--York equation, which is 
well-studied for the (3, 0; --1) CWB and asymptotically-flat cases.  By the artful construction 
above,
this is a {\sl quasilinear} elliptic equation, permitting the approaches outlined in C.2.2.  
It has been studied including most fundamental matter fields \cite{IOY}, and for Sobolev spaces matching those 
then used in the GR CP \cite{B76, CBY}.  

Note that the choice of using the scale--scalefree decomposition of the metric and the 
trace-tracefree and TT--TL decompositions of the extrinsic curvature are irreducible $n$-d 
generally-covariant choices, a decided advantage over the coordinate-dependent, ambiguous componentwise methods.  
This and the decoupling of the constraints in the conformal method are signature-independent.  
Some of the methods to solve the Codazzi equation can be used regardless of signature, but 
the usual study of the Lichnerowicz--York equation involves elliptic methods which are absolutely not 
generalizable to the $s = 1$ case.

One can attempt to preserve the maximal or CMC conditions away from the `initial' hypersurface 
$\Upsilon_0$, by solving lapse-fixing equations (LFE's, see I.2.10).    
Such choices of slicings deliberately prevent unnecessary focussing of geodesics 
\cite{CGLich, CGSYkin}, enhancing the practical longevity of the 
evolution (the opposite occurs for normal coordinates!).   
However, the conformal method is not absolutely general, 
for some spacetimes may not have any maximal or CMC slice to identify with $\Upsilon_0$ in the 
first place, while in others the maximal or CMC slicing cannot be maintained to 
cover the whole spacetime (see I.2.10).  Furthermore, results concerning this depend on the asymptotics 
assumed.\fn{It was these caveats, along with the reliance on $s = 0$ of York's method and 
restrictions on the values $R$ can take for some subcases of the Lichnerowicz--York equation, 
that made me suspicious of the supposed generality of the Magaard method.}  
However, in the usual 

\noindent(3, 0; --1) case, this method (and its variants) 
is widely accepted as a practical method by the numerical relativity community (see I.2.11).  
The LFE study to date has relied on them being elliptic, and thus 
cannot be simply generalized away from $s = 0$.  

Finally, for $s = 0$ there is the useful property that certain local data patches can be proven to 
suffice for the treatment of astrophysical problems, by building on the notion of DOD.  As promised, 
this sort of technique is also applicable to protect pieces of local data obtained by elimination 
methods such as that of Magaard.

\mbox{ }

\noindent{\bf 2.9.4 Further approaches and related mathematics used in (3, 0; --1) case}

\mbox{ }

\noindent{\bf 2.9.4.1 The conformal thin sandwich method}

\mbox{ }

\noindent This \cite{YCTS, Cook, CIY, PfeifferYork} is a combination of conformal and 
thin sandwich ideas.  There are 5 unknowns, taken to be $\beta_i$, $\psi$ and 
$\alpha_{\mbox{\scriptsize Y\normalsize}} \equiv \psi^6\alpha$, whereas $h_{ij}$ up to 
scale, $\dot{h}_{ij}^{\mbox{\scriptsize T\normalsize}} 
= \psi^{      -{   4    }    }\dot{h}_{ij}^{\mbox{\scriptsize T\normalsize}}$ and the 
hypersurface constant $K$ are taken to be knowns.    
As compared with the original thin sandwich scheme, ${\cal H}$ is now an equation for 
$\psi$ while there is a new equation for the scaled-up lapse.    
Then one writes the definition of 
$K^{\mbox{\scriptsize T\normalsize}}_{ij}$ in the form 
\be
K^{\mbox{\scriptsize T\normalsize}}_{ij} = \frac{\psi^{10}}
{2\alpha_{\mbox{\scriptsize Y\normalsize}}}
[(|\mbox{L}\beta)^{ij} - \dot{h}^{\mbox{\scriptsize T\normalsize}ij}] 
\label{KTdef2}
\ee
and one uses this in the Codazzi constraint (\ref{Acod}) to obtain an equation for $\beta_i$.  
One additionally 
has the Lichnerowicz--York equation (\ref{ndlich}) and 

\mbox{ }

\noindent
\be
-\psi^2D^2\alpha_{\mbox{\scriptsize Y\normalsize}} 
- 14\psi^2\pa_a\phi \pa^a\alpha_{\mbox{\scriptsize Y\normalsize}} 
- 42\pa_a\psi\pa^a\psi \alpha_{\mbox{\scriptsize Y\normalsize}} 
+ \frac{7}{4}\alpha_{\mbox{\scriptsize Y\normalsize}}\psi^{-6}
K^{\mbox{\scriptsize T\normalsize}} \circ K^{\mbox{\scriptsize T\normalsize}} 
- \frac{3}{4}\psi^2\alpha_{\mbox{\scriptsize Y\normalsize}} R 
= - \psi^{-2}\delta_{\check{\beta}}K 
\ee

\mbox{ }

One solves these as coupled equations.  One can then finally reconstruct the 
physical $\tilde{h}_{ij}$ and $\tilde{K}^{\mbox{\scriptsize T\normalsize}ij}$ (and hence 
$\tilde{K}^{ij}$) from (\ref{ecd}) and (\ref{KTdef2}).  The theoretical point of this 
formulation is that everything then scales correctly.  

\mbox{ }

\noindent{\bf 2.9.4.2 Conformal Superspace}

\mbox{ }

\noindent In the CWB case, York showed  that `$\frac{2}{3}$ of Superspace' may be taken to be Conformal 
\be
\mbox{Superspace (CS) 
\cite{York74} }
\mbox{\hspace{0.5in}}
\mbox{CS} = \frac{\mbox{Riem}}{\mbox{3-diffeomorphisms}\times\mbox{conformal transformations}} 
\mbox{\hspace{0.5in}} 
\ee
(see \cite{CGFischMon} for a more recent mathematical study).    
This is the restriction of Riem for which the momentum is transverse ($D_ip^{ij} = 0$) and traceless ($p = 0$).  

The quotienting out of conformal transformations is based on the decomposition 
of arbitrary 3-metrics $h_{ij}$ into their determinant $h$ and their scale-free part  
$h_{ij}^{\mbox{\scriptsize unit\normalsize}} \equiv h^{-\frac{1}{3}}h_{ij}$.
CS is essentially the geodesically-complete space arising in the split (\ref{55BLOCK}).  
There is accumulating evidence that a case can be made for 
CS \sl almost \normalfont corresponding to a representation of the space of true d.o.f's of GR.  
It turns out that to obtain this for GR one must adjoin a solitary 
\be
\mbox{global degree of freedom: 
the volume of the universe, }
\mbox{\hspace{0.7in}}
V=\int h^{\frac{1}{2}}\textrm{d}^{3}x 
\label{CGVolDef} 
\mbox{ } , 
\mbox{\hspace{0.7in}}
\ee
which interacts with the infinitely many local shape d.o.f's represented by ${h}^{\mbox{\scriptsize unit\normalsize}}_{ij}$.  
I investigate theories on CS and `CS+V' in III.2.

\mbox{ }

\noindent{\bf 2.9.4.3 3-d Conformal tensors and the York 1971 formulation}

\mbox{ }
   
\noindent For dimension $n>3$, the Weyl tensor exists and serves as a conformal curvature 
tensor.  
\be
\mbox{For $n = 3$ one may define instead the {\it Bach tensor} }
\mbox{\hspace{0.3in}}
b_{abc} = 2\left(D_{[c}R_{b]a} - \frac{1}{4}D_{[c}Rh_{b]a}\right)
\mbox{\hspace{0.3in}}
\ee
\be
\mbox{and form from it the {\it Cotton--York tensor density} }
\mbox{\hspace{0.5in}}
y^{ab} \equiv B^{ab} = -\frac{1}{2}\epsilon^{ade}h^{bf}b_{fde}
\mbox{\hspace{0.5in}}
\ee
which is a conformal curvature tensor.  

This is used first in York's earliest (1971) 
formulation of the GR IVP, which is cast in a form which looks closely-analogous to that of 
electromagnetism:
\be
D_bB^{ab} = 0 \mbox{ } , \mbox{ } \mbox{ } D_bE^{ab} = 0
\ee
\be
B = E = 0
\ee
for $E^{ab} = h^{\frac{1}{3}}p^{ab}$ the momentum density of weight $\frac{5}{3}$.  
However, instead of the physical gauge freedom of electromagnetism, {\sl what is physical here 
is the gauge fixed by solution of the original (vacuum maximal) Lichnerowicz equation}.  

\mbox{ }

\noindent{\bf 2.9.5 A new elimination method in terms of irreducibles}

\mbox{ }

\noindent The route to avoid Bruhat's non-covariance criticism and the ensuing ambiguities of 
procedure is to work not with components but with irreducibles.  The conformal method uses the 
scale--scalefree split of the metric and the TT--TL splits of the extrinsic curvature.
It is then declared that the $n + 1$ unknowns are the single scale of the metric and the 
TL part of the extrinsic curvature encapsulated in the $n$-vector potential $W_i$.  

With this hindsight I constructed an \it irreducible elimination method \normalfont. 
Write the constraints in terms of the trace-tracefree decomposition.  
Now consider the Gauss equation as an 
 
\be
\mbox{algebraic equation for the trace part $K$: } 
\mbox{\hspace{0.4in}}
K = \sqrt{\frac{n}{n - 1}}\sqrt{    K^{\mbox{\scriptsize T\normalsize}}\circ 
K^{\mbox{\scriptsize T\normalsize}} - \sigma R + 2\rho    } 
\mbox{ } ,
\mbox{\hspace{0.4in}}
\ee
and substitute this into the Codazzi equation to obtain
\be
D^iK^{\mbox{\scriptsize T\normalsize}}_{ij} - \sqrt{\frac{n - 1}{n}}D_j
\sqrt{    K^{\mbox{\scriptsize T\normalsize}}\circ K^{\mbox{\scriptsize T\normalsize}} 
- \sigma R + 2\rho    } = j_j 
\mbox{ } . 
\ee
Now treat this as an $n$ equations for the $n$ unknowns $K^{\mbox{\scriptsize TL\normalsize}}_{ab}$ encapsulated in the $n$-vector 
potential $W_i$:
\be
(K\delta^p_jh^{qr} - \delta^r_jK^{{\mbox{\scriptsize T\normalsize}}pq})
D_r
\left[
K^{\mbox{\scriptsize TT\normalsize}}_{pq} + 2
\left(
D_{(p}W_{q)} - \frac{1}{n}h_{pq}D_kW^k
\right)
\right]
= Kj^j - \frac{1}{2}D_j(\epsilon R + 2\rho)
\label{irredelcod}
\ee
where $K$ and $K^{\mbox{\scriptsize T\normalsize}}_{ij}$ are treated as functions of 
$W_l$, $\pa_mW_n$ and knowns 
[which may be easily written down using (\ref{ecd}) and (\ref{bigsplit})].  
See C.3 for more general consideration of this equation.  It suffices to say that it is 
quite a complicated equation, but then so is the thin sandwich equation or the conformal thin 
sandwich system.  

Here I note that if $x_1$ is declared to be an auxiliary IDV, then if two certain functions do not 
have any zeros in the region of interest, the system (\ref{irredelcod}) may be cast into second-order 
Cauchy--Kovalevskaya form.  Very quickly, isolating the relevant terms of the $u$-component
\be 
\mbox{of the Codazzi equation leads to }
\mbox{\hspace{0.65in}}  
\pa^2_1W_u = F_u(W_i, \pa_jW_k; \mbox{ knowns}) \mbox{ } , 
\mbox{\hspace{1in}}
\label{fstCK}
\ee
provided that division by $Kh^{11}$ is valid.  
The 1-component gives 
$$
2
\left(
Kh^{11}\frac{n - 1}{n} - K^{{\mbox{\scriptsize T\normalsize}}11}
\right)
\pa_1^2W_1 + 
\left(
Kh^{1u}\frac{n - 2}{n} - 2K^{{\mbox{\scriptsize T\normalsize}}1u}
\right)
\pa^2_1W_u
= F_1(W_i, \pa_jW_k; \mbox{ knowns})
$$
which upon use of (\ref{fstCK}) and provided that division by 
$Kh^{11}\frac{n - 1}{n} - K^{{\mbox{\scriptsize T\normalsize}}11}$ 
is valid gives 
\be
\pa_1^2W_1 = \bar{F}_1(W_i, \pa_jW_k; \mbox{ knowns}) \mbox{ } .
\ee
This provides a method of proof of the Campbell result which is less prone to Bruhat's two 
criticisms since its Problem of zeros is of the third kind so at least the eliminated system 
is always valid, and the system is built in an unambiguous generally-covariant manner.  It will 
still be prone to the information leak Problem if $s = 1$, but as a genuine GR IVP method ($s = 0$) 
it is protected by the DOD property.  

So whilst some aspects of the old thin sandwich scheme have resurfaced again in the 
modern conformal thin sandwich scheme, the above is an attempt to have some aspects of old 
elimination schemes resurface in a new method with enough of the old faults corrected to 
be of potential interest to numerical relativity.   

All quantities prescribed in my method are the physical ones.  Although it is an elimination method, 
it is carefully thought-out: the eliminated quantity has no prefactor and hence no associated 
Problem of zeros invalidating the eliminated system in certain places, and the procedure is 
uniquely-defined and coordinate-independent.  In my method, one is faced with a single, more 
difficult vector p.d.e in place of the simpler Codazzi vector p.d.e followed by solving the 
decoupled Lichnerowicz--York equation.  
%

\mbox{ }

\noindent{\bf 2.9.6 The IVP with fundamental matter and for alternative theories of gravity}

\mbox{ }

\noindent Minimally-coupled scalars offer no complications.  In the Einstein--Maxwell IVP 
one has to scale 
$\tilde{E}_i = \psi^{-6}E_i$ and 
$\tilde{\rho}_{\mbox{\scriptsize e\normalsize}} = 
\psi^{-6}{\rho}_{\mbox{\scriptsize e\normalsize}}$ 
so that the electromagnetic Gauss constraint (\ref{curemgau}) is conformally-invariant \cite{IOY}.  This idea 
is considerably generalizable, both to more complicated matter theories and to large classes of 
alternative theories of gravity \cite{IOY, IN}.  I note however 1) that the preservation of energy conditions 
is not always applicable (also mentioned in \cite{CIY}) if complicated enough matter is included 
and 2) that higher-derivative theories remain uninvestigated and may cause significant difficulties 
if required for numerical relativity through having a more complicated IVP than GR.

\subsection{Maximal and CMC slicings}

This section is for the (3, 0; --1) case and in terms of momenta to match its application in III.2.  
\be
\mbox{The maximal condition for an embedded spatial hypersurface is  } 
\mbox{\hspace{0.65in}}
p = 0 
\label{CGtraceless} 
\mbox{ } .
\mbox{\hspace{0.65in}}
\ee 
\be
\mbox{The CMC condition is  } 
\mbox{\hspace{1in}}
\tau_{\mbox{\scriptsize Y\normalsize}} \equiv \frac{2p}{3\sqrt{h}} = C(\lambda) \mbox{ } 
,\mbox{ a spatial constant} 
\label{Yorktime} 
\mbox{ } .
\mbox{\hspace{2in}}
\ee 
In GR, one regards (\ref{CGtraceless}) and (\ref{Yorktime}) 
as maximal and CMC {\sl gauge conditions} respectively. 

From the scaling of the tracefree extrinsic curvature in I.2.9.3, 
the tracefree momentum 
\be
\mbox{scales as }
\mbox{\hspace{2.2in}}
\tilde{p}^{\mbox{\scriptsize T\normalsize}ij} = \phi^{-4}p^{\mbox{\scriptsize T\normalsize}ij}
\mbox{\hspace{2.2in}}
\ee
since it is a density of weight 1.  The original (vacuum maximal) Lichnerowicz equation now 
\be
\mbox{takes the form }
\mbox{\hspace{1.3in}}
8\triangle\phi + M\phi^{-7} - R\phi = 0 \mbox{ } , \mbox{ } \mbox{ }
hM \equiv p \mbox{ } \circ \mbox{ } p \geq 0 
\mbox{ } , 
\mbox{\hspace{1.0in}}
\label{maxvaclich}
\ee 
whereas the vacuum CMC Lichnerowicz--York equation (\ref{ndlich}) is 
\be
8\triangle\phi + {\cal M}\phi^{-7} - R\phi  + m^2\phi^5 = 0 \mbox{ } , \mbox{ } \mbox{ } 
h{\cal M} \equiv p^{\mbox{\scriptsize T\normalsize}} \circ p^{\mbox{\scriptsize T\normalsize}} \geq 0
\label{cmcvaclich}
\ee

It is important to distinguish between a single initial use of a 
slicing condition in order to find consistent initial data and 
subsequent use of the slicing when the obtained initial data are
propagated forward. This is by no means obligatory. The EFE's are
such that once consistent initial data have been found they can be
propagated with freely specified lapse and shift. This is
precisely the content of 4-d general covariance. 

If one wishes to attempt to maintain the maximal gauge condition 
(\ref{CGtraceless}), $\alpha$ must satisfy 
\be
\mbox{the maximal {\it lapse-fixing equation (LFE) } }
\mbox{\hspace{1in}}
\frac{\alpha}{h} p \mbox{ } \circ \mbox{ }  p - \triangle\alpha = 0 
\mbox{ } .
\mbox{\hspace{3in}}
\label{CGMSE}
\ee
In this thesis, I call such equations as LFE's, 
because, since GR is not always be presupposed, I do not always work in a context where 
the notion of slicing of GR-like spacetime makes sense.  Being homogeneous, 
(\ref{CGMSE}) does not fix $\alpha$ uniquely but only up to global 
$\lambda$-reparameterization 
$\alpha \longrightarrow f(\lambda)\alpha$, 
where $f(\lambda)$ is an arbitrary monotonic function of $\lambda$.  

Similarly, to maintain the CMC gauge condition during the evolution, it is necessary to 
choose the lapse $\alpha$ in such a way that it satisfies the CMC LFE
\be
2
\left(
\frac{\alpha}{h}p\circ p - \triangle\alpha 
\right) 
- \frac{\alpha p^2}{2h} = B(\lambda) = \frac{\pa}{\pa\lambda}
\left(
\frac{p}{\sqrt{h}}
\right) 
\mbox{ } .
\label{CGLapseFixing}
\ee

The above lapse-fixing equations are for vacuum GR, following from the vacuum version of the 
first form of (\ref{trevK}).  The versions with matter are likewise easily obtained from the 
full equation (\ref{trevK}).  

Maintenance of CMC slicing yields a foliation that is extremely convenient in the case of 
globally hyperbolic spatially-CWB spacetimes.  The foliation is unique \cite{CGunique, MT}, 
and the value of $\lambda$ increases monotonically, either from 
$-\infty$ to $\infty$ in the case of a Big-Bang to Big-Crunch cosmological solution or from 
$-\infty$ to zero in the case of eternally expanding universes.  In the first case, the 
volume of the universe increases monotonically from zero to a maximum expansion, at which 
the maximal condition (\ref{CGtraceless}) is satisfied, after which it decreases 
monotonically to zero.  In CWB GR, the total spatial volume cannot be maintained constant 
except momentarily at maximum expansion, when 
$\tau_{\mbox{\scriptsize Y\normalsize}} = \frac{2p}{3\sqrt{h}} =0$.  Thus, in CWB GR the volume is a 
dynamical variable.  This corresponds to the maximal LFE not being soluble in this case 
(see C.1).  The above properties of $\tau_{\mbox{\scriptsize Y\normalsize}}$ allow its 
interpretation as a notion of time, the extrinsic {\it York time}
(see e.g \cite{POTlit2, BEAR}).  

The maximal and CMC gauges in GR exhibit the following gauge artifact known as the 
{\it collapse of the lapse}:  a blowup in $R$ (for example in gravitational collapse) tends 
to drive $\alpha$ to zero.  Thus these gauges are {\it singularity-avoiding}.  But not all 
singularities are avoided: Eardley and Smarr \cite{Eardley} found an example of 
Tolman--Bondi model in which gravitational collapse is too sudden (i.e the curvature profile 
is too spiked) so that the LFE does not manage to drive $\alpha$ to zero before the 
singularity is reached.  Such singularity-avoiding gauges are not necessarily a good 
numerical strategy to use \cite{BSrev} because they cause other difficulties though stretching 
the numerical grid.  The CMC gauge is also not always applicable: not all GR spacetimes are 
CMC sliceable, nor is a CMC slicing necessarily extendible to cover the maximal analytic 
extension of a spacetime \cite{CGSYkin}. 

Note that the above are only partial gauge fixings. The shift $\beta_i$ is as yet 
unspecified.  One numerically-useful choice for this is to use the shift given by the 
minimal distortion condition \cite{CGSYkin}.  Another choice (which in fact generalizes this) 
is to obtain both $\alpha$ and $\beta_i$ from solving the conformal thin sandwich.  

Finally, note that one is entirely free to use whatever gauge in GR.  In each case the gauge 
can be treated as above in terms of equations for $\alpha$ and $\beta_i$ (e.g see II.3.2 for the 
normal gauge, or \cite{CGSYkin} for the harmonic gauge, which turns out to be closely analogous 
to the electromagnetic Lorenz gauge).

\subsection{Numerical applications}

Solving the conformal IVP equations is central to the study of binary compact object data in 
numerical relativity.  Often simplified versions of these equations are considered 
(as explained in App C).  The simplest treatments predate York's work: the Misner 
\cite{Misnerdata} and Brill--Lindquist \cite{BLdata} multiple black hole data solve just the 
flat-space Laplace equation case.  Lindquist \cite{Lindquist63} also treated 
Einstein--Maxwell wormholes in the two flat-space Laplace equation case.  The Bowen--York 
data \cite{BY} solve another simple case (a system of flat-space Poisson equations).    
Use of the conformal thin sandwich formulation in data construction is currently popular 
\cite{BSrev}, partly because of its practical motivation, since unlike for the simplification 
$K_{ij} = 0$ which is trivial because it implies $p^{ij} = 0$ and momentum is conserved, 
$\dot{h}^{\mbox{\scriptsize T\normalsize}}_{ij} = 0$ is rather the condition 
for (quasi)equilibrium, which need not be maintained at later times.    

The motivation to consider more complicated data is physical and is to be tested by 
comparison with gravitational wave signals expected to be observed over the next few years 
\cite{CutThorne}.   The numerical relativity side of this will require reasonably long-lived 
simulations (relative to the orbital period) of evolution of realistic data.  The viability of 
use of the conformal method alone requires the tractability of gravity wave emission within this 
formalism, for which there is recent evidence \cite{Gowdy}.  Traditional binary black hole 
evolution simulations have been of head-on and hence axisymmetric collisions 
\cite{Smarrsim, Annisim}.  Modern work attempts to study more likely collisions following from 
the more general modern data above.  Excision of black hole interiors (see C.2.3) is currently 
favoured over use of singularity-avoiding gauges. Nevertheless, simulations to date have been 
numerically unstable on small timescales (see e.g \cite{BSrev}).  The study of binary neutron star 
evolution is currently more fruitful \cite{Shibata}. 

\mbox{ }

\mbox{ }

\noindent{\large{\bf 2.12 Further formulations of the EFE's} }

\mbox{ }

\noindent The EFE's have been studied in many guises for a number of different purposes.  
The guises may roughly be qualified 
by what geometrical object represents the gravitational field (metric or `bein'), 
by how this is split (a matter of signatures) 
and how this split is to be interpreted (a matter of prescriptions), 
by how the split equations are pieced together to make systems of particular forms, 
and by how extra variables may be introduced.  
Among the purposes are both practical and theoretical aspects of numerics, and quantization.  

\mbox{ }

\noindent{\bf 2.12.1 Metric formulations}

\mbox{ }

\noindent If one uses the metric, one may then apply the (3, 0; --1) ADM split.  
The resulting system (\ref{Vham} \ref{Vmom}, \ref{CGsmallevol}, \ref{CGBSWEL}) may then be 
modified to form the equivalent but numerically better-behaved {\it BSSN} system \cite{BSrev} 
by use of conformal IVP-like variables, adding constraints 
to the evolution equations and using $\check{\Gamma}^A$ as new variables.  One can also form larger 
`Einstein--Ricci' or `Einstein--Christoffel' \cite{AY} systems, seeking for particular kinds of 
hyperbolicity, which give good theorems to protect the numerics.  One could also use 
an `Einstein--Weyl' system with extra `electric' and `magnetic' Weyl tensor variables.  
One such formulation is the {\it threading formulation} \cite{Hawkingthread, Ellisthread} in which the fluid flow congruence (rather 
than hypersurfaces perpendicular to it) is treated as primary, on the grounds that the information 
available to us as observers is on incoming geodesics and not on some spatial surface.  
One has then a `deliberately incomplete' system if regarded from the foliation perspective.  
Weyl variables are also employed in \cite{SMS} (used in Part B).

\mbox{ }


\noindent{\bf 2.12.2 `Bein' formulations}
\be
\mbox{One can start again using `beins' ${\mbox{\sc e}_A}^{\bar{A}}$ such that } 
\mbox{\hspace{0.8in}}
g_{AB} = {\mbox{\sc e}_A}^{\bar{A}}\mbox{\sc e}_{B{\bar{A}}} 
\mbox{ } , 
\mbox{\hspace{1.5in}}
\ee

\noindent or spacetime spinors.  This is useful to accommodate Dirac fields and in the study of 
supergravity.  The unsplit spinor formulation is useful in the study of exact solutions.  One 
can again do a (3, 0; --1) split (see VI.4).  Note that there are then additional `frame rotation' constraints 
${\cal J}_{\mu\nu}$.  A somewhat different `bein' gives Ashtekar variables (see below), 
of importance in quantization attempts.  

\mbox{ }

\noindent{\bf 2.12.3 Ashtekar variables}

\mbox{ }

\noindent Pass from ($h_{ab}$, $p^{ab}$) to a SU(2) connection 
${\mbox{\tt A\normalfont}_a}^{\mbox{\scriptsize\tt AB\normalfont\normalsize}}$ and 
its conjugate momentum 
${\mbox{\tt E\normalfont}^a}_{\mbox{\scriptsize\tt AB\normalfont\normalsize}}$ 
[which is related to the 3-metric by 
$h_{ab} = - tr(\mbox{\tt E\normalfont}_a\mbox{\tt E\normalfont}_b)$].\fn{The capital 
typewriter indices denote the Ashtekar variable use of internal spinorial SU(2).  
$tr$ denotes the trace over these. $\mbox{\bf\tt D\normalfont\normalfont}_a$ is the SU(2) 
covariant derivative as defined in the first equality of (\ref{ashgauss}).}  
The constraints are
\be
\mbox{\bf\tt D\normalfont}_a{\mbox{\tt E\normalfont}^a}_{\mbox{\scriptsize\tt AB\normalfont\normalsize}} 
\equiv \pa_a{\mbox{\tt E\normalfont}^a}_{\mbox{\scriptsize\tt AB\normalfont\normalsize}}  +  
|[\mbox{\tt E\normalfont}_a, \mbox{\tt E\normalfont}^a]|_{\mbox{\scriptsize\tt AB\normalfont\normalsize}} 
= 0 
\label{ashgauss} 
\mbox{ } , 
\ee
\be
tr(\mbox{\tt E\normalfont}^a \mbox{\tt F\normalfont}_{ab}) = 0 
\label{ashmom} 
\mbox{ } , 
\ee
\be
tr(\mbox{\tt E\normalfont}^a\mbox{\tt E\normalfont}^{ab}\mbox{\tt F\normalfont}_{ab}) = 0 
\label{ashham} 
\mbox{ } .
\ee
(\ref{ashgauss}) arises because one is using a first-order formalism.   Note that in this 
Ashtekar formalism \cite{Ashtekar} it has the form of an SU(2) Yang--Mills constraint.  
(\ref{ashmom}) and (\ref{ashgauss}) are the polynomial forms that the momentum and 
Hamiltonian constraints respectively, where 
$\mbox{\tt F\normalfont}^{\mbox{\scriptsize\tt AB\normalfont\normalsize}}_{ab} 
\equiv 2\partial_{[a}\mbox{\tt A\normalfont}^{\mbox{\scriptsize\tt AB\normalfont\normalsize}}_{b]} 
+ |[\mbox{\tt A\normalfont}_a, \mbox{\tt A\normalfont}^a_b]|^{\mbox{\scriptsize\tt AB\normalfont\normalsize}}$ 
is the field strength corresponding to ${\mbox{\tt A\normalfont}_a}^{\mbox{\scriptsize\tt AB\normalfont\normalsize}}$. 
One can see that (\ref{ashmom}) is indeed associated with momentum since it is the condition for a vanishing 
Poynting vector.  The Hamiltonian constraint $(\ref{ashham})$ has no such clear-cut interpretation.  

It is important for the formulation that ${\mbox{\tt A\normalfont}_a}^{\mbox{\scriptsize\tt AB\normalfont}}$ 
is a self-dual connection, which makes this formulation specific to dimension 4.  One has more or less a map 
of the GR phase space into the Yang--Mills one, which is exploited below.  `More or less' means that the 
asymptotics used are different, and also that the GR phase space has been enlarged to include the degenerate 
metrics.  Finally, the Ashtekar variables formulation is of complex GR.  But one requires troublesome 
\it reality conditions \normalfont (see I.3.3.3) in order to recover real GR.  Nowadays so as to 
avoid reality conditions, one usually prefers to work not with Ashtekar's original complex variables but with 
Barbero's real variables \cite{Barbero}.  

\mbox{ }

\noindent{\bf 2.12.4 Formulations using different splits}

\mbox{ }

\noindent Finally, one need not stick to using a (3, 0; -1) or indeed ($r$, $s$; $\epsilon$) split.  
One could split with respect to null surfaces: the characteristic formulation \cite{Winicour}.  
This may be useful numerically; it is well-suited to the study of gravitational waves.  
One might use a mixture of Cauchy and characteristic formulations for this purpose.  The difficulty 
then is in how to match the two regions.  One could also split in the `2 + 2' way \cite{2+2}, 
in which the gravitational d.o.f's are isolated as 
the conformal 2-metric.  These are not yet fully explored options.  

\vspace{3in}

\section{Quantum physics}

\subsection{Finite systems}

This might be based on the postulates  

\noindent {\bf QM1} Associated with each physical state s of a system, there exists a 
{\it wavefunction} $\psi_{\mbox{\scriptsize s\normalfont}}$ which is a ray in a complex 
vector space {\sl V}.  

\noindent {\bf QM2} Classical quantities $A$ have associated hermitian operators $\hat{A}$.  
The form of $A$ in terms of the fundamental variables 
$q_{\mbox{\sffamily\scriptsize A\normalfont\normalsize}}$,
$p^{\mbox{\sffamily\scriptsize A\normalfont\normalsize}}$ is supposedly reflected 
by the form of $\hat{A}$ in terms of $\hat{q}_{\mbox{\sffamily\scriptsize A\normalfont\normalsize}}$,  
$\hat{p}^{\mbox{\sffamily\scriptsize A\normalfont\normalsize}}$.   

\noindent {\bf QM3} Measurements and probabilities: {\sl V} has an associated inner product 
(i.p) 

\noindent $<||>: \mbox{{\sl V}}\times\mbox{{\sl V}} \longrightarrow \Re$ 
so that {\sl V} is Hilbert.  This i.p admits a probabilistic interpretation:  

\noindent$<\psi|\hat{A}|\psi>$ is the expectation value of $\hat{A}$, and $<\psi_1|\psi_2>$ is the 
overlap probability or transition amplitude between states 1 and 2. The i.p is required to 
be positive-definite to permit the probabilistic interpretation, and normalizable.   
By the `Copenhagen' interpretation, making measurements supposedly entails collapse 
of the wavefunction from a probability distribution of states to a particular state, so that  
the measurement yields a single real number (the eigenvalue of that state).  

\noindent {\bf QM4} The wavefunction unitarily evolves in time.  
For a nonrelativistic quantum system this evolution is given by a {\it time-dependent 
Schr\"{o}dinger equation (TDSE)}
\be
\hat{\mbox{\sffamily H\normalfont}}|\psi> = i\pa_t|\psi> 
\mbox{ } . 
\ee
This can yield a {\it time-independent Schr\"{o}dinger equation (TISE)} for stationary systems   
\be
\hat{\mbox{\sffamily H\normalfont}}|\psi> = \mbox{\sffamily E\normalfont}|\psi>
\ee
\be
\mbox{by separation of variables.  
The Schr\"{o}dinger i.p is }
\mbox{\hspace{0.1in}}
<\psi_1|\psi_2> = \int_{\Re^3}d\sigma_{\bar{p}}\frac{1}{2i}\delta^{\bar{p}\bar{q}}\psi_1
\stackrel{\longleftrightarrow}{{\pa}_{\bar{q}}}\psi_2 
\mbox{ } . 
\mbox{\hspace{0.1in}} 
\ee

The canonical procedure is to pass from Poisson brackets to 
{\it equal-time commutation } 
\be
\mbox{relations} (ETCR's)
\mbox{\hspace{0.8in}}
|[\hat{q}_{\mbox{\sffamily\scriptsize A\normalfont\normalsize}} \mbox{ } , \mbox{ } \hat{p}^{\mbox{\sffamily\scriptsize B\normalfont\normalsize}} ]| 
= i{\delta_{\mbox{\sffamily\scriptsize A\normalfont\normalsize}}}^{\mbox{\sffamily\scriptsize B\normalfont\normalsize}} 
\mbox{ } , \mbox{ }  
|[\hat{q}_{\mbox{\sffamily\scriptsize A\normalfont\normalsize}} \mbox{ } , \mbox{ } \hat{q}_{\mbox{\sffamily\scriptsize B\normalfont\normalsize}} ]| 
=
|[\hat{p}^{\mbox{\sffamily\scriptsize A\normalfont\normalsize}} \mbox{ } , \mbox{ } \hat{p}^{\mbox{\sffamily\scriptsize B\normalfont\normalsize}} ]| 
= 0 
\mbox{ } . 
\mbox{\hspace{1in}}
\ee
I will always choose to use the {\it position representation} 
in setting up quantum operators:
\be
\hat{q}_{\mbox{\sffamily\scriptsize A\normalfont\normalsize}} 
= q_{\mbox{\sffamily\scriptsize A\normalfont\normalsize}} 
\mbox{ } , \mbox{ }  
\hat{p}^{\mbox{\sffamily\scriptsize A\normalfont\normalsize}} 
= -i\frac{\pa }{\pa q_{\mbox{\sffamily\tiny A\normalfont\normalsize}}}
\mbox{ } .
\ee
Obtain the Hamiltonian in terms of these and solve for the eigenspectrum and eigenfunctions (wavefunctions).  
To have the standard interpretation available, one must know which inner product to use.    

The {\it semiclassical approximation}, tied to Hamilton--Jacobi theory, gives a sometimes useful approximate 
view of wavefunctions peaking around classical trajectories.

\subsection{Infinite systems: quantum field theory}

\be
\mbox{If one attempts to use }
\mbox{\hspace{1.0in}}
<\psi_1|\psi_2> = \int_{\Re^3}d\sigma_{\bar{P}}\frac{1}{2i}\eta^{\bar{P}\bar{Q}}
\psi_1\stackrel{\longleftrightarrow}{{\pa}_{\bar{Q}}}\psi_2
\mbox{\hspace{1.0in}}
\ee
in Klein--Gordon theory, the probabilistic interpretation is ruined because 
$\eta^{\bar{P}\bar{Q}}$ is now indefinite.  Fortunately, if $\varsigma$ and 
$\pi_{\varsigma}$ are promoted to operators, one happens to obtain an acceptable i.p 
(this is now a {\sl many} particle interpretation).  
Quantize canonically by imposing the obvious ETCR's, obtaining 
$\hat{\mbox{\sffamily H\normalfont}}$ and solving it e.g in close analogy with an infinite 
collection of uncoupled harmonic oscillators.  

Considering instead Dirac theory adds little: the fields being fermions, 
one imposes ET{\sl anti}CR's, and being charged, one uses distinct oscillators for $e^-$ and $e^+$.  
But the quantum theory of the electromagnetic field is complicated by its constraint. 
Although electromagnetism is tractable, theories with constraints ${\cal C}_X$ generally 
lend themselves to ambiguity of procedure: should these be imposed classically (which may 
complicate the ETCR's), or quantum-mechanically as restrictions $\hat{{\cal C}}_X |\psi> = 0$ 
on the permissible wavefunctions?    The latter case generally suffers from 
$\hat{{\cal C}}_X|\psi> = 0 \not\Rightarrow |[\hat{{\cal C}}_X, \hat{{\cal C}}_Y]||\psi> = 0$ 
unless one is lucky with the operator ordering.  Interacting QFT's are treated as perturbative 
expansions in the coupling constants, leading to the QED, Weinberg--Salam and QCD successes 
mentioned in I.1.7.  

QFT has well-known technical difficulties.  Corrections due to loop contributions may destroy 
classical symmetries: {\it anomalies}.  Integrals may not be well-defined, requiring 
{\it regularization}.  Also, some theories require {\it renormalization} (counterterms in order 
to get finite answers), whereas others are unsatisfactory in being nonrenormalizable.  
More details of na\"{\i}ve renormalizability are used in IV.

\subsection{Approaches to quantum gravity}

\noindent\bf{3.3.1 Outline of need for and approaches to quantum gravity}\normalfont

\mbox{ }

\noindent Given quantum physics holds for three of the four fundamental forces and for matter, 
it is unseemly for the involvement of gravitation to cheat quantum physics (see \cite{Carliprev} for references).    
Without further development, the EFE's read 
$$
\left(
\begin{array}{c}
\mbox{classical geometry 
as curvature} 
\end{array}
\right) = 
\left(
\begin{array}{c}
\mbox{quantum matter expectation value} \\
\mbox{as energy-momentum}  
\end{array} 
\right) \mbox{ }.  
$$
A fully quantum description would be more satisfactory.  Whereas all everyday physics can be 
understood by neglecting either GR or QM (consider the spacetime curvature inside CERN or the 
Compton wavelength of Mercury), {\it Planckian} regions where neither are negligible are conceivable.  
This situation is thought to occur in the very early universe and in black hole physics (both 
associated with classical GR breakdown at singularities), as well as possibly in very fine 
detail of space or spacetime structure.
 
Quantizing gravity is hard; one talks mostly of programs.  Some have failed.  Others have 
serious theoretical difficulties (see below).  None has predicted anything that is testable 
so as to provide convincing evidence.  Even if some genuinely theoretically-flawless theory is 
established, what of inequivalent quantizations of the same scheme? And one can never dismiss the 
possibility of entirely different flawless theories being subsequently discovered.  Without 
confirmatory (and discriminatory) experiments, physics reverts from Galilean to Aristotelian, 
regardless of the blinding usage of modern mathematics.  A lot of research is going into 
phenomenology!

There are two traditional branching programs for quantizing gravity: the canonical 
(3.3.2) and the covariant (3.3.4).  Their woes are explained below.  Another classification 
of attempts is into `top-down' which take the known laws and attempt to deduce quantum 
gravitational behaviour, and `bottom-up' which guess at the constituent structure of nature 
and attempt to recover known laws.  The former aims at respectful extrapolation, whereas the 
latter may benefit from (perhaps new, perhaps rich) mathematical structure assumed, at the 
price of present unjustifiability of such assumptions.  Additionally, it may be incomplete or 
generalized QM or GR that is involved.  

Note that the above remain programs, as is made clear by the critical nature of many papers in the 
subject.  There are mixed-ontology and nonstandard versions of these programs, while other 
practitioners have had their reasons to consider completely different schemes 
\cite{Carliprev, qgravrevI00}.  With the full theories being hard, toy theories (3.3.3--4) are frequently 
used to voice both innovations and criticisms.  This thesis extends or stems from Barbour's 
attempt toward quantization, which has closed-universe, known-law, canonical, traditional 
variable connotations.  Innovation vies with criticism; old toys are used and new toys (or just 
possibly alternative theories) emerge.  

\mbox{ }

\noindent\bf{3.3.2 The canonical approach and the Problem of time}\normalfont

\mbox{ }

\noindent 
Note how QM uses external time in {\bf QM4} whereas there is no such thing in GR.\fn{QFT in 
spacetimes such as Minkowski relies on timelike killing vectors which in turn are only present 
due to high symmetry, whereas generic solutions of GR have no symmetry.}  
So what would ETCR's now be?  Should one seek to identify a time before or after quantization 
or not at all?  With its Hamiltonian being zero, GR looks rather like a (\sffamily E \normalfont = 0) 
TISE, so is it frozen rather than unitarily-evolving?  These are some aspects of the {\it 
Problem of Time}.  GR is constrained, so should one quantize it before or after 
constraining?\fn{A third option is to take ${\cal H}_i$ into account classically and 
${\cal H}$ into account quantum-mechanically: {\it Superspace quantization}.}   
There are also Problems with finding i.p's and observables, and operator ordering, 
regularization and defining measures, depending on what one attempts.\fn{In everyday QM, 
experiment would settle any such ambiguities and support the correctness of theoretical 
technicalities employed.} And these Problems are interlinked.  
Conceptually, why is nature classical to good approximation?  
Semiclassical Hamilton--Jacobi interpretations run into `what is a superposition of geometries?' 
Embeddability is surely lost.  
Is nature a spacetime foam leading to loss of causality and (or) permissible topology change 
\cite{Wheeler}?  Also, the `Copenhagen' connotations in {\bf QM3} of the centrality of 
measurement by classical observers cleanly external to the observed system make no sense 
for a closed quantum universe.  One usually gets round this one by using rather a 
{\it many-worlds interpretation} rather than a wavefunction collapse interpretation 
\cite{DeWitt, Hartlerev}.  
 
Choosing to work in traditional variables $h_{ab}(x)$, $p^{ab}(x)$, one requires a fixed choice of the 
3-space topology.  Choosing to quantize before constraining, I elevate these to operators and 
choose the configuration representation (analogue of the position representation)\fn{I always make 
this choice in this Thesis. I use `` " to denote heuristic rather than well defined tractable expressions.} 
\be
h_{ij}(x) \longrightarrow \hat{h}_{ij}(x) = h_{ij}(x) \mbox{ } , \mbox{ } 
p^{ij}(x) \longrightarrow \hat{p}^{ij}(x) = ``-i\frac{\delta}{\delta h_{ij}(x)}" 
\mbox{ } . 
\ee 
Furthermore one can choose to pass from (\ref{gpb}) to standard ETCR's
\be
|[ \hat{h}_{ab}(x) \mbox{} , \mbox{} \hat{p}^{cd}(y) ]| = i{\delta_{(ab)}}^{cd} \overline{\delta}^{(3)}(x, y) 
\mbox{ } .  
\ee

\be
\mbox{The momentum constraint can then be chosen to read} \hspace{0.4in} D_b\frac{\delta }{\delta h_{ab}(x)}|\Psi> = 0
\mbox{\hspace{0.4in}}
\ee 
(ordered with $h_{ij}$ to the left) which signifies that $|\Psi>$
depends on the 3-geometry alone.
The Hamiltonian constraint ${\cal H}$ becomes the Wheeler--DeWitt equation (WDE)\fn{This is  
written here with the $\hat{p}^{ij}$ to the right of the $\hat{h}_{ij}$ 
ordering, and the heuristics include a lack of consideration for regularization.} 
\be 
\hat{{\cal H}}|\psi> \equiv ``
\left(
G_{abcd}(h_{ab}(x))\frac{\delta^2}{\delta h_{ab}(x)\delta h_{cd}(x)} - \sqrt{h}(h_{ab}(x))R(h_{ab}(x))
\right)
"|\Psi> = 0 
\mbox{ } . 
\label{WDE}
\ee
This is manifestly timeless (it does not contain any $i\frac{\pa|\Psi>}{\pa t}$ or 
$\frac{\pa^2|\Psi>}{\pa t^2}$). One consequence of this is that there is no conventional 
Schr\"odinger or Klein--Gordon i.p.  One could treat (\ref{WDE}) as an equation 
on Superspace; the indefiniteness of the DeWitt supermetric then provides a `time'.  This 
cannot permit a Schr\"{o}dinger i.p due to the indefiniteness of the metric (as in Klein--Gordon theory).  
Furthermore for GR the Klein--Gordon implementation  
\be
<\Psi_1||\Psi_2>  = \frac{1}{2i}\prod_{x \in \Sigma}\int_{\Sigma}d\Sigma_{\hat{A}}
G^{\hat{A}\hat{B}}
\left(
\Psi_1\stackrel{\longleftrightarrow}{\pa_{\hat{B}} }\Psi_2
\right) 
\mbox{ } , 
\ee
is also blocked by the absence of several convenient features that Klein--Gordon theory 
happened to have.  GR has a lack of well defined --, + mode separation tied to nonstationarity and 
an indefinite potential $R$ (unlike $m\varsigma^2$) so that this i.p is not normalizable.  
Two other not generally successful \cite{POTlit1} interpretations of the WDE are the 
{\it semiclassical interpretation} (akin to Hamilton--Jacobi formulation, using Euclidean path 
integrals \cite{Hallirev}) and {\it third quantization} (in which the wavefunction of the 
universe itself is promoted to an operator). 

Suppose instead that one can perform a mythical canonical transformation 
\be
h_{ij}(x), p^{ij}(x) \longrightarrow Q_A(x), P^A(x) ; 
q^{\mbox{\scriptsize true\normalsize}}_{\mbox{\sffamily\scriptsize Z\normalfont\normalsize}}(x), 
p^{\mbox{\scriptsize true\normalsize}\mbox{\sffamily\scriptsize Z\normalfont\normalsize}}(x)  
\ee
which separates out the embedding variables from the true d.o.f's.  This is equivalent to 
finding an internal time (and exemplifies constraining before quantizing).  Then in these new variables 
one has by construction constraints which look as if they will form a TDSE once quantized 
\be
{\cal H}_{A}
( q^{\mbox{\scriptsize true\normalsize}}_{\mbox{\sffamily\scriptsize Z\normalfont\normalsize}}(x)
, p^{\mbox{\scriptsize true\normalsize}}_{\mbox{\sffamily\scriptsize Z\normalfont\normalsize}}(x) ; 
Q_A(x), P^A(x)) \equiv P_A(x) - P_A
( q^{\mbox{\scriptsize true\normalsize}}_{\mbox{\sffamily\scriptsize Z\normalfont\normalsize}}(x)
, p^{\mbox{\scriptsize true\normalsize}}_{\mbox{\sffamily\scriptsize Z\normalfont\normalsize}}(x) ; 
Q_A(x))= 0
\mbox{ } .
\ee
One can then employ the corresponding well-defined Schr\"{o}dinger i.p.  However it could be that these 
variables are nonunique and each choice leads to a distinct quantum theory.  Besides not one 
such canonical transformation has been found to date: none of the internal time candidates 
considered has been satisfactory.  These are intrinsic time (undeveloped \cite{POTlit1}),  
extrinsic time (e.g York time\fn{This is actually good as a time but its implementation 
soon turns sour in other ways -- see VIII.3} \cite{Kuchar80, POTlit2} 
or Einstein--Rosen time \cite{Kucharcyl}), and time associated with matter fields (e.g Gaussian 
reference fluid \cite{DeWitt62, KuTorre90, KI85b}).    

Finally, one could accept the timelessness of GR.  One could then try the 
{\it na\"{\i}ve Schr\"odinger }
\be
\mbox{{\it interpretation } (NSI) \cite{H84, HP, UW}, based on the i.p } 
\mbox{\hspace{0.4in}}
<\Psi_1|\Psi_2> = \int\Psi_1^*\Psi_2 
\mbox{ } .  
\mbox{\hspace{0.4in}}
\label{NSIIP}
\ee
This permits the use of certain relative probabilities, but is conventionally regarded as of 
limited use since it does not permit answers to questions of evolution.  It is further 
discussed in VIII.1--2.  An attempted refinement of this is the \it conditional probability 
interpretation \normalfont \cite{PaWoo}; the \it sum over histories interpretation \normalfont 
\cite{Hartlerev} may be viewed as another such, and has been further developed as the 
{\it generally-covariant histories formalism} \cite{IL, Savvidou, KoulKu, VKouletsis}.  
An unrelated timeless approach involves \it perennials \normalfont (evolving constants of the 
motion) \cite{RovAshStach}.  

\mbox{ }

One could approach canonical quantization instead using Ashtekar variables.  
For the original complex Ashtekar variables, the form of the constraints 
(\ref{ashgauss} -- \ref{ashham}) makes operator ordering Problems less severe.  
One then attempts to use the reality conditions to construct the i.p.  
However, Kucha\v{r} \cite{CGK93} argued that unlikely hopes were being placed on 
these reality conditions.  Real Ashtekar variables have since been adopted, which have 
not been comprehensively criticized.  The loop representation \cite{RovSmo, PG, Aetal} 
(which works for both complex and real Ashtekar variables) provides a natural regularization, 
and leads to a discrete spacetime picture \cite{RSspinnetworks}.  
Modern approaches include Thiemann's construction of a quantum spin dynamics 
Hamiltonian operator \cite{Thiemann, Thiemannpapers, Phoenix}, and the spin foam approach \cite{Perez}.  
Because degenerate triads are being allowed and then exploited to obtain these results, I am not 
sure whether one is actually still considering GR.  Furthermore, the recovery of semiclassical 
space or spacetime is turning out to be difficult.  

The canonical approaches are based on the idea that quantizing GR alone is possible.  Then matter 
can be `added on' \cite{Ashtekar, Thiemannmatter}.  This is a counterpoint to the stringy unification below.  Indeed, for all we 
experimentally know, nature might hinge on specifically 4-d, non-unified, non-supersymmetric properties.  
Gravitation might play a distinguished role not through unification but as a universal regulator 
\cite{ADM, Thiemannmatter}.   

\mbox{ }

\noindent\bf{3.3.3 Minisuperspace quantum cosmology and other toys}\normalfont

\mbox{ }

\noindent 
Minisuperspace (I.2.7.1) was conceived in the late 1960's by Misner \cite{VMisner} as a toy 
quantum cosmology.  It is however a gross truncation, amounting to entirely suppressing the 
momentum constraint, leading to finite-dimensionality, and thus neglecting most of GR's technical 
Problems.  It is not even a self-consistent truncation \cite{KucharRyan} in that the Taub model 
behaves differently from the Bianchi {\sl IX} model that contains it! And such truncations are 
{\sl not} QM solutions  since the uncertainty principle is disregarded in switching off 
inhomogeneous modes.  I use Minisuperspace as a preliminary testing ground.    
If a supposedly general statement fails to be true even for Minisuperspace... (see V.1, VI.1.3).

Quantum cosmology was revived in the 1980's \cite{HH, Hallirev,Vilenkin}.  
More realistic Midisuperspace models 
(simple infinite-d models) such as with Einstein--Rosen and Gowdy models 
(see e.g \cite{Kucharcyl, Gowdymodel})
have been studied.  Recently, Bojowald \cite{Bojowald} 
has begun to investigate Minisuperspace quantum cosmology arising from quantum spin dynamics.  
As this {\sl is} a full quantum theory, such truncations then actually do make 
quantum-mechanical sense; a discrete singularity-free picture is beginning to emerge. 

Minisuperspace is but one of many toys used to speculate about the behaviour of full GR.  Other 
toys include parameterized fields \cite{Dirac, KucharII, KucharIII}, relativistic particles in 
curved spacetime \cite{kuhaj} anharmonic oscillators \cite{anhHaj, Ashtekar},  
perturbations away from homogeneous models \cite{HallHaw}, 2+1 GR \cite{Carlipbook}, 
1+1 dilaton-gravity 
, and strings instead of 3-geometries \cite{KuTorre, KoulKu}.  
See II, III and VIII for more theories which may be regarded as providing toys (some of which 
are new): strong-coupled gravities, Barbour--Bertotti-type particle models, and 
privileged-slicing conformal theories.  Different toys are required to 
uncover different aspects of full quantum GR, whilst it should be borne in mind that no 
particular toy works well.  Indeed extrapolation from toys is dangerous since many 
proposals that are tractable for a toy are specifically tied to non-generalizable features 
of that toy.  

\mbox{ }

\noindent\bf{3.3.4 The covariant approach and unification}\normalfont

\mbox{ }

\noindent 
The idea is to split the metric into background (usually Minkowski) and small perturbation 
pieces: $g_{AB} = g_{AB}^0 + g_{AB}^1$.  This clearly has spacetime and fixed-background 
connotations.  One then tries to treat this as a QFT for a `spin-2 graviton' $g^{1}_{AB}$, 
which has particle physics scattering connotations.  This suffers from nonrenormalizability.  
One way around this is to alter the gravitational theory!  For example, in higher-derivative 
theory, renormalizability can be obtained but only at the price of nonunitarity \cite{Stelle}. 

Because the subject has moved on from here to be considered alongside unification, it is 
appropriate to first consider the unificatory input.  It has been suggested in particle 
physics to try to use in place of SU(3) $\times$ SU(2) $\times$ U(1) a single Grand Unified 
Theory (GUT) gauge group e.g SU(5) or SO(10).  One has then less free parameters, e.g just one 
fundamental coupling constant, as well as predictiveness through novel (but to date 
unverified) particle processes such as proton decay.  Good convergence of the coupling 
constants, the {\it hierarchy Problem} (explanation of why the GUT scale of convergence is so 
much greater than the electroweak scale), and superior renormalization, may be obtained by 
{\it supersymmetry}.  This is the hypothesis that each observed bosonic and fermionic species 
has respectively a fermionic or bosonic superpartner species.  Although none of these have 
ever been seen, if the hierarchy Problem is to be resolved in this way, the forthcoming 
generation of particle accelerators are predicted to see superparticles.  Supersymmetry may be 
incorporated by passing to a minimally supersymmetric standard model, or by considering 
supersymmetric GUT's.  

Now, local supersymmetry is one route\fn{The analogue of the Fierz--Pauli route to GR.} to 
{\it supergravity} \cite{sugy}.  This includes a massless spin-2 field identified as gravity, and 11-d is 
picked out by uniqueness arguments giving a single set of (super)particle multiplets.  One then 
has the Problem of accounting for 
the apparent 4-d world.  However, the old {\it compactification} 
idea of Kaluza--Klein theory\fn{This is a unification of electromagnetism and gravity 
by means of a 5-d geometry with one Planckian-radius cylindrical dimension.  It is considered 
to be a failure at the quantum level because of the particle mode excitations being of Planck 
energy and thus not realistic particles!  Witten \cite{WittenKK} suggested a larger 
Kaluza--Klein theory to accommodate strong and weak forces also; this can be done in 11-d, with 
more complicated compact topology than cylindrical, to reflect the standard model gauge group.} 
whereby unwanted dimensions are 
currently unprobed through being curled up small, was then revived as a possible reconciliation.  
For a long time supergravity was believed to be renormalizable, but unfortunately this fails at 
higher orders \cite{Bern}.

However, the failed string theory of the strong force was investigated in a new light, and the 
closed string spectrum was found to include\fn{Alongside this, it contains a non-minimally coupled scalar dilaton and an antisymmetric form-field.  
Gauge fields are to be included on the ends of open strings.} a massless spin-2 field, which was  
then conjectured to be the graviton, leading to a background-dependent, 26-d theory free of  
renormalization, anomaly and unitarity difficulties \cite{VGSW}.  Another use of supersymmetry 
was then found to cure its tachyon Problem, trading the 26-d for the 10-d of superstring 
theory.  But this was found to be nonunique (five string theories were found), and although 
perturbatively finite order-by-order, the sum of the orders nevertheless diverged.  11-d 
{\it M-theory} \cite{MtheoryW, MtheoryT} comprises relations ({\it dualities}) between the 
small and the large, the weakly-coupled and the strongly coupled, thereby connecting the 
five string theories and supergravity.  It is intended to be nonperturbative.  It does not 
as yet exist as a theory.  Another aspect of modern string/M theory is the inclusion of 
extended objects with more dimensions than strings: (mem){\it branes} \cite{Polchinski}.  

Here are some simple stringy regimes.  Lovelock gravity 
(p 65) arises as a correction to GR, 
a Born--Infeld theory\fn{The general Born--Infeld Lagrangian density is 
$\overline{\mbox{\sffamily L\normalfont}} = \overline{f}(F_{AB}, g_{AB})$, whereas the stringy case is the specific form 
$\overline{\mbox{\sffamily L\normalfont}} = \sqrt{\mbox{det}(g + F)}$.} arises as a correction to Einstein--Maxwell theory.  
To attempt particle phenomenology, traditionally the extra dimensions are compactified, 
typically into a Calabi--Yau space.  The Ho\v{r}ava--Witten model \cite{string1} has large extra 
dimensions of AdS (anti de Sitter) form; there are analogous braneworld toy models of this 
form discussed in Part B.  
 
One concern is that string theoretic phenomenology suffers from severe nonuniqueness    
-- there are very many ways of hiding the extra dimensions\fn{Although neglected, 
one way out would be to have alternative 4-d string theories (e.g \cite{FL} or maybe using 
Liouville strings). } (e.g types of Calabi--Yau space, 
or of embeddings with large extra dimensions), and the corresponding split-up of matter fields 
(including gauge group breaking) can also be done in many ways.  Thus one obtains actions with 
many scalar fields, along with $p$-form fields.  Another concern is whether some of the model 
actions are unlikely to reflect truly string-theoretic properties (also discussed in Part B).  

\mbox{ }

\vspace{9in}

\noindent \Huge{\bf A The 3-space approach}

\mbox{ }

\noindent\Huge\bf{II Answering Wheeler's question}\normalfont\normalsize

\mbox{ }

\noindent \it ``If one did not know the Einstein--Hamilton--Jacobi equation, how might one hope to derive it straight off 
from plausible first principles without ever going through the formulation of the Einstein field equations themselves?" 
\normalfont John Archibald Wheeler \cite{Wheeler} 

\mbox{ }

The GR Einstein--Hamilton--Jacobi equation is made by substituting 
$p^{ij} = \frac{\pa \mbox{\sffamily\scriptsize S\normalsize\normalfont}}{\pa h_{ij}}$ into 
the Hamiltonian constraint 
${\cal H}$.  This is a supplementary constraint equation (\ref{HJSUPPL}) rather than a 
Hamilton--Jacobi equation (\ref{HJEQ}) per se.\fn{It would become a Hamilton--Jacobi equation however 
were it reformulated in terms of an internal time.} 

Thus the first stage of answering the question above involves finding a derivation of the GR form of the  
Hamiltonian constraint 
\be
{\cal H} \equiv G_{ijkl}p^{ij}p^{kl} - \sqrt{h}R = 0 
\mbox{ } ,
\label{HaminHKT}
\ee
where $G_{ijkl} = \frac{1}{\sqrt{h}}\left(h_{i(k|}h_{j|l)} -\frac{1}{2}h_{ij}h_{kl}\right)$ is 
the DeWitt supermetric.  ADM got this form precisely by starting off with the 4-d EFE's and 
performing their 3 + 1 split followed by the passage to the Hamiltonian (third 
route).   Wheeler's question is about not only the reverse of ADM's work (fourth route), but 
also about whether first principles can be found upon which this is to rest.  
Because Wheeler listed six routes, Hojman, Kucha\v{r} and Teitelboim (HKT) called their 
derivation from deformation algebra first principles {\it the seventh route}.  I take 
``seventh route" to mean any route which leads to geometrodynamics regained from first 
principles.

In addition to the HKT approach (II.1), there is now also a different {\it 3-space approach (TSA) } 
started by Barbour, Foster and \'{O} Murchadha (BF\'{O}) (II.2),\fn{N.B I distinguish between 
TSA the program name and BF\'{O} the first attempt.  HKT is a consolidation; I hope to extract 
and publish a consolidation of the TSA from this thesis \cite{LAnderson}.} which 
entails a distinct ontology (see II.3).  In brief, HKT presuppose spacetime 
whereas BF\'{O} presuppose space alone.  

\mbox{ }  

\noindent{\Large{\bf 1 Hojman, Kucha\v{r} and Teitelboim's answer}}

\mbox{ }

\noindent The constraints ${\cal H}$ and ${\cal H}_i$ of GR close as the Dirac Algebra\fn{This was 
already presented in a smeared way as (\ref{mommom}--\ref{hamham}).} 
\be
\begin{array}{cc}
\{{\cal H}(x), {\cal H}(y)\} = {\cal H}^i(x)\delta_{,i}(x,y) + {\cal H}^i(y)\delta_{,i}(x,y)\\
\{{\cal H}_i(x), {\cal H}(y)\} = {\cal H}(x)\delta_{,i}(x,y)\\
\{{\cal H}_i(x), {\cal H}_j(y)\} = {\cal H}_i(y)\delta_{,j}(x,y) + {\cal H}_j(x)\delta_{,i}(x,y) \mbox{ } . 
\label{DiracAlgebra}
\end{array}
\ee
This was originally derived by Dirac for slices in Minkowski spacetime \cite{Dirac51} and then 
DeWitt \cite{DeWitt} established by brute force that it holds for general spacetimes.  
Then Teitelboim \cite{T73} showed that the Dirac Algebra may be geometrically interpreted as the 
embeddability condition.\fn{Equivalently one can talk of the evolution of the 3-geometry being 
path-independent (in Superspace), also known as foliation invariance.}   

The HKT idea \cite{HKT} is to consider the {\it algebra of deformations}
of a spacelike hypersurface,
\bea
\{              {\cal H}^{\mbox{\scriptsize d\normalsize}}(x),           {\cal H}^{\mbox{\scriptsize d\normalsize}}(y)          \} =
{\cal H}^{\mbox{\scriptsize d\normalsize}i}(x)\delta_{,i}(x,y) + {\cal H}^{\mbox{\scriptsize d\normalsize}i}(y)\delta_{,i}(x,y)\\
\label{DeformAlgebra1}
\{              {\cal H}_i^{\mbox{\scriptsize d\normalsize}}(x), {\cal H}^{\mbox{\scriptsize d\normalsize}}(y)                  \} = 
{\cal H}^{\mbox{\scriptsize d\normalsize}}(x)\delta_{,i}(x,y) \mbox{ } \mbox{ } \mbox{ }  \mbox{ } \mbox{ } \mbox{ } \mbox{ } \mbox{ } \mbox{ } \mbox{ } \mbox{ } \mbox{ } \\
\label{DeformAlgebra2}
\{              {\cal H}_i^{\mbox{\scriptsize d\normalsize}}(x), {\cal H}^{\mbox{\scriptsize d\normalsize}}_j(y)                \} = 
{\cal H}^{\mbox{\scriptsize d\normalsize}}_i(y)\delta_{,j}(x,y) + {\cal H}^{\mbox{\scriptsize d\normalsize}}_j(x)\delta_{,i}(x,y) 
\mbox{ } , 
\label{DeformAlgebra3}
\eea
as primary i.e as their plausible first principles.  
${\cal H}^{\mbox{\scriptsize d\normalsize}}$ is a pure deformation (see fig 8), whereas 
${\cal H}^{\mbox{\scriptsize d\normalsize}}_i$ is a stretching within the hypersurface itself.  
In doing so, HKT are following Wheeler's advice of presupposing embeddability into spacetime 
in order to answer his question, since HKT's first principles encapsulate embeddability.  HKT 
next demand the `representation postulate':  that the 
${\cal H}^{\mbox{\scriptsize trial\normalsize}}$,  
${\cal H}^{\mbox{\scriptsize trial\normalsize}}_i$ for a  
prospective gravitational theory close as the 
${\cal H}^{\mbox{\scriptsize d\normalsize}}$, ${\cal H}^{\mbox{\scriptsize d\normalsize}}_i$ 
do (i.e as the mathematical structure commonly known as the Dirac Algebra, but now regarded as 
{\sl emerging} as the deformation algebra).  
\begin{figure}[h]
\centerline{\def\epsfsize#1#2{0.4#1}\epsffile{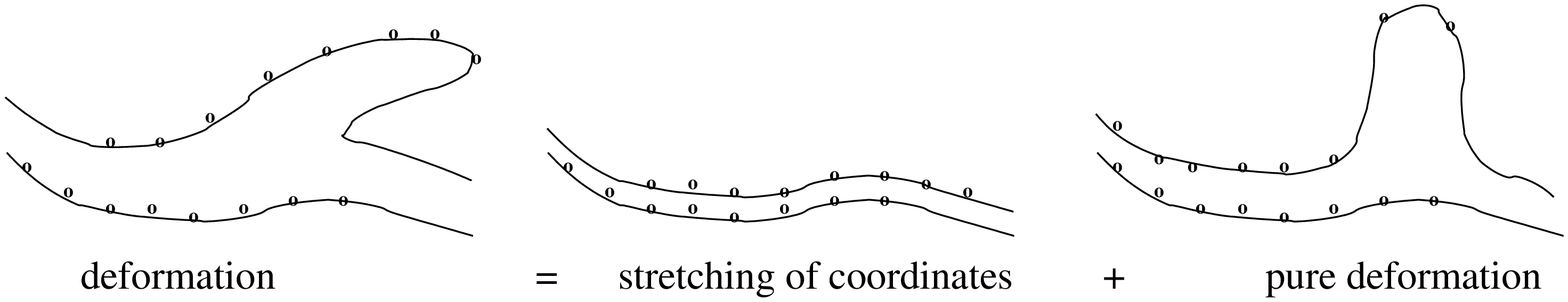}}
\caption[]{\label{1}
\footnotesize  
\normalsize}
\end{figure}

Their derivation spanned several years and a number of half-way-stage papers 
\cite{VKucharearly1, VKucharearly2, VKucharearly3}, in which various other assumptions were 
removed.  The further time-reversal assumption of \cite{HKT} was removed in \cite{Kuchar74}. 

The form of ${\cal H}_i$ essentially follows from it being a tangential deformation within each hypersurface, so one is quickly down to 
Wheeler's question about  ${\cal H}$ alone.  
From deformational first principles,
\be
\stackrel{\longrightarrow}{{\cal H}^{\mbox{\scriptsize d\normalsize}}(x^{\prime})} h_{ij}(x) = - 2K_{ij}(x)\delta(x, x^{\prime}),
\ee
from which it follows that any representation of ${\cal H}^{\mbox{\scriptsize d\normalsize}}$ 
must be ultralocal in $p^{ij}$.  From deformational first principles, 
(289) holds which 
implies ${\cal H}^{\mbox{\scriptsize d\normalsize}}$ is a scalar density of weight 1.  One can 
also deduce the form of (\ref{DeformAlgebra3}) from deformational first principles.  From this 
it follows by use of induction that ${\cal H}^{\mbox{\scriptsize d\normalsize}}$ is quadratic 
in $p^{ij}$ and also, in addition using Lovelock's theorem\fn{Thus the result is 
dimension-dependent.  Just as the Einstein--Hilbert action is a topological invariant in 
dimension $n = 2$ but leads to nontrivial theories for $n > 2$, the Gauss--Bonnet topological 
invariant in dimension $n = 4$ may be used as a nontrivial action for $n > 4$, and this sequence 
continues for all higher odd dimensions.  The gravity resulting from any linear combination of 
these actions obeys the same simplicities as GR.  Thus for $n > 4$ there are {\it Lovelock gravities} 
other than GR.  Teitelboim and Zanelli have shown that Lovelock gravity's constraints also 
close as the Dirac Algebra \cite{Lovelocksystem}.}, that the potential is 
$\sigma R + \Lambda$ for $\sigma = 1$ (it works equally well for $\sigma = - 1$ but one is then 
presupposing embeddability into a Euclidean signature geometry).  I emphasize that these proofs follow 
from the elements of spacetime structure present in the mathematics of the deformation algebra.  
I discuss an unsatisfactory residual assumption associated with these proofs in VI.1.5.  

Recollect that since the hope of pure geometrodynamics being by itself a total unified theory 
along the lines of RMW theory has largely been abandoned, asking about the form of ${\cal H}$ 
translates to asking about the form of $^{\Psi}{\cal H}$ which includes all the known fundamental 
matter fields, $\Psi$.  So given any seventh route to relativity from some first principles, 
one can assess whether these first principles are truly plausible by seeing if some form of 
them naturally extends to permit a route to relativity with all the known fundamental matter 
fields `added on'.  For HKT, the representation postulate idea extends additively (at least 
na\"{\i}vely) to matter contributions to ${\cal H}$ and ${\cal H}_i$.  In this way Teitelboim  
\cite{Teitelboim} included electromagnetism and Yang--Mills theory.  One must note 
however the absence of spin-$\frac{1}{2}$ fields from this list \cite{RMW2}, without which the 
HKT seventh route is not yet satisfactory by our criterion.  

Kouletsis \cite{VKouletsis} has recently worked on a variation on HKT's work (related to the 
generally-covariant histories formalism) which makes additional explicit use of spacetime structure 
so as to clarify the spacetime origin of HKT's postulates.  In this thesis, I do the opposite: 
rather than introducing yet more spacetime structure than HKT, I attempt to use less, along the lines 
of the next section.  

\mbox{ }

\noindent{\Large{\bf 2 Barbour, Foster and \'{O} Murchadha's answer:}}  

\vspace{.1in}  

\noindent{\Large{\bf the 3-Space approach (TSA)}}

\mbox{ }

\noindent Barbour, Foster and \'{O} Murchadha (BF\'{O}) \cite{BOF} require mere closure in place 
of closure as the Dirac Algebra.  Whereas  HKT's requirement that the Dirac Algebra be reproduced 
imports the embeddability into 4-d spacetime, BF\'{O} have been able to show that this is largely 
unnecessary.\fn{In this section, this means unnecessary for the purpose of 
retrieving vacuum GR.  Matter is considered in later chapters.} For they were 
able to derive GR from $3$-d principles alone, by use of Dirac's generalized Hamiltonian 
dynamics to exhaustively provide a highly restrictive scheme (see \cite{AB} or I.1.2.3).  Thus 
the two answers advocate distinct ontologies:  HKT assume the structure of 4-d spacetime 
(implicitly within the embeddability assumption rather than explicitly), whereas BF\'{O} adopt 
a TSA i.e they assume the structure of space alone.  

I begin by considering the `Relativity without Relativity' (RWR) paper in which BF\'{O}, 
starting from their 3-space point of view, obtained new insights into the origin of both 
special and general relativity.  Furthermore, when matter is `added on', the null cone 
structure and the Abelian gauge theory of electromagnetism are enforced and share a common 
origin.  The TSA also gives rise to new theories of evolving 3-geometries: 
strong gravity theories, and conformal gravity theories.  

\mbox{ } 

My first four contributions to this TSA program are:  first the tightening of RWR in this section 
based on \cite{AB, Sanderson, Vanderson} and additional original work, thus making the pure gravity TSA more 
complete and correct.  Second, the discovery and investigation of alternative strong and conformal  
theories of evolving 3-geometries (III and V).  Third, the inclusion of 
fundamental physical fields (IV), extending RWR to the case of many interacting 1-form fields 
in which case Yang--Mills theory is recovered, observing how matter couples to strong gravity, 
and demonstrating that the viability of conformal gravity is not threatened by preclusion of 
well-established fundamental bosonic fields.  I adopt a systematic approach in which it 
suffices to treat the matter terms piecemeal and independently of the gravitational working, 
thanks to working at the level of the general matter ELE.  All cases of 
interest can then be quickly retrieved from this working.  Fourth, I criticize the TSA as 
conceived in RWR (VI).  This is necessary for practical reasons and for including spin-$\frac{1}{2}$ fermion fields.  Then I provide a new interpretation of 
the TSA which may be viewed from within Kucha\v{r}'s `split spacetime framework' (SSF) 
(or hypersurface framework) \cite{KucharI, KucharII, KucharIII, KucharIV}, 
which furthermore permits a quick diagnostic for what 
can be easily put into a sufficiently broad-minded TSA formulation to replace RWR.      
Thus it is demonstrated that there is a TSA formulation which permits the 
inclusion of spin-$\frac{1}{2}$ fermions and indeed of a set of fundamental matter fields rich 
enough to contain the Standard Model, i.e the simplest collection of matter fields that 
suffices to agree with observations, at a classical level.  

More recent contributions of mine to the TSA program, on the relation to 
the relativity principle and to the equivalence principle and the accommodation of massive 1-form 
fields, appear in V and VII.  III is partly driven by preliminaries to quantization, which are revisited 
in the quantum TSA chapter VIII.  

\mbox{ }

\noindent{\large{\bf 2.1 The Barbour--Bertotti 1982 theory and best matching}} 

\mbox{ }

\noindent{\bf 2.1.1 Machian point-particle theories}

\mbox{ }

\noindent I discuss here the origin of the ideas behind the TSA.  
As these date back to the inception of Newtonian physics, 
it is appropriate to begin by considering them in the context of particle mechanics.  
The ideas are relational as opposed to absolute.  
They are rooted in Leibniz' \cite{LeibnizClarkecorr} `identity of indiscernibles' and were subsequently considered 
by Bishop Berkeley \cite{BishopBerkeley} and Mach \cite{Mach}.  Whereas the ideas are sound, it counted heavily against 
them that nobody was able to construct any theories which implemented them.  
However, Barbour and Bertotti found two such point particle theories \cite{BB77, BB82} 
(BB77 and BB82).  Whereas BB77 has experimental problems, BB82 does not, and its 
{\sl best matching} framework may furthermore be used to formulate gauge theories and 
theories of 3-geometries (i.e of gravitation from a geometrodynamical viewpoint).  
As Kucha\v{r} pointed out \cite{KuBacom}, the 3-geometry case of the BB82 model 
\it pretty much gives \normalfont GR in what is more or less identifiable with the BSW 
formulation and with a close analogue of the Jacobi principle.  So it gives a means of showing 
from a certain perspective that {\sl the direct implementation of Machian principles leads to 
the entirely orthodox gravitational physics of GR!}  However, Barbour and Bertotti perceived 
this as a disappointment, causing the program to stall for many years.  But in fact 
they had stopped just short of a large set of interesting results in traditional-variable 
geometrodynamics.  

The Machian ideas for point-particle theories are

\noindent{\bf R1} In particle dynamics, only the relative distances 
between the particles are physically-relevant.  

\noindent\bf R2 \normalfont Time is nothing but an arbitrary monotonic parameter $\lambda$ 
used to label the sequence of relative configurations that the universe passes through.  

Note that some 
relationalists only consider the first demand.  Temporal relationalism was 
emphasized alongside spatial relationalism by Mach \cite{Mach}.  As we shall see below, the common supposition that distant masses govern 
local inertia (c.f the motivation behind Brans--Dicke theory \cite{CGBD}, or arguments that 
in some sense GR possesses non-Machian solutions) only sometimes arises from our adopted 
starting point: for particles it happens in the BB77 implementation, but not in the BB82 one.    

Thus mathematically, given $n$ particles indexed by $(i)$ whose positions are described by 
$\mbox{\b{q}}_{(i)}$, spatiotemporal relationalists should demand actions that are invariant under the 
\be
\mbox{transformations of the `Leibniz group' }
\mbox{\hspace{0.3in}}
\mbox{\b{q}}_{(i)} \longrightarrow \mbox{\b{\b{A}}}(\lambda)\mbox{\b{q}}_{(i)} + \mbox{\b{g}}(\lambda) 
\mbox{ } , \mbox{ } \mbox{ } \mbox{\b{\b{A}}}(\lambda) \mbox{ orthogonal }
\mbox{\hspace{0.3in}}
\label{Leib1}
\ee
\be
\mbox{\hspace{2.3in}}
\lambda \longrightarrow f(\lambda) 
\mbox{ } .
\label{Leib2}
\ee
The transformations (\ref{Leib1}), which would suffice for purely spatial relationalists, form 
the Euclidean group Eucl i.e the kinematic group of Euclidean space \cite{Ehlers}. To be 
more precise, I take Eucl to comprise the group of {\sl small} isometries of Euclidean space.  
Note that it is from Eucl that the notions of distance and angle derive, which are the 
building-blocks for the relations.    

Invariance under these transformations is termed the \it kinematical principle of relativity 
\normalfont \cite{Weylkin}, which is notably different from the dynamical principle of 
relativity [invariance under the transformations (\ref{Gal1}) of the Galileo group].  
Newton's point of view on the mismatch between the invariances (\ref{Leib1}) of the 
interactions and (\ref{Gal1}) of the inertial forces was that it proves the existence of the 
absolute structures that he postulated \cite{Scholium}.  However, Mach conjectured that a 
completely relational physics of the whole universe could give effective local physics that is 
invariant under (\ref{Gal1}) alone.  Ehlers \cite{Ehlers} claimed that Leibnizian spacetime has 
insufficient structure\fn{Lange \cite{Lange} and Cartan \cite{Cartansp} spacetimes have more 
structure.  The former was not constructive in that no specific physical laws 
fitting the description of such a spacetime were postulated.  The latter has Newtonian mechanics 
merged with Newtonian gravity: as in GR, the gravitation provides the inertial frames.}.  
However this is not correct according to one interpretation, since Barbour and Bertotti then 
provided a satisfactory Leibnizian theory.  

The transformation (\ref{Leib2}) adjoined to (\ref{Leib1}) expresses the {\it reparameterization 
invariance} (RI) of the time label, in other words the absence of a meaningful absolute time.  The 
origin of Barbour's claim \cite{B94I, EOT} that the laws of physics are timeless is that they appear to be 
expressible in terms of only relative times, the content of which is nothing but relative

\noindent 
distances.  This may be illustrated as follows.  The RI Jacobi principle 
of Newtonian mechanics gives the shape of the dynamical orbit of the whole system, but if one 
is considering the universe as a whole, then none of the physically-relevant observers (i.e 
those internal to the system since by definition there is nothing external to the system) 
could notice any difference at all if the rates of all motions in the universe were to 
simultaneously double.  For example, if one considers the solar system (neglecting for the 
moment planet-planet perturbations) as a simple whole universe model, the answer to Jacobi's 
principle is then several confocal ellipses when represented in $\Re^3$.  Now, an internal 
observer may indeed allot a rate at which one of these ellipses is traversed, and use it to 
calibrate the rate at which the other ellipses are traversed.  Thus the relative rate at which 
planets orbit the sun remains a meaningful concept.  This amounts to using the motion of one 
planet as an internal clock with respect to which the motions of the other planets are measured.  
However, because we are interpreting this as a whole universe model, by definition there 
exists no clock external to the system with respect to which the motion of the planet-clock itself can 
be calibrated.  So it would make no difference to the study of the system if another internal 
observer chose to study the system by allotting a different rate to the planet-clock.  Note 
also how such use of a part of the motion as an internal clock is of limited 
accuracy.  Planet-planet perturbations are {\it not} negligible, so one must consider increasingly 
more extended configurations of planets in the quest for accuracy, until all the planets are 
included.   The only arbitrarily precise clock in the universe is the universe itself.  

The meaninglessness of performing translations and rotations on the whole $n$-particle 
universe is captured by working not on the (3$n$)-d configuration space 
$\mbox{\sffamily Q\normalfont}(n)$ of the $\mbox{\b{q}}_{(i)}$ but rather on 
the (3$n$ -- 6)-d ($n \geq 3$) quotient space 
\be
\mbox{$n$-particle Relative Configuration Space } 
\mbox{\sffamily RCS\normalfont}\mbox{($n$)} 
= \frac{\mbox{\sffamily Q\normalfont}{\mbox{($n$)}}}{\mbox{Eucl}} 
\mbox{ } .
\ee

The BB77 model \cite{BB77} is an attempt to work directly on this space, using the following 
implementations.  

\noindent {\bf RI[R2]} A Jacobi-type action is considered (to have manifest RI), and 
$$
\mbox{{\bf Direct[R1]} the action is built out of $r_{(i)(j)}$: }  
\mbox{\sffamily I\normalfont}_{\mbox{\scriptsize BB77\normalfont}} 
= \int \textrm{d}\lambda\sqrt{\mbox{-\sffamily V\normalfont}\mbox{\sffamily T\normalfont}} 
\mbox{ } , 
\mbox{\hspace{2in}}
$$
\be
\mbox{\sffamily V\normalfont} = - \sum_{    (i)<(j)    }\frac{    m_{  (i)  }m_{  (j)  }    }{    r_{  (i)(j)  }    }
\mbox{ } , \mbox{ } \mbox{ } 
\mbox{\sffamily T\normalfont} = \sum_{(i)<(j)}\frac{m_{(i)}m_{(j)}}{r_{(i)(j)}}
\left(
\frac{dr_{(i)(j)}}{d \lambda}
\right)^2 \mbox{ } . 
\ee
This is in fact an often-rediscovered theory, first found by Reissner \cite{Reissner} and 
then  by Schr\"{o}dinger \cite{Schrodinger25}, albeit BB77 obtained it on slightly different 
premises (spatiotemporal rather than purely spatial relationalism). This theory is interesting e.g 
Schr\"{o}dinger showed how it explains the perihelion precession of Mercury as well as GR 
does, and Barbour--Bertotti showed that it yields a new cosmology.  And it relates local inertia 
to distant masses; however mass anisotropy arises thus, which is experimentally unacceptable to very 
high precision (5 parts in $10^{23}$) \cite{Hughes, Drever}.

\mbox{ }  

In the BB82 model, they use 

\noindent {\bf BM[R1]} work indirectly on the RCS by considering the correction of the `bare' 
velocities  
\be
\dot{\mbox{\b{q}}}_{(i)} \longrightarrow 
\mbox{\ss}_{\mbox{\scriptsize k\normalsize}, \Omega} \mbox{\b{q}}_{(i)} \equiv
\dot{\mbox{\b{q}}}_{(j)} - \mbox{\b{k}} - \mbox{\b{$\Omega$} \scriptsize $\times$ \normalsize 
\b{q}}_{(j)} \mbox{ } , 
\ee
where the \b{k} and \b{$\Omega$} are auxiliary variables associated with the action on 
$q_{(i)}$ of the generators of the translations and the rotations.  This corresponds to keeping 
one configuration fixed and shuffling a second other one around by means of translations and 
rotations as a means of casting the second configuration into as similar a form as possible 
to the first configuration.   This procedure is called \it best matching (BM)\normalfont.  
The corrections are Lie derivatives

\noindent corresponding to draggings in the unphysical directions.  
Thus indeed no additional structure is used in this construction.
 
They furthermore use the implementation {\bf RI[R2]}, thus considering actions 
\be
\mbox{\sffamily I\normalfont}_{\mbox{\scriptsize BB82\normalfont}} 
= 2\int \textrm{d} \lambda \sqrt{ (\mbox{\sffamily E\normalfont}     -     \mbox{\sffamily V\normalfont}(\mbox{\b{q}}_{(k)}))\mbox{\sffamily T\normalfont}}
\mbox{ } , \mbox{ } \mbox{ } 
\mbox{\sffamily T\normalfont} = \sum_{(j) = (1)}^{(n)} \frac{  m_{(j)}  }{  2  }
\mbox{\ss}_{\mbox{\scriptsize k\normalsize}, \Omega}\mbox{\b{q}}_{(j)}  
\cdot
\mbox{\ss}_{\mbox{\scriptsize k\normalsize}, \Omega}{\mbox{\b{q}}}_{(j)}        \mbox{ } . 
\ee

\be
\mbox{ The particle momenta are } 
\mbox{\hspace{0.9in}}
\mbox{\b{p}}^{(i)} \equiv \frac{\pa \mbox{\sffamily L\normalfont}}{\pa \dot{\mbox{\b{q}}}_{(i)}} =  
\sqrt{      \frac{ \mbox{\sffamily E\normalfont}  -     \mbox{\sffamily V\normalfont}   } 
{   \mbox{\sffamily T\normalfont}   }      }
m_{(i)}\mbox{\ss}_{\mbox{\scriptsize k\normalsize}, \Omega}{\mbox{\b{q}}}^{(i)} 
\mbox{ } . 
\mbox{\hspace{1.2in}}
\label{BB82mom}
\ee
Now, the presence of a square root means that the Lagrangian is homogeneous of degree $1$ in 
the velocities. It is this homogeneity property that guarantees the RI and furthermore implies 
that the theory must satisfy at least one primary constraint as an identity by the following 
argument of Dirac's \cite{Dirac}.    The canonical momenta must be homogeneous of degree $0$.  
Thus they are functions of ratios of velocities alone, but there are only $n - 1$ of these, so that there 
is at least one relation between the $n$ momenta.  I now show how this occurs in the above 
type of action:
$$ 
\sum_{(i) = (1)}^{(n)}  \frac {\mbox{\b{p} \normalfont  }^{(i)} \cdot \mbox{\b{p} \normalfont  }^{(i)}} 
{m_{(i)}} = \sum_{(i) = (1)}^{(n)}  \frac{1}{m_{(i)}}
\left(
\sqrt{\frac{\mbox{\sffamily E\normalfont} - \mbox{\sffamily V\normalfont}}{\mbox{\sffamily T\normalfont}}} 
m_{(i)}\mbox{\ss}_{\mbox{\scriptsize k\normalsize}, \Omega}{\mbox{\b{q}}}_{(i)} 
\right)
\cdot
\left(
\sqrt{\frac{\mbox{\sffamily E\normalfont} - \mbox{\sffamily V\normalfont}}{\mbox{\sffamily T\normalfont}}}
m_{(i)}\mbox{\ss}_{\mbox{\scriptsize k\normalsize}, \Omega}{\mbox{\b{q}}}_{(i)} 
\right) 
$$
$$
= \frac{\mbox{\sffamily E\normalfont} - \mbox{\sffamily V\normalfont}}{\mbox{\sffamily T\normalfont}} 
\sum_{(i) = (1)}^{(n)} m_{(i)} 
\mbox{\ss}_{\mbox{\scriptsize k\normalsize}, \Omega}{\mbox{\b{q}}}_{(i)} 
\cdot
\mbox{\ss}_{\mbox{\scriptsize k\normalsize}, \Omega}{\mbox{\b{q}}}_{(i)}  
=\frac{\mbox{\sffamily E\normalfont} - \mbox{\sffamily V\normalfont}}{\mbox{\sffamily T\normalfont}}2\mbox{\sffamily T\normalfont} = 2(\mbox{\sffamily E\normalfont} - 
\mbox{\sffamily V\normalfont}) 
\mbox{ } . 
$$ 
Thus the square root form gives rise to one primary constraint by ``Pythagoras' theorem'': 
\be 
{\cal P} \equiv \sum_{(i) = (1)}^{(n)}  \frac {\mbox{\b{p} \normalfont  }^{(i)} \cdot  \mbox{\b{p} \normalfont  }^{(i)}} {2m_{(i)}} -(\mbox{\sffamily E\normalfont} 
- \mbox{\sffamily V\normalfont}) = 0 \mbox{ } .    
\label{BB82prim}
\ee 

Varying with respect to the auxiliaries \b{k} and \b{$\Omega$} respectively, one obtains that the total 
momentum and angular momentum of the whole $n$-particle universe must be zero, 
\be
\mbox{\b{${\cal M}$}} \equiv \sum_{(i) = (1)}^{(n)}\mbox{\b{p}}^{(i)} = 0 
\mbox{ } , \mbox{ } \mbox{ } 
\mbox{\b{${\cal L}$}} \equiv \sum_{(i) = (1)}^{(n)}\mbox{\b{q}}_{(i)} \mbox{\scriptsize $\times$ \normalsize } 
\mbox{\b{p}}^{(i)} = 0 \mbox{ } .
\label{BBcons}
\ee
\be
\mbox{The particle ELE's are }  
\mbox{\hspace{1.2in}}
\mbox{\ss}_{\Omega}\mbox{\b{p}}^{(i)}
= \sqrt{      \frac{    \mbox{\sffamily T\normalfont}     }{    \mbox{\sffamily V\normalfont}    }      }\frac{\pa \mbox{\sffamily V\normalfont}}{    \pa \mbox{\b{q}}_{(i)}    }
\mbox{\hspace{1.5in}}
\label{BB82EL}
\ee
so, by coupling (\ref{BB82mom}) and (\ref{BB82EL}), if one picks the unique distinguished 
choice of label time such that 
$\sqrt{\mbox{\sffamily T\normalfont}} = \sqrt{\mbox{\sffamily V\normalfont}}$, one recovers 
Newton's second Law (\ref{NII}).  This choice corresponds to the total energy of the universe 
also being zero.  So both the {\bf BM[R1]} and {\bf RI[R2]} implementations lead to 
constraints.  These happen to be preserved by the evolution equations provided that 
the potential is a function of the relative separations alone.    

\mbox{ } 

I consider it useful to view the above in other ways.  I {\sl show} here that the above 
auxiliaries are in fact not Lagrange multipliers but velocities associated with cyclic 
coordinates.  This necessitates a more careful account of how the variation is to be done.  
It also means then the kinetic term truly remains homogeneous of degree 2 in its velocities 
upon Eucl-BM, thus indeed not compromising the RI of the action.

I start with the implementation {\bf AF[R1]}: I write down my actions in the appropriate 

\noindent arbitrary frame.\fn{I have since seen that Lynden-Bell also thought along these lines \cite{LB}.}  
This in my opinion ought to underly BM: the BM auxiliaries are 
\sl velocities associated with the transformation between the stacked and arbitrary 
frames\normalfont.  Variation with respect to these auxiliaries then ensures that this choice of 
frame does 
not affect the physics (this is again an indirect implementation of {\bf R1}).  

For particle mechanics, writing down the action in an arbitrary frame requires use of 
\be
\mbox{`corrected coordinates' 
with respect to Eucl, }
\mbox{\hspace{1.1in}}
\mbox{\b{q}}_{(i)}^{\prime} = {q}_c^{\prime}\mbox{\b{e}}^{\prime c}_{(i)} 
= \stackrel{\longrightarrow}{{E}_{\mbox{\scriptsize l\normalsize}, \Theta}}\mbox{\b{q}}_{(i)}  
\mbox{ } , 
\mbox{\hspace{1.1in}}
\ee 
for $\mbox{\b{e}}^k$ the frame basis and $q_{(i)k}$ the components of the $(i)$th particle 
with respect to it, and for $E_{\mbox{\scriptsize l\normalsize}, \Theta}$ the coordinate 
transformations of Eucl and where the arrow indicates group action (which happens to be 
on the underline since the action on the basis $c$ index and the component $c$-index are 
opposite and cancel out).  

The potential should then be built to be Eucl-invariant from the start.  
Dealing with the kinetic term is trickier since $\frac{\pa}{\pa\lambda}$ is not even 
Eucl-covariant, due to the involvement of 
\be
\mbox{the frame itself: } 
\mbox{\hspace{0.8in}}
\frac{\pa \mbox{\b{q}}_{(i)}}{\pa\lambda} =  
\left(
\frac{\pa\mbox{\b{q}}_{(i)}}{\pa\lambda}
\right)
^{\prime}   +  \frac{\pa \stackrel{\longrightarrow}
                                  {{E}_{      \mbox{\scriptsize l\normalsize}, \Theta} }     }
                   {      \pa\lambda      }
\mbox{\b{q}}_{(i)}     = 
\left(\frac{\pa\mbox{\b{q}}_{(i)}}{\pa\lambda}
\right)
^{\prime}   +  
\stackrel{\longrightarrow}{{E}_{\dot{\mbox{\scriptsize l\normalsize}}, \dot{\Theta}} }
\mbox{\b{q}}_{(i)} 
\mbox{ } .
\mbox{\hspace{0.8in}}
\ee 
Thus use of an arbitrary frame in building the action amounts to passing from the bare stacked 
velocity to the unstacked, BM velocity 
$\left(
\frac{    \pa\mbox{\scriptsize\b{q}\normalsize}_{\mbox{\tiny(i)\normalsize}}    }
     {    \pa\lambda}
\right)
^{\prime} \equiv \mbox{\ss}_{\mbox{\scriptsize k\normalsize}, \Omega}\mbox{\b{q}}_{(i)}$ for 
$k = \dot{l}$ and $\Omega = \dot{\Theta}$.  This amounts to bringing in Eucl-generators which 
infinitesimally drag in all unphysical directions (corresponding here to the translationary 
and rotationary status of the whole universe).  

That one is treating each auxiliary not as a Lagrange multiplier 
$q_{\mbox{\scriptsize l\normalsize}}$ but as velocity $\dot{q}_{\mbox{\scriptsize c\normalsize}}$ 
corresponding to a cyclic coordinate  
$q_{\mbox{\scriptsize c\normalsize}}$ is a justifiable choice, by use of the following 
equivalent \it free endpoint variation \normalfont with respect to
$q_{\mbox{\scriptsize c\normalsize}}$.  Starting from the standard expression (\ref{ELDERIV}) 
$$
0 = \delta S = \int \textrm{d}\lambda
\left[
\frac{\pa \mbox{\sffamily L\normalfont}}{\pa q_{\mbox{\scriptsize c\normalsize}}} - \frac{\pa}{\pa \lambda}
\left(
\frac{\pa \mbox{\sffamily L\normalfont}}{\pa \dot{q}_{\mbox{\scriptsize c\normalsize}}}
\right)
\right]
\delta q_{\mbox{\scriptsize c\normalsize}} + 
\left[
\frac{    \pa\mbox{\sffamily L\normalfont}    }{    \pa \dot{q}_{\mbox{\scriptsize c\normalsize}}    }
\delta q_{\mbox{\scriptsize c\normalsize}}
\right]_{e_1}^{e_2} 
$$
$$
\Rightarrow 
\left.
\begin{array}{l}
\mbox{ } 
\frac{        \mbox{\normalsize $\pa$\sffamily L\normalfont}        }{        \mbox{\normalsize $\pa \dot{q}$\normalsize}_{\mbox{\scriptsize c\normalsize}}        } = \mbox{ const }  \\
\left.
\frac{        \mbox{\normalsize$\pa$\sffamily L\normalfont}         }{        \mbox{\normalsize $\pa \dot{q}$\normalsize}_{\mbox{\scriptsize c\normalsize}}         }
\right|
_{e_1, \mbox{ } e_2} = 0
\end{array}
\right\}  
\Rightarrow
\frac{\pa \mbox{\sffamily L\normalfont} }{\pa \dot{q}_{\mbox{\scriptsize c\normalsize}}} = 0 \mbox{ } , 
$$
where the first step uses (\ref{cyclicel}) since $q_{\mbox{\scriptsize c\normalsize}}$ is 
cyclic, and noting that $\delta q_{\mbox{\scriptsize c\normalsize}}$ no longer vanishes at 
the endpoints $e_1$ and $e_2$, and the second step uses the endpoint condition to fix the 
value of the constant $p_{\mbox{\scriptsize c\normalsize}}$.

I also note that RI Lagrangian Eucl-BM correction corresponds precisely to the 
Hamiltonian Dirac-appending of constraints according to 
$$
\mbox{\sffamily H\normalfont} = \mbox{\sffamily T\normalfont}\mbox{+}
\mbox{\sffamily V\normalfont}\mbox{+}\mbox{\b{k}}\cdot\sum_{(i)} \mbox{\b{p}}^{(i)} 
+ \mbox{\b{$\Omega$}}\cdot\sum_{(i)}\mbox{\b{q}}_{(i)} 
\mbox{\scriptsize $\times$ \normalsize} \mbox{\b{p}}^{(i)}
\mbox{$\longrightarrow$}
\mbox{\sffamily L\normalfont} 
= \sum_{(i)}\mbox{\b{p}}^{(i)}\cdot\dot{\mbox{\b{q}}}_{(i)}\mbox{--}\mbox{\sffamily T\normalfont}\mbox{--}
\mbox{\sffamily V\normalfont}\mbox{--}\mbox{\b{k}}\cdot\sum_{(i)} 
\mbox{\b{p}}^{(i)}\mbox{--}\mbox{\b{$\Omega$}}\cdot\sum_{(i)}\mbox{\b{q}}_{(i)}\mbox{\scriptsize $\times$ \normalsize}\mbox{\b{p}}^{(i)} 
$$
\be
= 
\sum_{i}(\dot{\mbox{\b{q}}}_{(i)} - \mbox{\b{k}} - \mbox{\b{$\Omega$}}\cdot \mbox{\b{q}}_{(i)} 
\mbox{ \scriptsize $\times$ \normalsize }\mbox{\b{p}}^{(i)}    )^2 
- \mbox{\sffamily T\normalfont}  - \mbox{\sffamily V\normalfont}      
= \mbox{\sffamily T\normalfont}(\ss_{\mbox{\scriptsize k,$\Omega$\normalsize}}\mbox{\b{q}}_{(i)}) 
- \mbox{\sffamily V\normalfont}(\mbox{\b{q}}_{(i)})
\longrightarrow 
\mbox{\sffamily L\normalfont}_{\mbox{\scriptsize J\normalsize}} 
= 2\sqrt{    \mbox{\sffamily T\normalfont}(\ss_{\mbox{\scriptsize k,$\Omega$\normalsize}}\mbox{\b{q}}_{(i)})
\mbox{\sffamily V\normalfont}(\mbox{\b{q}}_{(i)})    }
\ee
(presented in the \sffamily T \normalfont homogeneous quadratic in the $\dot{q}_{(i)}$ case), 
where the first three steps form a Legendre transform and the fourth step is the passage to the 
Jacobi form.  


\mbox{ }

\mbox{ }

\mbox{ }

\noindent{\bf 2.1.2 General best-matched reparameterization-invariant schemes}

\mbox{ }

\noindent Consider a collection {\sffamily Q} of objects 
$q_{\mbox{\sffamily\scriptsize A\normalfont\normalsize}}$ which are functions of position on 
some manifold and of label time $\lambda$.  Suppose there is a continuous group $G$ of small motions that 
have been

\noindent declared to be irrelevant to the physics of the objects.  This could include both the isometry 
(or kinematic) group of space and internal symmetries of the objects themselves.  Then the extension of the above 
Machian scheme is to demand 

\noindent \bf general R1 \normalfont that the action be invariant under $G$-transformations, and 

\noindent \bf R2 \normalfont be unaffected by any overall choice of time for the whole universe.

The first of these is implemented as follows. 
  
\noindent \bf BM[general R1] \normalfont would involve working not on the configuration space 
$\mbox{\sffamily Q\normalfont}({\mbox{\sffamily A\normalfont}})$ but rather
\be
\mbox{the quotient space }
\mbox{\hspace{1.6in}}
\mbox{\sffamily RCS\normalfont}
({\mbox{\sffamily A\normalfont}}) 
= \frac{\mbox{\sffamily Q\normalfont}({\mbox{\sffamily A\normalfont}})      }
       {G} 
\mbox{ } . 
\mbox{\hspace{2.1in}}
\ee
by the indirect means of correcting the bare velocities with auxiliaries corresponding to the freedom of 
moving in the G-directions.  Thus one passes to the `doubly degenerate' 
$\mbox{\sffamily Q\normalfont}({\mbox{\sffamily A\normalfont}}) \times G$ but then the constraints corresponding to 
variation with respect to the auxiliaries ensures that $G$ is physically irrelevant.  

I take this implementation to be built on an underlying implementation 

\noindent{\bf AF[general R1]}: 
the action is to be written in an arbitrary $G$-frame and so in terms of 
\be
\mbox{the `corrected coordinate' }
\mbox{\hspace{1.5in}}
{q}^{\prime}_{\alpha\mbox{\sffamily\scriptsize A\normalsize\normalfont}} = 
{{e}^{\prime}}_{\alpha}^{\beta}q^{\prime}_{\beta \mbox{\sffamily\scriptsize A\normalsize\normalfont}} = 
\stackrel{\longrightarrow}{L_{\mbox{\scriptsize a\normalfont}_{[i]}}}q_{\alpha\mbox{\sffamily\scriptsize A\normalsize\normalfont}}
\mbox{\hspace{2.3in}}
\ee 
where greeks are manifold-multi-indices, Greeks are internal or field-type multi-indices, 
${e}_{\alpha}^{\beta}$ is the obvious product of frame bases with respect to which the 
{\sffamily A}-th 
object has components $q_{\alpha \mbox{\sffamily\scriptsize A\normalsize\normalfont}}$, and  
$L_{\mbox{\scriptsize a\normalsize}_{[i]}}$ are the coordinate transformations of $G$ 
corresponding to generators associated with the auxiliary variables ${\mbox a}_{[i]}$, From this it 
follows that the auxiliaries involved are $G$-frame velocities.  

The second is implemented as 

\noindent \bf RI[R2] \normalfont: the use of manifestly RI actions  
(of the usual Jacobi type or some suitable generalization homogeneously linear in the 
velocities).  

According to {\bf AF[general R1]}, the potential should then be built to be $G$-invariant, to which end it is useful 
if a collection of $G$-invariant or $G$-covariant objects are known.  Such \it 
$G$-concomitants \normalfont are available for all the examples in this chapter (but not 
in III.2).   The kinetic term is trickier since $\frac{\pa}{\pa\lambda}$ is  not even 
$G$-covariant due to the involvement of the frame
\be
\mbox{itself:} 
\mbox{\hspace{0.3in}}
\frac{               \pa {q}_{\mbox{\sffamily\scriptsize A\normalfont\normalsize}}               }{            \pa\lambda              } =  
\left(
\frac{               \pa {q}_{\mbox{\sffamily\scriptsize A\normalfont\normalsize}}               }{            \pa\lambda              }
\right)
^{\prime}   
+ \sum_{                [{\mbox{\scriptsize i\normalsize}}] \in 
\mbox{ generators of $G$}(       \mbox{\sffamily A\normalfont}        )                       }
\frac{            \pa\stackrel{\longrightarrow}
                                 {{L}_{\mbox{\scriptsize a\normalsize}_{\mbox{\tiny[i]\normalsize}}} }              }{                 \pa\lambda                  }
{q}_{\mbox{\sffamily\scriptsize A\normalfont\normalsize}} = 
\left(
\frac{               \pa {q}_{\mbox{\sffamily\scriptsize A\normalfont\normalsize}}               }{            \pa\lambda              }
\right)
^{\prime}   
+  
\sum_{                [{\mbox{\scriptsize i\normalsize}}] }
                     \stackrel
                     {\longrightarrow}
                     {{L}_{\dot{\mbox{\scriptsize a\normalsize}}_{\mbox{\tiny [i]\normalfont}}} }    
{q}_{\mbox{\sffamily\scriptsize A\normalfont\normalsize}} 
\mbox{ } . 
\mbox{\hspace{0.3in}}
\ee   
Thus use of an arbitrary $G$-frame in building the action amounts to passing from the bare stacked 
velocity to the unstacked, BM velocity 
$\left(
\frac{\pa {q}_{\mbox{\sffamily\tiny A\normalfont\normalsize}} }{\pa \lambda}
\right)
^{\prime} \equiv \mbox{\ss}_{\mbox{\scriptsize a\normalsize}_{[i]}}
q_{\mbox{\sffamily\scriptsize A\normalfont\normalsize}}$
Thus one has brought in generators which infinitesimally drag in all unphysical directions.  
This corresponds to keeping one configuration fixed and shuffling a second one around by means 
of the physically-irrelevant symmetries of the collection of objects as a means of casting the second configuration 
into as similar a form as possible to the first configuration.  

Variation with respect to each $\mbox{a}_{[\mbox{\scriptsize i\normalsize}]}$ gives one 
secondary constraint.  For homogeneous quadratic actions, these are linear in the momenta.  
If these are satisfied, one has successfully passed from $\mbox{\sffamily Q\normalfont}
({\mbox{\sffamily A\normalfont}}) \times G$ to 
$\frac{\mbox{\sffamily\scriptsize Q\normalsize\normalfont}({\mbox{\sffamily\scriptsize A\normalsize
\normalfont}})}
{G}$. By Dirac's argument there still remains at 
least 1 primary constraint due to \bf R2\normalfont.  There may also be additional 
secondary constraints from applying the Dirac procedure to the constraints.  

This procedure may be interpreted as in the previous section with the  
$\mbox{a}_{[\mbox{\scriptsize i\normalsize}]}$ 
as auxiliary velocities 
corresponding to cyclic coordinates, \sffamily T \normalfont indeed remains RI.  The 
interconversion with Hamiltonian Dirac-appending continues to hold, at least while the 
appended terms are linear in the appending Lagrange multipliers. 

Whereas mechanics is not usually considered in the above RCS fashion (which 
reconciles Newtonian mechanics with Leibniz's ideas at a small global cost), it so happens 
that conventional electromagnetism and GR can be thought of as working along these lines.  
I discuss these in greater depth in App II.B (generalized to the Yang--Mills case) and in 
II.2.2 respectively.  

\mbox{ }

\noindent{\large{\bf 2.2 Relativity without relativity} }

\mbox{ }

\noindent{\bf 2.2.1 Local square roots}

\mbox{ }

\noindent The configuration space is 
$$
\mbox{\sffamily Q\normalfont}(\mbox{GR}) 
= \mbox{Riem} 
= \{\mbox{set of 3-metrics on a fixed topology, taken to be CWB} \}
$$
\be
\mbox{The RCS is }
\mbox{\hspace{1.0in}}
\mbox{\sffamily RCS\normalfont}({\mbox{\scriptsize GR\normalsize}}) \subset \{\mbox{Superspace}\} 
= \frac{\{\mbox{Riem}\}}{\{\mbox{3--Diffeomorphisms}\}} 
\mbox{ } . 
\mbox{\hspace{1.0in}}
\ee
though this is only part of the way to wards isolating a representation of the true 
dynamical d.o.f's of GR (the Hamiltonian constraint remains and is problematic).   

This is to be implemented indirectly by BM with respect to the 3-diffeomorphisms: 
\be
\dot{h}_{ij} \longrightarrow \mbox{\ss}_{{\xi}}h_{ij} \equiv 
\dot{h}_{ij} - \pounds_{\xi}h_{ij} = 
\dot{h}_{ij} - 2D_{(i}\xi_{j)}\mbox{ } . 
\ee
So for any two 3-metrics on 3-geometries $\Sigma_1$, $\Sigma_2$, this corresponds to keeping 
the coordinates of $\Sigma_1$ fixed whilst shuffling around those of $\Sigma_2$ until they are 
as `close' as possible to those of $\Sigma_1$.

The new ingredient in the RWR paper is to implement the lack of external time by 
using a RI action with a {\sl local square root ordering}.  For particle mechanics, 
this would mean using  
$\sum_{(i) = (1)}^{(n)}\sqrt{m_{(i)}\dot{\mbox{\b{q}}}_{(i)}\dot{\mbox{\b{q}}}_{(i)}}$ 
rather than the {\sl global square root ordering }
$\sqrt{\sum_{(i) = (1)}^{(n)} m_{(i)}\dot{\mbox{\b{q}}}_{(i)}\cdot\dot{\mbox{\b{q}}}_{(i)}}$.   
If one has a field theory in place of a finite number of particles, the sums are replaced by 
integrals.  For gravitation, BF\'{O} thus consider {\it BSW-type actions}, taken to be of form 
\be
\mbox{\sffamily I\normalfont}_{\mbox{\scriptsize BSW-type\normalsize}} 
= \int \textrm{d}\lambda \int \textrm{d}x^3 \sqrt{h}\sqrt{\mbox{\sffamily P\normalfont}}
\sqrt{\mbox{\sffamily T\normalfont}} 
\mbox{ } ,
\label{BSWTYPE}
\ee
as opposed to BB82's global square root actions.\fn{Following \cite{KuBacom}, mention of BSW 
was included in BB82 but such local square roots were not studied until BF\'{O}'s paper.}  Although the 
square root is local in that it sits inside the integral over all space, there is also a 
`little sum' over the indices hidden within the kinetic term \sffamily T \normalfont in 
(\ref{BSWTYPE}). 

The global analogue of GR (the original BB82 geometrodynamics), has the action 
\be
\mbox{\sffamily I\normalfont}^{\mbox{\scriptsize G\normalsize}}_{\mbox{\scriptsize BSW-type\normalsize}} 
= \int\textrm{d}\lambda\sqrt{\int\textrm{d}^3x
\sqrt{h}\mbox{\sffamily T\normalfont}^{\mbox{\scriptsize g\normalsize}}      }
\sqrt{\int\textrm{d}^3x\sqrt{h}R} 
\mbox{ } ,
\ee
whose global square root then gives one constraint only rather than the one constraint 
per space point of GR.  Note that this global alternative would arise from the ADM 
Lagrangian by the BSW procedure were one to assume $\alpha$ is independent of $x_i$.  

Note that the suggestion in II.2.1.1 about treating auxiliaries may be applied in 
GR to $\xi^i$: to have an action that is homogeneous of degree 1 in the velocities and hence 
reparameterization invariant, we can treat it as an auxiliary velocity to which free 
end-point variation is applied.  

One of Barbour's conceptual ideas behind the TSA is that such 
RI actions could be considered as geodesic principles on 
configuration space: the reduction of the physical problem of motion to the geometrical 
problem of finding the geodesics of the configuration space geometry.   This followed from the 
insight of homogeneous-quadratic Newtonian mechanics, for which use of Jacobi's principle 
reduces the problem of motion to a problem (conformal to) the well-defined, well-studied 
problem of finding the geodesics of the Riemannian configuration space geometry.  
But I dash his original hope that something similar would happen in GR in VI.1.  

The BSW formulation of GR and the Jacobi formulation of mechanics are similar in that in each 
case the presence of a square root means that the Lagrangian is homogeneous of degree $1$ in 
the velocities.  Thus primary constraints exist by Dirac's argument; 
specifically here the square roots give rise to 
primary constraints by `some version or other of Pythagoras'.

\mbox{ } 

The standard (global square root) Jacobi case is just the usual bare version of the previous 
section.  The global square root gives 1 constraint ${\cal P}_{\mbox{\scriptsize G\normalsize}}$ 
in total and then ELE's imply that 
$\dot{\cal P}_{\mbox{\scriptsize G\normalsize}} \approx 0$.  Also for the choice of 
$\lambda$ such that \sffamily T \normalfont = \sffamily E \normalfont 
-- \sffamily V \normalfont holds, the bare version of (\ref{BB82EL}) reduces to the form of 
Newton's Second Law.  This special $\lambda$ has the same properties as Newton's absolute 
time.  If there is no external time, then \sffamily E \normalfont = \sffamily T \normalfont 
+ \sffamily V \normalfont is to be interpreted not as energy conservation but as an emergent 
definition of Newtonian time.

In contrast, local square roots give many primary constraints: one per object summed over 
(1 per space point in the continuum case). Typically applying the Dirac procedure to many 
primary constraints will give many more secondary constraints which use up all the d.o.f's.  
This overconstraining due to the local square root led BF\'{O} to call it the `bad' ordering 
and the global square root the `good' ordering.
  
The local square root homogeneous quadratic mechanics in which each particle has its 
\be
\mbox{own potential would be } 
\mbox{\hspace{1in}}
S_{\mbox{\scriptsize Mech\normalsize}}^{\mbox{\scriptsize L\normalsize}} = 2\int \textrm{d}\lambda 
\sum_{  (i) = (1)  }^{(n)}\sqrt{(  \mbox{\sffamily E\normalfont}_{(i)} 
- \mbox{\sffamily V\normalfont}_{(i)}  )\mbox{\sffamily T\normalfont}_{(i)}} 
\mbox{ } ,
\mbox{\hspace{2in}}
\ee
\be
\mbox{which gives one constraint } 
\mbox{\hspace{1in}}
{\cal P}_{\mbox{\scriptsize L\normalsize}}^{(i)} \equiv \frac {\mbox{\b{p} \normalfont  }^{(i)} \cdot  \mbox{\b{p} \normalfont  }^{(i)}} {2m_{(i)}} 
- ( \mbox{\sffamily E\normalfont}_{(i)} - \mbox{\sffamily V\normalfont}_{(i)}) = 0    
\label{primmech}
\mbox{\hspace{2in}}
\ee
per particle, which then gives a tower of constraints rendering the theory inconsistent by 
using up all the d.o.f's.  A simple special case avoiding this can however be found by 
inspection of the primary constraint (\ref{primmech}): the noninteracting case in which 
each particle moves in its own potential alone. 

\mbox{ }

In the case of the BSW Lagrangian one obtains an infinity of primary constraints: 1 per 
space point.  As we show below however, despite GR having the bad ordering, it turns out not 
to be overconstrained.
\be
\mbox{ }\mbox{ The BSW action may be written as }
\mbox{\hspace{0.6in}}
\mbox{\sffamily I\normalfont}_{\mbox{\scriptsize BSW \normalsize}} 
= \int \textrm{d}\lambda \int \textrm{d}^3x \sqrt{h} \sqrt{R}\sqrt{\mbox{\sffamily T\normalfont}^{\mbox{\scriptsize g\normalsize}}} 
\mbox{ } , 
\mbox{\hspace{2in}} 
\label{ABSW} 
\ee
\be
\mbox{where the gravitational kinetic term $\mbox{\sffamily T\normalfont}^{\mbox{\scriptsize g\normalsize}}$ is given by }
\mbox{\hspace{0.3in}}                              
\mbox{\sffamily T\normalfont}^{\mbox{\scriptsize g\normalsize}} = 
\frac{1}{\sqrt{h}}G^{abcd}\mbox{\ss}_{\xi}{h}_{ab}\mbox{\ss}_{\xi}{h}_{cd} 
\mbox{ } , 
\mbox{\hspace{2in}}
\ee 
\be
\mbox{for }
\mbox{\hspace{2in}}
G^{abcd} = \sqrt{h}(h^{ac}h^{bd} - h^{ab}h^{cd})
\mbox{\hspace{2in}}
\ee 
the inverse of the DeWitt supermetric, and where I am now regarding it as an action constructed to meet 
appropriate relational demands {\bf general R1} and {\bf R2}, as emphasized by making use of $\xi^i$ 
for the relational auxiliary, and $N$ for the emergent lapse to distinguish these from the presupposed 
spacetime notions of shift and lapse, $\beta^i$ and $\alpha$.  I will also regard $\xi^i$ as some 
$\frac{\pa {x}^i}{\pa \lambda}$, where $x^i$ are the coordinates and I am building an $x^i$-invariant 
action; care  

\noindent
\be
\mbox{is needed particularly with } 
\frac{\pa}{\pa\lambda} \mbox{ } :
\mbox{\hspace{0.15in}}
\frac{\pa}{\pa\lambda}h_{ij}^{\prime} = \frac{\pa x^a}{\pa x^{\prime i}}\frac{\pa x^b}{\pa x^{\prime j}}
\left(
\frac{\pa h_{ab}}{\pa\lambda}  - 2D_{(a}\xi_{b)}
\right)
\mbox{ } , \mbox{ } \xi_a = \frac{\pa x_a}{\pa\lambda} 
\mbox{ } . 
\mbox{\hspace{0.6in}}  
\ee
\be
\mbox{The canonical momenta (defined at each space point) are }
\mbox{\hspace{0.1in}}
p^{ij} \equiv \frac{  \partial\mbox{\sffamily{L}\normalfont}  }{ \partial \dot{h}_{ij}  }
 = \sqrt{    \frac{  h  }{  2N  }    }G^{ijcd}\mbox{\ss}_{\xi}{h}_{cd} 
\label{GRmom}
\ee
for $2N \equiv \sqrt{\frac{\mbox{\sffamily\scriptsize T\normalsize\normalfont}^{\mbox{\tiny g\normalfont}}}{R}}$.
At each space point 
\be
G_{ijkl}p^{ij}p^{kl} = G_{ijkl}G^{ijcd}\frac{1}{2N}\mbox{\ss}_{\xi}h_{cd}G^{klab}\frac{1}{2N}(\mbox{\ss}{h}_{ab}  
= \sqrt{h}\frac{  \mbox{\sffamily T\normalfont}^{\mbox{\scriptsize g\normalsize}}  }{  (2N)^2  } 
= \sqrt{h}(\sigma R + \Lambda) 
\mbox{ } ,
\ee
\be
\mbox{so the local square root gives one primary constraint } 
\mbox{\hspace{0.3in}} 
{\cal H} \equiv G_{ijkl}p^{ij}p^{kl} - \sqrt{h}R = 0 
\mbox{ } . 
\mbox{\hspace{0.3in}}
\label{GRHam} 
\ee 
In addition, $\xi^i$-variation (treated as a Lagrange multiplier) of the BSW action leads to 
the momentum constraint ${\cal H}_i \equiv -2D_j{p^{j}}_i = 0$ as a secondary constraint.  It 
is unsurprising that we get this since in starting with a 3-space ontology we are allowed to 
provide the BM correction with respect to 3-diffeomorphisms.  Thus we get the momentum constraint quite 
on purpose.  Note also that (\ref{GRHam}) may be identified as the Hamiltonian constraint of 
GR, ${\cal H}$. Whereas the action (\ref{ABSW}) is associated with curves on the space 
Riem $\times $ $\Xi$, where $\Xi$ is the space of the $\xi^i$, if the momentum constraint can be 
solved as a p.d.e for $\xi^i$ (the thin sandwich conjecture of  I.2.9.1), the action will 
depend only on the curve in Superspace. This follows from the constraints being free of 
$\xi^i$, and the three components of the momentum constraint reducing the number of d.o.f's 
from the 6 of Riem to the $3$ per space point in a $3$-geometry.  
$$
\mbox{ }\mbox{The ELE's are (\ref{GRmom}) and } 
\mbox{\hspace{0.15in}}
\dot{p}^{ij}  = \frac{   \delta\mbox{\sffamily L\normalfont}   }{   \delta h_{ij}   } = \sigma\sqrt{h}N(h^{ij}R - R^{ij}) - \frac{2N}{\sqrt{h}}
\left(
p^{im}{p_m}^j -\frac{1}{2}p^{ij}p
\right) 
\mbox{\hspace{0.2in}}
$$
\be
\mbox{\hspace{1.75in}}
+ \sigma\sqrt{h}(D^iD^jN -h^{ij}{D}^2 N) + \pounds_{\xi}p^{ij} \mbox{ } .
\label{GReleq}
\ee
This is indeed GR, for which it is well-known by the contracted Bianchi identity (\ref{contbi}) that both constraints 
propagate without further secondary constraints arising.  Furthermore, in this thesis, 
the momentum constraints are automatically propagated as a further consequence of the action 
being deliberately constructed to be invariant under \it $\lambda$-dependent 
$3$-diffeomorphisms\normalfont. 

In the present case this momentum constraint propagation takes the form 
\be
\dot{{\cal H}}_i = - \frac{1}{N}D_i(N^2{\cal H}) + \pounds_{\xi}{\cal H}_i \mbox{ } .
\ee
The Hamiltonian constraint propagation takes the form
\be
\dot{\cal H} = -\frac{1}{N}D^j
\left(
N^2{\cal H}_j 
\right) 
+ \frac{Np}{2\sqrt{h}}{\cal H}
+ \pounds_{\xi}{\cal H} 
\mbox{ } .
\label{evolham}
\ee
(c.f I.2.6, II.1).  
Thus, unlike in mechanics, this system is not over-constrained.  From the 3 + 1 perspective, 
the reason that GR is not over-constrained is that it possesses a hidden foliation invariance.  Whereas at 
first glance, one would expect the BSW action to be invariant only with respect to the global 
reparameterization $\lambda \longrightarrow \lambda^{\prime}(\lambda)$ for $\lambda^{\prime}$ 
a monotonic arbitrary function of $\lambda$ (in accord with Noether's theorem), in fact the 
action is invariant under the far more general local transformation 
\be
\lambda \longrightarrow \lambda^{\prime}(\lambda),\mbox{  }
h_{ij}(x,\lambda) \longrightarrow h_{ij}(x,\lambda^{\prime}),
\mbox{ } \xi_i(x,\lambda) \longrightarrow
\frac{d\lambda^{\prime}}{d\lambda}\xi_i(x,\lambda) 
\mbox{ } . 
\ee
This is the foliation invariance of the 3 + 1 split of GR -- the hidden symmetry associated 
with the Hamiltonian constraint ${\cal H}$.  
In contrast mechanics has just the global $\lambda \longrightarrow \lambda^{\prime}(\lambda)$ 
invariance unless the aforementioned restriction is applied.

\mbox{ }

\noindent{\bf 2.2.2 Uniqueness of consistent BSW-type actions}

\mbox{ }

\noindent We know [by virtue of the contracted Bianchi identity (\ref{contbi})] that GR will 
work as a constraint algebra.  The question then is {\sl does anything else work?}  

BF\'{O}'s idea was to postulate 

\noindent
\bf BM[general R1] \normalfont :
the {\sl best matching} rule is used
to implement the {\sl{3-d}} diffeomorphism invariance
by correcting the bare metric velocities 
$\dot{h}_{ij} \longrightarrow \mbox{\ss}_{\xi}h_{ij} \equiv \dot{h}_{ij} - \pounds_{\xi}h_{ij}$.  
This rule will be applied in IV to the velocities corresponding 
to all the matter fields $\Psi_{\mbox{\sffamily\scriptsize A\normalsize\normalfont}}$ as well: 

\noindent$\dot{\Psi}_{\mbox{\sffamily\scriptsize A\normalsize\normalfont}} \longrightarrow 
\mbox{\ss}_{\xi}{\Psi}_{\mbox{\sffamily\scriptsize A\normalsize\normalfont}} \equiv 
\dot{\Psi}_{\mbox{\sffamily\scriptsize A\normalsize\normalfont}} 
- \pounds_{\xi}\Psi_{\mbox{\sffamily\scriptsize A\normalsize\normalfont}}$ so it is a universal 
rule.  
For everything in this chapter, this can be taken to arise from the corresponding  
\noindent{\bf AF[general R1]}.   

\noindent  
\bf RI[R2]\normalfont: a {\sl local square root reparameterization-invariant} implementation is used.  
The pure gravity actions considered are of Baierlein--Sharp--Wheeler (BSW) type 
with {\sffamily T} homogeneously quadratic in the velocities and 
ultralocal in the metric.  

\noindent My analysis below however differs from BF\'{O}'s since they 
missed out a number of possibilities (as explained below).  My trial BSW-type action is 
\be
\mbox{\sffamily I\normalfont}_{\mbox{\scriptsize BSW-type \normalsize}}
= \int \textrm{d}\lambda \int \textrm{d}^3x \sqrt{h} \sqrt{\sigma R + \Lambda}
\sqrt{\mbox{\sffamily T\normalfont}^{\mbox{\scriptsize g\normalsize}}_{\mbox{\scriptsize WY\normalsize}}} 
\mbox{ } , \mbox{ } \mbox{ } 
\mbox{\sffamily T\normalfont}^{\mbox{\scriptsize g\normalsize}}_{\mbox{\scriptsize WY\normalsize}} 
= \frac{1}{\sqrt{h}Y}G_{\mbox{\scriptsize W\normalsize}}^{abcd}
\mbox{\ss}{h}_{ab}\mbox{\ss}{h}_{cd} \mbox{ } ,
\label{VASBSW}
\ee
\be
\mbox{where}
\mbox{\hspace{1.4in}}
G^{ijkl}_{\mbox{\scriptsize W\normalsize}} \equiv \sqrt{h}(h^{ik}h^{jl} - Wh^{ij}h^{kl}) 
\mbox{ } , \mbox{ } \mbox{ }
W \neq \frac{1}{3} \mbox{ } , 
\mbox{\hspace{1.4in}}
\ee
is the inverse of the most general (invertible) ultralocal
supermetric (\ref{ULSUPER}), 
and without loss of generality $\sigma \in \{-1, 0, 1\}$.  More general potentials are discussed in II.2.2.5.       
\be
\mbox{ }\mbox{ Setting $2N \equiv \sqrt{          
\frac{        \mbox{\sffamily\scriptsize T\normalsize\normalfont}^{\mbox{\tiny g\normalsize}}_{\mbox{\tiny W\normalsize}}        }
{\sigma R + \Lambda}        }$, the gravitational momenta are }
p^{ij} \equiv \frac{\partial\mbox{\sffamily{L}\normalfont} }{ \partial\dot{h}_{ij}} =
\frac{\sqrt{h}Y}{2N}G^{ijcd}_{\mbox{\scriptsize W\normalsize}}\mbox{\ss}{h}_{cd} 
\mbox{ } .
\mbox{\hspace{0.1in}}
\label{wmom}
\ee
\be
\mbox{The primary constraint }
\mbox{\hspace{0.6in}}
{\cal H } \equiv \frac{Y}{\sqrt{h}}
\left( 
p \circ p - \frac{X}{2}p^2
\right) - \sqrt{h}(\sigma R \mbox{ } + \mbox{ } \Lambda)  = 0 
\mbox{\hspace{2in}} 
\label{VGRHam}
\ee
then follows merely from the local square-root form of the Lagrangian.  
In addition, $\xi^i$-variation leads to a secondary constraint which is the usual 
momentum constraint (\ref{Vmom}). 
$$
\mbox{ } \mbox{ The ELE's are }
\dot{p}^{ij} = \frac{\delta\mbox{\sffamily L\normalfont}}{\delta h_{ij}} =  \sqrt{h}Nh^{ij}(\sigma R + \Lambda) -\sqrt{h}\sigma NR^{ij}
- \frac{2NY}{\sqrt{h}}
\left(
p^{im}{p_m}^j -\frac{X}{2}p^{ij}p
\right)
$$
\be
\mbox{\hspace{0.5in}}
+ \sqrt{h}\sigma (D^iD^j N - h^{ij}D^2 N) + \pounds_{\xi}p^{ij} 
\mbox{ } . 
\label{wel}
\ee
The propagation of ${\cal H}$ then gives \cite{Sanderson} 
\be
\dot{{\cal H}} = \frac{Y\sigma}{N}D^i(N^2{{\cal H}_i})
+ \frac{(3X - 2)NpY}{2\sqrt{h}}{\cal H}
+ \pounds_{\xi}{\cal H} + \frac{2}{N}(1 - X)Y\sigma D_i\left(N^2 D^ip\right).
\label{mastereq}
\ee
The first three terms of this are related to existing constraints and thus vanish weakly in 
the sense of Dirac.  However note that the last term is not related to the existing 
constraints. 
\be 
\mbox{It has 4 factors which could conceivably be zero: }
\mbox{\hspace{0.4in}}
(1 - X)Y\sigma D_i(N^2D^ip) 
\mbox{ } .
\label{keyterm}
\mbox{\hspace{0.6in}}
\ee
Any of the first three factors being zero would be strong equations restricting the form of 
the ansatz.  The fourth factor might however lead to new constraints and thus vanish weakly.  

\mbox{ }

\noindent\bf{2.2.3 Interpretation of the consistency condition}\normalfont

\mbox{ }

\noindent The question posed in RWR -- and answered in the affirmative -- is whether GR can be 
derived solely from $3$-d arguments, that is, without any use of arguments involving  
$4$-d general covariance (spacetime structure).  The approach succeeds because of the need to 
propagate ${\cal H }$ acts as a powerful filter of viable theories, which are already strongly 
restricted by ${\cal H}_i$ being tied to 3-diffeomorphism invariance.  
This means that one has only 2 d.o.f's per space point to play with. The remarkable invariance 
at the end of I.2.2.1 does not usually hold for the generalization (\ref{VASBSW}) of the BSW 
action.  

We require the term (\ref{keyterm}) to vanish in order to have a consistent theory. Unless one 
wishes to have a theory with a privileged slicing, constraints must be independent of $N$.  
In this case the fourth factor requires $D_i p = 0 \Rightarrow \frac{p}{\sqrt{h}} = C(\lambda)$.  
But this new constraint must also propagate.  This leads to a 
nontrivial lapse-fixing equation (LFE) which (if soluble) gives a constant mean curvature 
(CMC) foliation.  The LFE is 
\be
\frac{\pa}{\pa\lambda}\left(\frac{p}{\sqrt{h}}\right) = B(\lambda) =  3\Lambda N + 2\sigma (NR - D^2N) 
+ \frac{(3X - 2)NYp^2}{2h} \mbox{ } ,
\label{SGslicingeq}
\ee
where $B(\lambda)$ is a spatial constant.   Note that for $\sigma \neq 0$ this is a nontrivial equation for the lapse 
$N$.
It is the standard CMC LFE (\ref{CGLapseFixing}) in the GR case ($\sigma = Y = W = 1$).   

So if neither $\sigma = 0$ nor $Y = 0$, the DeWitt ($W = X = 1$) supermetric of relativity is 
enforced, which is BF\'{O}'s `Relativity Without Relativity' result.  Note that the Lorentzian 
signature ($\sigma = 1$) of GR does not drop out of this working; one can just as well obtain 
Euclidean GR ($\sigma = -1$) in this way.  Earlier work of Giulini already noted that the 
$W = 1$ supermetric is mathematically special \cite{SGGiulini}.  

But there are alternatives arising from the second, third and fourth factors.  The study of 
any new alternatives arising this way is motivated by how these arise from an exhaustive route 
to GR.  Can we overrule them and thus arrive uniquely at GR, or do serious alternative 
theories arise?  In addition, these alternative theories have interesting properties which 
we use to motivate their study in their own right.  These motivations are presented at the 
start of the treatment of each theory in III and V.    

The $\sigma = 0$ branch works for any $W$.  This is a generalization of strong gravity 
\cite{SIsham, SHenneaux, STeitel, Pilatilit1, Pilatilit2, Pilatilit3} which is strong-coupled 
limit of GR in the $W = 1$ case.  The other theories are not related to GR, but are related to 
scalar--tensor theories; strong gravity theories are argued to be relevant near singularities 
(see III.1 for all of this).
The HKT program would discard the strong gravity theories since they are not a 
representation of the Dirac Algebra (although Kucha\v{r} \cite{Ku81JMP} and Teitelboim 
\cite{STeitel} did elsewhere study strong gravity).  However, the strong gravity theories meet 
the TSA's immediate criteria in being dynamically consistent theories of 3-geometries.  

The fourth factor of (\ref{keyterm}) leads to conformal theories 
\cite{conformal, CG, Kelleher, Kellehertheory, ABFKO}.  
These theories now have privileged slicings, which happen to 
correspond to those most commonly used in the GR IVP: maximal and CMC slices.  

For example, our \it conformal gravity \normalfont \cite{conformal, CG} has the action 
\bea
\mbox{\sffamily I\normalfont}_{\mbox{\scriptsize C\normalsize}} =
\int \textrm{d}\lambda
\frac{\int \textrm{d}^3x\sqrt{h}\phi^4
\sqrt{\sigma
\left(
R - \frac{8D^2\phi}{\phi}
\right) +
\frac{   \Lambda\phi^4   }{   V(\phi)^{  \frac{2}{3}}   } }
\sqrt{\mbox{\sffamily T\normalfont}^{\mbox{\scriptsize g\normalsize}}_{\mbox{\tiny C\normalsize}}}}
{  V(\phi)^{\frac{2}{3}}  } \mbox{ } ,
\mbox{   volume } V = \int \textrm{d}^3x\sqrt{h}\phi^6 \mbox{ } \\
\mbox{\sffamily T\normalfont}_{\mbox{\scriptsize C\normalsize}} = 
\frac{1}{\sqrt{h}}G_{(\mbox{\scriptsize W \normalsize} = 0)}^{abcd}
\left[
\mbox{\ss}_{\xi}{h}_{ab} + \frac{ 4\mbox{\ss}_{\xi}{\phi}}{\phi}h_{ab}
\right]
\left[
\mbox{\ss}_{\xi}{h}_{cd} + \frac{4\mbox{\ss}_{\xi}{\phi}}{\phi}h_{cd}
\right] \mbox{ } ,
\label{VBOaction}
\eea
which is consistent for $\sigma = 1$ because it circumvents the above argument about the 
fourth factor by independently guaranteeing a new slicing equation for the lapse.  Despite its 
lack of 4-d general covariance, conformal gravity is very similar to CWB GR 
in that CWB GR's dynamical d.o.f's may be taken to be 
represented by \cite{York72, York73, York74}
$$
\mbox{CS + V} \equiv 
\mbox{Conformal Superspace $+$ Volume} 
\mbox{\hspace{1.4in}}
$$
\be
\mbox{\hspace{0.6in}}
= \frac{\mbox{Riem}}
{  \mbox{Diff} \times
\{\mbox{ Volume-preserving conformal transformations}\}  }
\label{VCSplusV}
\ee
and conformal gravity arises by considering instead
\be
\mbox{Conformal Superspace}
= \frac{\mbox{Riem}}
{\mbox{Diff} \times 
\{\mbox{Conformal transformations}\}} 
\mbox{ } .
\label{VCS}
\ee
This has an infinite number of `shape' d.o.f's whereas there is only one volume d.o.f.  Yet 
removing this single d.o.f changes one's usual concept of cosmology (see V.2), and ought to change the 
Problems associated with the quantization of the theory (by permitting the use of a 
positive-definite inner product and a new interpretation for ${\cal H}$) \cite{CG}. 
One arrives at further 3-space {\it CS+V theories} if one chooses to work on
(\ref{VCSplusV}) instead of (\ref{VCS}) \cite{CG, ABFKO} whilst still retaining a
fundamental slicing.  There are yet other conformal theories (III.2.6, V.2.2).  

I included the $Y$, rather than scaling it to 1 like BF\'O, to make clear my recent insight into 
how a `Galilean' branch arises for $Y = 0$ i.e when the gravitational momenta completely vanish in the 
Hamiltonian constraint.  From this as well as the Lorentzian branch arising, it becomes 
clear that the condition that (\ref{keyterm}) vanishes is closely related to the choice of postulates that 
Einstein faced in setting up special and general relativity.  This point is best further 
developed in V once matter has been introduced. 

\mbox{ } 

To mathematically distinguish GR from these other theories, I use

\noindent
\bf TSA GR postulate 3 \normalfont: the theory does not admit privileged foliations 
and has signature $\epsilon = -\sigma = -1$. 

\noindent My future strategy will involve seeking to overrule the conformal alternatives on 
fundamental grounds, by thought experiments or by use of current astronomical data, 
which would tighten the uniqueness of GR as a viable 3-space theory \sl on physical 
grounds\normalfont.  If such attempts persistently fail, these theories will become 
established as serious alternatives to GR (see V.2).  

\mbox{ }

\noindent{\bf 2.2.4 Deducing BM from BSW alone}

\mbox{ }

\noindent \'{O} Murchadha \cite{SGNiall} and I \cite{Sanderson} separately considered the possibility of starting off with 
`bare velocities' $\dot{h}_{ij}$ rather than BM ones $\dot{h}_{ij} - \pounds_{\xi}h_{ij}$,  
to provide a variation on the 3-space theme in which the spatial relationalism is emergent 
rather than assumed.  My generalized version (including $\sigma$, $Y$, $W$ and $\Lambda$) 
works as follows.  I obtain a bare momentum 
\be
p^{ij} = \frac{\partial\mbox{\sffamily{L}\normalfont} }{ \partial\dot{h}_{ij}} =
\frac{\sqrt{h}}{2NY}G^{ijcd}_{\mbox{\scriptsize W\normalsize}}\dot{h}_{cd} \mbox{ } ,
\label{barewmom}
\ee
and the usual Hamiltonian-type constraint (\ref{VGRHam}) arises as a primary 
constraint from the local square root.  The temporary ELE's read
$$
\dot{p}^{ij} = \frac{\delta\mbox{\sffamily L\normalfont}}{\delta h_{ij}} =   \sqrt{h}Nh^{ij}(\sigma R + \Lambda) -\sqrt{h}\sigma NR^{ij}
- \frac{2NY}{\sqrt{h}}\left(p^{im}{p_m}^j -\frac{X}{2}p^{ij}p\right)
$$
\be
+ \sqrt{h}\sigma (D^iD^j N - h^{ij}D^2 N) \mbox{ } , 
\label{barewel}
\ee
which permit the evaluation of the propagation of ${\cal H}$ 
\be
\dot{{\cal H}} = -\frac{2\sigma Y}{N}D_a(N^2[D_b{p^{ab}} + (X - 1)D^ap]) + \frac{(3X - 2)Np}{2\sqrt{h}}{\cal H} 
\mbox{ } .
\ee
For neither $\sigma = 0$ nor $Y = 0$, I then obtain the secondary constraint
\be 
{\cal S}_a \equiv D_b{p^b}_a + (X - 1)D_a p = 0 \mbox{ } .
\ee 
But propagating this gives 
$$
\dot{{\cal S}}_a = 
\sqrt{h}(X - 1)
\left(
D_a
\left[
2\sigma(NR - D^2N) + 3N\Lambda + \frac{NY(3X - 2)p^2}{2h}
\right]  
- \frac{(3X - 2)NpY}{2h}D_ap
\right)
$$
\be
-\frac{1}{2N}D_a(N^2{\cal H}) 
\mbox{ } , 
\ee
so if constraints alone arise (rather than conditions on $N$ imposing preferred foliations), 
I require $X = 1$ and recover relativity since the secondary constraint $-2{\cal S}_i = 0$ becomes 
the momentum constraint\fn{That ${\cal H}_i$ is an integrability of ${\cal H}$ was already known to 
Teitelboim and Moncrief \cite{TM72, Teitelthesis}. } 
\be
{\cal H}_i \equiv -2D_j{p_i}^j = 0.  
\ee

One then argues that this `discovered' constraint may be `encoded' into the bare action by the 
introduction of an auxiliary variable $\xi^i$.  It is then this encoding that may be thought 
of as the content of BM.  This is similar to how the emergent gauge-theoretic Gauss constraints 
are treated in IV.1.   One then re-evaluates the momenta and ELE's, obtaining the 
GR ones (\ref{GRmom}) and (\ref{GReleq}) in place of the above temporary bare ones (\ref{barewmom}) 
and (\ref{barewel}).

For $\sigma = 0$ or $Y = 0$ no additional constraints arise.  Thus it is possible to `miss out' 
rather than `discover' constraints by this integrability method above (further discussed in 
III.1.2).    
Thus although the `bare' and BM schemes are equivalent for pure GR \cite{SGNiall},  
they are not in general equivalent \cite{Sanderson}.  Another example of this is 
`bare Barbour--Bertotti theory' in which \b{${\cal M}$} = \b{${\cal L}$} = 0 are not recovered, and hence is just 
Newtonian mechanics.  

\mbox{ }

\noindent\bf{2.2.5 Higher derivative potentials}\normalfont 

\mbox{ }

\noindent Using the potential \sffamily P \normalfont $= \sigma R + \Lambda$ assumed in 1.2.3 
amounts to applying a temporary 

\noindent{\bf TSA gravity simplicity 4}: the pure gravity action is constructed with at most second-order 
derivatives in the potential, and with a homogeneously quadratic BM kinetic term. 

\noindent Furthermore, BF\'{O} considered potentials that are more complicated scalar concomitants 
of the 3-metric $h_{ij}$ than the above:  \sffamily P\normalfont $ = R^n$ and 
\sffamily P\normalfont $= C_1R^2 + C_2R\circ R + C_3D^2R$ (the most general fourth-order 
curvature correction in 3-d because of the Gauss--Bonnet theorem).  Among these the 
potential of GR alone permits the Hamiltonian constraint to propagate.  Also, recently, 
\'{O} Murchadha \cite{Niall03} considered actions based on matrices $M_{ab}(x_i, \lambda)$ 
and their conjugates $P^{ab}(x_i, \lambda)$ which contain all terms in the former with the use 
of up to two derivatives and are ultralocal in the latter.  In this bare approach, he recovers 
the combination $R(M_{ab})$ for the form of the potential (along with the strong and conformal 
options), as singled out by the propagation of the local square root constraint.

The first difficulty with the above methods is that they are not as yet systematic 
order-by-order, whereas HKT were able to use induction to dismiss all alternative potentials 
to all orders.  Furthermore, HKT's proofs rest on the extra structure assumed, so similar 
proofs cannot be easily envisaged for the TSA.  The second difficulty is that 
other known theories are in fact implicitly excluded both in BF\'{O}'s and in HKT's work 
(further discussed in VI.1.5 along with other difficulties).  Finally, HKT's result is known 
to be dimension-dependent since Lovelock gravity also reproduces the Dirac Algebra \cite{Lovelocksystem}.

\mbox{ }

\noindent{\bf 2.2.6 The inclusion of matter} 

\mbox{ }

\noindent Just as Teitelboim \cite{Teitelthesis, Teitelboim} `added on' matter so as to strengthen HKT's answer 
(see App IV.C), we have strengthened BF\'{O}'s answer so.  Our first works \cite{BOF, AB} appear to 
give some striking derivations of the 
classical laws of bosonic physics.  Rather than being presupposed, both the null cone 
structure shared between gravitation and classical bosonic matter theories, and gauge theory,  
are enforced and share a common origin in the propagation of the Hamiltonian constraint.  
Thus electromagnetism and Yang--Mills theory are picked out.  Gauge theory arises in this 
picture because one discovers the gauge-theoretic Gauss constraint as a secondary constraint 
from the propagation of the Hamiltonian constraint.  This is then encoded by an auxiliary 
field which occurs as gauge-theoretic BM corrections to the velocities.  

The matter considered was subject to the 

\noindent{\bf TSA matter simplicity 5}: the matter potential has at most first-order derivatives and the kinetic 
term is  ultralocal and homogeneous quadratic in the velocities.  Apart from the homogeneity, 
this parallels Teitelboim's main matter assumptions.    

\noindent As explained in VI.2--3 there is a further tacit simplicity hidden in the 
`adding on' of matter.  This is linked to how including general matter can alter the 
gravitational part of the theory.  This issue is tied both to the relationship between 
the TSA and the {\bf POE} (see VII.2), and 
in my contesting BF\'{O}'s speculation that the matter results ``hint at partial unification"  
(see VI, VII).  

\mbox{ }

\noindent\Large{\bf{3 Spacetime or space?}}\normalsize

\mbox{ }

\noindent Having built up a theoretical 3-space scheme working in turn with point particles, 
gauge fields and 3-geometries, I now argue for its merit and compare it with Einstein's route, 
ADM's formulation and HKT's route.  

I begin by elaborating on Einstein's motivations for his route to GR and then 
explain how his route neither directly implements these nor is in accord with the 
dynamics-centred development of the rest of physics.  The first of these points is a relationalist's 
complaint for which RWR offers a peaceful reconciliation: the direct implementation \sl also 
\normalfont gives rise to GR. The second point has become quite a tension, as explained below.   

Einstein had several goals \cite{CGspacetime} during the years that he created 
special and general relativity.  The first, realized in special relativity, was to reconcile 
electromagnetism with the universal validity of the restricted relativity principle {\bf RP1}. 
In contrast to Lorentz \cite{Lorentz}, who explicitly sought a constructive theory \cite{CGconstructive} to 
explain the Michelson--Morley experiment and the relativity principle, Einstein was convinced 
that the quantum effects discovered by Planck invalidated such an approach \cite{CGspacetime}, 
p. 45. He ``despaired of the possibility of discovering the true laws by means of constructive 
efforts'' \cite{CGspacetime}, p. 53, and instead adopted {\bf RP1} as an axiomatic principle 
\cite{CGconstructive}. 
Notably this reconciliation involves adopting the physically-motivated \bf Lorentzian RP2 \normalfont in 
place of the absolute-time-motivated {\bf Galilean RP2}.  But whereas one has a new set of 
privileged frames (SR inertial rather than Newton inertial), 
 
\noindent 1) one is still giving distinct status to configurations which are relationally 
identical;

\noindent 2) there is still distinct unaccounted-for status being given to a privileged class 
of frames.

\noindent Einstein's further goals were to free 
physics of absolute space (`implement Mach's principle', see below), and to construct a field 
theory of gravitation analogous to Maxwell's electromagnetism.  
Encouraged by his treatment of {\bf RP1} as a principle to be adopted rather than a result to 
be derived, Einstein generalized it to the general relativity principle {\bf GRP}, 
according to which the laws of nature must take an identical form in all frames of reference. 
The {\bf GRP} was eventually implemented as the 4-d general covariance of a 
pseudo-Riemannian dynamical spacetime.

Whereas Einstein's route addresses 2) via the equivalence principle and the local vanishing of 
the connection which is in turn influenced by dynamical geometry that is itself influenceable, 
1) is apparently disregarded.  Also, in making spacetime the arena of dynamics, Einstein broke 
radically with the historical development of dynamics, in which the configuration space and 
phase space had come to play ever more dominant roles. Both of these played decisive 
roles in the discovery of quantum mechanics, especially the symplectic invariance of Hamiltonian 
dynamics on phase space.  Since then, Hamiltonian dynamics has also played a vital role in the 
emergence of modern gauge theory \cite{Dirac}. In fact, spacetime and the canonical dynamical 
approach have now coexisted for almost a century, often creatively but also not without tension.
This tension became especially acute when Dirac and Arnowitt, Deser and Misner (ADM) 
\cite{Dirac, ADM} reformulated the Einstein field equations as the constrained Hamiltonian 
dynamical system (\ref{Vham}, \ref{Vmom}, \ref{CGsmallevol}, \ref{CGBSWEL}) describing the 
evolution of Riemannian 3-metrics $h_{ij}$ Dirac was so impressed by the simplicity of the 
Hamiltonian formulation that he questioned the status of spacetime, 
remarking ``I am inclined to believe \dots that 4-d symmetry is not a fundamental 
property of the physical world" \cite{CGDiracPRC}.  Wheeler too was struck by this 
development \cite{Wheeler}, and contributed the thoughts and terminology of I.2.7.  

However, the hitherto unresolved status of the one remaining redundancy due to the Hamiltonian 
constraint, which can only be eliminated at the price of breaking the spacetime covariance of 
GR, is probably the reason why neither Dirac nor Wheeler subsequently made any serious attempt 
to free themselves of spacetime. In particular, Wheeler rather formulated the idea of embeddability, 
i.e, that Riemmanian 3-geometries always evolve in such a way that they can be embedded in a
4-d pseudo-Riemannian spacetime \cite{Wheeler}, which led Hojman, 
Kucha$\check{\textrm r}$ and Teitelboim \cite{HKT} to their seventh route to general 
relativity.  

Contrast this, and Einstein's indirect approach through generalization of \bf RP1 \normalfont 
to \bf GRP\normalfont, with the TSA first principles, which constructively implement   
Mach's idea of a relational dynamics!  Whereas Einstein tried to incorporate his 
interpretation\fn{Mach's principle was left open to diverse and misunderstood interpretations 
\cite{DOD, buckets}.} of Mach's principle into GR, that interpretation and GR were found not to be satisfactorily 
compatible.  {\sl But CWB GR is Machian in Barbour's (or Wheeler's) sense, in addition to being a 
spacetime theory.}

The primality of spacetime or space constitute two different ontological viewpoints.  
In presupposing embeddability, HKT assume the habitual spacetime viewpoint 
of relativists, whereas Barbour defends the space viewpoint.  In the former, there are two 
presuppositions about how 3-geometries stack: spatial stacking and foliation invariance.  
These presuppositions become manifest in Kucha\v{r}'s general 
decompositions of spacetime objects with respect to spatial hypersurfaces (see VI.2).  
In the latter, spatial stacking and RI are presupposed.    
One then sees that there are alternatives to the spacetime foliation invariance  
presupposition in the form of the alternative theories of III (even if these are distasteful), 
and it becomes interesting to see if these can realistically describe the universe.  
{\sl Not all Machian theories are spacetime theories!}  


Supportive arguments for BM theories have recently been put forward by 
Brown and Pooley \cite{RRICM} for mechanics and by Pooley \cite{RRII} for relativity.  
An argument against it (in mechanics) is that relationalists have it easy when 
$\mbox{\b{${\cal L}$}} = 0$ \cite{Earman89}.  But $\mbox{\b{${\cal L}$}} = 0$ may be argued to 
be a {\sl prediction} of BB82 theory: a restricted subset of Newtonian mechanics emerges, so 
the BB82 theory is good in the sense of being more restrictive and thus in principle more 
readily falsifiable than Newtonian mechanics.  I emphasize that within geometrodynamics the 
debate ends up being ``spacetime versus space" and not the full ``physical objects plus 
container versus objects alone" of the absolutist versus relationalist debate.  Reality {\sl is} 
ascribed to structure, but this is to the {\sl 3-geometry} and not to space(time) points. 

Before the discussion of spacetime versus space, I require the introduction of matter (IV, VI).  
I then show how the space scheme works in V.1 (including the emergence of the Relativity 
Principle), and how split space-time works in VI.2.  These are contrasted in VI.3.  
The status of the {\bf POE} in these schemes is explored in VII.2.  
 
The two viewpoints are suggestive of how one approaches quantum gravity and its 
Problem of Time (in VIII).  Internal time is based on belief in spacetime 
(as are formulations in terms of sum over histories, generally-covariant histories, spin foams 
and causal sets), whereas not specifically (spacetime) generally covariant canonical approaches and 
Barbour's hope in the na\"{\i}ve Schr\"{o}dinger interpretation are space-based. 
The question is then: was spacetime a historical accident in 
that, because of being a very fruitful approach in classical GR and in quantum particle 
physics in a Minkowskian background, we have been overly bedazzled by spacetime and 
consequently misled by it in our attempts to quantize gravity?  

Barbour also put forward his classical and quantum ideas in \cite{B94I, B94II} and the popular 
book \cite{EOT}.  
These have been critically discussed by Butterfield \cite{VButterfield} and Smolin 
\cite{VSmolin} largely from a quantum mechanical, philosophical perspective.  
In contrast, my criticisms in V, VI and VII concentrate on the later technical papers, 
from a classical, mathematical perspective.

\mbox{ }

\noindent\Large{\bf App II.A : Deriving the relativity without relativity result}\normalsize

\mbox{ }

\noindent For the arbitrary ultralocal supermetric, one can build up the following 
$\xi^i$-free parts of results:  
\be
\dot{h}_{ij} = \frac{2NY}{\sqrt{h}}
\left(
p_{ij} - \frac{X}{2}ph_{ij}
\right) \mbox{ } ,
\ee
\be
\dot{h}^{ij} = -\frac{2NY}{\sqrt{h}}
\left(
p^{ij} - \frac{X}{2}ph^{ij}
\right) \mbox{ } ,
\label{hupstairs}
\ee
\be
\dot{\sqrt{h}} = \frac{2 - 3X}{2}\sqrt{h}NYp 
\mbox{ } ,
\ee
\be
{    \dot{\Gamma}^t    }_{lm} = \frac{Y}{\sqrt{h}}
\left(
D_m
\left[    
N
\left(
{p_l}^t\mbox{--}\frac{X}{2}{h_l}^t   
\right) 
\right]
\mbox{+}D_l
\left[   
N
\left(
{p_m}^t\mbox{--}\frac{X}{2}p{h_m}^t  
\right)
\right]
\mbox{--} D^t
\left(  
N
\left[
p_{ml}\mbox{--}\frac{X}{2}ph_{ml}   
\right)
\right]
\right) 
\mbox{ } ,   
\label{CDOT}
\ee
\be
{\dot{\Gamma}^t}_{lt} = \frac{(2 - 3X)Y}{2\sqrt{h}}D_l(Np) 
\mbox{ } ,
\label{CCDOT}
\ee
$$
\sqrt{h}\dot{R}  = -2NY
\left(
p^{ij} - \frac{X}{2}ph^{ij}
\right)
R_{ij} + 2YD_jD_iNp^{ij} + 2(X - 1)YD^2Np 
$$
\be
+ \frac{2Y}{N}D_i(N^2D_j{p^{ij}}) 
+ \frac{2(X - 1)Y}{N}D^i(N^2D_ip) \mbox{ } .   
\label{RDOT}
\ee
These (along with the metric ELE) can be composed by the chain-rule to form the propagation of the Hamiltonian constraint, 
and hence weakly obtain the consistency condition \ref{keyterm} (which includes the RWR result), 
from the last term of (\ref{RDOT}).  One can similarly obtain the working for the propagation of the momentum constraint.  

The $\xi^i$ part of the Hamiltonian constraint propagation is simple once one realizes that 
$\pounds_{\xi}R$ is simply $\xi^i\pa_iR$ [which could also be derived longhand using  
the symmetry of $R_{ab}$, the Ricci lemma (\ref{Riccilemma}) and 
the contracted Bianchi identity (\ref{contbi})]. 
By similar techniques, the $\xi^i$-part of the propagation of the momentum constraint is  
$$
D_j(      \pounds_{\xi}p^{ij}      )\mbox{+}{\dot{\Gamma}^i}_{kj}|_{\xi}p^{jk}\mbox{--}\pounds_{\xi}(      D_jp^{ij}      ) 
= \xi_k(      D_jD^kp^{ij} - D^kD_jp^{ij}     ) 
+ p^{ij}(      D_jD_k{\xi^k} - D_kD_j{\xi^k}      )                          
\mbox{\hspace{0.4in}}
$$
$$
\mbox{ } \mbox{ } \mbox{ }  \mbox{ } \mbox{ }  
\mbox{ } \mbox{ } \mbox{ }  \mbox{ } \mbox{ } \mbox{ }  \mbox{ } \mbox{ } \mbox{ } 
+ p^{jk}(      D_jD^i{\xi_k} - D^iD_j\xi_{k}      )                                                                        
\mbox{\hspace{0.5in}}
$$
\be
\mbox{\hspace{2in}}
= \xi_k(      {{R_{kj}}^i}_dp^{dj} + {{R_{kj}}^j}_dp^{id}      ) + p^{ij}{{R_{kj}}^k}_d\xi^d 
+ p^{jk}{R^i}_{jkd}\xi^d = 0 
\mbox{ } .  
\label{liepid}
\ee

\noindent\bf\Large{App II.B 3-space approach and gauge theory}\normalsize\normalfont  

\mbox{ }

\noindent I go beyond BB82's treatment in including an arbitrary gauge group G rather than U(1), 
and in considering a local as well as a global ordering.  Given a configuration space 
that is a collection of 1-forms 
$A_i^{\mbox{\bf\scriptsize I\normalfont\normalsize}}$ (where {\bf I} running over 1 to 
{\bf K} are internal indices) which are functions of position on $\Re^3$ and of label time 
$\lambda$, and which are invariant under small   

\be
\mbox{G-transformations, I require {\bf general R1:} to work on the }
\mbox{\hspace{0.2in}}
{\mbox{\sffamily RCS\normalfont}}(A^{\mbox{\bf\scriptsize I\normalfont\normalsize}}) 
= \frac{Q(A^{\mbox{\bf\scriptsize I\normalfont\normalsize}})}{\mbox{G} \times \mbox{Eucl}} 
\mbox{ } . 
\mbox{\hspace{1.5in}}
\ee
N.B I could also (more standardly) proceed without involving Eucl.  

\noindent{\bf R2} The choice of time label is not to affect the system.   

I use the indirect implementation {\bf BM[general R1]} by passing to 
$\mbox{\sffamily Q\normalfont}(A^{\mbox{\bf\scriptsize I\normalfont\normalsize}}) 
\times \mbox{G} \times$ Eucl by best matching each velocity.  
The electromagnetic BM is with respect to the U(1) gauge transformations 
$
\dot{\mbox{\b{A}}} \longrightarrow \dot{\mbox{\b{A}}} - \mbox{\b{$\pa$}}\Lambda.  
$
So for any two 3-d 1-form fields on a flat space, one without loss of generality keeps one in a fixed gauge and 
changes the gauge of the other until the two 1-form fields are as close as possible.
Thus overall 
$
\dot{\mbox{\b{A}}}_{\mbox{\bf\scriptsize I\normalfont\normalsize}} \longrightarrow 
\mbox{\ss}_{\Lambda, \mbox{\scriptsize a\normalsize}, \mbox{\scriptsize b\normalsize}}
{\mbox{\b{A}}}_{\mbox{\bf\scriptsize I\normalfont\normalsize}} \equiv 
\dot{\mbox{\b{A}}}_{\mbox{\bf\scriptsize I\normalfont\normalsize}} 
- \mbox{\b{$\pa$}}\Lambda^{\mbox{\bf\scriptsize I\normalfont\normalsize}} 
+ \mbox{\sffamily g\normalfont}_{\mbox{\scriptsize c\normalsize}}[\Lambda, \mbox{\b{A}}]_{\mbox{\bf\scriptsize I\normalfont\normalsize}} 
- \mbox{\b{a}}_{\mbox{\bf\scriptsize I\normalfont\normalsize}}
- \mbox{\b{b} \scriptsize $\times$ \normalsize \b{A}}_{\mbox{\bf\scriptsize I\normalfont\normalsize}},
$
where $\mbox{\b{a}}^{\mbox{\bf\scriptsize I\normalfont\normalsize}}$ is 
\bf K \normalfont copies of \b{a}, and both 
$\mbox{\b{a}}^{\mbox{\bf\scriptsize I\normalfont\normalsize}}$ and \b{b} are independent 
of position on $\Re^3$.  

I first use the global square root reparameterization-invariant implementation {\bf RI[R2] } 
\be
\mbox{\sffamily I\normalfont}^{\mbox{\sffamily\scriptsize L\normalsize\normalfont}}_{\mbox{\scriptsize YM({\cal G})\normalfont}} = \int \textrm{d} \lambda 
\sqrt{                
\left(
{\mbox{\sffamily E\normalfont}}     -    \int\textrm{d}^3x         
F_{\bar{p}\bar{q}}^{\mbox{\bf\scriptsize I\normalfont\normalsize}}
F^{\bar{p}\bar{q}}_{\mbox{\bf\scriptsize I\normalfont\normalsize}}    
\right)
\int \textrm{d}^3x
\mbox{\ss}_{\Lambda, \mbox{\scriptsize a\normalsize}, \mbox{\scriptsize b\normalsize}}
{\mbox{\b{A}}}_{\mbox{\bf\scriptsize I\normalfont\normalsize}}
\cdot
\mbox{\ss}_{\Lambda, \mbox{\scriptsize a\normalsize}, \mbox{\scriptsize b\normalsize}}
{\mbox{\b{A}}}^{\mbox{\bf\scriptsize I\normalfont\normalsize}}
                      } \mbox{ } .  
\ee
Defining $2N \equiv \sqrt{    
\frac{        ({\mbox{\sffamily\scriptsize E\normalsize\normalfont}}   
- \int\mbox{\scriptsize d\normalsize}^3x(   \mbox{ \scriptsize\b{$\pa$}\normalsize } \times 
\mbox{\scriptsize\b{A}\normalsize}^{\mbox{\bf\tiny I\normalfont\normalsize}} + 
\mbox{\sffamily\scriptsize g\normalsize\normalfont}[\mbox{\scriptsize\b{A}\normalsize}, 
\mbox{\scriptsize\b{A}\normalsize}]^{\mbox{\bf\tiny I\normalfont\normalsize}}  )^2    )  }
{\int\mbox{\scriptsize d\normalsize}^3x
\mbox{\ss}_{\Lambda, \mbox{\tiny a\normalsize}, \mbox{\tiny b\normalsize}}
\dot{\mbox{\scriptsize\b{A}\normalsize}}_{\mbox{\bf\tiny I\normalfont\normalsize}}
\cdot
\mbox{\ss}_{\Lambda, \mbox{\tiny a\normalsize}, \mbox{\tiny b\normalsize}}
\dot{\mbox{\scriptsize\b{A}\normalsize}}^{\mbox{\bf\tiny I\normalfont\normalsize}}       }      }$, 
the field momenta are 
$\pi_{\mbox{\bf\scriptsize I\normalfont\normalsize}} = 2N^{\mbox{\scriptsize G\normalsize}}
\mbox{\ss}_{\Lambda, \mbox{\scriptsize a\normalsize}, \mbox{\scriptsize b\normalsize}}
{\mbox{\b{A}}}_{\mbox{\bf\scriptsize I\normalfont\normalsize}}.$
Variation with respect to the auxiliaries $\Lambda_{\mbox{\bf\scriptsize I\normalfont\normalsize}}(x)$, 
$\mbox{\b{a}}^{\mbox{\bf\scriptsize I\normalfont\normalsize}}$ and \b{b} respectively yields 
\be
\mbox{\hspace{1.5in}}
\mbox{the Yang--Mills Gauss constraint } 
\bar{\mbox{\b{D}\normalfont}}^{\mbox{\scriptsize G\normalfont}}\cdot\mbox{\b{$\pi$}}_{\mbox{\bf\scriptsize I\normalfont\normalsize}} = 0 
\mbox{ } , 
\mbox{\hspace{1.5in}}
\label{ymaux1}
\ee
\be
\mbox{\b{${\cal M}$}} 
= \int\textrm{d}^3x \sum_{\mbox{\bf\scriptsize I\normalfont\normalsize}} \mbox{\b{$\pi$}}^{\mbox{\bf\scriptsize I\normalfont\normalsize}} \mbox{ } ,
\label{ymaux2}
\ee
\be
\mbox{\b{${\cal L}$}} = \int\textrm{d}^3x \mbox{\b{A}}^{\mbox{\bf\scriptsize I\normalfont\normalsize}}  
\mbox{ \scriptsize $\times$ \normalsize} \mbox{\b{$\pi$}}_{\mbox{\bf\scriptsize I\normalfont\normalsize}} 
\mbox{ } .
\label{ymaux3} 
\ee 
There is additionally a primary constraint from the global square root: 
\be
{\cal P} \equiv \int \textrm{d}^3x
\left(
\mbox{\b{$\pi$}}^{\mbox{\bf\scriptsize I\normalfont\normalsize}} 
\cdot 
\mbox{\b{$\pi$}}_{\mbox{\bf\scriptsize I\normalfont\normalsize}} +   
F_{\bar{p}\bar{q}}^{\mbox{\bf\scriptsize I\normalfont\normalsize}}
F^{\bar{p}\bar{q}}_{\mbox{\bf\scriptsize I\normalfont\normalsize}}           
\right)
= \int{\textrm{d}^3x}\mbox{\sffamily E\normalsize} \mbox{ } . 
\ee

The ELE's are $\dot{\pi}^{\bar{p}}_{\mbox{\bf\scriptsize I\normalfont\normalsize}} 
= \bar{D}^{\mbox{\scriptsize G\normalsize}}_{\bar{q}}
(2N^{\mbox{\scriptsize G\normalsize}} F\normalfont^{\bar{p}\bar{q}}_{\mbox{\bf\scriptsize I\normalfont\normalsize}}) = 0.$ 
These guarantee the propagation of the constraints.  Now, it is required for the time-label 
\be
\mbox{$\lambda$ to be such that } 
\mbox{\hspace{1.1in}}
{\mbox{\sffamily E\normalfont}}  =  \int \textrm{d}^3x
\left[  
{      F_{\bar{p}\bar{q}}^{\mbox{\bf\scriptsize I\normalfont\normalsize}}
F^{\bar{p}\bar{q}}_{\mbox{\bf\scriptsize I\normalfont\normalsize}}      }      +
     {      
\mbox{\ss}_{\Lambda, \mbox{\scriptsize a\normalsize}, \mbox{\scriptsize b\normalsize}}
\dot{\mbox{\b{A}}}_{\mbox{\bf\scriptsize I\normalfont\normalsize}}
\cdot
\mbox{\ss}_{\lambda, \mbox{\scriptsize a\normalsize}, \mbox{\scriptsize b\normalsize}}
\dot{\mbox{\b{A}}}^{\mbox{\bf\scriptsize I\normalfont\normalsize}}
     }
\right]
\mbox{\hspace{1.1in}} 
\ee
in order to recover the Yang--Mills field equations.  One further does not usually correct 
with respect to Eucl, but \sl the standard form of gauge theory precisely corresponds to G-BM\normalfont.  

\mbox{ } 

I next consider the local square root ordering.
\be
\mbox{\sffamily I\normalfont}^{\mbox{\sffamily\scriptsize L\normalsize\normalfont}}_{\mbox{\scriptsize YM({\cal G})\normalfont}} 
= \int \textrm{d} \lambda \int \textrm{d}^3x   ({\mbox{\sffamily E\normalfont}}      
- F_{\bar{p}\bar{q}}^{\mbox{\bf\scriptsize I\normalfont\normalsize}}
F^{\bar{p}\bar{q}}_{\mbox{\bf\scriptsize I\normalfont\normalsize}})
\mbox{\ss}_{\Lambda, \mbox{\scriptsize a\normalsize}, \mbox{\scriptsize b\normalsize}}
\dot{\mbox{\b{A}}}_{\mbox{\bf\scriptsize I\normalfont\normalsize}}
\cdot
\mbox{\ss}_{\Lambda, \mbox{\scriptsize a\normalsize}, \mbox{\scriptsize b\normalsize}}
\dot{\mbox{\b{A}}}^{\mbox{\bf\scriptsize I\normalfont\normalsize}} \mbox{ } .
\ee
Defining $2N = \sqrt{                   
\frac{        \mbox{\scriptsize\sffamily E\normalfont\normalsize} 
-  F_{\bar{p}\bar{q}}^{\mbox{\bf\tiny I\normalfont\normalsize}}
F^{\bar{p}\bar{q}}_{\mbox{\bf\tiny I\normalfont\normalsize}}       }{                
\mbox{\ss}_{\Lambda, \mbox{\tiny a\normalsize}, \mbox{\tiny b\normalsize}}
\dot{\mbox{\scriptsize\b{A}\normalsize}}_{\mbox{\bf\tiny I\normalfont\normalsize}}
\cdot
\mbox{\ss}_{\Lambda, \mbox{\tiny a\normalsize}, \mbox{\tiny b\normalsize}}
\dot{\mbox{\scriptsize\b{A}\normalsize}}^{\mbox{\bf\tiny I\normalfont\normalsize}}
}                                  }$, 
the field momenta are now
$\mbox{\b{$\pi$}}_{\mbox{\bf\scriptsize I\normalfont\normalsize}} = 2N
\mbox{\ss}_{\Lambda, \mbox{\scriptsize a\normalsize}, \mbox{\scriptsize b\normalsize}}
\dot{\mbox{\b{A}}}_{\mbox{\bf\scriptsize I\normalfont\normalsize}}.$
As above, variation with respect to the auxiliary variables yields (\ref{ymaux1}), (\ref{ymaux2}) and 
(\ref{ymaux3}), whereas there 
\be
\mbox{is now one primary constraint per space point, }
\mbox{\hspace{0.6in}}
{\cal P}(\mbox{\b{x}}) \equiv 
\pi_{\bar{p}}^{\mbox{\bf\scriptsize I\normalfont\normalsize}}
\pi_{\mbox{\bf\scriptsize I\normalfont\normalsize}}^{\bar{p}} 
+ F_{\bar{p}\bar{q}}^{\mbox{\bf\scriptsize I\normalfont\normalsize}}
F^{\bar{p}\bar{q}}_{\mbox{\bf\scriptsize I\normalfont\normalsize}}        
= \mbox{\sffamily E\normalfont}
\mbox{\hspace{0.5in}}
\ee
due to the local square root.  The ELE's are now 
$\dot{\pi}^{\bar{p}}_{\mbox{\bf\scriptsize I\normalfont\normalsize}} = 
\bar{D}^{\mbox{\scriptsize G\normalsize}}_{\bar{q}}
\left(
2N                 
\bar{D}^{[\bar{q}}A^{\bar{p}]}_{\mbox{\bf\scriptsize I\normalfont\normalsize}}
\right).$
These propagate by standard energy--momentum conservation: $\dot{{\cal P}}$ gives  
\be
{\cal S}_{\bar{p}} \equiv (\mbox{\b{E}}_{\mbox{\bf\scriptsize I\normalfont\normalsize}} 
\mbox{\scriptsize $\times$ \normalsize} \mbox{\b{B}}^{\mbox{\bf\scriptsize I\normalfont\normalsize}})_{\bar{p}} 
= \pi_{\mbox{\bf\scriptsize I\normalfont\normalsize}}^{\bar{q}}
F_{\bar{q}\bar{p}}^{\mbox{\bf\scriptsize I\normalfont\normalsize}} = 
\pi_{\mbox{\bf\scriptsize I\normalfont\normalsize}}^{\bar{q}}
\left(
\pa_{\bar{p}}A_{\bar{q}}^{\mbox{\bf\scriptsize I\normalfont\normalsize}} 
- \pa_{\bar{q}}A_{\bar{p}}^{\mbox{\bf\scriptsize I\normalfont\normalsize}}  
+ {C^{\mbox{\bf\scriptsize I\normalfont\normalsize}}}_{{\mbox{\bf\scriptsize JK\normalfont\normalsize}}}
A_{\bar{q}}^{\mbox{\bf\scriptsize J\normalfont\normalsize}}A_{\bar{p}}^{\mbox{\bf\scriptsize K\normalfont\normalsize}}
\right) = 0 
\mbox{\hspace{0.1in}}
\ee
as a secondary, whose propagation involves no new constraints.  It is true that now this is locally 
rather than globally restrictive and thus only gives a small fragment of conventional electromagnetism 
or Yang--Mills theory.  Similar thoughts play a corrective role in the universal null cone result 
(see V.1).  This difficulty goes away in GR coupled to these matter fields because of the role of 
the momentum constraint.   Thus, GR permits more complete 
Machian matter field theories.  Finally, I note that the Hamiltonian constraints arise in place 
of labelling conditions that ensure the recovery of the standard form of the equations of physics.  
There are to be no such privileged labellings in GR!  

\vspace{9in}


\noindent\Huge\bf{III: Alternative TSA theories}\normalfont\normalsize

\mbox{ }

\noindent\Large{\bf 1 Strong gravity alternatives}\normalsize

\mbox{ }

\noindent `` \it ` Well, in our country' said Alice ... `you'd generally get to somewhere else if you run 
very fast for a long time' \normalfont  '' Lewis Carroll \cite{ATTLG}

\mbox{ }

\noindent\large{\bf 1.1 Introduction}\normalsize

\mbox{ }

\begin{figure}[h]
\centerline{\def\epsfsize#1#2{0.4#1}\epsffile{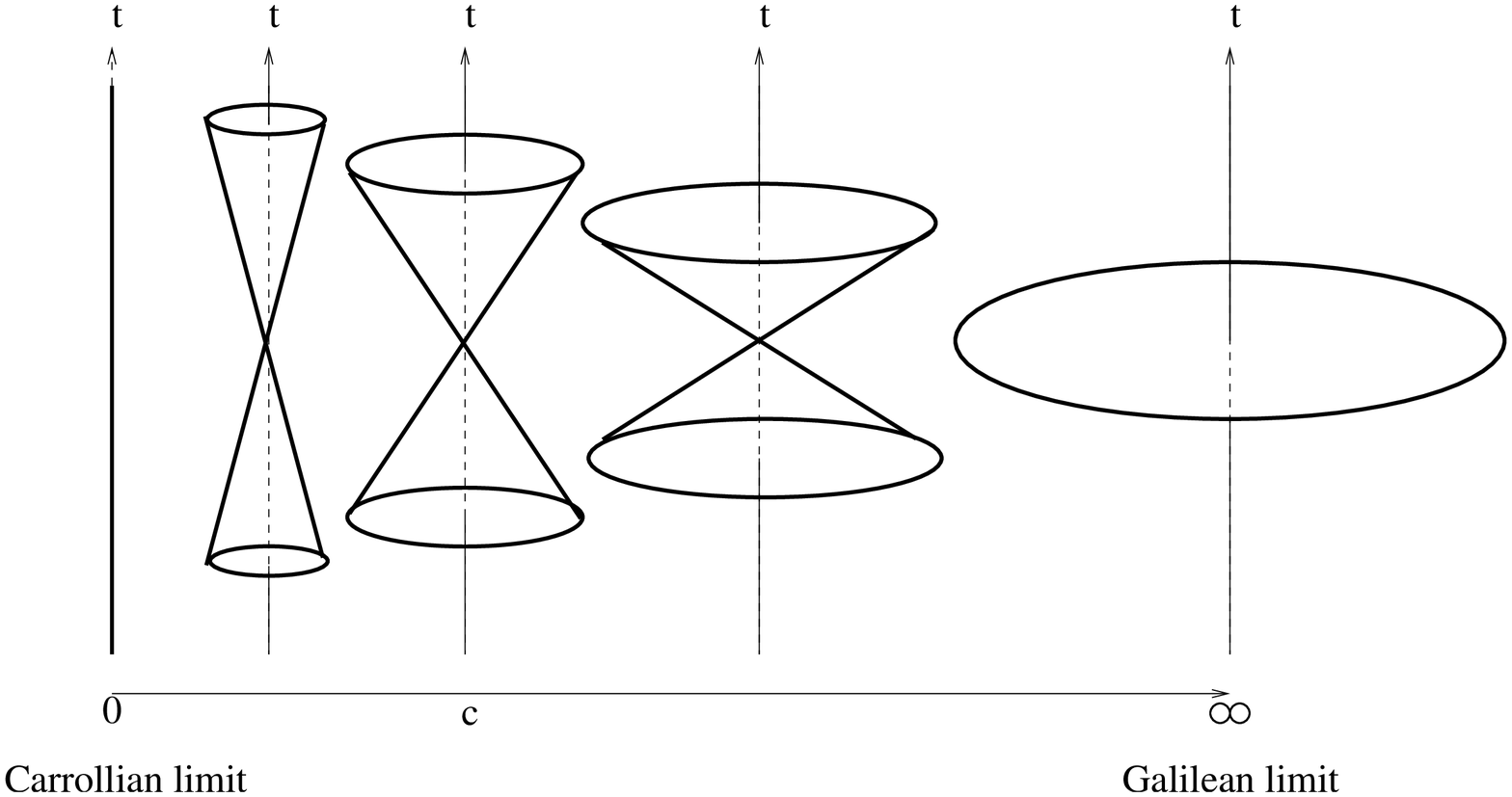}}
\caption[]{\label{2}
\scriptsize  
\normalsize}
\end{figure}
\noindent If one adopts Einstein's \bf RP1\normalfont, there is then actually a third 
choice\fn{Another source of choices might be to introduce somewhat more complicated structures 
such as in noncommutative geometry approaches.} (in addition to the Galilean $v_{prop} = \infty$ 
and the Lorentzian $v_{prop} =$ finite): the {\bf Carrollian RP2} \cite{STeitel} $v_{prop} = 0$.   
Whereas in the Galilean limit the Lorentzian light-cone becomes squashed into a plane of simultaneity 
due to the infinite speed of physical signals, in the Carrollian limit the Lorentzian 
light-cone becomes squeezed into a line as points become entirely isolated due to the zero 
speed of physical signals (see fig 9).  Thus the Carrollian universe consists of entirely 
isolated worldlines.  The most common interpretation of this is as ultralocal physics (no 
spatial derivatives). 
Klauder \cite{Klauderlit1, Klauderlit2} has studied ultralocal scalar field theory as a 
non-renormalizable but nevertheless controllable quantum field theory.  

{\it Strong gravity } is the regime of GR in which $K\circ K - K^2 \ll R$ which may be considered to 
be letting $G \longrightarrow \infty$, or alternatively letting 
$v_{\mbox{\scriptsize prop\normalsize}} = c \longrightarrow 0$ \cite{STeitel}.  
This was first considered by Isham \cite{SIsham} so as to be a starting point for a nonstandard perturbative 
theory of quantum gravity along the lines of Klauder's method above.  There are actually two 
forms of strong gravity
\be
K \circ K - K^2 = \Lambda
\label{5dofsg}
\ee
\be
\mbox{and } 
\mbox{\hspace{1.5in}}
K \circ K - K^2 = \Lambda \mbox{ , } D_iK^{ij} - D^jK = 0
\mbox{\hspace{3in}}
\label{2dofsg}
\ee
which are thus 5 d.o.f and 2 d.o.f theories of gravity respectively.  
I retain a `cosmological constant' $\Lambda$ to ensure the theory is nontrivial.  
Strictly, one is taking the $G \longrightarrow \infty$ limit while keeping 
$\frac{\Lambda}{G}$ constant to arrive at these theories.  Strong gravity is locally Carroll invariant just like GR is locally Lorentz invariant.  

Note how these are written in `split' form.  Now, strong gravity has a degenerate metric 
\be
g_{AB} = 
\left(
\begin{array}{cc}
0 & 0 \\
0 & h_{ij}
\end{array}
\right) \mbox{ } . 
\ee
This gives an unusual geometry, as studied by Henneaux \cite{SHenneaux}.  One needs to use a 
weaker notion of inverse than usual, leading to only a very limited notion of parallel 
transport.  The extrinsic curvature $K_{ab}$ becomes merely the $ab$ components of a 4-tensor 
$K_{AB}$ rather than being the whole of $K_{AB}$, and (\ref{2dofsg}) cease to be the $0A$ 
components of a 4-tensor equation (the EFE's).  There does not appear to be a clean 
interpretation of strong gravity as a 4-tensor theory -- spacetime structure does not appear 
here as it does in GR.  

What use is strong gravity?  It approximates GR near the cosmological singularity, since it
gives an independent Kasner universe at each spatial point, which is the conjectured 
behaviour of the general solution of GR (see I.2.7.1).  Thus it is a worthwhile regime to 
quantize \cite{Pilatilit1,Pilatilit2,Pilatilit3}.  The notion of strong gravity is related 
to the two most popular approaches to quantum gravity as follows.  It is analogous to the 
tensionless string \cite{SGstringyrefs11,SGstringyrefs12,SGstringyrefs13,SGstringyrefs14,SGstringyrefs15}, 
and it admits an Ashtekar variable formulation \cite{A88, SGHusain} (see III.1.5). 

In the TSA, we consider strong gravity as a dynamically-consistent theory of 
evolving 3-geometries on its own merit, obtained as an alternative to GR.  
My more careful analysis revealed that the 3-space approach gives rather a 1-parameter 
family of strong gravity theories.  Their discovery provides a \sl different \normalfont 
answer to Wheeler's question from the uniqueness of BF\'{O}, in the case of the strong-coupled 
limit $\sigma = 0$:  there is a consistent theory not only for the $W = 1$ DeWitt supermetric 
of the usual strong gravity, but also for any ultralocal invertible ($W \neq \frac{1}{3}$) 
supermetric.   Whereas the use of initially bare or 
initially BM velocities in the GR case of the TSA does not affect the final form of the emergent theory, in the strong 
gravity case the outcome {\sl is} affected.  
I furthermore discuss how they can be related (as limits relevant 
to the very early universe) to the well-known scalar-tensor theories of gravity (III.1.3), and used 
as toy models toward the study of conformal gravity (III.1.4).    
 
Like Minisuperspace, strong gravity is a GR regime amenable to quantization.  My theories 
provide an enlargement of this second arena, in which different permitted ranges of values of 
$W$ permits considerable mathematical differences \cite{SGGiulini}.  In particular, for 
$W < \frac{1}{3}$ one has theories with positive-definite i.p's. The study of these 
could broaden the understanding of the i.p Problem of quantum gravity (see VIII).  I furthermore 
discuss the possibility of the very early universe actually having a positive-definite i.p.  
I then show (III.1.5) that the Ashtekar variable formulation, of potentially great use 
in quantization, is not readily available for $W \neq 1$.  The strong gravities are studied as 
systems of equations in III.1.6.  

I also return to strong gravity in IV.1.4, where I investigate the coupling to it of fundamental 
matter.  This enables fruitful comparison with matter coupling in the GR case, leading to better 
understanding of some of the GR TSA results.  

\mbox{ }

\noindent\large{\bf 1.2 Strong gravity and the TSA: X is arbitrary}\normalsize

\mbox{ }

\noindent We are interested in finding consistent theories of evolving 3-geometries;
we use the TSA 
\be
\mbox{to construct them.  Consider then the BM RI action, } 
\mbox{ } \mbox{ }
\mbox{\sffamily I\normalfont}
= \int \textrm{d}\lambda \int \textrm{d}^3x \sqrt{h} \sqrt{\Lambda}
\sqrt{\mbox{\sffamily T\normalfont}^{\mbox{\scriptsize g\normalsize}}_{\mbox{\scriptsize W\normalsize}}}.
\mbox{\hspace{1in}}
\ee
The canonical gravitational momenta are given by (\ref{wmom}) with  
$2N =\sqrt{\frac{\mbox{\sffamily\scriptsize T\normalsize\normalfont}^{\mbox{\tiny g\normalsize}}_{\mbox{\tiny W\normalsize}}}{\Lambda}}$.

The strong gravity Hamiltonian constraint follows as a primary constraint from the local 
\be
\mbox{square root form of the Lagrangian: } 
\mbox{\hspace{0.5in}}
{\cal H}^{\mbox{\scriptsize strong\normalsize}}_{\mbox{\scriptsize W\normalsize}} \equiv \frac{1}{\sqrt{h}}\left(p\circ p - \frac{X}{2}p^2\right) - \sqrt{h}\Lambda = 0 
\mbox { } . 
\mbox{\hspace{0.5in}}
\ee
$\xi^i$-variation yields the usual secondary momentum constraint (\ref{Vmom}).

The ELE's are the obvious subcase of the ELE's (\ref{wel}):     
\be
\dot{p}^{ij} = \frac{    \delta\mbox{\sffamily L\normalfont}    }{    \delta h_{ij}    } =   \sqrt{h}Nh^{ij}\Lambda - \frac{2N}{\sqrt{h}}
\left(
p^{im}{p_m}^j -\frac{X}{2}p^{ij}p
\right)
+ \pounds_{\xi}p^{ij},
\ee
\be
\mbox{From these one can evaluate $\dot{{\cal H}}$: }
\mbox{\hspace{1in}}
\dot{{\cal H}} = \frac{Np(3X - 2)}{2\sqrt{h}}{\cal H} + \pounds_{\xi}{\cal H}  
\label{SGevolham} 
\mbox{ } ,
\mbox{\hspace{1in}}
\ee
which vanishes weakly in the sense of Dirac. 

This corresponds to the $\sigma = 0$ possibility of (\ref{mastereq}), for which any ultralocal 
supermetric is allowed.  The theory traditionally called strong gravity has $W = 1$ because it 
is obtained as a truncation of GR.  But I have now shown that there is a family of such 
theories as far as dynamical consistency is concerned.  Because $W = \frac{1}{3}$ is badly 
behaved, the family of dynamically-consistent theories naturally splits into $W > \frac{1}{3}$ 
and $W < \frac{1}{3}$ subfamilies.  Such $W < \frac{1}{3}$ theories should be simpler to quantize 
than more habitual $W > \frac{1}{3}$ theories (including GR), because the former have a 
positive-definite supermetric, removing difficulties in defining a suitable quantum i.p.  
A particularly simple example of such a dynamically-consistent theory 
is the $W = 0$ theory, for which the constraints are \mbox{ } \mbox{ } 
$p \circ p  = h\Lambda \mbox{ } , \mbox{ }  D_j{p^{ij}} = 0 
\mbox{ } .$

\noindent $W = 0$ may be of particular relevance, in part because conformal gravity has been formulated 
in terms of the $W = 0$ supermetric, and in part from string-theoretic considerations (see III.1.3).   
Conformal gravity arises because, in a conformal generalization of the
above working, the equivalent of the slicing equation (\ref{SGslicingeq}) is independently guaranteed to hold.
There is also a separate strong conformal theory \cite {CG} (see III.2.6).

Since for $\sigma = 0$ (\ref{mastereq}) reduces to (\ref{SGevolham}), the momentum constraint 
may no longer be seen to arise as an integrability condition.  This fact was already noted by 
Henneaux \cite{SHenneaux}.  Strong gravity thus provides a counterexample to the suggestion that 
all additional constraints need arise from the propagation of ${\cal H}$.  However, all the 
other constraints can be interpreted as arising in this way in the standard approach to GR 
(see VI.3--4): ${\cal H }_i$, the electromagnetic Gauss constraint ${\cal G}$, the 
Yang--Mills Gauss constraint ${\cal G}_{\mbox{\scriptsize\bf J\normalsize\normalfont}}$ and the 
`locally Lorentz' constraint ${\cal J}_{AB}$ from working in some first order formalism.

So strong gravity without a momentum constraint is also dynamically consistent \cite{SGRovelli}, 
and the new $W \neq 1$ strong gravities may be treated in this way too.  In fact, it is this 
treatment that corresponds to strictly taking the $G \longrightarrow \infty$ limit of GR (as opposed 
to Pilati's approach \cite{SGPilatiID} in which the momentum constraint is kept).  This is because the 
GR momenta are proportional to $G^{-1}$ \cite{A88}.  One could also start off with a
${\cal H}_i$, but use Ashtekar variables in place of the traditional ones.
In this case, the analogues of ${\cal H}_i$ and ${\cal J}_{AB}$
cease to be independent \cite{A88}.

If one starts off with `bare velocities' rather than BM ones, for $\sigma = 0$ one obtains 
\be
\dot{{\cal H}} =  \frac{(3X - 2)Np}{2\sqrt{h}}{\cal H} \mbox{ } .
\ee
which vanishes weakly, and so no further, secondary constraints emerge, and one has the 5-d.o.f 
`metrodynamical' strong gravity theories.  

 \mbox{ }

\noindent\large{\bf 1.3 On the meaning of the theories: application to scalar-tensor theories}\normalsize

\mbox{ }

\noindent The $X \neq 1$ departure from the DeWitt supermetric does not appear to affect 
Henneaux's study of the geometry.  Whereas these theories are no longer interpretable as 
truncations of GR, they do correspond to truncations of scalar-tensor theories (such as 
Brans--Dicke theory), in a region where the scalar field is a large constant.  The relations 
between the Brans--Dicke parameter $\omega$ and our coefficients $W$ and $X$ are shown in fig 10.  
I now discuss the possibility that a positive-definite ($W < \frac{1}{3}$ i.e 
$\omega < 0$) inner product can occur in our universe.   There is no point in considering 
Brans--Dicke theory, an action for which is 
\be
\mbox{\sffamily I\normalfont}_{\mbox{\scriptsize BD\normalfont}} = \int \textrm{d}^4x\sqrt{|g|}e^{-\chi}(\check{R} - \omega\pa_a\chi \pa^a\chi)   
\mbox{ } ,
\ee
since this has $\omega$ constant in space and time and we know from the recent analysis of Cassini 
data \cite{Cassini} that today $\omega \geq 20000$ \cite{Will}, corresponding to $W$ being very 
slightly larger than the GR value 1.  However, {\it general scalar-tensor theory} permits $\omega(\chi)$ so $\omega$ 
varies with space and in particular with time. So it could be that the very early universe had 
a very different value of $\omega$ from that around us today, since the bounds on $\omega$ 
from nucleosynthesis \cite{dampich} permit 
$\frac{\omega_{\mbox{\scriptsize nucleosynthesis\normalsize}}}{\omega_{\mbox{\scriptsize today\normalsize}}} \approx \frac{1}{25}$.  
The bounds from \cite{KiefGiu} are less strict but presumably applicable to a wider range of 
theories since the origin of the departure from $W \neq 1$ (i.e $\omega \neq \infty$) is there 
unspecified.  Furthermore, $\omega$ is attracted to the GR value at late times in scalar-tensor 
theories \cite{letter, paper}, so it need not have started off large.  One would expect 
$\omega$ of order unity in any fundamental scalar-tensor theory \cite{letter}.  For example 
$\omega = -1$ arises in low-energy string theory \cite{BarDab}.    

\begin{figure}[h]
\centerline{\def\epsfsize#1#2{0.6#1}\epsffile{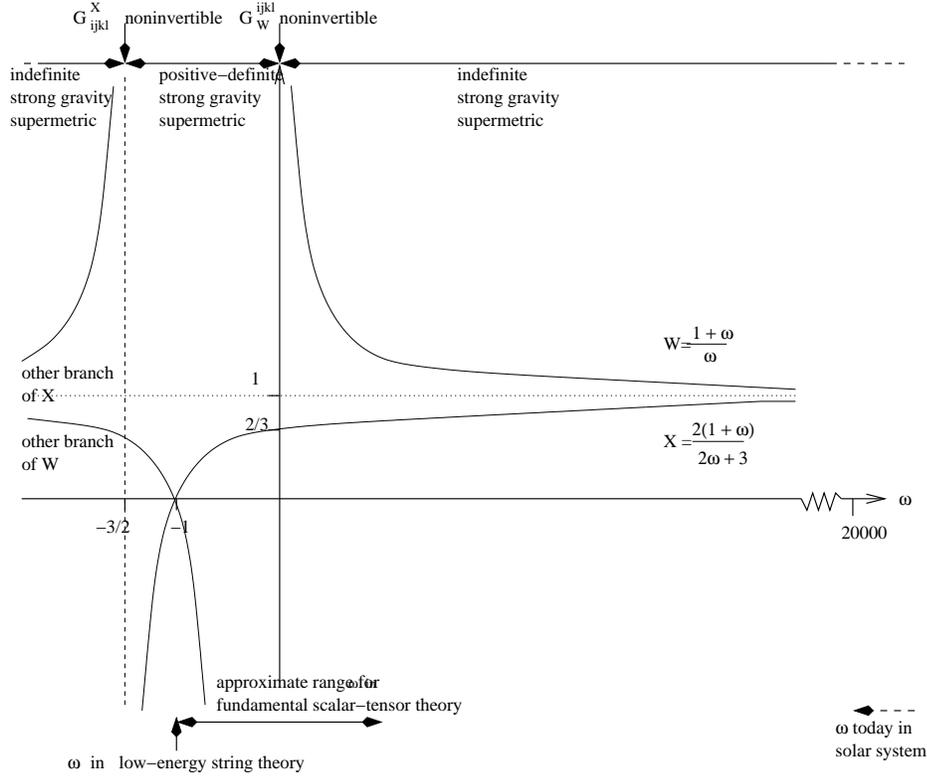}}
\caption[]{\label{3}
\footnotesize Although the Brans--Dicke parameter $\omega \geq 20000$ in the solar system today 
\cite{Cassini}, scalar-tensor theories permit $\omega$ to vary and it is suggestive that 
$\omega$ could have been smaller in the early universe.  If $\omega$ tends to or passes through 
zero, the corresponding strong gravity theory there would be considerably different from the 
strong gravity theory corresponding to GR due to the character of its supermetric. \normalsize}
\end{figure}

It is thus an open question whether $\omega$ at early times could have passed through the 
value 0.  This question is interesting for the quantum-mechanical reasons given in the next 
paragraph.  I first wish to clarify the role of my strong gravities in such a study.  They 
do \sl not \normalfont permit $W$ (and hence $\omega$) to change with 
time, so I am not suggesting to use these to investigate whether such a transition through 
$\omega = 0$ is possible.  But if such transitions are found to  be possible, the very early 
universe could then be described by scalar-tensor theories which have $\omega < 0$.  Then one 
of my strong gravity theories which behaves qualitatively differently from the usual $W = 1$ 
strong gravity corresponding to GR would be relevant as an approximation near the initial 
singularity.  I propose to study the possibility of having such a transition using the full 
scalar-tensor theories.  Unlike their strong gravity limits discussed above, for which 
this transition involves passing through a noninvertible supermetric,\fn{In fact, this 
$W = \frac{1}{3}$ supermetric is of the same form as the degenerate strong gravity 4-metric, 
so the pointwise geometry of Superspace for $W = \frac{1}{3}$ should be taken to be akin to 
Henneaux's geometry \cite{SHenneaux} of strong gravity spacetimes.} the full scalar-tensor 
theories are not badly-behaved as $\omega \longrightarrow 0$.  This is because despite the 
degeneracy of the tensor (`gravity') supermetric for $\omega = 1$, what counts for the full 
theory is the larger scalar-tensor supermetric.  Just like the GR supermetric may be 
represented by a $6 \times 6$ matrix,  the scalar-tensor supermetric may be represented by 
a $7 \times 7$ matrix where the new seventh index is due to the scalar field \cite{KM}:

\noindent
\be
\left(
\begin{array}{ll} 
\mbox{{\large G}}_{abcd}(h_{ab}) & \mbox{{\large G}}_{ab\chi}(h_{ab}, \chi) \\  
\mbox{{\large G}}_{\chi cd}(h_{ab}, \chi) & \mbox{{\large G}}_{\chi\chi}(h_{ab}, \chi) 
\end{array}
\right) = 
\left(
\begin{array}{ll}
G^{\mbox{\scriptsize X\normalsize}}_{abcd} & (X - 1)h_{ab} \\ 
(X - 1)h_{cd} & -(X - 1) \\
\end{array}
\right)
\label{hornedhelmet}
\ee  
Now, this is well-behaved at $\omega = 0$ because of its scalar-tensor cross-terms, due to 
which the degeneracy of the $6 \times 6$ block is not sufficient to cause the whole 
$7 \times 7$ scalar-tensor supermetric to be degenerate.  But in the approximation by which 
the theories of this subsection arise from scalar-tensor theories the scalar momentum is 
negligible, so one \sl is \normalfont then left with only the `gravity' supermetric.  

Thus in principle there could be different possible early universe behaviours which admit 3 
different sorts of strong gravity limits, corresponding to indefinite, degenerate and positive-definite i.p's.   
Each of these cases has a correspondingly different kind of natural candidate for the early-universe WDE i.p.  

\mbox{ } 
 
\noindent\large{\bf 1.4 Application to conformal gravity}\normalsize

\mbox{ }

\noindent Another application of the $W < \frac{1}{3}$ theories follows from 
conformal gravity (see III.2) being $W$-insensitive and thus expressible in terms of a 
$W = 0 < \frac{1}{3}$ supermetric. 
I will additionally show that a sequence of theories can be formed: $W = 0$ strong gravity, 
conformal strong gravity, conformal gravity, which could permit the isolated study of 
some of the novel features of classical and quantum conformal gravity (see V.2.1.3)

In answer to whether arguments from the II.1.3 are applicable to conformal gravity, 
begin by noting that there is no `expansion term' $p$ in conformal gravity.    Because this 
absence is due to $p = 0$ being separately variationally imposed (rather than due to $W = 0$ 
occurring for the vacuum theory), the presence of non-minimally-coupled-scalars or the related 
use of conformal transformations are unable to reintroduce a $p$ into conformal gravity.    
Consequently, conformal gravity cannot be included among the `wider range of theories' for which the 
less stringent bounds on $W$ mentioned in 1.3 are applicable.  Conformal gravity is a theory in 
which $W$ plays no role at all.  Presumably the classical and quantum study of conformal gravity on 
Superspace with $W <\frac{1}{3}$ and $W > \frac{1}{3}$ are equivalent once projected down to conformal 
Superspace.   Working out how this happens may be interesting and instructive, at least from a 
theoretical point of view. 

\mbox{ }

\noindent\large{\bf 1.5 Difficulty with implementation of Ashtekar variables}\normalsize

\mbox{ }

\noindent This section includes theories for which the i.p Problem of quantum GR is altered 
(if not ameliorated).  In the case of conformal gravity, its preferred foliation 
additionally represents an attempt to circumvent the Problem of time of quantum GR.  
One must however recall that these Problems of quantum gravity are always intertwined 
with other formidable Problems, which include operator ordering and regularization.  
At least in GR, Ashtekar variables have nice properties as regards these Problems (I.3.3.3).   
scalar-tensor theories or conformal gravity admit an analogue of Ashtekar variables.  
Indeed, how special is GR in admitting Ashtekar variables with their nice properties?   

In the strong-coupled limit of GR, the constraints 
(\ref{ashgauss}), (\ref{ashmom}) and (\ref{ashham}) become \cite{bombelli,SGHusain}
\be
[\mbox{\tt A\normalfont}_a, \mbox{\tt E\normalfont}^a] = 0 \mbox{ } ,
\label{sashgauss}
\ee
\be
tr({\tt E\normalfont}^m[\mbox{\tt A\normalfont}_m, \mbox{\tt A\normalfont}_i]) = 0 \mbox{ } ,
\label{sashmom}
\ee
\be
tr({\tt E\normalfont}^a{\tt E\normalfont}^b[\mbox{\tt A\normalfont}_a, \mbox{\tt A\normalfont}_b]) - h\Lambda = 0 \mbox{ } .  
\label{sashham}
\ee
By the cyclic property of the trace and use of (\ref{sashgauss}),  that 
(\ref{sashmom}) is redundant as claimed in III.1.1.  Furthermore, there is an equivalent form 
for the remaining constraints (\ref{sashgauss}) and 
\be
\mbox{(\ref{sashham}) \cite{bombelli}: } 
\mbox{\tt A\normalfont}_{[ab]} = 0 
\mbox{ } ,
\label{equivsashgauss}
\ee
\be
\mbox{\tt A\normalfont}_{ab}\mbox{\tt A\normalfont}_{cd}\underline{G}^{abcd} = \Lambda 
\label{equivsashham}
\ee
(for $\mbox{\tt A\normalfont}_{ab} \equiv tr({\tt A\normalfont}_a\sigma_b)$), which 
manifestly displays the dependence on the (now overall undensitized inverse) DeWitt 
supermetric $\underline{G}^{abcd}$.  I then investigate 
what happens when $G^{abcd}$ is replaced by $G_{\mbox{\scriptsize W\normalsize}}^{abcd}$.  
Notice how then the Hamiltonian constraint no longer contains a truncation of the natural 
object ${\tt F\normalfont}^{\mbox{\scriptsize\tt A\normalfont\normalsize}}_{ab} 
= 2\partial_{[a}{\tt A\normalfont}^{\mbox{\scriptsize\tt A\normalfont\normalsize}}_{b]} 
+ |[{\tt A\normalfont}_a, {\tt A\normalfont}_b]|^{\mbox{\scriptsize\tt A\normalfont\normalsize}}$ 
of SU(2) Yang--Mills theory, in 
correspondence with $W \neq 1$ strong gravity not being a natural truncation of the GR 
Hamiltonian constraint.  The constraint algebra closes.

For full scalar-tensor theory, I do not think Ashtekar variable with $W \neq 1$ will work.  
One has there the option of making conformal transformations to put scalar-tensor theory into a 
$W = 1$ form, but the conformal factor required then causes the constraints to be 
non-polynomial \cite{SGCapovilla}.  As for conformal gravity, one could as well write the 
theory with $W = 1$, but I see no way that conformal gravity's lapse-fixing equation 
(\ref{CGarbslice}) can be expressed polynomially.  All these theories' lacks could be viewed 
as a `second RWR result': the form of the supermetric in 
${\cal H}^{\mbox{\scriptsize trial\normalsize}}$ is fixed to be DeWitt's GR one if one 
requires passage to Ashtekar variables with their quantization-geared neatness.    

\mbox{ }

\noindent\large{\bf 1.6 PDE problems for strong gravity theories}\normalsize 

\mbox{ }

\noindent The 5 d.o.f version of the theories has a trivial thin sandwich formulation, as a consequence 
of having no thin sandwich equation.  Beware this simple state of affairs 
since it is reminiscent of Minisuperspace models.  
Also, it only works for the $\Lambda < 0$ case by the Problem of zeros corresponding to the 
thin sandwich approach.  But within this case there are globally no zeros, a decided 
improvement on the situation in GR.  
The condition $\Lambda < 0 $ has the knock-on effect of forcing $W > 0$ from the Gauss 
constraint.   
\be
\mbox{ } \mbox{ The IVP involves solving the constraint }
\mbox{\hspace{0.6in}}
K^{\mbox{\scriptsize T\normalsize}}\circ K^{\mbox{\scriptsize T\normalsize}} + \frac{1 - 3W}{3}K^2 
= \Lambda 
\mbox{ } .
\mbox{\hspace{1in}}
\label{III16}
\ee
No scalings are fixed in the `York IVP method' corresponding to these theories, because 
there is no Codazzi constraint to keep conformally-invariant and the conformally-transformed 
Gauss equation can contain no $|D\psi|^2$ terms because it has no $R$ in the first place.  
This latter point means that the IVP now involves no differential equations at all: the  
`York IVP method' is now an algebraic method.  
If one declares everything to be known except the scale of the metric, 
one can scale the various objects in the Gauss constraint as one pleases and explicitly
solve it algebraically for the scale.  The reason for the GR study's restriction to CMC slicings 
is also absent because there is no Codazzi constraint.

\mbox{ }

The 2 d.o.f version of the theories has a nontrivial thin sandwich formulation.  
As above, the Problem of zeros becomes the mere requirement that $ - \Lambda > 0 \Rightarrow W > 0$   
which is a global rather than local condition.  
Although many  of the counterexamples hold irrespective of $W$ and of the fixed sign of the potential,   
I have reasons to suspect the proofs will turn out to be $W$-dependent.  
\be
\mbox{ } \mbox{ The IVP now involves solving the constraints (\ref{III16}) and } 
\mbox{\hspace{0.3in}}
D_iK^{\mbox{\scriptsize T\normalsize}ij} + \frac{1 - 3W}{3}D^{j}K = 0 \mbox{ } .
\mbox{\hspace{0.3in}}
\ee
Now, one scaling is fixed in the `York IVP method', because there is a Codazzi 
equation to keep conformally-invariant, but again the conformally-transformed 
Gauss equation can contain no $|D\psi|^2$ terms because it has no $R$.  
So for $h_{ij}$ and $K^{\mbox{\scriptsize T\normalsize}ij}$ scaling as in (\ref{metscal}) and (\ref{otherscal}),  
$\eta = -2\zeta$.  According to one interpretation there should be no $\rho$ and $j^a$ as these terms become 
negligible in the strong-coupled gravity limit (this is distinct from the `Pilati limit' interpretation used in IV.1.4  
to illustrate other points).  In this case the full constraints have less terms in them 
than in GR.  The resulting IVP is not trivial: one starts off by solving a $j^i = 0$ momentum constraint.  
But then, in place of a Lichnerowicz--York p.d.e, one has 
\be
\psi^{2\zeta - 2\eta}K^{\mbox{\scriptsize T\normalsize}}\circ K^{\mbox{\scriptsize T\normalsize}} 
+ \frac{1 - 3W}{3}K^2 
= \Lambda
\label{easylich}
\ee
where we have demanded that $\Lambda$ does not scale so that it remains a cosmological 
constant in the physical frame.  
Then for example, using the usual but now unforced $\eta = 4$, this has 
\be
\mbox{the trivial explicit 
algebraic solution }
\mbox{\hspace{1.0in}}
\psi = 
\left(
\frac
{K^{\mbox{\scriptsize T\normalsize}}\circ K^{\mbox{\scriptsize T\normalsize}}}
{\frac{3W - 1}{3}K^2 + \Lambda}
\right)^{\frac{1}{12}} 
\mbox{ } .
\mbox{\hspace{1.7in}}
\label{easypsi}
\ee 

The maximal case is never sustainable, since the corresponding lapse-fixing equation 
is trivially only solved by frozen dynamics ($N = 0$), independently of what 
asymptotics the spaces have:
$
\dot{p} \approx 3N\Lambda \mbox{ } , \mbox{ } \mbox{ } \Lambda \neq 0 
\mbox{ } , \mbox{ } \mbox{ } N > 0 \mbox{ } .  
$
The CMC  case is sustainable, 
trivially soluble and trivial in evolutionary character:   
\be
B(\lambda) = \frac{\pa}{\pa\lambda}\left(\frac{p}{\sqrt{h}}\right) = 3N\Lambda 
+ \frac{p^2}{h}\frac{3X - 2}{2} \mbox{ } \Rightarrow N = \mbox{spatial constant} \mbox{ } .
\ee
Finally, note thus that the conformal thin sandwich approach to strong gravity just reduces to 
solving the Codazzi equation, now for the shift $\xi_i$ rather than the vector potential $W_i$.

\mbox{ }

What of the Cauchy problem for these equations?  The indefinite signature 
means that only derivatives with respect to label time are contained in the evolution 
equations.      
Thus the Cauchy problem is merely that for an o.d.e, and may be cast into the form 
$\ddot{h}_{ij} = F(h_{ij}, \dot{h}_{ij}; \xi_i; N)$
away from the bad value $W = \frac{1}{3}$.  

\mbox{ }

\noindent\Large{\bf 2 Conformal alternatives}\normalsize

\mbox{ }

\noindent\large{\bf 2.1 Introduction}\normalsize  

\mbox{ }

\noindent Recollect from II.3 that there is a tension between the spacetime and split  
spacetime geometrodynamical interpretations of GR.  The TSA would add to this tension if it 
leads to {\sl realistic} geometrodynamical theories in which the GR structure of spacetime is 
\sl not \normalfont emergent.   III.1 contains {\sl toy} examples of this, appropriate for 
extreme regimes.  This section is about attempts to build such theories which are realistic in 
everyday as well as extreme regimes.  

In the usual geometrodynamical interpretation of GR, the configuration space is the extension 
of Riem to Riem $\mbox{ }\times B \times P$, where $B$ is the space of shifts $\beta^i$ 
and $P$ is the space of lapses $\alpha$ (which are positive functions).  However, since 
$\alpha$ and $\beta^i$ have no conjugate momenta, the true gravitational d.o.f's of GR are 
contained in Riem.   They are furthermore subjected to ${\cal H}$ and ${\cal H}_i$. If, in the 
thin-sandwich problem for geometrodynamics, one could solve ${\cal H}_i$ in terms of the Lagrangian 
variables for $\beta^i$, then the theory would be defined on the quotient, Superspace.  However, 
because of ${\cal H}$, Superspace still contains one redundant d.o.f per space 
point.  Then GR is a BM RI theory whose true configuration space is contained within 
Superspace.  As part of this interpretation it is natural and equivalent to regard $N$ and 
$\xi^i$ (which become identified with $\alpha$, $\beta^i$) as velocities associated with 
cyclic coordinates rather than as Lagrange multiplier coordinates.  Also (I.2.9.4.2), one 
may take the true configuration space of CWB GR as close to being Conformal Superspace (CS).  In this section the TSA to 
geometrodynamics is tentatively extended to include scale invariance.  The resulting conformal 
gravity theory is a BM RI theory on CS; the role of the auxiliaries being velocities and not 
coordinates plays a role in deriving this from first principles.  Whereas conformal 
gravity, has an additional linear constraint $p = 0$, the theory still has 2 d.o.f's per space 
point.  To form a configuration space of CWB GR, one requires to adjoin the single global volume 
of the universe to CS, thus forming a space `CS+V'.  Theories on CS+V will also be developed.  

In III.2.2, I somewhat rework Barbour's scale-invariant particle dynamics model, to be used 
to help test out a number of conceptual and technical issues for scale-invariant 
geometrodynamics.  This model requires \sffamily E -- V \normalfont to be homogeneous of degree -- 2 
in the particles' relative positions, which does not disrupt usual physics by the trick of 
division by suitable powers of the moment of inertia.  I begin by treating the auxiliary $a$ 
responsible for scale invariance as a coordinate.\fn{Whereas for conformal gravity the first 
principles derivation is conceptually helpful, it is not clear how to do this for alternative 
theories considered later.  This is why I also provide the auxiliary coordinate 
interpretation.}  $a$-variation yields a new constraint 
$\sum_{(i) = (1)}^{(n)}q_{(i)}p^{(i)} = 0$, analogous to $p = 0$.  I next treat the 
auxiliaries as velocities and give a derivation of the scale-invariant particle dynamics 
theory from BM first principles.  This involves the compensatory auxiliary field giving the 
action an overall `banal' scale invariance, and the use of a free-endpoint variational 
principle.  

In III.2.3, I give a 2-auxiliary Lagrange multiplier coordinate ($\phi$ and $\theta$) 
formulation of conformal gravity.  $\theta$-variation yields the $p = 0$ constraint.  
The maintenance of this is guaranteed by the lapse-fixing equation (LFE) arising from 
$\phi$-variation.  For this LFE to be workable, one requires division by a suitable 
power of the volume of the universe, $V$.  For this to be workable, one must choose 
between e.g CWB or asymptotically flat {\sl before} the theory of gravity is declared.  

In III.2.4, we arrive at conformal gravity instead from the 3-space principles of demanding 
a consistent RI BM action where there is now both Diff-BM and Conf-BM.  The conformal gravity 
action is required to be homogeneous of degree zero in the auxiliary variable $\phi$, which is 
implemented by division by the suitable power of V.  We find that what 
was previously considered to be a second auxiliary coordinate $\theta$ is in fact closely 
related to the velocity corresponding to $\phi$.  However, there are still 2 separate equations arising 
from the auxiliaries by the free-endpoint variational principle.  
In III.2.5 we include a cosmological constant as a precursor of the inclusion of matter in IV.2.

In III.2.6 we work in the Hamiltonian formulation treating whichever auxiliaries arise as 
Lagrange multiplier coordinates.  We provide a range of CS and CS+V theories.  The idea is 
to explore the whether CWB GR in York's formulation is closely analogous to some BM theory in 
CS+V, in which given an initial point and an initial direction in CS+V, a unique dynamical curve is determined.  In 
both GR and {\it CS+V theory}, there is a unique definition of simultaneity given by CMC slicings 
(corresponding to York time).  
CS+V schemes may be the more promising ones as regards both reformulating GR and establishing less 
radical but more secure departures from GR.  Further theories which share some features with GR and conformal gravity 
are also presented, including the asymptotically-flat counterparts of conformal gravity and 
the CS+V theory.  Material supporting III.2.2--6 drawn from the GR conformal IVP method is 
provided in App III.2.A.   A comparison of conformal gravity with other conformally-invariant theories 
is provided in App III.2.B.  

\mbox{ }

Here is a summary of which aspects of these theories are considered further on in the thesis.  
Matter is included in IV.2: it turns out, as in RWR, that a universal null cone, 
electromagnetism and Yang--Mills theory are picked out.  There is a discussion in V.2: conformal 
gravity should be in agreement with the standard tests of GR, and strong differences will 
emerge in cosmology and in the quantum theory; these differences will be less pronounced for 
CS+V theories.  In V.2.2 I improve on the CS+V theory, and extend the list of conformal 
theories.  In V.2.3 I provide and an account of how conformal gravity and the CS+V theory work 
as p.d.e systems, from both the IVP--CP and thin sandwich perspectives, update the cosmology 
and provide a comparison of CS+V theories with GR.   The quantum treatment is updated in VIII.  

\mbox{ }

\noindent\large{\bf 2.2 Scale-invariant particle dynamics model}\normalsize

\mbox{ }

\noindent Here I rerun the Machian program for an $n$-particle mechanics with the additional 
supposition that \sl scale \normalfont should be relative.  Then one should demand invariance 
under the 
\be
\mbox{transformations }
\mbox{\hspace{0.1in}}
\mbox{\b{q}}_{(i)} \longrightarrow \mbox{\b{\b{A}}}(\lambda)\mbox{\b{q}}_{(i)} + \mbox{\b{g}}(\lambda) \mbox{ } , \mbox{ } \mbox{ } \mbox{\b{\b{A}}}(\lambda) 
\in \mbox{SO(3)} \times \Re^+ = \mbox{SL(3)} \mbox{ , the linear group }
\mbox{\hspace{1.5in}}
\label{genLeib1}
\ee
\be
\lambda \longrightarrow f(\lambda).
\label{genLeib2}
\ee
The group of the first of the above two transformations is now not the Euclidean group Eucl, 
but the \it similarity group \normalfont Sim = \{translations, rotations and dilations\}, 
which does not preserve lengths but does preserve angles and hence shapes.  

The relative configuration space is then the quotient with respect to this similarity group 
\be
\mbox{Shape Space } \mbox{ \sffamily SS\normalfont}(n) = \frac{ \mbox{\sffamily Q\normalfont}(n)}
{\mbox{Sim}} = \frac{ \mbox{\sffamily RCS\normalfont}(n)}{\mbox{dilations}} 
\mbox{ } .  
\ee
This is trivial for $n$ = 1, 2 and has dimension $3n - 7$ for $n \geq 3$.  For $n$ = 3, SS is 
Barbour's toy space \it Triangle Land \normalfont \cite{EOT}, the space of all triangular shapes. 

Barbour then works indirectly on SS by BM: replace bare velocities according to   
\be
\dot{\mbox{\b{q}}}_{(i)} \longrightarrow 
\mbox{\ss}_{\mbox{\scriptsize {k}, $\Omega$, $\theta$\normalsize}}{\mbox{\b{q}}}_{(i)}  \equiv 
\dot{\mbox{\b{q}}}_{(i)} - \mbox{\b{k}} - \mbox{\b{$\Omega$}} \mbox{ \scriptsize $\times$ \normalsize} 
\mbox{\b{q}}_{(i)} - \theta\mbox{\b{q}}_{(i)} 
\mbox{ } .  
\ee  
This involves a new dilational correction to the particle velocities; 
for the moment I consider bringing in this and the usual translational and rotational 
corrections by use of auxiliary coordinates.    
A RI action is as ever used to implement the invariance under 
the second 
\be
\mbox{transformation: }
\mbox{\hspace{0.2in}}
\mbox{\sffamily I\normalfont} = \int\textrm{d}\lambda \mbox{\sffamily L\normalfont}
  = \int\textrm{d}\lambda \sqrt{\mbox{\sffamily E\normalfont} - \mbox{\sffamily V\normalfont}}
\sqrt{\mbox{\sffamily T\normalfont}} 
\mbox{ } , \mbox{ } \mbox{ }
\mbox{\sffamily T\normalfont} = \sum_{(i) = (1)}^{(n)}m_{(i)}
\mbox{\ss}_{\mbox{\scriptsize {k}, $\Omega$, $\theta$\normalsize}}{\mbox{\b{q}}}_{(i)}\cdot
\mbox{\ss}_{\mbox{\scriptsize {k}, $\Omega$, $\theta$\normalsize}}{\mbox{\b{q}}}_{(i)} 
\mbox{ } .  
\mbox{\hspace{0.6in}}
\label{bolt}
\ee
\be
\mbox{The conjugate momenta are } 
\mbox{\hspace{0.7in}}
\mbox{\b{p}}^{(i)} \equiv \frac{\pa \mbox{\sffamily L\normalfont}}{\pa \mbox{\b{$\dot{q}$}}_{(i)}}
= \sqrt{\frac{\mbox{\sffamily E\normalfont} - \mbox{\sffamily V\normalfont}}{\mbox{\sffamily T\normalfont}}}m_{(i)}
\mbox{\ss}_{\mbox{\scriptsize {k}, $\Omega$, $\theta$\normalsize}}{\mbox{\b{q}}}_{(i)}
\mbox{ } . 
\mbox{\hspace{1in}} 
\ee
\be 
\mbox{One obtains from the square root form of the Lagrangian }
\mbox{\hspace{0.1in}}
\sum_{(i) = (1)}^{(n)}\frac{\mbox{\b{p}}_{(i)}\cdot\mbox{\b{p}}_{(i)}}{m_{(i)}} = \mbox{\sffamily E\normalfont} 
\mbox{ }-\mbox{ }\mbox{\sffamily V\normalfont} 
\mbox{ } . 
\mbox{\hspace{0.2in}}  
\label{sipdlsr}
\ee
Variation with respect to the auxiliaries then yields as multiplier equations the usual 2 BB82 conditions 
$\mbox{\b{${\cal M}$}} = \mbox{\b{${\cal L}$}} = \mbox{\b{0}}$ and furthermore the vanishing of the 
{\it dilational momentum}: 
\be
{\cal D} \equiv \sum_{(i) = (1)}^{(n)}\mbox{\b{q}}_{(i)} \cdot \mbox{\b{p}}^{(i)} = 0 
\mbox{ } .
\ee
This means that the moment of inertia of island 
universes described are thus conserved, which would not generally be true in Newtonian mechanics.      

The ELE's from variation with respect to $\mbox{\b{q}}_{(i)}$ are 
\be
\dot{\mbox{\b{p}}}^{(i)} \equiv \frac{\pa \mbox{\sffamily L\normalfont}}{\pa \mbox{\b{q}}_{(i)}} 
= -\frac{1}{2}\sqrt{    \frac{      \mbox{\sffamily T\normalfont}      }
                             {      \mbox{\sffamily E\normalfont} - \mbox{\sffamily V\normalfont}      }      
                   }
\frac{\pa \mbox{\sffamily V\normalfont}}{\pa \mbox{\b{q}}_{(i)}} 
+ \sqrt{     \frac{      \mbox{\sffamily E\normalfont} - \mbox{\sffamily V\normalfont}      }
             {      \mbox{\sffamily T\normalfont}      }
             }
\mbox{\ss}_{      \mbox{\scriptsize {k}, $\Omega$, $\theta$\normalsize}      }\mbox{\b{q}}_{(i)}       
\mbox{ } .
\ee
These maintain the constraints $\mbox{\b{${\cal M}$}} = \mbox{\b{${\cal L}$}} = \mbox{\b{0}}$  
as before, and also maintain ${\cal D} = 0$  provided that \sffamily V \normalfont -- \sffamily E \normalfont is 
homogeneous of degree $-2$:
\be
\dot{\cal D} \approx \sum_{(i) = (1)}^{(n)}\mbox{\b{p}}^{(i)}\cdot\mbox{\b{p}}^{(i)} 
+ \frac{1}{2} \sum_{(i) = (1)}^{(n)}q_{(i)} \cdot \frac{      \pa (\mbox{\sffamily V -- E \normalfont})      }
{      \pa \mbox{\b{q}}_{(i)}      } 
=  \frac{1}{2} 
\left(
2 + \sum_{(i) = (1)}^{(n)}\mbox{\b{q}}_{(i)} \cdot
\frac{      \pa       }{      \pa \mbox{\b{q}}_{(i)}      }
(\mbox{\sffamily V\normalfont} - \mbox{\sffamily E\normalfont})
\right) 
\mbox{ } ,  
\ee
where the second equality makes use of the primary constraint (\ref{sipdlsr}).  

\mbox{ }

At first sight, this homogeneity requirement is inconvenient because  
Coulomb's law (\ref{coulomb}) and Newton's law of gravitation (\ref{NLOG}) 
involve $\frac{1}{r}$ and not $\frac{1}{r^2}$ potentials.  
But there are actually reasonable ways of making $\frac{1}{r^2}$ potentials 
give effective physics similar to that for $\frac{1}{r}$ potentials.  
Barbour's favoured method to achieve this is by use of appropriate powers of the 
moment of inertia, which depends on $r$ but seldom changes much.  
Barbour's proposed action built along these lines is (\ref{bolt}) with
\be
\mbox{\sffamily V\normalfont} = 
\frac{        \sum  \sum_{  (i) < (j)  }\frac{             m_{(i)} m_{(j)}          }
                                             {           r_{(i)(j)}                 }              }
     {        \sqrt{   \sum \sum_{  (i) < (j)  } m_{(i)} m_{(j)} r_{(i)(j)}^2   }                  } 
\mbox{ } , \mbox{ }
\mbox{\sffamily E\normalfont} = \frac{      \mbox{\sffamily E\normalfont}_0                        }
                                     {       \sum\sum_{  (i) < (j)  }m_{(i)}m_{(j)}r_{(i)(j)}^2    }      
\mbox{ } . 
\ee

To aid the development of conformal gravity, I now discuss two further possible schemes 
for the above working.  First, to make proper sense of the action as a RI action, the 
auxiliaries associated with scaling, translation and rotation should all be regarded as 
the velocities associated with cyclic coordinates to which free-endpoint 
variation is applied.  Let the scaling auxiliary hitherto used be $\dot{\zeta} = \theta$.  
I also set $\zeta = ln a$, whereupon the action is  
\be
\mbox{\sffamily I\normalfont} = \int \textrm{d}\lambda 
\sqrt{\mbox{\sffamily V\normalfont}(\mbox{\b{q}}_{(i)})
\left|
\left|
a\sum_{(i) = (1)}^{(n)} 
\left(
\dot{\mbox{\b{q}}}_{(i)} + \mbox{\b{q}}_{(i)}\frac{\dot{a}}{a}
\right)
\right|
\right|^2} 
\mbox{ } , 
\ee
which corresponds to constructing the action in an arbitrarily-scaled frame (by passing to 
\be
\mbox{Foster and Barbour's corrected coordinate) }
\mbox{\hspace{1.3in}}
\mbox{\b{q}}_{(i)} \longrightarrow a\mbox{\b{q}}_{(i)} 
\mbox{\hspace{1.3in}}
\ee
\be 
\mbox{and writing the action as } 
\mbox{\hspace{0.8in}}
\mbox{\sffamily I\normalfont} = \int\textrm{d}\lambda\sqrt{\mbox{\sffamily V\normalfont}(a\mbox{\b{q}}_{(i)})
\left|
\left|
\sum_{(i) = (1)}^{(n)} 
\frac{\pa (a\mbox{\b{q}}_{(i)})}{\pa\lambda}
\right|
\right|^2} 
\mbox{ } . 
\mbox{\hspace{1in}}
\ee
The action then has the following `banal invariance' under scalings:
\be
\mbox{\sffamily I\normalfont} \longrightarrow \mbox{\sffamily I\normalfont} \mbox{ under } 
\left\{
\begin{array}{l}
\mbox{\b{q}}_{(i)} \longrightarrow b\mbox{\b{q}}_{(i)} \\
a \longrightarrow \frac{a}{b}
\end{array}
\right. 
\mbox{ } .
\label{banban}
\ee
To get the same field equations as before, consider $\delta b$ and $\delta\dot{b}$ as separate 
variations.  Then 
\be
\mbox{the primary constraint reads }
\mbox{\hspace{1.2in}}
\sum_{{(i)} = {(1)}}^{(n)}\mbox{\b{q}}_{(i)}\cdot\mbox{\b{p}}^{(i)} + ap^{\mbox{\scriptsize a\normalfont}} = 0 
\mbox{ } , 
\mbox{\hspace{1.2in}}
\ee
variation with respect to $a$ gives 
$\dot{p}^{\mbox{\scriptsize a\normalsize}} 
= \frac{\dot{a}}{a}p^{\mbox{\scriptsize a\normalsize}}$, 
and free-endpoint variation yields 
$p^{\mbox{\scriptsize a\normalsize}} = 0$, 
so ${\cal D} = 0$.
 
This second procedure may not be very general.  While it works for translations, dilations and indeed 
conformal transformations, it is at least obscured for the \it non-commuting 
\normalfont rotations.  

\mbox{ }



\noindent\large{\bf 2.3 Two auxiliary coordinate formulation of conformal gravity}\normalsize

\mbox{ }

\noindent I now present conformal gravity much as it was originally conceived by \'{O} Murchadha 
\cite{conformal, CG}, 
using two auxiliary coordinates to give and maintain the scale-invariance.  Conformal gravity 
has subsequently been explored from the single auxiliary corrected-coordinate perspective 
(III.2.4).  The style below is available further alternative theories in III.2.6, while it is 
not clear whether these possess a corrected coordinate formulation.  

We desire a theory on CS, i.e on Riem with restrictions due to $D_ap^{ab} = 0$ 
and $p = 0$.  The first restriction is familiar: it means that we are within Superspace.  
Thus one is to use corrections to the metric velocities that are Lie derivative with respect to an 
auxiliary $\xi^i$, variation with respect to which encodes $D_ap^{ab} = 0$  The 
second restriction is dealt with analogously:  
introduce a suitable correction to the metric velocities associated with some auxiliary 
$\theta$, variation with respect to which encodes $p = 0$.  On inspection, this requires the 
additional correction
\be
\dot{h}_{ij} - \pounds_{\xi}h_{ij} \longrightarrow \dot{h}_{ij} - \pounds_{\xi}h_{ij} - \theta h_{ij} 
\equiv{\mbox{\ss}}_{\xi, \theta}h_{ij}
\mbox{ } . 
\ee
$\theta$ can be interpreted as some $\dot{\zeta}$ in conjunction with free-endpoint variation so 
that the action is truly homogeneous of degree 1.  Thus one might consider an action (in analogy 
with the 
\be
\mbox{BSW formulation of GR) }
\mbox{\hspace{1.6in}}  
\mbox{\sffamily I\normalfont} = \int\textrm{d}\lambda\int\textrm{d}^3x\sqrt{h}
\sqrt{R \mbox{\sffamily T\normalfont}^{\mbox{\scriptsize g\normalsize}}_{\mbox{\scriptsize C\normalsize}}      } 
\mbox{ } ,
\mbox{\hspace{2in}}
\ee
\be
\mbox{for }
\mbox{\hspace{2.0in}}
\mbox{\sffamily T\normalfont}^{\mbox{\scriptsize g\normalsize}}_{\mbox{\scriptsize C\normalsize}} 
= G^{abcd}{\mbox{\ss}}_{\xi, \theta}h_{ab}{\mbox{\ss}}_{\xi, \theta}h_{cd}
\mbox{ } .
\mbox{\hspace{2.0in}}
\ee
\be
\mbox{The conjugate momenta are then }
\mbox{\hspace{1.1in}}
p^{ab} = \sqrt{\frac{hR}
{\mbox{\sffamily T\normalfont}^{\mbox{\scriptsize g\normalsize}}_{\mbox{\scriptsize C\normalsize}} }   }
G^{abcd}\equiv{\mbox{\ss}}_{\xi, \theta}h_{cd}
\mbox{ } .
\mbox{\hspace{2in}}
\ee
\be
\mbox{As ever, these are related by a primary constraint }
\mbox{\hspace{0.2in}}
\frac{1}{\sqrt{h}}
\left(
p \circ p - \frac{1}{2}p^2
\right) - \sqrt{h}R = 0 
\mbox{ } .
\mbox{\hspace{0.3in}}
\ee
\be
\mbox{and $\xi^i$-variation yields }
\mbox{\hspace{1.9in}}
D_ap^{ab} = 0 
\mbox{ } .
\mbox{\hspace{1.9in}}
\ee 
\be
\mbox{Now furthermore $\theta$-variation yields the maximal condition }
\mbox{\hspace{0.75in}}
p \equiv h_{ab}p^{ab} = 0 
\mbox{ } .
\mbox{\hspace{0.95in}}
\ee
which is the analogue of the vanishing of the dilational momentum 
${\cal D} = \sum_{(i) = (1)}^{(n)}\mbox{\b{q}}_{(i)}\cdot\mbox{\b{p}}^{(i)}$.  Note however that $p = 0$ is one constraint 
per space point, corresponding to the local conformal transformations,  
in contrast to the global scale transformations.  Just as ${\cal D}$ ensures moment of inertia conservation for 
island universes, $p = 0$ ensures the conservation of the volume for 
spatially compact universes.

When it comes to propagating this new condition, we know that in GR this is achieved where possible 
by solving the maximal LFE. Now, this can be successfully encoded by bringing in 
another auxiliary $\phi$ into the action as follows: use 
\be
\mbox{\sffamily I\normalfont} = \int\textrm{d}\lambda\int\textrm{d}^3x\sqrt{h}\phi^4\sqrt{
\left(
R - \frac{8D^2\phi}{\phi}
\right)
\mbox{\sffamily T\normalfont}^{\mbox{\scriptsize g\normalsize}}_{\mbox{\scriptsize C\normalsize}}    }
\label{primcg}
\ee
and $\phi$ variation gives the maximal LFE, in the distinguished representation (the one in 
which $\phi = 1$).  This distinguished representation is the one in which the theory being 
developed  most closely resemble the ADM split of GR.  In III.2.6 the ELE's, the LFE and the 
working that the constraints propagate is all checked to be independent of this choice.  

However the maximal LFE $\triangle N = NR$ is well-known to be insoluble in CWB GR (see C.1). 
This shortcoming of GR maximal slicing was bypassed by York's generalization to GR CMC slicing, 
which has a nicer CWB LFE (see C.1).  We bypass it here in a different way envisaged by 
\'{O} Murchadha: consider instead of (\ref{primcg}) an action in which division by some power 
of the volume of the universe occurs.  This was inspired by the Sobolev quotient used in the 
proof of the Yamabe result [(\ref{Sobquot}) of App III.2.A], and is analogous to the division 
by the moment of inertia in the particle model.  There is then the issue of what power of the 
volume should appear in the new action.  The answer is $\frac{2}{3}$.  One can arrive at this 
by arguing as in the next subsection that the action should be homogeneous of degree 0 in 
$\phi$ so as not to include volume among the d.o.f's (since otherwise the action would not be 
invariant under the constant scaling).  

Alternatively, one can argue that this particular power is necessary for consistency.  
Suppose the action is  

\noindent
\be
\mbox{\sffamily I\normalfont}^{\mbox{\scriptsize C\normalsize}}(n) = \int\textrm{d}\lambda \frac { \int \textrm{d}^3x\sqrt{h}\phi^4 
\sqrt{ R -   \frac{8D^2\phi}{\phi} } 
\sqrt{\mbox{\sffamily T\normalfont}^{\mbox{\scriptsize g\normalsize}}_{\mbox{\scriptsize C\normalsize}}  }    } {V^n }
\ee
for some $n$. Then variation with respect to $\phi$ yields the LFE, which is
\be
RN - D^2N = \frac{3}{2}n<NR>
\label{CGarbslice1} 
\ee
in the distinguished representation.   Then integrating this over space yields
$$
0 = \oint \sqrt{h}\textrm{d}S_aD^aN =  \int\textrm{d}^3x\sqrt{h}D^2N 
$$
\be
= \int \textrm{d}^3x\sqrt{h}RN 
- \frac{3}{2}n \int\textrm{d}^3x\sqrt{h(x)} 
\left( 
\frac{ \int \textrm{d}^3y\sqrt{  h(y)  }R(y)N(y) }{  V  }
\right) 
= \left(1 - \frac{3}{2}n\right)\int \textrm{d}^3x \sqrt{h}RN 
\mbox{ } ,
\label{CGConCheck}
\ee 
because of CWB, Gauss' theorem, the LFE, and that spatial integrals are spatial constants 
and can therefore be pulled outside further spatial integrals. 
Then the same integral inconsistency argument as for the maximal LFE holds unless $n=\frac{2}{3}$ 
(see App C).   

Thus the desired action is
\be
\mbox{\sffamily I\normalfont}^{\mbox{\scriptsize C\normalsize}} 
= \int\textrm{d}\lambda \frac { \int \textrm{d}^3x\sqrt{h}\phi^4 
\sqrt{ R -   \frac{8D^2\phi}{\phi} } \sqrt{\mbox{\sffamily T\normalfont}^{\mbox{\scriptsize g\normalsize}}_{\mbox{\scriptsize C\normalsize}}  }    } {V^{\frac{2}{3}} }
\label{CGBOCCtrial} 
\ee
\be
\mbox{with corresponding constraints } 
\mbox{\hspace{0.2in}}
-{\cal H}^{\mbox{\scriptsize C\normalsize}} \equiv \frac{\sqrt{h}}{V^{\frac{2}{3}}}R - \frac{\vc}{\sqrt{h}}p\circ p  = 0 \mbox{ } ,
\hspace{.5cm} D_bp^{ab} = 0 \mbox{ } ,
\hspace{.5cm} p = 0 \mbox{ } .
\label{CGcham}
\ee
\be
\mbox{and LFE }
\mbox{\hspace{2in}}
RN - D^2N = <NR>
\mbox{\hspace{2in}} 
\label{CGarbslice} 
\ee 
in the distinguished representation.  Whereas in the TSA to GR $N$ and $\xi^i$ are treated as 
freely specifiable gauge velocities, in this {\it conformal gravity} $N$ is fixed, and its role 
as gauge variable is taken over by $\phi$.  Since $\xi^i$ plays the same role in both theories, 
conformal gravity, like GR, has two d.o.f's per space point.  However, in contrast to GR, they 
are unambiguously identified as the two conformal shape d.o.f's of the 3-metric $h_{ij}$. 
Conceptually, this is a pleasing result, but it has a far-reaching consequence -- conformal 
gravity cannot be cast into the form of a 4-d generally covariant spacetime theory.  Because 
the lapse is fixed, absolute simultaneity and a preferred frame of reference are introduced.  
The reader may feel that this is too high a price to pay for a scale-invariant theory.  Of 
course, experiment will have the final word. However, one of the aims of pursuing this option is 
to show that conformal invariance already has the potential to undermine the spacetime covariance 
of GR.

\mbox{ }

\noindent \large{\bf 2.4 Single auxiliary Lagrangian formulation of conformal gravity}\normalsize 

\mbox{ }

\noindent My 3-space first principles treatment of conformal gravity is to consider an 
arbitrary-frame RI action.  The notion of `frame' now includes both Diff and 
Conf (all $\lambda$-dependent).  This will lead to Diff-BM and Conf-BM, i.e beginning with the 
highly redundant configuration space Riem, we add to that degeneracy 
by forming a Diff- and Conf-BM  action on 

\noindent $\mbox{Riem} \times \Xi \times P$, where $\Xi$ is the space of 
Diff-BM-encoding auxiliaries $\xi^i$ and $P$ is the space of suitably smooth positive 
Conf-BM-encoding auxiliaries $\phi$.  
\be
\mbox{ }\mbox{ Under Conf, the metric transforms according to } 
\mbox{\hspace{0.6in}}
h_{ij} \longrightarrow  \tilde{h}_{ij} = \omega^{4}h_{ij} 
\mbox{ } , 
\mbox{\hspace{1in}}
\label{CGConfTrans} 
\ee 
where the scalar $\omega$ is an arbitrary smooth positive function of the label $\lambda$ and 
of the position on the 3-space.  The fourth power of $\omega$ is the traditional one used for 
computational convenience, for the same reasons as in I.2.9.3.  
Then one has   

\noindent
\be
\tilde{h}^{ij} = \omega^{-4}h^{ij} 
\mbox{ } ,
\label{invom}
\ee
\be
\widetilde{\sqrt{h}} = \omega^6\sqrt{h} 
\mbox{ } , 
\label{CGPotTransLaw1} 
\ee
\be
\tilde{R} = \omega^{-4}\left(R - \frac{8D^2\omega}{\omega}\right) 
\mbox{ } .
\label{CGPotTransLaw2} 
\ee 

To work in the arbitrary conformal frame, one is to use the Foster--Barbour 
corrected coordinates 
\be
\bar{h}_{ij}=\phi^4h_{ij} \mbox{ } .  
\label{CGCoCo} 
\ee 
Note that there is then invariance under the pair of compensating transformations 
\be
h_{ab} \longrightarrow \omega^4 h_{ab} \mbox{ } , 
\label{splitban}
\ee
\be
\phi\longrightarrow {\frac{\phi}{\omega}} \mbox{ } ,
\label{CGConBanal}
\ee
which is the geometrodynamical generalization of the invariance under the 
banal transformation (\ref{banban}) of the particle model.  

Now, paralleling the derivation of (\ref{invom} -- \ref{CGPotTransLaw2}), (\ref{CGCoCo}) leads 
to  
\be
\bar{h}^{ij}=\phi^{-4}h^{ij} 
\mbox{ } ,
\label{invphi}
\ee
\be
\overline{\sqrt{h}} =\phi^6\sqrt{{h}},
\label{sqrtphi}
\ee
\be
\bar{R} = \phi^{-4}\left(R - \frac{8D^2\phi}{\phi}\right).
\label{Rphi}
\ee
\be
\mbox{Additionally, }
\mbox{\hspace{1in}}
\bar{{\mbox{\ss}_{\xi}h_{ab}}} = {\mbox{\ss}_{\xi}(\phi^4h_{ab})} \equiv \phi^4
\left(
{\mbox{\ss}_{\xi}h_{ab}} + {4}\frac{\mbox{\ss}_{\xi}\phi}{\phi}h_{ab}
\right) 
\mbox{\hspace{2in}}
\label{CGBabel}
\ee
which gives the Conf-BM metric velocity.

From (\ref{invphi}) and (\ref{CGBabel}) we can make a \sffamily T \normalfont that is Conf-invariant: 
\be
\mbox{\sffamily T\normalfont}^{\mbox{\scriptsize g\normalsize}}_{\mbox{\scriptsize C\normalsize}\mbox{\scriptsize W\normalsize}} 
\equiv \phi^{-8}G^{abcd}_{\mbox{\scriptsize W\normalsize}}
{\mbox{\ss}_{\xi}(\phi^4h_{ab})}{\mbox{\ss}_{\xi}(\phi^4h_{cd})}
\longrightarrow 
\tilde{\mbox{\sffamily T\normalfont}}^{\mbox{\scriptsize g\normalsize}}_{\mbox{\scriptsize C\normalsize}\mbox{\scriptsize W\normalsize}} 
= \mbox{\sffamily T\normalfont}^{\mbox{\scriptsize g\normalsize}}_{\mbox{\scriptsize C\normalsize}\mbox{\scriptsize W\normalsize}}  \mbox{ } . 
\label{CGzig}
\ee 
Note that we begin with an arbitrary $W$ in the inverse supermetric 

\noindent $G_{\mbox{\scriptsize W\normalsize}}^{abcd} = \sqrt{h}(h^{ac}h^{bd} - Wh^{ab}h^{cd})$. 
However, whereas $W = 1$ was found in RWR \cite{BOF} to be crucial for the consistency of GR, 
we will see that this apparent freedom does not play any role in conformal gravity.

\mbox{ }

One could likewise make a Conf-invariant potential density using the metric alone, 
from available conformally-invariant Bach tensor and the Cotton--York tensor of I.2.9.4.3, 
but the suitable combinations they give rise to are cumbersome, 
e.g 
\be
\left(
hb_{abc}b_{def}G^{abcdef}
\right)
^{\frac{1}{2}} 
\mbox{ } , \mbox{ }
h^{-\frac{1}{3}}
\left(
y^{ab}y^{cd}G^{\mbox{\scriptsize W\normalsize}}_{abcd}
\right)
^{\frac{1}{2}}
\mbox{ } , \mbox{ }
\left(
Y^{abcdef}b_{abc}b_{def}
\right)
^{\frac{1}{5}}
\ee
where $T^{abcdef} = \sum_{p \in S_6}K(p)t^{p(a)p(b)}t^{p(c)p(d)}t^{p(e)p(f)}$ 
for any tensor $t^{ab}$, where $K(p)$ are a priori free coefficients and $\mbox{S}_6$ 
is the permutation group of 6 objects.  

\mbox{ }

We instead resorted to using the auxiliary variable $\phi$, in addition to the metric, 
in order 
\be
\mbox{to build our potential, } 
\mbox{\hspace{1.5in}}
\phi^{-4}
\left( 
R - \frac{8D^2\phi}{\phi}
\right) 
\mbox{ } .
\mbox{\hspace{1.7in}}
\ee
Now in contrast to (\ref{CGzig}) and the situation in relativity \cite{BOF}, each term in the 
potential density part of a generalized BSW action changes under (\ref{splitban}). 
This has the consequence that the action of conformal gravity must, as a `conformalization' of the 
BSW action, depend not only on the velocity $\dot{\phi}$ of the auxiliary variable $\phi$ but 
also on $\phi$ itself. This parallels what was found for the gauge variable $a$ of the 
dilation-invariant particle action.  

So our action for pure (matter-free) conformal gravity is, upon additionally dividing by the 
homogenizing power of the volume as explained below: 
\be
\mbox{\sffamily I\normalfont}_{\mbox{\scriptsize \normalsize}}\mbox{=}\int\textrm{d}\lambda\int\textrm{d}^3x
\left(
\frac{\sqrt{h}\phi^6}{V}
\right)
\sqrt{
\left(
\frac{\vc}{\phi^4}
\right)
\left( 
R\mbox{--}\frac{8D^2\phi}{\phi}
\right)
                             } 
\sqrt{\mbox{\sffamily T\normalfont}^{\mbox{\scriptsize g\normalsize}}_{\mbox{\scriptsize C\normalsize}\mbox{\scriptsize W\normalsize}} }
\mbox{=}\int \textrm{d}\lambda \frac{\int\textrm{d}^3x\sqrt{h}\phi^4 \sqrt{R\mbox{--}\frac{8D^2\phi}{\phi}}
\sqrt{\mbox{\sffamily T\normalfont}^{\mbox{\scriptsize g\normalsize}}_{\mbox{\scriptsize C\normalsize}\mbox{\scriptsize W\normalsize}}  }    } {\vc } 
\label{CGSCGR} 
\ee 
\be
\mbox{where $V$ is the `conformalized' volume }
\mbox{\hspace{1.2in}}
V = \int \textrm{d}^3x \sqrt{h}\phi^6 
\mbox{ } .
\mbox{\hspace{2in}}
\label{CGVolumeDef}
\ee
This division by the volume implements homogeneity of degree zero in $\phi$.  
We next explain this homogeneity requirement and comment on this means of ours of 
implementing it.  
$$
\mbox{ } \mbox{ Let the Lagrangian density in (\ref{CGSCGR}) be } \mbox{ }
\mbox{\sffamily L\normalfont}
=\mbox{\sffamily L\normalfont}
\left(
h_{ij}, 
\frac{\textrm{d}h_{ij}}{\textrm{d}\lambda}, 
\phi,
\frac{\textrm{d}\phi}{\textrm{d}\lambda}
\right),
\hspace{.5cm}
\mbox{\sffamily I\normalfont}=\int\textrm{d}\lambda\int\textrm{d}^{3}x\mbox{\sffamily L\normalfont} 
\mbox{ } . 
\mbox{\hspace{0.3in}}
$$ 
Provided we ensure that $\mbox{\sffamily L\normalfont}$ is a functional of 
the corrected coordinates and velocities, it is bound to be invariant under the conformal transformations (\ref{CGConBanal}). This would be the case if we
omitted the volume $V$ in (\ref{CGSCGR}). But by the rules of BM suggested in 
III.2.2 the Lagrangian must also be invariant separately under all possible variations of 
the auxiliary variable. Now in this case the auxiliary $\phi$ is a
function of position, so one condition that $\mbox{\sffamily L\normalfont}$ 
\be 
\mbox{must satisfy is }
\mbox{\hspace{1.6in}}
\frac{\partial \mbox{\sffamily L\normalfont}}{\partial\phi}=0\hspace{.2cm}
\textrm{if}\hspace{.2cm} \phi=\textrm{spatial constant} 
\mbox{ } . 
\mbox{\hspace{2in}}
\label{CGHomCond}
\ee 
A glance at (\ref{CGSCGR}) shows that if $V$ were removed, (\ref{CGHomCond}) could not be 
satisfied. The action must be homogeneous of degree zero in $\phi$.\fn{This is the single 
symmetry property that distinguishes conformal gravity  
from GR, which for this reason \sl just \normalfont fails to be fully conformally 
invariant.} We exhibit this in the first expression in (\ref{CGSCGR}), 
in which the two expressions (\ref{sqrtphi}, \ref{Rphi}) that derive from
$\sqrt h$ and $\sqrt R$ in the BSW action have been multiplied by appropriate powers of 
$V$. (The second expression is more convenient for calculations.) 

A separate issue is the means of achieving homogeneity.  Barbour and \'{O} Murchadha 
originally attempted to achieve homogeneity by using not $R^{\frac{1}{2}}$ but 
$R^{\frac{3}{2}}$, since $\sqrt{h}R^{\frac{3}{2}}$ will satisfy (\ref{CGHomCond}). 
However, this already leads to an inconsistent theory even before one attempts 
`conformalization' (see II.2.2.3).  Since an ultralocal kinetic term 
$\mbox{\sffamily T \normalfont}$ has no conformal weight, another possibility would be, 
as mentioned above, to construct a conformally-invariant action by multiplying such a 
\sffamily T \normalfont  by a 3-d conformal concomitant scalar density.  Much hard 
work would be required to investigate whether any such possibilities yield consistent 
theories. Even if they do, they will certainly be far more complicated than conformal gravity.
 
We have therefore simply used powers of the volume $V$ in the style of the Yamabe work.  
This gives us {\it conformal gravity}.  The use of the volume has the added 
advantage that $V$ is conserved. This ensures that conformal gravity shares with the particle 
model the attractive properties it acquires from conservation of the moment
of inertia.  We extend this method (of using powers of $V$ to achieve homogeneity) 
in IV.2 to include matter coupled to conformal gravity.  It is the use of $V$ that necessitates our 
CWB assumption.  It is a physical assumption, not a mere mathematical 
convenience. It would not be necessary in the case of theories of the type considered in the 
previous paragraph.

\mbox{ } 

We must now find and check the consistency of the equations of conformal gravity.  The 
treatment of BM in the particle model in III.2.2 tells us that we must calculate the canonical 
momenta of $h_{ij}$ and $\phi$, find the conditions that ensure vanishing of the variation of 
the Lagrangian separately with respect to possible independent variations of the auxiliary $\phi$ and 
its velocity $v_{\phi}=\dot{\phi}$, and then show that these conditions, which involve the 
canonical momenta, together with the ELE's for $h_{ij}$ form a consistent set. This implements 
BM by the free-endpoint method.
\be
\mbox{ } \mbox{ The canonical momentum $p_{\phi}$ of $\phi$ is } 
\mbox{\hspace{0.35in}}
p_{\phi}\equiv \frac{      \partial \mbox{\sffamily L\normalfont}      }{       \partial v_{\phi}      } = 
\frac{      \sqrt{h}      }{      2N\vc      }G_{\mbox{\scriptsize W\normalsize}}^{abcd}
\mbox{\ss}_{\xi}(\phi^4h_{ab})\frac{4}{\phi}h_{cd} 
\mbox{ } , 
\mbox{\hspace{0.35in}}
\label{CGconfcanmom} 
\ee 
\be
\mbox{where 
$2N = \sqrt{            \frac{        \mbox{\sffamily\scriptsize T\normalsize\normalfont}
                                            ^{\mbox{\tiny W\normalsize}}
                                            _{\mbox{\tiny C\normalsize}}      }
                             {        R - \frac{8D^2\phi}{\phi}      }
           }$. 
The gravitational momenta are }
\mbox{\hspace{0.2in}} 
p^{ab} = \frac{      \sqrt{h}      }{      2N\vc      }G_{\mbox{\scriptsize W\normalsize}}^{abcd}
\mbox{\ss}_{\xi}(\phi^4h_{cd}) 
\mbox{ } ,
\mbox{\hspace{0.2in}}
\label{CGgravcanmom} 
\ee 
\be
\mbox{Thus one has the primary constraint } 
\mbox{\hspace{1.5in}}
p = \frac{\phi}{4}p_{\phi} 
\mbox{ } . 
\mbox{\hspace{1.5in}}
\label{CGprimconstr} 
\ee 
The primary constraint (\ref{CGprimconstr}) is a direct consequence of 
the invariance of the action (\ref{CGSCGR}) under the banal transformations (\ref{CGConBanal}). 

Independent variations $\delta\phi$ and $\delta v_{\phi}$ of $\phi$ and its velocity in the instantaneous Lagrangian that can
be considered are: 

\noindent 1) $\delta\phi$ is a spatial constant and $\delta v_{\phi}\equiv 0$. 

\noindent 2) $\delta\phi$ is a general function of position and $\delta v_{\phi}\equiv 0$.

\noindent 3) $\delta\phi\equiv 0$ and $\delta v_{\phi}\neq 0$ in an infinitesimal spatial 
region.

\noindent The possibility 1) has already been used to fix the homogeneity of 
$\mbox{\sffamily L\normalfont}$. Let us next consider 3). This tells us that
$\frac{\pa \mbox{\sffamily\scriptsize L\normalsize\normalfont}}{\pa v_{\phi}}=0$. But 
$\frac{\pa \mbox{\sffamily\scriptsize L\normalsize\normalfont}}{\pa v_{\phi}}\equiv p_{\phi}$, so we 
see that the canonical momentum of $\phi$ 
\be
\mbox{must vanish. Then the primary constraint (\ref{CGprimconstr}) gives } 
\mbox{\hspace{1in}} 
p = 0 
\mbox{ } . 
\mbox{\hspace{1in}}
\label{CGPisa}
\ee

As a result of this, without loss of generality we can set $W$, the coefficient of the trace, 
equal to zero, $W = 0$, in the generalization of the DeWitt supermetric used in (\ref{CGSCGR}). 
If conformal gravity proves to be a viable theory,
this result could be significant, especially for the quantization programme (see II.6), since 
it ensures that conformal gravity, in contrast to GR, has a 
$$
\mbox{positive-definite kinetic term.  
Indeed, in GR, since }
\mbox{\hspace{0.3in}}
\dot{h}_{ab} = \frac{2N}{\sqrt{h}}\left(p_{ab} - \frac{p}{2}h_{ab}\right) + 2D_{(a}\xi_{b)} 
\mbox{ } ,  
$$ 
the rate of change of $\sqrt h$ (which defines the volume element $\sqrt h\textrm{d}^3x$)  
is measured by $p$,
\be
\dot{\sqrt{h} } = - \frac{Np}{2} + \sqrt{h}D_a\xi^a 
\mbox{ } .
\ee
Therefore, in conformal gravity the volume element -- and with it the
volume of 3-space -- does not change and cannot make a contribution to the kinetic energy of 
the opposite sign to the contribution of the shape d.o.f's.  
Finally, we consider 2) and (including the use of 
\be
\mbox{$p_{\phi} = 0$) we obtain } 
\mbox{\hspace{0.4in}}
\phi^3N\left(R - \frac{7D^2\phi}{\phi}\right) - D^2(\phi^3N) = \phi^5
\left<
\phi^4N 
\left( R - \frac{8D^2\phi}{\phi} 
\right) 
\right> 
\mbox{ } , 
\mbox{\hspace{0.6in}}
\label{CGfullslicing} 
\ee
\be
\mbox{where we use the usual notion of {\it global average}: }
\mbox{\hspace{0.6in}}
<A> \equiv \frac{\int \textrm{d}^3x \sqrt{h} A} {\int \textrm{d}^3x\sqrt{h}} 
\mbox{ } .
\mbox{\hspace{0.8in}}
\ee
The lapse-fixing equation (LFE) (\ref{CGfullslicing}) 
has a status different from the constraints and it has no analogue in 
the scale-invariant particle model, or in any other gauge theory of which we are aware.  
We get this LFE and not a further homogeneity requirement such as in the particle model, 
because although the action is homogeneous in $\phi$ we are now dealing with a field theory in 
which there are spatial derivatives of $\phi$.   It is, however, a direct consequence of 
Conf-BM, and as explained below, it plays an important role in the mathematical structure of 
conformal gravity.  

Besides the trace constraint (\ref{CGPisa}), $p^{ab}$ must satisfy the primary constraint 
\be
-{\cal H}^{\mbox{\scriptsize C\normalsize}} \equiv \frac{\sqrt{h}\phi^4}{V^{\frac{2}{3}}}
\left( 
R - \frac{8D^2\phi}{\phi} 
\right)
- \frac{\vc}{\sqrt{h}\phi^4}p\circ p = 0
\label{CGfullham} 
\ee 
due to the local square-root form of the Lagrangian, and the secondary constraint 
\be
D_bp^{ab} = 0
\label{CGcmom} 
\ee 
from $\xi^{i}$-variation. Of course, (\ref{CGcmom}) is identical to the 
GR momentum constraint (\ref{Vmom}), while (\ref{CGfullham}) is very similar to the GR 
Hamiltonian constraint (\ref{Vham}). 

The ELE's for $h_{ij}$ are
$$
\mbox{\ss}_{\xi}p^{ab} =  \frac{\phi^4\sqrt{h}N}{\vc}
\left(
Rh^{ab} - \frac{4D^2\phi}{\phi} - R^{ab}
\right) 
- \frac{\sqrt{h}}{\vc}[  h^{ab}D^2(\phi^4N) - D^aD^b(\phi^4N)  ]
$$
$$ 
- \frac{2N\vc}{\sqrt{h}\phi^4}p^{ac}{p^b}_c 
+ \frac{8\sqrt{h}}{\vc}
\left(
\frac{1}{2}h^{ab}h^{cd} - h^{ac}h^{bd}
\right) 
\pa_c(\phi^3N)\pa_d\phi 
$$
\be
- \frac{2\sqrt{h}}{3\vc}\phi^6h^{ab}
\left<
\phi^4N
\left( 
R - \frac{8D^2\phi}{\phi}
\right)
\right> 
+ \frac{4}{\phi}p^{ab}\mbox{\ss}_{\xi}\phi 
\mbox{ } .
\label{CGGravEL}
\ee 
They can be used to check the consistency of the full set of equations, constraints and 
LFE of conformal gravity. To simplify these calculations and simultaneously 
establish the connection with GR, we go over to the distinguished representation 
in which $\phi = 1$ and $\xi^i = 0$. The three constraints that must be satisfied by the 
gravitational canonical momenta 
\be
\mbox{are }
\mbox{\hspace{1.2in}}
-{\cal H}^{\mbox{\scriptsize C\normalsize}} \equiv \frac{\sqrt{h}}{V^{\frac{2}{3}}}R 
- \frac{\vc}{\sqrt{h}}p \mbox{ } \circ \mbox{ } p  = 0,
\mbox{\hspace{.5cm}} D_bp^{ab} = 0 \mbox{ } ,
\mbox{\hspace{.5cm}} p = 0 \mbox{ } .  
\mbox{\hspace{2in}}
\ee
\be
\mbox{The LFE (\ref{CGfullslicing}) becomes }
\mbox{\hspace{1.3in}}
D^2N - NR = - <NR> \mbox{ } , 
\label{CGBMFLFC}
\hspace{1.3in}
\ee
and the ELE's are 
\be  
\dot{p}^{ab} =  \frac{\sqrt{h}N}{\vc}(Rh^{ab} - R^{ab}) -
\frac{\sqrt{h}}{\vc}(  h^{ab}D^2N - D^aD^bN  ) - \frac{2N\vc}{\sqrt{h}}p^{ac}{p^b}_c - \frac{2\sqrt{h}}{3\vc}h^{ab}<NR>.
\label{CGBMFGravEL}
\ee 

\mbox{ }

Since the volume $V$ is conserved, and its numerical value depends on a nominal lengthscale, 
for the purpose of comparison with the equations of GR we can set $V=1$. 
(This cannot, of course, be done before the variation that leads to the above equations, since 
the variation of $V$ generates forces. It is important not to confuse quantities on-shell and 
off-shell.) We see, setting $V=1$, that the similarity with GR in York's CMC slicing is strong.  
In fact, the constraints (\ref{CGcham}) are identical to the GR constraints and the CMC 
slicing condition at maximal expansion, and the ELE's differ only by the 
absence of the GR term proportional to $p$ and by the presence of the final force term. 
Finally, there is the LFE 
(\ref{CGBMFLFC}), which is an eigenvalue equation of essentially the same kind 
as the LFE (\ref{CGLapseFixing}) required to 
maintain York's CMC slicing condition (\ref{Yorktime}). 

In our view, one of the most interesting results of this work is the derivation of such 
LFE's directly from a fundamental symmetry requirement rather than as an 
equation which could be interpreted as maintaining a gauge fixing.  We now develop this point.  

This is part of the confirmation that we do have a consistent set of equations, constraints 
and LFE.  We show this in III.2.2.6, which amounts to demonstrating that if the 
constraints (\ref{CGcham}) hold initially, then they will propagate due to the ELE's 
(\ref{CGBMFGravEL}) \textit{and} the LFE (\ref{CGfullslicing}).  
The propagation of the vector momentum constraint is always unproblematic, being guaranteed by 
the 3-diffeomorphism invariance of the theory. The two constraints that could give difficulty 
are the quadratic and linear scalar constraints. In RWR \cite{BOF}, the propagation of the 
quadratic constraint proved to be a delicate matter and generated the new results of that 
paper.  However, in this section we are merely `conformalizing' the results of \cite{BOF}, 
and we shall see that the consistency achieved for the quadratic constraint in \cite{BOF} 
carries forward to conformal gravity. The only issue is therefore whether the new constraint 
$p = 0$ is propagated.  Now, we find that the form of $\dot{p}$ evaluated from the 
ELE's \sl is automatically guaranteed to be zero \normalfont by virtue of 
the LFE.  Thus the propagation of $p = 0$ is guaranteed rather than 
separately imposed.  It is in this sense that conformal gravity is not maintaining a 
gauge fixing.  

\mbox{ }

The same conformal gravity theory can also be reached using the `bare principles' of II.2.2.4, 
provided that one takes the $D_ip = 0$ branch of possibilities to propagate the Hamiltonian constraint.  
Notice how one gets two conditions to encode, and the encoding of the second of these (the LFE) 
does not look like any encodements we have seen before.  We will also see how conformal gravity 
can be formulated in a lapse uneliminated fashion in VI.  

Finally, we briefly mention two sources of variety.  First, we find that there is analogously 
a Euclidean conformal gravity and a \textit{strong conformal gravity} (which unlike in III.1 has 
without loss of generality $W = 0$), 
with action 
\be
\mbox{\sffamily I\normalfont} ^{\mbox{\scriptsize Strong\normalsize}} 
= \int \textrm{d}\lambda\frac{      \int   \textrm{d}^3x         \sqrt{h}\phi^6 \sqrt{\Lambda}
\sqrt{    \mbox{\sffamily T\normalfont}^{\mbox{\scriptsize g\normalsize}}_{\mbox{\scriptsize C\normalsize}}     }     }{V} 
\mbox{ } ,
\label{CGConStrongGrav}
\ee
which may be of use in understanding quantum conformal gravity.  Note that the power of the 
volume $V$ needed to make the action homogeneous of degree zero is here one, since now it has 
to balance only $\sqrt{h}$ and not the product $\sqrt{h}\sqrt{R}$.  This theory is simpler 
than conformal gravity in two ways: $\Lambda$ is less intricate than $R$, and the LFE for 
strong conformal gravity is $\Lambda N = <\Lambda N>$ so since $<\Lambda N>$ is a spatial 
constant, $N$ is a spatial constant.

Second, instead of constructing an action with a local square root, one could use instead a 
\textit{global} square root and thus obtain 
\be
\mbox{\sffamily I\normalfont} = \int \textrm{d}\lambda \frac{\sqrt{ \int \textrm{d}^3x \sqrt{h}\phi^2
\left(
R - \frac{8D^2\phi}{\phi}
\right)}
\sqrt{\int\textrm{d}^3x\sqrt{h}\phi^6\mbox{\sffamily T\normalfont}^{\mbox{\scriptsize g\normalsize}}_{\mbox{\scriptsize C\normalsize}}}}{\vc} 
\mbox{ } ,  
\ee
which gives rise to a single global quadratic constraint.  The above alternatives 
are cumulative: for both conformal gravity above and each theory below, we could consider 6 
variants by picking Euclidean, strong, or Lorentzian signature and a local or global square 
root.  The Lorentzian, local choices expanded throughout are the most obviously physical choice.  

\mbox{ }

\noindent\large{\bf 2.5 Integral conditions and the cosmological constant} \normalsize

\mbox{ }

\noindent The coupling of matter originally left BF\'{O} at a loss through 
Foster's demonstration of the appearance of integral inconsistencies in the actions they were 
considering.  The idea I used to resolve this problem comes from the analogous attempt to 
combine the actions of strong conformal gravity and conformal gravity in order to consider conformal gravity with a 
cosmological constant, $\Lambda$.  We give the combined action obtained by applying homogeneity 
of degree zero in $\phi$:  
$$
\mbox{\sffamily I\normalfont}^{\mbox{\scriptsize $\Lambda$\normalsize}} 
= \int\textrm{d}\lambda \int\textrm{d}^3x 
\left\{
\left( 
\frac{\sqrt{h}\phi^6}{V}
\right)
\sqrt{
\left(
\frac{\vc}{\phi^4}
\right)
\sigma
\left( 
R - \frac{8D^2\phi}{\phi}\right) + \Lambda } 
\sqrt{\mbox{\sffamily T\normalfont}^{\mbox{\scriptsize g\normalsize}}_{\mbox{\scriptsize C\normalsize}}     }
\right\} 
$$ 
\be 
= \int \textrm{d}\lambda \frac{\int \textrm{d}^3x\sqrt{h}\phi^4 \sqrt{\sigma 
\left( 
R - \frac{8D^2\phi}{\phi} 
\right) 
+ \frac{   \Lambda\phi^4   }{    V(\phi)^{  \frac{2}{3}}   } } 
\sqrt{\mbox{\sffamily T\normalfont}^{\mbox{\scriptsize g\normalsize}}_{\mbox{\scriptsize C\normalsize}}}} 
{ V(\phi)^{\frac{2}{3}}  } = \int \textrm{d}\lambda \frac{\bar{J}}{V^{\frac{2}{3}}} 
\mbox{ } ,
\label{CGBeast} 
\ee 
where we have also included a signature $\sigma$ to show how (\ref{CGBeast})
reduces to the strong conformal gravity action (\ref{CGConStrongGrav}) in the limit 
$\sigma \longrightarrow 0$.
That this is a correct way to proceed is emphasized below by the lack in this case of an integral 
inconsistency.

The conjugate momenta $p^{ij}$ and $p_{\phi}$ are given by (\ref{CGconfcanmom}) and 
(\ref{CGgravcanmom}) as before, but now with
\be
2N = \sqrt{ \frac{\mbox{\sffamily T\normalfont}^{\mbox{\scriptsize g\normalsize}}_{\mbox{\scriptsize C\normalsize}} } 
{   \sigma \left( R - \frac{ 8D^2 \phi }{ \phi } \right) 
+ \frac{  \Lambda\phi^4 }{   V(\phi)^{\frac{2}{3}}  }   } } 
\mbox{ } ,
\ee
and the primary constraint (\ref{CGprimconstr}) holds. Again, the end-point part of the 
$\phi$ variation yields $p_{\phi} = 0$, so $p = 0$, so without loss of generality $W = 0$, but 
now the rest of the $\phi$-variation gives a new LFE, 
\be
2\sigma(NR - D^2N) + \frac{3N\Lambda}{\vc} = \frac{\bar{J}}{V} + \frac{<N\Lambda>}{\vc}
\label{CGCClapsefix} 
\ee 
in the distinguished representation. For this choice of the action, there is indeed no 
integral inconsistency: 
$$
0 = 2\sigma\oint \sqrt{h}\textrm{d}S_aD^aN = 2\sigma\int\textrm{d}^3x\sqrt{h}D^2N
$$
\be
= \int \textrm{d}^3x\sqrt{h}
\left\{
2\sigma RN + \frac{3N\Lambda}{\vc} - 2N
\left(
\sigma R + \frac{\Lambda}{\vc}
\right)
 - \frac{N\Lambda}{\vc}
\right\}. 
\ee

The $\xi^{i}$-variation yields the usual momentum constraint (\ref{Vmom}), 
and the local square root gives the primary Hamiltonian-type constraint 
\be
-^{\Lambda}{\cal H}^{\mbox{\scriptsize C\normalsize}} \equiv \frac{\sqrt{g}}{V^{\frac{2}{3}}}\left(\sigma R + \frac{\Lambda}{V^{\frac{2}{3}}}\right) -
\frac{\vc}{\sqrt{h}}p\circ p = 0
\ee 
in the distinguished representation.  The ELE's are 
$$
\mbox{\ss}_{\xi}p^{ij} =  \frac{\sigma\sqrt{h}N}{V^{\frac{2}{3}}}(h^{ij}R - R^{ij}) +
\frac{\sqrt{h}\sigma}{\vc}(D^iD^jN - h^{ij}D^2N) - \frac{2N\vc}{\sqrt{h}}p^{im}{p^j}_m 
$$
\be
- \frac{\bar{J}\sqrt{h}}{V^{\frac{5}{3}}}h^{ij} +
\frac{\Lambda\sqrt{h}}{V^{\frac{4}{3}}}h^{ij}\left( N - \frac{<N>}{3}\right),
\label{CGLambdaEL} 
\ee 
where we have split the working up into pure conformal gravity and $\Lambda$ parts (the $\bar{J}$ here can also be split into the pure conformal gravity integrand and a $\Lambda$ part).
The $\Lambda$ part of the working for the constraint propagation presents no extra difficulties.  


Note that in conformal gravity the cosmological constant $\Lambda$ (just like its particle model analogue, the Newtonian
energy {\sffamily E\normalfont}) contributes to a conformally-induced cosmological-constant type force. The penultimate term in (\ref{CGLambdaEL}) is the final term
of pure conformal gravity in (\ref{CGBMFGravEL}), and the final term is induced by $\Lambda$. We will see in IV.2 
that matter also gives analogous contributions. The significance of this is discussed in V.2.

\mbox{ }

\mbox{ }

\noindent\large{\bf 2.6 Hamiltonian formulation and alternative theories}\normalsize

\mbox{ }

\noindent In this section, we examine the Hamiltonian formulation.  We exercise two
options in constructing theories such as GR and conformal gravity, so that this section is 
more general than IV.2.5.  First, while we have hitherto only considered the CWB case, 
we now also touch on the asymptotically-flat case.  Second, we also 
attempt to incorporate both maximal and CMC conditions.  

Our methods of doing so necessitate a 
discussion of whether the corresponding LFE's are fundamental or a gauge-fixing.  When we 
conformally-correct only the $\dot{h}_{ij}$, I think that one can interpret the slicing 
condition as pertaining to a particular slice and then either adopt the LFE which propagates 
it, or not, as a continued gauge choice.  When we conformally-correct both $h_{ij}$ and 
$\dot{h}_{ij}$, we find that the corresponding LFE is variationally guaranteed.  
This is the case in conformal gravity.  
I think such fully-conformally-corrected theories 
are the more complete ones.  Below I build up the Hamiltonian structure toward this completeness.    
 
\mbox{ }
 
We begin with the case that corresponds to working on Superspace.   Recollect how ADM and 
Dirac showed that the Hamiltonian for GR can be written as 
\be
\mbox{\sffamily H\normalfont} = 
\int\textrm{d}^3x(N{\cal H} + \xi^i{\cal H}_i)
\ee 
where $N$ and $\xi^i$ are regarded as 
arbitrary auxiliary Lagrange multiplier coordinates.  Now we wish to parallel this work on 
the space CS+V by working with CMC slices.  There is then little difference between the 
compact and asymptotically-flat cases.  Our treatment simply involves adding another 
constraint to the Dirac--ADM Hamiltonian:
\be
\mbox{\sffamily H\normalfont}^{\eta} 
= \int \textrm{d}^3x [N{\cal H} + \xi^i{\cal H}_i + (D^2\eta)p]
\label{CGham'}
\ee
and treating $\eta$ as another Lagrange multiplier.  The Laplacian is introduced here to 
obtain the CMC condition as the new constraint arising from $\eta$-variation: 
$D^2p = 0 
\Rightarrow \frac{p}{\sqrt{h}} = C(\lambda)$.  In addition we get the standard 
GR Hamiltonian and momentum constraints from $N$- and $\xi^i$-variation.  Now, we can choose 
to impose 
$\frac{\pa}{\pa\lambda}
\left(
\frac{p}{\sqrt{h}}
\right) 
= C(\lambda)$ 
which gives the CMC LFE.  Whereas one could reinterpret this as the study in the CMC gauge of 
the subset of (pieces of) GR solutions which are CMC foliable, we can also consider this to be 
a new theory with a preferred fundamental CMC slicing.\fn{By a preferred fundamental slicing, 
we mean a single stack of Riemannian 3-spaces.  This is not to be confused with GR, where 
there are an infinity of stacks between any two given spacelike hypersurfaces (the foliations), 
each of which is unprivileged.}    The latter interpretation 
is our first CS+V theory.  

This CS+V theory's evolution equations are (\ref{CGsmallevol}) and (\ref{CGBSWEL}) but picking 
up the extra terms $h_{ab}D^2\eta$ and $-p^{ab}D^2\eta$ respectively.  We have already dealt 
with the CMC constraint;     it turns out that we need to set $D^2\eta = 0$ to preserve the 
other constraints.  Therefore the CMC Hamiltonian is well-defined when one uses the 
distinguished representation $D^2\eta = 0$ and that $N$ satisfies the CMC LFE.
     
It is more satisfactory however to introduce a second auxiliary variable to conformally-correct 
the objects associated with the metric.  We used $(1 + D^2\zeta)^{\frac{1}{6}}$ in place of 
$\phi$ since this implements {\it the volume-preserving conformal transformations } VPConf: 
\be
\bar{V} - V = \int \textrm{d}^3x\sqrt{h}
\left[
(1+ D^2\zeta)^{\frac{1}{6}}
\right]
^6 
- \int \textrm{d}^3x\sqrt{h} = \int \textrm{d}^3x\sqrt{h}D^2\zeta 
= \oint \textrm{d}S_aD^a\zeta = 0 
\ee
in the case where the 3-space is CWB.  
Applying these corrections to the Hamiltonian 
\be
\mbox{ansatz}
\mbox{\hspace{1.6in}}
\mbox{\sffamily H\normalfont}^{\zeta\eta} 
= \int \textrm{d}^3x [N{\cal H}^{\zeta} + \xi^i{\cal H}_i + (D^2\eta)p] 
\mbox{ } ,
\mbox{\hspace{1.6in}} 
\label{CGdcham'}
\ee
\be 
{\cal H}^{\zeta} \equiv  \frac{1}{\sqrt{h}}
\left[
{        p^{\mbox{\scriptsize T\normalsize}}\circ p^{\mbox{\scriptsize T\normalsize}}        }
(1\mbox{+}D^2\zeta)^{\alpha} - \frac{    (1\mbox{+}D^2\zeta)^{\beta}p^2        }{        6        }
\right] 
- \sqrt{h}(1\mbox{+}D^2\zeta)^{\frac{2}{3}}
\left[
R - \frac{        8D^2(1\mbox{+}D^2\zeta)^{\frac{1}{6}}        }{        (1\mbox{+}D^2\zeta)^{\frac{1}{6}}        }
\right]  
\label{halfton} \mbox{ } , 
\ee
$$
\mbox{in order for $\zeta$-variation to encode the correct LFE }
\mbox{\hspace{0.4in}}
NR - D^2N + \frac{Np^2}{4h} = B(\lambda).
\mbox{\hspace{0.4in}}
$$
(in the distinguished representation $D^2\zeta = D^2\eta = 0$) to maintain the 
$\frac{p}{\sqrt{h}} = C(\lambda)$ given by $\eta$-variation, we require that 
$\alpha = -\frac{2}{3}$ and $\beta = \frac{4}{3}$.  
This corresponds to $\phi^{-4}$ and $\phi^8$ scalings, i.e a relative scaling by $\phi^{12}$, 
between the trace and tracefree terms.  Thus, (\ref{halfton}) happens to be the 
Lichnerowicz--York equation in $\phi$, subject to its 
solutions only being admitted if they are in VPConf,  
$\phi = (1 + D^2\zeta)^{\frac{1}{6}}$.  At this stage, we do not know why we are obtaining the 
same scalings as in York's work.  
   
\mbox{ }

We next try to extend the above workings to the case of the maximal condition.  This 
\be
\mbox{involves starting with a slightly different Hamiltonian, } 
\mbox{\hspace{0.2in}}
\mbox{\sffamily H\normalfont}^{\theta} = \int \textrm{d}^3x (N{\cal H} + \xi^i{\cal H}_i - \theta p) 
\mbox{ } .
\mbox{\hspace{0.2in}}
\label{CGpooras}
\ee  
$\theta$-variation gives us the maximal condition $p = 0$.  In the asymptotically-flat case, 
it makes 
\be
\mbox{sense to impose $\dot{p} = 0$ to arrive at the maximal LFE }
\mbox{\hspace{0.8in}}
D^2N = RN 
\mbox{ } , 
\mbox{\hspace{0.8in}}
\label{CGMSE2}
\ee
since this is an extremely well-behaved equation when taken in conjunction with the boundary 
condition $N \longrightarrow 1$ at infinity.  Again, $N$- and $\xi^i$-variation yield the 
standard GR Hamiltonian and momentum constraints respectively.  Now if we choose $\theta = 0$ 
in the evolution we preserve the Hamiltonian and momentum constraints.   Therefore 
$\mbox{\sffamily H\normalfont}^{\theta}$ can be interpreted as yielding standard GR with a 
choice of maximal gauge, or alternatively that this represents a new theory with a preferred 
fundamental maximal slicing: our first asymptotically-flat theory.

This does not work in the CWB case, since then now the only solution of (\ref{CGMSE2}) is 
$N \equiv 0$: frozen dynamics.  But if we use the volume of the universe (which amounts to 
moving away from GR), we are led to new theories.  First, consider
\be
\mbox{\sffamily H\normalfont}^{\theta \mbox{\scriptsize V\normalsize}} 
= \int \textrm{d}^3x(N{\cal H}^{\mbox{\scriptsize V\normalsize}} + \xi^i{\cal H}_i - \theta p) \mbox{ }, \mbox{  }
{\cal H}^{\mbox{\scriptsize V\normalsize}} \equiv \frac{V^{\frac{2}{3}}}{\sqrt{h}}p\circ p - \frac{\sqrt{h}}{V^{\frac{2}{3}}} R 
\mbox{ } .
\label{CGpoormanscg}
\ee
Then the $N$, $\xi^i$, $\theta$ variations yield the constraints ${\cal H}^V = 0$,  
${\cal H}_i = 0$ and $p = 0$.  Imposing 
\be
\mbox{$\dot{p} = 0$ yields a LFE }
\mbox{\hspace{1.5in}}
D^2N = RN - <RN> 
\mbox{ } .  
\mbox{\hspace{1.5in}}
\label{CGnav}
\ee
Whereas this could be regarded as a gauge fixing, the underlying theory is no longer GR.  The 
other interpretation is that of a new theory with a preferred fundamental maximal slicing -- 
a poor man's version of conformal gravity.  

Second, consider the use of two auxiliary conformal variables:  
\be
\mbox{\sffamily H\normalfont}^{\phi\theta \mbox{\scriptsize V\normalsize}} = \int \textrm{d}^3x
(N{\cal H}^{\mbox{\scriptsize C\normalsize}} + \xi^i{\cal H}_i - \theta p) \mbox{ } , \mbox{ } 
{\cal H}^{\mbox{\scriptsize C\normalsize}} \equiv \frac{V^{\frac{2}{3}}}{\phi^4\sqrt{h}}p\circ p
- \frac{\phi^4\sqrt{h}}{V^{\frac{2}{3}}} 
\left(
R - \frac{8D^2\phi}{\phi}
\right) \mbox{ } .  
\label{CGhamiltonian}
\ee
Then the $N$-, $\xi^i$- and $\theta$-variations yield the constraints 
${\cal H}^{\mbox{\scriptsize C\normalsize}} = 0$,  ${\cal H}_i = 0$ and $p = 0$.  Hamilton's 
equations are now the evolution equations 
$$
\mbox{\ss}_{\xi}h_{ab} =
\frac{2N\vc}{\sqrt{h}}p_{ab} - \theta h_{ab}
\mbox{ } ,
$$
$$
\mbox{\ss}_{\xi}p^{ab} = 
\frac{\sqrt{h}N}{\vc}
\left(
\frac{R}{2}h^{ab} - R^{ab}
\right) - \frac{\sqrt{h}}{\vc}(  h^{ab}D^2N - D^aD^bN  ) \\     +
\frac{N\vc}{2\sqrt{h}}h^{ab}p\circ p  
$$
\be 
- \frac{2N\vc}{\sqrt{h}}p^{ac}{p^b}_c - \frac{2}{3\vc}\left<NR\right>h^{ab} + 
\theta p^{ab}
\label{CGBSWELMOD}
\ee
in the distinguished representation $\phi = 1$.  Whilst $\phi$-variation yields (\ref{CGnav}), 
Hamilton's 
\be 
\mbox{equations give } 
\mbox{\hspace{1.5in}}
\dot{p} =  \frac{2\sqrt{h}}{\vc}(NR - <NR> - D^2N) 
\mbox{\hspace{1.5in}}
\ee
which is thus automatically zero due to (\ref{CGnav}).  This theory is (full) conformal 
gravity, as can be confirmed by Legendre transformation and BSW elimination \cite{BSW} to 
recover the Lagrangian of III.2.5.  Because the LFE that maintains the condition is also 
guaranteed, conformal gravity is definitely \sl not \normalfont interpretable as a gauge 
fixing.  

Before turning to these, we mention that there is a further (full rather than poor man's) 
maximal theory which arises from considering two conformal auxiliary variables in the 
asymptotically-flat case:
\be
\mbox{\sffamily H\normalfont}^{\phi\theta} = \int \textrm{d}^3x (N{\cal H}^{\phi} + \xi^i{\cal H}_i - \theta p) 
\mbox{ } , \mbox{ } 
{\cal H}^{\phi} \equiv \frac{1}{\phi^4\sqrt{h}}p\circ p - {\phi^4\sqrt{h}} 
\left(
R - \frac{8D^2\phi}{\phi}
\right) 
\mbox{ } .  
\label{CGascg}
\ee 
Then $N$, $\xi^i$, $\theta$ variation yield the constraints ${\cal H}^{\phi} = 0$,  
${\cal H}_i = 0$ and $p = 0$.  Now, $\phi$ variation gives the maximal LFE (\ref{CGMSE2}), and 
so automatically guarantees the propagation of the condition $p = 0$.  Note that 
${\cal H}^{\phi} = 0$ is the original Lichnerowicz equation (\ref{maxvaclich}).  Again, the 
Lichnerowicz equation carries the auxiliary variable which encode the LFE.  
In contrast, in conformal gravity, one must modify the corresponding Lichnerowicz 
equation by the introduction of volume terms to get an analogous scheme.  

Like conformal gravity, our full asymptotically-flat theory has no role for $p$ in its 
dynamics, but, unlike conformal gravity, it does not possess global terms.  We are less 
interested in this theory than in conformal gravity because it would not be immediately 
applicable to cosmology, on account of being asymptotically flat.  These two theories should 
be contrasted with our full CS+V theory, in which $p$ does play a role, which means that the 
standard GR explanation of cosmology is available.  We believe CS+V theory merits a full 
treatment elsewhere (started in V.2.2) as another potential rival to GR.

\mbox{ }
  
We now consider the preservation of the other conformal gravity constraints 
${\cal H}^{\mbox{\scriptsize C\normalsize}}$ and ${\cal H}_i$.  We find that we need to set 
$\theta = 0$.  Therefore (by comparison with GR) the only term 
\be
\mbox{we need to worry about is the }
\mbox{\hspace{1.35in}}
-\frac{2}{3\vc}<NR>h^{ab}
\mbox{\hspace{1.35in}}
\ee
term in the $\dot{p}^{ab}$ equation.  Since it is of the form 
$Ch^{ab}$ it clearly will not disturb the momentum constraint.  Therefore we need only worry 
about conserving the Hamiltonian constraint.  
\be
\mbox{This is quite straightforward.  We first 
realize that }
\mbox{\hspace{0.8in}}
\dot{\sqrt{h}} = \sqrt{h}D_i\xi^i 
\mbox{ } .
\mbox{\hspace{0.8in}}
\label{CGnog}
\ee
\be 
\mbox{Hence }
\mbox{\hspace{2.5in}}
\dot{V} = 0
\mbox{ } . 
\mbox{\hspace{2.5in}}
\label{CGnov} 
\ee
\be
\mbox{The only other term to worry about is }
\mbox{\hspace{1.2in}}
\frac{\vc}{\sqrt{h}}p\circ p \mbox{ } . 
\mbox{\hspace{1.2in}}
\ee
\be
\mbox{Varying $p^{ab}$ gives }
\mbox{\hspace{1.7in}}
-\frac{4}{3\sqrt{h}}<NR>p = 0 \mbox{ } . 
\mbox{\hspace{1.7in}}
\ee
Therefore the constraints are preserved under evolution.  

\mbox{ }

We now show that we can just as easily treat the Hamiltonian dynamics of conformal gravity in 
the general representation.  The LFE from $\phi$ variation is now 

\be
D^2(\phi^3N) = \phi^3N
\left(
R - \frac{7D^2\phi}{\phi}
\right) 
- \phi^5
\left<
\phi^4N 
\left( 
R - \frac{8D^2\phi}{\phi} 
\right) 
\right> \mbox{ } ,
\label{CGfullslicing2} 
\ee
whilst Hamilton's evolution equations are now 
$$
\mbox{\ss}_{\xi}h_{ab} =  \frac{2N\vc}{\phi^4\sqrt{h}}p_{ab} - \theta h_{ab}
\mbox{ } ,
$$
$$ 
\mbox{\ss}_{\xi}p^{ab} =  \frac{\phi^4\sqrt{h}N}{\vc}
\left(
\frac{R}{2}h^{ab} - R^{ab}
\right) 
$$
$$
+ \frac{N\vc}{2\phi^4\sqrt{h}}p\circ p h^{ab}  - \frac{\sqrt{h}}{\vc}(h^{ab}D^2(\phi^4N) - D^aD^b(\phi^4N)  ) + \theta p^{ab}       
-\frac{2N\vc}{\sqrt{h}\phi^4}p^{ac}{p^b}_c  
$$
\be
\mbox{\hspace{0.4in}}
+ \frac{8\sqrt{h}}{\vc}
\left(
\frac{1}{2}h^{ab}h^{cd}-h^{ac}h^{bd}
\right)
\pa_c(\phi^3N)\pa_d\phi - \frac{2\sqrt{h}}{3\vc}\phi^6h^{ab}
\left<
\phi^4N
\left( 
R\mbox{--}\frac{8D^2\phi}{\phi}
\right)
\right> 
\mbox{ } .
\label{CGham1}
\ee
We can compare these expressions to their Lagrangian analogues (\ref{CGgravcanmom}), 
(\ref{CGGravEL}) and we see they coincide if $\theta = \frac{4}{\phi}
\frac{\textrm{\scriptsize{d}}\phi}{\textrm{\scriptsize{d}}\lambda}$.  This will guarantee that 
the constraints are preserved by the evolution.  Alternatively, we could evolve the 
constraints using the Hamiltonian evolution equations (\ref{CGham1}) and discover that the
constraints propagate if and only if the lapse function, $N$, satisfies (\ref{CGfullslicing2}), 
the shift, $\xi$, is arbitrary, and $\theta$ satisfies  $\theta = \frac{4}{\phi}\mbox{\ss}_{\xi}\phi$.  
The $\phi$-variation gives the LFE (\ref{CGfullslicing}).  We emphasize that $\mbox{\ss}_{\xi}\phi$ 
is arbitrary in the full theories, unlike in the poor man's versions, where one ends up having 
to set the auxiliary ($\theta$ or $D^2\eta$) to zero.  

It is not obvious then that $\dot{p} = 0$ is guaranteed from Hamilton's equations, since what 
one immediately obtains is, weakly, 
$$
\dot{p} \approx  \frac{2N\sqrt{h}\phi^4}{V^{\frac{2}{3}}}\left(R - \frac{6D^2\phi}{\phi}\right) - \frac{2\sqrt{h}}{\vc}  D^2(\phi^4N)  
+ \frac{4\sqrt{h}}{\vc}h^{cd}\pa_c(\phi^3N)\pa_d\phi  
$$
\be
- \frac{2\sqrt{h}}{\vc}\phi^6
\left<
\phi^4N
\left(R - \frac{8D^2\phi}{\phi}
\right)
\right> 
\mbox{ } .  
\ee
We now require use of $D^2(\phi^4N) = \phi D^2(\phi^3N) 
+ 2h^{cd}\pa_c(\phi^3N)\pa_d\phi + \phi^3ND^2\phi$ 
to see that (\ref{CGfullslicing}) indeed guarantees $\dot{p} = 0$.

\mbox{ }

\noindent\large{\bf App III.2.A: Supporting material on conformal IVP method}\normalsize

\mbox{ }

\noindent{\bf Links between our alternative theories and York's work}

\mbox{ }

\noindent
With their preferred maximal and CMC slicings, our theories are written naturally in the 
conformal IVP language, and are best compared 
to GR via the York formulation of GR.  York's work thus contains many useful concepts and tricks 
(see I.2.9.3, I.2.10, which were largely given as an early section in ABF\'{O}).  
Rather than being exactly the same, I will show that some of our tricks share a common origin with York's in V.2.    
The notion of $CS$ and $CS + V$ themselves follow from York's work in I.2.9.4.2.  
York's work also offers as-yet unexplored options to some things, such as the style of treatment of matter in I.2.9.6 
and the possible alternative use of actions based on the 3-d metric concomitants of I.2.9.4.3.   
I furthermore study the closeness of both CS+V theory and conformal gravity to GR in both the 
traditional IVP and the thin sandwich formulation in V.2.3. I next outline the additional piece 
of conformal mathematics that inspired the division by the volume that led to conformal gravity.  

\mbox{ }

\mbox{ }

\noindent{\bf The Yamabe conjecture}

\mbox{ }

\noindent Yamabe's \cite{CGYamabe} manifestly-global conjecture is that all $n$-metrics are conformally equivalent 

\noindent to 
metrics of constant scalar curvature.  The proof for for $R \leq 0$ was relatively simple, 
but to prove the $R > 0$ case Schoen \cite{Schoen} required a new global idea: the Sobolev quotient 
\be
Q(\psi) = \frac{          \int_{\Sigma}
\left(        
|D\psi|^2 + \frac{      R\psi^2      }{      8      } 
\right)
\sqrt{h}\textrm{d}^3x          }
{          
\left(        \int_{\Sigma}\psi^{    {6}    }\sqrt{h}\textrm{d}^3x
\right)^{      \frac{1}{3}     }          }  
\label{Sobquot}
\ee
(presented here in 3-d).  
Note that the ELE corresponding to interpreting this as an action is the 
${\cal M} = 0$, $m = 1$ vacuum 
Lichnerowicz--York equation.  We are not concerned with what 
Schoen did with this quotient to prove the theorem, but rather with the idea of the quotient itself.  

The Yamabe conjecture is well-known in the GR IVP   
since the choice of constant curvature unphysical metrics simplifies the Lichnerowicz--York 
equation \cite{Yam, CIY}.  

\mbox{ }

\noindent{\bf Some Open Issues}

\mbox{ }

\noindent
In GR, the CMC condition is a gauge fixing without fundamental physical significance. 
The solutions of conformal gravity will strongly resemble solutions of GR in the CMC foliation 
at maximum expansion.  Some CS+V theories will resemble CMC-sliced GR even more closely.  
However, there are interpretational differences between those conformal 
theories that have a privileged slicing and reslicing-invariant GR.  First, there is a 
restriction in the solution space of GR since not all GR spacetimes are CMC sliceable, nor is a CMC slicing necessarily extendible to cover 
the maximal analytic extension of a given spacetime \cite{CGSYkin}.  Second, effects regarded 
as gauge artifacts in GR, such as the `collapse of the lapse' $N \longrightarrow 0$ in 
gravitational collapse (I.2.10) can have corresponding physical effects in 
privileged-CMC slicing theories.  

Can CS+V theory be formulated in terms of one auxiliary alone, like conformal gravity? 
Does our CS+V shed new light supporting York's contention that the two conformal 
shape d.o.f's of CS are, together with $V$, a representation of the true dynamical d.o.f's 
in CWB GR?  
Finally why does the Lichnerowicz--York equation arise 
from demanding the conformalized GR action to encode the LFE?  
Some of these issues are further pursued in V.2.

\mbox{ }

\noindent{\large\bf App III.2.B Other scale-invariant theories}\normalsize

\mbox{ }

\noindent There are two important differences between the manner in which conformal covariance 
is achieved in conformal gravity and the two best known earlier attempts to create conformally 
covariant theories: Weyl's 1917 theory \cite{CGWeyltheory} and 
Dirac's simplified modification of it \cite{CGDirac73}. First, both of these earlier theories 
are spacetime theories, and their conformal covariance leaves 4-d general 
covariance intact. In conformal gravity and in York's representation of GR, there is an 
element of the absolute.  Second, the conformal covariance is achieved in the theories of Weyl 
and Dirac through a compensating field that is conformally transformed with the gravitational 
field. In Weyl's theory, the compensating field is a 4-vector field that Weyl identified as the 
electromagnetic field until Einstein \cite{CGEinstein1} pointed out that atomic spectra 
would then be path-dependent, in contradiction with astronomical observations. 
Weyl later reinterpreted the idea of a compensating field in his 
effective creation of gauge theory \cite{CGWeylGauge}, but he never salvaged his original 
theory. In Dirac's simplification, the compensating field is the additional scalar field in 
Brans--Dicke theory \cite{CGBD}. This possibility has been exploited more recently in theories 
with a dilatonic field \cite{CGWetterich, buchdrag}. In contrast, conformal gravity has no 
physical compensating field; the variable $\phi$ (\ref{CGCoCo}) is a purely auxiliary gauge 
variable used to implement Conf-BM. This is therefore a more radical approach, in which full 
scale and conformal invariance of the gravitational field by itself is achieved.  In checking 
the literature on Weyl's theory, we came across Einstein's 1921 paper \cite{CGEinstein2}, in 
which he attempts to implement an idea similar to ours, albeit the implementation itself 
is very different.   Einstein follows Weyl in employing only ratios of the 4-metric 
components, but drops the idea of a compensating field.  Thus he is also aiming for a 
gravitational field that is scale invariant by itself.  The implementation Einstein proposed 
was very tentative, and we are not aware of anyone attempting to further develop it.

\vspace{9in}

\mbox{}

\noindent\Huge\bf{IV TSA: coupling of bosonic matter}\normalfont\normalsize

\mbox{ }

\noindent We now seek to convert the TSA to a route to relativity with matter 
`added on'.  
In IV.1.1 we consider the analysis of BF\'{O} concerning scalars and the apparent selection of 
electromagnetism among the possible single 1-form theories.  
In IV.1.2 we consider the theorem that underlaid my calculation with Barbour \cite{AB} about 
the apparent selection of Yang--Mills theory among the possible many interacting 1-form 
theories.  Consider this as a {\sl derivation} of the standard material provided  
for comparison in I.1.7.  

A tidier version of all the calculations for minimally coupled scalars and many 
interacting 1-forms is presented in IV.1.3 as following from the theorem.    
The strong gravity counterpart of these calculations \cite{Sanderson} is presented in IV.1.4, 
and compared with the GR results.  

The theorem is modified in IV.2.1 for use in conformal gravity, and a further theorem is built 
guaranteeing there is no integral inconsistency arising from the LFE that propagates the new 
$p = 0$ constraint, for the broad class of homogeneous actions considered.  
In IV.2.2 we show all the matter fields above can be coupled to conformal gravity.  
It is interesting that locally Lorentz-invariant physics drops out of a non-generally covariant 
theory.  I include two technical appendices IV.A on na\"{\i}ve renormalizability and IV.B on the 
Gell-Mann--Glashow theorem (which is used in both the TSA and the standard derivations 
of Yang--Mills theory).  App IV.C provides for comparison Teitelboim's derivation of 
Yang--Mills theory by the HKT route.  

For future reference of related material, the existence of suitable bosonic matter fields in 
the TSA is much strengthened in VI by an indirect method.  This however does not have the 
constructive status of the method of this chapter.  Following the criticism in VI and suitable 
reshaping of the whole program, I succeed in including also spin-$\frac{1}{2}$ fermions, in 
fact I include a full set of matter fields that suffice to describe nature as we know it.  
There are more matter issues in VII.  Overall, these later works show that the TSA is 
certainly not a unification and that the matter fields are picked out a great deal less 
uniquely than originally thought, although the TSA still does appear to be selective.

\mbox{ }

\noindent\Large{\bf 1 Matter in TSA formulation of GR}\normalsize

\mbox{ }

\noindent\large{\bf 1.1 Original BF\'{O} work}\normalsize

\mbox{ }

\noindent{\bf 1.1.1 Scalar fields}

\mbox{ }

\noindent Barbour, Foster and \'{O} Murchadha included a scalar field $\varsigma$ by considering 
the action 
\be 
\mbox{\sffamily I\normalfont}_{\mbox{\scriptsize BSW\normalsize}}^{\varsigma} 
= \int \textrm{d}\lambda \int \textrm{d}^3x \sqrt{h} 
\sqrt{        R + \mbox{\sffamily U\normalfont}^{\varsigma}        } 
\sqrt{        \mbox{\sffamily T\normalfont}^{\mbox{\scriptsize g\normalsize}} 
+ \mbox{\sffamily T\normalfont}^{\varsigma}        } 
\label{VARBSWAC}
\ee 
with the gravitationally BM scalar kinetic term\fn{NB the \sffamily U\normalfont's and 
\sffamily T\normalfont's in this chapter differ by constants from the standard spacetime 
forms.  This has no effect on results.  I leave them as they are to reflect the emergent 
character of the standard physics from the TSA point of view.} 
$\mbox{\sffamily T\normalfont}^{\varsigma} = (\mbox{\ss}_{\xi}{\varsigma})^2$ and the potential ansatz 
$\mbox{\sffamily U\normalfont}^{\varsigma} 
= -\frac{C}{4}h^{ab}\pa_a\varsigma\pa_b\varsigma + \sum_{(n)} A_{(n)}\varsigma^{(n)}$. 

Defining $2N \equiv \sqrt{      \frac{    
\mbox{\sffamily\scriptsize T\normalsize\normalfont}^{\mbox{\tiny g\normalsize}} 
+ \mbox{\sffamily\scriptsize T\normalsize\normalfont}^{\varsigma}    }{    R 
+ \mbox{\sffamily\scriptsize U\normalsize\normalfont}^{\varsigma}    }      } $, the conjugate momenta are given by 
(\ref{GRmom}) and
\be 
\pi \equiv \frac   {\partial\mbox{\sffamily{L}\normalfont}}
{\partial \dot{\varsigma}}  = \frac{\sqrt{h}}{2N} \mbox{\ss}_{\xi}\varsigma \mbox{ } .
\label{scalarmom}
\ee 
Then the local square root gives as a primary constraint a Hamiltonian-type constraint  
\be 
{}^{\varsigma}{\cal H } \equiv \sqrt{h}(R + \mbox{\sffamily U\normalfont}^{\varsigma}) 
- \frac{1}{\sqrt{h}}
\left(
p \circ p - \frac{1}{2}p^2 + \pi^2
\right)
 = 0 \mbox{ } . 
\ee 
${\xi}^i$-variation gives as a secondary constraint the momentum constraint 
\be 
^{\varsigma}{\cal H}_i \equiv -2D_j{p^{j}}_{i} + \pi\pa_i\varsigma = 0 
\mbox{ } . 
\label{SCSC}
\ee 
The ELE's are 
$$
\frac{\partial p^{ij}} {\partial \lambda} =
\frac{\delta\mbox{\sffamily L\normalfont}}{\delta h_{ij}} =  
\sqrt{h}N(h^{ij}R - R^{ij}) - \frac{2N}{\sqrt{h}}(p^{im}{p_m}^j -\frac{1}{2}p^{ij}p) +
\sqrt{h}( + D^jD^iN - h^{ij}D^2 N)
$$
\be
+ \frac{\sqrt{h}CN}{4}(\pa^i\varsigma\pa^j\varsigma -\pa_a\varsigma\pa^a\varsigma h^{ij})
+ \sqrt{h}N\sum_{(n)}A_{(n)}\varsigma^{(n)}h^{ij}+ \pounds_{\xi}p^{ij} \mbox{ } ,
\ee
\be
\frac{    \partial \pi    }{    \partial \lambda    } =
\frac{    \delta\mbox{\sffamily L\normalfont}    }{    \delta\varsigma    } =
\frac{    \sqrt{h}C    }{    2    }D^i(N\pa_i\varsigma)  
+ \sqrt{h}N\sum_{(n)}nA_{(n)}\varsigma^{n - 1} + \pounds_{\xi}\pi \mbox{ } . 
\label{phiELeq}
\ee

The constraint $^{\varsigma}{\cal H}$ contains the canonical propagation speed $C$ of the 
scalar field.  This can be read off the coupled form of (\ref{scalarmom}) and (\ref{phiELeq}):  
\be
\frac{   \partial   }{   \partial\lambda   }
\left\{
\frac{   \sqrt{h}   }{   N   }
\left( 
\frac{   \partial\varsigma   }{   \partial\lambda   }
- \pounds_{\xi}\varsigma
\right)
\right\} 
= \sqrt{h}C D_i(   N\pa^i\varsigma   ) + 2\sqrt{h}N\frac{   dV(  \varsigma  )   }{   d\varsigma   } 
+ 2\pounds_{\xi}\pi \mbox{ } .
\ee
In the TSA, 
a priori, $ C \neq 1$, which means there is no reason for the scalar field to obey the
same null cone as gravity.  However, propagating $^{\varsigma}{\cal H}$ gives
\be
^{\varsigma}\dot{{\cal H}} = 
- \frac{    D^i(N^2 \mbox{ }^{\varsigma}{\cal H}_i)    }{    N    }  
+ \frac{   Np \mbox{ }^{\varsigma}{\cal H}   }{   2   } 
+ \pounds_{\xi} (\mbox{ }^{\varsigma}{\cal H}) 
+ \frac{    (1 - C)   }{    N    }D^i(N^2\pi \pa^i\varsigma) \mbox{ } .
\ee
The theory has just one scalar d.o.f, so if the cofactor of $(1 - C)$ in the last term were 
zero, the scalar dynamics would be trivial.\fn{This argument requires further elaboration, 
which is provided in V.1.}  Thus 
$C = 1$: scalar fields must obey the null cone dictated by gravity.   The counterpart in the 
standard approach is that $C = 1$ due to \bf RP1\normalfont.  The point is that BF\'{O} \sl 
deduce \normalfont this as enforced by gravitation.  

Notice also that this scheme gives minimal coupling of the scalar field to gravity. 
However, there is one other possibility because the most general ultralocal kinetic term 
includes also a metric--scalar cross-term. This gives Brans--Dicke theory.  This was considered 
in the revised version of RWR, but I am not satisfied with that treatment since it is based on 
a conformal transformation (which does not preserve geodesics).  I present a satisfactory and 
in fact more illuminating treatment in VI.1.4.   

\mbox{ }

\noindent{\bf 1.1.2 A single 1-form field}

\mbox{ }

\noindent To include a single $1$-form field $A_a$, BF\'{O} considered the action
\be 
\mbox{\sffamily I\normalfont}^{\mbox{\scriptsize A\normalsize}}_{\mbox{\scriptsize BSW\normalsize}}
= \int \textrm{d}\lambda \int \textrm{d}^3x \sqrt{h} 
\sqrt{R + \mbox{\sffamily U\normalfont}^{\mbox{\scriptsize A\normalsize}}} 
\sqrt{\mbox{\sffamily T\normalfont}^{\mbox{\scriptsize g\normalsize}} 
+ \mbox{\sffamily T\normalfont}^{\mbox{\scriptsize A\normalsize}}}  
\ee 
for
$\mbox{\sffamily T\normalfont}^{\mbox{\scriptsize A\normalsize}} = 
h^{ab}\mbox{\ss}_{\xi}{A}_a\mbox{\ss}_{\xi}{A}_b$ 
the quadratic gravitationally best-matched kinetic term of $A_a$, and the potential ansatz 
$\mbox{\sffamily U\normalfont}^{\mbox{\scriptsize A\normalsize}} =
C_1D_bA_{a}D^bA^{a} + C_2D_bA_{a}D^aA^{b} + C_3D^a{A_a}D^b{A_b} + 
\sum_{(k)} B_{(k)}(A_aA^a)^{k} $.  The first part of this can be expressed
more conveniently for some purposes by using a generalized supermetric 
$C^{abcd} = C_1h^{ab}h^{cd} + C_2h^{ad}h^{bc} + C_3h^{ac}h^{bd}$.  

Defining $2N \equiv \sqrt{     \frac{    \mbox{\sffamily\scriptsize T\normalsize\normalfont}^{\mbox{\tiny g\normalsize}} 
+ \mbox{\sffamily\scriptsize T\normalsize\normalfont}^{\mbox{\tiny A\normalsize}}    }
{    R + \mbox{\sffamily\scriptsize U\normalsize\normalfont}^{\mbox{\tiny A\normalsize}}    }     }$, the conjugate momenta are given by 
(\ref{GRmom}) and 
\be 
\pi^i \equiv \frac   {\partial\mbox{\sffamily{L}\normalfont}}
{\partial \dot{A_i}}  = \frac{\sqrt{h}}{2N} \mbox{\ss}_{\xi} A^i \mbox{ } .
\label{1formmom}
\ee 
Then, the local square root gives as a primary constraint a Hamiltonian-type constraint 
\be 
^{\mbox{\scriptsize A\normalsize}}{\cal H } \equiv 
\sqrt{h}(R + \mbox{\sffamily U\normalfont}^{\mbox{\scriptsize A\normalsize}}) -  
\left(
p \circ p - \frac{1}{2}p^2 + \pi^a\pi_a
\right) 
= 0 \mbox{ } .
\ee 
$\xi^i$-variation gives as a secondary constraint the momentum constraint 
\be
^{\mbox{\scriptsize A\normalsize}}{\cal H}_i \equiv 
-2D_j{p^{j}}_{i} + \pi^{c}(D^j{A_{i}} - D_i{{A}^j}) - D_c{\pi^{c}}A_{i} = 0 \mbox{ } . 
\label{emmomcon}
\ee 

Then, propagating $^{\mbox{\scriptsize A\normalsize}}{\cal H }$ 
gives\fn{I do not provide the cumbersome ELE's since I am to build a 
method in the next section which does not require their explicit use.} 
$$
^{\mbox{\scriptsize A\normalsize}}\dot{{\cal H}} = 
- \frac{    D^i(N^2 \mbox{ }^{A}{\cal H}_i)    }{    N    }  
+ \frac{   Np \mbox{ }^{A}{\cal H}   }{   2   } 
+ \pounds_{\xi} \mbox{ }^{A}{\cal H} 
$$
$$
+ \frac{1}{N} 
\left\{
(4C_1 + 1)D_b(N^2\pi^aD^b{A_{a}}) + (4C_2 - 1)D_b(N^2\pi^aD_a{A^b}) + 4C_3D_a(N^2\pi^aD_b{A^b}) 
\right\}
$$
\be
- \frac{1}{N} D_b(N^2D_a{\pi^a}A^b) 
- \frac{1}{N} 
\left\{
N^2 D_a
\left(
p_{ij} - \frac{p}{2} h_{ij}
\right) 
D_dA_b 
(2A^iC^{ajbd} - A^aC^{ijbd}) 
\right\}  
\mbox{ } .
\ee
Now, the system has a priori 5 d.o.f's per space point, that is 2 geometric d.o.f's  
and the 3 d.o.f's of the 1-form field itself.  The constraints cannot include $N$, so the 
penultimate line includes a 3-vector of constraints multiplied by $\pa^aN$, which would take 
away all the 1-form d.o.f's, thus rendering a trivial theory, unless the  cofactor of 
$\pa^aN$ vanishes strongly.  This gives a nontrivial theory only for $C_1 = -C_2 = -\frac{1}{4}$,
$C_3 = 0$ and if there is a secondary constraint 
\be
{\cal G} \equiv D_a{\pi^a} = 0 \mbox{ } .
\label{emergau}
\ee 
The conditions on the $C$'s mean that the 1-form field obeys the null cone dictated by gravity, 
and furthermore that the derivative terms in $\mbox{\sffamily U\normalfont}_{\mbox{\scriptsize A\normalsize}}$ are
$-\frac{1}{4}|\mbox{\b{$\pa$}} \mbox{ \scriptsize $\times$ \normalsize} \mbox{\b{A}\normalfont}|^2$.  We identify the new 
constraint (\ref{emergau}) as the Gauss constraint of electromagnetism (\ref{M1}, \ref{curemgau}).   
$$
\mbox{ }\mbox{ In order to propagate the new constraint ${\cal G}$, first note that }
\mbox{\hspace{0.4in}}
\frac{\partial}{\partial \lambda}(D_a{\pi^a}) = D_a
\left(
\frac{\partial \pi^a}{\partial \lambda}
\right)
\mbox{\hspace{0.4in}}
$$
because $\pi^a$ is a $(1,0)$ density.  Thus, 
\be
\frac{\partial}{\partial \lambda}(D_i{\pi^i}) =  2\sqrt{h}D_a
\left\{
N\sum_{(k)}kB_{(k)}(A^aA_a)^{k - 1} A^i
\right\} 
+ \pounds_\xi(D_i{\pi^i})
\ee
where we have used $\mbox{\b{$\pa$}} \cdot \mbox{\b{$\pa$}} \mbox{ \scriptsize $\times$ \normalsize} \mbox{\b{A}} = 0$ 
and mixed partial equality to eliminate a further 
term 
\be
\frac{\sqrt{h}}{2}D_iD^b[N(D_b{A^i} - D^i{A_b})] \mbox{ } .
\ee
Again, one can argue that constraints cannot depend on $N$, and then that the only way of 
avoiding triviality of the 1-form field due to the terms in  $\pa_aN$ is to have all the 
$B_{(k)}$ be zero. In particular, $B_{(1)} = 0$ means that the 1-form field must be massless.  

Now, the allowed form  
$\mbox{\sffamily U\normalfont}^{\mbox{\scriptsize A\normalsize}} = -\frac{1}{4}D_bA_b(D^bA^a - D^aA^b)$ 
[c.f (\ref{Lemcssplit})], is invariant under the 
\be
\mbox{gauge transformation }
\mbox{\hspace{1.5in}}
A_a \longrightarrow A_a + \pa_a\Lambda 
\mbox{ } , 
\mbox{\hspace{2in}}
\ee
so we are dealing with a gauge theory.  
Note first how the gauge theory and the fixing of the light-cone to be equal to the 
gravity-cone arise together in the same part of the above calculation.  These are two aspects 
of the same consistency condition arising from the role of the momentum constraint in the 
propagation of the Hamiltonian constraint.  Second, because we have a gauge symmetry, if we 
introduce an auxiliary variable $\Phi$ into $\mbox{\sffamily T\normalfont}^{\mbox{\scriptsize A\normalsize}}$ such that 
variation with respect to it encodes ${\cal G}$, then we should do so according to  U(1)-BM.   
\be
\mbox{This uniquely fixes the form of 
$\mbox{\sffamily T\normalfont}^{\mbox{\scriptsize A\normalsize}}(A, \Phi)$ to be }
\mbox{\sffamily T\normalfont}^{\mbox{\scriptsize A\normalsize}} = (\dot{A}_a - \pounds_{\xi}A_a - \pa_a\Phi)
(\dot{A}^a - \pounds_{\xi}A^a - \pa^a\Phi)   
\label{Vcorrection}
\ee 
[c.f (\ref{Lemcssplit})].  Thus, if one identifies $\Phi$ as $A_{0}$, this
derivation forces \v{A} $= [A_0, \mbox{\b{A}}]$ 
to obey Maxwell's equations minimally-coupled
to gravity. Moreover, as pointed out by Giulini \cite{giulini} and reported in \cite{BOF2}, the 
massive (Proca) 1-form field does not fit into this TSA formulation despite being a perfectly 
good generally covariant theory. BF\'{O} originally took this to be evidence that the TSA 
does not yield all generally covariant theories.  This is further explored 
in later sections and chapters.  My main aim in this chapter is to show that the 
TSA does at least permit standard fundamental bosonic field theories.  

Finally, we mention that on attempting to couple $A_a$ to scalar fields by the inclusion of 
interaction terms, BF\'{O} showed similarly that demanding the propagation of 
$ ^{\mbox{\scriptsize A\normalsize}, \varsigma}{\cal H}$, and of any secondary constraints arising 
from it, leads to U(1) gauge theory minimally-coupled to GR \cite{BOF2}. We have thus a chain of 
successively more sophisticated theories, each arising from its predecessor by iteration of 
constraint propagation consistency. This provides a different means of deriving classical 
physics: Dirac's work, applied to BM RI actions, leads to a 
striking alternative to Einstein and Minkowski's $4$-d foundation of physics.  I will next 
build the above outline of Relativity without Relativity into an almost 
algorithmic formality, from which Yang--Mills gauge theory will emerge from allowing a general
collection of $1$-form fields to interact with each other.

\mbox{ }

\noindent\large{\bf 1.2 Matter workings from the perspective of a general theorem}\normalsize

\mbox{ }

\noindent We next find that Yang--Mills theory minimally coupled to GR emerges as one of a 
few possibilities allowed for quite a general ansatz for the 1-form fields' potential.  More 
precisely, the 1-form fields are again found to respect the gravitational null cone, to be fundamentally massless, 
and to have a Yang--Mills type mutual interaction.  The calculation if done in the above style 
is much messier than that for a single 1-form field.  We present it rather by use of a theorem 
that uses the general ELE's and permits a clear term-by-term computation of the 
propagation of the Hamiltonian constraint.  This theorem-based approach is good in being more 
systematic than the original RWR-type treatment.  

We conclude that, within the bosonic sector, the TSA yields the key features of 
the observed world. Gravity, the universal null cone, and gauge theory all arise in
essentially the same manner through the single mechanism of consistent Dirac-type constraint 
propagation applied to the interplay of BM with the local square root. Some field-theoretic 
issues are addressed in the conclusion.  

I consider $\mbox{\sffamily T\normalfont}^{\Psi}$ and $\mbox{\sffamily U\normalfont}^{\Psi}$ 
to consist of contributions from each of the matter fields $\Psi_{\mbox{\scriptsize\sffamily A\normalfont\normalsize}}$ present.  I then 
obtain the following formulae for the propagation of the Hamiltonian constraint.

\mbox{ }

\noindent \bf Theorem 1 \normalfont

\noindent i) For nonderivative coupled matter fields
$\Psi_{\mbox{\scriptsize\sffamily A\normalfont\normalsize}}$ with 
$\mbox{\sffamily T\normalfont}^{\Psi}$ homogeneously quadratic in
$\dot{\Psi}_{\mbox{\scriptsize\sffamily A\normalfont\normalsize}}$ and 
$\mbox{\sffamily U\normalfont}^{\Psi}$ containing at most
first-order derivatives, 
\be
-^{\Psi}\dot{{\cal H}} = \frac{1}{N}D_b 
\left\{
N^2
\left(
2G_{\mbox{\scriptsize\sffamily AB\normalfont\normalsize}}
\Pi^{\mbox{\scriptsize\sffamily B\normalfont\normalsize}}
\frac{ \pa \mbox{\sffamily U\normalfont}^{\Psi}    }
{    \pa(D_b\Psi_{\mbox{\scriptsize\sffamily A\normalfont\normalsize}})  } + \sigma
\left[
\Pi^{\mbox{\scriptsize\sffamily A\normalfont\normalsize}}
\frac{\delta    (\pounds_\xi\Psi_{\mbox{\scriptsize\sffamily A\normalfont\normalsize}}) }{\pa\xi_b} 
\right] 
\right)
\right\},
\ee 
where in this chapter and the next [ ] denotes the extent of applicability of the functional 
derivative within, $G_{\mbox{\scriptsize\sffamily AB\normalfont\normalsize}}$ is an invertible 
ultralocal kinetic metric and 
$\Pi^{\mbox{\scriptsize\sffamily A\normalfont\normalsize}}$ is the momentum conjugate to 
$\Psi_{\mbox{\scriptsize\sffamily A\normalfont\normalsize}}$.

\noindent ii) If, additionally, the potential contains covariant derivatives, then there is an 
extra 

\be
\mbox{contribution to i): } 
\mbox{ } \mbox{ } \mbox{ }
\frac{2\sqrt{h}}{N}D_b
\left\{
N^2
\left(
p_{ij} - \frac{X}{2}h_{ij}
\right)
\left(
\frac{\pa \mbox{\sffamily U\normalfont}^{\Psi}}{\pa{\Gamma^a}_{ic}}h^{aj} 
- \frac{1}{2}\frac{\pa \mbox{\sffamily U\normalfont}^{\Psi}}{\pa{\Gamma^a}_{ij}}h^{ac}
\right)
\right\}.
\ee

The proof offered here includes both GR ($\sigma = 1$, $W = 1$) and strong gravity 
($\sigma = 0$, $\Lambda \neq 0$, $W \neq \frac{1}{3}$ but otherwise arbitrary).  Result i) in 
the GR case is related to a result of Teitelboim \cite{Teitelthesis, Teitelboim} that the contributions of 
nonderivatively-coupled fields to the Hamiltonian and momentum constraints independently 
satisfy the Dirac Algebra (see App IV.C).  In the working below, this is reflected by our 
ability to split the working into pure gravity and matter parts.

Use of formulae i), ii) permits the $^{\Psi}\dot{{\cal H}}$ calculations to be done without 
explicitly computing each case's ELE's. This is because our derivation uses once 
and for all the \sl general \normalfont ELE's.

\mbox{ }

\noindent \bf Proof \normalfont The working can be split into its $\xi^i$-free and $\xi^i$ parts (the latter is not exhibited).  

\be
\mbox{i) For a homogeneous quadratic kinetic term }
\mbox{\hspace{0.7in}}
\mbox{\sffamily T\normalfont}^{\Psi}   = 
\mbox{\ss}_{\xi}{\Psi}_{\mbox{\scriptsize\sffamily A\normalfont\normalsize}} 
\mbox{\ss}_{\xi}{\Psi}_{\mbox{\scriptsize\sffamily B\normalfont\normalsize}} 
           G^{\mbox{\scriptsize\sffamily AB\normalfont\normalsize}}(h^{ij}) \mbox{ } ,
\mbox{\hspace{0.7in}}
\ee
\be
\mbox{the conjugate momenta are }
\mbox{\hspace{1.2in}}
\Pi^{\mbox{\scriptsize\sffamily A\normalfont\normalsize}} 
= \frac{\pa \mbox{\sffamily L\normalfont}    }{\pa\dot{\Psi}_{\mbox{\tiny\sffamily A\normalfont\normalsize}}} 
= \frac{\sqrt{h}}{2N}
{    }G^{\mbox{\scriptsize\sffamily AB\normalfont\normalsize}}
\mbox{\ss}_{\xi}{\Psi}_{\mbox{\scriptsize\sffamily B\normalfont\normalsize}} \mbox{ } .
\mbox{\hspace{1.2in}}
\ee
\noindent The $\xi^i$-variation gives as a secondary constraint the momentum constraint
\be 
-^{\Psi}{\cal H}_i \equiv 2D_j{p_i}^j
- \Pi^{\mbox{\scriptsize\sffamily A\normalfont\normalsize}}
\frac{\delta(\pounds_{\xi}\Psi_{\mbox{\scriptsize\sffamily A\normalfont\normalsize}})}
{\delta\xi^i}    = 0 
\label{genGRmom}
\ee 
and the local square root gives as a primary constraint a Hamiltonian-type constraint  
\be
-^{\Psi}{\cal H} \equiv 
\sqrt{h}(\sigma R + \Lambda + \mbox{\sffamily U\normalfont}^{\Psi})  
- \frac{ 1 }{  \sqrt{h}  }(p \circ p - \frac{X}{2}p^2 + 
G_{\mbox{\scriptsize\sffamily AB\normalfont\normalsize}}\Pi^{\mbox{\scriptsize\sffamily A\normalfont\normalsize}}
\Pi^{\mbox{\scriptsize\sffamily B\normalfont\normalsize}}) = 0.
\label{genGRham} 
\ee 
$$
\mbox{ Then } 
\mbox{\hspace{0.2in}}
-^{\Psi}\dot{{\cal H}} \approx 
\dot{\mbox{ }\sqrt{h}}(\sigma R + \Lambda + \mbox{\sffamily U\normalfont}^{\Psi}) -
\dot{
\left(
\frac{1}{\sqrt{h}}
\right)
}
\left(
p \circ p - \frac{X}{2}p^2 
+ G_{\mbox{\scriptsize\sffamily AB\normalfont\normalsize}}\Pi^{\mbox{\scriptsize\sffamily A\normalfont\normalsize}}
\Pi^{\mbox{\scriptsize\sffamily B\normalfont\normalsize}}
\right)     
\mbox{\hspace{1in}}
$$
$$
\mbox{\hspace{0.7in}}
+
{\sqrt{h}}(\sigma\dot{R} + \dot{\mbox{\sffamily U\normalfont}}^{\Psi}) -
\frac{2}{\sqrt{h}}
\left\{
\dot{p}^{ij}
\left(
p_{ij} - \frac{X}{2}ph_{ij}
\right) 
+ p^{ij}p^{kl}
\left(
\dot{h}_{ik}h_{jl} - \frac{X}{2}\dot{h}_{ij}h_{kl}
\right)
\right\}  
$$
\be
\mbox{\hspace{0.7in}}
- \frac{1}{\sqrt{h}}(2
\dot{\Pi}^{\mbox{\scriptsize\sffamily A\normalfont\normalsize}} 
{    }G_{\mbox{\scriptsize\sffamily AB\normalfont\normalsize}}\Pi^{\mbox{\scriptsize\sffamily B\normalfont\normalsize}} 
+  {    }\dot{G}_{\mbox{\scriptsize\sffamily AB\normalfont\normalsize}}\Pi^{\mbox{\scriptsize\sffamily A\normalfont\normalsize}}
\Pi^{\mbox{\scriptsize\sffamily B\normalfont\normalsize}}),
\ee
using the chain-rule on (\ref{genGRham}).  Now use the chain-rule on 
$\dot{\mbox{\sffamily U\normalfont}}^{\Psi}$, the ELE's 
$\dot{p}^{ij} = \frac{\delta \mbox{\sffamily\scriptsize L\normalsize\normalfont}}{\delta h_{ij}}$ and 
$\dot{\Pi}^{\mbox{\scriptsize\sffamily A\normalfont\normalsize}} 
= \frac{\delta \mbox{\sffamily\scriptsize L\normalsize\normalfont}}{\delta\Psi_{\mbox{\tiny\sffamily A\normalfont\normalsize}}}$, to obtain the first step below: 
$$
-^{\Psi}\dot{{\cal H}} = 
\frac{2 - 3X}{2}Np(\sigma R + \Lambda + \mbox{\sffamily U\normalfont}^{\Psi}) + 
\frac{2 - 3X}{2}Np
\left(
p \circ p - \frac{X}{2}p^2 + G_{\mbox{\scriptsize\sffamily AB\normalfont\normalsize}}
\Pi^{\mbox{\scriptsize\sffamily B\normalfont\normalsize}}
\Pi^{\mbox{\scriptsize\sffamily A\normalfont\normalsize}}
\right)                                                                                  
$$
$$
+ {\sqrt{h}\sigma}\dot{R} + {\sqrt{h}}
\left\{
\frac{\pa\mbox{\sffamily U\normalfont}^{\Psi}}{\pa\Psi_{\mbox{\scriptsize\sffamily B\normalfont\normalsize}}}
\dot{\Psi}_{\mbox{\scriptsize\sffamily B\normalfont\normalsize}} 
+ \frac{\pa\mbox{\sffamily U\normalfont}^{\Psi}}{\pa(D_b\Psi_{\mbox{\scriptsize\sffamily B\normalfont\normalsize}})}
\dot{(D_b\Psi_{\mbox{\scriptsize\sffamily B\normalfont\normalsize}})}
+ \frac{\pa\mbox{\sffamily U\normalfont}^{\Psi}}{\pa h_{ab}}\dot{h}_{ab}
\right\}                                                                                   
$$
$$
- 2
\left(
p_{ij}\mbox{--}\frac{X}{2}ph_{ij}
\right)
\left\{
\left[
\sigma N\frac{\delta R}{\delta{h_{ij}}}
\right] 
+ 
\left[
N\frac{\delta\mbox{\sffamily U\normalfont}^{\Psi}}{\delta{h_{ij}}}
\right] 
+ \frac{1}{4N}\frac{\pa\mbox{\sffamily T\normalfont}^{\Psi}}{\pa{h_{ij}}} 
\right\}
$$
$$
- \frac{4N}{{h}}
\left\{
p^{ij}p^{kl}
\left(
p_{ik}\mbox{--}\frac{X}{2}ph_{ik}
\right)
h_{jl}
\mbox{--}\frac{X}{2}
\left(
p_{ij}\mbox{--}\frac{X}{2}ph_{ij}
\right)
h_{kl}
\right\}
\mbox{--}{2}{ }G_{\mbox{\scriptsize\sffamily AB\normalfont\normalsize}}\Pi^{\mbox{\scriptsize\sffamily B\normalfont\normalsize}}
\left[
N\frac{\delta \mbox{\sffamily U\normalfont}^{\Psi}}{\delta\psi_{\mbox{\scriptsize\sffamily A\normalfont\normalsize}}} 
\right] 
- \frac{1}{\sqrt{h}}\mbox{}\dot{G}_{\mbox{\scriptsize\sffamily AB\normalfont\normalsize}}
\Pi^{\mbox{\scriptsize\sffamily A\normalfont\normalsize}}\Pi^{\mbox{\scriptsize\sffamily B\normalfont\normalsize}}
$$                                                                                     
$$
=  
\left\{
\frac{2 - 3X}{2}Np
\left(
\sigma R + \Lambda + ^{\Psi}U + 
p \circ p - \frac{X}{2}p^2 + 
G_{\mbox{\scriptsize\sffamily AB\normalfont\normalsize}}\Pi^{\mbox{\scriptsize\sffamily B\normalfont\normalsize}}
\Pi^{\mbox{\scriptsize\sffamily A\normalfont\normalsize}}
\right)                                                                                  
\right.
$$
$$
+ \sigma
\left(
\sqrt{h}\dot{R}-2
\left\{
p_{ij}\mbox-\frac{X}{2}ph_{ij}
\right\}
\left[
N \frac{\delta R}{\delta{h_{ij}}} 
\right]
\right)
$$
$$
\left.
-\frac{4N}{{h}}
\left(
p^{ij}p^{kl}
\left\{
p_{ik}-\frac{X}{2}ph_{ik}
\right\}
h_{jl}-\frac{X}{2}
\left\{
p_{ij}-\frac{X}{2}ph_{ij}
\right\}
h_{kl}
\right)                                                                                    
\right\}                                                                                  
$$
$$
+{\sqrt{h}}
\left\{
\frac{\pa \mbox{\sffamily U\normalfont}^{\Psi}}{\pa\Psi_{\mbox{\scriptsize\sffamily A\normalfont\normalsize}}}
\left(
\frac{2N}{\sqrt{h}}\Pi^{\mbox{\scriptsize\sffamily B\normalfont\normalsize}}
G_{\mbox{\scriptsize\sffamily AB\normalfont\normalsize}}
\right)
+ \frac{\pa\mbox{\sffamily U\normalfont}^{\Psi}}{\pa(D_b\Psi_{\mbox{\scriptsize\sffamily A\normalfont\normalsize}})}D_b
\left(
\frac{2N}{\sqrt{h}}\Pi^{\mbox{\scriptsize\sffamily B\normalfont\normalsize}}
G_{\mbox{\scriptsize\sffamily AB\normalfont\normalsize}}
\right) 
+ \frac{\pa\mbox{\sffamily U\normalfont}^{\Psi}}{\pa h_{ab}}
\frac{2N}{\sqrt{h}}
\left(
p_{ab} - \frac{X}{2}ph_{ab}
\right)
\right\}                                                                                   
$$
$$
- {2}
\left(
p_{ab} - \frac{X}{2}ph_{ab}
\right)
\frac{\pa\mbox{\sffamily U\normalfont}^{\Psi}}{\pa{h_{ab}}} N 
- \frac{1}{2N}
\left(
p_{ab} - \frac{X}{2}ph_{ab}
\right)
\frac{\pa{ }G^{\mbox{\scriptsize\sffamily AB\normalfont\normalsize}}}{\pa h_{ab}}
\dot{\Psi}_{\mbox{\scriptsize\sffamily B\normalfont\normalsize}}
\dot{\Psi}_{\mbox{\scriptsize\sffamily A\normalfont\normalsize}} 
$$
$$
- {2}{ }G_{\mbox{\scriptsize\sffamily AB\normalfont\normalsize}}
\Pi^{\mbox{\scriptsize\sffamily B\normalfont\normalsize}}N
\frac{\pa\mbox{\sffamily U\normalfont}^{\Psi}}
{\pa\psi_{\mbox{\scriptsize\sffamily A\normalfont\normalsize}}}
+ {2}G_{\mbox{\scriptsize\sffamily AB\normalfont\normalsize}}
\Pi^{\mbox{\scriptsize\sffamily B\normalfont\normalsize}}
D_b\left(N\frac{\pa\mbox{\sffamily U\normalfont}^{\Psi}} 
{\pa(D_b\Psi_{\mbox{\scriptsize\sffamily A\normalfont\normalsize}})}
\right)
- \frac{1}{\sqrt{h}}\frac{\pa{G}_{\mbox{\scriptsize\sffamily AB\normalfont\normalsize}}}
{\pa h_{ij}}\dot{h}_{ij}
\Pi^{\mbox{\scriptsize\sffamily A\normalfont\normalsize}}
\Pi^{\mbox{\scriptsize\sffamily B\normalfont\normalsize}}                                                       
$$
\be
\mbox{$\approx$}\frac{\sigma}{N}D_b
\left(
N^2
\left[
\Pi^{\Delta}\frac{\delta(\pounds_{\xi}\Psi_{\mbox{\scriptsize\sffamily A\normalfont\normalsize}})}{\delta\xi_b}                 
\right]
\right)
+ \sqrt{h}\frac{\pa\mbox{\sffamily U\normalfont}^{\Psi}}{\pa(D_b\Psi_{\mbox{\scriptsize\sffamily A\normalfont\normalsize}})}D_b
\left(
\frac{2N}{\sqrt{h}}\Pi^{\mbox{\scriptsize\sffamily A\normalfont\normalsize}}{ }
G_{\mbox{\scriptsize\sffamily AB\normalfont\normalsize}}
\right) 
+ {2}G_{\mbox{\scriptsize\sffamily AB\normalfont\normalsize}}\Pi^{\mbox{\scriptsize\sffamily B\normalfont\normalsize}}D_b
\left(
N\frac{\pa\mbox{\sffamily U\normalfont}^{\Psi}} {\pa(D_b\Psi_{\mbox{\scriptsize\sffamily A\normalfont\normalsize}})}
\right).
\ee 
In the second step above, I regroup the terms into pure gravity terms and matter terms, 
expand the matter variational derivatives and use the definitions of the momenta to eliminate
the velocities in the first three matter terms.  I now observe that the first and sixth matter 
terms cancel, as do the third and fourth.  In the third step I discard a term proportional to 
$^{\Psi}{\cal H}$, use the momentum constraint (\ref{genGRmom}), and the pure gravity working 
(N.B we assume that either $W = 1$ or $\sigma = 0$, so the pure gravity working does work!), 
and the definitions of the momenta to cancel the fifth and eight terms of step 2.  
Factorization of step 3 gives the result.  

ii) Now $-^{\Psi}\dot{{\cal H}}$ has 2 additional contributions in step 2 due to the presence 
of the connections: 
\be
{\sqrt{h}}\frac{\pa \mbox{\sffamily U\normalfont}^{\Psi}}
{\pa{\Gamma^a}_{bc}} \dot{\Gamma}^a {}_{bc} - {2}
\left(
p_{ij} - \frac{X}{2}ph_{ij}
\right)
\left[
\frac{\pa \mbox{\sffamily U\normalfont}^{\Psi}}{\pa{\Gamma^a}_{bc}}
\frac{\delta {\Gamma^a}_{bc}}{\delta h_{ij}}N
\right] \mbox{ } ,
\label{GRhalfwaycon}
\ee 
which, using (\ref{deltaGamma}), (\ref{CDOT}), 
integration by parts on the second term of (\ref{GRhalfwaycon}) and factorization yields ii). 
$\Box$

\mbox{ }

\noindent\large{\bf 1.3 Examples}\normalsize

\mbox{}

\noindent{\bf 1.3.1 Minimally-coupled scalar fields}

\mbox{ }

\noindent Application of formula i) to the minimally-coupled scalar field action (\ref{VARBSWAC}) 
immediately 
\be
\mbox{yields that }
\mbox{\hspace{1.8in}}
^{\varsigma}\dot{{\cal H}} \approx \frac{1 - C}{N}D^b(N^2\pi_{\varsigma} \pa_b\varsigma) \mbox{ } .
\mbox{\hspace{2.4in}}
\ee

\mbox{ }

\noindent\bf{1.3.2  K Interacting 1-form Fields}\normalfont

\mbox{ }

\noindent We consider a BSW-type action containing the a priori unrestricted 1-form fields 
$A_a^{\mbox{\bf\scriptsize I\normalsize\normalfont}}$, 
\be 
\mbox{\sffamily I\normalfont}_{\mbox{\scriptsize BSW\normalsize}}^{{\mbox{\scriptsize A\normalsize}}_{\mbox{\bf\tiny I\normalsize\normalfont}}} 
= \int \textrm{d}\lambda\int \textrm{d}^3x \sqrt{h} \mbox{{\sffamily L\normalfont}} (h_{ij},
\dot{h}^{ij}, A_i^{\mbox{\bf\scriptsize I\normalsize\normalfont}}           , 
\dot{A}^i_{\mbox{\bf\scriptsize I\normalsize\normalfont}}, N, \xi^i) 
= \int \textrm{d}\lambda \int\textrm{d}^3x \sqrt{h} \sqrt{R + 
\mbox{\sffamily U\normalfont}^{\mbox{\scriptsize A\normalsize}_{\mbox{\bf\scriptsize I\normalsize\normalfont}}}}
\sqrt{ \mbox{\sffamily T\normalfont}^{\mbox{\scriptsize g\normalsize}} + 
\mbox{\sffamily T\normalfont}^{\mbox{\scriptsize A\normalsize}_{\mbox{\bf\scriptsize I\normalsize\normalfont}}}} \mbox{ } .
\ee 
We use the most general homogeneous quadratic BM kinetic term 
$\mbox{\sffamily T\normalfont}^{\mbox{\scriptsize A\normalsize}_{\mbox{\bf\tiny I\normalsize\normalfont}}}$, 
and a general ansatz for the potential term 
$\mbox{\sffamily U\normalfont}^{\mbox{\scriptsize A\normalsize}_{\mbox{\bf\tiny I\normalsize\normalfont}}}$. 
We constructed these using both the inverse $3$-metric $h^{ab}$ 
and the
antisymmetric tensor density $\epsilon^{abc}$. 

We note that no kinetic cross-term $\dot{h}_{ab}\dot{A}_{{\mbox{\bf\scriptsize I\normalsize\normalfont}}c}$ 
is possible within this ansatz.  This is because the only way to contract 3 spatial indices is 
to use $\epsilon^{abc}$, and $\dot{h}_{ab}$ is symmetric. Then 
$\mbox{\sffamily T\normalfont}^{\mbox{\scriptsize A\normalsize}_{\mbox{\bf\tiny I\normalsize\normalfont}}}$ is 
\be
\mbox{unambiguously }
\mbox{\hspace{1.2in}}
\mbox{\sffamily T\normalfont}^{\mbox{\scriptsize A\normalsize}_{\mbox{\bf\scriptsize I\normalsize\normalfont}}} =
P_{\mbox{\bf\scriptsize IJ\normalsize\normalfont}}h^{ad}( \dot{A}_a^{\mbox{\bf\scriptsize I\normalsize\normalfont}} 
- \pounds_{\xi}A_a^{\mbox{\bf\scriptsize I\normalsize\normalfont}})( \dot{A}^{\mbox{\bf\scriptsize J\normalsize\normalfont}}_d 
- \pounds_{\xi}A^{\mbox{\bf\scriptsize J\normalsize\normalfont}}_d) 
\mbox{\hspace{1.6in}}
\ee
for 
$P_{\mbox{\bf\scriptsize IJ\normalsize\normalfont}}$ without loss of generality a symmetric 
constant matrix. We will assume that $P_{\mbox{\bf\scriptsize IJ\normalsize\normalfont}}$ is 
positive-definite so that the local flat-space limit quantum theory of 
$A^{\mbox{\bf\scriptsize I\normalsize\normalfont}}_a$ has a well-behaved inner product. 
In this case, we can take $ P_{\mbox{\bf\scriptsize IJ\normalsize\normalfont}} = 
\delta_{\mbox{\bf\scriptsize IJ\normalsize\normalfont}} $ by rescaling the 1-form fields.

We consider the most general 
$\mbox{\sffamily U\normalfont}^{\mbox{\scriptsize A\normalsize}_{\mbox{\bf\scriptsize I\normalsize\normalfont}}}$ 
up to first derivatives of $A_{{\mbox{\bf\scriptsize I\normalsize\normalfont}}a}$, and up to four spatial index 
contractions. This is equivalent to the necessary na\"{\i}ve power-counting requirement for the 
renormalizability of any emergent 4-d quantum field theory for $A_{{\mbox{\bf\scriptsize I\normalsize\normalfont}}a}$ (see App IV.A).  
Then $\mbox{\sffamily U\normalfont}^{\mbox{\scriptsize A\normalsize}_{\mbox{\bf\scriptsize I\normalsize\normalfont}}}$ has the form 
$$
\mbox{\sffamily U\normalfont}^{\mbox{\scriptsize A\normalsize}_{\mbox{\bf\scriptsize I\normalsize\normalfont}}} = 
O_{\mbox{\bf\scriptsize IK\normalsize\normalfont}}C^{abcd}
D_bA^{\mbox{\bf\scriptsize I\normalsize\normalfont}}_{a}D_dA^{\mbox{\bf\scriptsize K\normalsize\normalfont}}_{c} 
+ {B^{\mbox{\bf\scriptsize I\normalsize\normalfont}}}_{\mbox{\bf\scriptsize JK\normalsize\normalfont}}
\bar{C}^{abcd}D_bA_{{\mbox{\bf\scriptsize I\normalsize\normalfont}}a}A^{\mbox{\bf\scriptsize J\normalsize\normalfont}}_c A^{\mbox{\bf\scriptsize K\normalsize\normalfont}}_d 
+ I_{\mbox{\bf\scriptsize JKLM\normalsize\normalfont}}\bar{\bar{C}}^{abcd}
A^{\mbox{\bf\scriptsize J\normalsize\normalfont}}_aA^{\mbox{\bf\scriptsize K\normalsize\normalfont}}_bA^{\mbox{\bf\scriptsize L\normalsize\normalfont}}_cA^{\mbox{\bf\scriptsize M\normalsize\normalfont}}_d 
$$
\be
+ \frac{1}{\sqrt{h}}\epsilon^{abc}(Z_{\mbox{\bf\scriptsize IK\normalsize\normalfont}}
D_bA^{\mbox{\bf\scriptsize I\normalsize\normalfont}}_{a}A^{\mbox{\bf\scriptsize K\normalsize\normalfont}}_c + 
E_{\mbox{\bf\scriptsize IJK\normalsize\normalfont}}A^{\mbox{\bf\scriptsize I\normalsize\normalfont}}_a
A^{\mbox{\bf\scriptsize J\normalsize\normalfont}}_bA^{\mbox{\bf\scriptsize K\normalsize\normalfont}}_c) 
+ F_{\mbox{\bf\scriptsize I\normalsize\normalfont}}h^{ab}D_bA^{\mbox{\bf\scriptsize I\normalsize\normalfont}}_{a} 
+ M_{\mbox{\bf\scriptsize IK\normalsize\normalfont}}h^{ab}A^{\mbox{\bf\scriptsize I\normalsize\normalfont}}_aA^{\mbox{\bf\scriptsize K\normalsize\normalfont}}_b
\ee 
where $C$, $\bar{C}$ and $\bar{\bar{C}}$ are general ultralocal supermetrics, each 
with distinct coefficients. 
$O_{\mbox{\bf\scriptsize IK\normalsize\normalfont}}$, 
${B^{\mbox{\bf\scriptsize I\normalsize\normalfont}}}_{\mbox{\bf\scriptsize JK\normalsize\normalfont}}$, 
$I_{\mbox{\bf\scriptsize JKLM\normalsize\normalfont}}$, 
$Z_{\mbox{\bf\scriptsize IK\normalsize\normalfont}}$, 
$E_{\mbox{\bf\scriptsize IJK\normalsize\normalfont}}$, 
$F_{\mbox{\bf\scriptsize I\normalsize\normalfont}}$ and 
$M_{\mbox{\bf\scriptsize IK\normalsize\normalfont}}$
are constant arbitrary arrays.  W.l.o.g, 
$O_{\mbox{\bf\scriptsize IK\normalsize\normalfont}}$ and 
$M_{\mbox{\bf\scriptsize IK\normalsize\normalfont}}$ are symmetric, and 
$E_{\mbox{\bf\scriptsize IJK\normalsize\normalfont}}$ is totally antisymmetric.

Defining 
$2N \equiv \sqrt {\frac{\mbox{\sffamily\scriptsize T\normalsize\normalfont}^{\mbox{\tiny g\normalsize}} 
+ \mbox{\sffamily\scriptsize T\normalsize\normalfont}^{\mbox{\tiny A\normalsize}_{\mbox{\bf\tiny I\normalsize\normalfont}}}}
{R + \mbox{\sffamily\scriptsize U\normalsize\normalfont}^{\mbox{\tiny A\normalsize}_{\mbox{\bf\tiny I\normalsize\normalfont}}}}}$, 
the conjugate momenta are given by ($\ref{GRmom}$) and 
\be 
\pi_{\mbox{\bf\scriptsize I\normalsize\normalfont}}^i 
\equiv \frac{    \pa\mbox{\sffamily{L}\normalfont}    }
{    \partial \dot{A^{\mbox{\bf\scriptsize I\normalsize\normalfont}}_i}    }  
= \frac{\sqrt{h}}{2N} \mbox{\ss}_{\xi}{A_{\mbox{\bf\scriptsize I\normalsize\normalfont}}^i} \mbox{ } . 
\label{YMformmom}
\ee 
The local square root gives as a primary constraint the Hamiltonian constraint 
\be 
^{\mbox{\scriptsize A\normalsize}_{\mbox{\bf\scriptsize I\normalsize\normalfont}}}{\cal H } 
\equiv
\sqrt{h}(R + \mbox{\sffamily U\normalfont}_{\mbox{\scriptsize A\normalsize}_{\mbox{\bf\scriptsize I\normalsize\normalfont}}}) 
- \frac{1}{\sqrt{h}} 
\left(
p \circ p - \frac{1}{2}p^2 + \pi^{\mbox{\bf\scriptsize I\normalsize\normalfont}}_i\pi_{\mbox{\bf\scriptsize I\normalsize\normalfont}}^i
\right)
   = 0 \mbox{ } . 
\ee
$\xi^i$-variation gives as a secondary constraint the momentum constraint 
\be 
^{\mbox{\scriptsize A\normalsize}_{\mbox{\bf\scriptsize I\normalsize\normalfont}}}{\cal H}_i 
\equiv -2D_j{p^{j}}_{i} 
+ \pi^{{\mbox{\bf\scriptsize I\normalsize\normalfont}}c}
(D_i{A_{{\mbox{\bf\scriptsize I\normalsize\normalfont}}c}} 
- D_c{{A_{\mbox{\bf\scriptsize I\normalsize\normalfont}}}}_i) 
- D_c{\pi_{\mbox{\bf\scriptsize I\normalsize\normalfont}}^c}
A^{{\mbox{\bf\scriptsize I\normalsize\normalfont}}}_i = 0 \mbox{ } .   
\label{eagleeye}
\ee

By the theorem, the propagation of the Hamiltonian constraint is 
$$
^{A_{{\mbox{\bf\scriptsize I\normalsize\normalfont}}}}\dot{{\cal H}} = 
- \frac{    D^i(N^2 {}^{A_{{\mbox{\bf\tiny I\normalsize\normalfont}}}} {\cal H}_i)    }{    N    }  
+ \frac{   Np {}^{A_{\mbox{\bf\tiny I\normalsize\normalfont}}} {\cal H}   }{   2   } 
+ \pounds_{\xi} {}^{A_{\mbox{\bf\scriptsize I\normalsize\normalfont}}} {\cal H} \\
$$
$$
+ \frac{1}{N}
\left\{
(4C_1O^{\mbox{\bf\scriptsize IK\normalsize\normalfont}} 
+ \delta^{\mbox{\bf\scriptsize IK\normalsize\normalfont}})
D_b(N^2\pi_{\mbox{\bf\scriptsize I\normalsize\normalfont}}^aD^b{A_{{\mbox{\bf\scriptsize K\normalsize\normalfont}}a}}) 
+ (4C_2O^{\mbox{\bf\scriptsize IK\normalsize\normalfont}} 
- \delta^{\mbox{\bf\scriptsize IK\normalsize\normalfont}})
D_b(N^2\pi_{\mbox{\bf\scriptsize I\normalsize\normalfont}}^aD_a{A_{\mbox{\bf\scriptsize K\normalsize\normalfont}}^b}) 
\right.
$$
$$
\left.
+ 4C_3O^{\mbox{\bf\scriptsize IK\normalsize\normalfont}}
D_a(N^2\pi_{\mbox{\bf\scriptsize I\normalsize\normalfont}}^aD_b{A_{\mbox{\bf\scriptsize K\normalsize\normalfont}}^b})
\right\}    
- \frac{1}{N} D_b(N^2D_a{\pi_{\mbox{\bf\scriptsize I\normalsize\normalfont}}^a}A^{{\mbox{\bf\scriptsize I\normalsize\normalfont}}b}) 
$$
$$
+ \frac{2}{N} \bar{C}^{abcd}{B^{\mbox{\bf\scriptsize I\normalsize\normalfont}}}_{\mbox{\bf\scriptsize JK\normalsize\normalfont}}  
D_b(N^2 \pi_{{\mbox{\bf\scriptsize I\normalsize\normalfont}}a}A^{\mbox{\bf\scriptsize J\normalsize\normalfont}}_cA^{\mbox{\bf\scriptsize K\normalsize\normalfont}}_d) 
+ \frac{2}{N} \epsilon^{abc}Z_{\mbox{\bf\scriptsize IK\normalsize\normalfont}}D_b(N^2\pi^{\mbox{\bf\scriptsize I\normalsize\normalfont}}_aA^{\mbox{\bf\scriptsize K\normalsize\normalfont}}_c) 
$$
$$
+ \frac{2}{N} F^{\mbox{\bf\scriptsize I\normalsize\normalfont}}D^i(N^2\pi_{{\mbox{\bf\scriptsize I\normalsize\normalfont}}i}) 
- \frac{1}{N} O^{\mbox{\bf\scriptsize IK\normalsize\normalfont}}D_a
\left\{
N^2 (p_{ij} - \frac{p}{2} h_{ij}) D_dA_{{\mbox{\bf\scriptsize K\normalsize\normalfont}}b}
(2A_{\mbox{\bf\scriptsize I\normalsize\normalfont}}^iC^{ajbd} - A_{\mbox{\bf\scriptsize I\normalsize\normalfont}}^aC^{ijbd}) 
\right\}      
$$
$$
- \frac{1}{N} {B^{\mbox{\bf\scriptsize I\normalsize\normalfont}}}_{\mbox{\bf\scriptsize JK\normalsize\normalfont}}D_a 
\left\{ 
N^2 (p_{ij} - \frac{p}{2} h_{ij} ) A^{\mbox{\bf\scriptsize J\normalsize\normalfont}}_b 
A^{\mbox{\bf\scriptsize K\normalsize\normalfont}}_d 
( 2A_{\mbox{\bf\scriptsize I\normalsize\normalfont}}^i\bar{C}^{ajbd} - A_{\mbox{\bf\scriptsize I\normalsize\normalfont}}^a\bar{C}^{ijbd}) 
\right\}       
$$
\be
- \frac{1}{N} F^{\mbox{\bf\scriptsize I\normalsize\normalfont}} D_a
\left\{
 N^2 (p_{ij} - \frac{p}{2} h_{ij} )(2A_{\mbox{\bf\scriptsize I\normalsize\normalfont}}^ih^{aj} - A_{\mbox{\bf\scriptsize I\normalsize\normalfont}}^ah^{ij}) 
\right\} 
\mbox{ } .
\label{evolham5}
\ee 
We demand that $^{A_{\mbox{\bf\scriptsize I\normalsize\normalfont}}}\dot{{\cal H}}$ vanishes 
weakly.  Supposing that this does not automatically vanish, then we would require new constraints. 
However, we have at most 2 + 3\bf K \normalfont d.o.f's, so if we had 3\bf K \normalfont or 
more new constraints, the 1-form field theory would be trivial. 
Furthermore, all constraints must be independent of $N$. Thus, terms in $\pa^aN$ must be of 
the form $(\pa^aNV_{{\mbox{\bf\scriptsize I\normalsize\normalfont}}a})S^{\mbox{\bf\scriptsize I\normalsize\normalfont}}$ 
for the theory to be nontrivial [and we cannot have more than 3\bf K \normalfont independent (spatial) 
scalar constraint factors $S^{\mbox{\bf\scriptsize I\normalsize\normalfont}}$ in total]. Most of these scalars will vanish strongly, which 
means they will fix coefficients in the potential ansatz. Finally, (\ref{evolham5}) is such that 
all the non-automatically vanishing terms in $N$ are partnered by terms in $\pa^aN$.  So the 
above big restriction on the terms in $\pa^aN$ affects all the terms.

1) The first, second, third, sixth and seventh extra terms have no nontrivial scalar factors, 
thus forcing  
$O^{\mbox{\bf\scriptsize IK\normalsize\normalfont}}  
= \delta^{\mbox{\bf\scriptsize IK\normalsize\normalfont}}$, 
$C_1 = -C_2 = -\frac{1}{4}$, $C_3 = 0$, 
$Z_{\mbox{\bf\scriptsize IK\normalsize\normalfont}} = 0$ and
$F_{\mbox{\bf\scriptsize I\normalsize\normalfont}} = 0$ \mbox{ } .  

2) This automatically implies that the eighth and tenth terms also vanish. The conditions on 
the $C$'s correspond to the 1-form fields obeying the null cone dictated by gravity.  The only 
nontrivial possibilities for the vanishing of the ninth term are 
$\bar{C}_3 = 0$ and either 
$B_{{\mbox{\bf\scriptsize I\normalsize\normalfont}}({\mbox{\bf\scriptsize JK\normalsize\normalfont}})} = 0$ 
or $\bar{C_1} = -\bar{C_2} \equiv -\frac{\mbox{\sffamily\scriptsize g\normalsize\normalfont}}{4}$, say. In fact,
these are equivalent, by the following lemma.
\be
\mbox{{\bf Lemma } Given that $\bar{C}_3 = 0$, $B_{{\mbox{\bf\scriptsize I\normalsize\normalfont}}({\mbox{\bf\scriptsize JK\normalsize\normalfont}})} = 0 \Leftrightarrow \bar{C_1} = -\bar{C_2}$.} 
\mbox{ } \mbox{ } \mbox{ } \mbox{ } \mbox{ } \mbox{ } \mbox{ } \mbox{ } \mbox{ }
\mbox{ } \mbox{ } \mbox{ } \mbox{ } \mbox{ } \mbox{ } \mbox{ } \mbox{ } \mbox{ } 
\mbox{ } \mbox{ } \mbox{ } \mbox{ } \mbox{ } \mbox{ } \mbox{ } \mbox{ } \mbox{ } 
\mbox{ } \mbox{ } \mbox{ } \mbox{ } \mbox{ } \mbox{ } 
\mbox{\hspace{2in}}
\ee
{\bf Proof}: Since $\bar{C}_3 = 0$, the potential term in question has a factor of 
\be
\Lambda_{\mbox{\bf\scriptsize I\normalsize\normalfont}}^{ab}(\bar{C_1}, \bar{C_2}) = B_{\mbox{\bf\scriptsize IJK\normalsize\normalfont}}A^{\mbox{\bf\scriptsize J\normalsize\normalfont}}_cA^{\mbox{\bf\scriptsize K\normalsize\normalfont}}_d
(\bar{C_1}h^{ac}h^{bd} + \bar{C_2}h^{ad}h^{bc}) \mbox{ } . 
\label{Lambdadef}
\ee

\noindent$(\Leftarrow)$ The last factor in $\Lambda_{\mbox{\bf\scriptsize I\normalsize\normalfont}}^{ab}(\bar{C_1} , -\bar{C_1})$ 
is manifestly antisymmetric in $c$ and $d$, but 
$B_{{\mbox{\bf\scriptsize I\normalsize\normalfont}}({\mbox{\bf\scriptsize JK\normalsize\normalfont}})}
A^{\mbox{\bf\scriptsize J\normalsize\normalfont}}_cA^{\mbox{\bf\scriptsize K\normalsize\normalfont}}_d$ 
is symmetric in $c$ and $d$. So one can take 
$B_{{\mbox{\bf\scriptsize I\normalsize\normalfont}}({\mbox{\bf\scriptsize JK\normalsize\normalfont}})} = 0$ in this potential term.

\noindent$(\Rightarrow)$ $\Lambda_{\mbox{\bf\scriptsize I\normalsize\normalfont}}^{ab}
(\bar{C_1}, \bar{C_2}) = - \Lambda_{\mbox{\bf\scriptsize I\normalsize\normalfont}}^{ab}
(\bar{C_2}, \bar{C_1})$ by the antisymmetry in $B_{\mbox{\bf\scriptsize IJK\normalsize\normalfont}}$ and exchanging dummy 
internal and spatial indices in (\ref{Lambdadef}).  Hence 
$B_{\mbox{\bf\scriptsize IJK\normalsize\normalfont}}
A^{\mbox{\bf\scriptsize J\normalsize\normalfont}}_cA^{\mbox{\bf\scriptsize K\normalsize\normalfont}}_d
(\bar{C}_1 + \bar{C}_2)(h^{ac}h^{bd} + h^{ad}h^{bc}) = 0$.  Then the only nontrivial possibility 
is $\bar{C}_1 + \bar{C}_2 = 0$. \mbox{ } $\Box$

There is in fact another possibility for the vanishing of the ninth term,
\be
{\cal K}^{\mbox{\bf\scriptsize JK\normalsize\normalfont}} = 
A^{{\mbox{\bf\scriptsize J\normalsize\normalfont}}j}A^{{\mbox{\bf\scriptsize K\normalsize\normalfont}}i}
\left(
p_{ij} - \frac{p}{2}h_{ij}
\right) 
\approx 0
\ee
but we will now dismiss it.  Because the last factor is symmetric in $ij$, ${\cal K}^{\mbox{\bf\scriptsize JK\normalsize\normalfont}}$ is symmetric in \bf JK\normalfont, 
and so uses up  $\frac{\mbox{\bf K\normalfont}(\mbox{\bf K\normalfont + 1})}{2}$ d.o.f's, so the theory is trivial for 
\bf K \normalfont $\geq 5$.  Otherwise, we require this new constraint to propagate
$$
\frac{   \partial  {\cal K}^{\mbox{\bf\tiny JK\normalsize\normalfont}}   }    {   \partial\lambda   } 
=  - \frac{  Np  }{  2  }    {\cal K} ^{\mbox{\bf\scriptsize JK\normalsize\normalfont}}  + 
\pounds_{\xi}{\cal  K}^{\mbox{\bf\scriptsize JK\normalsize\normalfont}}  
+ \frac{  2N  }{  \sqrt{h}  }(   \pi_i^{\mbox{\bf\scriptsize J\normalsize\normalfont}}A_i^{\mbox{\bf\scriptsize K\normalsize\normalfont}} 
+ \pi_j^{\mbox{\bf\scriptsize K\normalsize\normalfont}}A_i^{\mbox{\bf\scriptsize K\normalsize\normalfont}}   )
\left(
p^{ij} - \frac{  p  }{  2  }h^{ij}
\right)  
$$
\be
+  A^{\mbox{\bf\scriptsize J\normalsize\normalfont}}_iA^{\mbox{\bf\scriptsize K\normalsize\normalfont}}_j
\left(   
\dot{p}^{ij} - \frac{1}{2}\dot{p}^{ij}p^{ab}h_{ab}
\right)  
- \frac{  N  }{  \sqrt{h}  } A^{\mbox{\bf\scriptsize J\normalsize\normalfont}}_jA^{{\mbox{\bf\scriptsize K\normalsize\normalfont}}j}p^{ab}
\left(  
{p}_{ab} - \frac{  p  }{  2  }h_{ab}
\right)  
\mbox{ } .  
\ee
But $\dot{p}^{ij}$ alone contains $\pa^iN$ terms, so there must be at least 3 more 
constraints, and we discover yet 3 more below, so this theory is trivial.  

3) So we are finally left with \bf K \normalfont new scalar constraint factors from 
the fourth and fifth terms, 
\be
{\cal G}_{\mbox{\bf\scriptsize J\normalsize\normalfont}} 
\equiv D_a{\pi_{\mbox{\bf\scriptsize J\normalsize\normalfont}}^a} 
- \mbox{\sffamily g\normalfont} B_{{\mbox{\bf\scriptsize IJK\normalsize\normalfont}}}
\pi^{\mbox{\bf\scriptsize I\normalsize\normalfont}}_aA^{{\mbox{\bf\scriptsize K\normalsize\normalfont}}a} \approx 0 \mbox{ } . 
\label{YaMiGau}
\ee

\mbox{ }

Next, we examine the evolution of this internal-index vector of new constraints.  
We do now require the $A_i^{\mbox{\bf\scriptsize I\normalsize\normalfont}}$ ELE, but we can 
simplify it by use of the restrictions imposed by $\dot{{\cal H}} \approx 0$ and with some 
regrouping of terms by dummy index exchanges, to obtain\fn{The corresponding $\xi^i$ part of 
the calculation uses 
$$
D_i(\pounds_{\xi}\pi^i_{\mbox{\bf\tiny I\normalsize\normalfont}}) 
- \pounds_\xi(D_i\pi^i_{\mbox{\bf\tiny I\normalsize\normalfont}}) 
= \xi_i(D_jD^i{\pi}^i_{\mbox{\bf\tiny I\normalsize\normalfont}} 
- D^iD_j{\pi^j}_{\mbox{\bf\tiny I\normalsize\normalfont}}) + 
\pi^i_{\mbox{\bf\tiny I\normalsize\normalfont}}(      D_i{D^j{\xi_j}} - D^j{D_i\xi_{j}}      ) 
= \xi_i{R^{ij}}_{jd}\pi^d_{\mbox{\bf\tiny I\normalsize\normalfont}} 
+ \pi^i_{\mbox{\bf\tiny I\normalsize\normalfont}}{{R_j}^{ij}}_d\xi^d = 0 \mbox{ } .  
$$}
$$
\frac{\partial \pi^{{\mbox{\bf\tiny J\normalsize\normalfont}}i}} {\partial \lambda} = 
\frac{\delta\mbox{\sffamily L\normalfont}}
{\delta A_{{\mbox{\bf\tiny J\normalsize\normalfont}}i}} = 
- 2\sqrt{h}O^{{\mbox{\bf\scriptsize JK\normalsize\normalfont}}}
D^b\{C_1(ND_b{A_{\mbox{\bf\scriptsize K\normalsize\normalfont}}^i}) 
+ C_2D^b(ND^i{A_{{\mbox{\bf\scriptsize K\normalsize\normalfont}}b}})\} 
$$
$$
+ \sqrt{h}\{ ND_bA^{\mbox{\bf\scriptsize I\normalsize\normalfont}}_{a}A_{Mc}(\bar{C}^{abci}
{B_{\mbox{\bf\scriptsize I\normalsize\normalfont}}}^{{\mbox{\bf\scriptsize MJ\normalsize\normalfont}}} 
+ \bar{C}^{abic}{B_{\mbox{\bf\scriptsize I\normalsize\normalfont}}}^{{\mbox{\bf\scriptsize JM\normalsize\normalfont}}}) -
D_b(NA^{\mbox{\bf\scriptsize M\normalsize\normalfont}}_cA^{\mbox{\bf\scriptsize K\normalsize\normalfont}}_d)
\bar{C}^{ibcd}{B^{\mbox{\bf\scriptsize J\normalsize\normalfont}}}_{{\mbox{\bf\scriptsize MK\normalsize\normalfont}}}\} 
$$
$$
+\sqrt{h}N(\bar{\bar{C}}^{ibcd}I^{\mbox{\bf\scriptsize JKLM\normalsize\normalfont}} +
\bar{\bar{C}}^{bicd}I^{\mbox{\bf\scriptsize KJLM\normalsize\normalfont}} 
+ \bar{\bar{C}}^{bcid}I^{\mbox{\bf\scriptsize KLJM\normalsize\normalfont}} +
\bar{\bar{C}}^{bcdi}I^{\mbox{\bf\scriptsize KLMJ\normalsize\normalfont}})
A_{{\mbox{\bf\scriptsize K\normalsize\normalfont}}b}A_{{\mbox{\bf\scriptsize L\normalsize\normalfont}}c}
A_{{\mbox{\bf\scriptsize M\normalsize\normalfont}}d} 
$$
\be
 +
3\epsilon^{ibc}E^{\mbox{\bf\scriptsize JNK\normalsize\normalfont}}
NA_{{\mbox{\bf\scriptsize K\normalsize\normalfont}}c}A_{{\mbox{\bf\scriptsize N\normalsize\normalfont}}b}  + 
2\sqrt{h}NM^{\mbox{\bf\scriptsize JK\normalsize\normalfont}}A_{\mbox{\bf\scriptsize K\normalsize\normalfont}}^i 
+ \pounds_{\xi}\pi^{{\mbox{\bf\scriptsize J\normalsize\normalfont}}i}
\mbox{}. 
\label{piELeq2}
\ee

Then, using the fact that $\pi^i_{\mbox{\bf\scriptsize I\normalsize\normalfont}}$ is a 
(1, 0)-density,  the propagation of ${\cal G}_{\mbox{\bf\scriptsize J\normalsize\normalfont}}$ 
gives
$$
\mbox{$\dot{{\cal G}_{\mbox{\bf\scriptsize I\normalsize\normalfont}}}$} = 
\pounds_{\xi}{\cal G}_{\mbox{\bf\scriptsize I\normalsize\normalfont}} 
-\frac{2N}{\sqrt{h}} \mbox{\sffamily g\normalfont} \pi^{\mbox{\bf\scriptsize K\normalsize\normalfont}}_i
\pi^{{\mbox{\bf\scriptsize I\normalsize\normalfont}}i}B_{{\mbox{\bf\scriptsize IJK\normalsize\normalfont}}} 
+ \frac{\sqrt{h}}{2}\mbox{\sffamily g\normalfont} D^bA^{{\mbox{\bf\scriptsize K\normalsize\normalfont}}i}
(D_bA_{{\mbox{\bf\scriptsize I\normalsize\normalfont}}i} 
- D_iA_{{\mbox{\bf\scriptsize I\normalsize\normalfont}}b})
{B^{\mbox{\bf\scriptsize I\normalsize\normalfont}}}_{{\mbox{\bf\scriptsize JK\normalsize\normalfont}}} 
$$
$$
+ \sqrt{h}D_i(NA^{d{\mbox{\bf\scriptsize K\normalsize\normalfont}}}A^{i{\mbox{\bf\scriptsize L\normalsize\normalfont}}}
A^{\mbox{\bf\scriptsize M\normalsize\normalfont}}_d)
\left\{ 
\bar{\bar{C_1}}(I_{\mbox{\bf\scriptsize JKLM\normalsize\normalfont}} + I_{\mbox{\bf\scriptsize KJML\normalsize\normalfont}} 
+ I_{\mbox{\bf\scriptsize LMJK\normalsize\normalfont}} + I_{\mbox{\bf\scriptsize MLKJ\normalsize\normalfont}})
\right.
$$
$$
\left.
+ \bar{\bar{C_2}}(I_{\mbox{\bf\scriptsize JKML\normalsize\normalfont}} + I_{\mbox{\bf\scriptsize KJLM\normalsize\normalfont}} 
+ I_{\mbox{\bf\scriptsize MLJK\normalsize\normalfont}} + I_{\mbox{\bf\scriptsize LMKJ\normalsize\normalfont}}) \\
+ 2\bar{\bar{C_3}}(I_{({\mbox{\bf\scriptsize JL\normalsize\normalfont}}){\mbox{\bf\scriptsize KM\normalsize\normalfont}}} 
+ I_{{\mbox{\bf\scriptsize KM\normalsize\normalfont}}({\mbox{\bf\scriptsize JL\normalsize\normalfont}})})
-  \frac{1}{2} \mbox{\sffamily g\normalfont}^2 
{B^{\mbox{\bf\scriptsize I\normalsize\normalfont}}}_{{\mbox{\bf\scriptsize JK\normalsize\normalfont}}}
B_{{\mbox{\bf\scriptsize IML\normalsize\normalfont}}}  
\right\}
$$
$$
- \frac{\sqrt{h}}{2}  \mbox{\sffamily g\normalfont}^2  N A^{{\mbox{\bf\scriptsize L\normalsize\normalfont}}i} 
A^{{\mbox{\bf\scriptsize M\normalsize\normalfont}}b} D_iA^{\mbox{\bf\scriptsize K\normalsize\normalfont}}_{b} 
({B^{\mbox{\bf\scriptsize I\normalsize\normalfont}}}_{{\mbox{\bf\scriptsize JK\normalsize\normalfont}}}
B_{\mbox{\bf\scriptsize ILM\normalsize\normalfont}} 
+ {B^{\mbox{\bf\scriptsize I\normalsize\normalfont}}}_{{\mbox{\bf\scriptsize JM\normalsize\normalfont}}}
B_{\mbox{\bf\scriptsize IKL\normalsize\normalfont}} 
+{B^{\mbox{\bf\scriptsize I\normalsize\normalfont}}}_{{\mbox{\bf\scriptsize JL\normalsize\normalfont}}}
B_{\mbox{\bf\scriptsize IMK\normalsize\normalfont}}) 
$$
$$
- {B^{\mbox{\bf\scriptsize Q\normalsize\normalfont}}}_{{\mbox{\bf\scriptsize JP\normalsize\normalfont}}} 
\mbox{\sffamily g\normalfont} A^{i{\mbox{\bf\scriptsize P\normalsize\normalfont}}}
A^{{\mbox{\bf\scriptsize K\normalsize\normalfont}}d}A^{\mbox{\bf\scriptsize L\normalsize\normalfont}}_i
A^{\mbox{\bf\scriptsize M\normalsize\normalfont}}_d
\left\{ 
\bar{\bar{C_1}}(I_{\mbox{\bf\scriptsize QKLM\normalsize\normalfont}} + I_{\mbox{\bf\scriptsize KLMQ\normalsize\normalfont}} 
+ I_{\mbox{\bf\scriptsize LKQM\normalsize\normalfont}} + I_{\mbox{\bf\scriptsize LKMQ\normalsize\normalfont}})  
\right.
$$
$$
\left.
+ \bar{\bar{C_2}}(I_{\mbox{\bf\scriptsize KQLM\normalsize\normalfont}}  + I_{\mbox{\bf\scriptsize KLQM\normalsize\normalfont}} 
+ I_{\mbox{\bf\scriptsize QKML\normalsize\normalfont}} + I_{\mbox{\bf\scriptsize KQML\normalsize\normalfont}}) \\
+ 2\bar{\bar{C_3}}(I_{{\mbox{\bf\scriptsize KM\normalsize\normalfont}}({\mbox{\bf\scriptsize QL\normalsize\normalfont}})} 
+ I_{({\mbox{\bf\scriptsize QL\normalsize\normalfont}}){\mbox{\bf\scriptsize KM\normalsize\normalfont}}} )  
\right\}                                                                                      
$$
$$
+ 3\epsilon^{ibc}
\left\{
E_{{\mbox{\bf\scriptsize JNK\normalsize\normalfont}}}D_i(NA^{\mbox{\bf\scriptsize K\normalsize\normalfont}}_cA^{\mbox{\bf\scriptsize N\normalsize\normalfont}}_b) 
+ \mbox{\sffamily g\normalfont} E_{\mbox{\bf\scriptsize QNK\normalsize\normalfont}} N 
A^{\mbox{\bf\scriptsize K\normalsize\normalfont}}_c A^{\mbox{\bf\scriptsize N\normalsize\normalfont}}_b 
A^{\mbox{\bf\scriptsize P\normalsize\normalfont}}_i 
{B^{\mbox{\bf\scriptsize Q\normalsize\normalfont}}}_{\mbox{\bf\scriptsize JP\normalsize\normalfont}} 
\right\}                                                                                       
$$
\be
+ 2\sqrt{h}
\left\{
M_{{\mbox{\bf\scriptsize JK\normalsize\normalfont}}}D_i(NA^{{\mbox{\bf\scriptsize K\normalsize\normalfont}}i}) 
- \mbox{\sffamily g\normalfont} M_{\mbox{\bf\scriptsize QK\normalsize\normalfont}} 
{B^{\mbox{\bf\scriptsize Q\normalsize\normalfont}}}_{\mbox{\bf\scriptsize JP\normalsize\normalfont}}
NA^{{\mbox{\bf\scriptsize K\normalsize\normalfont}}i}A^{\mbox{\bf\scriptsize P\normalsize\normalfont}}_i 
\right\} 
\mbox{ } .
\label{propagnew}
\ee 
In obtaining this result, we have used the following cancellations in addition to those which 
\be
\mbox{occur for the single 1-form case: }
\mbox{\hspace{0.85in}}
D_iD_b(NA_{\mbox{\bf\scriptsize M\normalsize\normalfont}}^iA_{\mbox{\bf\scriptsize K\normalsize\normalfont}}^b)
B^{\mbox{\bf\scriptsize JMK\normalsize\normalfont}} = 0 \mbox{ } ,
\mbox{\hspace{1.2in}}
\ee
\be
\mbox{\hspace{1.8in}}
D_i\{N(D^i{A_{{\mbox{\bf\scriptsize I\normalsize\normalfont}}b}} 
- D_b{A_{\mbox{\bf\scriptsize I\normalsize\normalfont}}^i})
A_{\mbox{\bf\scriptsize K\normalsize\normalfont}}^b\}
B^{{\mbox{\bf\scriptsize I\normalsize\normalfont}}({\mbox{\bf\scriptsize JK\normalsize\normalfont}})} = 0 \mbox{ } .
\ee
These follow from the last two indices of 
${B^{\mbox{\bf\scriptsize I\normalsize\normalfont}}}_{\mbox{\bf\scriptsize JK\normalsize\normalfont}}$ 
being antisymmetric.  

We next demand that (\ref{propagnew}) vanishes weakly.  Again, we will first consider the 
$\pa^aN$ terms.  For the theory to be nontrivial, the third, sixth and seventh 
non-automatically vanishing terms force us to have, without loss of generality, 
$I_{\mbox{\bf\scriptsize JKLM\normalsize\normalfont}} = 
{B^{\mbox{\bf\scriptsize I\normalsize\normalfont}}}_{\mbox{\bf\scriptsize JK\normalsize\normalfont}}
B_{\mbox{\bf\scriptsize ILM\normalsize\normalfont}}$, 
$\bar{\bar{C}}_2 = - \bar{\bar{C}}_1 =
\frac{\mbox{\sffamily g\normalfont}^2}{16}$, $\bar{\bar{C}}_3 = 0$,
$E_{\mbox{\bf\scriptsize JNK\normalsize\normalfont}} = 0$ 
and $M_{\mbox{\bf\scriptsize JK\normalsize\normalfont}} = 0$.  
This last condition means that, within this ansatz, fundamental interacting 1-form fields 
are massless.  We are then left with the first, second, fourth and fifth terms. The fourth 
term forces upon us 
\be
{B^{\mbox{\bf\scriptsize I\normalsize\normalfont}}}_{\mbox{\bf\scriptsize JK\normalsize\normalfont}}
B_{\mbox{\bf\scriptsize ILM\normalsize\normalfont}} + 
{B^{\mbox{\bf\scriptsize I\normalsize\normalfont}}}_{\mbox{\bf\scriptsize JM\normalsize\normalfont}}
B_{{\mbox{\bf\scriptsize IKL\normalsize\normalfont}}} + 
{B^{\mbox{\bf\scriptsize I\normalsize\normalfont}}}_{\mbox{\bf\scriptsize JL\normalsize\normalfont}}
B_{\mbox{\bf\scriptsize IMK\normalsize\normalfont}} = 0 \mbox{ } ,
\label{Jacobi}
\ee
which is the Jacobi identity (\ref{firstJac}), so the 
${B^{\mbox{\bf\scriptsize I\normalsize\normalfont}}}_{\mbox{\bf\scriptsize JK\normalsize\normalfont}}$'s 
are axiomatically the structure constants of some Lie algebra, $\bf {\cal A} $ \normalfont.  
Thus (\ref{YaMiGau}) can be identified with the Yang--Mills--Gauss constraint (\ref{flatYMgauss},\ref{curYMgau}).
Furthermore, the vanishing of the first term forces us to have
$B_{\mbox{\bf\scriptsize IJK\normalsize\normalfont}} = 
B_{[{\mbox{\bf\scriptsize I\normalsize\normalfont}}|{\mbox{\bf\scriptsize J\normalsize\normalfont}}|{\mbox{\bf\scriptsize K\normalsize\normalfont}}]}$, 
which means that the 
$B_{\mbox{\bf\scriptsize IJK\normalsize\normalfont}}$'s 
are totally antisymmetric.  The remaining terms are then automatically zero:
the second term vanishes trivially because of the new antisymmetry, and the fifth term vanishes
by the following argument.  Writing $S^{\mbox{\bf\scriptsize KF\normalsize\normalfont}}$ for the symmetric matrix 
$A^{\mbox{\bf\scriptsize K\normalsize\normalfont}}_iA^{i{\mbox{\bf\scriptsize F\normalsize\normalfont}}}$, 
and rearranging the (\ref{Jacobi}), 
\be
B_{\mbox{\bf\scriptsize IJK\normalsize\normalfont}}
{B^{\mbox{\bf\scriptsize I\normalsize\normalfont}}}_{\mbox{\bf\scriptsize MP\normalsize\normalfont}}
{B^{\mbox{\bf\scriptsize P\normalsize\normalfont}}}_{\mbox{\bf\scriptsize FG\normalsize\normalfont}}
S^{\mbox{\bf\scriptsize KF\normalsize\normalfont}}
S^{\mbox{\bf\scriptsize MG\normalsize\normalfont}} = 
(B_{\mbox{\bf\scriptsize IJM\normalsize\normalfont}} 
{B^{\mbox{\bf\scriptsize I\normalsize\normalfont}}}_{\mbox{\bf\scriptsize KP\normalsize\normalfont}} - 
{B^{\mbox{\bf\scriptsize I\normalsize\normalfont}}}_{\mbox{\bf\scriptsize JP\normalsize\normalfont}} 
B_{\mbox{\bf\scriptsize IKM\normalsize\normalfont}} )
{B^{\mbox{\bf\scriptsize P\normalsize\normalfont}}}_{\mbox{\bf\scriptsize FG\normalsize\normalfont}}
S^{\mbox{\bf\scriptsize KF\normalsize\normalfont}}
S^{\mbox{\bf\scriptsize MG\normalsize\normalfont}} \mbox{ } .  
\ee 
By means of a 
\bf KF \normalfont $\leftrightarrow$ \bf MG \normalfont dummy index change, 3 uses of 
antisymmetry and a ${\mbox{\bf I\normalfont}} \leftrightarrow$ \bf P \normalfont dummy index 
change, the second term on the right-hand side is equal to its negative and 
\be
\mbox{hence is zero:  }
\mbox{\hspace{1.5in}}
{B^{\mbox{\bf\scriptsize I\normalsize\normalfont}}}_{\mbox{\bf\scriptsize JP\normalsize\normalfont}} 
B_{\mbox{\bf\scriptsize IKM\normalsize\normalfont}} 
{B^{\mbox{\bf\scriptsize P\normalsize\normalfont}}}_{\mbox{\bf\scriptsize FG\normalsize\normalfont}}
S^{\mbox{\bf\scriptsize KF\normalsize\normalfont}}
S^{\mbox{\bf\scriptsize MG\normalsize\normalfont}} = 0 
\mbox{ } .
\mbox{\hspace{1.5in}}
\ee
$$
\mbox{Hence we have the symmetry }
B_{\mbox{\bf\scriptsize IJK\normalsize\normalfont}}
{B^{\mbox{\bf\scriptsize I\normalsize\normalfont}}}_{\mbox{\bf\scriptsize MP\normalsize\normalfont}}
{B^{\mbox{\bf\scriptsize P\normalsize\normalfont}}}_{\mbox{\bf\scriptsize FG\normalsize\normalfont}}
S^{\mbox{\bf\scriptsize KF\normalsize\normalfont}}
S^{\mbox{\bf\scriptsize MG\normalsize\normalfont}} = 
B_{{\mbox{\bf\scriptsize IJ\normalsize\normalfont}}({\mbox{\bf\scriptsize K\normalsize\normalfont}}|}
{B^{\mbox{\bf\scriptsize I\normalsize\normalfont}}}_{|{\mbox{\bf\scriptsize M\normalsize\normalfont}}){\mbox{\bf\scriptsize P\normalsize\normalfont}}}
{B^{\mbox{\bf\scriptsize P\normalsize\normalfont}}}_{\mbox{\bf\scriptsize FG\normalsize\normalfont}}
S^{\mbox{\bf\scriptsize KF\normalsize\normalfont}}
S^{\mbox{\bf\scriptsize MG\normalsize\normalfont}}
$$
but then the dummy index change \bf KF \normalfont $\leftrightarrow$ \bf MG \normalfont 
and antisymmetry in the third structure constant along with symmetry between the two 
$S^{\mbox{\bf\scriptsize AB\normalsize\normalfont}}$'s 
means that the right-hand side of the above is zero as required.

So the potential term must be [c.f (\ref{LYMcssplit})]
\be 
U_{\mbox{\scriptsize A\normalsize}_{\mbox{\bf\scriptsize I\normalsize\normalfont}}} = 
- \frac{1}{8}(D_bA^{\mbox{\bf\scriptsize I\normalsize\normalfont}}_{a} 
- D_aA^{\mbox{\bf\scriptsize I\normalsize\normalfont}}_{b} 
+ \mbox{\sffamily g\normalfont}{B^{\mbox{\bf\scriptsize I\normalsize\normalfont}}}_{\mbox{\bf\scriptsize JK\normalsize\normalfont}}
A^{\mbox{\bf\scriptsize J\normalsize\normalfont}}_aA^{\mbox{\bf\scriptsize K\normalsize\normalfont}}_b) 
(D^bA_{\mbox{\bf\scriptsize I\normalsize\normalfont}}^{a} - D^aA_{\mbox{\bf\scriptsize I\normalsize\normalfont}}^{b} 
+ \mbox{\sffamily g\normalfont} B_{{\mbox{\bf\scriptsize ILM\normalsize\normalfont}}}
A^{{\mbox{\bf\scriptsize L\normalsize\normalfont}}a}
A^{{\mbox{\bf\scriptsize M\normalsize\normalfont}}b}) \mbox{ } .  
\ee

\mbox{ }

We now investigate the meaning of totally antisymmetric structure constants $B_{\mbox{\bf\scriptsize IJK\normalsize\normalfont}}$\normalfont. 
This involves expanding on the account in I.1.7.3 of the standard approach 
to Yang--Mills theory in flat spacetime, where one starts with Lorentz and parity invariance, 
which restricts the Lagrangian to be 
$\mbox{\sffamily L\normalfont}^{\mbox{\scriptsize A\normalsize}_{\mbox{\bf\scriptsize I\normalsize\normalfont}}} =
-Q_{\mbox{\bf\scriptsize AB\normalsize\normalfont}}\check{F}^{\mbox{\bf\scriptsize A\normalsize\normalfont}}\circ
\check{F}^{{\mbox{\bf\scriptsize B\normalsize\normalfont}}}$.  Furthermore, one demands invariance  
$\delta \mbox{\sffamily{L}\normalfont} = 0$ 
\be 
\mbox{under the gauge transformation }
\mbox{\hspace{1in}}
\check{A}^{\mbox{\bf\scriptsize I\normalsize\normalfont}}_{A} \longrightarrow 
\check{A}^{\mbox{\bf\scriptsize I\normalsize\normalfont}}_{A} +
i\mbox{\sffamily g\normalfont}_{\mbox{\scriptsize c\normalsize}}
{f^{\mbox{\bf\scriptsize I\normalsize\normalfont}}}_{\mbox{\bf\scriptsize JK\normalsize\normalfont}}
\Lambda^{\mbox{\bf\scriptsize J\normalsize\normalfont}} \check{A}^{\mbox{\bf\scriptsize K\normalsize\normalfont}}_{A} 
\mbox{ } . 
\mbox{\hspace{1in}} 
\ee 
Using the standard result under a set of infinitesimal transformations parameterized by 
$\epsilon^A$ 
$$
\mbox{\cite{Weinberg} , }
\mbox{\hspace{2.0in}}
\delta \check{F}^{\mbox{\bf\scriptsize B\normalsize\normalfont}}_{AB} = 
\epsilon^{\mbox{\bf\scriptsize A\normalsize\normalfont}}{f^{\mbox{\bf\scriptsize B\normalsize\normalfont}}}_{{\mbox{\bf\scriptsize CA\normalsize\normalfont}}}
\check{F}^{\mbox{\bf\scriptsize C\normalsize\normalfont}}_{AB} 
\mbox{ } , 
\mbox{\hspace{2.0in}}
$$
\be
\mbox{$\delta \mbox{\sffamily L\normalfont} = 0$ yields
$Q_{\mbox{\bf\scriptsize AB\normalsize\normalfont}}
{f^{\mbox{\bf\scriptsize B\normalsize\normalfont}}}_{\mbox{\bf\scriptsize CD\normalsize\normalfont}}
\check{F}^{\mbox{\bf\scriptsize A\normalsize\normalfont}}\circ\check{F}^{{\mbox{\bf\scriptsize C\normalsize\normalfont}}} = 0$, 
which is equivalent to }
\mbox{\hspace{0.4in}} 
Q_{({\mbox{\bf\scriptsize A\normalsize\normalfont}}|{\mbox{\bf\scriptsize B\normalsize\normalfont}}}
{f^{\mbox{\bf\scriptsize B\normalsize\normalfont}}}_{{\mbox{\bf\scriptsize C\normalsize\normalfont}}|{\mbox{\bf\scriptsize D\normalsize\normalfont}})} = 0 
\label{a} 
\mbox{\hspace{0.4in}}
\ee 
by symmetry in the internal indices of $\check{F}^{\mbox{\bf\scriptsize A\normalsize\normalfont}}\circ \check{F}^{{\mbox{\bf\scriptsize C\normalsize\normalfont}}}$. 

For $Q_{\mbox{\bf\scriptsize AB\normalsize\normalfont}}$ positive-definite, there is the following theorem 
\cite{GMG, Weinberg} 

\mbox{ }

\noindent\bf{Gell--Mann Glashow Theorem }\normalfont : 
(\ref{a}) and the following two statements are equivalent.
\be 
\exists \mbox{ basis in which } f_{\mbox{\bf\scriptsize ABC\normalsize\normalfont}} = 
f_{[{\mbox{\bf\scriptsize ABC\normalsize\normalfont}}]} 
\label{b}
\ee 
\be 
\mbox{The corresponding algebra is a direct sum of compact simple and
U(1) Lie subalgebras}. 
\label{c} 
\ee 
The Lie algebra terminology used and the proof of the theorem are summarized in App IV.B.  
Also, (\ref{a}) $\Leftrightarrow$ (\ref{b}) $\Leftrightarrow \dot{{\cal G}}_{\mbox{\bf\scriptsize J\normalsize\normalfont}} \approx 0 $ 
in the usual flat spacetime canonical working.

Although we get ${\cal G}_{\mbox{\bf\scriptsize J\normalsize\normalfont}}$ to do the usual work, 
we arrive at the above from a different angle.  
We started with $3$-d 1-form fields on $3$-geometries, obtained 
${\cal H}$ as an identity and demanded that $\dot {\cal H} \approx 0$, which has forced us to 
have the secondary constraints ${\cal G}_{\mbox{\bf\scriptsize J\normalsize\normalfont}}$.  
But once we have the ${\cal G}_{\mbox{\bf\scriptsize J\normalsize\normalfont}}$, we can use 
$\dot{\cal G}_{\mbox{\bf\scriptsize J\normalsize\normalfont}} 
\approx 0  \Leftrightarrow$ (\ref{b}) $\Leftrightarrow$ (\ref{a}), so our scheme allows the 
usual restriction (\ref{c}) on the type of Lie algebra. We can moreover take (\ref{a}) to be
equivalent to the gauge 
\be 
\mbox{invariance of 
$\mbox{\sffamily U\normalfont}_{\mbox{\scriptsize A\normalsize}_{\mbox{\bf\scriptsize I\normalsize\normalfont}}}$ 
under } 
\mbox{\hspace{1.4in}}
A^{\mbox{\bf\scriptsize I\normalsize\normalfont}}_a \longrightarrow 
A^{\mbox{\bf\scriptsize I\normalsize\normalfont}}_a + i\mbox{\sffamily g\normalfont}_{\mbox{\scriptsize c\normalsize}} 
{f^{\mbox{\bf\scriptsize I\normalsize\normalfont}}}_{\mbox{\bf\scriptsize JK\normalsize\normalfont}}
\Lambda^{\mbox{\bf\scriptsize J\normalsize\normalfont}} A^{\mbox{\bf\scriptsize K\normalsize\normalfont}}_a 
\mbox{\hspace{1.4in}} \mbox{ } .
\ee
Thus, if we introduce {\bf K} auxiliary variables $\Phi^{\mbox{\bf\scriptsize K\normalsize\normalfont}}$ 
such that variation with respect to them encodes ${\cal G}_{\mbox{\bf\scriptsize K\normalsize\normalfont}}$ 
then we should do so according to G-BM.  In our emergent notation, this uniquely fixes the form of 
$\mbox{\sffamily T\normalsize}^{\mbox{\scriptsize A\normalsize}_{\mbox{\bf\tiny I\normalsize\normalfont}}}
(A_{{\mbox{\bf\scriptsize I\normalsize\normalfont}}a}, \Phi_{\mbox{\bf\scriptsize J\normalsize\normalfont}})$ to be 
[c.f (\ref{LYMcssplit})]
\be 
\mbox{\sffamily T\normalfont}^{\mbox{\scriptsize A\normalsize}_{\mbox{\bf\tiny I\normalsize\normalfont}}} 
= h^{ad}(\dot{A}^{\mbox{\bf\scriptsize I\normalsize\normalfont}}_a - \pounds_{\xi}A^{\mbox{\bf\scriptsize I\normalsize\normalfont}}_a -
\pa_a\Phi^{\mbox{\bf\scriptsize I\normalsize\normalfont}} + \mbox{\sffamily g\normalfont}
{B^{\mbox{\bf\scriptsize I\normalsize\normalfont}}}_{\mbox{\bf\scriptsize JL\normalsize\normalfont}}
A^{\mbox{\bf\scriptsize J\normalsize\normalfont}}_a\Phi^{\mbox{\bf\scriptsize K\normalsize\normalfont}})                    
(\dot{A}_{{\mbox{\bf\scriptsize I\normalsize\normalfont}}d} 
- \pounds_{\xi}A_{{\mbox{\bf\scriptsize I\normalsize\normalfont}}d} 
- \pa_d\Phi_{\mbox{\bf\scriptsize I\normalsize\normalfont}} + \mbox{\sffamily
g\normalfont}B_{\mbox{\bf\scriptsize ILM\normalsize\normalfont}}A^{\mbox{\bf\scriptsize L\normalsize\normalfont}}_d
\Phi^{\mbox{\bf\scriptsize M\normalsize\normalfont}}) \mbox{ } .  
\label{YMKE} 
\ee 
Finally, if we identify $\Phi^{\mbox{\bf\scriptsize K\normalsize\normalfont}}$ 
with $A_0^{\mbox{\bf\scriptsize K\normalsize\normalfont}}$, we arrive at Yang--Mills 
theory for $\mbox{\v{A}}^{\mbox{\bf\scriptsize K\normalsize\normalfont}} = [A_{0}^{\mbox{\bf\scriptsize K\normalsize\normalfont}}, 
\mbox{\b{A}}^{\mbox{\bf\scriptsize K\normalsize\normalfont}}]$, with coupling
constant $\mbox{\sffamily g\normalfont}$ and gauge group {\sc G } 
(corresponding to the structure constants ${B^{\mbox{\bf\scriptsize I\normalsize\normalfont}}}_{\mbox{\bf\scriptsize JK\normalsize\normalfont}}$). 
So this work constitutes a derivation, from $3$-d principles alone, of Yang--Mills theory minimally-coupled to GR.

\mbox{ }

One may argue also that the auxiliary variables should require to be Diff-BM.   
An investigation of this however shows that this is weakly equivalent to not 
applying BM to the auxiliary variables.  A number of interesting points arise from this line of thinking.  

First, consider the toy example of particles for which all velocities involved are Eucl-BM.    
From (\ref{trac}), (\ref{rotac}) the translational and rotational actions on each vector velocity are of the form 
\be
\stackrel{\longrightarrow}{{T}_{\mbox{\scriptsize k\normalsize}}}\mbox{\b{$\dot{v}$}} = \mbox{\b{$\dot{v}$}} 
- \mbox{\b{$\dot{k}$}} 
\mbox{ } ,
\ee
\be 
\stackrel{\longrightarrow}{{R}_{\Omega}}\mbox{\b{$\dot{v}$}} = \mbox{\b{$\dot{v}$}} 
- \mbox{\b{$\Omega$}} \mbox{ \scriptsize $\times$ \normalfont} \mbox{\b{$v$}} 
\mbox{ } .
\ee
\be
\mbox{So the BM for each particle could in fact take the form }
\mbox{\hspace{0.25in}}
\dot{\mbox{\b{q}}}_{(i)} \longrightarrow \dot{\mbox{\b{q}}}_{(i)} 
- \dot{\mbox{\b{a}}} -(\dot{\mbox{\b{b}}} - \dot{\mbox{\b{a}}}) \mbox{ \scriptsize $\times$ \normalfont} \dot{\mbox{\b{q}}}_{(i)} 
\mbox{ } .
\mbox{\hspace{0.25in}}
\ee
Then free-endpoint variations with respect to $\dot{\mbox{a}}$ and $\dot{\mbox{b}}$ yield 
$\mbox{\b{${\cal M}$}} + \mbox{\b{${\cal L}$}} = 0$ 
and $\mbox{\b{${\cal L}$}} = 0$ respectively, so things keep on working out as before.

What is and is not BM has a strong physical impact and is put in by hand.  Each field has 
associated charges that permit it to feel some forces but not others.  

In the current situation, the electromagnetic auxiliary $\dot{\Xi}$ picks up a Diff-BM 
correction: 
\be
\dot{A}_a \longrightarrow \dot{A}_a - \pounds_{\xi}A_a - \pa_a(\dot{\Xi} - \pounds_{\xi}\Xi) \mbox{ } .
\ee
Unlike above, there is an extra complication: this inclusion involves the previously cyclic 
coordinate itself appearing.  This is unproblematic: the EL equations, definition of momentum  
and the free endpoint equation ensure this works out.
Finally $\xi^i$-variation (or free-endpoint 
\be
\mbox{$s^i$-variation for $\dot{s}^i = \xi^i$) yields the weak equivalent }
\mbox{\hspace{0.8in}}
^A{\cal H}_i - {\cal G}\pa_i\Xi= 0
\mbox{\hspace{1in}}
\ee
of the momentum constraint.  Yang--Mills theory works along the same lines.  

\mbox{ }

\noindent{\bf 1.3.3 Discussion}

\mbox{ }

\noindent This work shows that the TSA can accommodate many examples of physical theories. We
can immediately write down a gravity-coupled formalism with the
SU(3) gauge group of the strong force, or with larger groups
such as SU(5) or O(10) of grand unified theories.
However, the work does not restrict attention to a single simple
gauge group, since it also holds for the direct sum (\ref{c}). As provided in App IV.B as a corollary
to the Gell-Mann--Glashow theorem, we 
can then rescale the structure constants of each U(1) or compact
simple subalgebra separately, which is equivalent to each
subalgebra having a distinct coupling constant \cite{Weinberg}.
The simplest example of this is to have ${B^{{\mbox{\bf\scriptsize I\normalsize\normalfont}}}}_{{\mbox{\bf\scriptsize JK\normalsize\normalfont}}}$ = 0, which
corresponds to \bf K \normalfont non-interacting copies of electromagnetism. Other
examples include the gauge bosons of (unbroken) SU(2) ${\times}$ U(1) electroweak theory and
of the SU(3) ${\times}$ SU(2) ${\times}$ U(1) Standard Model.  

We emphasize that our formalism cannot predict how many of these gauge fields there are in nature, 
nor what their gauge groups are.  This is to be expected once one accepts that the TSA 
is `adding on matter' rather than a unification (see also VI and VII).  
   
BF\'{O} showed that a scalar field, a 1-form field, and a 1-form field coupled to scalar 
fields all obey the same null cone as gravity.  In this section we have shown that this is
also true for {\bf K} interacting 1-form fields, thus providing more evidence for a universal null 
cone for the bosonic fields, derived entirely from $3$-d principles.  Investigation 
of the fermionic sector would tell us whether this null cone is indeed universal for all
the known fields of nature.  We also note that our formalism reveals that the universality of 
the null cone and gauge theory have a common origin resulting from the universal application 
of Diff-BM in conjunction with the need to propagate the 
quadratic Hamiltonian constraint.

In this TSA formulation, fundamental 1-form fields are not allowed to have mass. The 
only bosonic fields allowed to have mass are scalar fields. This would make spontaneous 
symmetry breaking a necessity if we are to describe the real world, since the weak 
$W^+$, $W^-$ and $Z$ bosons are massive. 
I then speculated that it might be that (Higgs) scalars alone are allowed to have mass in the TSA.    
However, I showed that strong gravity easily accommodates massive 1-forms (IV.1.4) and the 
spin-$\frac{1}{2}$ fermion mass term causes no trouble (VI.4).  Finally I figured out 
how to include massive 1-forms in another particular TSA formulation (VII).

We finally consider whether classical topological terms 
can be accommodated in the TSA.  Although it is not free of
controversy \cite{Peccei, PS}, t'Hooft's standard explanation of
the low energy QCD spectrum makes use of an extra topological
term 
\be
\frac{\Theta^2\mbox{\sffamily g\normalfont}
_{\mbox{\scriptsize strong \normalsize }}^2}{32{\pi}^2}
\epsilon_{ABCD}\check{F}^{AB}_{\mbox{\bf\scriptsize I\normalsize\normalfont}}\check{F}^{{\mbox{\bf\scriptsize I\normalsize\normalfont}}CD}
\ee 
in the classical Lagrangian \cite{Hooft}, to avoid the {\it U(1) Problem}.  This consists of a peak 
observed to be unsplit but theoretically expected to be split due to a U(1) symmetry.  The topological 
term resolves the U(1) Problem by breaking the U(1) symmetry.  Then the parameter $\Theta$ is 
constrained to be small ($|\Theta|< 10^{-9}$) by the non-observation of the neutron dipole 
moment \cite{nedm}. The inclusion of the topological term corresponds to dropping the parity-invariance of the Lagrangian. 
There is then a new {\it strong CP Problem}: if this symmetry is broken, why is it broken so weakly?  
The topological term is a total derivative \cite{Weinberg}. Nevertheless it makes a
contribution to the action when the QCD vacuum is nontrivial.

We argue also that we need not yet confront the accommodation of topological terms,  
because so far we are only describing a classical, unbroken, fermion-free world. But 
the need for the new term arises from QM considerations when massive quarks are present \cite{Peccei}.
On the long run, it is not clear to me whether spatial compactness and other topological features 
associated with curved spaces can affect the global notions used in particle physics.  

\mbox{ }

\noindent\large{\bf 1.4 Coupling matter to strong gravity}\normalsize
   
\mbox{ }

\noindent I now attempt to couple matter to the strong gravity TSA theories, following the 
procedure of BF\'{O}.  This enables comparison with the GR case, and leads to a better 
understanding of how the TSA works.  In particular 
\noindent 1) strong gravity theories impose an ultralocal structure rather than a Lorentz one, 
and they cause the breakdown of gauge theory, which reinforces BF\'{O}'s notion that the 
null cone and gauge theory have a common origin in GR. 2)
This helps clarify the central role of the differential Gauss laws of electromagnetism 
and Yang--Mills theory in the masslessness of 1-form fields in the TSA.   

\mbox{ }

\noindent{\bf 1.4.1 Scalar fields}

\mbox{ }

\noindent I include first a single scalar field by considering the action
\be
\mbox{\sffamily I\normalfont}^{(\mbox{\scriptsize strong\normalsize}),\varsigma}_{\mbox{\scriptsize BSW\normalsize}}
= \int \textrm{d}\lambda \int \textrm{d}^3x \sqrt{h} \sqrt{\Lambda + \mbox{\sffamily U\normalfont}^{\varsigma}}
\sqrt{\mbox{\sffamily T\normalfont}^{\mbox{\scriptsize g\normalsize}}_{\mbox{\scriptsize W\normalsize}} 
+ \mbox{\sffamily T\normalfont}^{\varsigma}} 
\mbox{ } , 
\ee
with $\mbox{\sffamily T\normalfont}^{\varsigma}$ and $\mbox{\sffamily U\normalfont}^{\varsigma}$ 
as before.  

The conjugate momenta are given by the usual expressions (\ref{GRmom}) and (\ref{scalarmom}), 
where now $ 2N = \sqrt {      \frac{    \mbox{\sffamily\scriptsize T\normalsize\normalfont}^{\mbox{\tiny g\normalsize}}_{\mbox{\tiny W\normalsize}} +
\mbox{\sffamily\scriptsize T\normalsize\normalfont}^{\varsigma}    }
{   \Lambda + \mbox{\sffamily\scriptsize U\normalsize\normalfont}^{\varsigma}    }     }$.  
The local square root gives the primary Hamiltonian constraint 
\be
^{\varsigma}{\cal H} \equiv  \frac{1}{\sqrt{h}}
\left(p \circ p - \frac{X}{2}p^2 + \pi^2\right) 
- \sqrt{h}(\Lambda + \mbox{\sffamily U\normalfont}^{\varsigma}) = 0 
\mbox{ } .
\ee
${\xi}^i$-variation gives the secondary momentum constraint (\ref{SCSC}).  

The constraint $^{\varsigma}{\cal H }$ contains the canonical propagation speed $\sqrt{C}$ of the 
scalar field.  A priori, this is unrestricted.  However, the propagation of the Hamiltonian 
constraint gives
\be
^{\varsigma}\dot{{\cal H}} = 
\frac{Np(3X - 2)^{\varsigma}{\cal H}}{2\sqrt{h}} 
+ \pounds_{\xi}\mbox{ }^{\varsigma}{\cal H} 
+ \frac{C}{N}D^i(N^2\pi D_i\varsigma) 
\mbox{ } .
\label{SGevolham2}
\ee

The theory has just one scalar d.o.f, so if the cofactor of $C$ in the last term were
zero, the scalar dynamics would be trivial. Thus I have derived that $C = 0$: the scalar field 
theory cannot have any spatial derivatives.  So, strong gravity necessarily induces the 
Carroll group structure on scalar fields present, thereby forcing them to obey ultralocal 
field theory.  This is analogous to how 
GR imposes the null cone structure on scalar fields present in \cite{BOF}.

I finally note that these results (and those in the next subsection) are unaffected by whether one
chooses to use the gravitationally `bare' instead of the gravitationally BM formulation.

\mbox{ }

\noindent{\bf 1.4.2 K interacting 1-form fields}

\mbox{ }

\noindent I consider a BSW-type action containing the a priori 
unrestricted 1-form fields $A_a^{\mbox{\bf\scriptsize I\normalsize\normalfont}}$, \bf I \normalfont = \bf 1 \normalfont  
to \bf K\normalfont,
\be
\mbox{\sffamily I\normalfont}^{\mbox{\scriptsize (strong)\normalsize}, 
\mbox{\scriptsize A\normalsize}_{\mbox{\tiny\bf I\normalfont\normalsize}} }_{\mbox{\scriptsize BSW\normalsize}}
= \int \textrm{d}\lambda \int \textrm{d}^3x \sqrt{h} \mbox{{\sffamily L\normalfont}}
(h_{ij}, \dot{h}^{ij}, A_i^{\mbox{\bf\scriptsize I\normalsize\normalfont}}, \dot{A}^i_{\mbox{\bf\scriptsize I\normalsize\normalfont}}, N, \xi^i)
= \int \textrm{d}\lambda \int \textrm{d}^3x \sqrt{h} \sqrt{        \Lambda
\mbox{+}\mbox{\sffamily U\normalfont}^{\mbox{\scriptsize A\normalsize}_{\mbox{\bf\scriptsize I\normalsize\normalfont}}}        }
\sqrt{        \mbox{\sffamily T\normalfont}^{\mbox{\scriptsize g\normalsize}}_{\mbox{\scriptsize W\normalsize}} 
+ \mbox{\sffamily T\normalfont}^{\mbox{\scriptsize A\normalsize}_{\mbox{\bf\scriptsize I\normalsize\normalfont}}}        } \mbox{ } .
\ee
where $\mbox{\sffamily T\normalfont}^{\mbox{\scriptsize A\normalsize}_{\mbox{\bf\scriptsize I\normalsize\normalfont}}}$ 
and $\mbox{\sffamily U\normalfont}^{\mbox{\scriptsize A\normalsize}_{\mbox{\bf\scriptsize I\normalsize\normalfont}}}$ are the ans\"{a}tze used before.

The conjugate momenta are given by (\ref{GRmom}) and  (\ref{YMformmom}) where now 
$2N \equiv \sqrt {      \frac{    
\mbox{\sffamily\scriptsize T\normalsize\normalfont}^{\mbox{\scriptsize g\normalsize}}_{\mbox{\tiny W\normalsize}} +
\mbox{\sffamily\scriptsize T\normalsize\normalfont}^{\mbox{\tiny A\normalsize}_{\mbox{\bf\tiny I\normalsize\normalfont}}}    }{    \Lambda + 
\mbox{\sffamily\scriptsize U\normalsize\normalfont}^{\mbox{\tiny A\normalsize}_{\mbox{\bf\tiny I\normalsize\normalfont}}}    }      }$.  
The local square root gives the primary Hamiltonian constraint,
\be
^{\mbox{\scriptsize A\normalsize}_{\mbox{\bf\tiny I\normalsize\normalfont}}}{\cal H } \equiv  
\frac{1}{\sqrt{h}}\left(p \circ p - \frac{X}{2}p^2 + 
\pi^{\mbox{\bf\scriptsize I\normalsize\normalfont}}_i\pi_{\mbox{\bf\scriptsize I\normalsize\normalfont}}^i\right) 
- \sqrt{h}(\Lambda + \mbox{\sffamily U\normalfont}^{\mbox{\scriptsize A\normalsize}_{\mbox{\bf\tiny I\normalsize\normalfont}}})  = 0 \mbox{ } .
\ee
I get the secondary momentum constraint (\ref{eagleeye}) by $\xi^i$-variation.  

From the strong subcase of the theorem, the evolution of the Hamiltonian constraint is then
$$
\dot{\cal H} = 
\frac{(3X - 2)Np{\cal H}}{2\sqrt{h}}
+ \pounds_{\xi}{\cal H}
$$
$$
- \frac{4}{N}O^{\mbox{\bf\scriptsize IK\normalsize\normalfont}} 
\left\{
C_1D_b(N^2\pi_{\mbox{\bf\scriptsize I\normalsize\normalfont}}^a
D^b{A_{{\mbox{\bf\scriptsize K\normalsize\normalfont}}a}}) + 
C_2D_b(N^2\pi_{\mbox{\bf\scriptsize I\normalsize\normalfont}}^aD_a{A_{\mbox{\bf\scriptsize K\normalsize\normalfont}}^b}) 
+ C_3D_a(N^2\pi_{\mbox{\bf\scriptsize I\normalsize\normalfont}}^aD_b{A_{\mbox{\bf\scriptsize K\normalsize\normalfont}}^b}) 
\right\}                           
$$
$$
- \frac{2}{N} \bar{C}^{abcd}{B^{\mbox{\bf\scriptsize I\normalsize\normalfont}}}_{{\mbox{\bf\scriptsize JK\normalsize\normalfont}}}  
D_b(N^2 \pi_{{\mbox{\bf\scriptsize I\normalsize\normalfont}}a}A^{\mbox{\bf\scriptsize J\normalsize\normalfont}}_c
A^{\mbox{\bf\scriptsize K\normalsize\normalfont}}_d)
+ \frac{2}{N} \epsilon^{abc}Z_{\mbox{\bf\scriptsize IK\normalsize\normalfont}}D_b(N^2
\pi^{\mbox{\bf\scriptsize I\normalsize\normalfont}}_aA^{\mbox{\bf\scriptsize K\normalsize\normalfont}}_c) 
+ \frac{2}{N}F^{\mbox{\bf\scriptsize I\normalsize\normalfont}}D^i(N^2\pi_{{\mbox{\bf\scriptsize I\normalsize\normalfont}}i}) 
$$
$$
+ \frac{1}{N} O^{\mbox{\bf\scriptsize IK\normalsize\normalfont}}D_a
\left\{
N^2 
\left(
p_{ij} - \frac{Xp}{2} h_{ij}
\right) 
D_dA_{{\mbox{\bf\scriptsize K\normalsize\normalfont}}b}
\left(
2A_{\mbox{\bf\scriptsize I\normalsize\normalfont}}^iC^{ajbd} - A_{\mbox{\bf\scriptsize I\normalsize\normalfont}}^aC^{ijbd}
\right) 
\right\}                            
$$
$$
+ \frac{1}{N} {B^{\mbox{\bf\scriptsize I\normalsize\normalfont}}}_{\mbox{\bf\scriptsize JK\normalsize\normalfont}}D_a 
\left\{
N^2 
\left(
p_{ij} - \frac{Xp}{2} h_{ij} 
\right) 
A^{\mbox{\bf\scriptsize J\normalsize\normalfont}}_b 
A^{\mbox{\bf\scriptsize K\normalsize\normalfont}}_d 
\left( 
2A_{\mbox{\bf\scriptsize I\normalsize\normalfont}}^i\bar{C}^{ajbd} 
- A_{\mbox{\bf\scriptsize I\normalsize\normalfont}}^a\bar{C}^{ijbd}
\right) 
\right\}                            
$$
\be
+ \frac{1}{N} F^{\mbox{\bf\scriptsize I\normalsize\normalfont}}D_a 
\left\{ 
N^2 
\left(
p_{ij} - \frac{Xp}{2} h_{ij}
\right)
(2A_{\mbox{\bf\scriptsize I\normalsize\normalfont}}^ih^{aj} 
- A_{\mbox{\bf\scriptsize I\normalsize\normalfont}}^ah^{ij}) 
\right\} 
\mbox{ } .
\label{SGevolSham}
\ee
I demand that $^{\mbox{\scriptsize A\normalsize}_{\mbox{\bf\scriptsize I\normalsize\normalfont}}}\dot{{\cal H}}$ 
vanishes weakly.  The first two terms vanish weakly by the Hamiltonian constraint, leaving us 
with nine extra terms.  
Because there are less than 3{\bf K} 1-form d.o.f's to use up, nontriviality dictates that most 
of these extra terms can 
only vanish strongly, that is by fixing coefficients in the potential ansatz.  Furthermore, I 
notice that all 
contributions to (\ref{SGevolSham}) are terms in $\pa^aN$ or are partnered by such terms.  
Since further constraints 
are independent of $N$, these terms in $\pa^aN$ are of the form 
$(\pa^aNV_{{\mbox{\bf\scriptsize J\normalsize\normalfont}}a})S^{\mbox{\bf\scriptsize J\normalsize\normalfont}}$, 
and nontriviality dictates that it must be the (spatial) scalar factors 
$S^{\mbox{\bf\scriptsize J\normalsize\normalfont}}$ 
that vanish.  I proceed in three steps.   

\noindent $1^{\prime}$) The first, second, third, fifth and sixth non-weakly-vanishing terms have no
nontrivial scalar factors, so we are forced to have $O^{\mbox{\bf\scriptsize IK\normalsize\normalfont}}  =
\delta^{\mbox{\bf\scriptsize IK\normalsize\normalfont}}$, $C_1 = C_2 = C_3 = 0$, 
$D_{\mbox{\bf\scriptsize IK\normalsize\normalfont}} = 0$ and $F_{\mbox{\bf\scriptsize I\normalsize\normalfont}} = 0$. 
The conditions on the $C$'s correspond to the 1-forms obeying the local Carroll structure.

\noindent $2^{\prime}$) This automatically implies that the seventh, eighth and ninth
terms also vanish.

\noindent $3^{\prime}$) The only nontrivial possibility for the vanishing of the fourth term is if
$B_{\mbox{\bf\scriptsize IJK\normalsize\normalfont}} = 0$, in which case the constraint algebra 
has been closed.

It is enlightening to contrast these (primed) steps with their (unprimed) counterparts from the GR case of IV.1.2.  

\noindent 1) is the same as $1^{\prime}$) except that $C_1 = - C_2 = -1/4$, which corresponds
to the 1-form fields obeying the local Lorentz light-cone structure 

\noindent 2) is the same as $2^{\prime}$) except that instead of the automatic vanishing of the eighth 
term, one is forced to take $B_{\mbox{\bf\scriptsize IJK\normalsize\normalfont}} = B_{{\mbox{\bf\scriptsize I\normalsize\normalfont}}
[{\mbox{\bf\scriptsize JK\normalsize\normalfont}}]}$, which is the start of the
imposition of an algebraic structure on the hitherto unknown arrays.

\noindent 3) One is now left with {\bf K} new \sl{nontrivial} \normalfont scalar constraints, which 
happen to form the Yang--Mills Gauss constraint (\ref{YaMiGau}).  
So the algebra is not yet closed, and the GR working is then substantially longer.

\mbox{ }

\noindent{\bf 1.4.3 Discussion of strong gravity matter-coupling results}

\mbox{ }

\noindent The above results help clarify some aspects of the TSA results for GR.
First, notice also how now that a family of supermetrics is allowed 
the matter dynamics is insensitive to a possible change of supermetric,
which is encouraging for the coupling of conformal gravity to matter fields.

Second, we can take further  
the view that local causal structure and gauge theory are manifestations of the same thing.
In the GR case, the universal light-cone and gauge theory come together from the $\dot{R}$ term in
$^{\mbox{\scriptsize A\normalsize}_I}\dot{{\cal H}}$,
whilst the absence of this in strong gravity ensures that the collapse of the null cone to the 
Carrollian line is accompanied by the breakdown of gauge theory: there is neither gauge 
symmetry nor a Gauss law.  In the GR case, the quantum-mechanics-inspired positive-definiteness assumed
of the 1-form kinetic matrix $P_{\mbox{\bf\scriptsize IJ\normalsize\normalfont}}$
then turns out to be necessary in the restriction of the choice of gauge group, so there would 
be a price to pay if one insisted instead on entirely classical assumptions.
In the strong gravity case,
the absence of emergent gauge structure means that there is no such price to pay
for using classical assumptions alone.
Provided that $P_{\mbox{\bf\scriptsize IJ\normalsize\normalfont}}$ is invertible, the outcome of steps 
$1^{\prime}$) to $3^{\prime}$) is unaltered.

Note through what happens above in the absence of the Gauss law that
it is specifically this characteristic of the 1-form theory that kills off its mass terms in the TSA of BF\'{O}, 
rather than some underlying principle for general matter.
This is a useful first insight into the status of mass in the 3-space approach to GR.  
It is also easy to demonstrate that the general derivative-free potential term built out of 1-forms  
persists coupled to strong gravity.    

I emphasize that our result concerning the breakdown of gauge theory is in particular a result 
about GR, although it clearly occurs for all our theories and the theories they approximate.  
In the strong-coupled gravity limit such as in the vicinity of the initial singularity, in this formalism 
dynamical consistency dictates that gauge theory breaks down in GR.   
Gauge interactions become impossible as one approaches such a regime.  
This appears not to be in accord with the view that gauge interactions persist in extreme regimes to form 
part of a unified theory with gravity, such as in string theory.  However, little is known about physics 
in such regimes, so this classical GR intuition might not hold.   If string theory can tame 
such singularities, the circumstances under which gauge theory breaks down according to GR 
might not occur.  However, it could even be that string theory breaks down in such a regime, 
since according to one interpretation, stringy matter could be a phase of some larger theory 
which breaks down in a high-energy phase transition \cite{SGstringyrefs2}.  Also, Carrollian 
regimes might arise in string theory under other circumstances, and exhibit different behaviour 
from the strong-coupled limit of GR coupled to gauge theory, 
as suggested by the recent Born--Infeld study \cite{SGGibbons}.

\mbox{ }

\noindent\Large{\bf 2 Coupling of matter to conformal gravity}\normalsize

\mbox{ }

\noindent It is an important test of the theoretical framework of conformal gravity to see 
whether it is capable of accommodating enough classical field
theories to be a viable description of nature.  Below we begin to show this is the case, in 
parallel with the previous section.  The universal null cone result is also obtained.  

\mbox{ }

\mbox{ }

\noindent\large{\bf 2.1 General theorems}\normalsize

\mbox{ }

\noindent We require a modified version of theorem 1 for use in conformal gravity, and also a 
new theorem relating homogeneity in $\phi$ to the propagation of the new constraint of 
conformal gravity, $p = 0$.  The forms of these below suffice for the construction of a useful range of classical field theories coupled 
to conformal gravity: we will demonstrate that the range of theories covered by these
theorems includes much of known classical bosonic physics coupled to conformal gravity.  
Furthermore, these theories are picked out from more general possibilities by exhaustive 
implementation of Dirac's demand for dynamical consistency.

Let $\Psi_{\mbox{\scriptsize\sffamily A\normalfont\normalsize}}$ be a set of matter fields that we wish to couple to conformal gravity, with 
potential term $\mbox{\sffamily U\normalfont}^{\Psi}$ and kinetic term
$\mbox{\sffamily T\normalfont}^{\Psi}$. We first decompose these as polynomials in the inverse
metric. This is because it is the power of the metric that determines the powers of $V$ that 
must be used to achieve the necessary homogeneity. Let $\Psi^{(n)}$ be the set of fields such
that these polynomials are of no higher degree than $n$. Thus 
\be
\mbox{\sffamily T\normalfont}^{\Psi^{(n)}} = \sum_{(k) = (0)}^{(n)}
\mbox{\sffamily T\normalfont}^{(k)}_{i_1j_1i_2j_2 ... i_kj_k}h^{i_1j_1} ... h^{i_kj_k} 
= \sum_{(k) = (0)}^{(n)}\mbox{\sffamily T\normalfont}_{(k)} \mbox{ } ,
\ee
\be
\mbox{\sffamily U\normalfont}^{\Psi^{(n)}} = \sum_{(k) = (0)}^{(n)}
\mbox{\sffamily U\normalfont}^{(k)}_{i_1j_1i_2j_2 ... i_kj_k}h^{i_1j_1} ... h^{i_kj_k} 
= \sum_{(k) = (0)}^{(n)}\mbox{\sffamily U\normalfont}_{(k)} \mbox{ } .
\ee
Then the following theorem guarantees that $p = 0$ is preserved by
the dynamical evolution.

\mbox{ }

\noindent\bf Theorem C1. \normalfont For matter fields $\Psi^{(n)}$, the conformal gravity plus 
matter action of the form 
\be
\mbox{\sffamily I\normalfont}^{\Psi^{(n)}} = \int \textrm{d}\lambda
\frac {      \int    \textrm{d}^3x\sqrt{h}\phi^4 \sqrt{ \sigma
\left( 
R - \frac{8D^2\phi}{\phi}
\right) 
+ \frac{ \phi^4 }{ V^{\frac{2}{3}} }\sum_{(k =    0)}^{(n)}
  \frac{  U_{(k)} V^{\frac{2k}{3}}  }{  \phi^{4k}  }   }
\sqrt{ \mbox{\sffamily T\normalfont}^{\mbox{\scriptsize g\normalsize}}_{\mbox{\scriptsize C\normalsize}} 
+ \sum_{(k = 0)}^{(n)}\frac{  \mbox{\sffamily T\normalfont}_{(k)}V^{\frac{ 2k }{    3 }}  }{ \phi^{4k}  }   } } 
{     V^{\frac{2}{3}}     }
\ee varied with free end points is guaranteed to have $\dot{p} =
0$ $\forall$ $n \in {\cal N}_0$.

Note how the powers of $V$ match
the powers of the inverse metric that are needed to make
3-diffeomorphism scalars from the matter fields of different
possible ranks.

\mbox{ }

\noindent\bf Proof \normalfont Vacuum conformal gravity works,
hence the theorem is true for $n = 0$.

\noindent Induction hypothesis: suppose the theorem is true for
some $n = q$.

\noindent Then, for $n = q + 1$, $\phi$ variation gives 
$$
0 = \frac{\delta \mbox{\sffamily S\normalfont}^{(q + 1)}}{ \delta\phi(x)} = 
\frac{\delta \mbox{\sffamily S\normalfont}^{(q)}}{\delta\phi(x)}
+ 4(2 - q)NV^{\frac{2(q - 1)}{3}}\mbox{\sffamily U\normalfont}_{(q + 1)} 
- \frac{q + 1}{N}\mbox{\sffamily T\normalfont}_{(q + 1)}V^{\frac{2}{3q}}
$$
\be
+ 4\int \textrm{d}^3x\sqrt{h}V^{\frac{2q - 5}{3}} 
\left( 
qN\mbox{\sffamily U\normalfont}_{(q + 1)} +  \frac{q + 1}{4N}V^{\frac{2}{3}}T_{(q + 1)} 
\right) 
\mbox{ } .
\label{CGline1}
\ee
Now, from $\dot{p}^{(q + 1)} = \dot{p}^{(q + 1)ij}h_{ij} + p^{(q +
1)ij}\dot{h}_{ij}$ and the metric ELE for $\dot{p}^{(q + 1)ij}$, 
$$
\dot{p}^{(q + 1)} =  \dot{p}^{(q)} + \int \textrm{d}^3x\sqrt{h}V^{\frac{2q    -5}{3}} 
\left( 
N\mbox{\sffamily U\normalfont}_{(q + 1)} 
+ \frac{   V^{  \frac{2}{3}  }   }{ 4N    }\mbox{\sffamily T\normalfont}_{(q + 1)} 
\right) 
+ 3N\sqrt{h}\mbox{\sffamily U\normalfont}_{(q + 1)}V^{  \frac{2(q -    2)}{3}  }
$$
\be
+ V^{  \frac{2(q - 2)}{3}  }\sqrt{h} 
\left( 
\frac{    V^{ \frac{2}{3} }   }{   4N   } 
\frac{   \delta \mbox{\sffamily T\normalfont}_{(q + 1)}   }{    \delta h_{ij}   } 
+ N\frac{   \delta \mbox{\sffamily U\normalfont}_{(q + 1)}   }{   \delta    h_{ij}   } 
\right)
h^{ij} 
\mbox{ } .
\ee
$$
\mbox{Hence, by (\ref{CGline1}) } \mbox{\hspace{0.4in}}
\dot{p}^{(q + 1)} =  V^{ \frac{2(q    -2)}{3} }\sqrt{h} 
\left\{
\frac{  V^{ \frac{2}{3} }  }{  4N  }
\left( 
\frac{\delta \mbox{\sffamily T\normalfont}_{(q + 1)}}{\delta h_{ij}}h^{ij}
 + (q + 1)\mbox{\sffamily T\normalfont}_{(q + 1)} 
\right)
\right.
\mbox{\hspace{0.4in}}
$$
$$
\left.
+ N 
\left( 
\frac{\delta \mbox{\sffamily U\normalfont}_{(q + 1)}}{\delta h_{ij}}h^{ij}\mbox{+}(q\mbox{+}1)U_{(q + 1)} 
\right) 
\right\}
= 0
$$
by the induction hypothesis and using that $\mbox{\sffamily U\normalfont}_{(q + 1)}$, 
$\mbox{\sffamily T\normalfont}_{(q + 1)}$ are homogeneous of degree $q + 1$ in $h^{ij}$. 
Hence, if the theorem is true for $n = q$, it is also true for $n = q + 1$.  But
it is true for $n = 0$, so it is true by induction $\forall$ $n \in {\cal N}_0$. $\Box$

As in IV.1.2, we will now consider $\mbox{\sffamily T\normalfont}^{\Psi}$ and 
$\mbox{\sffamily U\normalfont}^{\Psi}$ as being made up of contributions from each of the 
fields present.  We will label these fields, and the indices they carry, by capital Greek
indexing sets.  We then obtain the following formulae for the propagation of the 
conformal gravity Hamiltonian constraint.

\mbox{ }

\noindent \bf Theorem C2 \normalfont

\noindent i) For nonderivative coupled matter fields 
$\Psi_{\mbox{\scriptsize\sffamily A\normalfont\normalsize}}$ with 
$\mbox{\sffamily T\normalfont}^{\Psi}$ homogeneously quadratic in
$\dot{\Psi}_{\mbox{\scriptsize\sffamily A\normalfont\normalsize}}$ and 
$\mbox{\sffamily U\normalfont}^{\Psi}$ containing at most
first-order derivatives, 
\be
-^{\Psi}\dot{{\cal H}}^{\mbox{\scriptsize C\normalsize}} = \frac{1}{N}D_b 
\left\{
N^2
\left(
2G_{\mbox{\scriptsize\sffamily AB\normalfont\normalsize}}
\Pi^{\mbox{\scriptsize\sffamily A\normalfont\normalsize}}\frac{ \pa\mbox{\sffamily U\normalfont}^{\Psi}    }
{    \pa(D_b\Psi_{\mbox{\scriptsize\sffamily B\normalfont\normalsize}})  } + \sigma
\left[
\Pi^{\mbox{\scriptsize\sffamily A\normalfont\normalsize}}
\frac{\delta    (\pounds_\xi\Psi_{\mbox{\scriptsize\sffamily A\normalfont\normalsize}}) }
{\pa\xi_b} 
\right] 
\right)
\right\} 
\mbox{ } .
\ee 

\noindent ii) If, additionally, the potential contains covariant derivatives, then there is an 
extra contribution to i): 
\be
\frac{2\sqrt{h}}{N}D_b
\left\{
N^2p_{ij}
\left(
\frac{\pa\mbox{\sffamily U\normalfont}^{\Psi}}{\pa{\Gamma^a}_{ic}}h^{aj} 
- \frac{1}{2}\frac{\pa\mbox{\sffamily U\normalfont}^{\Psi}}{\pa{\Gamma^a}_{ij}}h^{ac}
\right)
\right\} 
\mbox{ } .  
\ee

The proof offered here includes both conformal gravity ($\sigma = 1$) and strong conformal 
gravity ($\sigma = 0$, $\Lambda \neq 0$).  Again, use of formulae i), ii) permits the 
$^{\Psi}\dot{{\cal H}}^{\mbox{\scriptsize C\normalsize}}$ calculations to be done without 
explicitly computing each case's ELE's. \hspace{4in}
\be
\mbox{\bf Proof \normalfont}
\mbox{i) For a homogeneous quadratic kinetic term }
\mbox{\hspace{0.2in}}
\mbox{\sffamily T\normalfont}^{\Psi}   = \mbox{\ss}_{\xi}{\Psi}_{\mbox{\scriptsize\sffamily A\normalfont\normalsize}}
                                          \mbox{\ss}_{\xi}{\Psi}_{\mbox{\scriptsize\sffamily B\normalfont\normalsize}} 
                                          G^{\mbox{\scriptsize\sffamily AB\normalfont\normalsize}}
\left(
\frac{\vc}{\phi^4}h^{ij}
\right)
\mbox{ } , 
\mbox{\hspace{0.4in}}
\ee
\be
\mbox{the conjugate momenta are } 
\mbox{\hspace{1in}}
\Pi^{\mbox{\scriptsize\sffamily A\normalfont\normalsize}} 
\equiv \frac{\pa \mbox{\sffamily L\normalfont}    }{\pa\dot{\Psi}_{\mbox{\scriptsize\sffamily A\normalfont\normalsize}}} = \frac{\sqrt{h}\phi^4}{2N\vc}
                {    }G^{\mbox{\scriptsize\sffamily AB\normalfont\normalsize}}
\mbox{\ss}_{\xi}{\Psi}_{\mbox{\scriptsize\sffamily B\normalfont\normalsize}} \mbox{ } . 
\mbox{\hspace{1in}}  
\ee
\be
\mbox{The $\xi^i$-variation gives the momentum constraint }
\mbox{\hspace{0.2in}}
-^{\Psi}{\cal H}^{\mbox{\scriptsize C\normalsize}}_i \equiv 2D_j{p_i}^j
- \Pi^{\mbox{\scriptsize\sffamily A\normalfont\normalsize}}
\frac{\delta(\pounds_{\xi}\Psi_{\mbox{\scriptsize\sffamily A\normalfont\normalsize}})}
{\delta\xi^i}    = 0 
\mbox{\hspace{0.2in}}
\label{CGgenconfmom}
\ee 
and the local square root gives a primary Hamiltonian-type constraint,   
\be
-^{\Psi}{\cal H}^{\mbox{\scriptsize C\normalsize}} \equiv 
\frac{  \sqrt{h}  }{   V^{  \frac{2}{3}  }   }(\sigma R + \mbox{\sffamily U\normalfont}^{\Psi})  
- \frac{ \vc  }{  \sqrt{h}  }(p \circ p + G_{\mbox{\scriptsize\sffamily AB\normalfont\normalsize}}\Pi^{\mbox{\scriptsize\sffamily A\normalfont\normalsize}}\Pi^{\mbox{\scriptsize\sffamily B\normalfont\normalsize}}) = 0
\label{CGgenconfham} 
\ee 
in the distinguished representation. Then 
\be 
-^{\Psi}\dot{{\cal H}}^{\mbox{\scriptsize C\normalsize}} \approx
\frac{\sqrt{h}}{V^{\frac{2}{3}}}(\sigma\dot{R} + \dot{\mbox{\sffamily U\normalfont}}^{\Psi}) -
\frac{2\vc}{\sqrt{h}}(\dot{p} \circ p + \dot{h}_{ik}p^{ij}{p^k}_j)
- \frac{\vc}{\sqrt{h}}(2\dot{\Pi}^{\mbox{\scriptsize\sffamily A\normalfont\normalsize}} 
{    }G_{\mbox{\scriptsize\sffamily AB\normalfont\normalsize}}\Pi^{\mbox{\scriptsize\sffamily B\normalfont\normalsize}} 
+  {    }\dot{G}_{\mbox{\scriptsize\sffamily AB\normalfont\normalsize}}\Pi^{\mbox{\scriptsize\sffamily A\normalfont\normalsize}}\Pi^{\mbox{\scriptsize\sffamily B\normalfont\normalsize}}),
\ee 
using the chain-rule on (\ref{CGgenconfham}) and using $\dot{h} = \dot{V} = 0$.  Now use the 
chain-rule on $\dot{\mbox{\sffamily U\normalfont}}^{\Psi}$, the ELE's 
$\dot{p}^{ij} = \frac{\delta \mbox{\sffamily\scriptsize L\normalsize\normalfont}    }
{\delta h_{ij}}$ and 
$\dot{\Pi}^{\mbox{\scriptsize\sffamily A\normalfont\normalsize}} 
= \frac{\delta \mbox{\sffamily\scriptsize L\normalsize\normalfont}    }
{\delta\Psi_{\mbox{\tiny\sffamily A\normalfont\normalsize}}}$, 
and $p = 0$ to obtain the first step below: 
$$
-^{\Psi}\dot{{\cal H}}^{\mbox{\scriptsize C\normalsize}} \approx 
\frac{\sqrt{h}\sigma}{V^{\frac{2}{3}}}\dot{R} + \frac{\sqrt{h}}{V^{\frac{2}{3}}}
\left\{
\frac{\pa\mbox{\sffamily U\normalfont}^{\Psi}}{\pa\Psi_{{\mbox{\scriptsize\sffamily A\normalfont\normalsize}}}}
\dot{\Psi}_{{\mbox{\scriptsize\sffamily A\normalfont\normalsize}}} 
+ \frac{\pa\mbox{\sffamily U\normalfont}^{\Psi}}
{\pa(D_b\Psi_{\mbox{\scriptsize\sffamily A\normalfont\normalsize}})}
\dot{(D_b\Psi_{\mbox{\scriptsize\sffamily A\normalfont\normalsize}})}
+ \frac{\pa\mbox{\sffamily U\normalfont}^{\Psi}}{\pa h_{ab}}\dot{h}_{ab}
\right\}                                                                                   
$$
$$
- {2\sigma}p^{ij}
\left[
\frac{\delta R}{\delta{h_{ij}}}N
\right] 
- {2}p^{ij}
\left[
\frac{\delta\mbox{\sffamily U\normalfont}^{\Psi}}{\delta{h_{ij}}} N
\right] 
- \frac{1}{2N}p^{ij}\frac{\pa\mbox{\sffamily T\normalfont}^{\Psi}}{\pa{h_{ij}}} 
- \frac{4N\vc}{{h}}p_{ik}p^{ij}{p^k}_j
$$
$$
- {2}{ }G_{\mbox{\scriptsize\sffamily AB\normalfont\normalsize}}\Pi^{\Gamma}
\left[
N\frac{\delta \mbox{\sffamily U\normalfont}^{\Psi}}{\delta\psi_{\mbox{\scriptsize\sffamily A\normalfont\normalsize}}} 
\right] 
- \frac{\vc}{\sqrt{h}}\mbox{}\dot{G}_{\mbox{\scriptsize\sffamily AB\normalfont\normalsize}}\Pi^{\mbox{\scriptsize\sffamily A\normalfont\normalsize}}\Pi^{\mbox{\scriptsize\sffamily B\normalfont\normalsize}}              \\
$$
$$
=  \left(\frac{\sqrt{h}\sigma}{V^{\frac{2}{3}}}\dot{R} - {2\sigma}p^{ij}
\left[
\frac{\delta R}{\delta{h_{ij}}} N
\right] 
- \frac{4N\vc}{{h}}p_{ik}p^{ij}{p^k}_j
\right)                                                                                   
$$
$$
+\frac{\sqrt{h}}{V^{\frac{2}{3}}}
\left\{
\frac{\pa\mbox{\sffamily U\normalfont}^{\Psi}}{\pa\Psi_{\mbox{\scriptsize\sffamily A\normalfont\normalsize}}}
\left(
\frac{2N\vc}{\sqrt{h}}\Pi^{\mbox{\scriptsize\sffamily B\normalfont\normalsize}}
G_{\mbox{\scriptsize\sffamily AB\normalfont\normalsize}}
\right)
+ \frac{\pa\mbox{\sffamily U\normalfont}^{\Psi}}{\pa(D_b\Psi_{\mbox{\scriptsize\sffamily A\normalfont\normalsize}})}D_b
\left(
\frac{2N\vc}{\sqrt{h}}\Pi^{\mbox{\scriptsize\sffamily B\normalfont\normalsize}}
G_{\mbox{\scriptsize\sffamily AB\normalfont\normalsize}}
\right) 
+ \frac{\pa\mbox{\sffamily U\normalfont}^{\Psi}}{\pa h_{ab}}
\left(
\frac{2N\vc}{\sqrt{h}}p_{ab} 
\right)
\right\}                                                                                   
$$
$$
- {2}p^{ij}\frac{\pa\mbox{\sffamily U\normalfont}^{\Psi}}{\pa{h_{ij}}} N 
- \frac{1}{2N}p^{ab}\frac{\pa{ }G^{\mbox{\scriptsize\sffamily AB\normalfont\normalsize}}}
{\pa h_{ab}}\dot{\Psi}_{\mbox{\scriptsize\sffamily A\normalfont\normalsize}}
\dot{\Psi}_{\mbox{\scriptsize\sffamily B\normalfont\normalsize}} 
$$
$$
- {2}{ }G_{\mbox{\scriptsize\sffamily AB\normalfont\normalsize}}
\Pi^{\mbox{\scriptsize\sffamily B\normalfont\normalsize}} 
N\frac{\pa\mbox{\sffamily U\normalfont}^{\Psi}}
{\pa\psi_{\mbox{\scriptsize\sffamily A\normalfont\normalsize}}}
+ {2}G_{\mbox{\scriptsize\sffamily AB\normalfont\normalsize}}
\Pi^{\mbox{\scriptsize\sffamily A\normalfont\normalsize}}
D_b\left(N\frac{\pa\mbox{\sffamily U\normalfont}^{\Psi}} 
{\pa(D_b\Psi_{\mbox{\scriptsize\sffamily B\normalfont\normalsize}})}
\right)
- \frac{\vc}{\sqrt{h}}\frac{\pa{G}_{\mbox{\scriptsize\sffamily AB\normalfont\normalsize}}}
{\pa h_{ij}}\dot{h}_{ij}\Pi^{\mbox{\scriptsize\sffamily A\normalfont\normalsize}}
\Pi^{\mbox{\scriptsize\sffamily B\normalfont\normalsize}}                                                       \\
$$
\be
\approx  \frac{\sigma}{N}D_b
\left(
N^2
\left[
\Pi^{\mbox{\scriptsize\sffamily A\normalfont\normalsize}}
\frac{\delta(\pounds_{\xi}\Psi_{\mbox{\scriptsize\sffamily A\normalfont\normalsize}})}
{\delta\xi_b}
\right]
\right)
+ \sqrt{h}\frac{\pa\mbox{\sffamily U\normalfont}^{\Psi}}{\pa(D_b\Psi_{\mbox{\scriptsize\sffamily A\normalfont\normalsize}})}D_b
\left(
\frac{2N}{\sqrt{h}}\Pi^{\mbox{\scriptsize\sffamily B\normalfont\normalsize}}{ }
G_{\mbox{\scriptsize\sffamily AB\normalfont\normalsize}}
\right) 
+ {2}G_{\mbox{\scriptsize\sffamily AB\normalfont\normalsize}}
\Pi^{\mbox{\scriptsize\sffamily B\normalfont\normalsize}}D_b
\left(
N\frac{\pa\mbox{\sffamily U\normalfont}^{\Psi}} {\pa(D_b\Psi_{\mbox{\scriptsize\sffamily A\normalfont\normalsize}})}
\right) 
\mbox{ } .
\ee 
In the second step above, we regroup the terms into pure gravity terms and matter terms, 
expand the matter variational derivatives and use the definitions of the momenta to eliminate
the velocities in the first three matter terms.  We now observe that the first and sixth matter 
terms cancel, as do the third and fourth.  In the third step we use the pure gravity working 
and the momentum constraint (\ref{CGgenconfmom}), and the definitions of the momenta to cancel 
the fifth and eight terms of step 2.  Factorization of step 3 gives the result.

ii) Now $-^{\Psi}\dot{{\cal H}}^{\mbox{\scriptsize C\normalsize}}$ has 2 additional 
contributions in step 2 due to the presence of the connections: 
\be
\frac{\sqrt{h}}{V^{\frac{2}{3}}}\frac{\pa \mbox{\sffamily U\normalfont}^{\Psi}}
{\pa{\Gamma^a}_{bc}} \dot{\Gamma}^a {}_{bc} - {2}p^{ij}
\left[
\frac{\pa \mbox{\sffamily U\normalfont}^{\Psi}}{\pa{\Gamma^a}_{bc}}
\frac{\delta {\Gamma^a}_{bc}}{\delta h_{ij}}N
\right] 
\mbox{ } ,
\label{CGhalfwaycon} 
\ee 
which, using (\ref{deltaGamma}) and  conformal gravity's slight modification of (\ref{CDOT}),
\be
\dot{\Gamma}^a {}_{bc} = \frac{\vc}{2\sqrt{h}}
\{
D_b(N{p_c}^a) + D_c(N{p_b}^a) - D^a(Np_{bc})
\} 
\mbox{ } ,
\ee 
integration by parts on the second term of (\ref{CGhalfwaycon}) and factorization yields ii). 
$\Box$

Although Theorem C1 does not consider potentials containing Christoffel symbols, in all the 
cases that we consider below (which suffice for the investigation of the classical bosonic
theories of nature) the propagation of $^{\Psi}{\cal H}^{\mbox{\scriptsize C\normalsize}}$ 
rules out all theories with such potentials. Thus it is not an issue whether such theories 
permit $p = 0$ to be propagated.

\mbox{ }

\noindent\large{\bf 2.2 Examples}\normalsize

\mbox{ }

\noindent In this section we take $\sigma = 1$ for Lorentzian (as opposed to
Euclidean or strong) conformal gravity.  We will also use $W = 0$
from the outset, and $\Lambda = 0$, so that we are investigating
whether our theory of pure conformal gravity is capable of
accommodating conventional classical matter theories and
establishing the physical consequences. We find that it does, and
that the known classical bosonic theories are picked out.

\mbox{ }

\noindent{\bf 2.2.1 Scalar fields }

\mbox{ }

\noindent The natural action to consider according to our prescription for
including a scalar field is 
$$
\mbox{\sffamily I\normalfont}^{\varsigma} = \int \textrm{d}\lambda \int \textrm{d}^3x
\left\{
\left(
\frac{\sqrt{h}\phi^6}{V}
\right) 
\sqrt{\left(\frac{\vc}{\phi^4}
\right) 
\left( 
R - \frac{8D^2\phi}{\phi} + \mbox{\sffamily U\normalfont}^{\varsigma}_{(1)} 
\right) 
+ \mbox{\sffamily U\normalfont}_{(0)}^{\varsigma} } 
\sqrt{\mbox{\sffamily T\normalfont}^{\mbox{\scriptsize g\normalsize}}_{\mbox{\scriptsize C\normalsize}} + 
\mbox{\sffamily T\normalfont}^{\varsigma}  }
\right\} 
$$ 
\be 
= \int \textrm{d}\lambda \frac{ 
\textrm{d}^3x\sqrt{h}\phi^4 \sqrt{ R - \frac{8D^2\phi}{\phi}  +
\mbox{\sffamily U\normalfont}_{(1)}^{\varsigma} + \frac{ \mbox{\sffamily U\normalfont}^{\varsigma}_{(0)}\phi^4 }
{   V(\phi)^{\frac{2}{3}}   } } 
\sqrt{\mbox{\sffamily T\normalfont}^{\mbox{\scriptsize g\normalsize}}_{\mbox{\scriptsize C\normalsize}} 
+ \mbox{\sffamily T\normalfont}^{\varsigma}}} {V(\phi)^{\frac{2}{3}}  } 
= \int \textrm{d}\lambda \frac{\bar{\mbox{\sffamily I\normalfont}}}{V^{\frac{2}{3}}} 
\mbox{ } , 
\label{CGBOSaction} 
\ee 
where, as in III.2.5, we give two different expressions to exhibit the homogeneity and to use 
in calculations. $\mbox{\sffamily U\normalfont}^{\varsigma}_{(0)}$ is an arbitrary function of 
${\varsigma}$ alone whilst $\mbox{\sffamily U\normalfont}^{\varsigma}_{(1)} = 
-\frac{C}{4}h^{ab}|\mbox{\b{$\pa$}} \varsigma |^2$.  

The conjugate momenta $p^{ij}$ and $\pi_{\phi}$ are given by
(\ref{CGgravcanmom}) and (\ref{CGconfcanmom}) but with
\be
2N = \sqrt{              
     \frac{          \mbox{\scriptsize\sffamily T\normalfont\normalsize}^{\mbox{\tiny g\normalsize}}_{\mbox{\tiny\sffamily C\normalfont\normalsize}} 
+ \mbox{\scriptsize\sffamily T\normalfont\normalsize}^{\mbox{\tiny $\varsigma$\normalsize}}         } 
{            R - \frac{     8D^2 \phi     }{     \phi     }  
+  \mbox{\scriptsize\sffamily U\normalfont\normalsize}^{\mbox{\tiny $\varsigma$\normalsize}}_{\mbox{\tiny (1)\normalsize}} 
+ \frac{    \mbox{\scriptsize\sffamily U\normalfont\normalsize}^{\mbox{\tiny $\varsigma$\normalsize}}_{(0)}
\phi^4     }{    V^{  \frac{2}{3}  }   }       }             } 
\mbox{ } ,
\ee
\be
\mbox{and additionally we have the momentum conjugate to $\varsigma$, }
\mbox{\hspace{0.6in}}
\pi = \frac{    \sqrt{h}\phi^4    }{    2NV^{  \frac{2}{3}  }    }\mbox{\ss}_{\xi}{\varsigma} 
\mbox{ } .
\mbox{\hspace{0.6in}}
\ee
As in the case of pure conformal gravity, we have the primary constraint (\ref{CGprimconstr}), 
and the end-point part of the $\phi$-variation gives $p_{\phi} = 0$, so that $p = 0$ by the 
primary constraint. But by construction (theorem C1) this action has the correct form to 
propagate the constraint $p = 0$ provided that the LFE 

\noindent
\be
    2(NR - D^2N) + \frac{3N\mbox{\sffamily U\normalfont}^{\varsigma}_{(0)}}{\vc} 
+ 2N \mbox{\sffamily U\normalfont}_{(1)}^{\varsigma} =
    \frac{1}{V^{\frac{5}{3}}}\int \textrm{d}^3x\sqrt{h}N
\mbox{\sffamily U\normalfont}_{(0)}^{\varsigma} + 
    \frac{\bar{\mbox{\sffamily I\normalfont}}}{V} 
\mbox{ } ,
\ee 
holds (in the distinguished representation), but this is guaranteed from the
rest of the $\phi$-variation.

\noindent The $\xi^i$-variation gives the secondary momentum constraint (\ref{SCSC})
\noindent whilst the local square root gives rise to a primary Hamiltonian-type constraint, 
which is 
\be
    -^{\varsigma}{\cal H}^{\mbox{\scriptsize C\normalsize}} \equiv
    \frac{\sqrt{h}}{V^{\frac{2}{3}}} \left(R + \mbox{\sffamily U\normalfont}^{\varsigma}_{(1)} +
    \frac{\mbox{\sffamily U\normalfont}^{\varsigma}_{(0)}}{\vc}\right) - \frac{\vc}{\sqrt{h}}(p \circ p +
    \pi^2) = 0
\ee 
in the distinguished representation.

Then, using formula i), the propagation of the Hamiltonian
constraint is, weakly, 
\be
^{\varsigma}\dot{\cal H}^{\mbox{\scriptsize C\normalsize}} \approx
\frac{(C - 1)}{N}D_b(N^2\pi\pa^b\varsigma) 
\mbox{ } .
\ee
Now, if the cofactor of $(C - 1)$ were zero, there would be a secondary constraint which would 
render the scalar field theory trivial by using up its d.o.f. Hence $C = 1$ is 
fixed, which is the universal light-cone condition applied to the scalar field.  This means 
that the null cone of gravitation is enforced even though the gravitational 
theory in question is not generally covariant in the spacetime sense.  

\mbox{ }

\noindent{\bf 2.2.2 1-Form fields}

\mbox{ }

\noindent According to our prescription, the natural action to include electromagnetism is 
$$ 
\mbox{\sffamily I\normalfont}^{\mbox{\scriptsize A\normalsize}}  
= \int\textrm{d}\lambda \int \textrm{d}^3x
\left\{
\left( 
\frac{  \sqrt{h}\phi^6  }{  V  }
\right) 
\sqrt{ 
\left( 
\frac{  \vc  }{  \phi^4  } 
\right) 
\left( 
R - \frac{ 8D^2\phi  }{  \phi  } 
\right) 
+ 
\left( 
\frac{    V^{\frac{4}{3}  }}{  \phi^8  } 
\right) 
\mbox{\sffamily U\normalfont}^{\mbox{\scriptsize A\normalsize}}   } 
\sqrt{ \mbox{\sffamily T\normalfont}^{\mbox{\scriptsize g\normalsize}}_{\mbox{\scriptsize C\normalsize}} +
\left( 
\frac{ \vc }{ \phi^4 } 
\right) 
\mbox{\sffamily T\normalfont}^{\mbox{\scriptsize A\normalsize}}} 
\right\} 
$$
\be 
= \int \textrm{d}\lambda \frac{ \int \textrm{d}^3x\sqrt{h}\phi^4 
\sqrt{        R - \frac{8D^2\phi}{\phi} +
\mbox{\sffamily U\normalfont}^{\mbox{\scriptsize A\normalsize}}    
\frac{   V(\phi)^{\frac{2}{3} } } {    \phi^4    }         } 
\sqrt{\mbox{\sffamily T\normalfont}^{\mbox{\scriptsize g\normalsize}}_{\mbox{\scriptsize C\normalsize}} + 
\mbox{\sffamily T\normalfont}^{\mbox{\scriptsize A\normalsize}}
\frac{ V(\phi)^{ \frac{2}{3} } } {  \phi^4  }}          } { V(\phi)^{ \frac{2}{3} } } 
= \int \textrm{d}\lambda \frac{\bar{\mbox{\sffamily I\normalfont}} }{V^{\frac{2}{3}}}, 
\label{CGBOCCaction} 
\ee 
for $\mbox{\sffamily U\normalfont}^{\mbox{\scriptsize A\normalsize}}$ and 
    $\mbox{\sffamily T\normalfont}^{\mbox{\scriptsize A\normalsize}}$  as in IV.1.2.  
We will first show that electromagnetism exists as a theory coupled to conformal gravity. We 
will then discuss how it is uniquely picked out (much as it is picked out in RWR \cite{BOF}), 
and how Yang--Mills theory is uniquely picked out upon consideration of \bf K \normalfont interacting 
1-form fields (much as it is picked out in \cite{AB}).  

Again, the conjugate momenta $p_{\phi}$ and $p^{ij}$ are given by (\ref{CGconfcanmom}) and 
(\ref{CGgravcanmom}) but now with
\be
    2N = \sqrt{
    \frac{\mbox{\sffamily T\normalfont}^{\mbox{\scriptsize g\normalsize}}_{\mbox{\scriptsize C\normalsize}} +
    \frac{\mbox{\sffamily\scriptsize T\normalsize\normalfont}^{\mbox{\tiny A\normalsize}}     
V(\phi)^{\frac{2}{3}}    }{\phi^4 } } {   \sigma 
\left( 
R - \frac{ 8D^2 \phi }{ \phi }
\right) 
+ \frac{\mbox{\sffamily\scriptsize U\normalsize\normalfont}^{\mbox{\tiny A\normalsize}}    
    V(\phi)^{\frac{2}{3}} }{\phi^4 } } } \mbox{ } ,
\ee
and additionally we have the momentum conjugate to $A_i$,  (\ref{1formmom}). 
By the same argument as in III.2.5, $p = 0$ arises and is preserved by a 
lapse-fixing equation, which is now 

\noindent
\be
2(NR - D^2N) + 
\left( 
N\mbox{\sffamily U\normalfont}^{\mbox{\scriptsize A\normalsize}} 
- \frac{      \mbox{\sffamily T\normalfont}^{\mbox{\scriptsize A\normalsize}}      }{4N} 
\right)
\vc + \frac{1}{V^{\frac{1}{3}}}\int\textrm{d}^3x\sqrt{h}
\left(
N\mbox{\sffamily U\normalfont}^{\mbox{\scriptsize A\normalsize}}  +
\frac{\mbox{\sffamily T\normalfont}^{\mbox{\scriptsize A\normalsize}}}{4N}
\right)
= \frac{\bar{I}}{V} \mbox{ } . 
\label{CGVECTORSLICE}
\ee 
The $\xi^i$-variation gives the secondary momentum constraint \ref{emmomcon},
whilst the local square root gives a primary Hamiltonian-type constraint, 
\be
    -^{\mbox{\scriptsize A\normalsize}}{\cal H}^{\mbox{\scriptsize C\normalsize}} \equiv 
\frac{\sqrt{h}}{V^{\frac{2}{3}}} 
\left(
\sigma R + {\mbox{\sffamily U\normalfont}^{\mbox{\scriptsize A\normalsize}}}{\vc}
\right) 
- \frac{\vc}{\sqrt{h}}
\left(p \circ p + \frac{1}{\vc}\pi_i\pi^i
\right) 
= 0
\ee  
in the distinguished representation.

Then, using formula i), the propagation of the Hamiltonian constraint is, weakly, 
\be
    -^{\mbox{\scriptsize A\normalsize}}\dot{\cal H}^{\mbox{\scriptsize C\normalsize}}
\approx \frac{1}{N} D^b 
\left\{
N^2 
\left( 
(1 - 4{C}^{\mbox{\scriptsize A\normalsize}})\pi^i(D_bA_i - D_iA_b) - A_bD_i\pi^i 
\right)
\right\}
\mbox{ } .
\ee 
Suppose the cofactor of $1 - 4{C}^{\mbox{\scriptsize A\normalsize}}$ is zero. Then we require 
$D_{[b}A_{i]} = 0$.  But this is three conditions on $A_i$, so the vector theory would be 
rendered trivial. Thus, exhaustively, the only way to obtain a consistent theory is to have 
the universal null cone condition ${C}^{\mbox{\scriptsize A\normalsize}} = \frac{1}{4}$ 
\be
\mbox{and 
the new constraint }
\mbox{\hspace{1.7in}}
{\cal G} \equiv D_a\pi^a = 0 
\mbox{ } , 
\mbox{\hspace{1.7in}}
\ee
which we identify as the electromagnetic Gauss constraint (\ref{M1}, \ref{curemgau}). The propagation of ${\cal G}$ 
\be
\mbox{is no further bother because the $A_i$ ELE }
\mbox{\hspace{0.6in}}
\mbox{\ss}_{\xi}\pi^i  = 2\sqrt{h}C^{\mbox{\scriptsize A\normalsize}}D_b(D^bA^i - D^iA^b)
\mbox{\hspace{0.6in}}
\label{CGemEL} 
\ee 
is free of $V$ and hence identical to that in the RWR case. Since the RWR argument for the 
propagation of ${\cal G}$ follows from (\ref{CGemEL}), this guarantees that the result also
holds in conformal gravity.

Since the 
potential is U(1) symmetric, we can finally encode this new constraint by making use of 
U(1)-BM , modifying the bare kinetic term by introducing an auxiliary 
\be
\mbox{variable $\Phi$:  }
\mbox{\hspace{1.1in}}
\mbox{\sffamily T\normalfont}^{\mbox{\scriptsize A\normalsize}}  = 
(\dot{A}_a - \pounds_{\xi}A_a - \pa_a\Phi)(\dot{A}^a - \pounds_{\xi}A^a - \pa^a\Phi) 
\mbox{ } .
\mbox{\hspace{1.1in}}
\ee

The following extensions of this working have been considered.

1) Additionally, replacing $\mbox{\sffamily U\normalfont}^{\mbox{\scriptsize A\normalsize}}$ by 
$C^{abcd}D_bA_a D_dA_c$ in the action preserves the correct form to guarantee $p = 0$ is maintained. We now have 
derivative coupling contributions also, so we need to make use of formula ii) of theorem 2 as 
well as formula i). Thus, weakly 
$$
    -^{\mbox{\scriptsize A\normalsize}}\dot{{\cal H}}^{\mbox{\scriptsize C\normalsize}}
\approx \frac{1}{N}D_b 
\left(
N^2 
\left\{
4C_1 + 1)\pi^aD^b{A_a} + (4C_2 - 1)\pi^aD_a{A^b} + 4C_3(N^2\pi^bD_a{A^a}) 
\right.
\right.
$$
\be
\left.
\left.
- D_a{\pi^a}A^b 
-  4p_{ij}D_{(d}A_{b)}
\left(
C^{ajbd}A^i - \frac{1}{2}C^{ijbd}A^a
\right)
\right\}
\right) 
\mbox{ } .
\ee
This has the same structure in $A_i$ as for the GR case [the overall $V^{-\frac{2}{3}}$ is 
unimportant, as is the replacement of the GR $(p_{ij} - \frac{p}{2}h_{ij})$ factors by 
$(p_{ij})$ factors here], so an argument along the same lines as that used in RWR will hold, 
forcing the Gauss constraint and $C_1 = - C_2 = -\frac{1}{4}$, $C_3 = 0$ (Maxwell theory).
\be
\mbox{ } \mbox{ } \mbox{ } \mbox{ 2) The changes }
\mbox{\hspace{1in}}
\mbox{\sffamily T\normalfont}^{\mbox{\scriptsize A\normalsize}}  \longrightarrow \
\mbox{\sffamily T\normalfont}^{\mbox{\scriptsize A\normalsize}_{\mbox{\bf\tiny I\normalsize\normalfont}}   }  
= h^{ij}(\dot{A}_i^{\mbox{\bf\scriptsize I\normalsize\normalfont}}    - \pounds_{\xi}A_i^{\mbox{\bf\scriptsize I\normalsize\normalfont}}   ) 
(\dot{A}_{j{\mbox{\bf\scriptsize I\normalsize\normalfont}}   } -    \pounds_{\xi}A_{j{\mbox{\bf\scriptsize I\normalsize\normalfont}}   })
\mbox{ } ,
\mbox{\hspace{1in}}
\label{CGTAISUB} 
\ee 
\be 
\mbox{\sffamily U\normalfont}^{\mbox{\scriptsize A\normalsize}}
\longrightarrow 
\mbox{\sffamily U\normalfont}^{\mbox{\scriptsize A\normalsize}_{\mbox{\bf\tiny I\normalsize\normalfont}}   }     
=  O_{\mbox{\bf\scriptsize IK\normalsize\normalfont}}C^{abcd}
D_bA^{{\mbox{\bf\scriptsize I\normalsize\normalfont}}}_{a}D_dA^{\mbox{\bf\scriptsize K\normalsize\normalfont}}_{c} 
+ {B^{\mbox{\bf\scriptsize I\normalsize\normalfont}}}_{\mbox{\bf\scriptsize JK\normalsize\normalfont}}
\bar{C}^{abcd}D_bA_{{\mbox{\bf\scriptsize I\normalsize\normalfont}}   a}
A^{\mbox{\bf\scriptsize J\normalsize\normalfont}}_c A^{\mbox{\bf\scriptsize K\normalsize\normalfont}}_d 
+ I_{\mbox{\bf\scriptsize JKLM\normalsize\normalfont}}\bar{\bar{C}}^{abcd}
A^{\mbox{\bf\scriptsize J\normalsize\normalfont}}_aA^{\mbox{\bf\scriptsize K\normalsize\normalfont}}_b
A^{\mbox{\bf\scriptsize L\normalsize\normalfont}}_cA^{\mbox{\bf\scriptsize M\normalsize\normalfont}}_d 
\label{CGUAISUB} 
\ee 
(for a priori distinct supermetrics $C$, $\bar{C}$, $\bar{\bar{C}}$) to the ansatz preserve 
the conformal properties, hence guaranteeing that $p = 0$ is maintained by the
lapse-fixing equation obtained by applying (\ref{CGTAISUB}, \ref{CGUAISUB}) to 
(\ref{CGVECTORSLICE}).  The new conjugate momenta are (\ref{YMformmom}).  
$\xi^i$-variation gives the secondary momentum constraint, (\ref{eagleeye})
and the local square root gives a primary Hamiltonian-type constraint,  
\be
   - ^{\mbox{\scriptsize A\normalsize}_{\mbox{\bf\scriptsize I\normalsize\normalfont}}}{\cal H}^{\mbox{\scriptsize C\normalsize}} \equiv 
\frac{\sqrt{h}}{V^{\frac{2}{3}}} 
\left(
\sigma R + {\mbox{\sffamily U\normalfont}^{\mbox{\scriptsize A\normalsize}_{\mbox{\bf\scriptsize I\normalsize\normalfont}}}}{\vc}
\right) 
- \frac{\vc}{\sqrt{h}}(p \circ p +  \frac{1}{\vc}\pi^{\mbox{\bf\scriptsize I\normalsize\normalfont}}_i
\pi_{\mbox{\bf\scriptsize I\normalsize\normalfont}}^i) = 0
\ee in the distinguished representation.

Using formulae i), ii) we read off that the propagation of the
Hamiltonian constraint is, weakly, 
$$
-{}^{{\mbox{\scriptsize A\normalsize}}_{\mbox{\bf\scriptsize I\normalsize\normalfont}}}\dot{{\cal H}}^{\mbox{\scriptsize C\normalsize}}
\approx \frac{1}{N} D_b 
\left( 
N^2 
\left\{
(4C_1O^{\mbox{\bf\scriptsize IK\normalsize\normalfont}} + \delta^{\mbox{\bf\scriptsize IK\normalsize\normalfont}})
\pi_{\mbox{\bf\scriptsize I\normalsize\normalfont}}^aD^b{A_{{\mbox{\bf\scriptsize K\normalsize\normalfont}}a}} 
+ (4C_2O^{\mbox{\bf\scriptsize IK\normalsize\normalfont}} - \delta^{\mbox{\bf\scriptsize IK\normalsize\normalfont}})
\pi_{\mbox{\bf\scriptsize I\normalsize\normalfont}}^aD_a{A_{\mbox{\bf\scriptsize K\normalsize\normalfont}}^b} 
\right.
\right.
$$
$$
 + 4C_3O^{\mbox{\bf\scriptsize IK\normalsize\normalfont}}\pi_{\mbox{\bf\scriptsize I\normalsize\normalfont}}^b
D_a{A_{\mbox{\bf\scriptsize K\normalsize\normalfont}}^a} -  
\left(
D_a{\pi_{\mbox{\bf\scriptsize K\normalsize\normalfont}}^a}A^{{\mbox{\bf\scriptsize K\normalsize\normalfont}}b} 
- 2 \bar{C}^{abcd}{B^{\mbox{\bf\scriptsize I\normalsize\normalfont}}}_{\mbox{\bf\scriptsize JK\normalsize\normalfont}}   
\pi_{{\mbox{\bf\scriptsize I\normalsize\normalfont}}a}A^{\mbox{\bf\scriptsize J\normalsize\normalfont}}_c
A^{\mbox{\bf\scriptsize K\normalsize\normalfont}}_d
\right)                                                                                    
$$
\be
\left.
\left.
- 2 O^{\mbox{\bf\scriptsize IK\normalsize\normalfont}}  p_{ij} 
D_{(d|}A_{{\mbox{\bf\scriptsize K\normalsize\normalfont}}|l)} 
(2A_{\mbox{\bf\scriptsize I\normalsize\normalfont}}^iC^{bjld} 
- A_{\mbox{\bf\scriptsize I\normalsize\normalfont}}^bC^{ijld})                        
- {B^{\mbox{\bf\scriptsize I\normalsize\normalfont}}}_{\mbox{\bf\scriptsize JK\normalsize\normalfont}} p_{ij} 
A^{\mbox{\bf\scriptsize J\normalsize\normalfont}}_l A^{\mbox{\bf\scriptsize K\normalsize\normalfont}}_d 
(    2A_{\mbox{\bf\scriptsize I\normalsize\normalfont}}^i\bar{C}^{bjld} 
- A_{\mbox{\bf\scriptsize I\normalsize\normalfont}}^b\bar{C}^{ijld})
\right\}
\right) 
\mbox{ } . 
\label{CGCYMhamprop} 
\ee
In the same sense as for the single 1-form case above, (\ref{CGCYMhamprop}) has the same 
structure as for the GR case, so the argument used in \cite{AB} will hold, forcing 
$$
    O^{\mbox{\bf\scriptsize IK\normalsize\normalfont}} = \delta^{\mbox{\bf\scriptsize IK\normalsize\normalfont}}, 
    C_1 = -C_2 = -\frac{1}{4}, C_3 = 0, \bar{C}_3 = 0 \mbox{ } , 
$$ 
\be 
B_{{\mbox{\bf\scriptsize I\normalsize\normalfont}}({\mbox{\bf\scriptsize JK\normalsize\normalfont}})} = 0
    \Leftrightarrow \bar{C}_1 = -\bar{C}_2 \equiv -\frac{\mbox{
    \sffamily g\normalfont}}{4}
    \label{CGanti}
\ee
(for some emergent coupling constant $\mbox{\sffamily g\normalfont}$) 
 and leaving the new constraint 
\be
    {\cal G}_{\mbox{\bf\scriptsize J\normalsize\normalfont}} \equiv
    D_a\pi^a_{\mbox{\bf\scriptsize J\normalsize\normalfont}} 
- \mbox{ \sffamily g\normalfont}B_{{\mbox{\bf\scriptsize IJK\normalsize\normalfont}}}
    \pi^{\mbox{\bf\scriptsize I\normalsize\normalfont}}_aA^{{\mbox{\bf\scriptsize K\normalsize\normalfont}}a} \mbox{ } .
\ee
Again as for the single vector field case, the $\pi^a_{\mbox{\bf\scriptsize J\normalsize\normalfont}}$ 
ELE is unchanged from the GR case. The action of the dot on 
$A_{{\mbox{\bf\scriptsize K\normalsize\normalfont}}a}$ gives no volume terms.  Hence the working
for the propagation of ${\cal G}_{\mbox{\bf\scriptsize J\normalsize\normalfont}}$ is unchanged 
from that in \cite{AB}, which enforces  
\be
    I_{\mbox{\bf\scriptsize JKLM\normalsize\normalfont}} 
= {B^{\mbox{\bf\scriptsize I\normalsize\normalfont}}}_{{\mbox{\bf\scriptsize JK\normalsize\normalfont}}}
B_{\mbox{\bf\scriptsize ILM\normalsize\normalfont}} 
\mbox{ } , \mbox{ } 
\bar{\bar{C}}_2 = - \bar{\bar{C}}_1
    = \frac{\mbox{ \sffamily g\normalfont}^2}{16} 
\mbox{ } , \mbox{ } 
\bar{\bar{C}}_3 = 0 \mbox{ } ,
\ee
\be 
{B^{\mbox{\bf\scriptsize I\normalsize\normalfont}}}_{\mbox{\bf\scriptsize JK\normalsize\normalfont}}
B_{\mbox{\bf\scriptsize ILM\normalsize\normalfont}} 
+ {B^{\mbox{\bf\scriptsize I\normalsize\normalfont}}}_{\mbox{\bf\scriptsize JM\normalsize\normalfont}}
B_{\mbox{\bf\scriptsize IKL\normalsize\normalfont}} 
+ {B^{\mbox{\bf\scriptsize I\normalsize\normalfont}}}_{\mbox{\bf\scriptsize JL\normalsize\normalfont}}
B_{\mbox{\bf\scriptsize IMK\normalsize\normalfont}} = 0 
\mbox{ } \mbox{ (Jacobi identity) },
\label{CGJacobi} 
\ee 
\be
    B_{{\mbox{\bf\scriptsize IJK\normalsize\normalfont}}} =
    B_{[{\mbox{\bf\scriptsize IJK\normalsize\normalfont}}]} 
\mbox{ } \mbox{ (total antisymmetry) }.
\label {totant} 
\ee 
From (\ref{CGanti}) and (\ref{CGJacobi}), it follows
that the $B_{\mbox{\bf\scriptsize IJK\normalsize\normalfont}}$ are the structure constants of some Lie
algebra, {\sc g}. From (\ref{totant}) and the
Gell-Mann--Glashow theorem \cite{GMG, Weinberg}, {\sc g} is the direct
sum of compact simple and U(1) subalgebras, provided that the
kinetic term is positive definite as assumed here. We can defend
this assumption because we are working on a theory in which even
the gravitational kinetic term is taken to be positive definite;
positive-definite kinetic terms ease quantization.

3) In the BF\'{O} formulation of the TSA,  mass terms are banned by the propagation of 
the Gauss laws.  Mass terms contain nontrivial powers of the volume;
however the above arguments can easily be extended to
accommodate them.  In
the many vector fields case, the effect of a mass term is to give
rise to a new term $\frac{2N}{\vc}M^{\mbox{\bf\scriptsize JK\normalsize\normalfont}}
A^i_{\mbox{\bf\scriptsize K\normalsize\normalfont}}$ in the 
ELE's, which contributes a term
$2M^{\mbox{\bf\scriptsize JK\normalsize\normalfont}}D_i
\left(
\frac{N}{\vc}A^i_{\mbox{\bf\scriptsize K\normalsize\normalfont}}
\right)$ 
to the propagation of ${\cal G}_{\mbox{\bf\scriptsize J\normalsize\normalfont}}$.  
For this to vanish, either $A^i_{\mbox{\bf\scriptsize K\normalsize\normalfont}} = 
0$ which renders the vector theory trivial, or $M^{\mbox{\bf\scriptsize JK\normalsize\normalfont}} = 0$.

\mbox{ }

\noindent\Large{\bf App IV.A: Na\"{\i}ve renormalizability}\normalsize

\mbox{ }

\noindent I use the below as part of the justification of the scope of the  
matter terms considered in IV and VI.  {\it Feynman diagrams} are schematic computational 
rules for each possible contribution to particle processes; path integral approaches are descended 
from such concepts.  Fig 11 has my propagator convention and depicts the terms used below.  
\begin{figure}[hhhh]
\centerline{\def\epsfsize#1#2{0.4#1}\epsffile{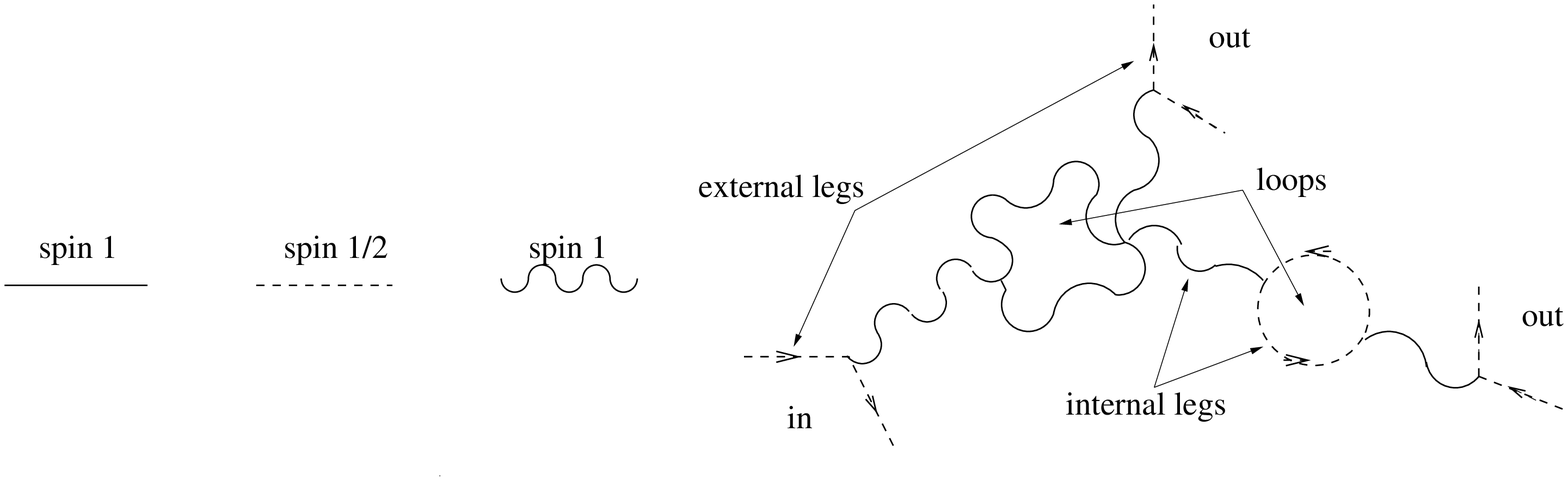}}
\caption[]{\label{4}
\scriptsize The {\it external legs} represent the input and output of the particle 
physics process (scattering), which although practical, has Minkowskian (open-universe) 
connotations if extrapolated. {\it Internal legs} and {\it loops} are also pictured. \normalsize}
\end{figure}

Now, for a boson-fermion theory with $(f + b)$-point interactions, the 
general Feynman diagram with 
$n(f, b)$ $(f + b)$-point vertices, 
$F_{\mbox{\scriptsize I\normalsize}}$ internal fermion legs, 
$F_{\mbox{\scriptsize E\normalsize}}$ external fermion legs, 
$B_{\mbox{\scriptsize I\normalsize}}$ internal boson legs, 
$B_{\mbox{\scriptsize E\normalsize}}$ external boson legs, and 
$L$ loops has the momentum space form 
$$
\left[
\int\frac{``\textrm{d}^4p"}{(2\pi)^4}
\right]
^{\sum_{f, b}n(f, b)}
\times
\left(
\frac{1}{p^2}
\right)
^{B_I}
\times
\left(
\frac{1}{p}
\right)
^{F_I}
\times
\left(
\begin{array}{c}
\delta-\mbox{functions due to} \\
\mbox{mass conservation}
\end{array}
\right) \mbox{ } .
$$
This has a superficial degree of divergence 
$D = 4L - 2B_{\mbox{\scriptsize I\normalsize}} - F_{\mbox{\scriptsize I\normalsize}}$.  
But by $L = B_{\mbox{\scriptsize I\normalsize}} + F_{\mbox{\scriptsize I\normalsize}}$, 
$\sum_{f}fn(f, b) = 2F_{\mbox{\scriptsize I\normalsize}} + F_{\mbox{\scriptsize E\normalsize}}$ 
and $\sum_{b}bn(f, b) = 2B_{\mbox{\scriptsize I\normalsize}} + B_{\mbox{\scriptsize E\normalsize}}$, 
$\frac{3f}{2} + b \leq 4$ for any vertex type $n(b, f)$, lest there be divergence.  Also, 
observed lepton number conservation requires fermion legs to occur in pairs.  

\begin{figure}[hhhh]
\centerline{\def\epsfsize#1#2{0.4#1}\epsffile{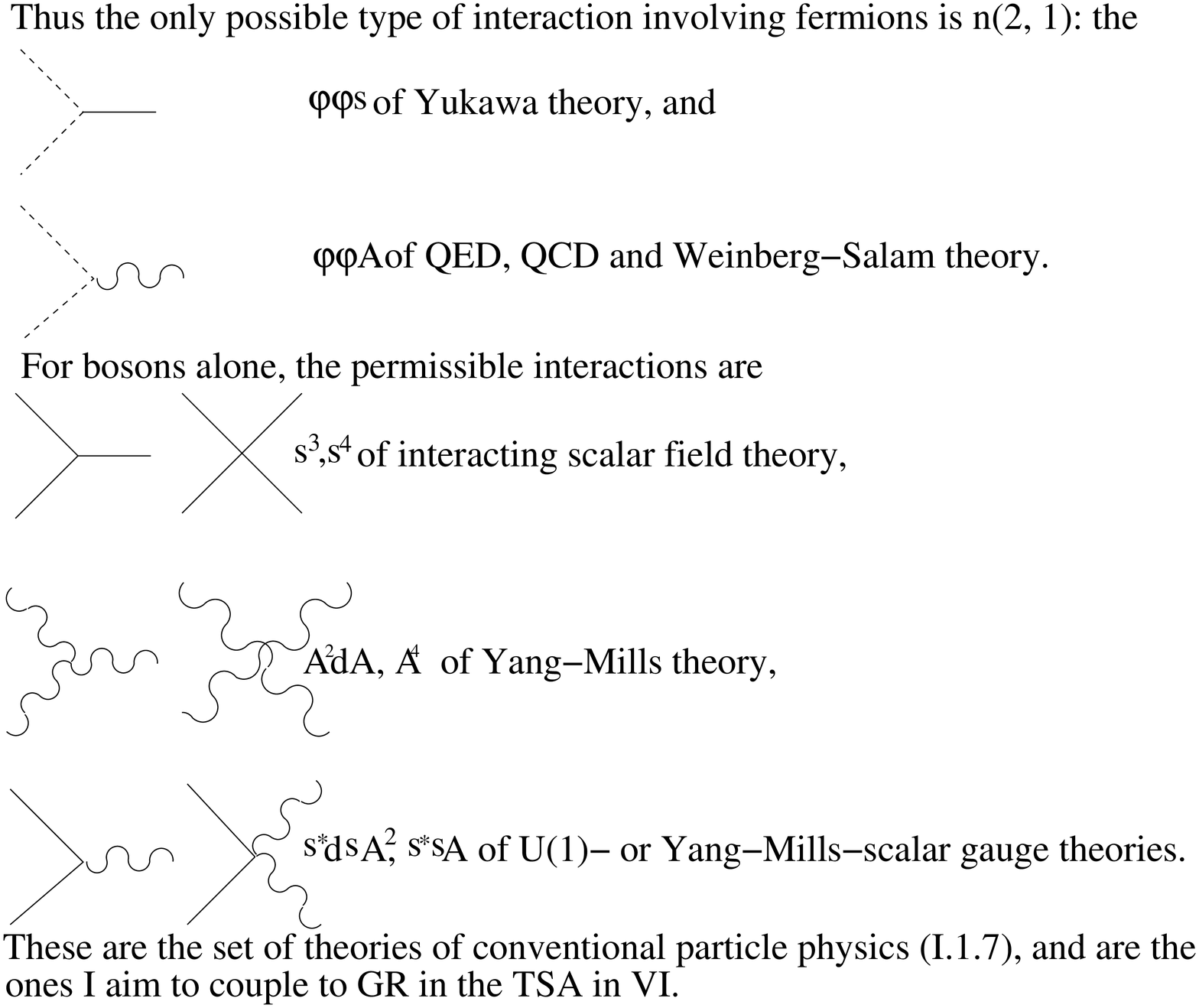}}
\end{figure}


\mbox{ }

\noindent\Large{\bf App IV.B: Gell-Mann--Glashow theorem}\normalsize

\mbox{ }

\noindent An algebra {\sc a} is a vector space {\sl V} equipped with a product 
$\mbox{\scriptsize $\Box$\normalsize} : \mbox{\sc a} \times \mbox{\sc a} \longrightarrow 
\mbox{\sc a}$.  A \it{subalgebra }\normalfont {\sc b} of {\sc a} is a vector subspace 
{\sl W} of {\sl V} which is also an algebra with product 
$\mbox{\scriptsize $\Box$\normalsize}$.  It is \it{invariant }\normalfont if 
${b} \mbox{ \scriptsize$\Box$ \normalsize} {a} \in \mbox{{\sc b}} 
\mbox{ } \forall \mbox{ } b \in \mbox{\sc b}, \forall \mbox{ } a \in \mbox{\sc a}$.  
A real \it{representation }\normalfont $\Gamma$ of {\sc a} is a map 
$\Gamma:\mbox{{\sc a}} \longrightarrow GL(n, \Re)$.  It is \it{reducible }\normalfont if 
$\exists$ some subspace {\sl U} in {\sl V} that is left invariant by $\Gamma$.  If it is not 
reducible, it is \it{irreducible}\normalfont.  It is \it{totally reducible }\normalfont if it 
can be written as a direct sum of reducibles.  

A Lie algebra {\sc g} is \it{Abelian }\normalfont if 
$|[g_1 , g_2]| = 0$  $\forall \mbox{ } g_1, \mbox{ } g_2$  $\in \mbox{\sc g}$.  
An Abelian algebra with 1 generator is a U(1) algebra.  
$\mbox{{\sc G}}$ is \it{simple }\normalfont if it contains no proper invariant subalgebras and 
it is not Abelian.  $\mbox{{\sc g}}$ is \it{semi-simple }\normalfont if it is the direct sum of 
simple Lie algebras.  A simple or semisimple Lie algebra is \it{compact }\normalfont if 
$tr(\tau_{\mbox{\bf\scriptsize A\normalsize\normalfont}}\tau_{\mbox{\bf\scriptsize B\normalsize\normalfont}}) 
= {f^{\mbox{\bf\scriptsize C\normalsize\normalfont}}}_{\mbox{\bf\scriptsize AD\normalsize\normalfont}}
{f^{\mbox{\bf\scriptsize D\normalsize\normalfont}}}_{\mbox{\bf\scriptsize BC\normalsize\normalfont}}$ 
is positive-definite.

\mbox{ } 

\noindent Now for the \bf{Proof }\normalfont of the Gell-Mann--Glashow theorem \cite{Weinberg, GMG}:

\noindent
(\ref{a}) $\Rightarrow$ (\ref{b}) trivially by basis change, provided that $Q _{AB}$ is 
positive definite, in which case we can define $(Q^{-\frac{1}{2}})_{AB}$.

\noindent
(\ref{b}) $\Rightarrow$ (\ref{c}) 

In the basis assumed in (\ref{b}) are matrices of the form 
$(\tilde{\tau}^{\mbox{\scriptsize A\normalsize}}_{\mbox{\bf\scriptsize A\normalsize\normalfont}})_{\mbox{\bf\scriptsize BC\normalsize\normalfont}} 
= - i \tilde{f}_{\mbox{\bf\scriptsize BCA\normalsize\normalfont}}$.\fn{A is for adjoint and R is for reducible.}  
Because these are imaginary and antisymmetric, they are by definition Hermitian. 
Thus we can use the simple fact that Hermitian matrices are either irreducible or totally reducible.  
.  In the irreducible case, by definition there is no proper 
W $\subset$ V that is left invariant by the 
${\tilde{\tau}^{\mbox{\scriptsize A\normalsize}}_{\mbox{\bf\scriptsize A\normalsize\normalfont}}}$, 
ie there is no set of less than K linearly-independent vectors 
$(v_R)_{\mbox{\bf\scriptsize B\normalsize\normalfont}}$ such that 
$\tilde{\tau}^{\mbox{\scriptsize A\normalsize}}_{\mbox{\bf\scriptsize A\normalsize\normalfont}}(v_R)_{\mbox{\bf\scriptsize C\normalsize\normalfont}}$ 
is a linear combination of the $(v_{\mbox{\scriptsize R\normalsize}})_{\mbox{\bf\scriptsize C\normalsize\normalfont}}$.  
As $(\tilde{\tau}^{{\mbox{\scriptsize A\normalsize}}}_{\mbox{\bf\scriptsize A\normalsize\normalfont}})_{\mbox{\bf\scriptsize BC\normalsize\normalfont}} 
\propto \tilde{f}_{\mbox{\bf\scriptsize ABC\normalsize\normalfont}}$, this means that there is no 
set of linear combinations $\mbox{\Large $\tau$\normalsize}_R = 
(v_{\mbox{\scriptsize R\normalsize}})_{\mbox{\bf\scriptsize C\normalsize\normalfont}}
\tilde{\tau}^{{\mbox{\scriptsize A\normalsize}}}_{\mbox{\bf\scriptsize C\normalsize\normalfont}}$ 
that is closed under $|[\mbox{ },\mbox{ }]|$ with all the 
$(\tilde{t}^{{\mbox{\scriptsize A\normalsize}}}_{\mbox{\bf\scriptsize A\normalsize\normalfont}})$.  But such a set would provide 
the generators of an invariant subalgebra.  Hence the absence of such a set means by definition that 
the Lie Algebra is simple.

In the totally reducible case, then there must be a suitable choice of the 
$(\tilde{t}^{{\mbox{\scriptsize A\normalsize}}}_{\mbox{\bf\scriptsize A\normalsize\normalfont}})$ that are 
block-diagonal supermatrices 
$(\tilde{t}^{{\mbox{\scriptsize A\normalsize}}}_{\mbox{\bf\scriptsize A\normalsize\normalfont}})_{{\mbox{\bf\scriptsize ME\normalsize\normalfont}},{\mbox{\bf\scriptsize NF\normalsize\normalfont}}} 
= (t^{{\mbox{\scriptsize A\normalsize}}({\mbox{\bf\scriptsize M\normalsize\normalfont}})})_{\mbox{\bf\scriptsize EF\normalsize\normalfont}}
\delta_{\mbox{\bf\scriptsize MN\normalsize\normalfont}} $, where the constituent 
$\tau^{{\mbox{\scriptsize A\normalsize}}({\mbox{\bf\scriptsize M\normalsize\normalfont}})}_{\mbox{\bf\scriptsize A\normalsize\normalfont}}$ 
submatrices are either irreducible or vanish.  Adopting this basis also for the Lie algebra gives structure 
constants of the form
\be
\tilde{f}_{{\mbox{\bf\scriptsize LC\normalsize\normalfont}},{\mbox{\bf\scriptsize MA\normalsize\normalfont}},{\mbox{\bf\scriptsize NB\normalsize\normalfont}}} 
= i(\tilde{\tau}^{{\mbox{\scriptsize A\normalsize}}}_{\mbox{\bf\scriptsize LC\normalsize\normalfont}})_{{\mbox{\bf\scriptsize MA\normalsize\normalfont}},{\mbox{\bf\scriptsize NB\normalsize\normalfont}}}  
= i(\tilde{\tau}^{{\mbox{\scriptsize A\normalsize}}({\mbox{\bf\scriptsize M\normalsize\normalfont}})}_{\mbox{\bf\scriptsize LC\normalsize\normalfont}})_{\mbox{\bf\scriptsize AB\normalsize\normalfont}}
\delta_{\mbox{\bf\scriptsize MN\normalsize\normalfont}}
\ee
But, since $\tilde{f}$ is totally antisymmetric, this is proportional also to 
$\delta_{\mbox{\bf\scriptsize LN\normalsize\normalfont}}$ and $\delta_{\mbox{\bf\scriptsize LM\normalsize\normalfont}}$.  
So for any representation $\tau^{({\mbox{\bf\scriptsize M\normalsize\normalfont}})}_{\mbox{\bf\scriptsize A\normalsize\normalfont}} 
\equiv \tau_{\mbox{\bf\scriptsize MA\normalsize\normalfont}}$ of the Lie algebra in this basis,
\be
|[\tau^{({\mbox{\bf\scriptsize M\normalsize\normalfont}})}_{\mbox{\bf\scriptsize A\normalsize\normalfont}} , 
\tau^{({\mbox{\bf\scriptsize N\normalsize\normalfont}})}_{\mbox{\bf\scriptsize B\normalsize\normalfont}} ]| 
= i\delta_{{\mbox{\bf\scriptsize MN\normalsize\normalfont}}}
f^{({\mbox{\bf\scriptsize M\normalsize\normalfont}})}_{{\mbox{\bf\scriptsize CAB\normalsize\normalfont}}}
\tau^{({\mbox{\bf\scriptsize M\normalsize\normalfont}})}_{\mbox{\bf\scriptsize C\normalsize\normalfont}} 
\ee
for $f^{({\mbox{\bf\scriptsize M\normalsize\normalfont}})}_{\mbox{\bf\scriptsize CAB\normalsize\normalfont}}$ 
real and totally antisymmetric in \bf CAB\normalfont.  
This is what we mean by an algebra being a direct sum of subalgebras, and furthermore, 
for each \bf M\normalfont,  
$\tau^{({\mbox{\bf\scriptsize M\normalsize\normalfont}})}_{\mbox{\bf\scriptsize A\normalsize\normalfont}}$ 
is either irreducible (corresponding to a simple subalgebra), or zero (corresponding to 
a U(1) subalgebra).  
\be
\mbox{ } \mbox{ Finally, $\forall \mbox{ } v_{\mbox{\bf\scriptsize A\normalsize\normalfont}} \in 
\Re^{\mbox{\bf\scriptsize K\normalsize\normalfont}}(\mbox{\bf M\normalfont})$, }
\mbox{\hspace{1.2in}}
Q^{({\mbox{\bf\scriptsize M\normalsize\normalfont}})}_{{\mbox{\bf\scriptsize AB\normalsize\normalfont}}} 
= -f^{({\mbox{\bf\scriptsize M\normalsize\normalfont}})}_{\mbox{\bf\scriptsize ACD\normalsize\normalfont}}
f^{({\mbox{\bf\scriptsize M\normalsize\normalfont}})}_{\mbox{\bf\scriptsize BCD\normalsize\normalfont}}
v_{\mbox{\bf\scriptsize A\normalsize\normalfont}}V_{\mbox{\bf\scriptsize B\normalsize\normalfont}} \geq 0
\mbox{\hspace{1.2in}}
\ee
 since by antisymmetry it is a sum of positive quantities.
Furthermore, strict equality occurs only if $v_{\mbox{\bf\scriptsize A\normalsize\normalfont}} = 0$.  For suppose not.  
Then $u_{\mbox{\bf\scriptsize A\normalsize\normalfont}}
\tau^{({\mbox{\bf\scriptsize M\normalsize\normalfont}})}_{\mbox{\bf\scriptsize A\normalsize\normalfont}}$ 
is an invariant Abelian subalgebra, which is a contradiction, since the 
$t^{({\mbox{\bf\scriptsize M\normalsize\normalfont}})}_{\mbox{\bf\scriptsize A\normalsize\normalfont}}$ 
form a simple Lie algebra.

\noindent
(\ref{c}) $\Rightarrow$ (\ref{a})

(\ref{c}) means that in some basis 
$T^{({\mbox{\bf\scriptsize M\normalsize\normalfont}})}_{\mbox{\bf\scriptsize A\normalsize\normalfont}} 
= \zeta_{{\mbox{\bf\scriptsize MA\normalsize\normalfont}},{\mbox{\scriptsize A\normalsize}}}\tau_{{\mbox{\scriptsize A\normalsize}}}$.  
To construct $Q_{\mbox{\bf\scriptsize AB\normalsize\normalfont}}$, take 
$Q_{{\mbox{\bf\scriptsize MA\normalsize\normalfont}},{\mbox{\bf\scriptsize NB\normalsize\normalfont}}} 
\equiv Q^{({\mbox{\bf\scriptsize M\normalsize\normalfont}})}_{\mbox{\bf\scriptsize AB\normalsize\normalfont}}
\delta_{{\mbox{\bf\scriptsize MN\normalsize\normalfont}}}$, which is an arbitrary real symmetric 
positive-definite matrix since each 
$Q^{({\mbox{\bf\scriptsize M\normalsize\normalfont}})}_{{\mbox{\bf\scriptsize AB\normalsize\normalfont}}}$ is.
Then (\ref{a}) follows for each simple subalgebra, using first the Jacobi identity and then antisymmetry: 
\be
Q^{({\mbox{\bf\scriptsize M\normalsize\normalfont}})}_{{\mbox{\bf\scriptsize DF\normalsize\normalfont}}}
{f^{({\mbox{\bf\scriptsize M\normalsize\normalfont}}){\mbox{\bf\scriptsize D\normalsize\normalfont}}}}_{{\mbox{\bf\scriptsize AB\normalsize\normalfont}}}
\mbox{=}{{f^{({\mbox{\bf\scriptsize M\normalsize\normalfont}}){\mbox{\bf\scriptsize C\normalsize\normalfont}}}}_{\mbox{\bf\scriptsize AD\normalsize\normalfont}}}
{f^{({\mbox{\bf\scriptsize M\normalsize\normalfont}}){\mbox{\bf\scriptsize D\normalsize\normalfont}}}}_{\mbox{\bf\scriptsize BE\normalsize\normalfont}}
{f^{({\mbox{\bf\scriptsize M\normalsize\normalfont}}){\mbox{\bf\scriptsize E\normalsize\normalfont}}}}_{\mbox{\bf\scriptsize CF\normalsize\normalfont}}
\mbox{--}{{f^{({\mbox{\bf\scriptsize M\normalsize\normalfont}}){\mbox{\bf\scriptsize C\normalsize\normalfont}}}}_{\mbox{\bf\scriptsize FD\normalsize\normalfont}}}
{f^{({\mbox{\bf\scriptsize M\normalsize\normalfont}}){\mbox{\bf\scriptsize D\normalsize\normalfont}}}}_{\mbox{\bf\scriptsize BE\normalsize\normalfont}}
{f^{({\mbox{\bf\scriptsize M\normalsize\normalfont}}){\mbox{\bf\scriptsize E\normalsize\normalfont}}}}_{\mbox{\bf\scriptsize CA\normalsize\normalfont}}
\mbox{=--}Q^{({\mbox{\bf\scriptsize M\normalsize\normalfont}})}_{\mbox{\bf\scriptsize DA\normalsize\normalfont}}
{f^{({\mbox{\bf\scriptsize M\normalsize\normalfont}}){\mbox{\bf\scriptsize D\normalsize\normalfont}}}}_{\mbox{\bf\scriptsize FB\normalsize\normalfont}}.
\ee
This is also trivially true for the U(1) subalgebras since their structure constants vanish. $\Box$.

\mbox{ }

\noindent\bf{Corollary}\normalfont:  One can always define the scale of the gauge fields such that 
$Q_{\mbox{\bf\scriptsize AB\normalsize\normalfont}} = \delta_{\mbox{\bf\scriptsize AB\normalsize\normalfont}}$.

\noindent\bf{proof }\normalfont : $|[Q_{\mbox{\bf\scriptsize AB\normalsize\normalfont}}, 
\tau^{\Gamma}_{\mbox{\bf\scriptsize A\normalsize\normalfont}}]| = 0$ for totally-antisymmetric structure constants.
All the $t^{\Gamma}_{\mbox{\bf\scriptsize C\normalsize\normalfont}}$ can be put into block diagonal form, 
with irreducible or zero submatrices along the diagonal.  Then by Schur's Lemma, 
$Q_{\mbox{\bf\scriptsize AB\normalsize\normalfont}}$ must also be block-diagonal, with blocks of the same 
size and position as in the $t^{\Gamma}_{\mbox{\bf\scriptsize A\normalsize\normalfont}}$, 
and with the submatrix in each block being proportional to the unit matrix.  So 
$Q_{\mbox{\bf\scriptsize AB\normalsize\normalfont}}$ has the form 
$Q_{{\mbox{\bf\scriptsize MA\normalsize\normalfont}},{\mbox{\bf\scriptsize NB\normalsize\normalfont}}} = 
\mbox{\sffamily g\normalfont}_{\mbox{\bf\scriptsize M\normalsize\normalfont}}
\delta_{\mbox{\bf\scriptsize MN\normalsize\normalfont}} 
\delta_{\mbox{\bf\scriptsize AB\normalsize\normalfont}}$ 
where the $\mbox{\sffamily g\normalfont}_{\mbox{\bf\scriptsize M\normalsize\normalfont}}$ 
are arbitrary real positive numbers.  Then we can always rescale the gauge fields so that 
these are all equal to 1. $\Box$

\mbox{ }

\noindent\Large{\bf App IV.C: Teitelboim's inclusion of matter into the HKT route}\normalsize

\mbox{ }

\noindent One wishes to include matter fields $\Psi$ in such a way that the resulting 
$^{\Psi}{\cal H}$ and $^{\Psi}{\cal H}_i$ close as the Dirac Algebra (\ref{DiracAlgebra}).  
Teitelboim notes and uses that minimally-coupled scalars, electromagnetism and 
Yang--Mills theory have $^{\Psi}{\cal H}$ ultralocal in $h_{ab}$.  
It so happens for all these examples that 
\be
^{\Psi}{\cal H} = 
{\cal H} + 
{}_{\Psi}{\cal H}
\mbox{ } \mbox{ and } \mbox{ }
^{\Psi}{\cal H}_i = 
{\cal H}_i +
{}_{\Psi}{\cal H}_i 
\ee
where $_{\Psi}{\cal H}$ and $_{\Psi}{\cal H}_i$ are the matter contributions, and that 
$_{\Psi}{\cal H}$ and $_{\Psi}{\cal H}_i$ 
separately obey the Dirac Algebra \cite{Teitelthesis, Teitelboim}.  This is the result mentioned in IV.1.2.  

The minimally-coupled scalar is trivially ultralocal.  
For a single 1-form to work in this way, it is enforced that $D^m\pi_m$ is physically irrelevant, 
and thus one quickly arrives at electromagnetism.  
For many 1-forms, ultralocality and the inclusion of the single 1-form case enforces 
${\cal G}_{\mbox{\bf\scriptsize I\normalsize\normalfont}}  \equiv 
D_a\pi^a_{\mbox{\bf\scriptsize I\normalsize\normalfont}} + \mbox{\sffamily g\normalfont}_{\mbox{\scriptsize c\normalsize}} 
{C^{{\mbox{\bf\scriptsize K\normalsize\normalfont}}}}_{\mbox{\bf\scriptsize IJ\normalsize\normalfont}}
A_i^{\mbox{\bf\scriptsize J\normalsize\normalfont}}\pi^i_{\mbox{\bf\scriptsize K\normalsize\normalfont}}$ 
to be physically irrelevant.  Now, by Teitelboim's extra assumption that the 
${\cal G}_{\mbox{\bf\scriptsize I\normalsize\normalfont}}$ generate an internal symmetry, the Poisson 
bracket of two ${\cal G}_{\mbox{\bf\scriptsize I\normalsize\normalfont}}$'s must be a combination 
of these ${\cal G}_{\mbox{\bf\scriptsize I\normalsize\normalfont}}$'s: 
$\{ {\cal G}(\Lambda^1), {\cal G}(\Lambda^2) \} = {\cal G}(\Omega)$.  This enforces both 
\be
{C^{\mbox{\bf\scriptsize C\normalsize\normalfont}}}_{\mbox{\bf\scriptsize AB\normalsize\normalfont}} 
= - {C^{\mbox{\bf\scriptsize C\normalsize\normalfont}}}_{\mbox{\bf\scriptsize BA\normalsize\normalfont}} 
\ee
\be
\mbox{and, via }
\mbox{\hspace{1.8in}}
\Omega^{\mbox{\bf\scriptsize C\normalsize\normalfont}} = 
{C^{\mbox{\bf\scriptsize C\normalsize\normalfont}}}_{\mbox{\bf\scriptsize AB\normalsize\normalfont}} 
\Lambda_1^{\mbox{\bf\scriptsize A\normalsize\normalfont}}\Lambda_2^{\mbox{\bf\scriptsize B\normalsize\normalfont}} 
\mbox{ } , 
\mbox{\hspace{2.2in}}
\ee
\be
\mbox{the Jacobi identity }
\mbox{\hspace{1.12in}}
{C^{\mbox{\bf\scriptsize I\normalsize\normalfont}}}_{\mbox{\bf\scriptsize JK\normalsize\normalfont}}
C_{\mbox{\bf\scriptsize ILM\normalsize\normalfont}} + 
{C^{\mbox{\bf\scriptsize I\normalsize\normalfont}}}_{\mbox{\bf\scriptsize JM\normalsize\normalfont}}
C_{{\mbox{\bf\scriptsize IKL\normalsize\normalfont}}} + 
{C^{\mbox{\bf\scriptsize I\normalsize\normalfont}}}_{\mbox{\bf\scriptsize JL\normalsize\normalfont}}
C_{\mbox{\bf\scriptsize IMK\normalsize\normalfont}} = 0 
\mbox{ } .
\mbox{\hspace{1in}}   
\ee
From the appearance of these, Teitelboim deduces that the theory has gauge symmetry.  
He argues this to be a consequence of embeddability.  
As for the TSA, I observe that the resulting Lagrangian contains a 
positive-definite 
combination of internal indices of $F_{\mbox{\bf\scriptsize I\normalsize\normalfont}}^{ab}$  
from which the GMG theorem tells me that the corresponding Lie algebra is furthermore a direct sum 
of U(1) and compact simple Lie subalgebras.  

\vspace{2in}

\mbox{ }

\noindent\Huge{\bf V TSA: Discussion and interpretation}\normalsize

\mbox{ } 

\noindent\Large{\bf 1 TSA versus the principles of relativity}\normalsize

\mbox{ }

\noindent I take Wheeler's question about the form of the Hamiltonian constraint seriously.  
Starting from the relational 3-space ontology, the TSA gives Hamiltonian-type constraints 
\be
`{\cal H}^{\mbox{\scriptsize trial\normalsize}} \equiv 
\sqrt{h}(\sigma R + \Lambda + \mbox{\sffamily U\normalfont}_{\Psi})  
- \frac{ 1 }{  \sqrt{h}  }
\left\{
Y
\left(
p \circ p - \frac{X}{2}p^2
\right)
+ G_{\mbox{\sffamily\scriptsize AB\normalfont\normalsize}}
\Pi^{\mbox{\sffamily\scriptsize A\normalfont\normalsize}}\Pi^{\mbox{\sffamily\scriptsize B\normalfont\normalsize}}
\right\} = 0 
\ee
as identities from RI.    
Consistency alone then dictates what options are available for 
${\cal H}^{\mbox{\scriptsize trial\normalsize}}$ -- the Dirac approach.  
I have included matter fields $\Psi_{\mbox{\sffamily\scriptsize A\normalfont\normalsize}}$ 
since I have found that conclusions are best made only once this is done.  
$\Psi_{\mbox{\sffamily\scriptsize A\normalfont\normalsize}}$ is s.t  
$\mbox{\sffamily T\normalfont}^{\Psi}$ is homogeneous quadratic in its velocities and 
$\mbox{\sffamily U\normalfont}^{\Psi}$ at worst depends on connections (rather than their 
derivatives). I then get the following master equation for the propagation of the 
Hamiltonian-type constraint: 
$$
\dot{{\cal H}}^{\mbox{\scriptsize trial\normalsize}} \approx  \frac{2}{N}D^a
\left\{
N^2
\left(
Y
\left\{
\sigma
\left(
D^bp_{ab} + \{X\mbox{--}1\}D_ap
\right)
+
\left(
p_{ij}\mbox{--}\frac{X}{2}ph_{ij}
\right)
\left(
\frac{\pa \mbox{\sffamily U\normalfont}^{\Psi}}{\pa {\Gamma^c}_{ia}}h^{cj}\mbox{--} 
\frac{1}{2}\frac{\pa \mbox{\sffamily U\normalfont}^{\Psi}}{\pa{\Gamma^c}_{ij}}h^{ac}
\right)
\right\}
\right.
\right.
$$
\be
\mbox{ } \mbox{ } \mbox{} \mbox{ } \mbox{ } \mbox{} \mbox{ } \mbox{ } \mbox{} \mbox{ } \mbox{ } \mbox{} \mbox{ } \mbox{ } \mbox{} \mbox{ } \mbox{ } \mbox{}
\left.
\left.
+ G_{\mbox{\sffamily\scriptsize AB\normalfont\normalsize}}\Pi^{\mbox{\sffamily\scriptsize A\normalfont\normalsize}}
\frac{\pa \mbox{\sffamily U\normalfont}^{\Psi}}{\pa(\pa_a\Psi_{\mbox{\sffamily\scriptsize B\normalfont\normalsize}})}
\right)
\right\}
\mbox{ } .
\label{sku}
\ee

The strategy tied to the {\bf Galilean RP2} for this is to declare that $Y = 0$.  This kills 
all but the last factor.  It would then seem natural to take 
$\Pi^{\mbox{\sffamily\scriptsize A\normalfont\normalsize}} = 0$, whereupon the fields are not 
dynamical.  They are however \sl not \normalfont trivial: they include fields obeying 
analogues of Poisson's law, or Amp\`{e}re's, which are capable of governing a wide variety of 
complicated patterns. One would then have an entirely nondynamical `Galilean' world.  Although 
this possibility cannot be obtained from a BSW-type Lagrangian (the \sffamily T \normalfont 
factor is badly behaved), this limit is unproblematic in the Hamiltonian description.  Of course, 
the Hamiltonian-type constraint ceases to be quadratic: 
\be
{\cal H}_{(\mbox{\scriptsize Y = 0\normalsize})} = \sigma R + \Lambda + \mbox{\sffamily U\normalfont}_{\Psi} = 0 
\mbox{ } .
\ee
Now one might still vary with respect to the metric, obtaining a multiplier equation in place of the ADM 
(or BSW) evolution equation, $N(h^{ij}R - R^{ij}) = h^{ij}D^2N - D^iD^jN$.  In vacuum the 
trace of this and ${\cal H} = R = 0$ leads to $D^2N = 0$ which in the absence of privileged 
vectors implies that $N$ is independent of position so that clocks everywhere march in step.  
Then also $R_{ij} = 0$.  The cosmological constant alone cannot exist in an unfrozen CWB 
world.  But the inclusion of matter generally breaks these results.  One might well however 
not vary with respect to the metric and consider the worlds with a fixed spatial background metric.  
This includes as a particular case the Hamiltonian study of the flat spatial background world 
in the local square root version of App II.B, but permits generalization to curved backgrounds. 
  
The strategy tied to the {\bf Carrollian RP2} is to declare that $\sigma = 0$.  One still has 
the penultimate term so presumably one further declares that 
$\mbox{\sffamily U\normalfont}^{\Psi}$  contains no connections (the possibility of 
connections is studied more satisfactorily in VI).  It is `natural' then to take the second 
factor of the last term to be 0 thus obtaining a world governed by Carrollian relativity.  

The strategy tied to the {\bf Lorentzian RP2L} is somewhat more colourful.  Use $0 = - 1 + 1$, 
reorder and invent a momentum constraint: 
$$
\dot{{\cal H}}^{\mbox{\scriptsize trial\normalsize}}\mbox{$\approx$}\frac{2D^a}{N} 
\left(
N^2
      \left\{
Y
             \left(
\sigma
                   \left\{
                          \left(
D^bp_{ab}\mbox{--}\frac{1}{2}
\left[
\Pi^{\mbox{\sffamily\scriptsize A\normalfont\normalsize}}
\frac{\delta\pounds_{\xi}\Psi_{\mbox{\sffamily\scriptsize A\normalfont\normalsize}}}{\delta\xi}
\right]
                          \right)
\mbox{+}\frac{1}{2}
\left[
\Pi^{\mbox{\sffamily\scriptsize A\normalfont\normalsize}}
\frac{\delta\pounds_{\xi}\Psi^{\mbox{\sffamily\scriptsize A\normalfont\normalsize}}}{\delta\xi}
\right]
                   \right)
            \right\}
\mbox{+}G_{\mbox{\sffamily\scriptsize AB\normalfont\normalsize}}\Pi^{\mbox{\sffamily\scriptsize A\normalfont\normalsize}}
\frac{\pa \mbox{\sffamily U\normalfont}^{\Psi}}{\pa(\pa_a\Psi_{\mbox{\sffamily\scriptsize B\normalfont\normalsize}})}
      \right.
\right.
$$
\be
\mbox{ } \mbox{ } \mbox{ } \mbox{ } \mbox{ } \mbox{ } \mbox{ } \mbox{ } \mbox{ } \mbox{ } \mbox{ } \mbox{ } \mbox{ } \mbox{ } \mbox{ } \mbox{ } \mbox{ } \mbox{ }
\left.
       \left.
            + Y\sigma(X - 1)D_ap
            + Y
            \left(
p_{ij} - \frac{X}{2}ph_{ij}
            \right)
            \left(
            \frac{\pa \mbox{\sffamily U\normalfont}^{\Psi}}{\pa {\Gamma^c}_{ia}}h^{cj} 
          - \frac{1}{2}\frac{\pa \mbox{\sffamily U\normalfont}^{\Psi}}{\pa{\Gamma^c}_{ij}}h^{ac}
            \right)
      \right\}
\right) 
\mbox{ } . 
\ee 
Now go for the orthodox general covariance option: that the third and fourth terms cancel, 
enforcing the null cone.  This needs to be accompanied by doing something about the fifth term.  
One can furthermore {\sl opt} for the orthodox $X = 1$: the recovery of embeddability into 
spacetime corresponding to GR (RWR result), or for the preferred-slicing but York-GR-like 
worlds of $D_ap = 0$.  Either will do: the recovery of locally-Lorentzian physics does not 
happen for generally-covariant theories alone!  One requires also to get rid of the connection 
terms but the Dirac procedure happens to do this automatically for our big ans\"{a}tze.  Thus 
GR spacetime arises alongside preferred slicing, Carrollian and Galilean worlds, in which 
aspects of GR-like spacetime structure are not recovered.  

\mbox{ }  

With the above in mind, a clarification is required as regards the previous use of exhaustive proofs.  
The ultralocal and nondynamical strategies for dealing with the last term of (\ref{sku}) are available in {\sl all} 
the above options.  It may not shock the reader that degenerate and dual-degenerate possibilities might 
coexist.  Indeed Carroll matter in the Galilean option permits a BSW Lagrangian... But in the Lorentzian 
case this means \bf RP1 \normalfont is not fully replaced! At the moment, we do derive that gravitation 
enforces a unique finite propagation speed, but the possibility of fields with infinite and 
zero propagation speeds is not precluded.\fn{The latter is the analogue of the `each particle 
moving in its own potential' bad-ordering mechanics example in II.2.2.1.}  Thus the objection 
that Newtonian mechanics and electromagnetism have different relativities is precisely not 
being countered!  So in this approach, if one were to observe an analogue of electrostatics 
(a Poisson law), or of magnetostatics, one could not infer that there is a missing displacement 
current (or any other  appropriate individual `Lorentzifications' of electrostatics and magnetostatics in the 
absence of a good reason such as Faraday's Law to believe in unifying these two analogue 
theories).  One would suspect that formulating physics in this way would open the door to analogue Aethers 
coexisting in a universe with Einstein's equations.  

In more detail, BF\'{O} dismissed this possibility as trivial from counting arguments.  
But these are generally misleading, since they do not take into account the geometry of the 
restrictions on the solution space.  It is true that if there are more conditions than degrees 
of freedom then there is typically no solution, but some such systems will 
nevertheless have {\sl undersized} and not empty solution spaces.  

As a first example of this, consider the flat spacetime single 1-form case of App II.B.  
The crucial term is then $(1 - C)\pi_iF^{ij}$.  The $C = 1$ option gives the 
universal light-cone, but the other factors could be zero in a variety of situations: 
they mean a vanishing Poynting vector: \b{E} {\scriptsize $\times$ }\b{B} = 0.  This includes 
\b{E} = 0 (a fragment of magnetostatics), \b{B} = 0 (a fragment of electrostatics) and 
$\mbox{\b{E}} \mbox{ } || \mbox{ } \mbox{\b{B}}$.  Each of these cases 
admits a number of solutions.  These include complicated patterns analogous to those which 
can occur in electrostatics and magnetostatics, which could not be described as trivial.  

As a second example, consider the single 1-form in homogeneous curved spacetimes.  
\b{$\pi$} = \b{E} = 0 imposes a severe but not total restriction on Minisuperspace.  
The Bianchi types {\sl IV}, {\sl V}, {\sl VI} (h $\neq -1$), {\sl VII} 
(h $= 0$), {\sl VIII}, {\sl IX} are banned outright, whereas the fields in Bianchi types 
{\sl II}, {\sl VI} (h = $-1$), {\sl VII} (h $\neq 0$) have less degrees of freedom than 
expected pointwise in Einstein--Maxwell theory.\fn{h is a further invariant for types {\sl VI} 
and {\sl VII} given by $(1 - \mbox{h})
{L^{\mbox{\tiny\bf A\normalfont\normalsize}}}_{\mbox{\tiny\bf BA\normalfont\normalsize}}
{L^{\mbox{\tiny\bf D\normalfont\normalsize}}}_{\mbox{\tiny\bf CD\normalfont\normalsize}} 
= - 2\mbox{h}
{L^{\mbox{\tiny\bf A\normalfont\normalsize}}}_{\mbox{\tiny\bf DB\normalfont\normalsize}}
{L^{\mbox{\tiny\bf D\normalfont\normalsize}}}_{\mbox{\tiny\bf AC\normalfont\normalsize}}$ 
for 
${L^{\mbox{\tiny\bf A\normalfont\normalsize}}}_{\mbox{\tiny\bf BC\normalfont\normalsize}}$ 
the structure constants of each Bianchi model's associated Lie Algebra.}  Nevertheless, solutions exist 
(see p 202 of \cite{MC}).  
The treatment of \b{B} = 0 is identical to that of \b{E} = 0 by dual rotation.  
\b{E} $\mbox{ } || \mbox{ }$ \b{B} also admits nontrivial solutions such as the charged Taub 
metric, or its generalization on p 195 of \cite{MC}.  These are not trivial models.  Thus one has indeed an 
undersized but still interesting solution space.  Now, these solutions could all be 
interpreted as belonging not just to Einstein--Maxwell theory, but also to a theory T with 
Einstein cones and distinct 

\noindent(even degenerate) cones belonging to a strange 1-form theory.  

However, despite these examples illustrating non-triviality, {\bf RP1} is safe.  
For, the theory T permits no macroscopic 1-form propagations, since $E_i = 0$ means no 
momentum, $B_i = 0$ means the theory is ultralocal so 1-form information does not propagate 
away from any point, and $E_i \mbox{ } || \mbox{ } B_i$ means that there is none of the 
mutual orthogonality that ensures the continued propagation of light in electromagnetism.  
In the absence of such propagation, the concept of a 1-form particle moving in a background 
solution of theory T makes no sense (since this is but an approximation to the field equations 
of theory T, which permit no 1-form propagation).  Thus such a 1-form is causally irrelevant, 
so the recovery of {\bf RP1 } from the TSA is not affected.   

Moreover, one does have a source of potentially nontrivial scenarios from this insight: 
such nonpropagating Carroll or non-($c=1$) Lorentz or Galileo fields could nevertheless be coupled 
via potential terms to propagating fields, leading to scattering of the propagating fields.  
Whether such unusual fields are capable of producing interesting 
theoretical cosmology results may deserve further investigation. 

\mbox{ }

\noindent\Large{\bf 2 Discussion of conformal theories}\normalsize

\mbox{ }

\noindent\large{\bf 2.1 Discussion from ABF\'{O} paper}\normalsize

\mbox{ }
  
\noindent The conformal branch of the TSA provides examples of theories of evolving 
3-geometries that do not fit together to form spacetime.  Nevertheless, the classical 
bosonic physics corresponding to these theories has the standard locally Lorentzian form.   
Thus the universal null cone can appear alongside preferred slicing rather than embeddability.  

The differences between the conformal-gravity--matter and GR--matter metric ELE's are small: 
the absence of a term containing the `expansion of the universe' $p$ and the presence of a 
global term such as 
\be -
h^{ab}\frac{\sqrt{h}}{3\vc}\left<N\left(2(R + U^{\varsigma}_{(1)}) + 3\frac{U^{\varsigma}_{(0)}}{\vc}\dots\right)\right> 
\ee 
where scalar matter has been included.  Such a global term mimics the effect of a small 
epoch-dependent cosmological constant. This global term is a `cosmological force' because it 
occurs in the ELE's with proportionality to $h^{ab}$, just like the cosmological constant 
contribution does in GR. We expect it to be epoch-dependent because it contains matter field 
contributions, which will change as the universe evolves.  The occurrence of this global term 
should be compared with the scale-invariant particle model, in which there is a universal 
cosmological force induced by all the familiar forces of nature such as Newtonian gravity and 
electrostatics. There, this cosmological force is extremely weak over solar system scales but 
has a decisive effect on cosmological scales, ensuring the conservation of the moment of 
inertia.  Our global term is an action-at-a-distance term which is the price to pay for our 
particular implementation of scale invariance.  Some of our other conformal theories do not 
have the global features; correspondingly they allow for the volume of the universe to be a 
meaningful concept.  Thus these other theories are closer to GR than conformal gravity in this 
respect.  

Conformal gravity represents a new approach to scale invariance.  Its construction shows how 
BM and constraint propagation are powerful tools for a different way of constructing theories.  
Conformal gravity also highlights the thought-provoking manner in which GR only just fails to 
be fully scale invariant.

\mbox{ }

\noindent{\bf 2.1.1 On the weak field limit}

\mbox{ }

\noindent We expect that conformal gravity will pass the the solar-system tests just as GR does.  
This is because, first, the expansion of the universe does not play a role on such small 
scales in GR so its absence will not affect the results. Second, at maximal expansion, a data 
set may be evolved by both the GR and conformal gravity equations. The difference between 
these two evolutions is well defined in Riem.  Since the first derivatives match up at maximal 
expansion, the difference between the evolutions is small.  For sure, the size of the 
difference will depend on the global terms. But these can be made small by a well-known 
construction as far as the finite-time evolution for a patch of initial data that is 
substantially smaller than the radius of the universe is concerned.  Such patches can be 
constructed to contain a simple model solar system:  a patch of spherically-symmetric weak 
field regime on which test particles and test light-rays move.  I have found nothing to date to 
contradict this assertion, briefly noting that the conformal gravity system is the same as the 
ADM one to first order.   
I have not pursued this very far because the currently unresolved cosmological difficulties 
below stand out as stronger grounds on which to question conformal gravity.  
Binary pulsar calculations, which in GR involve the second-order Ricci tensor \cite{Wald}, 
may be harder to check than solar system tests.  Additionally, in GR these are based on 
a tensors on Minkowski spacetime approximation, which may not be applicable to conformal 
gravity.  Finally, conformal gravity is the extreme 
in difference from GR of the conformal theories.  The closer to GR, the less doubt there is 
about the reproduction of GR results.

\mbox{ }

\noindent{\bf 2.1.2 Cosmology}

\mbox{ }

\noindent  On account of the strong evidence from the Hubble redshift, nucleosynthesis and the 
microwave background it does seem unlikely that conformal gravity will be able to supplant 
the Big Bang cosmology.  Crudely, as conformal gravity currently stands 
there is none of the first and the grounds on which a standard `hot' explanation of the other 
two would be based are questionable (although Barbour illustrates some simple points in 
\cite{CGPD}).  So when writing ABF\'{O}, we considered rather the potential theoretical value 
of conformal gravity as a foil to the Big Bang. Theorists concerned with achieving the deepest possible 
understanding of cosmology and the foundations of physics value alternative models,\fn{For 
example, in \cite{CGconscience} an eternal singularity is presented as an alternative 
explanation for the isotropy of the microwave background; in \cite{CGconscience2} it is shown 
that an anisotropic, Bianchi {\sl V} universe can account for the correct light element 
abundances.} even if they explain or mimic only part of the whole picture. This is akin to the 
role of Brans--Dicke theory in solar-system tests of GR.  

Conformal gravity may require a more sophisticated approach to cosmology.  Consider the 
Friedmann--Lema\^{\i}tre--Robertson--Walker (FLRW) homogeneous and isotropic cosmologies, 
which are the backbone of standard cosmology. As solutions in which nothing changes except 
size and homogeneous intensity, they are suspiciously trivial from a dynamical viewpoint.  
In a scale-invariant theory, the FLRW-type 
solutions are merely static points in the configuration space.  There has long been concern 
\cite{CGKrasinski} about the accuracy with which FLRW cosmologies approximate more physically 
realistic inhomogeneous solutions of GR under the assumption that it is the correct physical 
theory. Conformal gravity raises a more serious doubt: it gives the possibility of physics and 
small-scale gravity that is in close agreement with GR but which differs greatly from GR cosmologically 
(and quantum-mechanically).  In comparison, the Brans--Dicke/dilatonic modifications of GR 
\cite{CGWetterich, buchdrag} have not been found to significantly affect the key physical 
basis of Big Bang cosmology.  

Since conformal gravity has no dynamics analogous to the FLRW universes of GR, the only 
possible direct progress in its cosmology would be through the study of anisotropic and 
especially inhomogeneous solutions.  This is the opposite emphasis to the norm in classical 
and quantum cosmology.  In ABF\'{O} this was said to ``have some chance to throw up a radical 
new explanation of the redshift"\fn{This is not the only way out of difficulty though.  I 
explain in the next subsection why I currently rather favour more conventional explanations.}.  
This was based on it being known that in GR, in addition to Hubble redshift, a change in 
clumpiness (shape) of the universe can cause redshift.  The solar photons that reach us are 
redshifted by having to climb out of the solar gravitational potential well (gravitational 
redshift), and inhomogeneities cause similar effects in cosmology (the integrated 
Sachs--Wolfe and Rees--Sciama effects \cite{CGshapeshift2, CGshapeshift1}).  The 
scale-invariant particle model is suggestive in this respect. We speculated that the 
rearrangement of geometry and matter of an evolving universe can cause a similar redshift in 
conformal gravity. In such a case, it will not be due to differences in the gravitational 
potential between different points of space but between different epochs.  Now, the potential 
can be changed either by a change of scale or by a change of shape. In conformal gravity, the 
former is not available and so the latter would have to be the origin of the observed 
cosmological redshift.  Since the change of shape of the universe can be observed, this should 
lead to testable predictions.

Conformal gravity may also offer a different perspective on singularities.  The Big Bang 
itself is an initial singularity where the known laws of physics break down.  It is inevitable 
in GR by theorems of Hawking \cite{CGHawking}.  These require the expansion of past-directed 
normal timelike geodesic congruences to be positive everywhere on a given spatial hypersurface.  
The GR form of these theorems will not hold in conformal gravity since such a notion of 
expansion is no longer meaningful.\fn{ We do not know if other forms of singularity 
theorem hold. We cannot so easily dismiss results involving null and/or local expansion.  
Another source of trouble in adapting GR proofs for conformal gravity will be the lack of a 
4-d generally covariant equivalent to the EFE's.  If local singularities form, they will 
contribute to the global terms in conformal gravity {\sl everywhere}. This could be fatal in 
the particle model, but in conformal gravity there may be two ways out.  First, singularities 
may only contribute a finite amount once integrated.  Second, there may be a tendency to 
preclude singularities in our conformal theories by the `collapse of the lapse' becoming more 
than a gauge effect (see V.2.3.3).} In GR, the Hubble redshift interpretation forces one to 
admit the breakdown of known physics in our finite past, whilst in conformal gravity, the 
denial of such a breakdown requires a new interpretation of the Hubble redshift. 

Whereas our greatest interest is in whether conformal gravity can give us an alternative 
cosmology, our CS+V theory has a notion of universal expansion, so it will be much closer to 
GR both in agreeing with the standard cosmology and in not offering these new perspectives on 
nonsingularity and global cosmological forces.  I provide more cosmological ideas in V.2.3.2.  

\mbox{ }

\noindent{\bf 2.1.3 Brief quantum outline}

\mbox{ }

\noindent We finish with a simple discussion toward quantization of conformal 
gravity \cite{CG, Sanderson}.  We wonder about what r\^{o}le 
${\cal H}_{\mbox{\scriptsize C\normalfont}}$ now plays.  This and the fundamental LFE 
(\ref{CGfullslicing}) are nonstandard objects from the quantization perspective. The new 
global terms may also play a role.  Whereas in GR the DeWitt supermetric gives an indefinite 
i.p as a consequence of the sign of the expansion 
contribution to the kinetic energy, in conformal gravity the new $W = 0$ supermetric gives 
a positive-definite i.p, altering the status of the i.p Problem.  Our quantum program is 
attractive in that conformal gravity has a marginally smaller configuration space than GR 
(rather than some choice of additional structures).  Note that these features are widespread 
throughout the various conformal theories.  
    
The study of some of the novel features of quantum conformal gravity can be isolated by the 
sequential study of $W = 0$ strong gravity, strong conformal gravity and conformal gravity. 
The effect of using a positive-definite $W = 0$ supermetric can be tried out in $W = 0$ 
strong gravity.  Then the additional effect of introducing a volume and of the r\^{o}le of 
${\cal H}^{\mbox{\scriptsize C\normalsize}}$ can be tried out in strong conformal gravity, while the  
additional conformal gravity complication of a nontrivial integro-differential LFE is absent here, 
from my observation that its counterpart here leads to $N$ being a spatial constant.

We hope to use a `top-down' approach.  However, we start from space rather than spacetime for 
relational reasons \cite{BB82, B94I, EOT, BOF, CGPD} and to illustrate that it is potentially 
misguiding to always presuppose and generalize {\sl spacetime} structure.  We hope to 
quantize in the timeless na\"{\i}ve Schr\"{o}dinger interpretation favoured by Barbour 
\cite{B94I, B94II, EOT}. 

The Problems of quantizing gravity are hopelessly interrelated, so that adding to a 
partial resolution to tackle further Problems can spoil that partial resolution \cite{CGK93}.  
III.1.5 contains a further example of this malady: it is not to be expected that Ashtekar 
variable techniques \cite{Ashtekar}, with their simplification of operator ordering and their 
natural regularization, could be imported into conformal gravity.  Thus, quantization of 
conformal gravity will differ from, but not necessarily be easier than, quantization of GR.  
Should conformal gravity adequately describe the classical universe, its quantization program 
will become more important.  Even if this were not the case, we expect to further the 
understanding of quantization and of quantum GR by the study of such theories as toy models.  

\mbox{ }

\noindent\large{\bf 2.2 Further conformal alternatives}\normalsize

\mbox{ }

\noindent{\bf 2.2.1 First-principles formulation of CS+V theory}

\mbox{ }

\noindent Here are some well-founded {\sl candidates} for CS+V theory.  These 
particular candidates use a `Laplacian implementation' of volume-preserving conformal transformations 
(VPConf) like in that in III.2.6, but are built adhering to various more clear-cut first 
principles.  Note that we are also in the process of considering alternative implementations 
\cite{ABFKO}.  The {\sl finite} Laplacian implementation of the VPConf transformations in 
III.2.6, 
\be
h_{ab} \longrightarrow \tilde{h}_{ab} = (1 + D^2\xi)^{\frac{2}{3}}h_{ab}
\label{lrgtrans} 
\mbox{ } ,
\ee
is unsatisfactory, for \'{O} Murchadha and I independently found that these do not close to 
form a group.  Rather, 
\be
h_{ab} 
\stackrel{\zeta_1}{\longrightarrow}
\tilde{h}_{ab} 
\stackrel{\zeta_2}{\longrightarrow}
\tilde{\tilde{h}}_{ab} = 
\left(
1 + D^2\zeta_1 + \frac{        (1 + D^2\zeta_1)D^2\zeta_2 + \frac{1}{3}\pa_c\zeta_2\pa^cD^2\zeta_1     }{(1 + D^2\zeta_1)^{\frac{2}{3}}       }
\right)
^{\frac{2}{3}}h_{ab}
\ee
in terms of the Laplacian corresponding to the original metric, which is not of the form 

\noindent $(1 + D^2\Xi)^{\frac{2}{3}}h_{ab}$ for some $\Xi(\xi_1, \xi_2)$.  This is related to 
$D^2$ not being a conformally-covariant object.  
Moreover, there are no simple conformally-covariant differential operators acting on scalars 
that are plausible here.  


But (as presented in II), I realized that the form of BM is dictated by the {\sl infinitesimal} 
transformations associated with the generators.  Now, the 
\be
\mbox{clearly VP 
infinitesimal version of (\ref{lrgtrans}) }
\mbox{\hspace{0.8in}}
h_{ab} \longrightarrow \tilde{h}_{ab} = (1 + \frac{2}{3}D^2\xi)h_{ab}
\mbox{\hspace{0.8in}}
\ee
\be
\mbox{{\sl do} close to form a group: }
\mbox{\hspace{0.4in}}
h_{ab} 
\stackrel
{\zeta_1}
{\longrightarrow}
\tilde{h}_{ab} 
\stackrel
{\zeta_2}
{\longrightarrow}
\tilde{\tilde{h}}_{ab} = 
\left(
1 + \frac{2}{3}D^2(\zeta_1 +\zeta_2) + O(\zeta^2)
\right)
h_{ab} 
\mbox{ } . 
\mbox{\hspace{0.4in}}
\ee
Thus one should rather use this infinitesimal Laplacian implementation, which in fact 
reverts to earlier drafts of the paper \cite{CG}.  

I want an action such that part of the variation with respect to auxiliaries gives the CMC condition 
$\frac{p}{\sqrt{h}} = C(\lambda)$ and another part gives the correct LFE to 
propagate this.  I offer two approaches to this.  Both follow from my arbitrary frame 
principle.  The difference between them arises from there being two different kinds of 
covariance.

\mbox{ }

The `York style' is to write the action in the arbitrary conformal frame out of good 
conformally covariant objects, bearing in mind that the theory 
{\sl only has a temporary technically-convenient conformal gauge symmetry} 
(with respect to the bona fide metric scale-factor $\psi$), since the Lichnerowicz equation then 
gauge-fixes $\psi$ by specifically mapping to a particular point on the Conf orbit.\fn{Thus this use 
of gauge theory is different from the U(1), Yang--Mills and Diff uses of gauge theory.  
There, the choice of gauge is unphysical, whereas here it is physical because it is 
prescribed by an additional condition required for passage to the physical 3-metric.}  
Consequently there is no Conf BM in this approach: conformally-bare velocities are to be 
regarded as conformally covariant.  This conformal mathematics underlies both York's work and 
our attempt to write down a CS+V theory in III.2.6, and explains the underlying technical 
similarities.  In this approach, what might have been regarded as the trace and tracefree 
parts of a single tensor are rather regarded as distinct objects which are alloted distinct 
conformal rank as befits the formation of conformally-covariant derivatives.  In particular, 
a relative scaling of the 
$u^{\mbox{\scriptsize T\normalsize}}\circ u^{\mbox{\scriptsize T\normalsize}}$ and $u^2$ of 
$\phi^{12}$ arises.  This corrects the na\"{\i}ve mismatch in amount of $Np^2$ between the 
auxiliary variation LFE and the LFE required to propagate the CMC condition.\fn{I use 
$u \equiv tr(\mbox{\ss}_{\xi}h_{ij})$ and $u^{\mbox{\tiny T\normalsize}}_{ij} 
\equiv (\mbox{\ss}_{\xi}h_{ij})^{\mbox{\tiny T\normalsize}}$.} In this approach, there 
are two distinct auxiliaries, and the one encoding the CMC condition ($\eta$) 
should probably be regarded as a multiplier, not a best-matching.  The action is 
\be
\mbox{\sffamily I\normalsize} = 
\int d\lambda\int d^3x\sqrt{h}
\left(
1 + \frac{2\triangle\zeta}{3} 
\right)
\sqrt{
                                                \left( 
                                                R - \frac{4\triangle^2\zeta}{3}
                                                \right)                                       }
                                    \sqrt{      u^{\mbox{\scriptsize T\normalsize}}\circ u^{\mbox{\scriptsize T\normalsize}} 
                                                - \frac{2}{3}(1 - 2\triangle\zeta)
                                                \left(
                                                u + \triangle\eta
                                                \right)^2                                       }  
\mbox{ } .
\ee
This is (the infinitesimal version of) what was considered in III.2.6.  

\mbox{ }

The `Barbour style' is to consider an action with a true conformal symmetry, with scale factor 
$\omega$.  Then $\phi$ is neither the symmetry scale factor nor a bona fide metric scale-factor: 
it is an extraneous auxiliary whose role is to make the conformal symmetry true via its 
banal transformation property.      
Then VPConf-BM is indeed required\fn{It is in fact only a nontrivial correction to $u$ since 
$h^{\mbox{\tiny T\normalsize}}_{ij} = 0$.} to make $\dot{h}_{ij}$ into a good VPConf object.  
The natural action is then 
\be
\mbox{\sffamily I\normalfont} = \int\textrm{d}\lambda \int\textrm{d}^3x\sqrt{h}\phi^6
                                \sqrt{\phi^{-4}
                                \left(
                                R - \frac{8\triangle \phi}{\phi}
                                \right)} 
                                \sqrt{G^{ijkl}
                                \left(
                                \dot{h}_{ij} + \frac{4\dot{\phi}}{\phi}h_{ij}
                                \right)
                                \left(
                                \dot{h}_{kl} + \frac{4\dot{\phi}}{\phi}h_{kl}
                                \right)}
\ee
with $\phi = 1 + \frac{\triangle\zeta}{6}$, $\zeta$ infinitesimal, so that 
\be
\mbox{\sffamily I\normalsize} = 
\int d\lambda\int d^3x\sqrt{h}
\left(
1 + \frac{2\triangle\zeta}{3} 
\right)
\sqrt{
                                                \left( 
                                                R - \frac{4\triangle^2\zeta}{3}
                                                \right)                                       }
                                    \sqrt{      u^{\mbox{\scriptsize T\normalsize}}\circ u^{\mbox{\scriptsize T\normalsize}} - \frac{2}{3}
                                                \left(
                                                u + 2\frac{\pa (\triangle\zeta)}{\pa \lambda}
                                                \right)^2                                       }  
\mbox{ } .
\label{Barac}
\ee
\be
\mbox{The momenta are }
\mbox{\hspace{1.7in}}
\mbox{\tt p}^{ij} = p^{ij} + \frac{D^k(pD_k\zeta)}{3}h^{ij}
\label{CS+Vmom}
\mbox{\hspace{1.7in}}
\ee
\be
\mbox{for $p^{ij}$ the GR expression for momentum, and }
\mbox{\hspace{1.1in}}
p^{\zeta} = \frac{2}{3}\triangle p
\mbox{\hspace{1.9in}}
\ee
Then free-endpoint variation and the trace of (\ref{CS+Vmom}) give 
\be
\frac{p}{\sqrt{h}} = C(\lambda) \mbox{ } , \mbox{ } \mbox{\tt p} = p(1 + \triangle\zeta) \mbox{ i.e. }
\mbox{\tt p} = p\phi^6
\ee
so that the na\"{\i}ve $p$ is replaced by the scaled-up {\tt p}.

Note that (\ref{Barac}) does not have a relative scaling of $\phi^{12}$ 
between $u^{\mbox{\scriptsize T\normalsize}}\circ u^{\mbox{\scriptsize T\normalsize}}$ and $u^2$.  
Before, the exclusion of this led to a mismatch in the $Np^2$ terms between the $\zeta$-variation 
and $\frac{p}{\sqrt{h}} = C(\lambda)$ propagation LFE's, so one might expect (\ref{Barac}) to be 
inconsistent.  However, it turns out that $\zeta$-variation of (\ref{Barac}) has a new source of 
$Np^{2}$ terms.  
\be
\frac{\delta}{\delta\zeta}
\left[
\frac{\pa(\triangle\zeta)}{\pa\lambda}
\right] 
= \triangle
\left(
\frac{Np}{2\sqrt{h}}
\right)
\ee
by the $W$ = $Y$ = 1 cases of (\ref{hupstairs}) and (\ref{CDOT}).  This term occurs 
with cofactor numerically proportional to $\frac{p}{\sqrt{h}}$, and as this is constant, it can be 
taken inside $\triangle$, and happens to contribute just the right amount of $Np^2$ to the 
$\zeta$-variation LFE to obtain the CMC LFE.

Note also that a `$\phi^{12}$' relative scaling emerges in the primary constraint arising from 

\noindent squaring the momenta, due to the nonminimal coupling:    
\be
G_{ijkl}\mbox{\tt p}^{ij}\mbox{\tt p}^{kl} = 
\mbox{\tt p}^{\mbox{\scriptsize T\normalsize}} \circ \mbox{\tt p}^{\mbox{\scriptsize T\normalsize}} 
- \frac{\mbox{\tt p}^2}{6} = 
p^{\mbox{\scriptsize T\normalsize}} \circ p^{\mbox{\scriptsize T\normalsize}} - \frac{p^2}{6}(1 + 2\triangle{\zeta}) \mbox{ } .
\ee
\be
\mbox{This is }
\mbox{\hspace{1.4in}}
\phi^4\left(R - \frac{8\triangle\phi}{\phi}\right) = p^{\mbox{\scriptsize T\normalsize}} \circ p^{\mbox{\scriptsize T\normalsize}} - \phi^{12}\frac{p^2}{6} 
\mbox{ } ,
\mbox{\hspace{1.8in}}
\ee
which may be identified 
with the Lichnerowicz--York equation.  The above working could be seen as an alternative 
derivation of this equation.

Next, note that we came across the above via setting the 
$D^a(ND_ap)$ factor in (\ref{keyterm}) to 0, but they are also resliceable since $W = 1$ 
i.e the constraint algebra works just as well if CMC slices {\sl are not} chosen.  Thus 
I identify the above as not being alternative theories but new formulation of CWB GR in the 
CMC gauge.  However there is no longer any consistency reason for setting $W = 1$.  
So I can consider the arbitrary-$W$ versions of the two approaches above in order to 
obtain alternative theories of gravity which have genuine privileged slicings.  These follow 
from replacing the $-\frac{2}{3}$'s in the actions by $\frac{1- 3W}{2}$.  Thus, if one takes 
York's IVP mathematics more seriously than GR itself, then one is entitled to consider a range 
of privileged-slicing theories in addition to resliceable GR.

\mbox{ }

\noindent{\bf 2.2.2 Recovery of conformal gravity}   

\mbox{ }

\noindent The single-auxiliary formulation of conformal gravity is clearly in the 
`Barbour style'.  The 2-auxiliary formulation is not quite in the `York style', but can 
be easily made to comply with it since the scaling of $u^2$ is rendered irrelevant for 
conformal gravity because $p = 0$ arises.  
The action is then
\be
\mbox{\sffamily I \normalfont} = 
\int d\lambda
\frac{        \int d^3x\sqrt{h}\phi^6\sqrt{            \phi^{-4}
                                                       \left( 
                                                       R - \frac{8D^2\phi}{\phi}
                                                       \right)                             }
                                     \sqrt{            u^T\circ u^T - 
                                                       \left(
                                                       W - \frac{1}{3}
                                                       \right) 
                                                       \left(
                                                       u + \theta 
                                                       \right)
                                                       ^2
                                                       \frac{V^2}{\phi^{12}}                }                       }
     {                    V^{\frac{2}{3}}                                                                           }
\mbox{ } .    
\ee
Note also how $W = \frac{1}{3}$ (not $W = 0$) now clearly gives the simplest presentation (but 
this is the intractable degenerate case).  

\mbox{ }

\noindent{\bf 2.2.3 Further alternative conformal theories}   

\mbox{ }

\noindent The survey of alternative theories in III.2.6 is not exhaustive.   
First, recall that the trick of dividing through by the volume was adopted in the maximal CWB 
case to allow the LFE to work.  However, there is nothing stopping one also trying to divide through by 
the volume in the CMC CWB case, even though the LFE is already well-behaved 
without this correction.  This gives Kelleher's theory \cite{Kellehertheory, Kelleher}, 
\be
\mbox{\sffamily I\normalfont}_{\mbox{\scriptsize \normalsize}} = \int\textrm{d}\lambda 
\frac{      \int\textrm{d}^3x\sqrt{h}{\phi}^4\sqrt{R - \frac{8D^2\phi}{\phi}}
\sqrt{u^{\mbox{\scriptsize T\normalsize}} \circ u^{\mbox{\scriptsize T\normalsize}} 
- \frac{2}{3}\frac{V^2}{{\phi}^{12}}(u - D^2\theta)^2}     }{      V^{\frac{2}{3}}      }
\mbox{ } . 
\label{Kelltheo}
\ee
\be 
\mbox{This turns out to be restrictive: }
\mbox{\hspace{1.4in}}
\tau_{\mbox{\scriptsize Y\normalsize}} = \mbox{{\sl fully} constant} 
\mbox{ } ,
\mbox{\hspace{2in}}
\ee
is enforced giving an interesting (almost certainly too) restrictive cosmological scheme 
(see V.2.3.2).  

Another alternative theory is my

\be
\mbox{\sffamily I\normalfont} = \int\textrm{d}\lambda \int\textrm{d}^3x\sqrt{h}
\dot{\phi}^4\sqrt{      R - \frac{8D^2\dot{\phi}}{\dot{\phi}}      }
\sqrt{u^{\mbox{\scriptsize T\normalsize}} \circ u^{\mbox{\scriptsize T\normalsize}} 
- \frac{2}{3}\dot{\phi}^{-12}(u - D^2\theta)^2}
\label{Andertheo}
\ee
for $\dot{\phi}$ varied standardly rather than freely flapping.  Then consistency requires 
the York time 

\noindent to have the freedoms of Newtonian absolute time: 
\be
\tau_{\mbox{\scriptsize Y\normalsize}} = C\lambda + D \mbox{ } : \mbox{ } C \mbox{ } ,  \mbox{ } 
D \mbox{ {\sl fully} constant   }
\ee
$$
\mbox{from $\frac{p}{\sqrt{h}} = F(\lambda \mbox{ alone})$ and }
\dot{F}(\lambda) = \frac{\pa }{\pa\lambda}
\left(
\frac{p}{\sqrt{h}}
\right) 
= -2
\left[
D^2N - N
\left(
R + \frac{p^2}{4h}
\right) 
\right]
= Q(x \mbox{ alone}) 
\mbox{ }   
\mbox{\hspace{2in}} 
$$
where the last step is by the auxiliary variation.  Hence $\dot{F} = Q$, a total constant 
(and $C = \frac{2Q}{3}$).  Thus I name this theory {\it Newton--York absolute time theory}.  
It is intermediate in restrictiveness between GR and Kelleher's theory.  I found that all the
other combinations of types of auxiliary did not yield any further distinct consistent theories. 

Note that all bar CS+V and CG are 2-auxiliary `York style' theories, with their $\phi^{12}$ 
relative scalings.  I am not sure whether the choice of 
implementation (out of the above two, the Laplacian one, or the one in \cite{ABFKO}) 
affects precisely which `CS+V' configuration space is involved.  We consider this well 
worth investigating, given the connection between CS+V and a representation of the CWB GR 
d.o.f's.  

\mbox{ }

\noindent\large{\bf 2.3 Discussion and interpretation}\normalsize

\mbox{ }

\noindent{\bf 2.3.1 Conformal gravity and CS+V theory as PDE systems}

\mbox{ }

\noindent{\bf 2.3.1.1 Traditional thin sandwiches}

\mbox{ }

\noindent For conformal gravity, there is no Problem of zeros globally for all of its 
solutions in the classic thin sandwich approach since the potential is of fixed sign.  The 
impasse may then be instead whether such fixed sign potentials are capable of describing 
moderately complicated astrophysics.  For `York-style' CS+V theories, differences in $W$ should  
again alter the behaviour of the thin sandwich formulation.  For both conformal gravity and 
the CS+V theories, the auxiliary and its $\lambda$-derivative (or the second auxiliary) 
become involved so that one has not the usual sort of 3-equation thin sandwich system, but 
rather an extended 5-equation system.  

Here is the conformal gravity vacuum traditional thin sandwich system for $\xi$, $\phi$ 
and $\dot{\phi}$:  
$$
D^a
\left[
\sqrt{                  \frac{    R - \frac{8D^2\phi}{\phi}   }{       h^{ik}h^{jl}    
\left(
\dot{h}_{ij} - 2D_{(i}\xi_{j)} + 4\frac{\dot{\phi} - \xi^p\pa_{p}\phi}{\phi}h_{ij}
\right)
\left(
\dot{h}_{kl} - 2D_{(k}\xi_{l)} + 4\frac{\dot{\phi} - \xi^q\pa_{q}\phi}{\phi}h_{kl}
\right)
     }                       }
\right.
$$
\be
\left.
\times
\left(
\dot{h}_{ab} - 2D_{(a}\xi_{b)} + 4\frac{\dot{\phi} - \xi^r\pa_{r}\phi}{\phi}h_{ab}
\right)
\right] 
= 0
\mbox{ } , 
\ee
$$
h^{ij}\dot{h}_{ij} - 2D^i\xi_i + 4\frac{\dot{\phi} - \xi^i\pa_i\phi}{\phi} = 0 
\mbox{ } , 
$$ 
$$
\phi^3 
\sqrt{                  \frac{
(h^{ik}h^{jl}    
\left(
\dot{h}_{ij} - D_{(i}\xi_{j)} + 4\frac{\dot{\phi} - \xi^p\pa_{p}\phi}{\phi}h_{ij}
\right)
\left(
\dot{h}_{kl} - D_{(k}\xi_{l)} + 4\frac{\dot{\phi} - \xi^q\pa_{q}\phi}{\phi}h_{kl}
\right)
     }{    R - \frac{8D^2\phi}{\phi}   }                       }
\left(
R - \frac{7D^2\phi}{\phi}
\right) 
$$
$$
- D^2
\left(
\phi^3
\sqrt{                  \frac{
(h^{ik}h^{jl}    
\left(
\dot{h}_{ij} - 2D_{(i}\xi_{j)} + 4\frac{\dot{\phi} - \xi^p\pa_{p}\phi}{\phi}h_{ij}
\right)
\left(
\dot{h}_{kl} - 2D_{(k}\xi_{l)} + 4\frac{\dot{\phi} - \xi^q\pa_{q}\phi}{\phi}h_{kl}
\right)
     }{    R - \frac{8D^2\phi}{\phi}   }                       }
\right)
$$
\be
= \phi^5
\left<
\phi^4
\sqrt{                  \frac{
(h^{ik}h^{jl}    
\left(
\dot{h}_{ij}\mbox{--}2D_{(i}\xi_{j)}\mbox{+}4\frac{\dot{\phi}\mbox{--}\xi^p\pa_{p}\phi}{\phi}h_{ij}
\right)
\left(
\dot{h}_{kl}\mbox{--}2D_{(k}\xi_{l)}\mbox{+}4\frac{\dot{\phi}\mbox{--}\xi^q\pa_{q}\phi}{\phi}h_{kl}
\right)
     }{    R - \frac{8D^2\phi}{\phi}   }                       }
\left( 
R\mbox{--}\frac{8D^2\phi}{\phi} 
\right) 
\right> 
\mbox{ } .
\ee
The Barbour-style CS+V traditional thin sandwich system for $\xi$, $\zeta$ and $\dot{\zeta}$ is 
messy once I substitute for $N(\xi, \zeta,\dot{\zeta})$ so I do not give it explicitly.  
One feature is that the 3-vector equations are now fifth order whereas the LFE is sixth order, 
both due to the $D^4\zeta$ in the potential.  The $W = 1$ case is an as-yet  
untested scheme for a (piece of CWB) GR.  Were it to have good p.d.e properties, it would be of 
considerable direct interest both in numerical relativity and in quantization.  But it is 
plausible it will pick up difficulties from the traditional thin sandwich subsystem it 
includes.  However, allowing for other values of $W$ gives elliptic thin sandwich operators (see C.3).   Thus it may be 
so it may be that some working (toy?) worlds emerge from such a study, in which some 
of Wheeler's old classical and quantum hopes could be tried out.  

\mbox{ }

\noindent{\bf 2.3.1.2 Conformal thin sandwiches}

\mbox{ }

\noindent 
I next consider the `York style' $W = 1$ CS+V conformal thin sandwich.   This is messy if 
written using $1 +D^2\zeta$ for $\phi$, so I keep this as a later admissibility restriction 
to be imposed on the $\phi$ solutions.  Collecting together the $W = 1$ CS+V LFE, 
$D_ip^{ij} = 0$ and the `$W = 1$ CS+V Lichnerowicz--York' equation, the conventional GR 
conformal thin sandwich of I.2.9.4.1 arises once the equations are re-expressed in terms of 
$\alpha_{\mbox{\scriptsize Y\normalsize}}$ (s.t $\tilde{\alpha}_{\mbox{\scriptsize Y\normalsize}} 
= \phi^6{\alpha}_{\mbox{\scriptsize Y\normalsize}}$ rather than $\tilde{N} = \phi^2N$), and 
the Lichnerowicz--York equation along with the product rule is used in the LFE to eliminate 
$D^2\phi$.  

Next I consider the conformal gravity conformal thin sandwich formulation.  Collect the 
conformal gravity LFE, $D_{i}p^{ij} = 0$  and the `conformal gravity Lichnerowicz equation' 
(\ref{CGfullham}), write in terms of $\alpha_{\mbox{\scriptsize Y\normalsize}}$, and use the 
product rule along with (\ref{CGfullham}) to eliminate $D^2\phi$ in the LFE, thus obtaining:
$$
-\phi^2D^2\alpha_{\mbox{\scriptsize Y\normalsize}} - 14\phi^2\pa_a\phi \pa^a\alpha_{\mbox{\scriptsize Y\normalsize}} 
- 42\pa_a\phi\pa^a\phi \alpha_{\mbox{\scriptsize Y\normalsize}} \mbox{\hspace{1.8in}}
$$
\be
+ \frac{7\alpha_{\mbox{\scriptsize Y\normalsize}}}{4\phi^{6}}K^{\mbox{\scriptsize T\normalsize}} 
\circ K^{\mbox{\scriptsize T\normalsize}} - \frac{3}{4}\phi^2\alpha_{\mbox{\scriptsize Y\normalsize}} R 
\mbox{ } = \mbox{ } V^{\frac{1}{3}}\phi^4\int \sqrt{h}d^3x\alpha_{\mbox{\scriptsize Y\normalsize}} K^{\mbox{\scriptsize T\normalsize}}\circ K^{\mbox{\scriptsize T\normalsize}}
\mbox{ } ,
\ee
\be
D_i
\left(
\frac{\psi^{10}}{2\alpha_{\mbox{\scriptsize Y\normalsize}}   }
[(|L\beta)^{ij} - \dot{h}^{\mbox{\scriptsize T\normalsize}ij}]
\right) 
= 0 
\mbox{ } ,
\ee
\be
\frac{\phi^4}{V^{\frac{2}{3}}}
\left( 
R - \frac{8D^2\phi}{\phi} 
\right)
- \frac{\vc}{\phi^4} K^{\mbox{\scriptsize T\normalsize}}\circ K^{\mbox{\scriptsize T\normalsize}} = 0
\mbox{ } .
\ee 
This differs from GR's conformal thin sandwich and is thus a starting-point for conformal 
gravity predictions to compete with GR's as regards colliding compact objects.  Note however 
that some things will be shared: for moment of time-symmetry data (see App C, I.2.11), GR and 
the above are identical. Just as well numerical relativity is developing away from such overly 
simple cases!  Note that one can also build arbitrary-$W$ CS+V theory conformal thin 
sandwiches as a testbed for numerical relativity.  
 
\mbox{ }

\noindent{\bf 2.3.2.3 IVP--CP formulation}

\mbox{ }

\noindent 
For conformal gravity, one would first solve the 4 IVP equations for $\phi$ and the 
longitudinal vector potential $W_i$, then redeclare $\phi$ and $h_{ij}$ so that $\phi = 1$ and 
then tackle the evolution problem in the $\phi$-distinguished representation.  As the IVP's 
are identical in all relevant ways to the GR IVP, they are therefore as well-posed.  One then 
faces a CP step coupled to the solution of the LFE.  As a procedure this is not novel: its 
analogue in GR is the maintenance of a (partial) gauge-fixing in the to study of the evolution 
equations.  The LFE's in question here are also technically close to the usual 
ones in theoretical numerics.  The novelty is in some of LFE's being {\sl enforced}. 

It may be problematic that GR's p.d.e theorems are strictly tied to the harmonic gauge, which 
may be of limited computational value.  This becomes harsher for the above conformal 
theories since now there is no apparent right to use harmonic coordinates, so there are no 
known theorems at all.  

\mbox{ }

\noindent{\bf 2.3.2 Cosmology}

\mbox{ }

\noindent Whereas inhomogeneity can produce redshift, it is unlikely to produce anywhere near 
as much redshift (around 6) as is required to account for observations of distant galaxies.  
The standard cosmology furthermore makes use of much larger redshifts than this so as to 
explain features believed to be of earlier\fn{In the case of element abundances, the observed 
abundance is explained both in terms of early-universe nucleosynthesis and the more 
contemporary influence of astrophysics.} origin than galactic structures, such as the 
microwave background or the abundances of the light elements.  The GR cosmological effects 
mentioned above are capable of producing fluctuations in redshift but are small compared to 
the Hubble redshift.  It is also worth noting that in GR the maximum gravitational redshift from 
light emerging from a star is 2 \cite{Wald}.  It remains to be checked whether these results also occur within conformal gravity.  
Finally if one were to attempt to explain larger 
redshifts cumulatively from clumpy effects, one would be faced with unobserved anisotropy 
correlated with the presence of intervening galactic structures.  


I rather propose some simpler, more conventional solutions to this cosmological difficulty.    
First, note that if the universe is described by one of our conformal theories, it need not be 
conformal gravity.  The CS+V theories permit the standard GR cosmology if we so wish (though 
one might choose an appreciably different $W$).  This might be regarded as being too similar 
to GR to generate interest as an alternative cosmology.  However, there are theories of 
`intermediate restrictiveness' between conformal gravity (no standard cosmology) and the CS+V 
theories (roughly the same as the standard cosmology).  These examples {\sl do} achieve some of 
the goals of alternative cosmologies beyond the illustrative value of the nonexistence of 
standard cosmology in conformal gravity as originally conceived.   

For example, in Kelleher's theory \cite{Kelleher}, redshift is clearly available ($p =$ fully 
constant rather than conformal gravity's $p = 0$), and its form is most rigid compared to GR 
(where $p$ can vary with cosmic time.).  Furthermore, the isotropic cosmology ansatz rigidly 
leads to a scale-factor $a = t^{\frac{1}{4}}$ independently of the matter content: 
\be
\mbox{totally constant} = \frac{p}{\sqrt{h}} = V^{\frac{4}{3}}\frac{\dot{a}}{a} \propto a^3\frac{da}{dt} 
\mbox{ } .
\ee
Both the rigidity of this form and its insensitivity to the matter are completely unlike GR.  
From a GR perspective this looks like a flat universe with a peculiar but physically-admissible 
equation of state, and yet in truth it is a closed universe corresponding to a conformal theory.  
Of course, $a = t^{\frac{1}{4}}$ does not fit current observations well, and corresponds to a 
universe believed to be younger than some of its constituent stars!  Nevertheless,  
this example deserves further investigation to delineate which of its predictions agree with GR 
and which differ.  It may be seen as too rigid to be realistic.  To this I offer three ways out 
(some of which may be combined), which I leave as a future project.  

First, there is no guarantee that this $a = t^{\frac{1}{4}}$ is a stable attractor of 
inhomogeneous solutions.  It may be that the theory permits other more typical asymptotic 
behaviours to which there corresponds no exact solution.  This would furbish more 
interesting alternative cosmology examples.  

Second, Newton--York absolute time theory is somewhat less rigid.  Had it been possible to 
include volume division into this theory, the subsequent cosmology would have a dust-like 
$a = t^{\frac{1}{2}}$.  Unfortunately, the combination of this theory with volume division turns out to be 
Kelleher's theory again, because volume division interferes with the viability of the CMC bypass 
of the integral inconsistency (see C.1).  I am yet to investigate the cosmology of Newton--York 
absolute time theory itself. 


Third, the means of coupling the matter may not yet have been determined.  Note that the 
couplings in \cite{Kelleher}, \cite{CG} and \cite{IOY} are all different.  In fact, if I 
Conf-BM the matter in conformal gravity, I get a Hubble redshift:
for each species {\sffamily A}, replacing 
$\dot{\Psi}_{\mbox{\sffamily\scriptsize A\normalsize\normalfont}}$ with
\be
\frac{\pa}{\pa\lambda}
\left( 
\frac{        \Psi_{\mbox{\sffamily\scriptsize A\normalsize\normalfont}}
              \phi^{        w_{\mbox{\sffamily\tiny A\normalsize\normalfont}}        }           }
     {        V^{      \frac{    w_{\mbox{\sffamily\tiny A\normalsize\normalfont}}       }
                            {      6      }
                } 
     }
\right) 
                       = 
\left(
\frac{\phi}{V^{\frac{1}{6}}} 
\right)
^{               w_{\mbox{\sffamily\tiny A\normalsize\normalfont}}                }
\left(      
\dot{\Psi}_{\mbox{\sffamily\scriptsize A\normalsize\normalfont}} 
+ w_{\mbox{\sffamily\scriptsize A\normalsize\normalfont}}
\frac{\dot{\phi}}{\phi}    \Psi_{\mbox{\sffamily\scriptsize  A\normalsize\normalfont}}      
\right) 
\mbox{ } \Rightarrow \mbox{ }
p = - \sum_{\mbox{\sffamily\scriptsize  A\normalsize\normalfont}}
w_{\mbox{\sffamily\scriptsize A\normalsize\normalfont}}
\Phi_{\mbox{\sffamily\scriptsize  A\normalsize\normalfont}}
\pi^{\mbox{\sffamily\scriptsize A\normalsize\normalfont}} 
\mbox{ } .
\ee
Are there rigid rules determining matter coupling or is it down to taste?  
Isenberg--\'{O} Murchadha--York type scalings \cite{IOY} in a `Barbour style' formulation 
would give Hubble expansion, furbishing conformal gravity with a more conventional cosmology.  
The form of the matter coupling may also change the rigidities of the resulting cosmologies 
in Kelleher and Newton--York absolute time theories.  

\mbox{ }

\noindent{\bf 2.3.3 CS+V theory: interpretation and possible tests} 

\mbox{ }

\noindent 
I arrive at arbitrary-$W$ `York-style' conformal theories since embeddability is no longer necessary 
for consistency.  But then $W = 1$ and $W \neq 1$ have very different interpretations.  For, 
whereas for $W = 1$ the theory can be resliced away from the CMC stacking and the constraints 
still close for embeddability, this is not true for $W \neq 1$.  

For the $W = 1$ CS+V theory, the use of a CMC stack of hypersurfaces is thus ultimately a 
{\sl gauge choice}, which is available provided that the LFE is soluble.  It is a {\sl partial} 
gauge choice since the point-identification (shift) between hypersurfaces in the stack is still 
unspecified.  That the LFE encodes this gauge choice means that one is automatically provided 
with a partially {\sl gauge-fixed} action.  Any pathology in this CMC gauge might then go away 
under the valid procedure of reslicing so as to be in another gauge.  

But $W \neq 1$ CS+V theories are not just written to favour a particular slicing or possess 
a privileged slicing.  They are not generally resliceable because this leads to inconsistency.  
Thus these describe stacks of CMC slices and not pieces of GR-like spacetime.  As a result of 
this, pathologies of the stack of CMC slices become real effects since reslicing to avoid these 
is not possible.    

Thus while $W = 1$ CS+V theory is just (a restriction of) CMC-sliced GR, $W \neq 1$ 
CS+V theories are quite distinct at a conceptual level.    

To complete the picture, the choice of $W = 1$ slicing is available in conformal gravity, so 
one can use it to pass to other slicings by embeddability.  But unlike in CS+V, these new 
slicings {\sl remember} the privileged $p = 0$ slicing, since the volume of these slices gets 
incorporated into the field equations. 

\mbox{ }

Not being able to reslice leads to differences, at least in principle.  Suppose one has access to a 
compact object whose curvature profile permits (GR-inspired) collapse of the lapse to occur 
well outside its horizon.  Then in a GR world, one could send an observer past where the lapse 
collapses, and as nothing physical occurs there and the observer is still safely away from the 
horizon, the observer can `return to Earth' and report that $W \neq 1$ CS+V theory has been 
falsified.  But in a $W \neq 1$ CS+V theory world, the observer would have become frozen forever 
where the lapse collapses and thus would not be able to return.  Note that this is somewhat 
similar to the frozen star concept which predated the GR notion of black holes, except that 
the freezing could be occurring {\sl outside} the horizon.  Although $W \neq 1$ CS+V theory 
could therefore be an improvement as regards strong cosmic censorship (the occurrence of 
singularities {\it at all}), there is also the GR-inspired possibility in the Eardley--Smarr 
example: that sufficiently steep curvature profiles generate too slow a collapse of the lapse 
to avoid singularities.   Also, the collapse of the lapse would not save one from other 
non-curvature blowup pathologies usually regarded as singularities.  

$W \neq 1$ CS+V theory ought to also be testable much as Brans--Dicke theory is, by solar 
system tests.   While plain `arbitrary-$W$ GR' was suggested as another useful testbed for GR 
\cite{KiefGiu} (c.f the use of Brans--Dicke theory), this is not much good because it's 
inconsistent (by \cite{SGGiulini} or the RWR result).  What I have demonstrated however is that 
{\sl the idea of this `arbitrary-$W$ GR' can be salvaged because it is a consistent theory 
provided that it is treated as a (non-resliceable) stack of CMC hypersurfaces, in which case 
it becomes CS+V theory}.  Thus I provide a 1-parameter family of theories to test against (not 
just extreme but also) everyday GR.  Moreover, by the nature of the conformal mathematics in 
which they are so naturally expressed, they should be easily useable as testbeds for 
theoretical numerical relativity.  This should require but minor modifications of existing 
codes.  Conformal gravity could also be used/tested in this way.  {\sl Numerical relativity 
uses conformal mathematics, not necessarily any notion of embeddability into GR-like spacetime.  
The alternative theories of this section may be seen as arising from taking this conformal 
mathematics in its own right as possibly a serious alternative to GR itself.} Finally, $W = 1$ 
CS+V theory may be seen as a formulation of GR proper, and thus still be directly useful (both 
conceptually and as a tool) in theoretical numerical relativity even if the suggested 
alternatives to GR are dismissed or heavily bounded by future compact-object observations and 
analysis. 

\vspace{8in}

\mbox{ }  

\noindent\Huge\bf{VI TSA: criticism}\normalfont\normalsize

\mbox{ }

\noindent I sharpen the understanding of what the TSA is because of interest in why the 
impressive collection of results in the GR case above arises in BF\'{O} and AB.  I seek for tacit 
simplicity postulates, survey which assumptions may be weakened and assess the thoroughness 
and plausibility of BF\'{O} and AB's  principles, results and conjectures.  I 
thus arrive at a number of possible TSA variations.  I stress that this is not just about 
improving the axiomatization.  One must be able to find a version that naturally accommodates 
spin-$\frac{1}{2}$ fermions coupled 1) to GR if the TSA is to provide a set of plausible first 
principles for GR 2) To conformal gravity if this is to be a viable alternative.  

In VI.1, I argue that the use of BSW-type actions in the implementation of {\bf R2} 
is problematic.  First, Barbour's use of it draws inspiration from the analogy between it and 
the Jacobi formulation of mechanics. But in VI.1.1 I point out that the Jacobi formulation 
itself has limitations and a significant generalization.  Furthermore in VI.1.2 I point out 
that the analogy itself is inexact in several important respects.  There is then a `conformal  
Problem' and a `notion of distance Problem' (VI.1.3).  Second, should the notion of `BSW-type 
theories' not include all the theories that permit the BSW elimination process itself?  But 
when I sketch this to include fermions in VI.1.4, I find that not the BSW form but rather 
{\sl its} significant generalization is obtained.  Thus the inclusion of fermions will severely complicate the use 
of exhaustive proofs such as those in BF\'{O} and AB.  Another example of such complications 
is how particles are to be included alongside fields in RWR.  I furthermore point out in 
VI.1.5 that the usual higher derivative theories are not necessarily being excluded by BF\'{O}.  
These last two subsections include discussion of their HKT counterparts.

In VI.1.6, I formalize the second point above by showing that I could just as well use 
lapse-uneliminated but lapse eliminable actions for GR and conformal gravity.  For GR, these 
actions may be studied within Kucha\v{r}'s split spacetime framework (SSF)  
\cite{KucharI, KucharII, KucharIII, KucharIV}.  
This framework brings attention to \it tilt \normalfont and \it derivative coupling \normalfont 
complications in general (VI.2.1), which are however absent for the minimally-coupled scalar, 
and `accidentally absent' for the Maxwell and Yang--Mills 1-forms, which are what the TSA picks out.  
But tilt is present for the massive analogues of these 1-forms. 
I deduce the relation between tilt and the existence of a generalized BSW form.  In VI.2.2 I  
counter BF\'{O}'s hope that \sl just \normalfont the known fundamental matter fields are being 
picked out by the TSA, by showing that the massless 2-form is also compatible.  
In VI.2.3, I find alternative reasons why the Maxwell 1-form is singled out by BF\'{O}, 
from the point of view of the hypersurface framework.  I end by explaining the 
complications that would follow were one to permit derivative-coupled 1-forms.  

In VI.3, I point out that it is consistent to take the bosonic sector of nature to be far 
simpler than 4-d generally covariant spacetime structure might have us believe.  Out of the tilt, derivative coupling and best-matching 
kinematics of spacetime, the entirely spatial best-matching kinematics suffices to write down 
actions for GR coupled to a full enough set of bosonic fields to describe nature.    

In VI.4.1, I show how all these results also hold true upon inclusion of spin-$\frac{1}{2}$ 
fermions.  VI.4.2 lists further research topics for fermions in the light of the advances made.  

\mbox{ }

\noindent\Large{\bf 1 Problems with the use of BSW actions}\normalsize

\mbox{ }

\noindent\large{\bf 1.1 Insights from mechanics}\normalsize 

\mbox{ }

\noindent Barbour uses the analogy between the BSW formulation of GR (\ref{VBashwe}) and the 
Jacobi formulation of classical mechanics (in I.1.2.2).  I consider below limitations both on 
the Jacobi formulation, and limitations on the applicability of this analogy in the next 
subsection.    

The Jacobi formulation has a catch: the conformal factor is not allowed to have zeros.  If it 
does then the conformal transformation is only valid in regions where there are no such zeros.  
These zeros are physical barriers in mechanics.  For they correspond to zero kinetic energy by 
the conservation of energy equation.  As the configuration space metric is positive-definite, 
this means that the velocities must be zero there, so the zeros cannot be traversed.

The working for the Lagrangian homogeneously quadratic in its velocities (\ref{QUADRATIC}) 
leads to a configuration space geometry that is conformal to a
Riemannian geometry, which may be useful since Riemannian geometry is well-understood. 
Let $\mbox{\sffamily L\normalfont}(q_{\mbox{\sffamily\scriptsize C\normalsize\normalfont}}, 
\dot{q}_{\mbox{\sffamily\scriptsize A\normalsize\normalfont}})$ be instead a completely general 
function.  Then\footnote{Here, Newtonian time is $t$ whilst $\lambda$ is a parameter.  
Dot is used for $\frac{\pa} {\pa t}$ in mechanics workings and dash for $\frac{\pa} {\pa \lambda}$. 
\sffamily A\normalfont-indices take values 1 to $n$, $\hat{\mbox{\sffamily A\normalfont}}$ 
indices take values 1 to $n$ and $t$, and $\bar{\mbox{\sffamily A\normalfont}}$ indices take 
values 1 to $n - 1$.}  
\be 
\mbox{\sffamily I\normalfont} = \int_{\lambda_1}^{\lambda_2}\mbox{\sffamily L\normalfont}
\left(
q_{\mbox{\sffamily\scriptsize C\normalsize\normalfont}}, \frac{q^{\prime}_{\mbox{\sffamily\scriptsize A\normalsize\normalfont}}}{t^{\prime}}
\right)
t^{\prime}\mbox{d}\lambda \equiv  
\int_{\lambda_1}^{\lambda_2}\mbox{\sffamily l\normalfont}(q_{\hat{\mbox{\sffamily\scriptsize C\normalsize\normalfont}}}, 
q^{\prime}_{\hat{\mbox{\sffamily\scriptsize D\normalsize\normalfont}}})\textrm{d}\lambda 
\ee
\be
\mbox{may be modified to }
\mbox{\hspace{1.6in}}
\mbox{\sffamily I\normalfont}_{\mbox{\scriptsize J\normalsize}} 
= \int_{\lambda_1}^{\lambda_2}{\mbox{\sffamily R\normalfont}}(q_{\mbox{\sffamily\scriptsize C\normalsize\normalfont}}, 
q^{\prime}_{\mbox{\sffamily\scriptsize A\normalsize\normalfont}})\textrm{d}\lambda 
\mbox{\hspace{2in}}
\label{VJacgen}
\ee
by Routhian reduction, where ${\mbox{\sffamily R \normalfont}} = F$, some homogeneous 
linear function of the $q^{\prime}_{\mbox{\sffamily\scriptsize A\normalsize\normalfont}}$ 
\cite{Lanczos}.  For example, $F$ could be a \it Finslerian metric function \normalfont from 
which we could obtain a Finslerian metric $f_{_{\mbox{\sffamily\scriptsize AB\normalsize\normalfont}}} 
= \frac{1}{2}\frac{  \pa^2  }{  \pa q^{\prime}_{\mbox{\sffamily\tiny A\normalsize\normalfont}} 
\pa q^{\prime}_{\mbox{\sffamily\tiny B\normalsize\normalfont}}  }F^2$, provided that $F$ 
obeys further conditions \cite{Vfinsler} including the nondegeneracy of $f_{\mbox{\sffamily\scriptsize AB\normalsize\normalfont}}$.  
So in general the `geometrization problem' of reducing the motion of a mechanical system to a 
problem of finding geodesics involves more than the study of Riemannian geometry.

To some extent, there is conventional freedom in the choice of configuration space geometry.   
This is because standard manoeuvres can alter whether it is Riemannian.  This is because 
one is free in how many redundant configuration variables to include, and in the character of 
those variables (for example whether they all obey second-order ELE's).

As a first example of this, consider the outcome of the Routhian reduction of 
(\ref{QUADRATIC}) for without loss of generality $q_n$ cyclic,
\be
{\mbox{\sffamily R\normalfont}}(q_{\mbox{\scriptsize\sffamily A\normalfont\normalsize}}, 
\dot{q}_{\mbox{\scriptsize\sffamily A\normalfont\normalsize}}) = \frac{1}{2}
\left( 
M^{\mbox{\scriptsize\sffamily AB\normalfont\normalsize}}  
- \frac{M^{{\mbox{\scriptsize\sffamily A\normalfont\normalsize}} n}
M^{{\mbox{\scriptsize\sffamily A\normalfont\normalsize}} n}}{M^{nn}}
\right)
\dot{q}_{\mbox{\scriptsize\sffamily A\normalfont\normalsize}}
\dot{q}_{\mbox{\scriptsize\sffamily B\normalfont\normalsize}} 
+ \frac{c^nM^{{\mbox{\scriptsize\sffamily A\normalfont\normalsize}} n}}{M^{nn}}
\dot{q}_{\mbox{\scriptsize\sffamily A\normalfont\normalsize}} 
- \bar{\mbox{\sffamily V\normalfont}} 
\mbox{ } ,
\ee
where $\bar{\mbox{\sffamily V\normalfont}}$ is a modified potential.  So removing 
redundancy in a Riemann-geometrizable action by Routhian reduction can lead to a 
non-Riemann-geometrizable system, on account of the penultimate `gyroscopic term' 
\cite{Lanczos}, which is linear in the velocities.  I consider the reverse of this procedure as 
a possible means of arriving at Riemannian geometry to describe systems with linear and 
quadratic terms.  I observe that if the linear coefficients depend on configuration variables, 
then in general the quadratic structure becomes contaminated with these variables.

As a second example, higher-than-quadratic systems may be put into quadratic form by 
\it Ostrogradsky reduction \normalfont \cite{Vostrograd}, at the price of introducing extra
configuration variables.

\mbox{ }

\noindent{\bf 1.2 Criticism of BSW--Jacobi analogy}

\mbox{ } 

\noindent As explained in I.2.5, GR is an already-parameterized theory.  This is made manifest by writing 
it in its BSW form (\ref{VBashwe}).  Although this looks similar to the Jacobi action in 
mechanics, there are important differences.  Recollect from I.2.7.1 that the GR configuration 
space is infinite-d; with redundancies, one can consider it to be Superspace.  The DeWitt 
supermetric is defined on Superspace \sl pointwise\normalfont.  By use of DeWitt's 2-index to 
1-index map (\ref{2TO1}), it may be represented by a $6\times 6$ matrix, which is $( - + + + + + )$ 
and thus indefinite.  The special case Minisuperspace (I.3.3) has a 
$( - + + )$ minisupermetric, thus also indefinite.  In contrast mechanics has a 
positive-definite kinetic term.  

Also, the BSW action has the `bad' ordering, whereas the Jacobi action has the `good' ordering. 
Below, I first consider Minisuperspace, for which this extra complication does not arise, 
since by homogeneity the `good' Jacobi and `bad' BSW orderings are equivalent.

Finally, BSW's work led to the thin sandwich conjecture (I.2.9.1) the solubility of which 
features as a caveat in BF\'{O}'s original paper.  Being able to pose this conjecture for a 
theory amounts to being able to algebraically eliminate the lapse from its Lagrangian.  
This implies that the theory is timeless in Barbour's sense \cite{B94I, EOT}.  Note that the 
extension of the conjecture to include fundamental matter fields has only recently begun 
\cite{thin sandwich3}.  Such investigations are required to assess the robustness of the 
conjecture to different theoretical settings, to see if in any circumstances it becomes 
advantageous to base numerical relativity calculations on the thin sandwich algorithm.    

\mbox{ }

\noindent\large{\bf 1.3 Lack of validity of the BSW form}\normalsize

\mbox{ }

\noindent{\bf 1.3.1 Conformal Problem of zeros}

\mbox{ }

\noindent In perfect analogy with mechanics, vacuum GR has a conformally-related 
line element $\mbox{d}\tilde{s}^{2} = (\Lambda + \sigma R)\mbox{d}s^2$, for which the motion 
associated with (\ref{VBashwe}) is geodesic \cite{VDeWitt70}.  But the observation in 
mechanics that such conformal transformations are only valid in regions where the conformal 
factor is nonzero
still holds for GR.  It is true that the details are different,
due to the indefiniteness of the GR supermetric.  This causes the zeros to be spurious rather 
than physical barriers \cite{VMisner}.  For whilst a zero $z$ of the potential corresponds to a 
zero of the kinetic term by virtue of the Hamiltonian constraint, this now means that the 
velocity need be null, not necessarily zero, because of the indefiniteness.  Thus, rather than 
grinding to a halt, the motion may continue through $z$ `on the Superspace null cone', 
which is made up of perfectly reasonable Kasner universes.  Nevertheless, the conformal 
transformation used to obtain geodesic motion is not valid, so it is questionable whether the 
BSW form is a `geodesic principle', if in general it describes conformally untransformed 
\sl non-geodesic curves \normalfont for practical purposes.

To illustrate that the presence of zeros in the potential term is an important occurrence in 
GR, note that Bianchi IX spacetimes have an infinity of such zeros as one approaches the 
cosmological singularity.  And these spacetimes are important because of the BKL conjecture 
(see I.3.3).  The above argument was originally put forward by Burd and Tavakol \cite{VTB} to 
argue against the validity of the use of the `Jacobi principle' to characterize chaos in GR 
\cite{VSydlowski}.  Our point is that this argument holds against \sl any \normalfont use, 
BF\'{O}'s included, of the BSW form in Minisuperspace models of the early universe in GR.  

The way out of this argument that I suggest is to abstain from the self-infliction of spurious 
zeros by not performing the conformal transformation in the first place, thus abandoning the 
interpretation of the BSW form as a geodesic principle in GR.  Conformal gravity however is 
distinct from GR and has no cosmological singularity, so arguments based on the BKL conjecture 
are not applicable there.  Conformal gravity's zeros are real as in mechanics, because  
$\mbox{\sffamily T\normalfont}^{\mbox{\scriptsize g\normalsize}}_{\mbox{\scriptsize C\normalsize}}$ is positive-definite, and Barbour and \'{O} Murchadha 
use this to argue that topologies with $R < 0$ at any point are not allowed \cite{conformal}.  

\mbox{ }

\noindent{\bf 1.3.2 The BSW form is an unknown notion of distance}

\mbox{ }

\noindent BF\'{O} called the local square root ordering `bad' because it gives one constraint per space 
point, which would usually render a theory trivial by overconstraining due to the ensuing 
cascade of secondary constraints.  Yet GR contrives to survive this because of its hidden 
foliation invariance \cite{BOF}.  However Giulini \cite{thin sandwich3} has pointed out 
another reason why the local square root ordering is bad: it does not give rise to known 
geometry.  Below, we extend his finite-d counterexample to the geometry being Finslerian.

The BSW form as a notion of distance provides as the `full metric' on Superspace 
\be
\frac{1}{2}\frac{\pa^2
(\mbox{\sffamily I\normalfont}_{\mbox{\scriptsize BSW\normalsize}}^2)}{\pa v^A(u)\pa v^B(w)}
\mbox{=}
\left[
\tilde{G}_{\hat{A}\hat{C}}(u)\tilde{G}_{\hat{B}\hat{D}}(w)\mbox{+}2\delta^{(3)}(u, w)
\left(
\frac{\mbox{\sffamily I\normalfont}_{\mbox{\scriptsize BSW\normalsize}}}
{\sqrt{\tilde{G}_{\hat{B}\hat{D}}v^{\hat{B}}v^{\hat{D}} }}
\tilde{G}_{\hat{A}[\hat{B}}\tilde{G}_{\hat{C}]\hat{D}}
\right)
(u)
\right]
\hat{v}^{\hat{A}}\hat{v}^{\hat{C}},
\ee
where $v^{\hat{A}} \equiv \dot{h}^{\hat{A}} = \dot{h}^{ab}$ by DeWitt's 2-index to 1-index map 
(\ref{2TO1}) 
and where hats denote unit `vectors'.  So in general, $\tilde{G}_{\hat{A}[\hat{B}}\tilde{\hat{G}}_{\hat{C}]\hat{D}} = 0$ is 
a sufficient condition for the full metric to be degenerate and hence not Finsler (Giulini's 
particular example had a 1-d $v^{\hat{A}}$ so this always occurred).  But if 
$\tilde{G}_{\hat{A}[\hat{B}}\tilde{G}_{\hat{C}]\hat{D}} \neq 0$, the full metric is not a function 
(it contains delta functions and integrals).  So using the BSW form as a notion of distance leads to 
unknown geometry, so there is no scope for the practical application of the BSW form as a 
geodesic principle.

This is to be contrasted with the global square root, for which the above procedure gives 
instead (semi)Riemannian geometry.  For Minisuperspace, the local square root working presented 
does indeed collapse to coincide with this global square root working, and the resulting 
(semi)Riemannian geometry is of considerable use in Minisuperspace quantum cosmology 
\cite{VMisner}.  Whereas conformal gravity is still plagued by the notion of distance 
difficulty, the GR issue of the sheaves (I.2.8.1) is resolved by privileged foliation.
I should emphasize that I do not know of any reason why degeneracy due to stratification and 
degeneracy due to sheaves are at all related.  What I am sure about is that the badness of the 
local square root geometry has nothing to do with this since this badness is already there for 
field theories with simple configuration spaces.  I conclude that it is not appropriate to use 
the BSW form as a geodesic principle on Superspace.  

\mbox{ }

\noindent\large{\bf 1.4 The fermionic contribution to the action is linear}\normalsize

\mbox{ }

\noindent 
Since the kinetic terms of the bosons of nature are also quadratic in their velocities, I can 
use the modifications
\be
\mbox{\sffamily T\normalfont}^{\mbox{\scriptsize g\normalsize}}    \longrightarrow 
\mbox{\sffamily T\normalfont}^{\mbox{\scriptsize g\normalsize}} + 
\mbox{\sffamily T\normalfont}^{\mbox{\scriptsize B\normalsize}} \mbox{ } ,
\mbox{ } \Lambda +  \sigma R \longrightarrow \Lambda + \sigma R + 
\mbox{\sffamily U\normalfont}^{\mbox{\scriptsize B\normalsize}}
\ee
to accommodate bosonic fields $B_{\mbox{\sffamily\scriptsize A\normalfont\normalsize}}$ in a 
BSW-type action,
\be
\mbox{\sffamily I\normalfont}^{\mbox{\scriptsize B\normalsize}} = \int \textrm{d}\lambda
\int \textrm{d}^3x \sqrt{h} \sqrt{\Lambda + \sigma R + 
\mbox{\sffamily U\normalfont}^{\mbox{\scriptsize B\normalsize}}} \sqrt{
\mbox{\sffamily T\normalfont}^{\mbox{\scriptsize g\normalsize}} + 
\mbox{\sffamily T\normalfont}^{\mbox{\scriptsize B\normalsize}}}
\mbox{ } .
\label{VBSWboson}
\ee
This local square root encodes the correct Hamiltonian constraint for the gravity--boson 
system.  

Although the pointwise Riemannian kinetic metric is larger than the DeWitt supermetric, in the 
case of minimally-coupled matter it contains the DeWitt supermetric as an isolated block:
\be
\left(
\begin{array}{ll}
G_{\hat{A}\hat{B}}(h_{ab}) & 0 \\
0 & H^{\mbox{\scriptsize Matter\normalsize}}_{\Lambda\Sigma}(h_{ab})
\end{array}
\right) 
\mbox{ } .
\label{Vmin}
\ee 
If this is the case, it makes sense to study the pure gravity part by itself, which is a 
prominent feature of almost all the examples hitherto studied in the TSA.  I identify this as 
a tacit simplicity requirement, for without it the matter d.o.f's interfere with the 
gravitational ones, so it makes no sense then to study gravity first and then `add on' matter
In Brans--Dicke (BD) theory, this is not immediately the case: this is an example in which 
there are gravity-boson kinetic cross-terms $\mbox{{\large G}}_{ab\chi}$ 
[see (\ref{hornedhelmet})] in the pointwise
Riemannian kinetic metric unlike in (\ref{Vmin}). 
The TSA form is then 
$$
\mbox{\sffamily I\normalfont} = \int\textrm{d}\lambda\int\textrm{d}^3x\sqrt{h}
\sqrt{      e^{-\chi}(R - \omega|\pa\chi|^2)
\left[
\left(
h^{ac}h^{bd} - \frac{X -2}{3X - 4}h^{ab}h^{cd}
\right)
\mbox{\ss}_{\xi}h_{ab}\mbox{\ss}_{\xi}h_{cd} \mbox{\hspace{1in}} 
\right.
}
$$
\be
\overline{\mbox{\hspace{1in}} 
\left.
+ \frac{4}{3X  -4}h^{ab}\mbox{\ss}_{\xi}h_{ab}\mbox{\ss}_{\xi}\chi 
+ \frac{3X - 2}{(3X - 4)(X - 1)}\mbox{\ss}_{\xi}\chi\mbox{\ss}_{\xi}\chi  
\right]  }
\mbox{ } .
\ee
Thus, the metric and dilatonic fields form \sl together \normalfont a theory of gravity with 
3 d.o.f's.   Compared to the more disturbing examples in VI.2.4, this is a mild example of 
derivative coupling.\fn{Furthermore, if one admits the conformal redefinition 
of the metric and scalar d.o.f's, blockwise isolation of the form (\ref{Vmin}) is permitted, 
thus freeing the theory from derivative coupling.}

I now begin to consider whether and how the 3-space formulation can accommodate 
spin-$\frac{1}{2}$ fermionic fields, $F_{\mbox{\sffamily\scriptsize A\normalfont\normalsize}}$.  
Following the strategy employed above for bosons, the BSW working becomes 
$$
\mbox{\sffamily I\normalfont}^{\mbox{\scriptsize F\normalsize}} 
= \int \textrm{d}\lambda\int \textrm{d}^3x \sqrt{h}NL(h_{ab}, \dot{h}_{ab}; \xi_i; N; 
F_{\mbox{\sffamily\scriptsize A\normalfont\normalsize}}, 
\dot{F}_{\mbox{\sffamily\scriptsize A\normalfont\normalsize}}) 
$$
\be 
\mbox{\hspace{0.7in}} 
= \int \textrm{d}\lambda \int \textrm{d}^3x\sqrt{h}
\left[
N
\left(
\Lambda + \sigma R + 
\mbox{\sffamily U\normalfont}^{\mbox{\scriptsize F\normalsize}}
+ \frac{      \mbox{\sffamily T\normalfont}^{\mbox{\scriptsize g\normalsize}}
(\mbox{\ss}_{\xi}h_{ij})      }{      4N^2      }
\right)
+ \mbox{\sffamily T\normalsize}^{\mbox{\scriptsize F\normalsize}}
(\dot{F}_{\mbox{\scriptsize\sffamily A\normalfont\normalsize}})
\right]
\label{VSFtrue}
\ee
because $\mbox{\sffamily T\normalfont}^{\mbox{\scriptsize F\normalsize}}$ is linear in 
$\dot{F}_{\mbox{\scriptsize\sffamily A\normalfont\normalsize}}$.\fn{I show in VI.2 that the 
algebraic dependence on $N$ emergent from such decompositions requires rigorous justification.  
I provide this for (\ref{VSFtrue}) in VI.4.1}  Then the usual trick for eliminating $N$ does not touch 
$\mbox{\sffamily T\normalfont}^{\mbox{\scriptsize F\normalsize}}$,
which is left outside the square root:
\be
\mbox{\sffamily I\normalfont}^{\mbox{\scriptsize F\normalsize}} = \int \textrm{d}\lambda
\int \textrm{d}^3x \sqrt{h}
\left(
\sqrt{
\Lambda + \sigma R + \mbox{\sffamily U\normalfont}_{\mbox{\scriptsize F\normalsize}}}\sqrt{
\mbox{\sffamily T\normalfont}^{\mbox{\scriptsize g\normalsize}}} +
\mbox{\sffamily T\normalfont}^{\mbox{\scriptsize F\normalsize}}
\right). 
\label{Vactualfermi}
\ee
The local square root constraint encodes the correct gravity-fermion Hamiltonian constraint
\be
^{\mbox{\scriptsize F\normalsize}}{\cal H}
\equiv -\sqrt{h}(\Lambda + \sigma R + 
\mbox{\sffamily U\normalfont}_{\mbox{\scriptsize F \normalsize}})
+ \frac{1}{\sqrt{h}}
\left(
p\circ p - \frac{X}{2}p^2\right) = 0 
\mbox{ } .
\ee
I postpone the issue of the precise form of the best matching to be used until VI.4.1  My 
concern in this section is the complication of the configuration space geometry due to the 
inclusion of fermions.

For now the elimination procedure is analogous not to the Jacobi working but rather to its 
generalization (\ref{VJacgen}).  So even the pointwise geometry of the gravity-fermion 
configuration space is now compromised:
$\sqrt{\Lambda + \sigma R + 
\mbox{\sffamily U\normalfont}^{\mbox{\scriptsize F\normalsize}}}\sqrt{
\mbox{\sffamily T\normalfont}^{\mbox{\scriptsize g\normalsize}}} + 
\mbox{\sffamily T\normalfont}^{\mbox{\scriptsize F\normalsize}}$
could sometimes be a Finslerian metric function.  
By allowing (\ref{VSFtrue}), one is opening the door to all sorts of
complicated possible actions, such as

\noindent
1) $^k\sqrt{ G^{\Sigma_1 ... \Sigma_{k}}\dot{q}_{\Sigma_1} ... 
\dot{q}_{\Sigma_{k}} }$.

\noindent
2) Arbitrarily complicated compositions of such roots, powers and sums.

\noindent
3)More generally, 
$K_{\mbox{\sffamily\scriptsize Q\normalfont\normalsize}}\dot{q}^{\mbox{\sffamily\scriptsize Q\normalfont\normalsize}}$,
where $K_{\mbox{\sffamily\scriptsize Q\normalfont\normalsize}}$  is allowed to be an
arbitrary function of not only the $q_{\mbox{\scriptsize\sffamily Q\normalfont\normalsize}}$ 
but also of the {\sffamily Q} - 1 independent ratios of the velocities.

\noindent
4) The above examples could all be Finslerian or fail to be so
by being degenerate.    
They could also fail to be Finslerian if the
$K_{\mbox{\scriptsize\sffamily Q\normalfont\normalsize}}$ are permitted to be \sl functionals \normalfont
of overall degree 0 in the velocities,
which we can take to be a growth of the local-global square root ambiguity.

Here are some examples of these.  

\noindent i) Inclusion of particles: for 1 particle,  
$$
\mbox{\sffamily I\normalfont}_{\mbox{\scriptsize (1)}}\mbox{=}
\int\textrm{d}^4x\sqrt{|g|}\delta^{(3)}
\left(
x_j, \mbox{ } x^{(1)}_k
\right)
\sqrt{g_{AB}\dot{x}^A\dot{x}^B}\mbox{=}
\int\textrm{d}\lambda\int\textrm{d}^3x\sqrt{h}N\delta^{(3)}
\left(
x_j, \mbox{ } x^{(1)}_k
\right)
\sqrt{        h_{ab}\dot{x}^a\dot{x}^b\mbox{--}\dot{t}^2         }
$$
\be
=
\int\textrm{d}\lambda\sqrt{h_{ab}\dot{x}^{(1)a}\dot{x}^{(1)b} - \dot{t}^2         }
\ee
by use of proper time.  We could have  (i) = (1) to (n) of these, in which case want one root 
\be
\mbox{per particle (bad ordering) } 
\mbox{\hspace{0.8in}}
\mbox{\sffamily I\normalfont}_{\mbox{\scriptsize (1, 2, ... n)\normalsize}} = \int\textrm{d}\lambda\sum_{(i) = (1) }^{(n)}
\sqrt{h_{ab}\dot{x}^{(1)^a}\dot{x}^{(1)b} - \dot{t}^2 }
\mbox{\hspace{0.8in}}
\ee
\be
\mbox{to recover separate constraints}
\mbox{\hspace{1.35in}}
p_{(i)}^2 - E_{(i)}^2 = - E_{0(i)}^2
\mbox{\hspace{1.35in}}
\ee
for each particle from each square root in the sum.  
In a RWR context, one would Diff-BM the particle velocities (which act on everything) 
\be
\mbox{\sffamily I\normalfont}(n)_{\mbox{\scriptsize Diff\normalsize}} = 
\int\textrm{d}\lambda
\left[   
\int
\textrm{d}^3x\sqrt{h}
\left(
\sqrt{R\mbox{\sffamily T\normalfont}^{\mbox{\scriptsize g\normalsize}}}
\right)
+ \sqrt{h} \sum_{(1) = (i)}^{(n)}\sqrt{h_{ab}\mbox{\ss}_{\xi}{x}^{(1)a}\mbox{\ss}_{\xi}{x}^{(1)^b} 
- \mbox{\ss}_{\xi}{t}^2 }
\right] 
\mbox{ } . 
\ee
Then one has a momentum constraint with a standard vector and scalar contribution to it.

\noindent ii) If without loss of generality the first $m$  $(\leq n)$ particles are to be charged, the linear piece 
\be
\int\textrm{d}\lambda\sum_{(j) = (1)}^{(m)}(\dot{x}_{(j)}^aA_a - \dot{t}A_{\perp})
\ee
is to be added.  Note that the particles 
{\it are not} to be U(1)-BM; the well-known correction for charged 
particles is to the momenta and not the velocities, and this arises just so for the above form.  

\noindent iii) The phenomenological matter contribution $\xi^ij_i$ occurs linearly outside the 
BSW square root [c.f (\ref{tsac})].  Note however that if I study this system in more detail 
(i.e in terms of fundamental matter), then I find that this term really belongs inside the square 
root. For if one admits 
\be
\mbox{that }
\mbox{\hspace{1.8in}} 
p \mbox{ } \circ \mbox{ } \dot{h} - N({\cal H} + 2\sqrt{h}\rho) - \xi^i({\cal H}_i - \sqrt{h}j_i)
\mbox{\hspace{1.8in}}
\ee
should really have a contribution from matter momenta, and uses (\ref{EMTen}) for fundamental 
$\rho$ and $j^a$ (done here for the scalar field example), then one has 
\be
p \mbox{ } \circ \mbox{ } \dot{h} + \pi\varsigma - N
\left(
{\cal H} + \sqrt{h}[\frac{(\delta_{\xi}\varsigma)^2}{\alpha^2} 
+ |\pa\varsigma^2| + m_{\varsigma}\varsigma^2] 
\right) 
- \xi^i({\cal H}_i + \delta_{\xi}\varsigma\pa_i\varsigma)
\ee
\be
\mbox{which wraps up to form }
\mbox{\hspace{1.75in}}
\sqrt{h}\alpha
\left(
R + \mbox{\sffamily U\normalsize}^{\varsigma} 
+ \frac{      \mbox{\sffamily T\normalsize}^{\mbox{\scriptsize g\normalsize}} 
            + 4\mbox{\sffamily T\normalsize}^{\varsigma}      }
       {      4\alpha^2      }
\right)
\mbox{\hspace{1in}}
\ee
\be
\mbox{rather than }
\mbox{\hspace{2.5in}}
\sqrt{h}\alpha
\left(
R - 2\rho + \frac{      \mbox{\sffamily T\normalsize}^{\mbox{\scriptsize g\normalsize}}     }
                 {      4\alpha^2      }
\right)
+ \xi^ij_i
\mbox{\hspace{1.8in}}
\ee
so that its RI form is $\sqrt{AB}$ rather than $\sqrt{AB} + C$ .  

\noindent iv) I will eventually admit some ugly 1-form theories to the TSA via horrendous 
combinations of type 2) in VII.1.2.  

We would therefore need to modify the RI BSW implementation of principle 
\bf R2 \normalfont to a general BSW implementation capable of including fermions.  This amounts to 
dropping the requirement of the matter field kinetic term being homogeneously quadratic in its 
velocities, thus bringing the \bf TSA matter simplicity 5 \normalfont into alignment with Teitelboim's 
assumptions.  
We note that with increasing generality the possibility of uniqueness proofs becomes more 
remote.  Although some aims of the TSA such as a full derivation of the universal 
null cone would require some level of uniqueness proofs for spin-$\frac{1}{2}$ fermions, my 
strategy in this chapter is to show that spin-$\frac{1}{2}$ fermions coupled to GR do possess a 
3-space formulation and also to point out that the uniqueness results may have to be 
generalized in view of the generalization of the BSW form required in this section.   

Could one not choose to geometrize the gravity-fermion system as a Riemannian geometry instead, 
by use of the reverse of Routhian reduction?  But the coefficients of the linear fermionic 
velocities in the Einstein--Dirac system contain fermionic variables, so the resulting
Riemannian geometry's coefficients would contain the fermionic variables in addition 
to the metric.  This is a \it breach of the DeWitt structure\normalfont, 
since it means that contact is lost with DeWitt's study of the configuration space of pure GR 
\cite{DeWitt, VDeWitt70}.  So this choice also looks highly undesirable.

For 40 years the natural accommodation of spin-$\frac{1}{2}$ fermions in geometrodynamics 
has been problematic (see I.2.7.2, II.1).   So this is a big demand 
on the TSA, and one which must be met if the TSA is truly to describe nature.  My demands 
here are less than Wheeler's in \cite{RMW2}: I am after a route to relativity with all 
matter `added on' rather than a complete unified theory.  The HKT route appears also to be 
incomplete at this stage:  Teitelboim was unable to find a hypersurface deformation 
explanation for spin-$\frac{1}{2}$ fermions \cite{Teitelboim}.  
Thus when this work was started, all forms of the seventh route to relativity were 
incomplete with respect to the inclusion of spin-$\frac{1}{2}$ fermions.  
In VI.4.1, I point out the natural existence of GR--spin-$\frac{1}{2}$ theory 
within the TSA.  

\mbox{ }

\noindent\large{\bf 1.5 Higher derivative theories}\normalsize

\mbox{ }

\noindent I now argue against the significance of the preclusion of higher derivative theories 
by BF\'{O}.  For the precluded theories are easily seen \sl not \normalfont to be 
the usual higher derivative theories.  There are two simple ways of noticing this.  
First, the primary constraints encoded by the BF\'{O} theories
with arbitrary $\mbox{\sffamily V\normalfont}(h_{ij}, h_{ij,k}, ...)$ will always be of the form
\be
\sqrt{h}{\cal H} = -\sqrt{h}\mbox{\sffamily V\normalfont} + \frac{1}{\sqrt{h}}
\left(
p \circ p - \frac{X}{2}p^2
\right) 
= 0 
\mbox{ } ,
\ee
while the usual higher derivative theories give something more complicated.  Second, BF\'{O}'s 
theories have fourth-order terms in their potentials but their kinetic terms remain quadratic 
in the velocities, whilst the usual higher derivative theories' kinetic terms are quartic in 
the velocities.  This mismatch of derivatives between $\mbox{\sffamily T\normalfont}$ and 
$\mbox{\sffamily V\normalfont}$ for $\mbox{\sffamily V\normalfont} \neq \sigma R + \Lambda$, 
overrules the theories from within the generally covariant spacetime framework, 
so BF\'{O} are doing nothing more than 
general covariance can do in this case.  On the other hand, the potentials BF\'{O} consider are not absurd 
given their ontology; such potentials were specifically inquired about by Wheeler in 
\cite{Wheeler}.

It is not clear whether the usual higher derivative theories could be written in some 
generalized BSW form.  The form would either be considerably more complicated than that of 
pure GR or not exist at all.  Which of these is actually true should be checked case by case.  
I consider this to be a worthy project in its own right by the final comment in VI.1.2, 
since this problem may be phrased as `for which higher derivative theories can the thin 
sandwich formulation be posed?'  To illustrate why there is the possibility of nonexistence,
consider the simplest example, $\check{R} + x \check{R}^{2}$ 
theory.  The full doubly-contracted Gauss equation 
\be
\mbox{is }
\mbox{\hspace{1.2in}}
\check{R}
= R - \sigma(K\circ K - K^2) + 2\sigma D_a(n^b D_bn^a - n^a D_bn^b)
\mbox{\hspace{1.2in}}
\ee
and, whereas one may discard the
divergence term in the 3 + 1 split of $\check{R}$, in the 3 + 1 split
of $\check{R}^{2}$, this divergence is multiplied by 
$\check{R}$ and so cannot similarly be discarded.
So it is unlikely that the elimination of $N$ will be \sl algebraic \normalfont
in such theories, which is a requirement for the BSW procedure.
Were this algebraic elimination possible, we would get more complicated 
expressions than the local square root form from it.
Indeed, higher derivative theories are known to have considerably more
complicated canonical formulations than GR \cite{Vhdl1, Vhdl2};
it is standard to treat them by a variant of Ostrogradsky reduction
adapted to constrained systems \cite{Vhdl2}.

It is worth commenting that HKT's derivation of
${\cal H}$ being quadratic in its momenta
and containing at most second derivatives may also be interpreted as tainted,
since it comes about by restricting the gravity to have two d.o.f's, as opposed to e.g. the 
three of $\check{R} + x \check{R}^{2}$ theory or of 
Brans--Dicke theory.  Thus I do not foresee that any variant of the seventh route to 
relativity will be able to find a way round the second-order derivative assumption of the 
other routes.

\mbox{ }

\noindent\large{\bf 1.6 Lapse-uneliminated actions}\normalsize

\mbox{ }

\noindent
I have explained why the interpretation of the BSW form
as a geodesic principle is subject to considerable complications,
and that it may obscure which theories are permitted or forbidden
in the TSA.  I next show that the use of the manifestly lapse-eliminated BSW form,
and consequently the Problems with its interpretation,
may be circumvented by the use of lapse-uneliminated but {\sl lapse eliminable} actions 
(just as the Jacobi and Euler--Lagrange interpretations of mechanics are 
equivalent).
It is easy to show that the equations of motion that follow from
the lapse-uneliminated 3 + 1 `ADM' Lagrangian (\ref{VBSWmethod})
are weakly equivalent to the BSW ones:
$$
\left(\frac{\partial p^{ij}} {\partial \lambda}\right)_{\mbox{\scriptsize 
ADM\normalsize}}
= \sqrt{h}N
\left(
h^{ij}\frac{\sigma R + \Lambda}{2} - \sigma R^{ij}
\right)
 - \frac{2N}{\sqrt{h}}\left(p^{im}{p_m}^j -\frac{X}{2}p^{ij}p\right)
$$
$$
\mbox{  }\mbox{  } + \frac{N}{2\sqrt{h}}h^{ij}
\left(
p\circ p - \frac{X}{2}p^2 
\right)
+ \sigma\sqrt{h}(D^iD^jN -h^{ij} D^2 N) + \pounds_{\xi}p^{ij}
$$
$$
= \sqrt{h}N[h^{ij}(\sigma R\mbox{+}\Lambda)\mbox{--}\sigma R^{ij}] - 
\frac{2N}{\sqrt{h}}(p^{im}{p_m}^j -\frac{X}{2}p^{ij}p)
$$
$$
+ \sigma\sqrt{h}( D^iD^jN - h^{ij}D^2 N)
+ \pounds_{\xi}p^{ij}
- \frac{N}{2}h^{ij}
\left[
\sqrt{h}(\sigma R + \Lambda)
- \frac{1}{\sqrt{h}}
\left(
p\circ p - \frac{X}{2}p^2 
\right)
\right]
$$
\be
= 
\left(
\frac{\partial p^{ij}}{\partial \lambda}
\right)_{\mbox{\scriptsize BSW \normalsize}}
+ \frac{N}{2}h^{ij}{\cal H} 
\mbox{ } ,
\label{VBSWtoADM}
\ee
and similarly when matter terms are included.  I use arbitrary $\sigma$ and $W$ 
above to simultaneously treat the GR and strong gravity cases.
The ADM propagation of the Hamiltonian 
\be
\mbox{constraint is slightly simpler than 
the BSW one, }
\mbox{\hspace{0.6in}}
\dot{{\cal H}} = \frac{\sigma}{N}D^i(N^2{\cal H}_i) + \pounds_{\xi}{\cal H}
\mbox{\hspace{0.8in}}
\ee
for $W = 1$ or $\sigma = 0$,
where it is understood that the evolution is carried out by the
ADM ELE's or their strong gravity analogues.

I next check that using uneliminated actions does not damage the conformal branch of the TSA.  
The conformal gravity action (\ref{VBOaction}) is equivalent to
\be
\mbox{\sffamily I\normalfont} = \int \textrm{d}\lambda
\frac
{\int \textrm{d}^3x\sqrt{h}N\phi^4
\left[
\sigma
\left(
R - \frac{ 8 D^2 \phi }{ \phi }
\right)
+ \frac{\Lambda\phi^4}{V^{\frac{2}{3}}(\phi)}
+ \frac{\mbox{\sffamily\scriptsize T\normalsize\normalfont}_{\mbox{\tiny C\normalsize}}}{4N^2}
\right]}
{V(\phi)^{\frac{2}{3}}}
\ee
where the lapse is 
$N = \frac{1}{2}
\sqrt{    \frac{   \mbox{\sffamily\scriptsize T\normalsize\normalfont}_{\mbox{\tiny C\normalsize}}   }
{   \sigma 
\left(
R - \frac{ 8 D^2 \phi }{ \phi }
\right)
+ \frac{  \Lambda\phi^4  }{  V^{\frac{2}{3}}  }   }    }$.
The following equivalent of (\ref{VBSWtoADM}) holds:
\be
\left(\frac{\partial p^{ij}}
{\partial\lambda}\right)_{\mbox{\scriptsize N-uneliminated\normalsize}} =
\left(\frac{\partial p^{ij}}
{\partial\lambda}\right)_{\mbox{\scriptsize N-eliminated\normalsize}}
+ h^{ij}
\left(
\frac{N{\cal H}^{\mbox{\scriptsize C\normalsize}}}{2}
- \frac{\sqrt{h}\phi^6}{3V}\int \textrm{d}^3x 
{   N {\cal H}^{\mbox{\scriptsize C\normalsize}}   } 
\right) 
\mbox{ } ,
\label{VCBSWtoCADM}
\ee
\be
\mbox{for }
\mbox{\hspace{1.2in}}
{\cal H}^{\mbox{\scriptsize C\normalsize}} \equiv
- \frac{  \sqrt{h}\phi^4   }{   V^{  \frac{2}{3}  }   }
\left[
\sigma
\left( 
R - \frac{8D^2\phi}{\phi}
\right)
+ \frac{\Lambda\phi^4}{V^{\frac{2}{3}}}
\right]
+ \frac{V^{\frac{2}{3}}}{\sqrt{h}\phi^4}p\circ p 
\mbox{\hspace{1.2in}}
\ee
the conformal gravity equivalent of the Hamiltonian constraint.

In the next section, I develop a strategy involving the study of lapse-uneliminated actions.
This represents a first step in disentangling Barbour's no time \cite{B94I, B94II, EOT}
and no scale \cite{CGPD, CG} ideas.  It also permits me to investigate which standard 
theories exist according to the other TSA rules, by inspection of formalisms of these 
theories.  Where possible, one could then eliminate the lapse to cast theories into TSA form.  
This is not always possible: some perfectly good theories appear to have no actions which are 
simultaneously best-matched and BSW reformulable (in having an algebraically-eliminable 
lapse).  Thus the uneliminated form can be used to help test whether the TSA is or can be made 
to be a satisfactory scheme for all of nature.

I furthermore next use this lapse-uneliminated formulation to interpret the GR branch of the 
TSA within Kucha\v{r}'s {\it hypersurface} or {split spacetime framework} (SSF), which has 
striking interpretational consequences.

\mbox{ }

\noindent\Large{\bf 2 The 3-space approach and the split spacetime framework}\normalsize

\mbox{ }

\noindent\large{\bf 2.1 Kucha\v{r}'s hypersurface or split spacetime framework}\normalsize

\mbox{ }

\noindent In his series of four papers, Kucha\v{r} considers the deformation of a hypersurface \cite{KucharI}, 
the kinematics of tensor fields on the hypersurface \cite{KucharII}, the dynamics of the fields split 
with respect to the the hypersurface \cite{KucharIV}, and the geometrodynamics of the fields \cite{KucharIV}.  
The fields are decomposed into perpendicular and tangential parts, as in I.2.1 (N.B $\epsilon = -1$ 
in this section).  I am mainly concerned with 1-forms, for which the decomposition is
$A_{A} = n_{A}A_{\perp} + e^a_{A}A_a$. A deformation at a point $p$ of a hypersurface $\Sigma$ may 
be decomposed as in (fig 12). 
\begin{figure}[h]
\centerline{\def\epsfsize#1#2{0.4#1}\epsffile{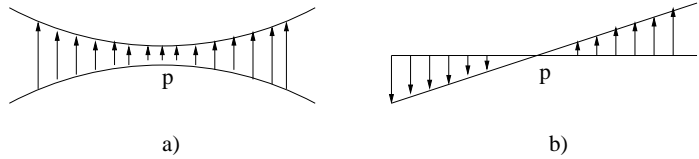}}
\caption[]{\label{TO7.ps}
\scriptsize a)the \it translation \normalfont part is such that $\alpha(p) \neq 0$, $[\pa_a\alpha](p) = 0$.  
 b) \it tilt \normalfont part is such that $\alpha(p) = 0$, $[\pa_a\alpha](p) \neq 0$.
\normalsize}
\end{figure}

I follow Kucha\v{r}'s use of first-order actions.  For the 1-form, this amounts to rewriting 
$$
\mbox{the second-order action } 
\mbox{\hspace{1.1in}}
\mbox{\sffamily I\normalfont}_{\mbox{\scriptsize A\normalsize}}
= \int \textrm{d}^4x \sqrt{|g|} \mbox{\sffamily L\normalfont}(A_{A}, \nabla_{B}A_{C}, g_{DE})
\mbox{\hspace{1.1in}}
$$
$$
\mbox{by setting }
\lambda^{AB} = \frac{\pa \mbox{\sffamily L\normalfont}}{\pa (\nabla_{B}A_{A})} \mbox{ and using the 
Legendre transformation } 
\mbox{\hspace{2in}}
$$
$(A_{A}, \nabla_{B}A_{C}, \mbox{\sffamily L\normalfont}) \longrightarrow (A_{A}, \lambda_{CB}, L)$,
where the {\it Lagrangian potential} is

\mbox{ }

\noindent 
$L = [\lambda^{AB}\nabla_{B}A_{A} - \mbox{\sffamily L\normalfont}](A_{A}, \lambda_{BC}, g_{DE})$.
Then the `hypersurface Lagrangian' is
\be
\delta_{\beta}  \mbox{\sffamily I\normalfont}^{\mbox{\scriptsize A\normalsize}}
= \int_{\Sigma}\textrm{d}^3x(\pi^{\perp}\delta_{\check{\beta}}A_{\perp}
+ \pi^a\delta_{\check{\beta}}A_a -
\alpha \mbox{ }_{\mbox{\scriptsize A\normalsize}}{\cal H}^{\mbox{\scriptsize o\normalsize}}
- \beta^a { }_{\mbox{\scriptsize A\normalsize}}{\cal H}^{\mbox{\scriptsize o\normalsize}}_a)
\label{Vcovectoraction}
\ee
where the $A_i$-contribution to the momentum constraint 
$_{\mbox{\scriptsize A\normalsize}}{\cal H}^{\mbox{\scriptsize o\normalsize}}_a$ 
is obtained from 
\be
\frac{\pa}{\pa\lambda} \equiv \delta_{\alpha} = \delta_{\check{\beta}} - \pounds_{\beta}
\ee
(see fig 13) integrating by parts where necessary, and the $A_i$-contribution to the 
Hamiltonian constraint on a fixed background 
$_{\mbox{\scriptsize A\normalsize}}{\cal H}^{\mbox{\scriptsize o\normalsize}}$ may be further 
decomposed into its translation and tilt parts,\fn{The subscript $_{\not-}$ denotes the tilt 
part and the subscript $_{\mbox{\scriptsize t\normalsize}}$ denotes the translational part.}
\be
_{\mbox{\scriptsize A\normalsize}}{\cal H}^{\mbox{\scriptsize o\normalsize}}  =
{}_{\mbox{\scriptsize A\normalsize}}{\cal H}^{\mbox{\scriptsize o\normalsize}}_{\mbox{\scriptsize t\normalsize}}  
+ {_{\mbox{\scriptsize A\normalsize}}{\cal H}^{\mbox{\scriptsize o\normalsize}}_{\not-}} 
\mbox{ } .
\label{Vhamdec}
\ee
The translational part 
$_{\mbox{\scriptsize A\normalsize}}{\cal H}^{\mbox{\scriptsize o\normalsize}}_{\mbox{\scriptsize t\normalsize}}$ 
may contain a term $2 {}_{\mbox{\scriptsize A\normalsize}}P^{ab}K_{ab}$ due to the possibility of 
\it derivative coupling \normalfont of the metric to the 1-form, whilst the remainder of 
$_{\mbox{\scriptsize A\normalsize}}{\cal H}^{\mbox{\scriptsize o\normalsize}}_{\mbox{\scriptsize t\normalsize}}$ 
is denoted by $_{\mbox{\scriptsize A\normalsize}}{\cal H}_{\mbox{\scriptsize t\normalsize}}$: 
\be
_{\mbox{\scriptsize A\normalsize}}{\cal H}^{\mbox{\scriptsize o\normalsize}}_{\mbox{\scriptsize t\normalsize}}  =
{}_{\mbox{\scriptsize A\normalsize}}{\cal H}_{\mbox{\scriptsize t\normalsize}} + 2 {}_{\mbox{\scriptsize 
A\normalsize}}P^{ab}K_{ab} 
\mbox{ } .
\label{Vtransdec}
\ee 
\begin{figure}[h]
\centerline{\def\epsfsize#1#2{0.6#1}\epsffile{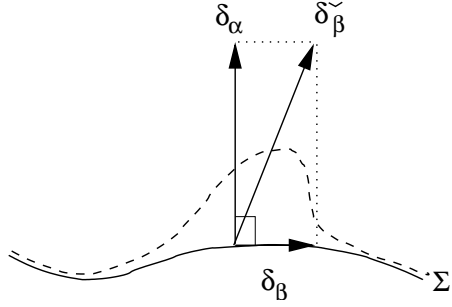}}
\caption[]{\label{TO8.ps}
\footnotesize The change along an arbitrary deformation of the hypersurface $\Sigma$ 
is split according to $\delta_{\check{\beta}} = \delta_{\alpha}+ \delta_{\beta}$.  
Kucha\v{r} shows that $\pounds_{\beta} = \delta_{\beta}$ when acting on spatial tensors 
in Sec 7 of \cite{KucharI} and Sec 3 of \cite{KucharIII}\normalsize}
\end{figure}

\mbox{ } 

\noindent\large{\bf 2.2 SSF as a TSA tool}\normalsize 

\mbox{ }

\noindent I will show that BF\'{O} original TSA is complete for the bosons of nature, and more.  

For the 1-form field, using the decomposition 
\be
\lambda^{AB} = \lambda^{\perp\perp}n^{A}n^{B} + 
\lambda^{a\perp}e^{A}_a n^{B} 
+ \lambda^{\perp b}n^{A}e^{B}_b + 
\lambda^{ab}e^{A}_ae^{B}_b
\ee
and 
$\lambda^{a\perp} = \pi^a, \lambda^{\perp\perp} = \pi^{\perp}$ 
(by the definition of canonical momentum), one obtains 
\be
_{\mbox{\scriptsize A\normalsize}}{\cal H}_{\mbox{\scriptsize t\normalsize}} = L + \sqrt{h}
(\lambda^{\perp a}D_a A_{\perp} - \lambda^{ab}D_aA_b) 
\mbox{ } .
\ee
\be
\mbox{One also requires }
\mbox{\hspace{0.9in}}
_{\mbox{\scriptsize A\normalsize}}P^{ab} = \frac{\sqrt{h}}{2}( - 
A^{(a|}\lambda^{\perp|b)}
+ A_{\perp}\lambda^{(ab)} - A^{(a}\pi^{b)}) 
\mbox{ } . 
\mbox{\hspace{0.9in}}
\label{V1formderiv}
\ee
For the 1-form, $\lambda^{\perp a}$ and $\lambda^{ab}$ play the role of Lagrange multipliers;
one would then use the corresponding multiplier equations to attempt to eliminate the 
multipliers from (\ref{Vcovectoraction}).  In my examples below, $A_{\perp}$ will also occur 
as a multiplier, but this is generally not the case.

The above sort of decomposition holds for any rank of tensor field.   
${\cal H}^{\mbox{\scriptsize o\normalsize}}_{\not-}$, $P^{ab}$ and $\pounds_{\beta}$ are 
universal for each rank, whereas ${\cal H}_{\mbox{\scriptsize t\normalsize}}$ contains $L$, 
which has further details of the particular field in question.  These three universal features 
are the kinematics due to the presupposition of spacetime.  The $\pounds_{\beta}$ contribution 
is `shift kinematics', whilst the tilt contribution is `lapse kinematics'.

The point of Kucha\v{r}'s papers is to construct very general consistent matter theories by 
presupposing spacetime and correctly implementing the resulting kinematics.  I show below that 
in not presupposing spacetime, BF\'{O} are attempting to construct consistent theories by using 
shift kinematics (Diff-BM) alone, and thus attempting to deny the presence of any 
`lapse kinematics' in nature.  This turns out to be remarkably successful for the bosonic 
theories of nature.     

I begin by noting that nonderivative-coupled fields are a lot simpler to deal with than 
derivative-coupled ones.  I then ask which fields are included in this simpler case, in which 
the matter fields do not affect the gravitational part of the Hamiltonian constraint so that 
the gravitational momenta remain independent of the matter fields.  Thus it became clear to 
me that this is a \sl tacit assumption \normalfont in almost all of BF\'{O}'s work.

\noindent \bf TSA Gravity--Matter Simplicity 0 \normalfont : the implementation of `adding on' matter is for 
matter contributions that do not interfere with the structure of the gravitational theory.

\noindent This amounts to the absence of Christoffel symbols in the matter Lagrangians, 
which is true of minimally-coupled scalar fields ($D_a\varsigma = \pa_a\varsigma$) and of Maxwell and 
Yang--Mills theories and their massive counterparts (since $D_aA_b - D_bA_a = \pa_aA_b - \pa_bA_a$).  
Thus it suffices to start off by considering the nonderivative-coupled case on the grounds that it 
includes all the fields hitherto thought to fit in with the BF\'{O} scheme, and also the massive 
1-form fields.    

Consider then the Proca 1-form.  Its Lagrangian is
\be
\mbox{\sffamily L\normalfont}^{\mbox{\scriptsize A\normalsize}}_{\mbox{\scriptsize Proca\normalsize}}
= - \nabla_{[A}A_{B]}\nabla^{[A}A^{B]} - 
\frac{m^2}{2}A_{A}A^{A} 
\mbox{ } ,
\label{Vlagproca}
\ee
\be
\mbox{with corresponding Lagrangian potential }
\mbox{\hspace{0.6in}}
L = -\frac{1}{4}\lambda^{[AB]}\lambda_{[AB]} + 
\frac{m^2}{2}A_{A}A^{A} 
\mbox{ } .
\mbox{\hspace{0.6in}}
\ee
\be
\mbox{Whereas $_{\mbox{\scriptsize A\normalsize}}{\cal H}_{\not-}^{\mbox{\scriptsize o\normalsize}}$ 
has in fact been completed to a divergence, }
_{\mbox{\scriptsize (A)\normalsize}}{\cal H}^{\mbox{\scriptsize o\normalsize}}_{\not-} = A^aD_a\pi^{\perp} 
+ A_{\perp} D_a\pi^a
\mbox{\hspace{2in}}
\ee
suffices to generate the tilt change of $A_{\perp}$ and $A_a$ for the universal 1-form 
(see Sec 6 of \cite{KucharIII}).  The first term of this vanishes since $\pi^{\perp} = 0$ by 
antisymmetry for the 1-forms described by (\ref{Vlagproca}).
Also $_{\mbox{\scriptsize A\normalsize}}P^{ab} = 0$ by antisymmetry so
\be
_{\mbox{\scriptsize A\normalsize}}{\cal H}^{\mbox{\scriptsize o\normalsize}} =
\sqrt{h}\left[ -\frac{1}{4}\lambda_{ab}\lambda^{ab} + \frac{1}{2h}\pi_a\pi^a
+ \frac{m^2}{2}(A_aA^a - A_{\perp}^2) - \lambda^{ab}A_{[a,b]} \right] + 
A_{\perp}D_a\pi^a 
\mbox{ } .
\label{VProcaham}
\ee
\be
\mbox{by (\ref{Vhamdec}, \ref{Vtransdec}).  
The multiplier equation for $\lambda_{ab}$ gives } 
\mbox{\hspace{0.35in}}
\lambda_{ab} = -2 D_{[b}A_{a]} \equiv B_{ab} 
\mbox{ } .
\mbox{\hspace{2in}}
\label{Vlme}
\ee
\be
\mbox{For $m \neq 0$, the multiplier equation for $A_{\perp}$ gives }
\mbox{\hspace{0.8in}}
A_{\perp} = - \frac{1}{m^2\sqrt{h}} D_a\pi^a 
\mbox{ } , 
\mbox{\hspace{2in}}
\label{Vlightening}
\ee
and elimination of the multipliers in (\ref{VProcaham}) using (\ref{Vlme}, \ref{Vlightening}) gives
\be
_{\mbox{\scriptsize A\normalsize}}{\cal H}^{\mbox{\scriptsize o\normalsize}} = \frac{1}{2\sqrt{h}}\pi_a\pi^a 
+ \frac{\sqrt{h}}{4}B_{ab}B^{ab}
+ \frac{m^2\sqrt{h}}{2}A_aA^a + \frac{1}{2m^2\sqrt{h}}(D_a\pi^a)^2 
\mbox{ } ,
\label{Vunel}
\ee
which is non-ultralocal in the momenta.  I note that this does nothing to eliminate 
the remaining term in the tilt: the Proca field has nonzero tilt.  

But, for $m = 0$, the $A_{\perp}$ multiplier equation gives instead the Gauss constraint of 
\be
\mbox{electromagnetism }  
\mbox{\hspace{2in}}
{\cal G} \equiv D_a\pi^a \approx 0 
\mbox{ } .  
\mbox{\hspace{2in}}
\label{VGaussweak}
\ee
This would not usually permit $A_{\perp}$ to be eliminated from (\ref{Vunel}) but the final 
form of $_{\mbox{\scriptsize A\normalsize}}{\cal H}^{\mbox{\scriptsize o\normalsize}}$ 
\be
\mbox{for $m = 0$ is }
\mbox{\hspace{1in}} 
_{\mbox{\scriptsize A\normalsize}}{\cal H}^{\mbox{\scriptsize o\normalsize}} = \frac{\sqrt{h}}{4}B^{ab}B_{ab} + 
\frac{1}{2\sqrt{h}}\pi_a\pi^a + A_{\perp}(D_a\pi^a \approx 0) 
\mbox{ } , 
\mbox{\hspace{1in}}
\ee
so the cofactor of $A_{\perp}$ in (\ref{Vcovectoraction}) weakly vanishes by (\ref{VGaussweak}), 
so $A_{\perp}$ may be taken to `accidentally' drop out.
This means that the tilt of the Maxwell field may be taken to be zero.
The tilt is also zero for the metric and for the scalar field.  
So far all these fields are allowed by BF\'{O} and have no tilt, 
whereas the so-far-disallowed Proca field has tilt.  

\mbox{ }

I can begin to relate this occurrence to the BSW or generalized BSW implementation of 
\bf R2 \normalfont.  For, suppose an action has a piece depending on 
$\pa_a\alpha$ in it.  Then the immediate elimination of $\alpha$ from it {\sl is not 
algebraic}, so the procedure of BSW is not possible.  By definition, the tilt part of 
the Hamiltonian constraint is built from the $\pa_a\alpha$ contribution using integration by 
parts.  But, for the $A_{\perp}$-eliminated Proca Lagrangian, this integration by parts 
gives a term that is non-ultralocal in the momenta, $(D_a\pi^a)^2$, which again contains 
$\pa_a\alpha$ within.  Thus, for this formulation of Proca theory, one cannot build a 
TSA Einstein--Proca action to start off with.   Of importance, this difficulty with spatial 
derivatives was not foreseen in the simple analogy with the Jacobi principle in mechanics,
where there is only one independent variable.  

The above argument requires refinement from the treatment of further important physical 
examples.  This is a fast method of finding matter theories compatible with the TSA by 
the following argument.  If there is no derivative coupling and if one can arrange for 
the tilt to play no part in a formulation of a matter theory, then all that is left of the 
hypersurface kinematics is the shift kinematics, which is {\bf BM[general R1]}.    
But complying with hypersurface kinematics is a guarantee for consistency for established 
spacetime theories so in these cases BM suffices for consistency.

\mbox{ }

First, I consider \bf K \normalfont interacting 1-forms 
$A_a^{\mbox{\bf\scriptsize K\normalsize\normalfont}}$
with Lagrangian\fn{By $\mbox{\sffamily g\normalfont}
{f^{\mbox{\bf\tiny A\normalsize\normalfont}}}_{\mbox{\bf\tiny BC\normalsize\normalfont}}$ 
I strictly mean $\mbox{\sffamily g\normalfont}^{\cal A}
f^{\mbox{\bf\tiny A\normalsize\normalfont}}_{{\cal A}\mbox{\bf\tiny BC\normalsize\normalfont}}$ where ${\cal A}$ indexes each gauge subgroup in a direct product.  
Then each such gauge subgroup can be associated with a distinct coupling constant $\mbox{\sffamily g\normalfont}^{\cal A}$.}.  
\be
\mbox{\sffamily L\normalfont}^{\mbox{\scriptsize A\normalsize}_{\mbox{\bf\tiny M\normalsize\normalfont}}}_{\mbox{\scriptsize massive YM\normalsize}}
\mbox{=--}
\left(
\nabla_{[A}A^{\mbox{\bf\scriptsize A\normalsize\normalfont}}_{B]}
+ \frac{\mbox{\sffamily g\normalfont}_{\mbox{\scriptsize c\normalfont}}}{2}
{f^{\mbox{\bf\scriptsize A\normalsize\normalfont}}}_{\mbox{\bf\scriptsize BC\normalsize\normalfont}}
A^{\mbox{\bf\scriptsize B\normalsize\normalfont}}_{B}
A^{\mbox{\bf\scriptsize C\normalsize\normalfont}}_{A}
\right)
\left(
\nabla^{[A}A_{\mbox{\bf\scriptsize A\normalsize\normalfont}}^{B]}
+ \frac{\mbox{\sffamily g\normalfont}_{\mbox{\scriptsize c\normalfont}}}{2}
{f_{\mbox{\bf\scriptsize ADE\normalsize\normalfont}}}A^{\mbox{\bf\scriptsize D\normalsize\normalfont}B}
A^{\mbox{\bf\scriptsize E\normalsize\normalfont}A}
\right)
- \frac{m^2}{2}A_{A{\mbox{\bf\scriptsize M\normalsize\normalfont}}}
A^{A{\mbox{\bf\scriptsize M\normalsize\normalfont}}} 
\mbox{ } .
\label{VlagYMmass}
\ee
$$
\mbox{I define } \mbox{ } 
\lambda^{AB}_{\mbox{\bf\scriptsize M\normalsize\normalfont}}
= \frac{\pa \mbox{\sffamily L\normalfont}}
{\pa (\nabla_{B}A^{\mbox{\bf\tiny M\normalsize\normalfont}}_A)} 
\mbox{ }
\mbox{ and the corresponding Lagrangian potential is }
\mbox{\hspace{2in}}
$$
\be
L = -\frac{1}{4}\lambda^{[AB]}_{\mbox{\bf\scriptsize M\normalsize\normalfont}}
\lambda^{\mbox{\bf\scriptsize M\normalsize\normalfont}}_{[AB]}
- \frac{\mbox{\sffamily g\normalfont}_{\mbox{\scriptsize c\normalfont}}}{2}
f_{\mbox{\bf\scriptsize BDE\normalsize\normalfont}}
A_{B}^{\mbox{\bf\scriptsize D\normalsize\normalfont}}
A_{A}^{\mbox{\bf\scriptsize E\normalsize\normalfont}}
\lambda^{AB{\mbox{\bf\scriptsize B\normalsize\normalfont}}}
+ \frac{m^2}{2}A_{B{\mbox{\bf\scriptsize M\normalsize\normalfont}}}
A^{A{\mbox{\bf\scriptsize M\normalsize\normalfont}}} 
\mbox{ } .
\ee
The overall tilt contribution is now the sum of the tilt contributions of the individual fields, so  
$_{(\mbox{\scriptsize A\normalsize}_{\mbox{\bf\tiny M\normalsize\normalfont}})}{\cal H}^{\mbox{\scriptsize o\normalsize}}_{\not-}
= A^{\mbox{\bf\scriptsize M\normalsize\normalfont}}_{\perp}D_a
\pi^a_{\mbox{\bf\scriptsize M\normalsize\normalfont}}$
suffices to
generate the
tilt change.
Again,
$_{{\mbox{\scriptsize A\normalsize}}_{\mbox{\bf\tiny M\normalsize\normalfont}}}P^{ab} = 0$
by antisymmetry so 
$$
_{\mbox{\scriptsize A\normalsize}_{\mbox{\bf\tiny M\normalsize\normalfont}}}
{\cal H}^{\mbox{\scriptsize o\normalsize}}
=  \sqrt{h}\left[ -\frac{1}{4}\lambda^{\mbox{\bf\scriptsize 
M\normalsize\normalfont}}_{ab}
\lambda_{\mbox{\bf\scriptsize M\normalsize\normalfont}}^{ab}
+ \frac{1}{2h}\pi^{\mbox{\bf\scriptsize M\normalsize\normalfont}}_a
\pi_{\mbox{\bf\scriptsize M\normalsize\normalfont}}^a
+ \frac{m^2}{2}(A^{\mbox{\bf\scriptsize M\normalsize\normalfont}}_a
A_{\mbox{\bf\scriptsize M\normalsize\normalfont}}^a -
A^{\mbox{\bf\scriptsize M\normalsize\normalfont}}_{\perp}
A_{\perp{\mbox{\bf\scriptsize M\normalsize\normalfont}}}) -
\lambda_{\mbox{\bf\scriptsize M\normalsize\normalfont}}^{ab}
A^{\mbox{\bf\scriptsize M\normalsize\normalfont}}_{[a,b]}\right]  
$$
\be
+ A^{\mbox{\bf\scriptsize M\normalsize\normalfont}}_{\perp}{\mbox{\bf 
D\normalfont}}_a
\pi_{\mbox{\bf\scriptsize M\normalsize\normalfont}}^a -
\frac{\mbox{\sffamily g\normalfont}_{\mbox{\scriptsize c\normalfont}}}{2}
f_{\mbox{\bf\scriptsize MPQ\normalsize\normalfont}}
(\sqrt{h}\lambda^{ab{\mbox{\bf\scriptsize M\normalsize\normalfont}}}
A^{\mbox{\bf\scriptsize P\normalsize\normalfont}}_bA^{\mbox{\bf\scriptsize 
Q\normalsize\normalfont}}_a +
2 \pi^{\mbox{\bf\scriptsize M\normalsize\normalfont}}
A^{\mbox{\bf\scriptsize P\normalsize\normalfont}}_{\perp}
A^{\mbox{\bf\scriptsize Q\normalsize\normalfont}}_a) 
\ee
by (\ref{Vhamdec}, \ref{Vtransdec}).  
The multipliers are
$\lambda^{ab}_{\mbox{\bf\scriptsize M\normalsize\normalfont}}$
and $A_\perp^{\mbox{\bf\scriptsize M\normalsize\normalfont}}$,
with corresponding
multiplier equations
\be
\lambda^{\mbox{\bf\scriptsize M\normalsize\normalfont}}_{ab}
= -2 D_{[b}A^{\mbox{\bf\scriptsize M\normalsize\normalfont}}_{a]}
\equiv B^{\mbox{\bf\scriptsize M\normalsize\normalfont}}_{ab} 
\mbox{ } ,
\ee
\be
A_{\mbox{\bf\scriptsize M\normalsize\normalfont}\perp} = - 
\frac{1}{m^2\sqrt{h}}
\mbox{\bf D\normalfont}_a\pi^a_{\mbox{\bf\scriptsize 
M\normalsize\normalfont}}
\equiv - \frac{1}{m^2\sqrt{h}}(D_a\pi^a{\mbox{\bf\scriptsize 
M\normalsize\normalfont}}
+ \mbox{\sffamily g\normalfont}_{\mbox{\scriptsize c\normalfont}}
f_{\mbox{\bf\scriptsize LMP\normalsize\normalfont}}
\pi^{\mbox{\bf\scriptsize L\normalsize\normalfont}a}
A_a^{\mbox{\bf\scriptsize P\normalsize\normalfont}})
\ee 
for $m \neq 0$.
I thus obtain the eliminated form
\be
_{{\mbox{\scriptsize A\normalsize}}_{\mbox{\bf\tiny M\normalsize\normalfont}}}
{\cal H}^{\mbox{\scriptsize o\normalsize}}
= \frac{1}{2\sqrt{h}}
\pi_{\mbox{\bf\scriptsize 
M\normalsize\normalfont}a}\pi^{\mbox{\bf\scriptsize 
M\normalsize\normalfont}a}
+ \frac{\sqrt{h}}{4}B_{\mbox{\bf\scriptsize M\normalsize\normalfont}ab}
B^{\mbox{\bf\scriptsize M\normalsize\normalfont}ab}
+ \frac{m^2\sqrt{h}}{2}
A_{\mbox{\bf\scriptsize M\normalsize\normalfont}a}A^{\mbox{\bf\scriptsize 
M\normalsize\normalfont}a}
+ \frac{1}{2m^2\sqrt{h}}
(\mbox{\bf D\normalfont}_a\pi^{\mbox{\bf\scriptsize 
M\normalsize\normalfont}a})
(\mbox{\bf D\normalfont}_b\pi_{\mbox{\bf\scriptsize 
M\normalsize\normalfont}b}) 
\mbox{ } .
\ee
and the massive Yang--Mills field is left with nonzero tilt.  For $m = 0$, the second multiplier 
\be
\mbox{equation gives instead the Yang--Mills Gauss constraint } 
\mbox{\hspace{0.5in}}
{\cal G}^{\mbox{\bf\scriptsize M\normalsize\normalfont}}
\equiv {\mbox{\bf D\normalfont}}_a\pi^{\mbox{\bf\scriptsize 
M\normalsize\normalfont}a} \approx 0 
\mbox{ } . 
\mbox{\hspace{2in}}
\label{VYMGaussweak}
\ee
In this case, the tilt is nonzero, but the Yang--Mills Gauss constraint 
`accidentally' enables the derivative part of the tilt to be converted into 
an algebraic expression, which then happens to cancel with part of the Lagrangian potential: 
\be
_{{\mbox{\scriptsize A\normalsize}}_{\mbox{\bf\tiny M\normalsize\normalfont}}}{\cal H}^{\mbox{\scriptsize o\normalsize}} 
= \frac{\sqrt{h}}{4}
B^{ab}_{\mbox{\bf\scriptsize M\normalsize\normalfont}}
B_{ab}^{\mbox{\bf\scriptsize M\normalsize\normalfont}} 
+ \frac{1}{2\sqrt{h}}
\pi^{\mbox{\bf\scriptsize M\normalsize\normalfont}}_a
\pi^a_{\mbox{\bf\scriptsize M\normalsize\normalfont}} 
+ A^{\mbox{\bf\scriptsize M\normalsize\normalfont}}_{\perp}
(D_a\pi^a_{\mbox{\bf\scriptsize M\normalsize\normalfont}} 
+ \mbox{\sffamily g\normalfont}
f_{\mbox{\bf\scriptsize LMP\normalsize\normalfont}}
\pi^{\mbox{\bf\scriptsize L\normalsize\normalfont}a}
A_a^{\mbox{\bf\scriptsize P\normalsize\normalfont}} \approx 0) 
\mbox{ } .
\ee

\mbox{ }

Second, I consider U(1) 1-form--scalar gauge theory [c.f (\ref{U1covderiv}--\ref{U1gautheo})]
\be
\mbox{\sffamily L\normalfont}_{\mbox{\scriptsize U\normalsize}(1)}^{\varsigma, \varsigma^*, \mbox{\scriptsize A\normalsize}}
= - \nabla_{[A}A_{B]}\nabla^{[A}A^{B]} +
(\pa_{A}\varsigma - ieA_{A}\chi)(\pa^{A}\varsigma^* + ieA^{A}\varsigma^*) - 
\frac{m_{\varsigma}^2}{2}\varsigma^*\varsigma 
\mbox{ } .
\label{VlagU(1)}
\ee
$$
\mbox{Now, in addition to $\lambda^{AB}$, I define } \mbox{ } 
\mu^{A} = \frac{\pa\mbox{\sffamily L\normalfont}}{\pa (\nabla_{A}\varsigma)} \mbox{ }
\mbox{ and } \mbox{ }
\nu^{A} = \frac{\pa\mbox{\sffamily L\normalfont}}
{\pa (\nabla_{A}\varsigma^*)} \mbox{ } , \mbox{so the Lagrangian }
\mbox{\hspace{2in}}
$$
\be
\mbox{potential is }
L = -\frac{1}{4}\lambda^{[AB]}\lambda_{[AB]} + 
\frac{m^2}{2}A_{A}A^{A}
+ \mu^{A}\nu_{A} - ieA_{A}(\varsigma^*\nu^{A} - 
\varsigma\mu^{A}) + \frac{m_{\varsigma}^2}{2}\varsigma^*\varsigma 
\mbox{ } .
\mbox{\hspace{0.5in}}
\ee
$_{\mbox{\scriptsize (A)\normalsize}}{\cal H}^{\mbox{\scriptsize o\normalsize}}_{\not-} =   A_{\perp} 
D_a\pi^a$ still
suffices to generate the tilt (as scalars contribute no tilt), 
$$
\mbox{$_{\varsigma, \varsigma^*, \mbox{\scriptsize A\normalsize}}P^{ab} = 0$, and }
_{\varsigma, \varsigma^*, \mbox{\scriptsize A\normalsize}}{\cal H}^{\mbox{\scriptsize o\normalsize}}_{\mbox{\scriptsize t\normalsize}} =  \sqrt{h}
\left[ 
- \frac{1}{4}\lambda_{ab}\lambda^{ab} + \mu_a\nu^a 
+ \frac{1}{h}
\left(
\frac{1}{2}\pi_a\pi^a 
- \pi_{\varsigma}\pi_{\varsigma^*}
\right) 
+ \frac{m_{\varsigma}^2}{2}\varsigma^*\varsigma 
\right.
\mbox{\hspace{2in}}
$$
\be
\mbox{\hspace{2in}}
\left.
- ie\left( A_a[\varsigma^*\nu^a - \varsigma\mu^a] 
- \frac{A_{\perp}}{\sqrt{h}}[\varsigma^*\pi_{\varsigma^*} - \varsigma\pi_{\varsigma}]\right)   
\right] 
\mbox{ } . 
\mbox{\hspace{2in}}
\label{VunelUH}
\ee
The $\lambda_{ab}$ multiplier equation is (\ref{Vlme}) again, whilst the 
$A_{\perp}$ multiplier equation is now 
\be
{\cal G}_{\mbox{\scriptsize U\normalsize\normalfont}(1)}
\equiv D_a\pi^a + ie(\varsigma^*\pi_{\varsigma^*} - \varsigma\pi_{\varsigma}) = 0 
\mbox{ } ,
\label{VsourceGauss}
\ee
which can be explained in terms of electromagnetism now having a fundamental source.  
In constructing $_{\varsigma, \varsigma^*, \mbox{\scriptsize A\normalsize}}{\cal H}^{\mbox{\scriptsize o\normalsize}} $ 
from (\ref{Vhamdec}, \ref{Vtransdec}, \ref{VunelUH}), I can convert the tilt to an algebraic expression by the sourced Gauss law (\ref{VsourceGauss}) 
which again happens to cancel with a contribution from 
$$
\mbox{the Lagrangian potential: }
\mbox{\hspace{0.4in}}
_{\varsigma, \varsigma^*, \mbox{\scriptsize A\normalsize}}{\cal H}^{\mbox{\scriptsize o\normalsize}} 
= -\lambda^{ab}A_{[a,b]} - (\mu^a + \nu^a)\varsigma_{,a} + { }_{(A)}{\cal H}^{\mbox{\scriptsize o\normalsize}}_{\not-} 
+ _{A,\varsigma,\varsigma^*}{\cal H}^{\mbox{\scriptsize o\normalsize}}_{\mbox{\scriptsize t\normalsize}} 
\mbox{\hspace{0.4in}}
$$
$$ 
\mbox{\hspace{2in}}
= 
\left[  
\frac{1}{4}B_{ab}B^{ab} - \mu_a\nu^a 
+ \frac{1}{h}
\left(
\frac{1}{2}\pi_a\pi^a 
- \pi_{\varsigma}\pi_{\varsigma^*}
\right) 
+ \frac{m_{\varsigma}^2}{2}\varsigma^*\varsigma
\right]
\mbox{\hspace{2in}}
$$
\be
\mbox{\hspace{2in}}
+ A_{\perp}[D_a\pi^a + ie(\varsigma^*\pi_{\varsigma^*} - \varsigma\pi_{\varsigma}) \approx 0] 
\mbox{ } . 
\mbox{\hspace{2in}}
\ee  
It is not too hard to show that the last two `accidents' also `accidentally' conspire together to 
wipe out the tilt contribution in Yang--Mills 1-form--scalar gauge theory (\ref{Gcovderiv}, \ref{Ggauscalar}, \ref{Varda}) used to obtain broken 
SU(2) $\times$ U(1) bosons for the electroweak force.  
This theory is also obviously nonderivative-coupled.

\mbox{ }

I now present a more general treatment about the occurrence of these `accidents'.  They arise from 
eliminating $A_{\perp}$ from its multiplier equation.  For this to make sense, $A_{\perp}$ must 
{\sl be} a multiplier, thus $\pi^{\perp} = 0$.  Then for general $L$, the multiplier equation is
\be
\frac{\pa L}{\pa A_{\perp}} +D_a\pi^a = 0 
\mbox{ } .
\label{multipleq}
\ee
Then the requirement that $A_{\perp}D_a\pi^a + L$ be independent of $A_{\perp}$ on using 
(\ref{multipleq}) means that $- A_{\perp}\frac{\pa L}{\pa A_{\perp}} + L$ is independent of 
$A_{\perp}$.  Thus the `accidents' occur whenever the Lagrangian potential is linear in $A_{\perp}$.

\mbox{ }

The above two examples have broadened my understanding, and lead to the following precise  
reformulation of the BSW-type version of \bf RI[R2]) \normalfont within the SSF as 

\noindent \bf RI-castable[R2] \normalfont: I use lapse-uneliminated actions homogeneously 
quadratic in their velocities and permit only those for which the matter contributes a weakly 
vanishing tilt.

\noindent I can combine this with the generalization of dropping the homogeneously quadratic 
requirement, in anticipation of the inclusion of spin-$\frac{1}{2}$ fermions.

So for Einstein--Maxwell theory, Einstein--Yang--Mills theory, and their corresponding 
scalar gauge theories, 
1) the absence of derivative coupling guarantees that they can be coupled to GR without disrupting 
its canonical structure as tacitly assumed by BF\'{O}.  
2) The absence of tilt guarantees that the resulting coupled theories can be put into BSW form.  
Because the theories have homogeneously quadratic kinetic terms, this is the most straightforward 
notion of BSW form.   
3) Now, the SSF guarantees consistency if all the required kinematics are included.  
But the only sort of kinematics left is BM.  Thus, all these theories are guaranteed to exist as 
TSA theories.  

These workings begin to show (if one presupposes spacetime), what sorts of obstacles 
in Kucha\v{r}'s split spacetime ontology might be regarded as responsible for the uniqueness 
results for bosonic matter when one starts from BF\'{O}'s 3-d ontology.    

\mbox{ }

\noindent\large{\bf 2.3 The TSA allows more than the fields of nature}\normalsize

\mbox{ }

\noindent I have described how the fields hitherto known to be permitted by the TSA may be 
identified within the SSF.  These fields all have just BM kinematics and no other 
significant universal feature (tilt or derivative coupling).  Are these fields then the known 
fundamental matter fields, which somehow have less universal kinematic features than generally 
covariant spacetime structure would lead one to expect?  This question may be subdivided as 
follows.  Does the TSA single out \sl only \normalfont the known fundamental matter fields?  
Does the TSA single out \sl all \normalfont the known fundamental matter fields?  Kucha\v{r} 
makes no big deal about the simplified form weakly equivalent to his decomposition of the 
electromagnetic field, because it does not close to reproduce the Dirac Algebra (see 
Secs 11-12 of \cite{KucharIII}); it only does so modulo ${\cal G}$.  He takes this to be an inconvenience, one 
which can be got round by adhering to the form directly obtained from the decomposition, 
whereas BF\'{O} take it as a virtue that the simplified form `points out' the new constraint, 
${\cal G}$, as an integrability condition.  

The first question can be answered by counterexample.  One should interpret the question as 
coarsely as possible; for example one could argue that the TSA is not capable of restricting 
the possibility of Yang--Mills theory to the gauge groups conventionally used to describe 
nature, or that by no means is massless 1-form--scalar gauge theory guaranteed to occur in 
nature.  Rather than such subcases or effects due to interaction terms, I find it more 
satisfactory to construct a distinct matter theory which is not known to be present in nature.  
The last subsection has put me into a good position to do this.
\be
\mbox{ } \mbox{ Consider the 2-form $\Phi_{AB}$ Lagrangian }
\mbox{\hspace{0.4in}}
\mbox{\sffamily L\normalfont}^{\Phi} =
- \nabla_{[C}\Phi_{AB]}\nabla^{[C}\Phi^{AB]}
- \frac{m^2}{2}\Phi_{AB}\Phi^{AB} 
\mbox{ } , 
\mbox{\hspace{0.3in}}
\ee
$$
\mbox{define } \mbox{ }
\lambda^{ABC} = \frac{\pa \mbox{\sffamily L\normalfont}}{\pa (\nabla_{C}\Phi_{AB})} 
\mbox{ } \mbox{ and use the Legendre transformation to obtain the}
\mbox{\hspace{1in}}
$$
\be
\mbox{Lagrangian potential }
\mbox{\hspace{1.5in}}
L = -\frac{1}{4}\lambda^{[ABC]}\lambda_{[ABC]}
+ \frac{m^2}{2}\Phi_{AB}\Phi^{AB} 
\mbox{ } .
\mbox{\hspace{2in}}
\ee
Then $_{(\Phi)}{\cal H}^{\mbox{\scriptsize o\normalsize}}_{\not-} = 2\Phi_{\perp b}D_a\pi^{ab}$
suffices to generate the 2-form tilt and $_\Phi P^{ab} = 0$ by antisymmetry.
The multipliers are $\lambda^{abc}$ and $A_{\perp a}$ with corresponding 
multiplier equations 
\be 
\mbox{$\lambda_{abc} = -2 D_{[b}\Phi_{ab]} \equiv B_{abc}$} \mbox{ } \mbox{ and, for 
$m \neq 0$, }  \mbox{ }
\Phi^b_{\perp} = - \frac{1}{m^2\sqrt{h}} D_a\pi^{ab} 
\mbox{ } , 
\mbox{\hspace{2in}}
\ee
which may be used to eliminate the multipliers,
giving rise to the non-ultralocal form
\be
_{\Phi}{\cal H}^{\mbox{\scriptsize o\normalsize}} = \frac{\sqrt{h}}{4}B^{abc}B_{abc} + 
\frac{3}{4\sqrt{h}}\pi^{ab}\pi_{ab}
+\frac{7}{8m^2\sqrt{h}}h_{bd}
(D_a\pi^{ab})(D_c\pi^{cd}) + \frac{m^2}{2}\Phi_{ab}\Phi^{ab} 
\mbox{ } .
\ee

\be
\mbox{ } \mbox{ But for $m = 0$, the $\Phi_{\perp b}$ multiplier constraint is }
\mbox{\hspace{0.8in}}
{\cal G}^b \equiv D_a\pi^{ab} \approx 0
\label{V2Gaussweak}
\mbox{\hspace{2in}} 
\ee
\be 
\mbox{and }\mbox{\hspace{1.6in}}
_{\Phi}{\cal H}^{\mbox{\scriptsize o\normalsize}} = \frac{h}{4}B^{abc}B_{abc} + \frac{3}{4\sqrt{h}}\pi^{ab}\pi_{ab} 
+ 2\Phi_{\perp b}(D_a\pi^{ab} \approx 0) 
\mbox{ } .
\mbox{\hspace{2in}}
\ee
So the massless 2-form's tilt is zero and this leads to the elimination of $\Phi_{\perp b}$ 
by the same sort of `accident' that permits $A_{\perp}$ to be eliminated in electromagnetism.    
So, for this massless 2-form, BM is equivalent to all the spacetime hypersurface kinematics, 
and as this guarantees closure, we deduce that there exists a resulting TSA theory 
starting with
\be
\mbox{\sffamily I\normalfont}^{\Phi} = \int \textrm{d}\lambda \int \textrm{d}^3x\sqrt{h}\sqrt{R + D_{[c}\Phi_{ab]}D^{[c}\Phi^{ab]}}
\sqrt{\mbox{\sffamily T\normalfont}^{\mbox{\scriptsize g\normalsize}} + h^{ab}h^{cd}(\dot{\Phi}_{[ab]} - \pounds_{\xi}\Phi_{ab})(\dot{\Phi}_{[cd]} - \pounds_{\xi}\Phi_{cd}) } 
\mbox{ } ,
\ee 
which leads to the enforcement of (\ref{V2Gaussweak}), which is subsequently encoded by the 
introduction of an auxiliary variable $\Theta_b$.  This working should also hold for any $p$-form 
for $p \leq d$, the number of spatial dimensions.  Yet only the $p = 1$ case, electromagnetism, is 
known to occur.  This is evidence against BF\'{O}'s speculation that the TSA 
``hints at partial unification'' of gravity and electromagnetism, since these extra unknown fields would also be 
included as naturally as the electromagnetic field.  Note also that the ingredients of low-energy 
string theory are getting included rather than excluded: $p$-forms, the dilatonic coupling...
These are signs that the TSA is not as restrictive as BF\'{O} originally hoped.  

The second question must be answered exhaustively.  It is the minimal requirement for 
the TSA to be taken seriously as a description of nature.  The TSA gives gravity, electromagnetism 
and Yang--Mills theories including those of the strong and electroweak nuclear forces.  
One may argue that disallowing fundamental Proca fields is unimportant -- use spontaneous symmetry 
breaking -- but the inability at this stage to treat the weak bosons directly is disturbing.  The 
next project is the inclusion of spin-$\frac{1}{2}$ fermions (see VI.4),  in order to complete the 
TSA for the theories of the simplest free fundamental fields that can account for nature (though 
recollect that this is subject to the simplicity in I.2.3.4).   I then follow by investigating all 
the interactions involved in the Standard Model \cite{PS}.  

\mbox{ }

\noindent\large{\bf 2.4 Derivative coupling and the 3-space 1-form ansatz}\normalsize

\mbox{ }

\noindent In their study of 1-forms, BF\'{O} used a BSW-type action with potential term 
\be
\mbox{\sffamily U\normalsize}_{\mbox{\scriptsize A\normalsize}} = C^{abcd}D_bA_{a}D_dA_{c} + \frac{M^2}{2}A_aA^a,
\label{vfPansatz}
\ee
which is natural within their 3-space ontology.  They then obtain 
$^{\mbox{\scriptsize A\normalsize}}{\cal H}$ and $^{\mbox{\scriptsize A\normalsize}}{\cal H}_i$ 
in the usual 3-space way (from the local square root and from $\xi^i$-variation).  Then the 
propagation of $^{\mbox{\scriptsize A\normalsize}}{\cal H}$ enforces 
$C_1 = -C_2$ , $C_3 = 0$ and also ${\cal G}$, whose propagation then enforces $M = 0$.  
Having thus discovered that a new (Abelian) gauge symmetry is present, 
${\cal G}$ is then encoded by the corresponding U(1)-BM,  
by introduction of an auxiliary velocity $\Theta$ [c.f (\ref{Vcorrection})].  
Identifying $\Theta = A_0$, this is a derivation of Einstein--Maxwell theory 
for $\check{A}_{A} = [A_0, A_i]$.

I find it profitable to also explain this occurrence starting from the 4-d ontology 
of the SSF.  The natural choice of 1-form potential and kinetic terms 
\be
\mbox{would then arise from the decomposition of } \mbox{ }
\mbox{\sffamily L\normalfont}^{\mbox{\scriptsize A\normalsize}} =
- C^{ABCD}\nabla_{B}A_{A}\nabla_{D}A_{C} - \frac{M^2}{2}A_AA^A 
\mbox{ } .
\mbox{\hspace{0.2in}}
\label{Vdostrons}
\ee
Using the following set of four results from Sec 2 of \cite{KucharII}, 
\be
\nabla_bA_{\perp} = D_bA_{\perp} - K_{bc}A^c 
\mbox{ } ,
\label{Vderivproj1}
\ee
\be
\alpha\nabla_{\perp} A_{a} = - \delta_{\check{\beta}} A_a - \alpha K_{ab}A^b - A_{\perp}\pa_a\alpha  
\mbox{ } ,
\label{Vderivproj2}
\ee
\be
\nabla_b A_a = D_bA_a - A_{\perp}K_{ab} 
\mbox{ } ,
\label{Vderivproj3}
\ee
\be
\alpha\nabla_{\perp}A_{\perp} = - \delta_{\check{\beta}}A_{\perp} - A^a\pa_a\alpha  
\mbox{ } , 
\label{Vderivproj4}
\ee
$$
\mbox{I obtain that }
\mbox{\hspace{1.3in}}
\mbox{\sffamily L\normalfont}^{\mbox{\scriptsize A\normalsize}}_{\mbox{\scriptsize 3+1\normalsize}} 
= -(C_1 + C_2 + C_3)
\left(
\frac{\delta_{\check{\beta}}A_{\perp} + A^a\pa_a\alpha}{\alpha}
\right)
^2
\mbox{\hspace{1.3in}}
$$
$$
+ C_1
\left[
\left(
\frac{ \delta_{\check{\beta}}A_a + A_{\perp}\pa_a\alpha}{\alpha} + K_{ac}A^c
\right)^2
+ (D_aA_{\perp} - K_{ac}A^c)^2
\right] 
$$
$$ 
+ 2C_2
\left(
\frac{ \delta_{\check{\beta}}A_a + A_{\perp}\pa_a\alpha}{\alpha} + K_{ac}A^c
\right)
\left(
D^aA_{\perp} - {K^a}_{c}A^c
\right)
$$
$$
- 2C_3
\left(
\frac{\delta_{\check{\beta}}A_{\perp} + A^c\pa_c\alpha}{\alpha}
\right)
\left(
D^aA_a - A_{\perp}K
\right) 
$$
\be
- C^{abcd}(D_bA_a - A_{\perp}K_{ab})(D_dA_c - A_{\perp}K_{cd})
- \frac{M^2}{2}(A_aA^a - A_{\perp}A^{\perp}) 
\mbox{ } .
\label{Vstarstar}
\ee
Then, if one adopts the spacetime ontology and then imports BF\'{O}'s 3-space assumptions 
into it, one finds the following `spacetime explanations' for BF\'{O}'s uniqueness results.  

First, BF\'{O}'s tacit assumption that addition of a 1-form $A_a$ does not affect the
3-geometry part of the action can be phrased as there being no derivative coupling,
$_{\mbox{\scriptsize A\normalsize}}P^{ab} = 0$, which using (\ref{V1formderiv}) implies 
that $\lambda^{(ab)} = 0$, $\pi^b = -\lambda^{\perp b}$.  
Since $\lambda^{AB} = -2C^{ABCD}\nabla_{D}A_{C}$, this {\sl by itself} implies 
$C_1 = - C_2$, $C_3 = 0$.

In \cite{Vanderson}, I wrote that Barbour would argue that $A_{\perp}$ is a velocity, since 
it occurs as a

\noindent correction to velocities in the kinetic term \cite{CGPD}\fn{When writing the 
thesis, I derived that this is essentially correct, but care is needed for the reason given 
here:  one could just as well have products of a velocity with coordinates.} (following from 
its auxiliary status, just as $N$ and $\xi^i$ are velocities),it makes sense for the 3-space 
ansatz to contain no $\mbox{\ss}_{\xi}A^{\perp}$.  But from (\ref{Vstarstar}), this by itself 
is also equivalent to $C_1 = - C_2$, $C_3 = 0$ from the 4-d perspective.  Also, inspecting 
(\ref{Vstarstar}) for Maxwell theory reveals that
\be
\mbox{\sffamily L\normalfont}^{\mbox{\scriptsize A\normalsize}}_{\mbox{\scriptsize em(3+1)\normalsize}} 
= \frac{C_1}{N^2}[\delta_{\check{\beta}}A_a - D_a(-\alpha A_{\perp})]^2 - C_1D^bA^a(D_bA_a - D_aA_b) 
\mbox{ } .
\ee
So in fact $\Theta = - \alpha A_{\perp}$, so $A_{\perp}$ itself is not a velocity.
Notice in contrast that the issue of precisely what $\Theta$ is does not arise in the TSA because 
it is merely an auxiliary velocity that appears in the last step of the working.

One argument for the 3-space 1-form field ansatz is simplicity: consideration of a 3-geometry 
and a single 3-d 1-form leads to Maxwell's equations.   However, I argue that in the lapse 
uneliminated form, provided that one is willing to accept the additional kinematics,  I can 
extend these d.o.f's to include a dynamical $A_{\perp}$.   The TSA is about \sl not \normalfont 
accepting kinematics other than BM,  but the spacetime hypersurface framework enables one to 
explore what happens when tilt and derivative-coupling kinematics are `switched on'.  Working 
within the SSF, if $A_{\perp}$ is allowed to be dynamical, there is derivative coupling,  and 
consistency would require the presence of 2 further bunches of terms, with coefficients 
proportional to $C_1 +  C_2$ and to $C_3$. The first bunch consists of the following sorts
$$
\mbox{of terms: }
\mbox{\hspace{0.45in}}  
D^bA^aA_{\perp}\delta_{\check{\beta}}h_{ab}\mbox{ , }
A^b\left(D^aA_{\perp} - A_{\perp}\frac{\pa^a\alpha}{\alpha}\right)\delta_{\check{\beta}}h_{ab} 
\mbox{ , }
\frac{1}{\alpha}h^{ab}A^c\delta_{\check{\beta}}A_a\delta_{\check{\beta}}h_{bc}\mbox{ , } 
\mbox{\hspace{0.45in}}
$$
\be
A^bA^dh^{ac}\delta_{\check{\beta}}h_{ab}\delta_{\check{\beta}}h_{cd}\mbox{ ,  }
A_{\perp}A_{\perp}h^{ac}h^{bd}\delta_{\check{\beta}}h_{ab}\delta_{\check{\beta}}h_{cd} 
\mbox{ } .
\label{Vfirstbunch}
\ee
The second bunch consists of the following sorts of terms:
\be
h^{ab}\left(A_{\perp}D^cA_c + A^c\frac{\pa_c\alpha}{\alpha}\right)\delta_{\check{\beta}}h_{ab} 
\mbox{ , }
\frac{1}{\alpha}h^{ab}A_{\perp}\delta_{\check{\beta}}A_{\perp}\delta_{\check{\beta}}h_{ab} 
\mbox{ , }
A_{\perp}A_{\perp}h^{ab}h^{cd}\delta_{\check{\beta}}h_{ab}\delta_{\check{\beta}}h_{cd} 
\mbox{ } .
\label{Vsecondbunch}
\ee
The na\"{\i}ve blockwise Riemannian structure of the configuration space of GR and 

\noindent nonderivative-coupled bosonic fields (\ref{Vmin}) can get badly
broken by derivative coupling (c.f Kucha\v{r} IV.5).  Either of the above bunches by 
itself exhibits all the unpleasant configuration space features mentioned in VI.1.4:
the first two terms of (\ref{Vfirstbunch}) are linear and hence the geometry is not 
Riemannian,  the third is a metric-matter cross-term, and the last two terms breach 
the DeWitt structure; likewise the first term of (\ref{Vsecondbunch}) is linear, the 
second is a cross-term and the third is a breach of the DeWitt structure.  If the 
DeWitt structure is breached in nature, then the study of pure canonical gravity and 
of the isolated configuration space of pure gravity are undermined.  Whereas there is 
no evidence for this occurrence, I have argued in I.2.3.4 that some forms of derivative 
coupling are only manifest in experimentally-unexplored high-curvature regimes.  

In the SSF, if $A_{\perp}$ were dynamical, then it would not be a Lagrange multiplier, 
and so it would not have a corresponding multiplier equation with which the tilt could be
`accidentally' removed, in which case there would not exist a corresponding BSW form 
containing $A_{\perp}$.  This argument however is not watertight, because it does not 
prevent some other BSW form from existing since variables other than $A_{\perp}$ could be 
used in attempts to write down actions that obey the 3-space principles.  As an example of 
such an attempt, I could use the $\alpha$-dependent change of variables to $A_0$ to put 
Proca theory into BSW form.  In this case I believed when I wrote \cite{Vanderson} that 
the attempt fails as far as the TSA is concerned, because $A_0$ features as a non-BM 
velocity in contradiction with the adopted implementation of 
principle \bf general R1\normalfont.\fn{But also 
see VII.1 for an amendment, and for further counterexamples of uniqueness once one allows 
form 2) on 
p 146.}  This shows however that criteria for whether a matter theory can be 
coupled to GR in the TSA are unfortunately rather dependent on the formalism used for the 
matter field.  The TSA would then amount to attaching particular significance to formalisms 
meeting its description.  This is similar in spirit to how those formalisms which close 
precisely as the Dirac Algebra are favoured in the hypersurface framework and the HKT and 
Teitelboim \cite{Teitelboim} articles.  In both cases one is required to find at least one 
compatible formalism for all the known fundamental matter fields.  

\mbox{ }

\noindent{\Large{\bf 3 Connection between space and spacetime viewpoints}}

\mbox{ }

\noindent Presupposing 3-space (taken to mean 3-geometry, not the 3-metric), 
the universal kinematics is Diff-BM.  The split (\ref{Vhamdec}, \ref{Vtransdec}) of 
$_{\mbox{\scriptsize A\normalsize}}{\cal H}^{\mbox{\scriptsize o\normalsize}}$
or perhaps more simply the equations 
(\ref{Vderivproj1}, \ref{Vderivproj2}, \ref{Vderivproj3}, \ref{Vderivproj4})
(and their analogues for higher-rank tensors (see e.g Sec 9 of \cite{KucharIII}), sum up the 
position of Diff-BM within the SSF.  In this latter case, there are three sorts of 
universal kinematics for tensor fields:  BM, tilt and derivative coupling.  All three 
of these are required in general in order to guarantee consistency (if the spacetime 
theory works) and Kucha\v{r}'s papers are a recipe for the computation of all the 
terms required for this consistency.  Thus in GR where it is available, the SSF is 
powerful and advantageous as a means of writing down consistent matter theories.  If 
conformal gravity is regarded as a competing theory to GR, it makes sense therefore to 
question what the 4-geometry of conformal gravity is, and whether its use could lead to a 
more illuminating understanding of matter coupling than offered by the TSA.  I am  thus 
free to ask how special GR is in admitting a constructive kinematic scheme for coupled 
consistent tensorial matter theories.

As BF\'{O} formulate it, the TSA denies the primary existence of the lapse.  But I have 
demonstrated that whether or not the lapse is eliminated does not affect the mathematics, 
so I prefer to think of the TSA as denying `lapse kinematics'.  BF\'{O}'s use of BSW forms 
does lead to a more restrictive scheme than the split spacetime framework, but I have 
demonstrated in VI.2 that this restriction can be understood in terms of when the split 
spacetime framework has no tilt.  Furthermore, I have unearthed the tacit \bf TSA gravity-matter simplicity 0 \normalfont 
and have rephrased this and the generalized BSW form of 
\bf RI[R2] \normalfont as nonderivative coupling and the no tilt condition  
respectively within the SSF.   

Working in the SSF with lapse-uneliminated actions with only shift (Diff-BM) kinematics  
has the additional advantage that I can immediately turn on and hence investigate the 
mathematical and physical implications of the tilt and derivative-coupling kinematics.  
Nevertheless, {\sl it is striking that BM kinematics suffice to describe all of the known 
fundamental bosonic fields coupled to GR.}  The absence of other kinematics includes the 
absence of the derivative-coupled theories whose presence in nature would undermine the study 
of pure canonical gravity of DeWitt and others.  I see my above work as support for this study.  
The less structure is assumed in theoretical physics, the more room is left for predictability.  
Could it really be that nature has less kinematics than the GR SSF might have us believe?  

Following from the debate of BM versus bare `discover and encode' methods in II.2.2 and III.1.2, 
I have by now demonstrated the gauge-theoretic gauge constraints to be integrabilities of ${\cal H}$.  
This permits the following alternative to starting with the BM implementation of {\bf general R1}: 

\noindent \bf Integrability[general R1] \normalfont :  start with a 3-d action with
bare velocities.  ${\cal H}$ can be deduced immediately from the action, and demanding 
$\dot{{\cal H}} \approx 0$ leads to a number of other constraints.  These are all
then to be encoded by use of auxiliary variables.

\noindent  This has the immediate advantage of treating Diff-BM on the same footing as the encoding of
Gauss constraints (G-BM).  In VII.1 I also consider achieving the same footing but rather 
by moving in the {\sl other} direction.  

So far, at least the bosonic sector of nature appears to be much simpler than the GR SSF might suggest, 
and the GR branch of the TSA may be formulated in two hitherto equivalent ways (making use of 
either the \bf BM \normalfont or the \bf Integrability \normalfont implementation of {\bf general R1}).  
I next consider both of these for spin-$\frac{1}{2}$ fermions.  

\mbox{ }

\noindent\Large{\bf 4 Spin-$\frac{1}{2}$ fermions and the TSA}\normalsize

\mbox{ }

\noindent Whereas it is true that the spinorial laws of physics may be rewritten in terms of tensors 
\cite{PenRind}, the resulting equations are complicated and it is not clear if and how 
they may be obtained from action principles.  Thus one is almost certainly compelled to 
investigate coupled spinorial and gravitational fields by attaching local flat frames to our 
manifolds.

There are two features I require for the study of the spin-$\frac{1}{2}$ laws of nature 
coupled to gravity.  First, I want the study to be clear in terms of shift and lapse 
kinematics, given my success in this chapter with this approach.  However, one should expect 
the spinors to have further sorts of kinematics not present for tensor fields.  Second, I  
want to explicitly build SO(3, 1) (spacetime) spinors out of SO(3) (spatial) ones.\fn{This is 
standard use of representation theory, based on the accidental Lie algebra relation 

\noindent SO(4)$\cong$ SO(3) $\bigoplus$ SO(3), which depends on space being 3-d.  This relation is a 
common source of tricks in particle physics and quantum gravity.  By SO(3, 1) and SO(3) spinors, I 
strictly mean spinors corresponding to their universal covering groups, SL(2, C) and SU(2) 
respectively.  I am not yet concerned in this thesis with the differences betwen SO(4) and SO(3, 1) 
from a quantization perspective, which render Euclidean quantum programs easier in some 
respects \cite{Thiemann, Perez}.}    

I begin here by looking at the first feature in the following 4-component spinor formalism.  
Introduce Dirac's suited triads, denoted by 
$\mbox{\sc e}_{\bar{A}}^{B}$, which obey 
$\mbox{\sc e}_{\bar{0}a} = 0$, 
$\mbox{\sc e}_{\bar{0}0} = - \alpha$, 
$\mbox{\sc e}_{\bar{A}}^{B} \mbox{\sc e}_{\bar{C}B} = \eta_{\bar{A}\bar{C}}$ and 
$\mbox{\sc e}_{\bar{A}B} \mbox{\sc e}^{\bar{A}}_{C} = g_{BC}$.
The spacetime {\it spinorial covariant derivative} is 
\be
\mbox{\hspace{0.2in}}
\nabla^{\mbox{\scriptsize s\normalsize}}_{\bar{A}}\psi 
= \psi_{,\bar{A}} - \frac{1}{4}\Omega_{\bar{R}\bar{S}\bar{A}}\gamma^{\bar{R}}\gamma^{\bar{S}}\psi 
\mbox{ } , 
\ee
\be
\mbox{where }
\mbox{\hspace{1.8in}}
\Omega_{\bar{R}\bar{S}\bar{P}} = (\nabla_{B}\mbox{\sc e}_{\bar{R}a})
                                            \mbox{\sc e}^{A}_{\bar{S}} \mbox{\sc e}^{B}_{\bar{P}} 
\label{4scdef}
\mbox{\hspace{2.2in}}
\ee
is the spacetime {\it spin connection }   
The spatial spinorial covariant derivative is
\be
\mbox{\hspace{2.2in}}
D^{\mbox{\scriptsize s\normalsize}}_{\bar{p}}\psi = \psi_{,\bar{p}} - \frac{1}{4}\omega_{\bar{r}\bar{s}\bar{p}}\gamma^{\bar{r}}\gamma^{\bar{s}}\psi 
\mbox{ } , 
\mbox{\hspace{2in}}
\ee
\be
\mbox{where } 
\mbox{\hspace{2in}}
\omega_{\bar{r}\bar{s}\bar{p}} = (D_{b}\mbox{\sc e}_{\bar{r}a})\mbox{\sc e}^{a}_{\bar{s}} \mbox{\sc e}^{b}_{\bar{p}}
\label{3scdef}
\mbox{\hspace{2.4in}}
\ee
is the spatial spin connection.  

Next split the spacetime spin connection, following Henneaux's treatment \cite{Hen} which has 
the virtue of keeping track of the geometrical significance of each of the pieces.  I supply 
each piece with contracting gamma matrices as suits its later application.  As
\be
\omega_{(\bar{P}\bar{Q})\bar{R}} = 0
\mbox{ } , 
\label{spantisym}
\ee
there are 4 components in its decomposition. $\omega_{\bar{p}\bar{q}\bar{r}}$ may be used as 
it is, to form the 3-d spinorial covariant derivative. 
By (\ref{3scdef}), suited tetrad properties, the Dirac algebra (\ref{Diracalgebra}) and 
$K_{ab} = K_{ba}$, $K_{a\perp} = 0$ (see I.2.1.3), 
\be  
\gamma^{\bar{r}}\gamma^{\bar{q}}\omega_{\bar{0}\bar{q}\bar{r}} = - K 
\mbox{ } . 
\label{womble}
\ee
\be
\mbox{By (\ref{3scdef}), (\ref{metcon}) in (\ref{Chris}) and use of suited tetrad properties, }
\mbox{\hspace{0.5in}}
\omega_{\bar{p}\bar{0}\bar{0}} = - \frac{\pa_{\bar{p}}\alpha}{\alpha} 
\mbox{ } . 
\mbox{\hspace{2in}}
\ee
\be
\mbox{By (\ref{3scdef}), }
\mbox{\hspace{1.7in}}
\omega_{\bar{r}\bar{s}\bar{0}} = \pa_{\bar{0}}\mbox{\sc e}_{[\bar{r}l}{\mbox{\sc e}^l}_{\bar{s}]} 
- {\alpha}(\pounds_{\beta}\mbox{\sc e}_{\bar{s}a}){\mbox{\sc e}^a}_{\bar{r}} 
\mbox{ } . 
\mbox{\hspace{1.7in}}
\label{ursa}
\ee
$$
\mbox{ } \mbox{ Then, using (\ref{4scdef}), (\ref{spantisym}) and (\ref{3scdef}), }
\mbox{\hspace{0.5in}}
\gamma^{\bar{A}}\nabla^{\mbox{\scriptsize s\normalsize}}_{\bar{A}}\psi 
= \gamma^{\bar{0}}\nabla^{\mbox{\scriptsize s\normalsize}}_{\bar{0}}\psi 
+ \gamma^{\bar{l}}\nabla^{\mbox{\scriptsize s\normalsize}}_{\bar{l}}\psi 
\mbox{\hspace{1in}}
$$                     
$$
= \gamma^{\bar{0}}\nabla_{\bar{0}}\psi + 
\left(
\gamma^{\bar{l}}\pa_{\bar{l}}\psi 
- \frac{1}{4}\gamma^{\bar{l}}\omega_{\bar{m}\bar{n}\bar{l}}\gamma^{\bar{m}}\gamma^{\bar{n}}\psi 
- \frac{1}{4}\gamma^{\bar{l}}\omega_{\bar{m}\bar{0}\bar{l}}\gamma^{\bar{m}}\gamma^{\bar{0}}\psi 
- \frac{1}{4}\gamma^{\bar{l}}\omega_{\bar{0}\bar{n}\bar{l}}\gamma^{\bar{0}}\gamma^{\bar{n}}\psi
- \frac{1}{4}\gamma^{\bar{l}}\omega_{\bar{0}\bar{0}\bar{l}}\gamma^{\bar{0}}\gamma^{\bar{0}}\psi
\right)
$$
\be
= \gamma^{\bar{0}}\nabla^{\mbox{\scriptsize s\normalsize}}_{\bar{0}}\psi 
+ \gamma^{\bar{l}}D^{\mbox{\scriptsize s\normalsize}}_{\bar{l}}\psi 
+ \frac{1}{2}\gamma^{\bar{0}}\gamma^{\bar{l}}\gamma^{\bar{m}}\omega_{\bar{0}\bar{m}\bar{l}}\psi 
\mbox{ } .
\label{threeterms} 
\ee
\be
\mbox{Then the first term may be replaced by } 
\mbox{\hspace{0.5in}}
\alpha\nabla_{\bar{0}}\psi 
= \dot{\psi} 
- \pounds^{\mbox{\scriptsize s\normalsize}}_{\beta}\psi 
- \pa_{\mbox{\scriptsize R\normalsize}}\psi 
+ \frac{1}{2}\pa_{\bar{r}}\alpha\gamma^{\bar{0}}\gamma^{\bar{r}} 
\mbox{\hspace{1in}}
\ee
by splitting (\ref{ursa}) into two pieces; the first of these is directly geometrically 
meaningful, whereas the second is geometrically meaningful when combined with 
$\pounds_{\beta}$:
\be
\mbox{\hspace{0.2in}}
\pa_{\mbox{\scriptsize R\normalsize}}\psi 
\equiv \frac{1}{4}\mbox{\sc e}^i_{[\bar{r}}\dot{\mbox{\sc e}}_{\bar{s}]i}\gamma^{\bar{r}}\gamma^{\bar{s}}\psi    
\mbox{\hspace{1in}}
\mbox{ (triad rotation correction) ,}
\label{VRottcod}
\ee 
\be
\mbox{and  } 
\mbox{\hspace{0.6in}}
\pounds^{\mbox{\scriptsize s\normalsize}}_{\beta}\psi \equiv \beta^i\pa_i\psi 
- \frac{1}{4}\mbox{\sc e}_{[\bar{r}|}^i\pounds_{\beta}
             \mbox{\sc e}_{|\bar{s}]i}\gamma^{\bar{s}}\gamma^{\bar{r}}\psi 
\mbox{\hspace{1.05in}}
\mbox{ (Lie derivative) .}
\mbox{\hspace{1.2in}}
\label{Vsplied}
\ee 
Thus the tensorial Lie derivative $\pounds_{\beta}\psi = \beta^i\pa_i\psi$ is but a piece of 
the spinorial Lie derivative (\ref{Vsplied}) \cite{VGH, VKosmann}.  The second term in (\ref{threeterms}) is already in 
clear-cut spatial form, while the last term is just $-\frac{\gamma^{\bar{0}}K}{2}$, by (\ref{womble}).  Thus
\be
\sqrt{|g|}\bar{\psi}\gamma^{\bar{A}}\nabla^{\mbox{\scriptsize s\normalsize}}_{\bar{A}}\psi 
= i\sqrt{h}{\psi}^{\dagger}
\left[
\alpha\gamma^{\bar{0}}\gamma^{\bar{l}}D^{\mbox{\scriptsize s\normalsize}}_{\bar{l}}\psi 
+ \frac{\alpha K}{2}\psi + \alpha_{,\bar{l}}\gamma^{\bar{0}}\gamma^{\bar{l}}\psi  
- (\dot{\psi} - \pounds^{\mbox{\scriptsize s\normalsize}}_{\beta}\psi 
- \pa_{\mbox{\scriptsize R\normalsize}}\psi)
\right].  
\label{Vspinsplit}
\ee

Next, although derivative coupling (second term) and tilt (third term) appear to be present 
in (\ref{Vspinsplit}), G\'{e}h\'{e}niau and Henneaux \cite{VGH} observed that these simply 
cancel out in the Dirac field contribution to the Lagrangian density, 
$$
\sqrt{|g|}\mbox{\sffamily L\normalfont}_{\mbox{\scriptsize D\normalsize}} = 
\sqrt{|g|}
\left[
\frac{1}{2}(\bar{\psi}\gamma^{\bar{A}}\nabla^{\mbox{\scriptsize s\normalsize}}_{\bar{A}}\psi
- \nabla^{\mbox{\scriptsize s\normalsize}}_{\bar{A}}\bar{\psi}\gamma^{\bar{A}}\psi) 
- m_{\psi}\bar{\psi}\psi
\right]
$$
$$
= \frac{i\sqrt{h}}{2}
\left(
\psi^{\dagger}
\left[
\alpha\gamma^{\bar{0}}\gamma^{\bar{l}}D^{\mbox{\scriptsize s\normalsize}}_{\bar{l}}\psi 
+ \frac{\alpha K}{2}\psi + \alpha_{,\bar{l}}\gamma^{\bar{0}}\gamma^{\bar{l}}\psi  
- (\dot{\psi} - \pounds^{\mbox{\scriptsize s\normalsize}}_{\beta}\psi 
- \pa_{\mbox{\scriptsize R\normalsize}}\psi)
\right]
- 
\right.
$$
$$
\left.
\left[
\alpha\gamma^{\bar{0}}\gamma^{\bar{l}}D^{\mbox{\scriptsize s\normalsize}}_{\bar{l}}\psi^{\dagger} 
+ \frac{\alpha K}{2}\psi^{\dagger} 
+ \alpha_{,\bar{l}}\gamma^{\bar{0}}\gamma^{\bar{l}}\psi^{\dagger}  
- (\dot{\psi}^{\dagger} - \pounds^{\mbox{\scriptsize s\normalsize}}_{\beta}\psi^{\dagger} 
- \pa_{\mbox{\scriptsize R\normalsize}}\psi^{\dagger})
\right] 
\right)
- \sqrt{h}\alpha m_{\psi}\bar{\psi}\psi
$$
$$
= 
\frac{i\sqrt{h}}{2}
\left[
\psi^{\dagger}\alpha\gamma^{\bar{0}}\gamma^{\bar{l}}D^{\mbox{\scriptsize s\normalsize}}_{\bar{l}}\psi  
- \psi^{\dagger}(\dot{\psi} - \pounds^{\mbox{\scriptsize s\normalsize}}_{\beta}\psi 
- \pa_{\mbox{\scriptsize R\normalsize}}\psi)
- \alpha\gamma^{\bar{0}}\gamma^{\bar{l}}(D^{\mbox{\scriptsize s\normalsize}}_{\bar{l}}\psi^{\dagger})\psi  
- (\dot{\psi}^{\dagger} - \pounds^{\mbox{\scriptsize s\normalsize}}_{\beta}\psi^{\dagger} 
- \pa_{\mbox{\scriptsize R\normalsize}}\psi^{\dagger})\psi 
\right]
$$
\be
- \sqrt{h}\alpha m_{\psi}\bar{\psi}\psi 
\mbox{ } 
\label{Vfermilag}
\ee
[c.f the second form in (\ref{Diraclag2})].

Whilst Nelson and Teitelboim \cite{Vfermi} do not regard such a choice of absence of 
derivative-coupling as a deep simplification (they adhere to the HKT school of thought and the 
simplification is not in line with the hypersurface deformation algebra), the  
G\'{e}h\'{e}niau--Henneaux formulation is clearly encouraging for the TSA.  For, using BSW 
procedure on the 3+1 split Einstein--Dirac Lagrangian,
one immediately obtains a generalized \bf RI-castable\normalfont-type action:  
$$
\mbox{\sffamily I\normalfont}^{\psi, \bar{\psi}}_{\mbox{\scriptsize ED\normalsize}} = \int \textrm{d}\lambda
\int \textrm{d}^3x \sqrt{h}
\left[
\sqrt{      \Lambda + \sigma R + 
\psi^{\dagger}\gamma^{\bar{0}}\gamma^{\bar{l}}D^{\mbox{\scriptsize s\normalsize}}_{\bar{l}}\psi  
- \gamma^{\bar{0}}\gamma^{\bar{l}}(D^{\mbox{\scriptsize s\normalsize}}_{\bar{l}}\psi^{\dagger})\psi  
- m_{\psi}\bar{\psi}\psi       }
\sqrt{\mbox{\sffamily T\normalfont}^{\mbox{\scriptsize g\normalsize}}} 
\right.
$$
\be
\left.
- \psi^{\dagger}(\dot{\psi} - \pounds^{\mbox{\scriptsize s\normalsize}}_{\xi}\psi - \pa_{\mbox{\scriptsize R\normalsize}}\psi)
+ (\dot{\psi}^{\dagger} - \pounds^{\mbox{\scriptsize s\normalsize}}_{\xi}\psi^{\dagger} 
- \pa_{\mbox{\scriptsize R\normalsize}}\psi^{\dagger})\psi 
\right] 
\mbox{ } .
\label{specificfermi}
\ee  

Thus one knows that 
$$
\mbox{\sffamily I\normalfont}^{\psi, \bar{\psi}}_{\mbox{\scriptsize TSA(ED)\normalsize}} 
= \int \textrm{d}\lambda\int \textrm{d}^3x \sqrt{h}
\left[
\sqrt{      \Lambda + \sigma R + 
\psi^{\dagger}\gamma^{\bar{0}}\gamma^{\bar{l}}D^{\mbox{\scriptsize s\normalsize}}_{\bar{l}}\psi  
- \gamma^{\bar{0}}\gamma^{\bar{l}}(D^{\mbox{\scriptsize s\normalsize}}_{\bar{l}}\psi^{\dagger})\psi  
- m_{\psi}\bar{\psi}\psi       }
\sqrt{\mbox{\sffamily T\normalfont}^{\mbox{\scriptsize g\normalsize}}(\mbox{\ss}_{\xi}h_{ij})} 
\right.
$$
\be
\left.
- \psi^{\dagger}\mbox{\ss}_{\xi, \mbox{\scriptsize R\normalsize}}{\psi} + \mbox{\ss}_{\xi, \mbox{\scriptsize R\normalsize}}{\psi}^{\dagger}\psi 
\right] 
\mbox{ }  
\label{specificfermi3}
\ee  
[a particular case of the form (\ref{Vactualfermi})], will work as a spatial ontology 
starting-point for Einstein--Dirac theory.   Note that in addition to the more complicated 
form of the Diff-BM correction to the Dirac velocities, there is also a triad rotation 
correction (\ref{VRottcod}).  Thus 
either of the following arguments should be adopted.  The {\bf BM[general R1]} implementation 
should be generalized to accommodate this additional, natural geometric correction: given two 
SO(3) spinor bundles $\Sigma_1$, $\Sigma_2$, the (full spinorial) drag shufflings of 
$\Sigma_2$ (keeping $\Sigma_1$ fixed) are accompanied by the rotation shufflings of the triads 
glued to it.   The triad rotation correction then leads to a further `locally Lorentz' constraint 
${\cal J}_{\bar{\mu}\bar{\nu}}$ \cite{VDiracrec}.  Alternatively, it can be checked that this 
constraint will be picked up by the {\bf Integrability[general R1]} implementation, since it is an 
integrability condition.  This check may be implied to be verified by Nelson and Teitelboim's 
work \cite{Vfermi}: both ${\cal H}_i$ and ${\cal J}_{\bar{\mu}\bar{\nu}}$ are indeed integrability 
conditions for ${\cal H}$.  For in terms of Dirac brackets (\ref{Dibra}) $\{\mbox{ }, \}^*$, starting from 
${\cal H}$, $\{{\cal H}, {\cal H}\}^*$ gives ${\cal H}_i$ and then we can form 
$\{{\cal H}, {\cal H}_i\}^*$ which gives ${\cal J}_{\bar{A}\bar{B}}$ 
(and ${\cal H}$) so one has recovered all the constraints as integrability conditions for 
${\cal H}$.  Also, one does not recover ${\cal H}$ if one starts with ${\cal H}_i$ or 
${\cal J}_{\bar{A}\bar{B}}$, so in some sense ${\cal H}$ is privileged.  However, this does 
highlight my other caveat for the integrability idea: one might choose to represent the constraint 
algebra differently by mixing up the usual generators.  For example, a linearly-related set of 
constraints is considered in \cite{Vfermi}, for which the integrability of any of the constraints 
forces the presence of all the others.   My defense against this is to invoke again that we only 
require one formulation of the TSA to work, so we would begin with the quadratic constraint 
${\cal H}$ nicely isolated.  

Also note I succeed in including the 1-form--fermion interaction terms of the Einstein--Standard 
Model theory: 
\be
\mbox{\sffamily g\normalfont}^{\cal A}\tau_{{\cal A}\mbox{\scriptsize\bf I\normalfont\normalsize}}
\bar{\psi}\gamma^{\bar{\beta}}
\mbox{\sc e}^{A}_{\bar{\beta}}A^{\mbox{\scriptsize\bf I\normalfont\normalsize}}_{A}\psi
\label{Vstar}
\ee
where ${\cal A}$ takes the values U(1), SU(2) and SU(3).  The decomposition of these into 
spatial quantities is trivial.  No additional complications arise from the inclusion of 
such terms, since  1) they contain no velocities so the definitions of the momenta are 
unaffected (this includes there being no scope for derivative coupling) 2) they are part of 
gauge-invariant combinations, unlike the Proca term which breaks gauge invariance and 
significantly alters the Maxwell canonical theory.  In particular, the new terms clearly 
contribute linearly in $A_{\perp}$ to the Lagrangian potential, so by the argument at the end 
of VI.2.2, an `accident' occurs ensuring that tilt kinematics is not necessary.  Also, clearly 
the use of the form (\ref{Vfermilag}) is compatible with the inclusion of the interactions 
(\ref{Vstar}) since, acting on $\bar{\psi}$ the gauge correction is the opposite sign.  So 
our proposed formulation's combined Standard Model matter Lagrangian is\fn{Here 
$\mbox{\sffamily L\normalfont}_{\mbox{\tiny YM\normalsize}}$
is given by the $m = 0$, SU(3) $\times$ SU(2) $\times$ U(1) version of (\ref{VlagYMmass}) and 
one would need to sum the square bracket over all the known fundamental fermionic species.}    
\be
\mbox{\sffamily L\normalfont}_{\mbox{\scriptsize SM\normalsize}} = 
\frac{1}{2}
\left[
(\bar{\psi}\gamma^{\bar{A}}
(\nabla^{\mbox{\scriptsize s\normalsize}}_{\bar{A}} 
- \mbox{\sffamily g\normalfont}_{\cal A}\tau^{\cal A}_{\mbox{\scriptsize\bf I\normalfont\normalsize}}
\mbox{\sc e}^{B}_{\bar{A}} A^{\mbox{\scriptsize\bf I\normalfont\normalsize}}_{B})\psi
- (\nabla^{\mbox{\scriptsize s\normalsize}}_{\bar{A}} 
+ \mbox{\sffamily g\normalfont}_{\cal A}\tau^{\cal A}_{\mbox{\scriptsize\bf I\normalfont\normalsize}}
\mbox{\sc e}^{B}_{\bar{\lambda}}A^{\mbox{\scriptsize\bf I\normalfont\normalsize}}_{A})\bar{\psi}\gamma^{\bar{B}}\psi
\right] 
- m_{\psi}\bar{\psi}\psi 
+ \mbox{\sffamily L\normalfont}_{\mbox{\scriptsize YM\normalsize}} 
\mbox{ } .  
\ee
which can be adjoined to 
$\mbox{\sffamily L\normalfont}_{\mbox{\sffamily\scriptsize 3+1(GR)\normalsize\normalfont}}$, 
and successfully subjected to the BSW procedure, thus announcing that an enlarged version of 
(\ref{specificfermi3}) will serve as a spatial ontology starting-point.  
  
There is also no trouble with the incorporation of the Yukawa interaction term 
$\varsigma\bar{\psi}\psi$ which could be required for some fermions to pick up mass from a 
Higgs scalar.  Thus the Lagrangian for all the known fundamental matter fields can be built 
by assuming BM kinematics and that the DeWitt structure is at the outset respected.  The 
thin sandwich conjecture can be posed for all these fields coupled to GR.  The classical 
physics of all these fields is timeless in Barbour's sense.  

It remains to cast the action (\ref{specificfermi3}) in entirely spatial terms, as it still has 
remnants of spacetime in its 
appearance: it is in terms of 4-component spinors and Dirac matrices.  However, recall that the 
Dirac matrices are built out of the Pauli matrices associated with SO(3), and choosing to work 
in the chiral representation (\ref{chiralgamma}), the 4-component spinors may be treated as D 
and L SO(3) 2-component spinors (\ref{chiralspinor}).  Thus a natural formulation of 
Einstein--Dirac theory in terms of 3-d  objects exists.  To accommodate neutrino (Weyl) fields, 
one would consider a single $\psi_{\mbox{\scriptsize L\normalsize}}$ SO(3) spinor, i.e set its pair 
$\psi_{\mbox{\scriptsize D\normalfont}}$ and its mass to zero before the variation is carried 
out.  Whilst we are free to accommodate all the known fundamental fermionic fields in the TSA, 
one cannot predict the number of Dirac and Weyl fields present in nature nor their masses nor 
the nongravitational forces felt by each field.  

\mbox{ }

\noindent{\bf Suggested Further Work}

\mbox{ }

\noindent
Here are some suggestions for further work given the above progress.  Consider lapse eliminated 
or lapse-eliminable actions including 
$\mbox{\sffamily U\normalfont}^{\mbox{\scriptsize F\normalsize}}$ and 
$\mbox{\sffamily T\normalfont}^{\mbox{\scriptsize F\normalsize}}$ built from spatial first 
principles using SO(3) spinors.  Obtain ${\cal H}$ and treat its propagation exhaustively to 
obtain constraint algebras.  Does the universal null cone result still hold ?  Is 
Einstein--Dirac theory in any sense picked out?  How does the thin sandwich conjecture for 
Einstein--Dirac theory behave?   On coupling a 1-form field, do these results hold for 
Einstein--Maxwell--Dirac theory?  On coupling {\bf K} 1-form fields, do they hold for 
Einstein--Yang--Mills--Dirac theories such as the Einstein--Standard Model?  There is also the 
issue of whether conformal gravity can accommodate spin-$\frac{1}{2}$ fermions.    

It is worth considering whether any of my ideas for generalizing the TSA extend to canonical 
supergravity \cite{Vsugy1, Vsugy2, Vsugy3, Vsugy4, Vsugy5, Jacsugy}, or to the minimal 
supersymmetric standard model  This could be seen as a robustness test for our ideas and 
possibly lead to a new formulation of supergravity, or as a precaution of looking if our way 
of seeing the world is compatible with physics that could conceivably soon become 
observationally established (see I.3.3).  The supergravity constraint algebra is not known 
well enough\fn{The most recent confirmation that we are aware of that this work remains 
incomplete is on p. 96 of \cite{Vsugy5}.} to comment whether the new supersymmetric constraint 
${\cal S}_{\bar{A}}$ arises as an integrability condition for ${\cal H}$.  Note however that 
Teitelboim was able to treat ${\cal S}_{\bar{\mu}}$ as arising from the square root of ${\cal H}$
\cite{Vsqrtteitel}; however this means that the bracket of ${\cal S}_{\bar{A}}$ and its conjugate 
gives ${\cal H}$, so it is questionable whether the supergravity ${\cal H}$ retains all of the 
primary importance of the GR ${\cal H}$.

Finally, given the competition from \cite{BOF, Vanderson, LAnderson} and this thesis it would be 
interesting to see whether the HKT or Kouletsis formulations can be made to accommodate 
spin-$\frac{1}{2}$ fermions, and also to refine Teitelboim's GR-matter postulates to the level of these 
formulations' pure GR postulates.  



\noindent\Huge{\bf VII More on TSA matter schemes}\normalsize

\mbox{ }

\noindent
In this chapter, I use the most general {spacetime} (rather than TSA) single 
1-form ansatz.  I then use the arguments of VI.2 to pass to TSA formulations 
where possible.  I carefully show how the TSA {\sl does} permit a formulation 
of massive 1-form fields.  This completes the dispelling of the possibility that the 
TSA has something to say about the origin of mass.  This and further more 
complicated examples show that electromagnetism is not particularly being picked 
out by the TSA's relational principles, thus also dispelling any ``hints at partial 
unification''.  I also use the single 1-form example to smooth out the treatment 
of lapse-eliminability in  VI.  I finally consider in the light of my examples 
what the TSA has to say about the Principle of Equivalence ({\bf POE}).   I make much 
use below of the following mathematics.

\mbox{ }

\noindent{\bf More on multipliers and cyclic coordinates}

\mbox{ }

\noindent 
I am dealing with fields, so spatial derivatives may enter the multiplier 
and cyclic equations (\ref{lmel}, \ref{cyclicel}) rendering them p.d.e's.  
Sometimes this can be avoided through integration by parts.  

In some cases, elimination from a multiplier equation is not possible.  
This occurs for {\sffamily L} linear in the multiplier (so the multiplier 
equation does not contain the multiplier itself),  for {\sffamily L} 
homogeneous in the multiplier (so the multiplier equation contains the 
multiplier as a pure factor), or if the multiplier equation, be it p.d.e or 
algebraic, is insoluble or not explicitly soluble for the multiplier.  
The insoluble case would mean the system is inconsistent.  It may also not 
be possible to eliminate using the cyclic 
equation (\ref{cyclicel}).  This is more delicate since it is accompanied by passage to the 
Routhian.  It is not possible to do this if {\sffamily L} is linear in the 
cyclic velocity so that (\ref{cyclicel}) is independent of it, or insoluble 
or not explicitly soluble.   Finally, note that some of the above features 
depend on the choice of variables.  

\mbox{ }

\noindent\Large{\bf 1 Update of TSA with a single 1-form} \normalsize 

\mbox{ }

\noindent{\bf 1.1 A means of including Proca theory}

\mbox{ }

\noindent  
Note that whether a theory can be cast into TSA form can only be treated formalism by formalism.    
I explain here how to obtain a formalism in which Proca theory is allowed.  
The tricks used do not suffice to put all other theories I considered into 
TSA form.  Thus the TSA retains some selectivity, although I demonstrate 
here via including Proca and various other theories that the TSA is less 
selective than previously assumed.  I tie this selectivity to the {\bf POE} in 
VII.2.  

Of what use is the inclusion of Proca theory?  It is then easy to see how to include 
massive Yang--Mills theory as a phenomenological theory of what weak bosons 
look like today, and Proca theory itself appears phenomenologically e.g in 
superconductivity.  Whereas these applications are quite peripheral, it is 
nevertheless reassuring that one need not abandon the TSA to do phenomenology.  

Here is another way of looking at electromagnetism.  The `accident' method of 
VI.1.2 `lets go' of the constraint; fortunately it is `caught again' because 
it arises as an integrability, but one would not generally expect this to be 
the case.  One could rather avoid the tilt by {\sl redefining variables} 
according to $A_{\perp} \longrightarrow A_0 = - \alpha A_{\perp}$.  Then one never 
`lets go' of the constraint.  

Do Proca theory just like the above.  Whereas this was previously objected to on 
p 158, notice that my observation on 
p 115-6 also holds ($A_0$ can be taken to be a 
BM velocity if one so wishes as this only changes the equations weakly), which 
removes the objection.  Thus the following is a TSA formulation of Einstein—-Proca 
theory  
\be
\mbox{\sffamily I\normalfont}^{\mbox{\scriptsize A\normalsize}, \Xi}_{\mbox{\scriptsize TSA(EP)\normalsize}} 
= \int\textrm{d}\lambda \int\textrm{d}^3x\sqrt{h}
\sqrt{       D_{[a}A_{b]}D^{[a}A^{b]}\mbox{+}m^2 A_aA^a\mbox{+}R    }
\sqrt{       (\dot{A}_a\mbox{--}\pounds_{\xi}\mbox{--}\pa_a(\dot{\Xi}\mbox{--}\pounds_{\xi}\Xi)^2m^2 A_0^2
\mbox{+}\mbox{\sffamily T\normalfont}_{\mbox{\scriptsize g\normalfont}}   }    
\mbox{ } .
\ee

Also note that the Proca constraint is second-class. It then makes no sense by 
definition to work immediately with constraint propagation.  Rather, the way to 
proceed is to obtain the Proca constraint, use it on the other constraints to 
eliminate $A_0$ and then one has constraints which close.   

Why did we not hit on Proca theory before? Before we were doing `discover and 
encode' as regards non-gravitational auxiliaries.  This does not work here, 
ostensibly because the Proca constraint is second-class, while `discover and 
encode' is a procedure whereby gauge theory (associated with first class 
constraints \cite{HTbook}) is emergent.  I require rather the more broad-minded approach in 
which all the auxiliaries are treated on the same footing by being present in 
the action from the outset.  This works here by starting with actions for 
3-geometries together with one scalar and one 1-form matter fields.  To have 
Proca theory, that scalar then turns out to be the above auxiliary.

Similarly, gauge theory breaking down near singularities (IV.1.4.3) is actually 
a permissible choice rather than an inescapable necessity.  For, if one chooses 
to adopt the action resulting from the `discover and encode' TSA formulation, 
then gauge theory breaks down, while if one chooses to start from the outset 
with scalars alongside gauge variables then it does not break down.  

\mbox{ } 

\noindent{\bf 1.2 Further nonuniqueness examples}

\mbox{ }

\noindent
First, choose to assume that there is no fundamental underlying theory so that  
local flat spacetime na\"{\i}ve renormalizability makes sense.  This puts a stringent 
bound on how many terms can be in the 1-form ansatz.  
\be
\mbox{ } \mbox{ The ansatz is}
\mbox{\hspace{0.4in}} 
\mbox{\sffamily L\normalfont}^{\mbox{\scriptsize A\normalsize}} = C^{ABCD}\nabla_B A_A\nabla_DA_C 
 + \bar{C}^{ABCD}\nabla_B A_A A_DA_C + mA^2 + qA^4
\mbox{\hspace{0.2in}}
\ee
(the other combinations are total derivatives or zero by symmetry--antisymmetry).  
Note that I already split the first term in VI.2.4.  The second term has in effect only one piece  
$bA^2g^{AB}\nabla_AA_B$ (including use of integration by parts).  The last two terms are trivial 
to decompose.  

Using (\ref{Vderivproj1}--\ref{Vderivproj4}), the second term's tilt contribution vanishes by parts.  
Suppose the first term is present but not the second.  
Then the action is TSA-castable if $C_1 = - C_2$ and $C_3 = 0$ either by the $A_{\perp}$ multiplier 
equation `accident' or by the $A_0$.  If the second term is also present the first of these options 
is blocked because $A_{\perp}$ is then not a multiplier, but the $A_0$ option survives thanks to 
additional integration by parts.  The third and fourth terms are compatible with both options.  

Thus I have found two classes of single 1-form theories which I can cast 
into TSA form:  
\be
\mbox{\sffamily L\normalfont}^{\mbox{\scriptsize A\normalsize}}_1 = a\nabla_{[A}A_{B]}\nabla^{[A}A^{B]} +   mA^2 + qA^4 
\mbox{ } , 
\ee
\be
\mbox{\sffamily L\normalfont}^{\mbox{\scriptsize A\normalsize}}_2 = 
\mbox{\sffamily L\normalfont}^{\mbox{\scriptsize A\normalsize}}_1 + bA^2\nabla_AA^A 
\mbox{ } .  
\label{theo2}
\ee 

The first is Proca theory if $q = 0$.  If $q \neq 0$, its $A_0$ formulation has a 
Lagrangian density of type 
\be
\sqrt{g}\mbox{\sffamily L\normalfont}^{\mbox{\scriptsize A\normalsize}} = \sqrt{h}\alpha
\left(
A + \frac{B}{\alpha^2}  +  \frac{C}{\alpha^4}
\right) 
\mbox{ } ,
\ee
so the $\alpha$-multiplier equation is $A\alpha^4 - B\alpha^2 - 3C = 0$, so by the quadratic formula, 
\be 
\mbox{\sffamily L\normalfont}^{\mbox{\scriptsize A\normalsize}}_{\mbox{\scriptsize TSA(1)\normalsize}} = 
\left(
\frac{2A}{B \pm \sqrt{B^2 + 12AC}}
\right)^{\frac{3}{2}}
\left[
\frac{      B(B \pm \sqrt{B^2 + 12AC})}{A} + 4C
\right] 
\mbox{ } .
\label{N4}
\ee
Compared to $\mbox{\sffamily L\normalfont}^{\mbox{\scriptsize A\normalsize}} = \alpha
\left(
A + \frac{B}{\alpha}
\right)
$ which gives $\mbox{\sffamily L\normalfont} = 2\sqrt{AB}$, the above is far more 
complicated [an example of my case 2) in VI.1.4] but nevertheless a valid TSA 
presentation.  Thus  $A^4$-theory was excluded by BF\'{O} on simplicity grounds rather 
than for fundamental reasons.   

For $a = 0$, the second class coupled to GR is linear as regards $\alpha$ 
(because it is in terms 
\be
\mbox{of $A_{\perp}$) and thus gives a TSA theory of form }
\mbox{\hspace{1.75in}}
\mbox{\sffamily L\normalfont} = 2\sqrt{AB} + D 
\mbox{\hspace{1.75in}}
\ee
where $D$ is the linear kinetic term.  This is similar in layout to the TSA formulation 
with spin-$\frac{1}{2}$ fermions, except that here tilt elimination does not happen to 
also eliminate the derivative coupling terms.  

Next, choose to assume there is a fundamental underlying theory rendering na\"{\i}ve 
renormalizability irrelevant.  Are there then any other sorts of TSA theories with a 
single 1-form?  

Clearly adjoining any polynomial in $A^2$ to the spacetime Maxwell Lagrangian will do, 
and the TSA forms just get nastier.  A messy Cardano formula is required if some amount 
of $A^6$ term is furthermore present.  If terms as high as  $A^{10}$ are present, 
despite being algebraic the $\alpha$-multiplier equation is not generally explicitly exactly 
soluble for $\alpha$ by Galois' well-known result.  

I also considered Born—-Infeld theory.  The easiest non-Maxwell Born—-Infeld theory 

\noindent[{\sffamily L} = $(F \circ F)^2$] also leads to the TSA form (\ref{N4}), while  the Born—-Infeld 
theory particular to string theory also gives a $\alpha$-multiplier equation not generally explicitly
exactly soluble for $\alpha$.  

Thus the TSA in fact admits a broad range of single 1-form theories.  

\mbox{ }

\noindent\Large{\bf 2 TSA and the principle of equivalence}\normalsize

\mbox{ }

\noindent 
Curved spacetime matter field equations are locally Lorentzian if they contain no worse than 
Christoffel symbols [by applying (\ref{Christrans})].  The gravitational field equations 
are given a special separate status in the {\bf POE} (`all the laws of physics bar gravity').   
However, derivatives of Christoffel symbols, be they from double derivatives or 
straightforward curvature terms multiplied by matter factors, cannot be eliminated 
likewise and are thus {\bf POE}-violating terms.  

Let me translate this to the level of the Lagrangians I am working with.  If the 
Lagrangian may be cast as  functionally-independent of Christoffel symbols, its 
field equations clearly will not inherit any, so the {\bf POE} is satisfied.  If the 
Lagrangian is a {\sl function} of the Christoffel symbols, then by the use of 
integration by parts in each Christoffel symbol's variation, generally derivatives 
will appear acting on the cofactor Christoffel symbols, leading to {\bf POE}-violating 
field equations.  A clear exception is when the Lagrangian is a {\sl linear} 
function of the Christoffel symbols.  Lagrangians unavoidably already containing 
matter-coupled Christoffel symbol derivatives lead to {\bf POE}-violating field equations.  

The first 1-form Lagrangian above contains no Christoffel symbols by antisymmetry 
(as do Yang--Mills theory and the various bosonic gauge theories).   Linearity 
gives a guarantee of protection to Dirac theory, in a different way from the 
fortunate rearrangement in VI: despite being derivative-coupled this behaves according 
to the {\bf POE} by this linearity.  This means also holds for my second class of 1-form theories 
above.  On the other hand, the excluded 1-form theories have Lagrangians nonlinear in the 
Christoffel symbols.  Thus the TSA and the {\bf POE} are acting in a similar way as regards 
the selection of admissible theories.  I conjecture that (possibly subjected to some 
restrictions)  the TSA leads to the {\bf POE}.   

Crudely, tilt and derivative coupling come together in the spacetime split, and tilt 
tends to prevent TSA  formulability. Crudely, tilt and derivative coupling originate 
in spacetime Christoffel symbol terms, which are {\bf POE}-violating.   At a finer level, 
1) I know tilt and derivative coupling need not always arise together by judicious 
construction otherwise (my second class of 1-form theory example).  This could potentially 
cause discrepancies between TSA formulability and obedience of the {\bf POE} from derivative-coupled 
but untilted examples. 2) Christoffel-linear actions are not {\bf POE}-violating; moreover in 
the examples considered [Dirac theory, the standard interacting theories related to 
Dirac theory, and theory (\ref{theo2}) ] this coincides with unexpected TSA 
formulability.  2) also overrules the given example of 1) from becoming a counterexample 
to the conjecture.      

There is one limitation I am aware of within the examples I considered.  I showed that 
Brans--Dicke (BD) theory, whose Lagrangian contains Christoffel symbol derivatives 
multiplied by matter terms (the $e^{-\chi}\check{R}$ term), happens to have derivative 
coupling but no tilt.  Hence this is another example of 1), but now BD theory is TSA 
formulable but {\bf POE}-violating.  Thus one should first classify {\bf POE}-violating Lagrangians 
into e.g ones which merely containing covariant derivatives and ones which contain matter coupled 
to the curvature scalar, and then have a conjecture only about the former.  BD theory was in fact 
the first theory considered in the TSA in connection with the {\bf POE} (see \cite{BOF} and now 
my conformally-untransformed working in VI.1.4).  But then I found all those other theories in VI 
and above have TSA fomulability and the {\bf POE} occurring together.  I suggest a further systematic 
search for (counter)examples should be carried out: including many 1-forms, higher derivative 
gravity terms, torsion.  …    

\mbox{ }

\noindent{\bf 2.1 On the origin of gauge theory }

\mbox{ }

\noindent 
Traditionally, gauge theory arises in flat spacetime by gauging a la Weyl (I.1.7).  One 
postulates Lorentz-, gauge- and parity-invariance in finding the dynamics of the gauge 
fields themselves.  Observation then requires some of the gauge theories to be broken a 
la Higgs.  However I emphasize instead that gravitation may be doing some of this for us.  
Non-Maxwellian curl leading derivative terms in flat spacetime are just as 
Lorentz-invariant as Maxwellian curl ones, but accepting that we live in a curved GR 
spacetime,\fn{This includes the {\bf POE} holding perfectly; the argument is unaffected by 
replacing this part of GR by the {\bf POE} holding up to somewhat past our current stringent 
observational limit on {\bf POE} violation.} the former {\sl could not locally arise in the first 
place since they are {\bf POE} violators}.  Thus both gauge theory and its broken form may be 
seen as arising from GR rather than separate gauge and gauge-breaking postulates.    

This viewpoint does not just relate to how TSA geometry--matter results come about.  I 
find Teitelboim also came across it: ``our efforts for preserving path independence have 
led us to gauge invariance" \cite{Teitelthesis}.  But, as the previous paragraph indicates, 
I also see it plainly in the unsplit spacetime formulation of GR, which makes me suspect 
that this point is older and has been forgotten.  

One issue is whether anything new could be inspired by this point of view.  For example, 
might theories along the lines of my second class of 1-form theory, which obeys the {\bf POE} via 
Christoffel-linearity rather than absence of Christoffel symbols through the Maxwellian curl 
leading derivatives, also be present in nature? This might underly cosmological mysteries or 
future particle physics experiments.  
N.B whether this is overruled on other grounds does not affect the viability of the above 
viewpoint on the origin of gauge theory.  

\vspace{3in}

\mbox{ }

\noindent\Huge\bf{VIII Toward a quantum TSA?}\normalfont\normalsize

\mbox{ }

\noindent This chapter describes the present state of work toward quantization, 
which underlies a fair amount of preceding material.  Recollect that spacetime 
and space are suggestive of different 
quantization approaches.  

\mbox{ }

\noindent{\Large{\bf 1 Barbour's suggestion for quantization}}

\mbox{ }

\noindent Barbour favours the timeless na\"{\i}ve Schr\"{o}dinger Interpretation (NSI) 
of quantum gravity.\fn{This idea was originally Hawking's \cite{H84} , and also used by 
him and Page \cite{HP}, and by Unruh and Wald \cite{UW} who coined its name.}  This is 
about the probability of this or that universe configuration occurring, based on the 
\be
\mbox{i.p } 
\mbox{\hspace{2.1in}}
<\Psi_1|\Psi_2> = \int\Psi_1^*\Psi_2 
\mbox{ } .  
\mbox{\hspace{2.1in}}
\ee
Whereas Barbour had hopes of rendering the GR configuration space 
tractable in a geodesic principle formulation, I have shown in VI.1.3 that the GR 
configuration space curve study is a lot nastier than determining Riemannian geodesics 
(and conformal gravity retains at least some aspects of this nastiness).  

Let me illustrate with simple toys that the geodesic principle was a good conceptual 
idea.  For a 3-body system, the configuration space is the Triangle Land, the space of triangles 
{$\mbox{T}_i$}.  Here typical 
NSI-tractable issues are asking what P(T equilateral) or P(T isosceles) are [compare Hawking and 
Page asking what P(large flat universe) and P(inflation) are \cite{HP}].  I expect the 
configuration space metric would often play an important role in answering these.  
Gergely has studied the relevant geometry \cite{Gergely}.  Detailed comparison with 
Superspace has not been done (and as Triangle Land has Superspace-like features such as 
stratification and singular barriers, this may be an interesting project).  For 
Minisuperspace, Misner wrote \cite{VMisner} that he eventually came to understand that 
the considering the configuration space metric (the minisupermetric) was often crucial at both a 
classical and a quantum level (but beware the zeros!).  

My above criticism should be taken as a lack of extendibility of insight depriving one of 
tools, but not a fault.  But the NSI should be surveyed for faults.  Barbour and I think that 
its `non-normalizability leading to relative probabilities only' feature is not a fault but a 
limitation (be it technical or actually physically-realized), so I do not worry about this here.  
But I do see the following as faults \cite{POTlit1, POTlit2}. 

\noindent{Fault 1 } No questions of {\sl becoming}, such as what is P(isosceles evolves toward 
equilateral) or P(Bianchi cosmologies isotropize toward FLRW cosmologies), are answerable.  

\noindent{Fault 2 } In the NSI, the WDE equation is unaccounted for; if accounted 
for, there is incompatibility: the NSI i.p  does not respect the WDE solution space.    
%
%

\mbox{ }

Barbour then proposes the  

\noindent{\bf Time Capsule Conjecture} for TISE-type equations, 

\noindent 1) there exist configurations with `memories' or `diaries' or `fossils' or `records' 
from which information about history might be inferrable in an actually timeless world.  

\noindent 2)  Furthermore, TISE-type equations might {\sl favour} such configurations.    

\noindent 3)  This might be driven by the asymmetric shape of the configuration space.  

This conjecture is an extrapolation of the behaviour of a toy TISE, the one which Mott 
\cite{Mott} used to explain the occurrence of alpha particle tracks in bubble chambers.  These 
track configurations {\sl are} time capsules, and prima facie very unlikely configurations but 
nevertheless often visible in the laboratory.   The target of the conjecture is the WDE viewed 
as a TISE.  

Note that 1) is compatible with the NSI, and is a novel means of cutting down 
fault 1, but 2) clearly takes one outside the NSI because the WDE is invoked.    
Thus the NSI i.p needs to be replaced, and my main concern is whether this replacement 
by a WDE-compatible i.p is generally possible.  I explain below that some of 
the theories in this thesis (viewed as toys), are rather nicer than GR as regards the i.p Problem 
and might just be related to a method of use in full quantum gravity.   Also note that whereas 
elements of the above conjecture have begun to be substantiated for simple semiclassical and Minisuperspace 
WDE examples \cite{Castagnino, Hallimott}, these works use consistent histories rather than the NSI.  

\mbox{ }

\noindent\Large{\bf 2 Wheeler--DeWitt approach}\normalsize

\mbox{ }

\noindent{\bf 2.1 Strong gravities }

\mbox{ }

\noindent I must first fix the manifold structure of space.  I promote $h_{ab}$, $p^{ab}$ 
to quantum operators 
\be
\mbox{and adopt the configuration representation }
\mbox{\hspace{0.6in}}
\hat{h}_{ab}(x) = h_{ab}(x) \mbox{ } , \mbox{ } \hat{p}^{ab} = -i\frac{\delta}{\delta h_{ab}(x)} 
\mbox{ }.  
\mbox{\hspace{0.6in}}
\ee
\be
\mbox{I then choose the constraints to read }
\mbox{\hspace{1.2in}}
D_b\frac{\delta }{\delta h_{ab}(x)}|\Psi> = 0   
\mbox{\hspace{1.2in}}
\ee
(ordered with $h_{ij}$ to the left) which signify that $|\Psi> = |\Psi(\mbox{3-geometry})>$ alone, and 
the arbitrary-$W$ strong gravity WDE 
\be
``
\left[
G^{\mbox{\scriptsize X\normalsize}}_{ijkl}(h_{ab}(x))
\frac{\delta}{\delta h_{ij}(x)} \frac{\delta}{\delta h_{kl}(x)} 
- \sqrt{h}(h_{ab}(x))\Lambda
\right]
"|\Psi> = 0 
\mbox{ } .  
\ee

Next one requires an i.p.  As in the argument for GR (I.3.3.3), consider first a 
Schr\"{o}dinger i.p in attempting to resolve the i.p Problem.  Now, unlike GR, in 
the $W < \frac{1}{3}$ cases this provides an immediate resolution of the i.p Problem, 
since now a probabilistic interpretation {\sl is} admitted, straight from the 
supermetric's positive-definiteness.  Thus I have direct analogues of ({\sffamily E } $= 0$) 
TISE's with constant potential.  This i.p should be normalizable, and as an analogue of 
`first quantization', it does not require a `+ --' split of states.

For $W > \frac{1}{3}$ the Schr\"{o}dinger i.p does not admit a probabilistic interpretation, 
straight off from the supermetric's indefiniteness.  So contemplate a Klein--Gordon i.p instead.  
Now unlike GR the potential is of a fixed sign, and thus directly analogous with the standard 
Klein--Gordon equation (or the tachyonic one, depending on the sign of the $\Lambda$ `mass').  
Thus the Klein--Gordon type i.p ought to be normalizable.  There is then the issue of the 
`+ --' split of states.  This follows in pure strong gravity theory because there is a 
Superspace Killing vector for all $W$, which is furthermore timelike for each 
$W > \frac{1}{3}$.  The use of this Killing vector, unlike in GR, is now acceptable since it 
adequately respects the $W > \frac{1}{3}$ strong gravities' much simpler potential 
term.    
   
NB if $R$-perturbations are to be much smaller than $\Lambda$, they cannot alter the sign of 
the potential.  Thus there is hope for the above procedure to be applicable to GR and a range of 
scalar-tensor theories somewhat away from the extreme regime.   This would closely parallel Isham 
and Pilati's perturbative quantization idea of expanding about strong gravity, in which a GR 
regime is recovered away from the singularity \cite{SIsham,Pilatilit1,Pilatilit2, Pilatilit3}.  
At the very least, I have opened up the $W \neq 1$ extension of this study, of relevance to 
quantum scalar-tensor theories.  But note that one may have to confront Pilati's full GR quantization 
stumbling block: obtaining a quantum expression for the $R$-perturbation.      

\mbox{ }

\noindent{\bf 2.2 Conformal gravity}

\mbox{ }

\noindent Again I fix the manifold structure of space.  I promote $h_{ab}$, $\phi$, $p^{ab}$ 
to quantum operators 

\be
\mbox{and adopt the position representation }
\mbox{\hspace{0.15in}}
\hat{h}_{ab}(x) = h_{ab}(x) 
\mbox{ } ,  \mbox{ } 
\hat{\phi} = \phi \mbox{ } , \mbox{ }  
\hat{p}^{ab}(x) = -i\frac{\delta}{\delta h_{ab}(x)} 
\mbox{ } .  
\mbox{\hspace{0.15in}}
\ee
\be
\mbox{I then choose the constraints to read }
\mbox{\hspace{0.45in}}
D_b\frac{\delta }{\delta h_{ab}(x)}|\Psi> = 0 
\mbox{ } , \mbox{ } h_{ab}\frac{\delta }{\delta h_{ab}(x)}|\Psi> = 0   
\mbox{\hspace{0.45in}}
\ee
(ordered with $h_{ij}$ to the left), which signify that 
$|\Psi> = |\Psi(\mbox{conformal 3-geometry})>$ alone,
and the conformal gravity WDE 
\be
``
\left[
\frac{V^{\frac{2}{3}}h^{ik}(x)h^{jl}(x)}{\sqrt{h}(x)\phi^4(x)}\frac{\delta^{2}}{\delta h_{ij}(x)\delta h_{kl}(x)}
\mbox{--}\frac{\sqrt{h}(h_{ab}(x))\phi^4(x)}{V^{\frac{2}{3}}}
\left(
R(h_{ij}(x))\mbox{--}8\frac{D^2\phi(x)}{\phi(x)}
\right)
\right]
" 
|\Psi> = 0 
\mbox{ } .   
\ee

Now the choice of some $W < \frac{1}{3}$ (e.g the $W = 0$ above) permits the use of Schr\"{o}dinger i.p, and 
makes quantum conformal gravity directly analogous to a {\sffamily E } $= 0$ TISE.  

Quantum conformal gravity has the additional issue of whether the LFE should be 
imposed classically or quantum-mechanically.  I would opt to treat it classically, akin to a  
gauge-fixing condition.  Tentatively, one has a timeless universe described by a 
TISE-like equation, nevertheless a preferred $N$ emerges as the solution of the LFE, which (being monotonic) 
may be used to order configurations along curves in configuration space.  The privileged foliation 
ties the further multiple choice aspect of the POT\fn{This is the possibility that the quantum 
theory could depend on which choice of time function is made \cite{POTlit1, POTlit2}.} and the 
sheaf aspect of the configuration space degeneracy to the existence and uniqueness properties of 
the elliptic LFE, for which many tools are likely to be available.  The global POT aspect\fn{This is 
the possibility that no time functions exist globally \cite{POTlit1, POTlit2}.} also becomes tied 
to the properties of the LFE.  These can be contemplated at a far more trivial level for strong conformal gravity.  
The foliation-dependence of quantization POT aspect is rendered irrelevant by the theory admitting a preferred 
foliation; this Problem rather becomes whether preferred slicing theories are viable descriptions of any part 
of our universe.  The $\hat{{\cal H}}|\psi> = 0 \not\Rightarrow 
|[\hat{{\cal H}}, \hat{{\cal C}}_Y]||\psi> = 0$ POT aspect remains unchallenged.  

Note that the configuration space CS is better behaved than Superspace in being geodesically-complete 
(c.f I.2.8.1, I.2.9.4.2).  Also note that it remains unexplored whether the linear constraints can be 
dealt with classically in strong and conformal gravities; in the latter case this would 
presumably entail `Conformal Superspace quantization'  rather than `Superspace quantization'.  
One interesting point is whether arbitrary-$W$ theories are perhaps more than toys, since 
varying-$W$ universes could be viable.  This is inspired by scalar--tensor theory's tendency 
to GR-like late-universe behaviour.  I speculate that something similar might happen in some sort of CS+V 
theories, permitting the emergence of GR-like spacetime from a privileged-sliced early universe 
regime.  It remains to be seen if such a regime could take $W$ values with canonical theory 
markedly distinct from GR in quantum-cosmologically relevant regimes.  Note that it follows from 
\cite{KM} that full scalar--tensor theory itself can posses unpleasant ultrahyperbolic supermetric 
regions in addition to hyperbolic supermetric regions, but not elliptic supermetric regions. 

I finally note that the above TISE's are unlike the many wave-equation-like toys habitually used in 
the quantum gravity literature.  The corresponding toys would now rather be Helmholtz-type linear 
elliptic equations (see C.1).  The potential sign then becomes important for good behaviour; the correct sign 
is clearly enforced in the gravitational theories due to ${\cal H}$: the kinetic term's positivity 
determines the potential term's sign.   Helmholtz-type equations on manifolds are known to have 
intricately-patterned shape-dependent solutions (see \cite{Levin98} for beautiful examples 
of patterns in compact universe models).  The superspace analogue of this might just help with 3) 
of Barbour's conjecture.  

\mbox{ }

\noindent\Large{\bf 3 Other approaches}\normalsize

\mbox{ }

\noindent First for the old and the new.  Following from the classical work in III.1.6 and V.2.2.3, 
it is likely that Wheeler's quantum-mechanically motivated thin sandwich scheme is realized in some 
of the alternative 3-space theories.  Again a small $R$-perturbation should not disrupt any pure strong gravity 
theory that works in this way.  On the other hand, I have already argued that Ashtekar variables-type 
approaches do not work in III.1.5.  

\mbox{ }

\noindent{\bf 3.1 Internal time approach}

\mbox{ }

\noindent Second, I consider the search for an internal time, motivated by a (typically GR) 
spacetime ontology.  
Decompose the traditional variables into $(
h^{\mbox{\scriptsize unit\normalsize}}_{ij} \equiv 
h^{      -  \frac{1}{3}        }h_{ij}, \sqrt{h} ; 
p^{\mbox{\scriptsize T\normalsize}ij} , p)$ and perform the canonical transformation to 
$(h^{\mbox{\scriptsize unit\normalsize}}_{ij}, 
\tau_{\mbox{\scriptsize Y\normalsize}} ;p^{\mbox{\scriptsize T\normalsize}ij} , \sqrt{h})$.  
Note that the definition of $\tau_{\mbox{\scriptsize Y\normalsize}}$ as the coordinate conjugate 
to $\sqrt{h}$ is $W$-independent.   
Then treat ${\cal H}$ classically as an equation for the scale factor (now related to $\sqrt{h}$).  
Having solved this, ${\cal H}$ may be replaced by the equation 
\be
\mbox{ } \mbox{ }\sqrt{h}(x) = \sqrt{h}(x, h^{\mbox{\scriptsize unit\normalsize}}_{ij}, \tau_{\mbox{\scriptsize Y\normalsize}}; 
p^{\mbox{\scriptsize T\normalsize}ij}) 
\mbox{ } .
\label{mythical}
\ee
Since $\sqrt{h}$ is now a momentum, its corresponding coordinate $\tau_{\mbox{\scriptsize Y\normalsize}}$ (the York time) 
serves as a good internal time function, as quantizing and adopting a position representation yields a 
\be
\mbox{TDSE }
\mbox{\hspace{1.6in}}
i\frac{\delta}{\delta \tau_{\mbox{\scriptsize Y\normalsize}} }|\Psi> = 
- \widehat{\sqrt{h}}
\left({h}^{\mbox{\scriptsize unit\normalsize}}_{ij}(x), 
\tau_{\mbox{\scriptsize Y\normalsize}}; 
-i\frac{      \delta     }{      \delta h^{\mbox{\scriptsize unit\normalsize}}_{ij}(x)          }  
\right)  |\Psi>
\mbox{ } .  
\mbox{\hspace{1.5in}}
\ee
Then the corresponding i.p solves the i.p Problem.  Unfortunately in GR ${\cal H}$ becomes the nonlinear Lichnerowicz p.d.e, 
which is not exactly soluble so one does not know how to explicitly construct 
$\sqrt{h}( 
h^{\mbox{\scriptsize unit\normalsize}}_{ij}(x), \tau_{\mbox{\scriptsize Y\normalsize}}; 
p^{\mbox{\scriptsize T\normalsize}ij}(x))$ \cite{Kuchar80}.

But for the strong gravity theories, ${\cal H}_{\mbox{\scriptsize strong\normalsize}}^{\mbox{\scriptsize W\normalsize}}$ 
is just an algebraic equation (\ref{easylich}), solved by (\ref{easypsi}).  Thus, (\ref{mythical}) 
is explicitly 
\be
\sqrt{h} = \sqrt{\frac{h^{\mbox{\scriptsize unit\normalsize}}_{ik}h^{\mbox{\scriptsize unit\normalsize}}_{jl}
p^{\mbox{\scriptsize T\normalsize}ij} p^{\mbox{\scriptsize T\normalsize}kl}      }{\frac{3X - 2}{6}
{\tau}^2 + \Lambda  }    }  
\mbox{ }.  
\ee  
\be
\mbox{This gives the TDSE }
\mbox{\hspace{0.7in}}
i\frac{\delta}{\delta {\cal T}_{\mbox{\scriptsize Y\normalsize}} }|\Psi> = -
\sqrt{{h^{\mbox{\scriptsize unit\normalsize}}_{ik}h^{\mbox{\scriptsize unit\normalsize}}_{jl}
\frac{\delta^2}{\delta h^{\mbox{\scriptsize unit\normalsize}}_{ij} \delta h^{\mbox{\scriptsize unit\normalsize}}_{kl}}  }    }  |\Psi>
\mbox{ },   
\mbox{\hspace{1.2in}}
\ee
where I have straightforwardly rescaled the internal time used.  
Note that the Hamiltonian (RHS operator) is of the form $\sqrt{D^2}$ for $D^2$ the functional Laplacian (containing a  
positive-definite pointwise supermetric which aids its well-definedness).  

Thus York time is a practically-realizable internal time in strong gravities.  The $W = 1$ case of this 
was alluded to by Pilati in \cite{Pilatilit1}.  He also observed that it is 
York time that crucially leads to the previous section's Klein--Gordon approach's Killing vector, and 
also that this conformal mathematics will break down if an $R$-perturbation is included.  
Thus I cannot see the above internal time use surviving away from the singularity.  

\mbox{ } 

For some conformal gravity theories, the use of York time is as for GR.  It is a good 
internal time in principle but in each case one requires the exact solution of a Lichnerowicz-type 
equation for the scale factor in order to make progress.  In others, like in conformal gravity 
and Kelleher's theory, York time is {\it not} a good internal time even in principle, since 
it is frozen.  Also, if $N$ is emergent, one has less use for another internal time candidate $\tau$.

\vspace{2in}

\mbox{ }

\noindent\Huge\bf{B Higher-dimensional spacetime?}\normalfont\normalsize

\mbox{ }

\noindent This largely separate Part makes a start \cite{ATlett, ATpap, Rio2} 
on applying 3 + 1 split ideas to the study of higher-d spacetimes with {\sl large} extra dimensions.  
%
Recollect from I.3.3.4 that these possibly play a role in string theory phenomenology.  
They involve most of the physics being confined or closely-bound to lower-d {\it braneworlds} 
surrounded by a higher-d {\it bulk} spacetime.   Some models studied are directly relevant to string theory 
and M-theory: Ho\v{r}ava--Witten theory \cite{string1}, heterotic string phenomenology 
\cite{string2} and the related ekpyrotic scenario \cite{Turok}.  However, many models studied 
\cite{stdbr1, stdbr2, stdbr3, stdbr4, stdw1, stdw2, SMS} are in fact formulated within 
the framework of (higher-d) general relativity (GR).  Among these are:

\noindent 1) the {\it second Randall--Sundrum scenario} \cite{stdw2} in which 
the graviton is tightly-bound to the brane by the curvature due to the 
warping of the bulk metric, which is pure anti de Sitter (AdS).  

\noindent 2) A more general scheme of {\it Shiromizu, Maeda and Sasaki (SMS) } 
\cite{SMS}, in which the 4-d Einstein field equations 
(EFE's) are replaced by 4-d ``braneworld EFE's''  
(BEFE's), which are not closed since there is a `dark energy' 
Weyl tensor term, knowledge of which requires solving also for the bulk. The perceived 
interpretational difficulties due to this constitute the `Weyl Problem'. 
The other main distinctive feature of SMS's BEFE's is the presence of a term quadratic 
in the braneworld energy-momentum, which arises from the junction condition used 
\cite{Lanczos23, Darmois27, jns, SMS}. 

Two important questions arise in consideration of such models.  

\noindent{\bf Question 1:} how should such models be built and interpreted consistently within the 
framework of the 5-d EFE's?  This would require a careful underlying choice of conceptually-clear 
general framework, in the sense we discuss below.  

\noindent{\bf Question 2:} what exactly is the connection between such models and the 
underlying string or M theory?  More precisely, to what extent can the agreements 
or otherwise of predictions of such models with observations be taken as support for 
(or disagreement with) such theories?  We concentrate on question 1 here, making a comparative 
study of the general schemes that have been employed in the literature in order to 
construct bulks which surround {\it thin matter sheets} such as branes or domain walls.  
Two broad schemes have been proposed to 
construct bulks: (3, 1; 1) constructions \cite{311lit, starapp, VW} starting from 
information on a (3, 1) spacetime hypersurface (usually taken to be the brane we 
hypothetically live on), and (4, 0; --1) constructions \cite{401SS, 401N} starting 
by the construction of data on a (4, 0) spatial hypersurface.  
However, we first emphasize that one should grasp the fundamental arguments and results 
before specializing to the thin matter sheet models.  

We begin by arguing in B.1.1 against ($n$, 1; 1) constructions.  In B.1.2--3 we add to the 
arguments in I.2.3.2 and I.2.9.2 against the usefulness of the Campbell--Magaard embedding result 
both as a theoretical pretext and as a constructor of higher-dimensional spacetimes.    
Having argued for the superiority of York's and not Magaard's data procedure,  
we investigate the extent to which this is adaptable to $s = 1$, $\epsilon = 1$ (B.1.4).   
Other data construction techniques are considered in B.1.5.  We further consolidate the 
viewpoint against 

\noindent($n$, 1; 1) schemes by arguing in B.1.6 against a suggested virtue of these: 
their use to remove singularities \cite{sing1, sing2}.    

We then introduce thin matter sheets in B.2, and study the (3, 1; 1) schemes with thin matter 
sheets, recollecting the derivation of the junction conditions (B.2.1), and showing how the 
SMS formulation (B.2.2) may be reformulated in a large number of ways using 
geometrical identities that interchange which terms are present in the braneworld EFE's.  
This is illustrated by the formulation in B.2.3 which directly parallels the GR CP 
formulation and thus makes no explicit use of the Weyl term, and by formulations in B.2.4 in 
which the quadratic term has been re-expressed entirely in terms of derivatives off the brane.  
These formulations are used to clarify a number of aspects of the `Weyl Problem', in particular 
to argue for the study of the full brane-bulk system.  Further aspects of such formulations are 
discussed in B.2.5.  The implications of material in B.1 in the case of thin matter sheets, 
are discussed in B.2.6--7.  

In B.3, we continue the study of the ($n$, 0; --1) scheme favoured by our arguments, 
in the presence of both thin and thick (i.e finitely thin) matter sheets.  
We begin in B.3.1 by providing a hierarchy of very difficult general thin 
matter sheet problems in which the main difficulties stem from low differentiability and 
details about the asymptotics.  Within this class of problems we identify how 
the far more specific scenarios currently studied emerge as more tractable cases, 
and thus identify which as-yet unjustified assumptions such studies entail.  
The IVP step, of use in the study of how braneworld black holes extend into the bulk (the `pancake' 
versus `cigar' debate \cite{311lit}), involves less assumptions.  
Thus we restrict attention to the ($n$, 0) data construction problem, 
which we apply to thin matter sheets in B.3.2 and more straightforwardly to 
thick matter sheets in B.3.3.  We then conclude as regards question 1.

\mbox{ }

\noindent\Large{\bf 1 Sideways problem in absence of thin matter sheets}\normalsize

\mbox{ }

\noindent With the advent of schemes to construct bulks 
\cite{311lit, 401SS, 401N, starapp, Gregory, Wesson03}, the old questions regarding 
identification of sensible input and output for p.d.e systems (I.2.3.1--2, I.2.8--10) become 
relevant in a new context.
\begin{figure}[h]
\centerline{\def\epsfsize#1#2{0.4#1}\epsffile{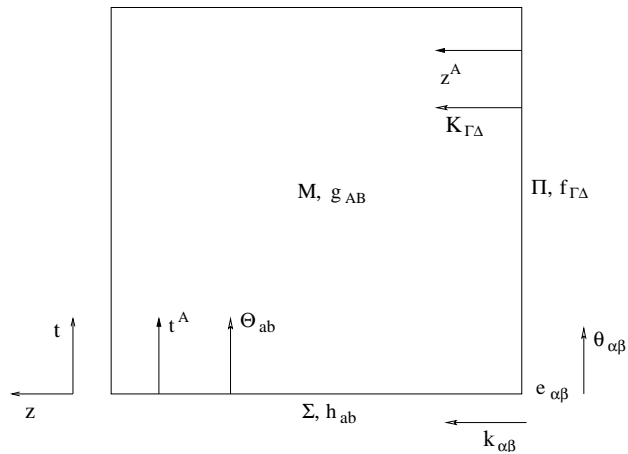}}
\caption[]{\label{TO9.ps}
\footnotesize  When we simultaneously talk of several splits in Part B, 
we use the notation presented here.  The white arrows represent the 
extrinsic curvatures associated with each of the embeddings.\normalsize}
\end{figure}
We principally use the arbitrary ($r$, $s$; $\epsilon$) ADM--type split (\ref{ADMs}), for  
which further specific notation is given in fig 14.  Discussion of some of the literature 
\cite{Wesson1, Wesson2} requires another metric split, the arbitrary ($r$, $s$; $\epsilon$) 
KK--type split  
\be
\begin{array}{ll}
g_{CD} =
\left(
\begin{array}{ll}
\epsilon\Phi^2 & \epsilon\Phi^2a_{d}\\ \epsilon\Phi^2a_{c} &  h_{cd} + \epsilon\Phi^2a_{c}a_{d}
\end{array}
\right)
\mbox{  } , \mbox{so that}  \mbox{  }
g^{CD} =
\left(
\begin{array}{ll}
a_{m}a^{m} + \epsilon\Phi^2 & - a^{d} \\ - a^{c} &  h^{cd} 
\end{array}
\right)
\end{array} 
\mbox{ } . 
\label{KKs}
\ee
Note first that for KK theory itself, this split is (3, 1; 1), with $h_{cd}$, $a_{c}$, 
$\Phi$ held to be independent of the added dimension's coordinate $z$.  It is this 
{\it cylindricity condition} that implicitly permits $a_{c}$ to be interpreted as the classical 
electromagnetic potential.  More recent generalizations of this scheme, called `noncompact 
KK theory' \cite{Wesson1, Wesson2}, have involved large extra dimensions in place of 
cylindricity, by permitting $h_{cd}$, $a_{c}$ ,$\Phi$ to depend on $z$.  In some simple 
instances, $a_{c}$ is held to be 0.  But clearly for $a_c = 0$ and $\beta_c = 0$, the two 
splits (\ref{ADMs}) and (\ref{KKs}) are identical.\fn{This requires the identification 
$\alpha \leftrightarrow \Phi$.  More generally, a KK split is the \sl inverse \normalfont 
of an ADM split with the identifications $\alpha \leftrightarrow \frac{1}{\Phi}$, 
$\beta_i \leftrightarrow - a_i$.} So there is already a vast literature on embeddings with 
a large extra dimension since this is what is usually treated in the GR CP. This fact is not  
exploited in \cite{Wesson1, Wesson2}, where rather the Campbell--Magaard result supposedly 
plays an important role.  But from the far larger and more careful GR CP and IVP literature, 
one can easily infer that such a use is technically and conceptually flawed.  

Note that the above GR CP analogue suggests at a simple level that (3, 1; 1) methods such as 
that of SMS might be viewed as \it z-dynamics \normalfont i.e ``dynamics"  with the coordinate of 
the new dimension, $z$, as IDV.  Such a ``dynamical" interpretation does not 
follow from KK theory, since there $h_{cd}$ was taken to be independent of $z$.   Whereas a way of 
relating SMS's BEFE's and `noncompact KK theory' has been pointed out \cite{PDL01}, it is surely far more revealing to 
relate the steps leading to SMS's BEFE's to the GR CP and IVP literature as we do in B.2.  

\mbox{ }

\noindent\large{\bf 1.1 Bad or unexplored behaviour of sideways analogue of GR CP}\normalsize

\mbox{ }

\noindent I.2.3.2 collects signature-independent results, including the Campbell--Magaard 
result.  Although this is identified as an arbitrary-signature CP and IVP result and already 
heavily criticized in the Introduction, I now further criticize it  
because of recent suggestions of applications of this result both in `noncompact KK' theory 
\cite{RTZ, Wesson1, Wesson2} and in braneworld bulk construction \cite{AL, ADLR, Wesson03}.  
Some of these criticisms hold more broadly against the other (3, 1; 1) constructions of bulks.   
My concern in this section is that the ($r$, 1; 1) counterparts of important GR CP results 
become bad and unexplored.  Further poor behaviour of Magaard's sideways IVP is in B.1.2.  
I sum up against the Campbell result in B.1.3.  

Most of the important results of the GR CP (c.f I.2.3.3) 
turn out to be entirely dependent on the lower-d signature $s = 0$.  In other words, the 
choice of methods which properly respect the difference between space and time is absolutely 
crucial.  

The results in question arise due to the need in proper mathematical physics for 
well-posedness and not just existence and uniqueness; for CP's the domain of dependence (DOD) 
property is part of this, in order to have a sensible notion of causality.   In our view 
causality can effectively be studied only in settings where the ``independent dynamical variable'' 
is a bona fide time.  One reason for this is that given information on an arbitrarily-thin 

\noindent (3, 1) 
hypersurface,the (4, 1)-d DOD is negligible because of this thinness (see fig 15).  Other 
reasons are discussed below.   
\begin{figure}[h]
\centerline{\def\epsfsize#1#2{0.5#1}\epsffile{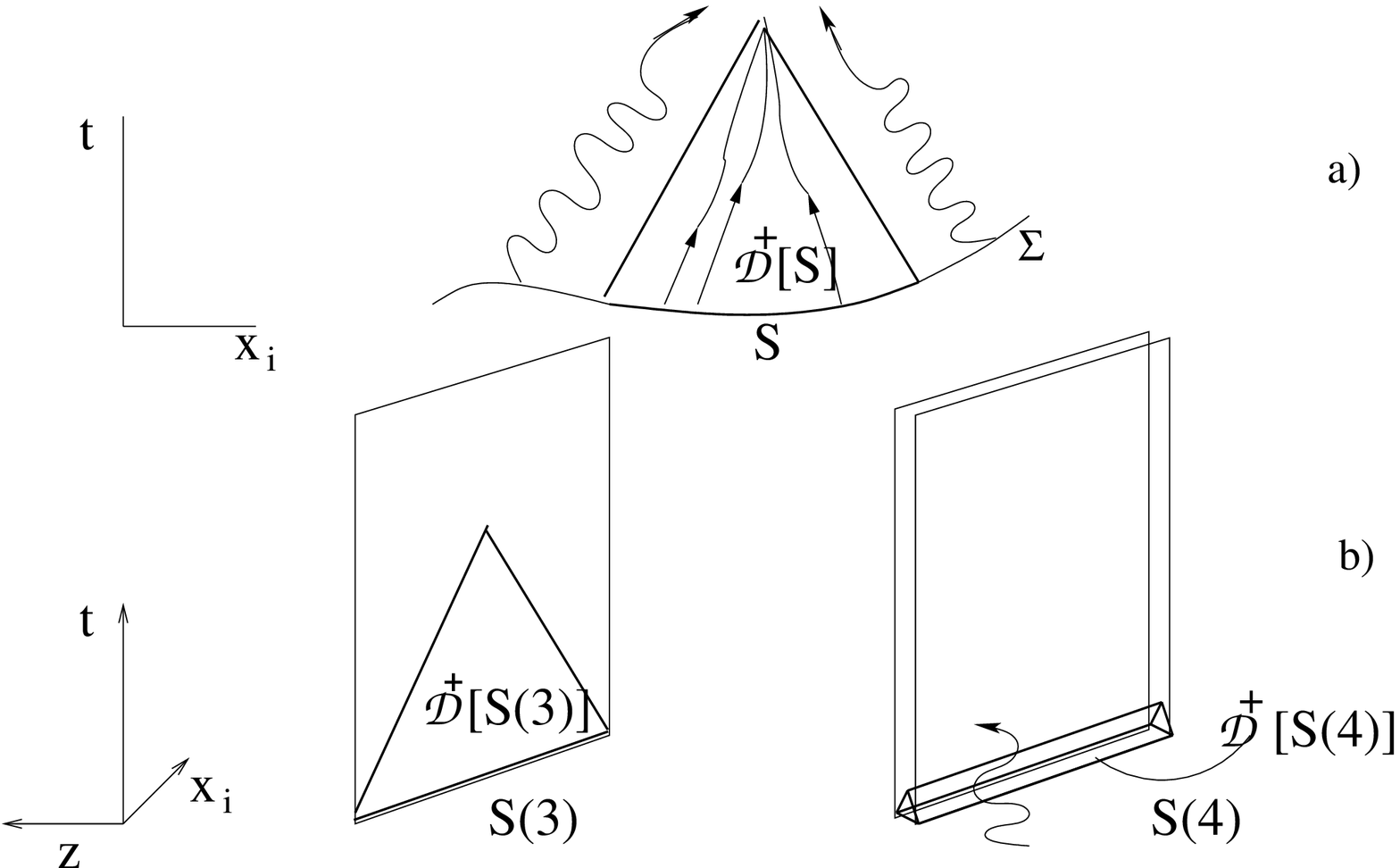}}
\caption[]{\label{TO10.ps} \footnotesize Extending the 4-d DOD figure 6 [reproduced as a) above].\normalsize 

\noindent \footnotesize If one claims to know the 4-d DOD on-brane, one is precluding the influence of   
the extra dimension, thus making it largely redundant.  The 5-d DOD of a thin sheet is small 
because of the thinness of the sheet.  Thus predictability breaks down very quickly.  
This corresponds to the 4-d branewold EFE system not being closed.\normalsize}
\end{figure}

The ($r$, 1; 1) problem is a complicated \it sideways Cauchy problem\normalfont.  For these 
problems, well-posedness theorems remain notably undeveloped.\fn{The few simple results known 
for flat spacetime sideways wave equation problems are collected in \cite{AS}; these results 
might serve as a starting point for the study of the much more complicated nonlinear sideways 
``GR CP" system.  So little is known about ultrahyperbolic equations that we deem it not 
sensible to talk about ($r$, $s$; $\epsilon$) procedures  at present for $s > 1$ or $s = 1$, 
$\epsilon = - 1$.  We argue that $s = 1$ $\epsilon = 1$ is hard enough!} There is quite simply 
no established way to proceed.  There is no known sideways ($s = 1$) analogue of Leray's 
theorem.   There are no function spaces known to be appropriate for sideways Cauchy problems.  
We can explain however why the applicability of Sobolev spaces to the GR CP does not carry over 
to sideways Cauchy problems.  First there is no sideways notion of DOD to make the construction.  
Second even if we assume the higher-d DEC holds, it would not give an inequality because the 
perpendicular vector $z^A$ in now spacelike (fig 16).  Third, we obtain a difference of 
squares rather than the sums in (\ref{sums}), so the equivalent of the energy method's use of 
Sobolev norms is simply of no use to control the ``evolution" given the data.  
\begin{figure}[h]
\centerline{\def\epsfsize#1#2{0.4#1}\epsffile{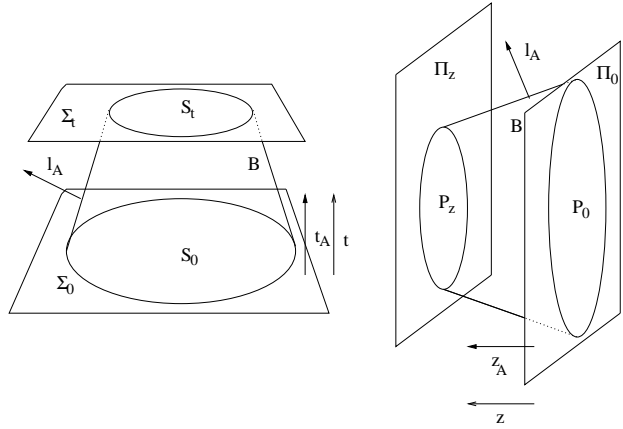}}
\caption[]{\label{TO11.ps} \footnotesize The bucket-shaped construction of Fig 7, and its meaningless 
sideways counterpart.\normalsize}
\end{figure}

\mbox{ }

\noindent\large{\bf 1.2 Extra difficulty with sideways Campbell--Magaard}\normalsize

\mbox{ }

\noindent The (r, 1; 1) Campbell--Magaard method encounters the following difficulty.  
The use of the strip $0 \leq x_1 < \eta$   as ``evolutionary" data is generally invalidated by 
the information leak construction in fig 17, unless one has had the luck to construct a full 
global data set.  But this would require the data construction to encounter no zeros nor to run 
into asymptotic difficulties.  Thus, in fact signature does play a role even for the 
Campbell--Magaard result.  
\begin{figure}[h]
\centerline{\def\epsfsize#1#2{0.4#1}\epsffile{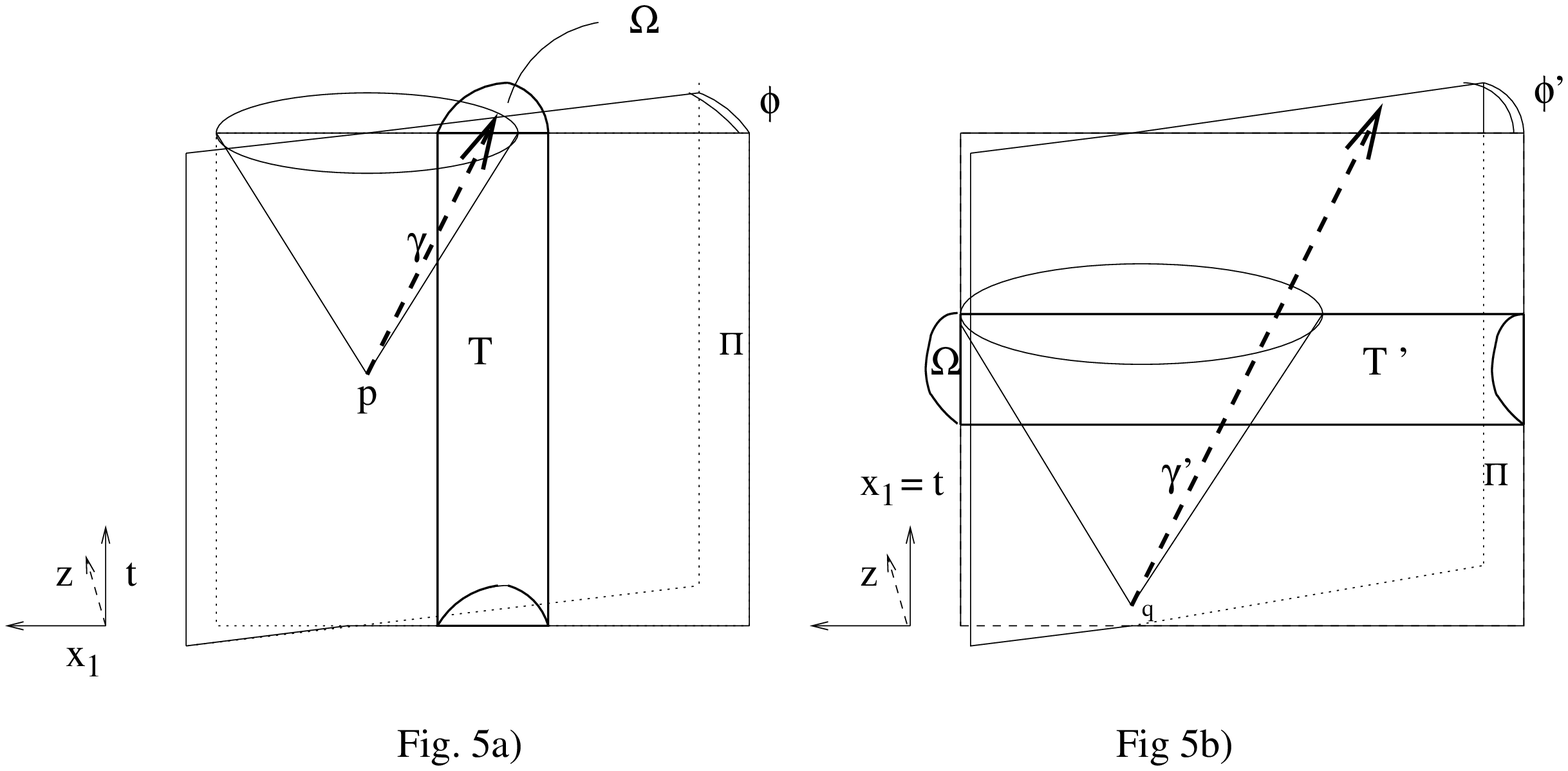}}
\caption[]{\label{TO12.ps} \footnotesize \noindent a) $\Pi$ is a ($r$, 1) (i.e timelike) 
hypersurface containing a data strip $T$ built by the Magaard method from prescribed 
`data for the data' on a timelike $x_1 = 0$. The region $\Omega$ is that part of the 
($r + 1$, 1) spacetime supposed to be entirely controlled by $T$ 
i.e for which the Campbell method can validly provide the evolution.  Typically $T$ will not cover all of $\Pi$ 
in the $x_1$-direction, so there will be nearby points like $p$ outside $T$.  Consider the future light-cone 
with apex $p$.  No matter how thin in the $z$-direction one considers $\Omega$ to be, it is always pierced by causal curves 
such as $\gamma$ on or in the light-cone, by taking $\gamma$  to be at a sufficiently slender angle $\phi$ to $\Pi$.  
Thus information can leak into $\Omega$ from elsewhere than $P$, which is a contradiction.  Therefore, in parts at least, 
$\Omega$ is arbitrarily thin in the $z$-direction. 
One can envisage that sometimes there will be enough points like $p$ that there is no region $\Omega$ at all.  \normalsize 

\noindent \footnotesize b) Suppose $x_1 = 0$ is spacelike and is a section of the entire hypersurface 
$\Pi$ and $T^{\prime}$ is the data strip built from it by the Magaard method.  
There is now no room for points like $p$! However, typically $T^{\prime}$ will not cover all of $\Pi$ in the 
$x_1$-direction, so there will be nearby points like $q$ in the causal past $J^-(P^{\prime})$ not in $P^{\prime}$.  
No matter how thin one considers $\Omega$ to be, it is pierced by causal curves $\gamma^{\prime}$ from $q$ 
\sl which do not pass through $P^{\prime}$ \normalfont, by taking these to be at a sufficiently slender angle $\phi^{\prime}$.  
So again, in parts at least, $\Omega$ is arbitrarily thin and sometimes there will be enough points like $q$ 
that there is no region $\Omega$ at all. Small pieces of timelike hypersurfaces need not hold useful information for 
hyperbolic-type systems! \normalsize}
\end{figure}

\mbox{ } 

\noindent\large{\bf 1.3 Arguments against Campbell's theorem: conclusion}\normalsize

\mbox{ }

\noindent After the above criticism, it is appropriate to conclude here about the 
Campbell--Magaard result.  
Although one can {\sl still} state a form of Campbell--Magaard in these terms 
(about constructing an embedding 
for a small region to be determined outside which all prior knowledge of the metric is 
forsaken), this only permits a rigorous theorem at the price of exposing its application 
to building bulks as weak and undesirable.  The result is also weak because of 
technical inappropriateness as described in I.2.3.2, and because of lack of further supporting 
results (I.2.3.3 and B.1.2). Its application to construct physically-meaningful embeddings of known 4-d worlds 
is furthermore rendered questionable because it is so readily generalizable to all manners of 
matter-containing bulks, among which there are insufficient criteria for discrimination.  

Some of these poor features are Magaard--specific, and can be avoided by realizing 
Campbell's theorem otherwise.  My Campbell-method 2 realization in I.2.9.2.3 is at 
least for small regions {\sl known beforehand}.  My Campbell--irreducible method 
realization in I.2.9.5 is at least for a system that is everywhere valid albeit the 
existence and uniqueness may break down in some regions, and is based on adequately 
covariant mathematics.   Thus these superior forms of Campbell's theorem are 
`about as badly behaved as the thin sandwich'.  But a strong theorem would be for 
{\sl all} regions, at least in many well-defined and substantial cases, like in York's work.  
A strong theorem would be for adequate function spaces and supported by further 
well-posedness theorems.   

About its application to `noncompact KK theory', subject to the above restrictions, 
it is indeed always possible to embed small pieces of spacetime with a given metric 
into vacuum spacetime with one extra dimension.  It is then claimed that the Campbell--Magaard  
theorem provides existence for the theory's $\check{R}_{AB} = 0$ equation to cover geometrization 
of any kind of matter \cite{Wesson2}, just like KK theory is a geometrization of electromagnetic 
matter.  As the Campbell--Magaard theorem relies entirely on the analytic functions which are 
inappropriate for relativistic theories, one must doubt that it is capable of 
providing support for any theory.  Also, as the Campbell--Magaard theorem also holds (trivially) 
for whichever other choice of analytic functional form of higher-dimensional energy-momentum 
tensor, it in no way picks out vacuum spacetimes.  Furthermore, this approach considers 
only 1-component (`induced') matter; counting degrees of freedom shows that it cannot be 
extended to many important cases of fundamental matter. Whereas allowing for more extra 
dimensions could improve similar situations \cite{Wesson2, WittenKK}, {\sl unification requires 
geometrization of the fundamental matter laws themselves}, whilst this `induced matter' 
approach only geometrizes {\sl solutions} of the EFE's coupled to matter of unspecified field 
dependence.\fn{For clarity, compare KK theory proper, in which the electromagnetic 
potential \sl and Maxwell's equations \normalfont are geometrized, as the $f_{\Gamma z}$ 
portion of the 5-metric and the $G_{\Gamma z}$ equations of the KK split respectively, 
along with a scalar field.  `Noncompact KK theory' could geometrize a more general vector 
field than the electromagnetic potential in addition to the generalization of the KK scalar 
field that is all that is usually considered.  Now, this vector field is no known vector field 
of nature for its field equations does not in general correspond to that of any known vector 
field (although clearly the field equation contains the inhomogeneous vacuum Maxwell 
equations as a subcase since KK theory would be included in this way within 
`noncompact KK theory'). 
For sure, this vector field (or even more obviously the scalar field) is not capable of 
being a simultaneous geometrization of the individual vector fields of the standard model.
If one were to treat these fields collectively, one throws away all that is gained by 
keeping their classical identities separate:  the results of Weinberg--Salam theory 
and QCD that are obtained then by quantization.}

Finally I note that the Campbell--Magaard theorem does not appear to actually be used in any 
of its claimed applications to constructions: the messy Magaard part of the prescription is 
simply ignored (and even replaced by what turn out to be rudiments of the conformal method, 
to which I next turn).  Unlike what is claimed in \cite{Wesson03} and elsewhere, without 
such a specific prescription, all one is doing is applying standard embedding mathematics, and 
not some old but little-known result to modern scenarios.  

\mbox{}

\noindent\large{\bf 1.4 What parts of the conformal IVP method survive? }\normalsize

\mbox{ }

\noindent Having argued much in their favour in I.2.9, to what extent conformal methods can be 
adapted both to heuristic and to general constructions for ($r$, 1; $\epsilon$) data ?

I treated the choice of the $n$-d generally-covariant, conformally-invariant trace-tracefree 
and TT-TL splits, and the subsequent decoupling of the Gauss and Codazzi equations in I.2.9.3, 
in a manifestly signature-independent fashion.  The defocussing property of the underlying 
choices of slicing is signature-independent.    
  
But the study of the usual (3, 0; --1) Lichnerowicz equation has been specifically elliptic.  
Investigation of whether the $s = 1$ `wave Lichnerowicz equation' has good existence and 
uniqueness properties could be interesting.  The natural setting for this is as a Cauchy 
problem (for even the 2-d wave equation is ill-posed as a Dirichlet problem).  Assuming that 
there exists a Cauchy surface in the (3, 1) spacetime sense, one can attempt forward and 
backward evolution to produce a global data set (assuming also that the decoupled procedure 
for finding ${K}^{\mbox{\scriptsize T\normalsize}}_{\Gamma\Delta}$ also yields a global 
solution).  Whilst this `wave Lichnerowicz' procedure for the data could conceivably produce 
global data sets in some subcases and thus avoid the onset of information leak difficulties, 
the other difficulties described in B.1.2 remain, since the next stage is still to be a 
sideways Cauchy problem.  For example, ultimately the `wave Lichnerowicz' problem would inherit 
the function space difficulty of the sideways CP that is to follow it.  So we do not think 
the `wave Lichnerowicz equation' is likely to be suitable as a general method. We favour 
instead the (4, 0; --1) approach in B.3.  

That said, the `wave Lichnerowicz equation' may still serve as the basis of a useful heuristic 
method.  We consider this in B.2.6.  Note also that the outline of the carefully thought-out 
simplifications and tricks for the Lichnerowicz and Codazzi equations listed in C.2.2 hold 
regardless of signature. Considerations of $j_i = 0$, cancellation of Lichnerowicz equation 
terms, conformal flatness, maximality, and of the `Bowen--York' system of $n$ + 1 flat-space 
linear equations are all available.

Also the study of the CMC LFE has been specifically elliptic, so the theory of existence 
of CMC slicings \cite{CGSYkin, CMClit} is likely to be signature-dependent.  

\mbox{ }

\noindent\large{\bf 1.5 Other methods and formulations}\normalsize

\mbox{ }

\noindent We note that the thin sandwich procedure of I.2.8, I.2.9.1 is 
($r$, $s$; $\epsilon$)-independent.  Thus we pose the thin sandwich conjecture for general 
($r$, $s$; $\epsilon$).  This is also considered for thin matter sheets in B.2.6.    
This is worth a try because of the concern with the sideways Cauchy problem being a bad prescription.  
Note that this suggestion signifies that the lower-d spacetime is a carrier of information 
about the extra dimension, which is now spacelike rather than a time.  Many of the aspects of 
the Problem of Time will arise again in this 

\noindent context.  

The conformal thin sandwich scalings are again signature-independent, giving me another possible 
bulk construction formulation.  Again however, the usual conformal thin sandwich mathematics is 
elliptic, so I do not know if this idea will work in the new sideways context.  

A (3, 1; 1) Hamiltonian formulation is not provided  
because although it 
is insensitive to $\epsilon$ \cite{KucharII} (just a few sign changes), and to dimensional increase 
(just a few traces), it {\sl is} sensitive to $s$ 
through the intimate involvement of the space of geometries.  
Whilst the usual Hamiltonian formulation is based on the space of Riemannian 
geometries (Superspace) \cite{Wheeler, DeWitt}, the space of semi-Riemannian geometries 
is reported to be not even Hausdorff \cite{Haus}.  

\mbox{ }

\noindent\large{\bf 1.6 Applying (r, 1; 1) methods to remove singularities}\normalsize

\mbox{ }

\noindent We have argued above in favour of (4, 0; --1) schemes over (3, 1; 1) schemes.  
However, it is often suggested that (3, 1; 1) schemes may be useful since in certain particular 
examples (3, 1) spacetime singularities are `removed' by embedding into nonsingular (4, 1) 
spacetimes \cite{sing1,sing2}.  Whereas embeddings are undoubtedly useful tools in the 
study of (3, 1) spacetimes (for the purpose of algebraic classification \cite{mathemb, MC}), 
we ask what is the status of their use to remove cosmological singularities.  Is it 
mathematically rigorous, generally applicable and physically meaningful?

By `singular' we mean geodesically-incomplete \cite{sclass}.  In the study of singularities, 
timelike and null geodesics (t.g's and n.g's) are the only objects of fundamental importance 
since they are the curves privileged by the free travel of massive and massless particles 
respectively.  The other objects associated with the study of singularities arise 
rather as a matter of convenience.  

Such objects include the expansion and shear of geodesic congruences, and curvature scalars.  
We focus first on these 
(B.1.6.1), explaining how they arise in the study of singularities and then demonstrating how 
the embedding and embedded versions of these objects can be very different (B.1.6.2). We then 
discuss the more important issue of geodesic incompleteness in the context of embeddings 
(B.1.6.3).  An example is used to illustrate some of these points and others (B.1.6.4).  A 
lack of rigour in singularity-removal claims is uncovered in (B.1.6.5).  We return to the 
case with thin matter sheets in B.2.7.  

\mbox{ }

\noindent{\bf 1.6.1 Secondary objects in the study of singularities}   

\mbox{ } 

\noindent For ($q$, 1)-d cosmology, smooth congruences of past-directed normal t.g's with 
normalized tangents denoted by $u^a$ are considered.  The decomposition 
\be
{\cal B}_{ab} \equiv D_au_b = \frac{\theta}{q}h_{ab} + \sigma_{ab} \mbox{ } \mbox{ } + \omega_{ab}     
\label{Bsplit}
\ee
provides the {\it expansion} $\theta \equiv h^{ab}{\cal B}_{ab}$ and {\it shear}  
$\sigma_{ab} \equiv {\cal B}_{(ab)}^{\mbox{\scriptsize T\normalsize}}$.  The {\it twist} 
$\omega_{ab} =  {\cal B}_{[ab]}$ is zero for the normal congruences considered here, in which 
case ${\cal B}_{ab}$ is an extrinsic curvature $\theta_{ab}$.  The corresponding normal 
Raychaudhuri equation [c.f (\ref{normRay})] can now be considered.   Although for 
$\theta_0 > 0$ this would usually mean that a caustic develops, under certain global 
conditions a contradiction about the existence of conjugate points arises.  Singularity 
theorems are thus obtained; in a cosmological context the simplest\fn{There are a number of 
other cosmological singularity theorems \cite{HE, Wald} not only because some have weaker 
assumptions, but also because one wants to be able to treat a number of pathological cases.  
One such is the Milne universe:  although this na\"{\i}vely looks like a cosmology with its 
focusing of geodesics normal to $t =$const as $t \longrightarrow 0$, it is merely a region of 
Minkowski spacetime. So one would not want to include the Milne universe among the singular 
spacetimes.  The way out of this is to demand that physically-meaningful cosmologies contain 
matter. This makes sense because the problematic physics that may be associated with geodesics 
focusing is the possible pile-up of matter travelling along these geodesics leading to 
infinite densities. One succeeds in not including the Milne universe by use of singularity 
theorems hinging on the `timelike genericity condition' $R_{abcd}u^au^b \neq 0 \mbox{ }
\mbox{ at least one point along each timelike geodesic}$, since the Milne universe is flat.}  
of these is (Hawking, theorem 1 of \cite{CGHawking}). 

\noindent  For globally-hyperbolic (3, 1) GR spacetimes obeying the strong energy condition and such that 
$\theta = C \geq  0$ on some smooth (spacelike) Cauchy surface $\Sigma$, then no past-directed 
timelike curve from $\Sigma$ can have greater length than $\frac{3}{|C|}$.  

\noindent By the definition of Cauchy surface, all past-directed t.g's are 
among these curves and are thus incomplete, so the spacetime is singular.  
Such theorems generalize to ($q$, 1) spacetimes (provided that these obey 
analogous energy conditions) in the obvious way since Raychaudhuri's 
equation clearly behaves in the same way for all $q$ and the global part 
of the arguments uses topological space methods that do not care about dimension.  

It must be noted that the singularity theorems are about existence whilst saying nothing about the nature of the singularity.    
Ellis and Schmidt began to classify singularities according to their properties \cite{sclass}, 
a difficult study which may never be completed \cite{Clarke}.   
Below we consider only the most elementary type of genuine singularities: \it curvature singularities\normalfont, 
for which at least one spacetime curvature scalar such as $\check{R}$ or $\check{R}_{AB}\check{R}^{AB}$ blows up.    

\mbox{ }

\noindent{\bf 1.6.2 Relating the embedded and embedding secondary objects}

\mbox{ }

\noindent Here we explain how knowledge of Gauss' hypersurface geometry renders it unsurprising 
that some singular spacetimes are `embeddable' in nonsingular ones.  
This is because the behaviour of the higher- and lower-d objects used in the study of 
singularities clearly need not be related.    As a first example, consider expansion.  
In the normal case, 
\be
\theta_{\alpha\beta} = -\frac{1}{2\alpha}\frac{\pa e_{\alpha\beta}}{\pa t} 
\mbox{ } \mbox{ and } \mbox{ } \Theta_{ab} = -\frac{1}{2\alpha}\frac{\pa h_{ab}}{\pa t}
\ee 
\be
\mbox{from which follows }
\mbox{\hspace{1.70in}}
\theta = \frac{n - 1}{n}\Theta \mbox{ } - \mbox{ } \Theta^{\mbox{\scriptsize T\normalsize}}_{\perp\perp} 
\mbox{ } .
\mbox{\hspace{1.75in}}
\ee
So a blowup in  ${\theta}$ need not imply a blowup in $\Theta$.    
Thus the (4, 1) spacetime perspective on focusing of geodesics 
can be completely different from the perspective on some (3, 1) hypersurface.  
Fig 18 shows how one's (3, 1) notion of expansion generally corresponds 
to expansion and shear from the perspective of an embedding (4, 1) spacetime.       
In particular, what look like caustics or singularities in the (3, 1) spacetime could well correspond to 
no such feature in some surrounding (4, 1) spacetime. 

\begin{figure}[h]
\centerline{\def\epsfsize#1#2{0.4#1}\epsffile{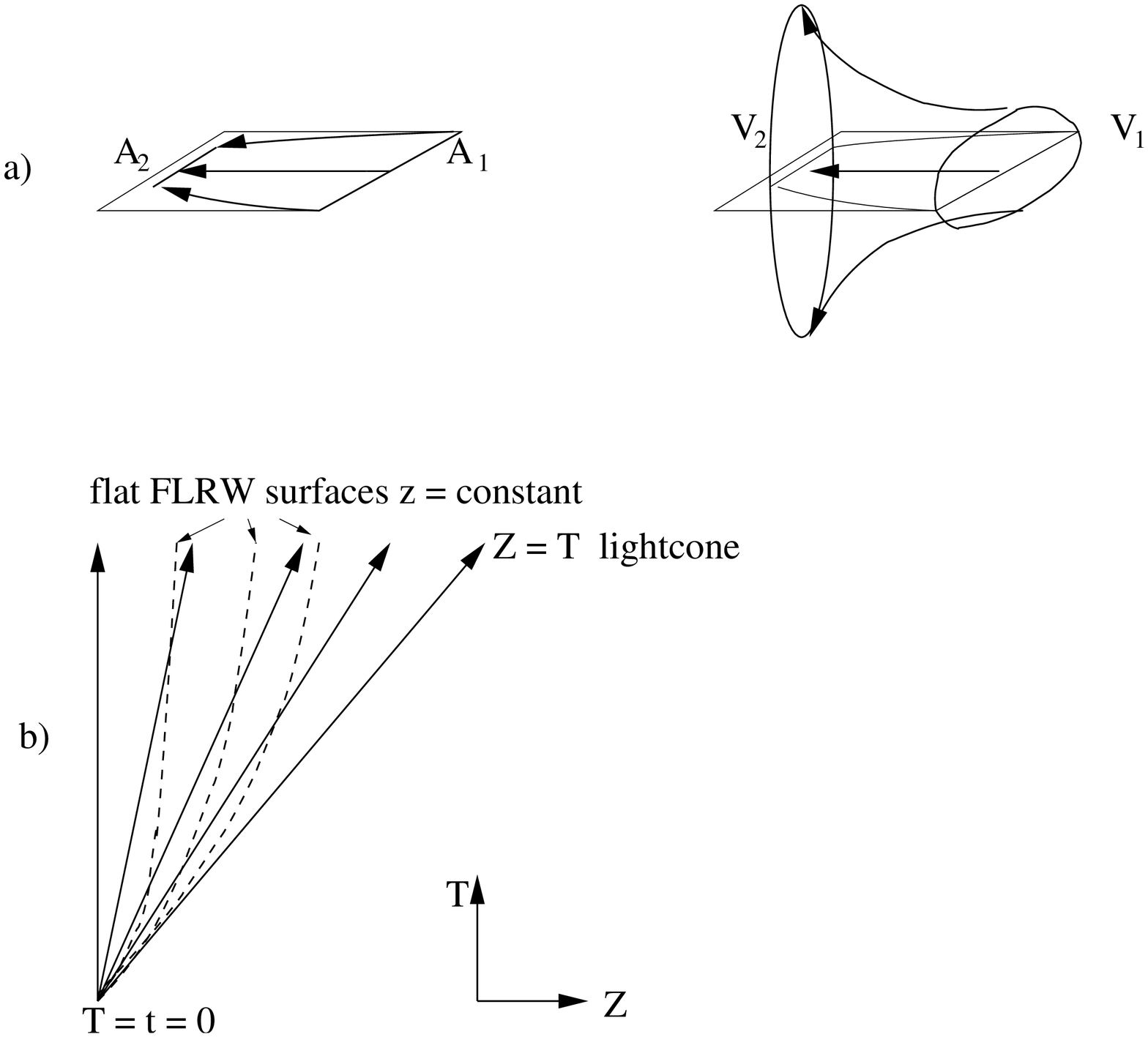}}
\caption[]{\label{TO13.ps} \footnotesize a) A simple illustration of how 4-d expansion may be interpreted as being mainly due to shear in a 5-d embedding manifold.\normalsize  

\noindent \footnotesize b) Diagram of the embedding of flat FLRW universes into Minkowski spacetime in the standard 5-d Minkowski coordinates $(T, X_1, X_2, X_3, Z)$.  
The curved surfaces are the FLRW spacetimes.  As one approaches $T = 0$ (corresponding to $t = 0$ in FLRW coordinates), each of these surfaces becomes 
tangent to the light cone (characteristics of Minkowski).  The foliation by these surfaces also becomes bad here because the surfaces intersect.  
Note also that the FLRW geodesics move within each of the surfaces whereas the Minkowski geodesics clearly pierce these surfaces.  Thus the 4-d and 
5-d geodesics in this example are not the same.\normalsize}
\end{figure}

As a second example, consider curvature scalars, in particular the Ricci scalar.
Preliminarily, Gauss' outstanding theorem is a clear indication that extrinsic curvature can 
compensate for differences between higher- and lower-d intrinsic curvatures.
 Our simple idea is to consider the implications for rigorous embedding mathematics of the 
generalization of this to the case where lower-d curvature scalars become infinite. 
From the form 
\be
\mbox{(\ref{Agauss}) for the (3, 1; 1) Gauss constraint, } 
\mbox{\hspace{0.5in}}
2\rho = 2\check{R}_{\perp\perp} - \check{R} = \frac{3}{4}{K}^2 - {K}^{\mbox{\scriptsize T\normalsize}} \circ {K}^{\mbox{\scriptsize T\normalsize}}  - R
\mbox{\hspace{0.5in}}
\label{singrem}
\ee 
clearly  -- at least for some (3, 1) worlds which have Ricci-scalar curvature 
singularities $R \longrightarrow \infty$ -- there will be surrounding (4, 1) worlds in which $\check{R}$ 
(and the 5-d $\rho$) remain finite.  For it could be that 
${K}^{\mbox{\scriptsize T\normalsize}} \circ {K}^{\mbox{\scriptsize T\normalsize}} - \frac{3}{4}{K}^2 \longrightarrow -\infty$ 
suffices to compensate for $R \longrightarrow \infty$. 
Thus it could be shear, expansion or both that dominate the compensation for $R \longrightarrow \infty$.  
If this involves $|{K}| \longrightarrow \infty$, it means a (4, 0) caustic forms in the (4, 1) 
spacetime corresponding to the (3, 1) spacetime's singularity.  
Yet this is not the only nonsingular possibility; if $R \longrightarrow -\infty$  
it may be the shear that compensates, including cases in which  the blowup is pure shear.    

Finally, clearly the higher-d singularity theorems hold so even if one were to 
succeed in excising singularities from a lower-d model, one would typically 
expect singularities to occur elsewhere in the resultant higher-d models. The 
nature of these higher-d singularities may not be the same as that of the excised 
lower-dimensional ones.  So, interestingly, by allowing extra dimensions, one would 
be even less certain of the character of singularities.

\mbox{ }

\noindent{\bf 1.6.3 embeddings and geodesic incompleteness}

\mbox{ }

\noindent The fundamental importance of geodesics is problematic for embeddings since in 
general the (3, 1)-d geodesics are not included among those of an embedding (4, 1)-d spacetime.  
There is then the dilemma of which of these sets of curves should be the physically 
privileged ones and thus be the set of curves whose extendibility is in question.  

If one wishes to postulate (4, 1)-d GR, in addition to the (4, 1)-d EFE's one must surely
require that the matter follows the (4, 1)-d geodesics (c.f I.1.5).  Then the issue of extendibility 
of the original (3, 1)-d geodesics becomes irrelevant since they are unprivileged curves 
in the (4, 1)-d spacetime.  One would rather replace them with (4, 1)-d geodesics, which 
might be viewed as unsatisfactory from a GR ontology.  Note that this is at the price of 
altering what the effective (3, 1)-d physics is: often it will be of affine-metric type and 
thus in principle distinguishable from ordinary (3, 1)-d GR physics.  This may contend with 
the recovered (3, 1)-d physics being as close to the usual (3, 1)-d physics as is widely 
claimed in the braneworld literature.   

Suppose instead that one postulates that both (3, 1)- and (4, 1)-d geodesics play a part in 
braneworld models, as the paths of brane-confined and bulk-traversable matter (or gravity waves) 
respectively. 
It is then not clear at all what is meant by a singularity -- exactly which curves are 
supposed to be incomplete?  

Finally, (3, 1) incompleteness might correspond to a (3, 1) hypersurface that 
becomes tangent to the characteristics of the bulk equations in which case t.g's and n.g's are 
being forced to exit that (3, 1) hypersurface. Also, the (3, 1)-d 
geodesics might be of (4, 1)-d geodesics which are extendible only by replacing a piece of 
the original foliation with a new one that extends into what was originally regarded as the 
extra dimension.  

\mbox{ }

\noindent{\bf 1.6.4 Embedding flat FLRW in Minkowski}

\mbox{ }

\noindent The following example illustrates many of the above points and leads us to further comments. 
Consider a (4, 1) spacetime with metric (see e.g \cite{sing1})
\be
g_{AB} = \mbox{diag}(g_{tt}, \mbox{ } e_{\alpha\beta}, \mbox{ } 
g_{zz}) = \mbox{diag}
\left(
-z^2, \mbox{ } t^{  \frac{2}{q}  }z^{  \frac{2}{1 - q}  }\delta_{\alpha\beta}, 
\mbox{ } \frac{q^2t^2}{(1 - q)^2}
\right) \mbox{ } , \mbox{ } \mbox{ } q > 1.
\label{funnymink}
\ee 
This is simple to treat because: 

\noindent 1) If we foliate it with constant $z$ hypersurfaces, a portion of each $z =$ const hypersurface has induced on it a flat FLRW cosmology.  
In particular, the coordinates have been chosen so that the $z = 1$ hypersurface is the FLRW cosmology with scale factor $t^{\frac{1}{q}}$. We restrict attention to 
$q \leq 3$ due to the DEC.  For $q \neq 2$ there is a (3, 1) Ricci scalar curvature singularity ($q = 2$ is the radiation universe, whence $R = S = 0$).  

\noindent 2) The (4, 1) spacetime is in fact Minkowski.

First, foliating the flat FLRW cosmology with constant $t$ surfaces,  
$\theta = \frac{3}{q t} \longrightarrow \infty$ as $t \longrightarrow 0$, so there is (3, 0; --1) 
focusing as the Big Bang is approached.  
Furthermore, \sl only \normalfont focusing occurs: $\theta^{\mbox{\scriptsize T\normalsize}}_{\alpha\beta} = 0$.  Next foliate 
the (4, 1) spacetime with constant $t$ surfaces.  The (4, 1) spacetime contains many FLRW worlds on constant-$z$ surfaces.  Build a congruence 
by collecting the FLRW geodesics on each $z =$ const slice.  Then the (4, 0; --1) expansion ${\Theta} = \frac{q + 3}{ztq}$ also blows up as $t \longrightarrow \infty$, 
but there is also a blowup of the corresponding 

\noindent (4, 0; --1) shear:
\be
\Theta^{\mbox{\scriptsize T\normalsize}}_{ab} \equiv \frac{q - 1}{4qzt}\mbox{diag}(-e_{\alpha\beta}, \mbox{ } 3g_{zz}) \mbox{ } .
\ee
Also, whilst still both $\theta \longrightarrow \infty$ and $\Theta \longrightarrow \infty$ in this case, we know that the former viewed 
from within the $z = 1$ (3, 1) hypersurface corresponds to a genuine (3, 1) singularity, whilst the latter in (4, 1) would be a mere caustic.  

Second, consider the $z = 1$ slice of the (4, 1) spacetime.  Here the (3, 1; 1) expansion and shear are 
\be
{K} = \frac{q - 4}{q t} \mbox{ }, \mbox{ } 
{K}^{\mbox{\scriptsize T\normalsize}}_{\Gamma\Delta} = -\frac{1}{4t}\mbox{diag}(3, \mbox{ } t^{\frac{2}{q}}\delta_{\gamma\delta}).
\ee
Thus for the physical range, both blow up as $t \longrightarrow 0$.  So the spacetime includes a $z = 0$, $t = 0$ point, at which there is a (4, 0) caustic.  
These blow-ups combine in ${K}^{\mbox{\scriptsize T\normalsize}} \circ {K}^{\mbox{\scriptsize T\normalsize}} - \frac{3}{4}{K}^2$ to cancel $R$ for all values of 
$q$; for $q = 2$ the shear and expansion contributions exactly cancel each other.  The unphysical $q = 4$ case is an example of a pure shear blowup.

Although the congruence two paragraphs up is built to naturally include the geodesics of all the included FLRW universes, these turn out not to be the (4, 1) geodesics 
(nor pieces of them).  Rather, by considering the coordinate transformation between (\ref{funnymink}) and standard (4, 1)-d 
Minkowski coordinates, it is easy to show that the Minkowski geodesics pierce the $z =$ constant surfaces that are the FLRW universes 
(more significantly in the early universe than today).  As $t \longrightarrow 0$ the $z=$ constant surfaces approach the null cone so the example is an illustration of 
the foliating (3, 1) hypersurfaces becoming tangent to the characteristics at the point of interest.  Also the foliation breaks down as $t \longrightarrow 0$ 
because the family of hypersurfaces of constant $z$ intersect at $t = 0$.  All these points are 
illustrated in Fig 18b).                               

Finally, the example cannot be taken to be typical by the (4, 1) version of genericity.  
Indeed, we know that not many (3, 1) spacetimes can be even locally embedded into 

\noindent (4, 1) Minkowski \cite{MC}.  
The value of particular examples is limited to exhibiting \sl possible \normalfont behaviours.  
One would require this to be upgraded to the study of  large classes of examples to assess \sl probable \normalfont behaviours.  
Indeed, one would expect higher-d cosmological models to exhibit a range of behaviours distinct from that of the Big Bang (including nonsingularness).  
This follows from knowledge of particular examples (from the algebraic study) of inhomogeneous cosmologies in standard GR \cite{CGKrasinski}.  
We especially note that if highly-symmetric cosmologies are haphazardly embedded, 
one would often expect the resulting higher-d models to be less symmetric i.e 
more anisotropic and inhomogeneous than the original lower-d models
[as na\"{\i}vely suggested by the appearance of (4, 0; --1) shear from models containing no (3, 0; --1) shear].  
So the resulting higher-d models are expected occasionally to exhibit more unusual behaviour than the lower-d models from which they arose,  
and this includes the possibility of nonsingular cosmologies.  Indeed, we would expect that an increase of dimension increases the variety of possible 
behaviours in inhomogeneous cosmology.  But the bottom line in the study of 4-d and 5-d singularities is the same: one is ultimately interested 
in generic behaviour and this requires more sophisticated means of study than building individual particular examples.     

\mbox{ }

\noindent{\bf 1.6.5 Nonrigorousness of singularity removal by embedding}

\mbox{ }

\noindent So, there are arguments against the use of embedding results toward making general 
statements about singularity removal.  Some of our arguments against the 
use of the Campbell--Magaard result (some of which also hold against the use of further 
embedding theorems) are relevant to this context.  In particular, given a spacetime, there are so many possible 
embeddings.  Were nature higher-d, why should it choose a particular nonsingular 
embedding out of an infinity of (nonsingular or singular) embeddings that mathematically 
exist?  The more extra dimensions are present, the more severe this nonuniqueness is.  
This makes singularity removal by such embeddings physically-questionable.  The 
Campbell--Magaard scheme, and other embedding schemes such as the 10-d Minkowski 
embedding theorem \cite{mathemb}), are \sl local\normalfont.  This localness includes 
the notion that the theorem is only applicable to a small region of the original 
manifold.  We question whether this need include the approach to a singularity, since 
these are edges of the lower-d spacetime and thus have no neighbourhood.  
\sl Singularities are global features of spacetime\normalfont.  However even some of the 
global embedding theorems (which require very many extra dimensions) are stated for the 
analytic functions \cite{mathemb}.  But low differentiability may be typical in the 
approach to singularities \cite{Clarke}.  It becomes much harder to make any general 
statements once one accepts that spacetime is not analytic! 

One may view at least some (3, 1) singularities as projective effects due to taking a 
badly-behaved foliation (for example a foliation that becomes tangent to the 
characteristics), manifested by the formation of caustics or pure shear blow-ups.  
The question is why there is confusion in reversing this projection procedure to embed 
(3, 1) singularities into nonsingular (4, 1) spacetimes.  If an embedding of the 
($r$, $s$; $\epsilon$) type such as those considered in this section is to be used, 
it is mandatory that the region of hypersurface to be embedded be entirely of one 
signature $s$ (the Cauchy--Kovalevskaya theorem demands a {\sl nowhere-characteristic} 
lower-d manifold).  Also, from the lower-d perspective, the singularity cannot be 
included in the set of points on which data are prescribed because it is an edge of 
spacetime, not a point of spacetime.  Were one to try and include it by providing data 
`right up' to that edge, one would expect that the data would become badly-behaved e.g 
some components of $K_{\Gamma\Delta}$ could be forced to be infinite.  Under 
the various circumstances above, embedding theorems become inapplicable to reconstruct 
the higher-d spacetime.

\mbox{ }

\noindent\Large{\bf 2 (r, 1; 1) methods with thin matter sheets}\normalsize 

\mbox{ }
 
\noindent The \it warpfactor \normalfont split \cite{stdw2} 
\be
g_{CD} = 
\left(
\begin{array}{ll}
\Phi^2(x^{\Pi}, z) & 0 \\ 0 & W(x^{\Pi}, z)f_{\Gamma\Delta}(x^{\Pi})
\end{array}
\right)
\mbox{ } , \mbox{ } \mbox{ } W(x^{\Pi}, 0) = 1
\label{wan}
\ee

\noindent
is a simple subcase of the $z$-dynamics scheme, in which the metric is allowed to 
$z$-evolve only in its scale, away from the $z = 0$ hypersurface where it is taken to be known.   
Then 
\be
\mbox{(\ref{ecd}) leads to } 
\frac{\pa f^{1/2}}{\pa z} = -f^{1/2}\Phi {K} 
\mbox{ which gives the equation } 
\mbox{\hspace{0.7in}}
\frac{\pa lnW}{\pa z} = -\Phi {K}
\label{wftr}
\mbox{\hspace{0.7in}}
\ee
for the warpfactor. For example, using the ansatz ${K}_{\Gamma\Delta} = Cf_{\Gamma\Delta}$ 
in (\ref{wftr}) gives an exponential Randall--Sundrum type warpfactor.  Whereas 
the split (\ref{wan}) does not cover very many cases, our scheme exhibits generalizations for 
it: to permit the whole metric to evolve and to recognize the gauge freedom in 
$\beta_{\Gamma}$, which should ideally be used to separate coordinate effects 
from true physics in the spirit of \cite{CGSYkin}.  

Full, overtly $z$-dynamical schemes are used in particular examples for domain walls 
\cite{Gregory}, braneworld black holes (such as for the pancake or cigar bulk horizon shape 
problem \cite{311lit}) and in models for braneworld stars \cite{starapp, VW}. Whereas in the 
Randall--Sundrum model the higher- and lower-d cosmological constants balance out leaving vacuum 
(in fact Minkowski spacetime) on the brane, more generally a brane would consist of a thin sheet 
of matter -- a junction.  We study such [Shiromizu--Maeda--Sasaki (SMS)-type] braneworlds 
below,\fn{This section is based on \cite{ATlett} and \cite{ATpap}.} starting first however with 
a careful recollection of where the underlying junction conditions come from.   

\mbox{ }

\noindent\large{\bf 2.1 Origin of the junction conditions}\normalsize

\mbox{ }

\noindent Assume that we have a ($D$ -- 1)-d thin matter sheet in a $D$-d bulk with one extra 
spatial dimension ($\epsilon = 1$).  Our discussion follows \cite{MTW} most closely whilst 
keeping the unraised index positions of Israel's work \cite{jns}.

Whereas the requirement of well-defined geometry dictates that the metric is 
continuous across the thin matter sheet yielding the junction condition (j.c)
\be
[f_{\Gamma\Delta}]^+_- \equiv f_{\Gamma\Delta}^+ - f_{\Gamma\Delta}^- = 0 
\mbox{ } ,
\label{jcf}
\ee
discontinuities in certain derivatives of the metric are permissible.
Consider then the 3 projections of the Einstein tensor $\check{G}_{AB}$.  
We use the $\epsilon = 1$ cases of the Codazzi and Gauss constraints (\ref{cod}) 
and (\ref{gauss}) for $\check{G}_{\Gamma \perp}$ and $\check{G}_{\perp\perp}$ respectively.  
For $\check{G}_{\Gamma\Delta}$, the following construction is used.

One begins by writing down the contracted Gauss equation (\ref{contG}) 
and subtracts off  $\frac{1}{2}f_{\Gamma\Delta} $ times the doubly-contracted Gauss equation 
--(\ref{Gpp}): 
\be
\begin{array}{c}
\check{R}_{\Gamma\Delta}                                            \\
- \frac{\check{R}}{2}f_{\Gamma\Delta}
\end{array}
\begin{array}{c}
- \\ + 
\end{array}
\begin{array}{c}
\check{R}_{\perp \Gamma \perp \Delta} = \\
\check{R}_{\perp\perp}f_{\Gamma\Delta} 
\end{array}
\begin{array}{c}
 R_{\Gamma\Delta}                \\
 -\frac{R}{2}f_{\Gamma\Delta}
\end{array}
\begin{array}{c}
- \\ +
\end{array}
\begin{array}{c}
{K} {K}_{\Gamma\Delta} + { {K}_{\Gamma}} ^{\Lambda}{K}_{\Delta\Lambda} \\
\frac{{K}^2 - {K} \circ {K}}{2}f_{\Gamma\Delta} 
\mbox{ } .
\end{array}
\label{bfbwefes}
\ee 
The following steps are then applied.  

\noindent Step 1: The Ricci equation (\ref{thirdproj}) 
is used to remove all the $\check{R}_{\perp\Gamma\perp\Delta}$.  

\noindent Step 2: The contracted Ricci equation (\ref{tpcont}) 
is used to remove all the  $\check{R}_{\perp\perp}$.   Thus   
\be
\left(
\begin{array}{c}
\check{R}_{\Gamma\Delta}                               \\
- \frac{\check{R}}{2}f_{\Gamma\Delta}
\end{array}
\right)
\begin{array}{c}
- \\ +
\end{array}
\begin{array}{c}
\left[
\frac{
\delta_{\check{\beta}} {K}_{\Gamma\Delta} -  D_{\Gamma} D_{\Delta}\alpha
}
{\alpha} + {{K}_{\Gamma}}^{\Pi} {K}_{\Delta\Pi}
\right]                                                         
\\  
\left[
\frac{\delta_{\check{\beta}} {K} -  D^2\alpha}{\alpha} - {K} \circ {K} 
\right] 
f_{\Gamma\Delta}
\end{array}
\mbox{ } \mbox{ } \mbox{ }
 =
\left(
\begin{array}{c}
R_{\Gamma\Delta} \\
-\frac{R}{2}f_{\Gamma\Delta}
\end{array}
\right)
\begin{array}{c}
- \\ +
\end{array}
\begin{array}{c} 
{K} {K}_{\Gamma\Delta} + { {K}_{\Gamma}} ^{\Lambda}{K}_{\Delta\Lambda} 
\\ 
\frac{{K}^2 - {K} \circ {K}}{2}f_{\Gamma\Delta} 
\mbox{ } .  
\end{array}
\label{step0}
\ee
This is then rearranged to form the ``GR CP'' geometrical identity (\ref{gid})
\be
\check{G}_{\Gamma\Delta} = G_{\Gamma\Delta} - {K} {K}_{\Gamma\Delta} + 2{ {K}_{\Gamma}} ^{\Lambda}{K}_{\Delta\Lambda}   
                                                                     + \frac{{K}^2\mbox{+}{K} \circ {K}}{2}f_{\Gamma\Delta}
                                                                     + \frac{\delta_{\check{\beta}} {K}_{\Gamma\Delta} -  D_{\Gamma} D_{\Delta}\alpha 
                                                                     - (\delta_{\check{\beta}} {K} -  D^2\alpha)f_{\Gamma\Delta}} {\alpha}.
\label{GRCPid}
\ee
$$\mbox{Then performing} \mbox{ } \begin{array}{c} \mbox{lim} \\ \epsilon 
\longrightarrow 0\end{array} \int_{-\epsilon}^{+\epsilon}\check{G}_{AB}dz \mbox{ } 
\mbox{one obtains the j.c's} 
\mbox{ } \mbox{ }\mbox{ } \mbox{ }\mbox{ } \mbox{ }\mbox{ } \mbox{ }\mbox{ } \mbox{ }\mbox{ } \mbox{ }\mbox{ } \mbox{ }\mbox{ } \mbox{ }
\mbox{ } \mbox{ }\mbox{ } \mbox{ }\mbox{ } \mbox{ }\mbox{ } \mbox{ }\mbox{ } \mbox{ }\mbox{ } \mbox{ }\mbox{ } \mbox{ }\mbox{ } \mbox{ }
\mbox{ } \mbox{ }\mbox{ } \mbox{ }\mbox{ } \mbox{ }\mbox{ } \mbox{ }\mbox{ } \mbox{ }\mbox{ } \mbox{ }\mbox{ } \mbox{ }\mbox{ } \mbox{ }
\mbox{ } \mbox{ }\mbox{ } \mbox{ }\mbox{ } \mbox{ }\mbox{ } \mbox{ }\mbox{ } \mbox{ }\mbox{ } \mbox{ }\mbox{ } \mbox{ }\mbox{ } \mbox{ }
\mbox{ } \mbox{ }\mbox{ } \mbox{ }\mbox{ } \mbox{ }\mbox{ } \mbox{ }\mbox{ } \mbox{ }\mbox{ } \mbox{ }\mbox{ } \mbox{ }\mbox{ } \mbox{ } $$
\be
[\check{G}_{\perp\perp}]_-^+ = 0 \mbox{ } , \mbox{ } \mbox{ } [\check{G}_{\Gamma\perp}]_-^+ = 0 
\mbox{ } ,
\label{jc0}
\ee
\be
[\check{G}_{\Gamma\Delta}]_-^+ =  [{K}_{\Gamma\Delta} - f_{\Gamma\Delta}{K}]^+_- 
\mbox{ } .   
\label{primjn}
\ee
The derivation of this last equation makes use of normal coordinates (in which case the 
hypersurface derivative $\delta_{\check{\beta}}$ becomes the normal derivative 
$\frac{\pa}{\pa z}$) and the rearrangement to the `Israel' geometrical identity 
\be
\check{G}_{\Gamma\Delta} = G_{\Gamma\Delta} + {K} {K}_{\Gamma\Delta} + 2{ {K}_{\Gamma}} ^{\Lambda}{K}_{\Delta\Lambda}   
                                                                     + \frac{{K}^2 + {K} \circ {K}}{2}f_{\Gamma\Delta}
                                                                     + \frac{\pa}{\pa z}
\left(
{K}_{\Gamma\Delta} -  f_{\Gamma\Delta}{K} 
\right) 
\label{Israel}
\ee
via the definition of extrinsic curvature (\ref{ecd}) to form the complete normal derivative 
$$
\frac{\pa}{\pa z}({K}_{\Gamma\Delta} - {f}_{\Gamma\Delta}{K}) \mbox{ }.
$$

\noindent Step 3: One then further assumes that the (4, 1)-d EFE's hold.  If one then 
$$
\mbox{additionally uses the thin matter sheet energy-momentum }
\check{Y}_{AB} = 
\begin{array}{c}
\mbox{lim} \\ 
\epsilon \longrightarrow 0
\end{array}
 \int_{-\epsilon}^{+\epsilon} \check{T}_{AB} dz,  
$$
\be
\mbox{one obtains the j.c's } 
\mbox{\hspace{1in}}
0 = {Y}_{\perp\perp} \mbox{ } , \mbox{ } \mbox{ } 0 = {Y}_{\Gamma\perp} 
\mbox{ } ,
\mbox{\hspace{2in}}
\label{jc0M}
\ee
\be
[{K}_{\Gamma\Delta}]_-^+ =  
\left(
{Y}_{\Gamma\Delta} - \frac{{Y}}{D - 2}f_{\Gamma\Delta}
\right)
\label{prez2}
\ee
(performing a trace-reversal to obtain the last equation).

\mbox{ }

\noindent\large{\bf 2.2 SMS's braneworld EFE's}\normalsize

\mbox{ }

\noindent We next recall the method SMS use to obtain their BEFE's \cite{SMS}.    
They begin by forming the (3, 1)-d Einstein tensor $ G_{\Gamma\Delta}$ just like we 
obtain (\ref{bfbwefes}) above.
SMS then apply three steps to this equation.

\noindent Step S1: Using the definition of the Weyl tensor, 
$\check{R}_{\perp \Gamma \perp \Delta }$  is replaced by the electric part of the Weyl tensor, 
$\check{E}_{\Gamma\Delta} \equiv \check{W}_{\perp \Gamma \perp \Delta} $ and extra terms built 
from the projections of $\check{R}_{AB}$.  

\noindent Step S2 ( = Step 3 of the above subsection): 
The (4, 1)-d EFE's are then assumed, which permits one to exchange all 
remaining projections of $\check{R}_{AB}$ for (4, 1)-d energy-momentum terms.  
Only when this is carried out does (\ref{bfbwefes}) become a system of field 
equations rather than of geometrical identities.    
We refer to the field equations at this stage as ``timelike hypersurface EFE's'' (THEFE's),\fn{This choice of name reflects their 
superficial resemblance to the (3, 1)-d EFE's, although as discussed below, 
this resemblance is {\sl only} superficial.} as opposed to the 
braneworld EFE's which arise at the next stage.  

\noindent Step S3: A special subcase of THEFE's are braneworld EFE's (BEFE's), which are obtained in normal coordinates by choosing the (thin) braneworld energy-momentum 
tensor ansatz 

\be
\check{T}_{AB} = \check{Y}_{AB}\delta(z) - \Lambda g_{AB} \mbox{ } , \mbox{ } \mbox{ } 
\check{Y}_{AB} \equiv (T_{AB} - \lambda f_{AB} ) \mbox{ } , \mbox{ } \mbox{ } T_{AB}z^A = 0 
\mbox{ } , 
\label{BWEM}
\ee
where $T_{\Gamma\Delta}$ is the energy-momentum of the matter confined to the brane. This is a specialization due to the specific presence of 
(3, 1) and (4, 1) cosmological constants $\lambda$ and $\Lambda$, and by $\Lambda$ being the only bulk contribution.  See B.2.5 
for more about this.  

One then adopts the j.c's (\ref{jcf}), (\ref{jc0}), and (\ref{prez2}) with the additional 
supposition of $Z_2$ symmetry:\fn{The sign difference between (\ref{prez2}) and SMS's paper is due 
to our use of the opposite sign convention in the definition of extrinsic curvature.
We compensate for this in subsequent formulae by also defining $K_{ab} = -K_{ab}^+$ rather than $+K_{ab}^+$.} 
\be
- {K}_{\Gamma\Delta} \equiv {K}_{\Gamma\Delta}^+ = - {K}_{\Gamma\Delta}^- 
\mbox{ } \mbox{ } \Rightarrow \mbox{ } \mbox{ }
{K}_{\Gamma\Delta} = - \frac{G_5}{2}
\left(
\check{Y}_{\Gamma\Delta} - \frac{\check{Y}}{3}f_{\Gamma\Delta}
\right) 
= - \frac{G_5}{2}
\left(
T_{\Gamma\Delta} - \frac{{T - \lambda}}{3}f_{\Gamma\Delta}
\right) 
\mbox{ } , 
\label{41jc}
\ee
where the (4, 1)-d gravitational constant $G_5$ has been made explicit.
Then SMS's BEFE's 
\be
\mbox{read }
\mbox{\hspace{1.8in}}
G_{\Gamma\Delta} = L^{\mbox{\scriptsize SMS\normalsize}}_{\Gamma\Delta} + Q^{\mbox{\scriptsize SMS\normalsize}}_{\Gamma\Delta} - \check{E}_{\Gamma\Delta} 
\mbox{ } , 
\mbox{\hspace{2in}} 
\ee
where $Q_{\Gamma\Delta}(T)$ and $L_{\Gamma\Delta}(T)$ are the terms quadratic in, and linear together with zeroth order in $T_{\Gamma\Delta}$
respectively, given by 
\be
Q^{\mbox{\scriptsize SMS\normalsize}}_{\Gamma\Delta} = G_5^2
\left[ 
\frac{T}{12}T_{\Gamma\Delta} - \frac{1}{4}T_{\Gamma\Pi}{T^{\Pi}}_{\Delta} + 
\left(
\frac{T \circ T}{8} - \frac{T^2}{24}
\right)
f_{\Gamma\Delta}
\right] 
\label{BEFE} 
\mbox{ } ,
\ee
\be
L_{\Gamma\Delta}^{\mbox{\scriptsize SMS\normalsize}} = -\frac{G_5}{2}
\left(
\Lambda + \frac{G_5}{6}\lambda^2
\right)
f_{\Gamma\Delta} + 
\frac{G^2_5}{6}\lambda T_{\Gamma\Delta} 
\mbox{ } .
\ee
As opposed to the (3, 1)-d EFE's, SMS's BEFE's are not closed since they contain 
the unspecified electric part of the Weyl tensor $\check{E}_{\Gamma\Delta}$. 
Although it also contains 15 equations, the SMS BEFE--Gauss--Codazzi system
is not equivalent to the (4, 1)-d EFE's: indeed SMS write down further third-order 
equations for the ``evolution'' away from the timelike brane of 
$\check{E}_{\Gamma\Delta}$, by use of the $z$-derivative of the contracted Gauss 
equation (\ref{contG}), Bianchi identities and the Ricci equation (\ref{thirdproj}).  
This then involves the magnetic part\fn{As the 5-d alternating tensor has 5 indices, 
one has two choices for the number of indices in what is to be taken to be the 
definition of the magnetic Weyl tensor.  The above is the 3-indexed definition; the 
other possible definition is 
$\check{H}^{\Lambda\Delta} = \epsilon^{\Lambda\Delta\Gamma\Sigma\Pi}\check{B}_{\Gamma\Sigma\Pi}$.} 
of the Weyl tensor 
$\check{B}_{\Lambda\Delta\Gamma} \equiv \check{W}_{\perp \Gamma\Delta\Lambda}$, 
the ``evolution" of which follows from further Bianchi identities.  
This full brane-bulk SMS system is then closed. 

\noindent Step 4: In practice, however, instead of the difficult treatment of this 
third-order system, other practitioners have often worked on the SMS BEFE's alone.  
This involves the ad hoc prescription of the functional form of  
$\check{E}_{\Gamma\Delta}$ (sometimes taken to be zero\fn{It is zero for example 
in the references 5 of \cite{ATlett}, as well as in the restricted setting of an AdS bulk.}).  
In fact $\check{E}_{\Gamma\Delta}$ is often first decomposed according to a standard procedure 
\cite{MWBH, Maartensdec}.  Because the original functional form is
unknown, the functional forms of each of the parts defined by the decomposition is 
also unknown.\fn{The unknowns are sometimes kept, for example the `Weyl charge' for black 
holes in \cite{BH}.}  Some of these parts are set equal to zero whereas other parts are 
restricted by unjustified but convenient ans\"{a}tze.  In particular, the anisotropic 
stress part $\check{P}_{\Gamma\Delta}$ is sometimes set to 0 (see e.g the references 7 in 
\cite{ATlett}), or otherwise restricted (see the references 8 in \cite{ATlett}).  The radiative perfect fluid 
part is often kept, but is then argued to be small (despite containing an unknown factor) in 
the circumstances arising in the inflationary \cite{MWBH} and perturbative \cite{Maartensdec} 
treatments.  
Having dealt with
$\check{E}_{\Gamma\Delta}$ in one of the above ways, the form (\ref{BEFE}) of
$Q_{\Gamma\Delta}^{\mbox{\scriptsize SMS\normalsize}}$ is then often taken to
be uniquely defined and the starting-point of many works on brane cosmology
\cite{Langlois1, otherpapers2, otherpapers11, otherpapers12, otherpapers13, Coley11, Coley12,Coley13}.

However, SMS's procedure is far from unique.  It turns out that there are many reformulations of the BEFE arising from geometrical identities.  
Each has a distinct split of the non-Einsteinian BEFE terms into `bulk' and `brane' terms.  Whereas all these formulations are clearly equivalent, their 
use helps clarify how to interpret SMS's braneworld.  Were one to truncate the `bulk' terms in each case (in direct analogy with the usual practice of throwing away 
the Weyl term), then the BEFE's obtained in each case would generally be inequivalent.        

\mbox{ }

\noindent\large{\bf 2.3 Ambiguity in the formulation of the BEFE's}\normalsize

\mbox{ }

\noindent
The Weyl term in SMS's BEFE's has been the subject of much mystery.  Our aim is 
not to argue about what functional form the Weyl term may take (e.g zero 
everywhere) in particular solutions of the system.  Rather we show that different 
formulations of the full brane-bulk system exist which -- although completely 
equivalent -- lead to very different splits of the BEFE's non-Einsteinian terms 
into `brane' and `bulk' terms.  Some of these reformulations have no explicit 
Weyl terms present in their BEFE's.  We first remove some misconceptions as to 
how SMS's procedure leads to a BEFE containing a Weyl term.  Does it have 
anything to do with the modelling of braneworld scenarios?  No, for the Weyl term 
is already in the SMS THEFE before the braneworld energy-momentum ansatz is 
invoked.  Furthermore,  all the procedures used in SMS's method are independent 
of signature and dimension.  Thus this issue of a Weyl term must have already 
arisen long ago in the study of the GR CP.  So why is there no manifest Weyl term 
piece in the GR CP formulation of the EFE's?  The answer is simple, and has 
already been used in I.2.2 and B.2.2: in the ``GR CP'' and Israel procedures, 
\sl one uses the Ricci equation (\ref{thirdproj}) to remove the 
$\check{R}_{\perp \Gamma \perp \Delta}$ term. \normalfont If there is no early use 
of the Ricci equation, one is left with $\check{E}_{\Gamma\Delta}$ in the THEFE's, 
which requires later use of the Ricci equation to ``evolve'' it.  

So there is a choice as to whether one formulates the BEFE's with or without an 
explicit Weyl term.  In the usual treatment of the split of the EFE's (I.2.2), 
one does not use an explicit Weyl term.  In the (4, 1)-d case, this gives a 
well-understood system of 15 p.d.e's in the variables $f_{\Gamma\Delta}$.  
The option of using an explicit Weyl term gives a considerably larger, more 
complicated system of p.d.e's with variables $f_{\Gamma\Delta}$, 
$\check{E}_{\Gamma\Delta}$ and $\check{B}_{\Gamma\Delta\Lambda}$. 

Although this SMS system looks somewhat like the threading formulation (I.2.12.1), 
it does not appear to be tied to the idea of deliberate incompleteness.  The use 
of this idea in the 
threading formulation is clearly tied to signature-specific 
physical reasons which do not carry over to the signature relevant to SMS's 
equations.  We also note that some other third-order reformulations of the 
$(3, 0; -1)$ split of the EFE's are sometimes used to seek to cast the EFE's into 
hyperbolic forms that manifestly have theorems associated with them \cite{RF}.  
Whereas this is precisely the sort of result that is spoiled by considering 
instead a sideways split\fn{Having argued that the second-order sideways 
``GR CP'' is not known to be well-posed, we should add that it is unlikely that 
third-order formulations are likely to be better-behaved in this respect.}, it 
serves to illustrate that what at first seems a `mere reformulation' of a set of 
equations can in fact be used to prove highly nontrivial theorems.  So, whereas 
similar complicated formulations have been used elsewhere in the GR literature, 
SMS's unstated motivation to have a complicated formulation does not coincide 
with the motivation elsewhere in the literature.  Below we bring attention to 
many reformulations of SMS's system, so we ask: what is the motivation for the 
original SMS formulation?  Should the use of some simpler second-order 
formulation be preferred?  Is SMS's formulation or any other third- or 
second-order formulation singled out by good behaviour, either in general or for 
some particular application?   

\mbox{ }

Also, from first principles the SMS procedure to obtain their BEFE is  
quite complicated. For, since they use the j.c obtained by the Israel procedure,  
their procedure actually entails beginning with the whole Israel procedure (Steps 1 to 3 of 
Sec 4.1), and then choosing to reintroduce $\check{R}_{\Gamma\perp\Delta\perp}$ and 
$\check{R}_{\perp\perp}$ by reverse application of Steps 1 and 2.  This is followed by the Weyl 
rearrangement (Step S1), the use of the EFE's (Step S2) and the  substitution of the j.c  
into the extrinsic curvature terms in the braneworld ansatz (Step S3).  
However, despite being complicated, all is well with SMS's scheme since 
any BEFE's obtained by other such combinations of careful procedures will always be equivalent 
because the different steps are related by geometrical identities.  

Step 4 however is not an instance of being careful as it is a truncation.    
Our first point is that whereas in SMS's formulation the non-Einsteinian terms in the BEFE 
might be regarded as a bulk-like $\check{E}_{\Gamma\Delta}$ and a term quadratic in the brane 
energy-momentum, in other formulations the content of these two terms can be mixed up.  
In general, BEFE's contain a group of non-Einsteinian `bulk' terms we denote by $B_{\Gamma\Delta}$ 
(which include both Weyl terms and  
normal derivatives of objects such as the extrinsic curvature), and a group of non-Einsteinian 
`brane' terms that depend on the brane energy-momentum.  
Thus any temptation to discard the Weyl term in the SMS formulation 
(on the grounds that it involves the unknown bulk over which one has no control) 
should be seen in the light that 
if one considered instead a reformulation, 
then there would be a similar temptation to discard the corresponding 
`bulk' term, which would generally lead to something \sl other \normalfont than the Weyl term being discarded.  
Thus for each formulation, the corresponding truncation of the `bulk term' would result in 
inequivalent residual `braneworld physics'.  This is because there are geometrical identities 
that relate `bulk' and `brane' terms, so that the splits mentioned above are highly non-unique 
and thus not true splits at all.  We take this as a clear indication that any such truncations
should be avoided in general.  
Instead, the full system must be studied.

Our second point is that each possible bulk spacetime may contain some hypersurface on  
which a given $B_{\Gamma\Delta}$ vanishes.  Then if one identifies this hypersurface with the 
position of the brane, one has a solution of the full brane-bulk system and not a truncation.  
For example, in any conformally-flat bulk, by definition $\check{W}_{ABCD} = 0$ and therefore 
$B^{\mbox{\scriptsize SMS\normalsize}}_{\Gamma\Delta} = -\check{E}_{\Gamma\Delta} = 0$ 
on all hypersurfaces.  Thus any of these could be identified as 
a brane to form a genuine (rather than truncated) $-\check{E}_{\Gamma\Delta} = 0$ braneworld.  
From this, we can see that \sl the SMS formulation is particularly well adapted for the study of 
conformally-flat bulks such as pure AdS\normalfont.  This motivates SMS's formulation as regards this 
common application.  However, also consider repeating the above procedure   
with some $B_{\Gamma\Delta} \neq - \check{E}_{\Gamma\Delta}$.  This would correspond to a 
genuine (rather than truncated) braneworld model with distinct braneworld physics from that given 
by SMS's particular quadratic term.  
Note that given a model with some $B_{\Gamma\Delta} = 0$,
the BEFE formulation for which $B_{\Gamma\Delta}$ is the bulk term is particularly well adapted for the study of that model.  
Thus different formulations may facilitate the study of braneworlds with different 
braneworld physics.  In the context of conformally-flat spacetimes, it is probably true that the 
$-\check{E}_{\Gamma\Delta} = 0$ braneworlds outnumber the braneworlds for which any other (or 
even all other) $B_{\Gamma\Delta} = 0$ since these other conditions 
are not automatically satisfied on all embedded hypersurfaces.  
Rather, each of these other conditions constitutes a difficult geometrical problem, 
somewhat reminiscent of the question of which spacetimes contain a maximal or CMC slice 
\cite{MT}.  However, generic spacetimes are not conformally-flat.  
For a generic spacetime, we see no difference between the status of the condition 
`$-\check{E}_{\Gamma\Delta} = 0$ on some hypersurface' and the condition `any other particular 
$B_{\Gamma\Delta} = 0$ on some hypersurface'.  Because braneworlds constructed in 
each of these cases have a different residual quadratic term and thus a propensity to have distinct  
braneworld physics, and because we do not know how frequently each of these cases occur,  
we question whether anything inferred from conformally-flat models with the SMS quadratic term 
need be typical of the full SMS brane-bulk system.  
Confirmation of this would require study of the full range of difficult geometrical problems 
$B_{\Gamma\Delta} = 0$, and the construction of concrete examples of 
non-SMS quadratic term braneworld models together with the assessment of whether their 
braneworld physics is conceptually and observationally acceptable.  
  
\mbox{ }

For the moment we study what is the available range of reformulations and thus of 
$B_{\Gamma\Delta}$.  To convince the reader that such reformulations exist,  
we provide a first example before listing all the steps which are available for 
reformulating the BEFE's.      

Assume we do not perform all the steps implicit within SMS's work but rather just the Israel 
steps to obtain the j.c and then use it in the field equation (\ref{Israel}) that gave rise to 
it (as done in \cite{ATlett}), or (as done below) use it in the ``GR CP'' field equation 
following from (\ref{GRCPid}).  
In other words, why not apply the braneworld ansatz to e.g the Israel or ``GR CP'' formulations 
rather than to the SMS formulation?    In the ``GR CP'' case we then obtain
\be
G_{\Gamma\Delta} = L_{\Gamma\Delta} + Q_{\Gamma\Delta} + B_{\Gamma\Delta} \mbox{ , with }
\label{CPBEFE}
\ee 
\be
Q_{\Gamma\Delta} = -\frac{G_5^2}{72}
\left[
36{T_{\Gamma}}^{\Pi}T_{\Pi\Delta} - 18TT_{\Gamma\Delta} + (9T\circ T + T^2)f_{\Gamma\Delta}
\right]
\label{QCP}
\ee
\be
L_{\Gamma\Delta} = \frac{G_5^2}{9}(T \lambda  - 2\lambda^2) f_{\Gamma\Delta}   
+ G_5[T_{\Gamma\Delta} - (\lambda + \Lambda) f_{\Gamma\Delta}]
\label{LCP}
\ee
\be
B_{\Gamma\Delta} = f_{\Gamma\Delta}\frac{\pa K}{\pa z} - \frac{\pa K_{\Gamma\Delta}}{\pa z}.  
\label{BCP}
\ee
This example serves to illustrate that choosing to use a different formulation can cause the 
`brane' quadratic term $Q_{\Gamma\Delta}(T)$ to be different.  Also note that this formulation 
makes no explicit use of the Weyl term.  Thus this BEFE, along with the Gauss and Codazzi 
constraints, forms a small second-order system, in contrast with the much larger third-order 
SMS system.   

\mbox{ }

Now we further study the list of steps \cite{ATlett} which may be applied in the construction 
of BEFE's.  

\noindent Steps S1 and 3 together mean that the Weyl 
`bulk' term $\check{E}_{\Gamma\Delta}$ is equivalent to 
the Riemann `bulk' term together with matter terms.  
This swap by itself involves no terms which are 
quadratic in the extrinsic curvature.

\noindent Step 1 says that the Riemann `bulk' term is equivalent 
to the hypersurface derivative of the extrinsic 
curvature together with a ${K}_{\Gamma\Pi}{{K}^{\Pi}}_{\Delta}$ term.

\noindent Steps S1 and 3 together say that the hypersurface derivative  
of the trace of the extrinsic curvature is equivalent to a matter term together with 
a ${K} \circ {K}$  term.  

Furthermore, one can use both Steps 2-3 and Step 1, on
arbitrary proportions 

\noindent(parameterized by $\mu$ and $\nu$)
of $\check{R}_{\perp \Gamma\perp \Delta}$ and of $\check{R}_{\perp\perp}$:

$$
G_{\Gamma\Delta} = \check{G}_{\Gamma\Delta} 
- (1 + \nu)\check{R}_{\perp \Gamma\perp \Delta} 
+ (1 - \mu)\check{R}_{\perp\perp}f_{\Gamma\Delta} 
\mbox{\hspace{1in}}
$$
$$
+ \frac{1}{\alpha}
\left[
\nu(\delta_{\check{\beta}} {K}_{\Gamma\Delta} -  D_{\Gamma} D_{\Delta}\alpha)  
+ \mu(\delta_{\check{\beta}}  {K} -  D^2\alpha)f_{\Gamma\Delta}
\right] 
$$
\be
\mbox{\hspace{1.2in}} 
+ {K} {K}_{\Gamma\Delta} + (\nu - 1) {{K}_{\Gamma}}^{\Pi} {K}_{\Delta\Pi} 
- \frac{{K}^2}{2}f_{\Gamma\Delta}  +  \left(\frac{1}{2} - \mu\right) {K} \circ {K}f_{\Gamma\Delta} 
\mbox{ } .
\label{2param}
\ee
This introduces freedom in the coefficients of the
${K}_{\Gamma\Pi}{{K}^{\Pi}}_{\Delta}$ and ${K} \circ {K}$ contributions to 
the quadratic term $Q_{\Gamma\Delta}({K})$ in the THEFE's.  
We next find further freedom in $Q_{\Gamma\Delta}$ by choice of the objects to be regarded as primary.  

We are free to choose a `bulk' term described by hypersurface derivatives 
$\delta_{\check{\beta}} $ (which are partial derivatives $\frac{\pa}{\pa z}$ 
in normal coordinates) of objects related to the extrinsic curvature ${K}_{\Gamma\Delta}$ 
by use of the metric tensor (including its inverse and determinant $f$).  
The underlying reason for doing this is that it is just as natural to treat such an object, 
rather than the extrinsic curvature itself, as primary (see below for examples).  

Upon careful consideration, there are three separate ways such objects can be related 
to the extrinsic curvature: raising indices, removing a portion of the trace by defining 
$_{\eta}{K}_{\Gamma\Delta} \equiv {K}_{\Gamma\Delta} - \eta{K}f_{\Gamma\Delta}$, 
and densitizing by defining $^{\xi}{K}_{\Gamma\Delta} \equiv f^{\xi} \times {K}_{\Gamma\Delta}$.  
The hypersurface derivatives of these objects are related to those of the extrinsic curvature by
\be
\delta_{\check{\beta}} {K}_{\Gamma\Delta} = \delta_{\check{\beta}}  
(f_{\Gamma\Pi}{K^{\Pi}}_{\Delta}) = f_{\Gamma\Pi}\delta_{\check{\beta}}  
{{K}^{\Pi}}_{\Delta} - 2\alpha {K}_{\Gamma\Pi}{{K}^{\Pi}}_{\Delta} 
\mbox{ } ,
\label{i}
\ee
\be
\delta_{\check{\beta}}{K}_{\Gamma\Delta} = 
\delta_{\check{\beta}}{K}^{\mbox{\scriptsize $\eta$\normalsize}}_{\Gamma\Delta} + \eta
\left(\delta_{\check{\beta}}{K}f_{\Gamma\Delta} - 2\alpha{K}{K}_{\Gamma\Delta}
\right) 
\mbox{ } ,
\label{64}
\ee    
\be
\delta_{\check{\beta}}{K}_{\Gamma\Delta} = (\mbox{}f)^{-\xi_1}\delta_{\check{\beta}}(\mbox{}f^{\xi_1}{K}_{\Gamma\Delta}) + 2\alpha\xi_1{K}{K}_{\Gamma\Delta} 
\mbox{ } .
\label{65}
\ee  
Further useful equations arise from the traces of these:  
\be 
\delta_{\check{\beta}} {K} = f^{\Gamma\Delta} \delta_{\check{\beta}}{K}_{\Gamma\Delta} + 2\alpha {K} \circ {K} 
\mbox{ } ,
\label{63}
\ee
\be
\delta_{\check{\beta}}{K} = \frac{1}{1 - 4\eta}[f^{\Gamma\Delta}\delta_{\check{\beta}}\mbox{}_\eta{K}_{\Gamma\Delta} 
- 2\alpha\eta({K}^2 - {K}\circ{K})] 
\mbox{ } , \mbox{ } \mbox{ } \eta \neq \frac{1}{4} 
\mbox{ } , 
\label{new}
\ee
\be
\delta_{\check{\beta}}{K} = (\mbox{}f)^{-\xi_2}\delta_{\check{\beta}}(\mbox{}f^{\xi_2}{K}) + 2\alpha\xi_2{K}^2 
\mbox{ } , 
\label{66}
\ee
where the $\delta_{\check{\beta}}{K}$ in (\ref{new}) and (\ref{66}) has been 
obtained via (\ref{63}).  

The following examples of $_{\eta}^{\xi}{K}_{\Gamma\Delta}$ illustrate that the use of 
such objects is entirely natural:  $_{{1/4}}^{0}{K}_{\Gamma\Delta}$ 
is the ${K}^{\mbox{\scriptsize T\normalsize}}_{\Gamma\Delta}$ commonly used in the IVP literature,  
and $_{0}^{{1/2}}{K}_{\Gamma\Delta}$ appears in the guise of forming the complete 
normal derivative in the Israel procedure.  
Also, the ``gravitational momenta'' are $p_{\Gamma\Delta} \equiv  - _{1}^{{1/2}}{K}_{\Gamma\Delta}$.

The above thorough consideration of possible `bulk' terms permits all four THEFE 
terms homogeneously quadratic in the extrinsic curvature to be changed 
independently.  One may think that we have a redundancy in providing 8 ways to 
change only 4 coefficients.  However, one can afford then to lose some of the 
freedoms by making extra demands, of which we now provide four relevant examples.  
First, one could further demand that there is no Weyl term in the THEFE's (as 
discussed in B.2.4).  Second, unequal densitization of ${K}_{\Gamma\Delta}$ 
and ${K}$ (i.e $\xi_1 \neq \xi_2$) corresponds to interpreting the fundamental 
variable to be some densitization of the metric rather than the metric itself.  
Whereas this is again a common practice (for example the scale-free metric of the 
IVP literature is $f^{\mbox{\scriptsize unit\normalsize}} \equiv 
{   f^{  -\frac{1}{n} }   }{   f_{\Gamma\Delta}   }$ in 
dimension $n$), the use of such an object as fundamental variable does appear to 
complicate the isolation of the (3, 1) Einstein tensor truly corresponding to 
this fundamental variable.  Thus this option is not pursued in this thesis.  Third, 
one may start by declaring that one is to use particular well-known primary 
objects (such as the ``gravitational momenta'') and still desire to be left with 
much freedom of formulation.   Fourth, one could declare that one is to use the 
raised objects given by (\ref{63}), in which case the further ability to change 
coefficients by use of (\ref{64}) is lost, since moving a Kronecker delta rather 
than a metric through the derivative clearly generates no terms quadratic in the 
extrinsic curvature.  

\mbox{ }

\noindent\large{\bf 2.4 Examples of formulations of BEFE's with no quadratic terms}\normalsize

\mbox{ }

\noindent
As a consequence of the above freedoms, there are many formulations in which all four
coefficients vanish, and hence $Q_{\Gamma\Delta}({K}) = 0$.       
From this it follows that $Q_{\Gamma\Delta}(T)$ is zero [and it is easy to show that all 
instances of $Q_{\Gamma\Delta}(T) = 0$ follow from $Q_{\Gamma\Delta}({K}) = 0$].  
Thus it suffices to seek for cases of THEFE's with $Q_{\Gamma\Delta}({K}) = 0$ 
to obtain all cases of BEFE formulations that have no quadratic term.  
We now motivate these formulations and then choose to exhibit three that comply with 
some of the extra demands in the previous paragraph.

The diversity of `brane'-`bulk' splits ensures that truncations such as Step 4 
produce all possible combinations of quadratic terms as residual `braneworld physics'.  
Alternatively, one may suspect that there might be solutions to the full brane-bulk system 
that just happen to have a particular $B_{\Gamma\Delta} = 0$ on some hypersurface 
which is then identified as a brane.   We speculate that each of these situations will often 
lead to different answers to questions of physical interest.  Whereas most FLRW perfect 
fluid models with equation of state $P = (\gamma - 1)\rho$ arising thus will be similar, 
differences will be more salient in models with more complicated equations of state, 
in perturbations about FLRW  (as started in \cite{Maartensdec, SLMW, MB} in the SMS-adapted 
case) and in anisotropic models (as started by \cite{Coley11, Coley12, Coley13, Coleyconj} 
in the SMS-adapted case).  These in turn constitute natural frameworks to seriously justify 
the late-time emergence of FLRW behaviour and the likelihood of inflation 
\cite{otherpapers11, otherpapers12, otherpapers13} as well as the study of singularities 
\cite{Coleyconj, MB} on the brane.  We emphasize that, 
for a satisfactory study of whether any particular full brane-bulk (as opposed to truncated) case 
leads to any differences from the hitherto-studied $-\check{E}_{\Gamma\Delta} = 0$ case, 
one would require a full brane-bulk solution explicitly constructed to satisfy some $B_{\Gamma\Delta} = 0$ 
on some hypersurface within a particular given bulk spacetime.  
Since we currently have no such example, our arguments currently only 
support the far simpler idea that truncation should be avoided.    

We illustrate that in different formulations, the truncation of the corresponding `bulk' terms 
can lead to 
big differences in the residual `braneworld physics', without any of the above 
lengthy calculations. We do this by formulating the `bulk' part so that there is no 
corresponding $Q$ term at all.  Thus these truncations give  the `$\rho$' of standard FLRW 
cosmology rather than the `$\rho$ and $\rho^2$' of brane cosmology  
\cite{Langlois1, MWBH, Maartensdec}.  As a result whether we have a `$\rho$ and $\rho^2$' 
brane cosmology depends on the choice of formulation.    So we argue that since the SMS 
procedure followed by truncating the Weyl term is a hitherto unaccounted-for choice out of 
many possible procedures, then adopting the particular 
homogeneous quadratic term of SMS (often taken as the 
starting-point of brane cosmology) appears to be unjustified.  Rather, we conclude that no 
particular truncation should be privileged as the act of truncation imprints undesirable 
arbitrariness into the study of the truncated system.  Whereas the (3, 1)-d trace 
of $\check{E}_{\Gamma\Delta}$ is zero and thus might\fn{This is if one treats the geometric 
content of the Weyl tensor as an effective or `induced' energy--momentum.  People often also 
assume that this ought to behave like a perfect fluid, although  Maartens' decomposition of 
$\check{E}_{\Gamma\Delta}$ \cite{Maartensdec}, permits a more general treatment.} 
phenomenologically look like pure radiation fluid to observers on the brane, other bulk 
characterizations would typically not look like a pure radiative fluid.  This may open up 
phenomenological possibilities.  

Also, before further study of SMS's full (untruncated, third-order)
system is undertaken, some of the reformulations along the lines
suggested in B.2.3 might turn out to be more tractable.
In particular those reformulations which fully eliminate
$\check{R}_{\perp \Gamma\perp \Delta}$ by the early use of the Ricci equation are already
closed as second-order systems. These include the Israel formulation in \cite{ATlett},
the ``GR CP'' formulation (\ref{CPBEFE}--\ref{BCP}), and our second and
third examples below, which contain neither a Weyl term nor a quadratic term.

For our first example, we take as the primary object the antidensitized extrinsic curvature
$\underline{K}_{ab} \equiv \frac{   K_{ab}   }{   \sqrt{ h  }   }$
so that the `bulk' term is (partly) a combination of this object's normal derivatives. 
The corresponding BEFE's are: 
\be
G_{\Gamma\Delta} = {L}_{\Gamma\Delta} + B_{\Gamma\Delta}
\mbox{ } , 
\ee
\be
L_{\Gamma\Delta} = \frac{\check{T}_{\Gamma\Delta}}{3} + \frac{1}{6}\left(5\check{T}_{\perp\perp} - 
\check{T}\right)f_{\Gamma\Delta} \mbox{ } , \mbox{ } \mbox{ }
B_{\Gamma\Delta} = -2\check{E}_{\Gamma\Delta} + \sqrt{\mbox{}f}
\left(
\frac{\pa {\underline{K}_{\Gamma\Delta}}}{\pa z}
- \frac{1}{2}f^{\Gamma\Delta}
\frac{\pa \underline{K}_{\Gamma\Delta}}{\pa z} f_{\Gamma\Delta}
\right)
\ee
(where we have chosen to remove all projections of $\check{R}_{AB}$ by the EFE's).

To derive this, take (\ref{2param}) in normal coordinates.  
Choose to convert all of the 
$$
\frac{\pa {K}}{\pa z} \mbox{ } \mbox{ into } \mbox{ } 
f^{\Lambda\Sigma}\frac{\pa {K}_{\Lambda\Sigma}}{\pa z} \mbox{ } \mbox{ by (\ref{63}): }
\mbox{\hspace{4in}}
$$
$$
G_{\Gamma\Delta} = \check{G}_{\Gamma\Delta} - (1 + \nu)\check{R}_{\perp \Gamma\perp \Delta} 
+ (1 - \mu)\check{R}_{\perp\perp}f_{\Gamma\Delta} 
+ \left(
\nu \frac{\pa {K}_{\Gamma\Delta}}{\pa z}  
+ \mu\frac{\pa {K}}{\pa z}f_{\Gamma\Delta}
\right) 
$$
\be
+ 
{K} {K}_{\Gamma\Delta} + (\nu - 1) {{K}_{\Gamma}}^{\Pi} {K}_{\Delta\Pi} 
- \frac{{K}^2}{2}f_{\Gamma\Delta}  +  (\frac{1}{2} + \mu) {K} \circ {K}f_{\Gamma\Delta} 
\mbox{ } .   
\ee
Now choosing the primary object to be some densitized $^{\xi}K_{\Gamma\Delta}$ by (\ref{65}) we have 
$$
G_{\Gamma\Delta} =   \check{G}_{\Gamma\Delta} - (1 + \nu)\check{R}_{\perp \Gamma\perp \Delta} 
+ (1 - \mu)\check{R}_{\perp\perp} f_{\Gamma\Delta}
+\frac{1}{f^{\xi}}
\left(
\nu \frac{\pa \mbox{}^{\xi}K_{\Gamma\Delta})}{\pa z}  
+ \mu f^{\Lambda\Sigma}\frac{\pa \mbox{}^{\xi}K_{\Lambda\Sigma})}{\pa z}  
\right) 
$$
\be
+ 
(1 + 2\nu\xi){K}{K}_{\Gamma\Delta} + (\nu - 1) {{K}_{\Gamma}}^{\Pi} {K}_{\Delta\Pi} 
+\left(2\mu\xi - \frac{1}{2}\right){K}^2f_{\Gamma\Delta}  
+  \left(\frac{1}{2} + \mu\right) {K} \circ {K}f_{\Gamma\Delta} 
\mbox{ } ,
\ee
so clearly $\mu = -\frac{1}{2}$, $\nu = 1$ and the antidensitization choice of 
weighting $\xi = -\frac{1}{2}$ ensure that $Q_{\Gamma\Delta}({K}) = 0$   

The following examples arose from asking if it is possible to find examples in which neither 
$Q_{\Gamma\Delta}(T)$ nor $\check{E}_{\Gamma\Delta}$ feature.  We found the following BEFE's: 
\be
G_{\Gamma\Delta} = L_{\Gamma\Delta} + B_{\Gamma\Delta} 
\mbox{ } ,
\ee
\be
L_{\Gamma\Delta} = \check{T}_{\Gamma\Delta} + \frac{1}{2}
\left(
\check{T}_{\perp\perp}- \frac{\check{T}}{3}
\right)f_{\Gamma\Delta} 
\mbox{ } , \mbox{ } \mbox{ }
B_{\Gamma\Delta} = \frac{1}{\sqrt f}
\left( 
\frac{1}{2}\frac{\pa \overline{K}}{\pa z} f_{\Gamma\Delta} - 
\frac{\pa {\overline {K}}^{\Pi}\mbox{}_{\Gamma}}{\pa z} f_{\Pi\Delta} 
\right)
\ee
by considering as our primary object the densitized extrinsic curvature with one index raised, 
${\overline {K}}^a\mbox{}_b \equiv \sqrt{\mbox{}h}{{K}^a}_b$.  

We obtained these BEFE's by arguing as follows.  In order for the BEFE's to contain no Weyl 
term, $\nu$ is fixed 

\noindent to be $-1$.  Then the only control over 
${K}_{\Gamma\Pi}{{K}^{\Pi}}_{\Delta}$ is from raising by (\ref{i}).  
It is easy to show that this raising must be applied to the whole 
$\frac{\pa{K}_{\Gamma\Delta}}{\pa z}$ in order for the coefficient of 
${K}_{\Gamma\Pi}{{K}^{\Pi}}_{\Delta}$ to be zero.  
Then using $_{\eta}{{K}^{\Lambda}}_{\Delta}$ does not change any terms quadratic in the 
extrinsic curvature.  Also, use of distinct densities for ${K}$ and 
${{K}^{\Lambda}}_{\Delta}$ does not appear to make sense since both quantities are related 
to ${K}_{\Gamma\Delta}$ by a single use of the inverse metric.  
Although all these restrictions make the outcome unlikely, the use of (\ref{65}) and (\ref{66}) 
alone suffices to obtain the above example:
$$
G_{\Gamma\Delta} = \check{G}_{\Gamma\Delta} + (1 - \mu)\check{R}_{\perp\perp}f_{\Gamma\Delta} + f^{-\xi}
\left(
- \frac{\pa\mbox{}^{\xi}{K}_{\Gamma\Delta}}{\pa z}  
+ \mu\frac{\pa {K}}{\pa z}f_{\Gamma\Delta}
\right) 
$$
\be
+  (1 - 2\xi)                              {K} {K}_{\Gamma\Delta} 
+ (2\xi\mu - \frac{1}{2})                {K}^2f_{\Gamma\Delta}  
+ (\frac{1}{2} - \mu)                    {K} \circ {K}f_{\Gamma\Delta} 
\mbox{ } ,
\ee
which has no quadratic terms if $\mu = 1/2$ and $\xi = 1/2$ (`densitization' weight).

Another possibility is  to replace $^{\xi}{{K}^{\Lambda}}_{\Delta}$ by 
$_{\eta}^{\xi}{{K}^{\Lambda}}_{\Delta}$.  Although this does not immediately 
do anything about the quadratic terms, if we also convert a portion 
$\pi$ of $\frac{\pa\mbox{}^{\xi}{K}}{\pa z}$ into 
$\frac{\pa\mbox{}_{\eta}^{\xi}{K}_{\Lambda\Sigma}}{\pa z}f^{\Lambda\Sigma}$ we obtain 
$$
G_{\Gamma\Delta} = \check{G}_{\Gamma\Delta} + (1 - \mu)\check{R}_{\perp\perp}f_{\Gamma\Delta} 
\mbox{\hspace{3.7in}}
$$
$$
+ f^{-\xi}
\left(
- \frac{\pa\mbox{}^{\xi}_{\eta}{{K}^{\Lambda}}_{\Delta}}{\pa z}  
+ (\mu - \eta)  
\left[  
\frac{\pi}{1 - 4\eta}  \frac{\pa \mbox{}^{\xi}_{\eta}{K}_{\Lambda\Sigma}}{\pa z}f^{\Lambda\Sigma} 
+ (1 - \pi)\frac{\pa\mbox{}^{\xi}{K}}{\pa z} 
\right] 
f_{\Gamma\Delta}
\right) 
\mbox{\hspace{0.5in}}
$$
\be
\mbox{\hspace{0.4in}}
+ (1 -2\xi)                                                                                     {K} {K}_{\Gamma\Delta}
+ \left[2\xi\mu - \frac{1}{2} - \frac{2\pi\eta(\mu - \eta)}{1 - 4\eta}\right]                   {K}^2f_{\Gamma\Delta} 
+  \left[\frac{1}{2} - \mu + \frac{2\pi\eta(\mu - \eta)}{1 - 4\eta}\right]                      {K} \circ {K}f_{\Gamma\Delta} 
\mbox{ } , 
\ee
which requires $\xi = \frac{1}{2}$, whereupon the two remaining equations become 
identical: 
\be
\mu - \frac{1}{2} = \frac{2\pi\eta(\mu - \eta)}{1 - 4\eta} 
\mbox{ } ,
\ee
which clearly has many solutions.  A particularly neat one is to take $\eta = 1$ so that the primary 
objects are `gravitational momenta' and $\pi = 1$ so that only two normal derivative terms 
appear in the `bulk' term.  Then $\mu = \frac{7}{10}$ so the BEFE's read 
\be
G_{\Gamma\Delta} = L_{\Gamma\Delta} + B_{\Gamma\Delta} 
\mbox{ } , 
\ee
\be
L_{\Gamma\Delta} = \check{T}_{\Gamma\Delta} + \frac{3\check{T}_{\perp\perp} - \check{T}}{10} f_{\Gamma\Delta} 
\mbox{ } , \mbox{ } \mbox{ }
B_{\Gamma\Delta} = \frac{1}{\sqrt f}
\left( 
\frac{\pa {p^{\Pi}}_{\Gamma}  }{\pa z} f_{\Pi\Delta} 
- \frac{1}{10}\frac{\pa p_{\Lambda\Sigma}}{\pa z}f^{\Lambda\Sigma} f_{\Gamma\Delta} 
\right).
\ee

Of course, it would make sense to particularly investigate the difficult geometrical problem 
`$B_{\Gamma\Delta} = 0$ on some embedded hypersurface' for such $B_{\Gamma\Delta} = 0$ 
corresponding to no quadratic terms, since by the same arguments as above, such a model 
would be sure to give braneworld physics distinguishable from that hitherto studied.

\mbox{ }

\mbox{ }

\noindent\large{\bf 2.5 Two further comments about building SMS-type braneworlds}\normalsize

\mbox{ }

\noindent 
First, so far we have talked in terms of the $\check{Y}_{\Gamma\Delta} = 
T_{\Gamma\Delta} - \lambda f_{\Gamma\Delta}$ split of the matter contribution 
to relate our work as clearly as possible to its predecessors in the literature.
However, from the outset \cite{SMS} it was pointed out that this split is not unique.  
On these grounds we would prefer to work with the unique trace-tracefree 
split in which all the $\lambda f_{\Gamma\Delta}$ contributes to the trace part.  
The (4, 0) version of this split is used in B.3.  

Second, given a fixed type of bulk energy-momentum such as the $\check{T}_{AB} = 0$ 
of `noncompact KK theory' or the $\check{T}_{AB} = \Lambda g_{AB}$, then 
establishing an embedding requires the existence of a suitable compensatory characterization 
of the bulk geometry.  The GR line of thought would be to only take results within 
such schemes seriously if they are robust to the addition of bulk matter fields. 
Of course privileged choices of bulk could arise from further theoretical
input.   We argue below that the theoretical arguments behind some privileged choices 
in the literature for the bulk energy-momentum are not convincing enough to anchor 
strongly credible physical predictions.  One would rather require rigorous and general 
theoretical input following directly from some fundamental theory such as string theory.  

We now argue about the $\check{T}_{AB} = \Lambda g_{AB}$ choice of bulk 
(an example of which is pure AdS).  This is clearly also always possible given the premises of a 
particular case of the generalized Campbell--Magaard result, but we have argued that this is not 
significant. We rather discuss the argument for pure AdS bulks from string theory. 
This is not generally justifiable since firstly, bulk gravitons are permitted so the bulk geometry 
would generically contain gravity waves. Secondly, bulk scalars ought to be permitted since they 
occur along with the graviton in the closed string spectrum \cite{Polchinski}.  The content of 
the closed string spectrum thus places interesting restrictions on bulk matter rather than completely 
abolishing it.  From the perspective of 5--d GR, evidence for the stability of vacuum or AdS bulks 
(and of any resulting physical predictions) to the introduction of suitable bulk fields would 
constitute important necessary support for such models and their predictions.  

Finally note that use of arbitrary smooth bulk $\check{T}_{AB}$ does not affect the form of the junction conditions since
only the thin matter sheet contribution to 
$\check{T}_{AB}$ enters these.  The conclusion of our second point is that there is no good reason not to explore at least certain kinds of bulk matter in order to have 
a more general feel for how these thin matter sheet models behave \cite{scalarbulk}.   

\mbox{ }

\noindent\large{\bf 2.6 Sideways York and thin-sandwich schemes with thin matter sheets}\normalsize

\mbox{ }

\noindent 
Although (3, 1; 1) methods which build 
higher-d bulks about the privileged (3, 1) worlds by ``z-dynamics'' look tempting at 
the simplest level, the more advanced issues from B.1 (of causality, and of 
well-posedness not being known) hold regardless of the presence of thin matter sheets.  
The issue of rough function spaces will become particularly relevant in the study of 
sufficiently general situations involving the evolution of thin matter sheets (see B.3.1).  

Beyond the objections in B.1.3, the Campbell--Magaard scheme is furthermore of limited 
use in models with thin matter sheets because the junction condition imposes restrictions 
on ${K}_{\Gamma\Delta}$ which prevents these being subdivided into the knowns and 
unknowns of Magaard's method.   

We next consider the (3, 1) version of the York method applied on the thin matter sheet, 
but recall that even if this does provide data sets, one is next confronted with the 
difficulties of the (3, 1; 1) ``evolution" scheme.   To date (4, 1) worlds built from 
(3, 1) ones have relied on very simple specialized ans\"{a}tze, such as 

\noindent A) $z$-symmetric surfaces ${K}_{\Gamma\Delta} = 0$ with known metric 
$f_{\Gamma\Delta}$, whereupon the vacuum Codazzi equation is automatically satisfied and 
then $R = 0$ is required from the Gauss constraint.   

\noindent B) ${K}_{\Gamma\Delta} = Cf_{\Gamma\Delta}$ with known $f_{\Gamma\Delta}$, 
for example to obtain the Randall--Sundrum bulk \cite{stdw2} or slightly more general 
solutions \cite{AL}.  Now the maximal subcase ${K} = 0$ of the CMC condition is a 
generalization of A), whereas the full CMC condition itself is a generalization of B).  
Moreover, now the metric is to be treated as only known up to scale.  Thus one would 
generally only know the full metric of each model's (3, 1) world once the `wave 
Lichnerowicz equation' for the embedding 
of this world into the (4, 1) world is solved.  So one loses the hold from the outset 
on whether each model will turn out to contain an interesting (3, 1) world.  
Nevertheless, some of the (3, 1) worlds will turn out to be of interest.  Furthermore, one should 
question the sensibleness of any ideas involving the prescription of full 
(3, 1) metrics if the most general technique available fails to respect such a prescription.  This 
point is more significant for 
(3, 1) data than for (3, 0) data because conformally-related metrics 
have different non-null geodesics.  For (3, 0) data no physical significance 
is attached to spatial geodesics, but for (3, 1) there are timelike geodesics which are 
physically interpreted as paths of free motion of massive particles.  So a 
$(\tilde{M}, \tilde{h}_{ab})$ spacetime which is conformally related to $(M, h_{ab})$ 
is different physically (for example one could violate energy conditions the other one does 
not violate). One can get out of this difficulty by either attaching no physical significance 
to one's inspired guesses for $(M, h_{ab})$  or by hoping for unobservably tiny 
deviations between the geodesic curves of the two geometries.  

In the specific case of thin matter sheets, by the j.c's, A) implies that 
$\check{Y}_{\Gamma\Delta} = 0$, whilst  B) implies that $\check{Y}$ is a hypersurface 
constant. The maximal condition implies that $\check{Y} = 0$ 
whilst the CMC one implies that $\check{Y}$ is a hypersurface constant.  
So whereas the maximal and CMC ans$\ddot{\mbox{a}}$tze are more general than 
A) and B) respectively, they are nevertheless restricted in this braneworld application.  
Notice, however, that a number of interesting cases are 
included: vacuum, radiative matter and electromagnetic matter 
are all among the $\check{Y} = 0$ spacetimes.

It is important to note that unlike the usual GR application, the choice of a 
hypersurface to be a brane is not a choice of slicing because localized 
energy--momentum is to be pinned on 
it. Almost all reslicings would fail to isolate this energy-momentum 
on a single slice.  We know of no good reason why the brane should be CMC 
nor what value the CMC should take on it. However, at least this is a well-defined 
notion and it is substantially simpler to solve for than in general because of the decoupling 
of the Gauss and Codazzi constraints.  

As regards possible use of either thin or thick forms of the thin sandwich conjecture
to treat branes, we first distinguish between thin sandwiches between 2 nearby branes
and thin sandwiches which have a brane on one side and an undistinguished hypersurface on the other.
One should be aware that non-intersection requirements may be different in these two cases,
and also different from that of the original thin sandwich setting of 2 unprivileged spacelike
hypersurfaces.
Second, one would have to take $\check{Y}_{\Gamma\Delta}$ as unknown until
it can be deduced from the $\check{K}_{\Gamma\Delta}$ evaluated from the thin sandwich procedure.
Finally we caution that thin sandwich schemes need not always exist.
They require the ``lapse'' to be algebraically-eliminable,
which for example is not the case for the analogous $\Phi$ of the KK split.

\mbox{ }

\noindent\large{\bf 2.7 (n, 1; 1) singularity removal and thin matter sheets}\normalsize

\mbox{ }

\noindent 
First, there is less scope for it occurring along the lines of (\ref{singrem}).  
For, as it is the energy-momentum on the brane, $\check{Y}_{\Gamma\Delta}$ is presumably 
finite for a nonsingular (4, 1) world so ${K}_{\Gamma\Delta}$ is finite and so 
cannot cause blowups in 
${K}^{\mbox{\scriptsize T\normalsize}}\circ{K}^{\mbox{\scriptsize T\normalsize}} - \frac{3}{4}{K}^2$.
However, blowups in this quantity 
can still occur if the inverse metric is badly-behaved (corresponding to $f$ = 0).
Note however that the application of the Cauchy--Kovalevskaya theorem to GR requires this 
not to be the case everywhere within the region of applicability.

Second, one would of course require the investigation of more elaborate 
(e.g `warped') embeddings than the example discussed above in order to 
investigate whether some embeddings lead to 5-d geodesics that exit the brane.
If the (3, 1)-d geodesics on the  brane are not included 
among those of the (4, 1)-d bulk, it is not clear at all what is meant 
by `singular' since one then would have to simultaneously consider some notion of 
extendibility for two congruences of privileged curves of different dimensionality.

Under the decomposition (\ref{ADMs}), the geodesic equation (\ref{afgeoeq}) 
becomes 
\be
\ddot{x}^{\Gamma} + \check{\Gamma^{\Gamma}}_{\Delta\Sigma}\dot{x}^{\Delta}\dot{x}^{\Sigma} 
- 2{{K}^{\Gamma}}_{\Delta}\dot{x}^{\Delta}\dot{x}^{\perp} = 0 \mbox{ } ,  
\label{geo1}
\ee
\be
\ddot{x}^{\perp} + {K}_{\Delta\Sigma}\dot{x}^{\Delta}\dot{x}^{\Sigma} = 0 \mbox{ } .
\label{geo2}
\ee
if one upholds the use of normal coordinates.  In order for (\ref{geo1}) to reduce 
in all cases to the (3, 1) geodesic equation on the brane, one requires 
$\dot{x}^{\perp} = 0$ or ${{K}^{\Gamma}}_{\Delta} = 0$ everywhere 
on the brane. By (\ref{geo2}), maintenance of $\dot{x}^{\perp} = 0$ 
along all geodesics is impossible unless ${K}_{\Delta\Sigma} = 0$.  
Thus in all cases ${K}_{\Delta\Sigma} = 0$ is required on the brane.  
But this means that ${{K}_{\Delta\Sigma}}^+ = {{K}_{\Delta\Sigma}}^-$ 
or equivalently $\check{Y}_{\Delta\Sigma} = 0$ i.e the absence of a brane.  
Thus in SMS-type braneworlds {\sl the}

\noindent{\sl (3, 1) geodesics need not be among the (4, 1) 
geodesics.}    

One is trying to model confined matter.  But the presence of brane-confined fields then means 
that brane-field particles fall off the brane if these follow (4, 1) geodesics, which cannot 
be a healthy state of affairs.   If one rather postulates that these follow (3, 1) geodesics, 
then 

\noindent A) there is {\bf POE} violation going on: bulk-traversable and brane-confined matter 
test-bodies undergo distinct free fall, which might have observable consequences.   

\noindent B) the on-brane causal structures due to brane-confined photons and due to 
bulk-traversable gravitons would generally be distinct.  Thus the new, serious conceptual 
difficulties in B.1.6.3 about the definition of singularities are indeed relevant to brane cosmology.  

\mbox{ }

\noindent\Large{\bf 3 (n, 0; --1) methods with thin matter sheets}\normalsize

\mbox{ }

\noindent\large{\bf 3.1 Hierarchy of problems and their difficulties}\normalsize

\mbox{ }

\noindent We are here concerned with the study of ($n$ + 1)-d GR models with thin matter sheets 
using only well--studied, well--behaved mathematical techniques which also make good physical 
sense and are applicable to the full EFE's.  In our view a good way of achieving this is via the 
($n$, 0) IVP, followed where possible by the heavily-protected ($n$, 0; --1) CP.  

We provide a hierarchy of modelling assumptions together with associated difficulties, and argue 
that evolution is more problematic than data construction.  We concentrate on the latter, making 
use of attractive features of the usual-signature York method in the new thin mater sheet setting. 
The differences between this setting and that of numerical relativity (c.f I.2.9--11, App C) lead 
to the following difficulties.   Firstly, certain thin matter sheet models require particularly 
rough mathematics.  Secondly, they require novel asymptotics and boundary conditions (b.c's).       

Note that the increase in dimensionality itself does not affect much the data construction nor 
the study of well-posedness and stability of the GR CP, although there is some 
dimensional dependence when the rougher function spaces come into use, due to the contrast 
between the Hughes--Kato--Marsden theorem's requirement that the Sobolev class of the metric 
to be no rougher than $H^{n + 1}$ 
(p 37) and that in    
all dimensions, the presence of thin matter sheets means that the metric is rougher 
than $C^2$ (the functions with continuous second derivatives) because of the jump in the Riemann 
tensor at the sheet.  This corresponds to the metric being too rough to belong to the Sobolev class $H^4$.  
Thus HKM's mathematics, the strongest used to date for both the GR IVP and CP \cite{CBY}, is not 
generally powerful enough to deal with thin matter sheets.

More powerful mathematics is thus desirable, and more may well appear over 
the next few years (see the program starting with \cite{Klainerman}).  Realizations of part of the 
hierarchy of models below would be an interesting application for this sort 
of mathematics, which further motivates the study of increasingly rough Sobolev spaces, since the 
higher the dimension the greater the improvement to the established HKM mathematics 
is required for a full study of thin matter sheets.  We must also mention that in spaces 
rougher than $H^4$ there is no guarantee of geodesic uniqueness \cite{HE}, so complete 
conventional physical sensibleness is generally lost no matter what rougher Sobolev class 
results are proved. This shortcoming is clearly dimension-independent 
(from the dimension-independence of the form of the geodesic equation).  
 
\mbox{ }

The asymptotics difficulty \cite{AsADS} follows from supposing that `AdS' bulks  are desirable for 
string-inspired scenarios. The study of `AdS' bulks by relativists would entail the study of 
{\sl asymptotically AdS bulks}, to permit a more general study of disturbances close to the thin 
matter sheet.  Whereas we will for the moment seek to avoid the function space difficulty, 
the asymptotics difficulty will remain in the `tractable scenarios'. In the usual GR, the use of AdS 
asymptotics for the application of the York method to small-scale astrophysics was neglected because of 
the negligible effect of a cosmological constant on such small scales and because of DOD arguments.
The braneworld application is however substantially different, so one might in this case have to study 
the York method with `AdS asymptotics'.  This could affect the tractability of the Lichnerowicz equation, 
and also whether the existence of the crucial maximal or CMC slices is as commonplace for these new 
spacetimes as it was for the (3, 1) GR ones.  Furthermore, because of the interest in infinite planar 
branes, the asymptotics is \sl directional\normalfont, `perpendicular' to the brane. 
However, in general there is no such notion of perpendicularity in GR.  Whereas one can locally define 
geodesics and draw hypersurfaces perpendicular to them, this procedure in general eventually breaks down, 
for example due to (spatial) caustic formation.  

We start our hierarchy by envisaging the most general situation possible for thin matter sheets within 
the framework of GR.  It is entirely legitimate to construct a (4, 0) initial 
data snapshot with whatever shape of thin matter sheet, but it is not legitimate 
to assume that any features of this are maintained over time unless it can be shown that the full 
evolution equations robustly maintain these features. Strictly, this is an 
\it initial-boundary problem\normalfont, such as occurs for water waves or for the 
surface of a star.  This is a very hard and quite new problem in GR \cite{RF, FN}.
Thus the evolution step is particularly hard both in its full generality and in the 
justifiability of simplifications such as a fixed boundary.  

Consider a thin matter sheet in an asymmetric (not $Z_2$ symmetric) bulk.  
One would ideally want to follow objects that crash into and possibly disrupt the thin matter sheet.  
One can then imagine asymmetric crashes which might punch though the thin 
matter sheet.  The sheet could thicken or disperse with time. It could emit a 
significant amount of gravitational radiation, which could moreover include gravitational shock 
waves that spread out the $H^3$ character that initially pertained only to the thin matter 
sheet and not to the smooth surrounding bulk.  That is, the $H^3$ character of a thin 
matter sheet data set $J$ could typically spread to the whole of its causal future ${\cal J}^+(J)$ (fig 19). 
\begin{figure}[h]
\centerline{\def\epsfsize#1#2{0.4#1}\epsffile{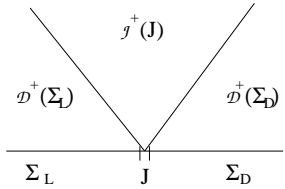}}
\caption[]{\label{TO14.ps} \footnotesize The spread of the $H^3$ region of a snapshot $\Sigma_L \bigcap J \bigcap \Sigma_R$ 
containing the $H^3$ junction $J$ between the 2 spacelike bulk pieces $\Sigma_L$ and $\Sigma_R$ 
upon which the prescribed data are smooth or analytic.  Then the evolutions in the domains of 
dependence ${\cal D}^+(\Sigma_L)$ and ${\cal D}^+(\Sigma_R)$ are smooth or analytic, but the 
causal future of $J$ will typically be $H^3$. \normalfont  }
\end{figure}

In simple words, why should the junction remain unscathed?  Why should the bulk in the immediate 
vicinity of the junction at later times be simple, smooth or known?  Addressing these questions 
goes beyond the reach of present-day techniques.  Rather we note that some usually-tacit 
assumptions are used to definitely avoid these difficulties, and that we do not know whether any 
of these are justifiable in the physics governed by the GR evolution equations. One can presuppose 
a thin matter sheet exists at all times to prevent it being created or destroyed.  One can allow 
only bulks with regular thin matter sheet boundaries to preclude shock waves.  One could also 
preclude asymmetric crashes by presupposing $Z_2$ symmetry.

Even if discontinuous emissions are precluded, one can imagine smooth emissions and absorptions 
(symmetric or asymmetric) whereby the bulk interacts with the thin matter sheet.  This would 
entail material leaking off or onto each side of the thin matter sheet.  This can be precluded 
by the presupposition that the thin matter sheet's energy-momentum resides at all times on the 
sheet, which is encapsulated by the well-known `equation of motion' of the thin matter sheet 
\cite{jns, MTW}. This is a significant restriction on the dynamics of thin matter sheets in GR, 
often carried out in the name of tractability but which may not be realistic.  

In the Randall--Sundrum scenario, the brane can be placed anywhere in the bulk without affecting 
it.  We strongly suspect that this is not a typical feature in GR-based braneworlds, nor is it 
desirable since it dangerously marginalizes the ontological status of the bulk.  We take this 
feature to be too specialized to be included in the developments below.  
   
\mbox{ }

We close this section by briefly discussing modelling assumptions which avoid some of the above 
difficulties.  We then implement these in two particular classes of tractable problems.  The 
desirability of each modelling assumption strategy and of other modelling assumptions from a 
string-theoretic perspective is discussed in B.3.4.

I) The choice of $Z^2$ symmetry may well look and indeed be arbitrary from a GR point of view 
and indeed from a string-theoretic point of view (see the end of this section).  But without such a choice one 
simply cannot establish any specific embeddings if there are thin matter sheets (in the GR study 
of stars one requires the absence of surface layers to perform matchings). Thus one studies the 
restricted case with $Z^2$ symmetry, in which one integrates only up to the b.c provided by the 
j.c. In the IVP part one can assume whatever configuration of thin matter sheets. Thus this step 
is more justifiable than the subsequent evolution, which may require some of the above ad hoc 
assumptions as to the existence and good behaviour of the thin matter sheet at all future times. 
So for the moment we are merely after the construction of (4, 0) initial data up to the junction. 
By themselves these data are useful in addressing issues such as the shape of the extensions of 
black hole horizons into the bulk, but the staticity and stability of these configurations remain 
unaddressed in the as-yet intractable evolutionary step of the problem \cite{401SS}.   

II) One could choose instead to work with thick matter sheets i.e ones that are finitely rather 
than infinitely thin \cite{Gregory}. Then the above troubles with the function spaces need not 
occur, since the walls would then not be rough in the above sense, although they could still have 
the tractable level of roughness that is able to accommodate astrophysical objects just as in 
ordinary GR.  It is then easier to envisage the study of the evolutionary step of the thick matter 
sheet problem quite high up our hierarchy, since this more closely resembles the stellar surface 
problem.     
      
III) One could investigate closed shells rather than open sheets in order not to require 
directional asymptotics.  

\mbox{ }
  
\noindent\large{\bf 3.2 Thin matter sheet IVP}\normalsize

\mbox{ }

\noindent
We consider the $j_b = 0$ case.  Then by definition an extrinsic curvature 
$\Theta^{\mbox{\scriptsize T\normalsize}ab} = \Theta^{\mbox{\scriptsize TT\normalsize}ab}$ 
suffices in order to solve the Codazzi constraint.  We begin by considering the (4, 0) case of 
the Lichnerowicz--York equation (\ref{ndlich}).  The $j^a \neq 0$ case, while important, lies beyond the 
scope of this thesis.    

We work with the unphysical line element $ds^2 = dz^2 + e^{w}dx_{\alpha}dx^{\alpha}$ for 
some known trial function $w = w(r, z)$, so as to model a $S^2$-symmetric object such as a 
black hole or a star on the brane.  For the $j^a= 0$ case, the IVP essentially reduces to the 
solution of the Lichnerowicz equation for some $\psi = \psi(r, z)$.  We begin with the 
$T_{ab} = 0$ case of (the spatial projection of) (\ref{BWEM}).  We first find the b.c's for 
the Lichnerowicz equation, then provide algorithms and then comment on the underlying mathematical 
physics.    

Note that the b.c's are to be imposed on the physical metric and then written in terms of the 
unphysical quantities to have b.c's in terms of the objects we start off with and the unknown 
conformal factor $\psi$.  

The inner radial b.c is commonly established using an isometry (c.f C.2.3).   
The Neumann reflection b.c case appropriate for a stellar source holds just as well 
here as in C.2.3.  The case appropriate for a black hole is however distinct.  
Now, part of $r = 0$ is singular and is approached by excision and an
inversion-in-the-sphere isometry about some throat at radius $a$ of the black hole 
within the apparent horizon [see fig  20a)].  But for braneworld black holes one does 
not know how far the black hole extends into the bulk.  Indeed, the main point of the study 
is to find this out (pancakes versus cigars).  Our idea is then to guess a profile 
$r = f(z)$ along which to excise.  About each point on $r = f(z)$ an isometry in the 
corresponding 2-sphere may be applied [fig 20 b)], leading to an 
\be
\mbox{inner Robin 
condition }
\mbox{\hspace{1.2in}}
\left.
\left[
\frac{\pa \psi }{\pa r} + \frac{1}{2f(z)}\psi
\right]
\right|
_{r = f(z)} = 0.
\mbox{\hspace{2in}}
\ee
The profile could be chosen so that it matches up smoothly with the $r = 0$ and $z = 1$ boundaries.  
Once the problem is solved, one can find out whether this profile was a good choice or not 
[fig 20 c)].  By picking a 1-parameter family of $r = f(z)$ curves, one could then iterate until a 
satisfactory profile is found.  
\begin{figure}[h]
\centerline{\def\epsfsize#1#2{0.4#1}\epsffile{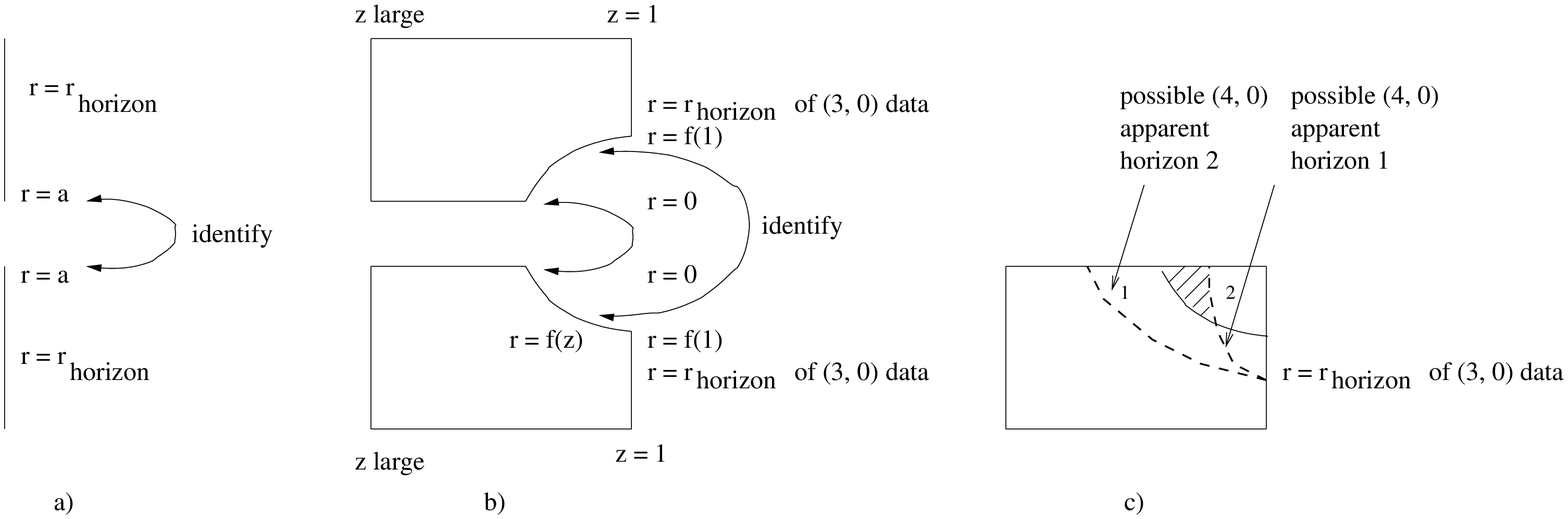}}
\caption[]{\label{TO15.ps}\footnotesize a) Excision of the inner region in (3, 0) black hole data construction by an 
inversion-in-the-sphere isometry identification between two copies of the black hole.  
This is along the lines of the method of images in electrostatics.\normalsize    

\noindent \footnotesize b) Excision of the inner region in (4, 0) braneworld black hole data construction 
would involve making a guess $r = f(z)$ for the excision region and then performing 
inversion-in-the-sphere isometries pointwise to identify two copies of the black hole.\normalsize      

\noindent \footnotesize c) One is then to numerically solve for the data.  One can then find the apparent 
horizon.  If this intersects $r = f(z)$ (horizon 1) then the data may be bad in that the 
part of the shaded portion of the excised region may be in causal contact with its surroundings.  
If the apparent horizon extends too far past the end of the excision region (horizon 2), one 
might worry that the singularity also extends past the excised region in which case some of the 
data prescribed along $r = 0$ should not have been prescribed since it lies on the singularity.\normalsize  }
\end{figure}

For the new inner-$z$ ($z = 1$) b.c, we impose the j.c's (\ref{jcf}) and (\ref{prez2}), 
which now read 
\be
\tilde{e}_{\alpha\beta}^+ = \tilde{e}_{\alpha\beta}^-
\mbox{ } , \mbox{ } \mbox{ } \mbox{ }
\tilde{k}_{\alpha\beta}^+ - \tilde{k}_{\alpha\beta}^- = -G_5\left(\tilde{Y}_{\alpha\beta} - \frac{\tilde{Y}}{2}\tilde{i}_{\alpha\beta}\right)
\label{scndjn}
\ee
(N.B these are imposed on the physical quantities).  Imposing $Z_2$ symmetry, the second of 
these becomes 
\be
\left. 
\left[
\frac{\pa \psi}{\pa z} + \frac{\psi}{2}\left(   \frac{\pa w}{\pa z} - \frac{G_5\tilde{\lambda}}{2}\psi 
\right)
\right] 
\right|
_{z = 1} = 0 
\label{nonlinbc}
\ee
by use of the definition of the physical $\tilde{K}_{\alpha\beta}$ in normal coordinates and 
then factorizing out the $e_{\alpha\beta}$.  This $Z_2$ isometry b.c is more complicated than 
usual because of the presence of the thin matter sheet.

That the $Z_2$ symmetry is termed a reflection appears to generate a certain amount of confusion.  
This is because there is also the reflected wave notion of reflection, taken to give a Neumann b.c 
(such as in for the inner-radius b.c for a stellar source).  In the case of Shiromizu and Shibata 
(SS) \cite{401SS} (see also \cite{401N}), who furthermore treat our known $w$ as their unknown, 
and our unknown $\psi$ as the perfect AdS $\psi \propto \frac{1}{z}$, it has the nice feature that 
the other 2 terms in (\ref{nonlinbc}) cancel leaving a Neumann b.c 
\be
\left. \frac{\pa w}{\pa z} \right|_{z=1} = 0,
\label{funnyref}
\ee 
which might look like a reflection in both of the above senses.  However, this is only possible for 
$\psi \propto \frac{1}{z + q(r)}$ which may be somewhat restrictive.  Note also that (\ref{funnyref}) 
is an unusual notion of reflection in this context, since it is of the conformally-untransformed 
(3, 0)-metric, whereas the natural notion of reflection isometry would be for the 
conformally-transformed (4, 0)-metric.  Thus we conclude that the presence of the notion of reflection 
(\ref{funnyref}) in SS's work is happenstance: it is separate from the $Z_2$ notion of reflection 
encapsulated in (\ref{scndjn}) and no analogue of it need hold in more general situations.  

Our b.c. (\ref{nonlinbc}) contains a reflection part $\frac{\pa\psi}{\pa z}$; the other part being a 
(nonlinear) absorption. This has the following implications.  First, the presence of absorption terms 
is interesting since pure reflection is the boundary condition of a perfect insulator in the 
potential theory of heat.  This suggests that pure reflection b.c's such as for pure AdS have 
built-in non-interaction of the bulk with its bounding brane, whereas our b.c may lead to (evolution 
models) with brane-bulk interactions.  Whereas neither the planar symmetry in (4, 1) spacetime of the 
underlying thin matter sheet nor the $S^2$ symmetry within the (3, 1) thin matter sheet of a simple 
astrophysical object generate gravity waves, we expect that spherical bumps on approximately planar 
branes in (4, 1) spacetimes to admit gravity waves, potentially giving rise to instabilities in the 
approximately AdS bulk braneworlds in which the branes contain astrophysical objects.  Our b.c could 
permit the modelling of such interactions.  Second the nonlinear absorption (much as arises in the 
theory of heat with temperature-dependent conductivity) is likely to complicate both analytic and 
numerical treatments of our b.v.p.

The first point above is one reason to favour our approach over that of SS (subject to overcoming the 
complications due to the second point).  Here is a further reason why our approach should be favoured 
on the long run.  Whereas one could try to generalize SS's work by keeping the notion that $w$ be 
unknown and $\psi$ known but not $\frac{1}{z}$, whereupon our b.c (\ref{nonlinbc}) is interpreted as 
an inhomogeneous Neumann b.c in $w$, the main trouble with SS's work is that there is no good reason 
for their method to be directly generalizable away from its restrictive assumption $\Theta_{ij} = 0$.  
Whereas treating $w$ not $\psi$ as the unknown may give nicer b.c's, it is not tied to a method known 
to be amenable to less trivial solutions of the Codazzi equation than 
$\Theta^{\mbox{\scriptsize T\normalsize}}_{ij} = 0$.  Our proposed method seeks a workable extension 
to the significantly more general case  
$\Theta^{\mbox{\scriptsize T\normalsize}}_{ij} = \Theta^{\mbox{\scriptsize TT\normalsize}}_{ij}$, and 
speculatively to completely general $\Theta_{ij}$.    

Another nice feature in SS's work is that the matter is scaled so that the linearized p.d.e obtained 
is a Poisson equation.  This p.d.e is simply invertible to obtain detailed asymptotics 
\cite{TanGarr, 401SS}.  For our scheme, the linearized equation (\ref{linlich}) is of Helmholtz-type 
rather than Poisson, complicating such a procedure.    We merely demand instead that 
$\psi \longrightarrow \frac{1}{z}$ for $z$ large wherever possible so that our models are 
`directionally asymptotically AdS'.  As for large $r$, at least on the brane, one can impose asymptotic 
flatness as $r \longrightarrow \infty$.  It is less clear what one should prescribe in this respect off 
the brane.  A more detailed study of the asymptotics for our b.v.p's should be required as part of the 
actual construction of examples of data sets.\fn{Particularly because numerical integration is done on 
large but finite grids, subleading order asymptotics are helpful.}  

\mbox{ }

Our proposed b.v.p is thus the mixed, nonlinear problem 
$$
\triangle\psi = \mbox{Nonlin}(\psi) \mbox{ } , \mbox{ } \mbox{ } 
$$
$$
\frac{\pa \psi}{\pa z} + \frac{\psi}{2}
\left(
\frac{\pa w}{\pa z} - \frac{G_5\tilde{\lambda\psi}}{2}
\right) 
= 0 \mbox{ for } z = 1 \mbox{ and } r \geq f(1) 
\mbox{ } ,  \mbox{ } \mbox{ }
\psi \longrightarrow \frac{1}{z} \mbox{ as } z \longrightarrow \infty 
\mbox{ } , \mbox{ } \mbox{ } 
$$
$$
\left. 
\frac{\pa \psi}{\pa r}
\right. 
= 0 \mbox{ for } r = 0 \mbox{ and } z \geq f^{-1}(0) 
\mbox{ } , \mbox{ } \mbox{ } 
\psi \longrightarrow 1 \mbox{ } \mbox{as} \mbox{ } r \longrightarrow \infty \mbox{ } ,
$$
\be
\frac{\pa \psi }{\pa r} + \frac{\psi}{2a} = 0 \mbox{ on } r = f(z) \mbox{ } .
\ee
One can use simplifications i) and/or ii) of C.2.2  on this and still have a more general 
case than that of SS.  Simplification i) might be plausible because 5-d AdS clearly admits 
conformally-flat spatial sections [$ds^2 = \frac{1}{z^2}(l^2dz^2 +dx_{\alpha}dx^{\alpha}$)] or 
maybe not, since it is regarded with suspicion in the (3, 1) GR 2-body problem \cite{BSrev}.  
The trick at the end of C.2.2 cannot be used to obtain particularly simple examples due to 
the following argument.  Unless one puts $\rho \propto \Lambda$ (constant) and has this maintained 
by non-scaling (i.e $a = 2$), it is overwhelmingly probable that the emergent $\tilde{\Lambda}$ is 
not constant.  Thus our freedom in $a$ is used up to ensure that $\tilde{\Lambda}$ is constant.  
Contrary to what York assumes in the ordinary GR context, we must permit $\rho < 0$ since our application 
requires a negative bulk cosmological constant, which further complicates the analysis of which 
cases are guaranteed to be well-behaved and renders $\rho$ cancellation with $m^2$ unavailable.  

Our algorithm for solving this problem is as follows:  

\noindent
i)   Prescribe the following unphysical quantities: the metric $h_{ab}$ and matter density $\rho$.    

\noindent
ii)  Pick any suitable $\Theta^{\mbox{\scriptsize TT\normalsize}ij}$ to solve the Codazzi 
constraint.  It follows that $M$ is known.     

\noindent
iii) Thus we can attempt to solve our b.v.p for the Lichnerowicz equation to obtain $\psi$.    

\noindent
iv) Then we can compute the physical bulk metric $\tilde{h}_{ab}$ and induced brane metric 
$\tilde{e}_{\alpha\beta}$ of our snapshot.  

This assumes that we have correctly guessed the profile.  One would now check whether this is 
the case by solving for the apparent horizon.  If this is unsatisfactory [fig 20c)] then one 
would repeat the algorithm with adjusted profile.  

For nonvacuum branes ($T_{ab} \neq 0$), a similar argument enforcing $a = 2$ holds. The j.c may 
now be split into a trace b.c part,

\noindent
\be
\left.
\left[
\frac{\pa \psi}{\pa z} + \frac{\psi}{2e}
\left(
\frac{ \pa e}{\pa z} + \frac{G_5Y\psi}{6}
\right)
\right]
\right|
_{z = 1} = 0
\ee
and a restriction on the tracefree part of the matter on the brane, which is 
\be
Y_{\alpha\beta}^{\mbox{\scriptsize T\normalsize}} = 0
\label{95}
\ee
for the metric ansatz and coordinate choice used.  
Our algorithm now becomes

\noindent
i)  One now requires the prescription of $h_{ac}$ everywhere and of $Y$.    

\noindent
ii) Pick any suitable $\Theta^{\mbox{\scriptsize TT\normalsize}ij}$; hence 
$M = \Theta^{\mbox{\scriptsize T\normalsize}}\circ\Theta^{\mbox{\scriptsize T\normalsize}}$ 
is known.  

\noindent
iii)Attempt to solve our b.v.p for the Lichnerowicz equation.   

\noindent
iv) Then we can compute $\tilde{h}_{ab}$ and $\tilde{e}_{\alpha\beta}$.  In this particular case, 
the simultaneous imposition of normal coordinates and an isotropic line element forces the 
restriction (\ref{95}) on the braneworld matter.  This is because one is applying too many 
coordinate restrictions; in this light the habitual practice of working in braneworlds using 
normal coordinates may be seen as a poor choice of gauge. If the above coordinate choices are not 
simultaneously made, the algorithm would contain further nontrivial equations [in place of 
(\ref{95})] to solve before the braneworld matter content can be deduced.

We assume that this nonvacuum application is for a star or clump of dust on the brane in which case 
there is no need to excise a corner with a profile as was done above to deal with the black hole 
singularity.  

Notice the lack of control of the physical metrics characteristic of theoretical numerical 
relativity.  Our non-scaling of the matter at least gives visible control over the matter.  
The sensibleness of doing this is in fact tied to the negativity of the bulk $\rho$: for 
$\rho \geq 0$ and $a \geq -1$, Lin receives a negative contribution with its tendency to encourage 
ill-posedness, whereas for predominantly $\rho \leq 0$, the danger \cite{CBY} lies in $a \leq -1$. 

A simpler example of b.v.p set up along the lines we suggest is that of Nakamura, Nakao and Mishima 
\cite{401N}.  Their metric is cylindrical not spherical and they have a slightly simpler version of 
(\ref{nonlinbc}).  Their case is still time-symmetric.  The linearized equation is then precisely 
Helmholtz, facilitating its inversion and the consequent more detailed knowledge of the asymptotics.  

We also propose the 2-brane version of the above, in which the $z \longrightarrow \infty$ condition 
is replaced by another `parallel' brane boundary at $z = 1 + l$. This problem is close to that 
considered by Piran and Sorkin \cite{401lit2}.  In contrast however, their study involves a 
periodically-identified fifth dimension to which there does correspond an isometry-based 
\be
\mbox{reflection Neumann condition} 
\mbox{\hspace{1.7in}}
\left. 
\frac{\pa\psi} {\pa z} 
\right
|_{z = 1} = 0 
\mbox{ } .  
\mbox{\hspace{1.7in}}
\ee

\mbox{ }

We can provide a local (in function space) uniqueness proof for our proposed type of b.v.p,  at the 
same level as that for standard GR black hole data in \cite{BY}.  In the conformally-flat case, 
suppose that $\psi_1 = \psi_2 + u$ for $u$ small.  Then the homogeneous linearized b.v.p in $u$ is 
applicable.  We then have 
$$
\int_{\Sigma}(|D u|^2 + \mbox{Lin}u^2)d^4x = \int_{\Sigma} (|D u|^2 + u\triangle u)d^4x =  
\oint_{\pa\Sigma} u\frac{\pa u}{\pa n}dS  
\mbox{\hspace{1in}}
$$
\be
\mbox{\hspace{1in}}
=\int_{z = Z} u\frac{D u}{\pa z}dS + \int_{r = R} u\frac{\pa u}{\pa r}dS 
 - \int_{z = 1 + \epsilon} u\frac{\pa u}{\pa z}dS 
- \int_{r = \eta} u\frac{\pa u}{\pa r}dS 
\ee
by Green's theorem.  For $u \longrightarrow 0$ at least as fast as $\frac{1}{z}$, the first 
surface integral tends to zero.  For $u \longrightarrow 0$ at least as fast as $\frac{1}{r}$  the 
second surface integral tends to zero.  The fourth integral is zero by the Neumann reflection b.c.  
Upon applying the linearization of the nonlinear b.c (\ref{nonlinbc}),
\be
\left.
\left[
\frac{\pa u}{\pa z} + \frac{u}{2}
\left(
\frac{\pa w}{\pa z} - G_5\tilde\lambda\psi_1
\right)
\right]
\right|
_{z=1} = 0, 
\label{linrob}
\ee
the third surface integral is non-positive provided that the restriction $\frac{\pa w}{\pa z} \geq 0$ 
holds.  Thus since Lin $\geq 0$, u must be zero.  For the non-conformally flat case, this argument 
requires $R \geq 0$.  The argument is not seriously changed in the black hole case when a corner is 
excised by a profile with a Robin condition on it.

Stronger existence and uniqueness proofs for our proposed problems are complicated by four factors:    
unboundedness, (albeit of the most benign kind), the boundary having corners, mixed b.c's 
(i.e a piecewise prescription on the boundary) and a portion of these being nonlinear.  The first is 
however of the simplest kind and the second and third are commonplace.  The last is less usual but 
perhaps not so bad either because by 
\be
\triangle \psi = \frac{1}{\sqrt{\mbox{\scriptsize \normalsize}h}}\frac{\pa} {\pa x_i}
\left(
\sqrt{\mbox{}h}h^{ij}\frac{\pa\psi}{\pa x_j}
\right)
\ee
the problem can be written in divergence form, for which b.v.p's with nonlinear b.c's  are treated 
in Ch. 10.2 of \cite{LU}.  The trouble is that the treatment there, unlike here, is neither for 
mixed b.c's nor for an unbounded region with corners.  Thus, the strongest point we can make at 
present is that there is good hope of obtaining existence and uniqueness results even for quite 
rough function spaces by more-or-less conventional, entirely rigorous mathematics for our proposed 
problem.  This should be contrasted with the hopeless state of affairs with sideways prescriptions!  
Once one begins to get good numerical results, it becomes worthwhile to explicitly prove the 
well-posedness of the method used, in order to support those results and the ongoing production of 
more such results. In this particular case, these results would serve to support and understand the 
crucial numerical step iii) in the above algorithms. 

To permit the even greater generality required to have momentum flows ($j^a \neq 0$), by our route 
one is forced to forfeit the DEC control over the matter.  This is because at least the $\Lambda$ 
part of $\rho$ cannot scale in accord with the scaling of $j^a$.  We leave this further development, 
which additionally requires posing and solving the b.v.p following from the Codazzi equation, for a 
future project.

\mbox{ }

\noindent\large{\bf 3.3 Thick matter sheet IVP}\normalsize

\mbox{ }

\noindent Many of the difficulties with evolution discussed above stem from the thinness of the 
matter sheets causing function-space-related problems.  But ordinary physics should not be 
sensitive to which function space is used.  One would hope that models with thick matter sheets 
would be more amenable to study, be good approximations to the thin matter sheets and in any case 
could be closer to reality than thin matter sheets.  

Bonjour, Charmousis and Gregory \cite{Gregory} have used the (2, 1; 1) version of the general split (\ref{gauss}), 
(\ref{cod}), (\ref{evK}) with scalar matter and consider both thick walls and the thin-wall limit
perturbatively in the presence of gravity. They further specialize to the case of a spherical 
domain wall, which they show collapses.  It would be worthwhile if this sort of example, which 
combines the modelling assumptions II) and III), is investigated using the 

\noindent($n$, 0; --1) split subject to flat and to AdS asymptotics.  

For the particular example mentioned above, the matter profile is sigmoidal.
However, to approximate a thin brane, one would want instead a hump profile, 
as indicated in fig 21.  One could still use an inner b.c at $z = 1$, where now $k_{ij} = 0$ so 
that the linear Robin b.c
\be
\left. 
\left[
\frac{\pa \psi}{\pa \lambda} + \frac{\pa w}{\pa z}\frac{\psi}{2}
\right]
\right|
_{z = 1} = 0
\label{thickrob}
\ee
holds. 
One could in fact consider 2 problems: the flat sheet with directional asymptotics 
and the spherical sheet with a suitable inversion-in-the-sphere in place of the above inner b.c.  
Both of these problems avoid the main obstacles to existence and uniqueness proofs by 
possessing linear b.c's. The second problem is also invertible to an inner problem 
which is on a bounded region with no corners, for which results in \cite{LU} apply.  
\begin{figure}[h]
\centerline{\def\epsfsize#1#2{0.4#1}\epsffile{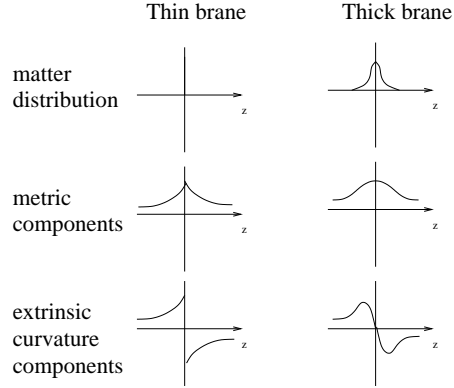}}
\caption[]{\label{soupdragon.ps}
\noindent \footnotesize The metric, extrinsic curvature and matter profiles for thin branes and thick branes.\normalsize} 

\end{figure}

These problems are for `plain' thick branes as opposed to branes containing spherically-symmetric objects. 
The first problem ought to more easily admit this extension.  
One might wish to study `plain' thin branes too, e.g in the context of branes in relative motion.  

Finally, suppose instead one attempted to use something like Magaard's method to construct 
(4, 0) thin-matter-sheet data.  Then one benefits from $s = 0$ and the identification of 
$x_1 = const$ with the $z = 1$ brane prescribes the topology, removes some of the sources of 
nonuniqueness and the brane's energy-momentum endows physical significance upon the (3, 0) 
covariance of the elimination procedure.  In the thin matter sheet case, one is blocked by 
the nonanalyticity of $\rho$, but at least some thick matter sheet models could be built in this way. 

\mbox{ }

\mbox{ }

\noindent\large{\bf 3.4 Discussion: modelling assumptions and stringy features}\normalsize

\mbox{ }

\noindent
We did not consider the possibility of a higher codimension \cite{cod2} which could substantially 
alter the nature of the embeddings used.  For models with 2 or more times, it is simply not 
possible to stick entirely to our suggestion to build on established mathematical physics.  
Finally, our study could be complicated by having more than the Einstein tensor in the bulk.  
For example, one can have a nontrivial Lovelock tensor term \cite{Lovelocktensor, Lovelockbranes},  
or one could have genuine higher-derivative terms.\fn{A starting-point for this study might 
be the case-by-case 4-d work of \cite{Vhdl2}.}  Whilst our incipient results -- the 
correspondence between ($n$, 0; --1) and ($n$ -- 1, 1; 1) schemes at the simplest level, or our 
BEFE ambiguity -- will have counterparts in these theories,  these theories' equations are much 
larger than the EFE's, and their CP/IVP is much less well-studied than that of GR.  Higher 
derivatives or non-standard bulk matter might lead to the violation of the energy conditions 
assumed in this thesis.  
  
We next discuss which features of GR-based braneworld models are desirable for string theory.  
First for some caveats.  We introduced the SMS braneworld as a GR generalization of the 
Randall--Sundrum scenario, moreover one that was as yet not sufficiently general for the 
purposes of GR in which the nature and existence of thin matter sheets at all times might not 
be a representative presupposition.  It could be that it is other aspects of the 
Randall--Sundrum scenario that are of interest in string theory since it also has particle 
physics aspects to it and because it is a toy model of Ho\v{r}ava--Witten theory.  One may 
prefer to consider only these stringy scenarios rather than Randall--Sundrum or SMS.   
Also a great source of tension between string theorists and relativists is that the latter 
believe GR is suggestive of the importance of background--independent theories, whilst there 
is as yet no background--independent formulation of string theory available.   Such ideas as 
conserved charges, flat sheets and simple bulks arise from assuming Minkowski or AdS 
backgrounds.  Since GR is background-independent, one is interested there in generic 
solutions to the full field equations.  This makes the notion of asymptotics necessary and 
subtle, and not necessarily compatible with flat sheets nor conserved charges.  From a general 
perspective it is doubtful whether highly-specialized solutions such as Minkowski or AdS are 
likely to reflect the gravitational physics of the universe, unless it can be shown that large 
classes of generic solutions behave likewise or are attracted toward such solutions.  

String theory conventionally favours many extra dimensions, although it is not clear how many of these 
should be large and how many compactified.  Among the corrections to the Einstein tensor 
mentioned above, string theorists might favour the Lovelock correction, 
since this is the only first-order correction for a heterotic string \cite{GS}; an infinite 
series of further such correction terms is predicted.  String theorists might favour very flat 
walls on the grounds that these could be stabilized by being the lowest-energy carriers of 
{\it BPS charge} \cite{Polchinski, Turok}.   These arguments would rule against the use of highly symmetric 
bulk manifolds/orbifolds, although string theorists might try to invoke them as fixed 
backgrounds so as to perturb about them.    There may be different reasons for stability in GR 
and string theory.  For example, in contrast to string-theoretic arguments favouring 
reflection-symmetric orbifolds, a GR domain wall could be stable precisely because the (3, 1) 
bulk is different on each side, as the wall separates domains in which symmetry is differently 
broken.  On the whole, the string theorist would agree that bulks could be complicated by the 
presence of bulk matter from the closed string multiplet (I.3.3.5).  At least in some models, 
gauge fields would be expected to occur only as fields confined to branes since they belong 
to the open string multiplet and some branes are where open strings end \cite{Polchinski}.    
As strictly, the branes are of Planckian thickness, so it is not a disaster that thick branes are 
favourable toward rigorous mathematics for GR-based scenarios (this is recycling the argument in footnote 32).    
Of course, one could worry that they are thin enough for quantum gravitational effects to be important. 

Finally, we summarize the aspects of question 2 that should be asked as the partial 
answers to question 1 \normalfont begin to be built up.  Does the introduction of stringy features 
change the outcome to any of issues?  And so, are GR-based braneworlds adequate or typical 
frameworks for string theory? What scenarios should one explore in order to furnish string 
theory with reasonably unique predictions?

\vspace{8in}

\mbox{ }

\noindent\Huge{\bf C Appendix on elliptic equations}\normalsize

\mbox{ }

\noindent 
Let $u$ take 1 to $U$, $v$ take 1 to $V$ and $w$ take 1 to $W$.  
Let $\Omega$ be a multi-index running over the integers $\omega_1$ to $\omega_W$, 
$\#\Omega = \sum_{w = 1}^{W}\omega_w$ and 
${\cal D}_{\Omega} \equiv 
\frac{      \pa^{\omega_1 ... \omega_{\mbox{\tiny W\normalsize}}      }                  }
     {      \pa (x^1)^{\omega_1} ... \pa (x^w)^{\omega_{\mbox{\tiny W\normalsize}}}      }$.  
The system of $U$ p.d.e's for $V$ unknowns $y^v$ of $W$ variables ($x^w$) 
\be
\sum_{\Omega = 0}^{max\#\Omega}A^{\Omega u}_{v}(x_w){\cal D}_{\Omega}y^{v} = 0
\ee 
\be 
\mbox{is {\it elliptic} if its {\it principal symbol}}
\mbox{\hspace{1.0in}}  
\sigma(k_d) \equiv \sum_{\Omega = max\#\Omega}A^{\Omega u}_{v}(x_w){k}_{\Omega} 
\mbox{\hspace{1.0in}}
\ee
is positive-definite and invertible.  

Many elliptic equations are associated with the GR IVP.  
Below I build up to these starting from the simplest elliptic equations, 
which emerge anyway as special cases in the GR IVP.    
The purpose of using elliptic formulations is that they lend themselves well to 
general well-posedness proofs.  But one may need to bear in mind in practice that they are computationally 
expensive (since, being instantaneous equations, their solution involves the whole numerical 
grid).

\mbox{ }

\noindent\large{\bf 1 Linear elliptic equations}\normalsize

\mbox{ } 

\noindent 
The Poisson equation in flat space (e.g Newtonian gravity or electrostatics)   
\hspace{0.1in} $\triangle_{\mbox{\scriptsize F\normalsize}}A = B$ \hspace{0.1in}
is typically considered in a region $\Omega$ and supplemented by b.c's on $\pa\Omega$.  
Typical b.c choices include
\be
\mbox{Dirichlet : }
\left.
A
\right|_{\pa\Omega} = K
\mbox{ } ,
\label{Diric}
\ee
\be
\mbox{Neumann : }
\left.
\frac{\pa A}{\pa\mbox{\b{n}}}
\right|_{\pa\Omega} = L \mbox{ } \mbox{ for \b{n} the normal to }\pa\Omega 
\mbox{ } ,
\ee
\be
 \mbox{Robin : }
\left.
\left(
\frac{\pa A}{\pa\mbox{\b{n}}} + fA
\right)
\right|_{\pa\Omega} = M \mbox{ } 
\mbox{ } ,
\ee
\be 
\mbox{mixed : } 
\left.
\frac{\pa A}{\pa\mbox{\b{n}}}
\right|_{\pa\Omega_1} = L \mbox{ } , \mbox{ }
\left.
A
\right|_{\pa\Omega_2} = M \mbox{ }, \mbox{ } \pa\Omega = \pa\Omega_1 \amalg \pa\Omega_2 
\mbox{ } .
\label{mixed}
\ee
These easily have good existence and uniqueness properties via Green's theorems 
It is profitable also to prove uniqueness for these from the {\it maximum principle} (see e.g \cite{CH}), 
a method which affords substantial generalization.  Whereas the above problems are furthermore 
well-posed, the Cauchy data choice 
\be
\left.
A
\right|_{\pa\Omega} = K 
\mbox{ } , \mbox{ } 
\left.
\frac{\pa A}{\pa\mbox{\b{n}}}
\right|_{\pa\Omega} = L \mbox{ } , \mbox{ }
\ee 
includes {\it Hadamard's example}, which fails to exhibit continuous dependence on the data.   
Solution methods for the b.v.p's (\ref{Diric}--\ref{mixed}) include fundamental solution 
techniques and the method of images.  

The above treatment can straightforwardly be extended to (generalized inhomogeneous Helmholtz) 
equations of form 
\be 
(M^{ij}\pa_i\pa_j + Q^i\pa_i + S)A = B \mbox{ } ,  \mbox{ } M^{ij} \mbox{ } \mbox{ positive definite and invertible}
\mbox{ } .
\label{genlinell}
\ee
A case of this is the following curved space equation:  
\be
(h^{ij}\pa_i\pa_j  - \Gamma^k\pa_k)N  =  \triangle N = NO + P 
\mbox{ } ,
\label{helmics}
\ee
where $h^{ij}$ is regarded as some inverse $n$-space metric and $\Gamma^k = h^{ij}{\Gamma^k}_{ij}$.  
These typically appear in this thesis for the CWB case.  Then, the above types of proof 
benefit from the additional fact that all surface integrals vanish:
\be
0 = \oint_{\pa\Omega}D_iN\textrm{d}S^i =\int\textrm{d}\Omega \triangle N =
\int\textrm{d}\Omega(NR + O) \Rightarrow \not{\exists} \mbox{ } N 
\mbox{ if } NR + O \mbox{ is of fixed sign.}
\ee
For the last step, suppose there is some 
point $x_0$ at which the integrand $I(x_0) = \epsilon > 0$.  Then for $I$ 
continuous, $|I(x) - I(x_0)| < \frac{\epsilon}{2} \mbox{ } \forall \mbox{ } |x - x_0| 
< \delta$, so $I(x) > \frac{\epsilon}{2} \mbox{ } \forall \mbox{ } |x - x_0| < \delta$, so 
$\int \textrm{d}^3xI(x) > K\delta^3\epsilon >0$, where $K$ is some positive constant, which is 
a contradiction.

The above nonexistence is the {\sl integral inconsistency} of III.2.  
A well-known case is non-maintainability of maximal slicing in CWB GR, tied to the LFE 
$\triangle^2N = NR$. This works out because $R$ is positive-definite from ${\cal H}$ and the 
lapse $N$ is strictly positive by definition, so the integrand is positive and hence cannot vanish.  

This is bypassed in GR by using CMC slicing instead.  Then the LFE is rather 

\noindent 
$\triangle N = N\left(R + \frac{p^2}{4h}\right) + C(\lambda)$, and $C(\lambda)$ may be 
taken to be negative to ruin the above argument.  This is also the means by which 
CS+V theory and Newton--York absolute time theory work.  Conformal gravity and Kelleher's theory work 
instead by having LFE's of a special integro-differential form 
$\triangle^2N = \overline{NQ} \equiv NQ - <NQ>$ 
arising from adopting volume-divided actions.  Then as $\int\textrm{d}\Omega\overline{NQ}$ is trivially 0, 
the integral inconsistency becomes irrelevant by construction.  
Finally, note that if both tricks are applied simultaneously, 
$\overline{C} = 0$ so the CMC resolution is not then possible (this is why volume-divided 
Newton--York absolute time theory is just Kelleher's theory again).  

\mbox{ }

\noindent\large{\bf 2 Lichnerowicz--York equation}\normalsize

\mbox{ }

\noindent The arbitrary-dimensional (but most definitely $s = 0$!) Lichnerowicz--York equation 
is a nonlinear elliptic p.d.e
$$
\triangle\psi = \mbox{Nonlin}(\psi) \equiv - c\psi(R - M\psi^c + m^2\psi^b - 2\rho\psi^a)
\mbox{ } ,
$$
\be
b = \frac{4}{n - 2} \mbox{ } , \mbox{ } c = \frac{1}{(1 - n)b} \mbox{ } , \mbox{ } 
a = 
\left\{ 
\begin{array}{ll} 
\mbox{arbitrary}             & j^b   =  0 \\
\frac{1}{2c}                 & j^b \neq 0 
\end{array} 
\right.
\mbox{ } .  
\label{Nonlinlich}
\ee
Note that the nonlinearity is of the relatively mild form known as \it quasilinearity \normalfont \cite{LU}
since $|\pa\psi|^2$ does not appear (due to the artful construction on p 47).  
A number of theorems apply to equations of this form.  

\mbox{ }

\noindent{\bf 2.1 Linearization}

\mbox{ }

\noindent First linearize it by setting 
$
\psi = \psi_0 + \varepsilon\psi_1$, 
$\rho = \rho_0 + \varepsilon\rho_1 
$
in (\ref{Nonlinlich}) and equate the $O(\varepsilon)$ terms to obtain   
$$
 [\triangle - \mbox{Lin}(\psi_0)] \psi_1 = \mbox{Inh}(\psi_0) 
\mbox{ } , \mbox{ } \mbox{ } 
$$
\be
\mbox{Lin}(\psi_0) = -c[R -(1 + c)M\psi^c_0 + m^2(1 + b)\psi^b_0 - 2(1 + a)\rho\psi_0^a ]
\mbox{ } , \mbox{ } \mbox{ }
\mbox{Inh}(\psi_0) = 2c\rho_1\psi^{1 + a}_0 
\mbox{ } .
\label{linlich}
\ee
If asymptotically 
$
h_{ab} \longrightarrow h^{\mbox{\scriptsize A\normalsize}}_{ab}$ 
and $\psi_0 \longrightarrow \psi_0^{\mbox{\scriptsize A\normalsize}}, 
$
one obtains  
$[\triangle^{\mbox{\scriptsize A\normalsize}} - \mbox{Lin}(\psi_0^{\mbox{\scriptsize A\normalsize}})]\psi_1 = 
\mbox{Inh}(\psi_0^{\mbox{\scriptsize A\normalsize}})$ 
from which the asymptotic behaviour of $\psi$ for the full equation may be obtained.  

NB this is an equation of the form (\ref{genlinell}).  Many b.v.p's for such equations are 
well-studied \cite{CH, ellbooks, LU}.  To give an idea of the techniques available, 
consider the Dirichlet problem: (\ref{genlinell}) in $\Sigma$ together with b.c  $\psi_1 = f$ on $\pa\Sigma$.    
One then proves existence from uniqueness \cite{LU}, by setting up a sequence of problems 
interpolating between the flat-space Laplace equation and (\ref{genlinell}).\fn{A word 
of caution as regards the applicability of this to (\ref{helmics}) equations: 
as $h^{ij}$ and $\Gamma^i$ have geometrical significance, (\ref{helmics})
equation is on an underlying $n$-manifold $\Sigma$ (generally with boundary).  
I thus caution that if $\Sigma$ cannot be deformed to flat space, 
the interpolation methods between the $\Re^n$ Laplacian on a portion of $\Re^n$ and our operator 
cannot apply.  To avoid this trouble, one could study a sufficiently small 
region of $\Sigma$ (as $\Sigma$ is a manifold and hence locally $\Re^n$ or restrict the 
allowed topology of $\Sigma$ to be the same as that of a portion of $\Re^n$.}  

\mbox{ }

\noindent{\bf 2.2 The full equation and its simplifications}

\mbox{ }

\noindent One can then interpolate between quasilinear elliptic equations such as the full Lichnerowicz 
equation and their linearizations.  Then one has topological theorems about the fixed points of suitable 
maps, such as Leray--Schauder degree theory \cite{LU}, to prove existence of the solutions to 
the quasilinear elliptic equation's Dirichlet problem.  Similar techniques are available for 
further b.v.p's \cite{LU}.  

Thus in handling the Lichnerowicz equation, one has an enhanced chance of having a well-behaved b.v.p 
if the corresponding Lin $\geq 0$. To this end, notice that $M$ and $m^2$ are always positive whereas 
the troublesome $R$ is of variable sign.  

One may furthermore try three simplifications:

\noindent i) Conformal flatness, which replaces $\triangle$ by the flat-space 
$\triangle_{\mbox{\scriptsize F\normalsize}}$ and wipes out the arbitrary-sign $R$ term.

\noindent ii) Maximality: $m = 0$ removes the highest-order term of Nonlin($\psi$). 

\noindent iii) Setting $M = 0$. If this is combined with ii), 
one has {\it moment of time symmetry }
$K_{ij} = 0$, which is regarded as overly simple in the standard applications of 
numerical relativity as it corresponds to the absence of gravitational momentum on 
the slice $\Sigma$.  

\noindent One can sometimes also use a trick: to scale $\rho$ like $M$, $R$ or $m^2$ 
and then fix it so as to cancel with this term.  

In the $j^a \neq 0$ case, the cancellation trick above is not available if one wishes 
the procedure to respect the DEC.  Simplification i) is convenient in this case.    

\mbox{ }

\noindent{\bf 2.3 Boundary conditions}

\mbox{ }

\noindent I consider these here for 3-d single spherical compact object data, in support of the braneworld 
black hole application in B.3.

The inner radial b.c, is commonly established using an isometry.  For example, for a stellar source,  
one could  use $\tilde{h}_{ab}(x_{\alpha}) = \tilde{h}_{ab}(-x_{\alpha})$ 
$\Rightarrow$ 
$\left. \frac{\pa \tilde{h}_{ab}}{\pa x_{\alpha}}\right| _{x_{\alpha} = 0} = 0$ 
$\Rightarrow$ 
$\left. \frac{\pa\tilde{h}_{ab}(x_{\alpha})}{\pa r}\right|_{r = 0}$ leading component-by-component to 
the homogeneous Neumann b.c (the so-called reflection b.c) 
\be 
\left. 
\frac{\pa \psi}{\pa r}
\right|
_{r = 0} = 0,
\label{irbc}
\ee  
and the restriction $ \left. \frac{\pa w}{\pa r} \right| _{r = 0} = 0$ on the valid form of the known function.

For a black hole, $r = 0$ is singular.  One excises this by use of an inversion-in-the-sphere 
isometry about some throat at radius $a$ of the black hole within the apparent horizon [see fig 20a)]. 
\fn{One has no choice but to work with apparent horizons rather than event horizons 
if one is given a single spatial slice.  At least the apparent horizon lies within the event horizon in GR.}  
This gives an inner Robin condition \cite{BY} 
\be
\left.
\left[
\frac{\pa \psi }{\pa r} + \frac{1}{2a}\psi
\right]
\right|_{r = a} = 0.
\ee

Asymptotic flatness provides a natural outer radius ($r \longrightarrow \infty$) Dirichlet b.c.  
Would like subleading terms because of working on a finite grid in practice.
Sophisticated analysis might rather use a mode-selecting outer Robin condition.

\mbox{ } 

\mbox{ }

\mbox{ }

\noindent\large{\bf 3 Techniques for solving the momentum constraint}\normalsize

\mbox{ }

\noindent 
The moment of time symmetry simplification $K_{ij} =0$ means no momentum constraint need be considered.  
This alongside conformal flatness and vacuum assumptions gives the simplest conformal method IVP: the 
remaining 
Lichnerowicz equation reduces to a flat-space Laplace equation, while its electrovacuum counterpart is two 
Laplace equations (the above and the electrostatic one). But moment of time symmetry corresponds to absence 
of gravitational momentum.  More general data are thus desirable.  In order to have a nonzero $j^b$, one 
furthermore requires $K_{ij}$ to have a longitudinal part.  The simplest longitudinal method is when 
the 
metric is conformally-flat.  Then the `Hodge decomposition' $W_i = V_i + \pa_i U$ of the vector potential 
reduces the momentum constraint to $n$ + 1 linear scalar Poisson 
\be
\mbox{equations \cite{BY} }
\mbox{\hspace{1.4in}}
\triangle_{\mbox{\scriptsize F\normalsize}}U = -\frac{n - 2}{2(n - 1)}\pa_jV^j \mbox{ } , 
\mbox{ } \mbox{ }  \triangle_{\mbox{\scriptsize F\normalsize}}V_i = -\epsilon j_i 
\mbox{ } .
\mbox{\hspace{1.4in}}
\ee
which are what is solved for e.g Bowen--York data.

The full momentum constraint is a well behaved vector elliptic equation.  
Recollect that it is decoupled from the Lichnerowicz equation by virtue of the CMC slicing.  
The conformal thin sandwich method requires all the constraints and the LFE to be treated 
together as a system.  In both these cases the momentum constraint's good elliptic properties 
help ensure the whole system works out.  

In contrast to these fortunately elliptic formulations for the momentum constraint, 
its traditional sandwich counterpart (\ref{thinsaneq}) is only elliptic in small regions, so the traditional 
thin sandwich is not thereby furbished with powerful theorems.   The Einstein--Maxwell extension 
of this system (including coupling to the electromagnetic Gauss constraint) is interesting, 
since the overall symbol becomes somewhat different \cite{thin sandwich3}.    

I note that my `Codazzi equation' (\ref{irredelcod}) is more complicated than usual.  
I do not expect this equation will be more than elliptic in small regions.  
The involvement of various values of $W$ in theories in this thesis 
will directly affect the ellipticity properties of their thin sandwich formulations. 
That this alters the nature of the principal symbol may be easily inferred from \cite{York83}.

\vspace{4in}

\mbox{ }

\mbox{ }

\noindent\Large{\bf Useful list of acronyms}\normalsize

\mbox{ }

\noindent AF arbitrary frame

\noindent BEFE braneworld Einstein field equation

\noindent BM best matching, or best matched 

\noindent CS conformal Superspace

\noindent CMC constant mean curvature

\noindent CWB compact without boundary

\noindent DEC dominant energy condition

\noindent DOD domain of dependence

\noindent ELE Euler--Lagrange equations

\noindent IDV ``independent dynamical variable''

\noindent LFE lapse-fixing equation

\noindent POE principle of equivalence

\noindent POT Problem of time

\noindent RCS relative configuration space

\noindent RI reparameterization invariant 

\noindent RP relativity principle

\noindent RWR relativity without relativity

\noindent SSF split spacetime framework

\noindent THEFE timelike hypersurface Einstein field equation

\noindent TSA 3-space approach

\noindent WDE Wheeler--DeWitt equation


\begin{thebibliography}{99}

\scriptsize

\bibitem{stdbr1}              K. Akama, Lect. Notes Phys. \bf 176 \normalfont 267 (1982), hep-th/0001113.

\bibitem{nedm}                I.S. Altarev et al, Phys. Lett. \bf B276 \normalfont 242 (1992).

\bibitem{AS}                  K.A. Ames and B. Straughan, \it Non-standard and Improperly Posed Problems \normalfont (Academic Press, San Diego 1997).

\bibitem{AY}                  A. Anderson and J.W. York, Phys. Rev. Lett. {\bf 82} 4384 (1999), gr-qc/9901021.

\bibitem{Sanderson}           E. Anderson, Gen. Rel. Grav. {\bf 36} 255, gr-qc/0205118.

\bibitem{Vanderson}           E. Anderson, Phys. Rev. {\bf D68} 104001 (2003), gr-qc/0302035.  

\bibitem{Rio1}                E. Anderson, ``Spacetime or Space and the Problem of Time", 
                              to appear in proceedings of 2003 Marcel Grossmann meeting, gr-qc/0312037.

\bibitem{Rio2}                E. Anderson, ``Geometry, P.d.e's and Foundations for GR-based Braneworlds", 
                              to appear in proceedings of 2003 Marcel Grossmann meeting, gr-qc/0402010.

\bibitem{LAnderson}           E. Anderson, ``Leibniz--Mach foundations for GR and fundamental physics", 
                              to appear in {\it Progress in General Relativity and Quantum Cosmology 
                              Research} (Nova, New York), gr-qc/0405022.  

\bibitem{AB}                  E. Anderson and J.B. Barbour, 
                              Class. Quantum Grav. \bf 19 \normalfont 3249 (2002), gr-qc/0201092.

\bibitem{CG}                  E. Anderson, J.B. Barbour, B.Z. Foster and N. \'{O} Murchadha, 
                              Class. Quantum Grav. {\bf 20} 157 (2003), gr-qc/0211022.

\bibitem{ABFKO}               E. Anderson, J.B. Barbour, B.Z. Foster, B. Kelleher and N. \'{O} Murchadha, 
                              ``A first-principles derivation of York scaling and the Lichnerowicz--York equation", 
                              gr-qc/0404099. 

\bibitem{AL}                  E. Anderson and J.E. Lidsey, Class. Quant. Grav. \bf 18\normalfont 4831 (2001), gr-qc/0106090.

\bibitem{ADLR}                E. Anderson, F. Dahia, J.E. Lidsey and C. Romero, J. Math. Phys {\bf 44} 5108 (2003), gr-qc/0111094.

\bibitem{ATlett}              E. Anderson and R. Tavakol, Class. Quant. Grav. {\bf 20} L267 (2003), gr-qc/0305013.

\bibitem{ATpap}               E. Anderson and R. Tavakol, gr-qc/0309063.  

\bibitem{Annisim}             P. Anninos et al, Phys. Rev. Lett. {\bf 71} 2851 (1993), gr-qc/9309016.  

\bibitem{stdbr3}              N. Arkani-Hamed, S. Dimopoulos and G. Dvali, Phys. Lett. \bf B429 \normalfont 263 (1998), hep-ph/9803315.

\bibitem{ADM}                 R. Arnowitt, S.Deser and C.W. Misner, in \it{Gravitation: an Introduction to Current Research} \normalfont ed L. Witten (Wiley, New York 1962).

\bibitem{Vfinsler}            G.S. Asanov, \it Finsler Geometry, Relativity and Gauge Theories \normalfont (Reidel, Dordrecht 1985).

\bibitem{Ashtekar1}           A. Ashtekar, Phys. Rev. Lett. {\bf 57} 2244 (1986).

\bibitem{Ashtekar2}           A. Ashtekar, Phys. Rev. {\bf D 36} 1587 (1987). 
 
\bibitem{A88}                 A. Ashtekar, {\it New perspectives in canonical quantum gravity} (Bibliopolis, Napoli 1988).  

\bibitem{Ashtekar}            A. Ashtekar (notes prepared in collaboration with R. Tate), \it Lectures on Nonperturbative Canonical Gravity \normalfont (World Scientific, Singapore 1991).


\bibitem{Aetal}               A. Ashtekar et al, J. Math. Phys. {\bf 36} 6456 (1995), gr-qc/9504018. 

\bibitem{BSW}                 R.F. Baierlein, D. Sharp and J.A. Wheeler, Phys. Rev. \bf 126 \normalfont 1864 (1962).

\bibitem{Barbero}             J.F. Barbero, Phys. Rev. {\bf D 51} 5507 (1995), gr-qc/9410014. 

\bibitem{B74}                 J.B. Barbour, Nature \bf 249 \normalfont 328 (1974); 
                              errata \bf 250 \normalfont 606.  

\bibitem{DOD}                 J.B. Barbour, {\it Absolute or Relative Motion? Vol 1: 
                              The Discovery of Dynamics} (Cambridge University Press, Cambridge, 1989).  

\bibitem{B94I}                J.B. Barbour, Class. Quantum Grav. \bf 11 \normalfont 2853 (1994).

\bibitem{B94II}               J.B. Barbour, Class. Quantum Grav. \bf 11 \normalfont 2875 (1994).

\bibitem{B95}                 J.B. Barbour, ``GR as a perfectly Machian theory'', in \cite{buckets}. 

\bibitem{B99}                 J.B. Barbour, in \it The Arguments of Time\normalfont, ed. J. Butterfield (Oxford University Press, Oxford 1999).

\bibitem{EOT}                 J.B. Barbour, {\it The End of Time} (Oxford University Press, Oxford 1999).


\bibitem{CGPD}                J.B. Barbour, Class. Quantum Grav. \textbf{20}, 1543 (2003), gr-qc/0211021.

\bibitem{BB77}                J.B. Barbour and B. Bertotti, Nuovo Cim. \bf B38 \normalfont 1 (1977).

\bibitem{BB82}                J.B. Barbour and B. Bertotti, Proc. Roy. Soc. Lond. \bf A382 \normalfont 295 (1982).

\bibitem{BOF}                 J.B. Barbour, B.Z. Foster and N. \'{O} Murchadha, Class. Quantum Grav. \bf 19 \normalfont 3217 (2002), gr-qc/0012089.

\bibitem{BOF2}                J.B. Barbour, B.Z. Foster and N. \'{O} Murchadha, forthcoming; is in v1 of preprint of \cite{BOF}.  

\bibitem{conformal}           J.B. Barbour and N. \'{O} Murchadha, gr-qc/9911071.

\bibitem{buckets}             \it Mach's principle: From Newton's Bucket to Quantum Gravity\normalfont, 
                              ed. J.B. Barbour and H. Pfister (Birkh\"{a}user, Boston 1995).


\bibitem{BarDab}                  See, for example, J.D. Barrow and M.P. Dabrowski, in Conference Proceedings, ISMC Potsdam, Germany, 30 March - 4 April 1998, gr-qc/9706023.

\bibitem{thin sandwich2}      R. Bartnik and G. Fodor, Phys. Rev. {\bf D48}, 3596 (1993). 

\bibitem{asymbranes}          R.A. Battye, B. Carter, A. Mennim and J.P. Uzan, Phys. Rev. \bf D64 \normalfont 124007 (2001), hep-th/0105091.

\bibitem{BSrev}               T.W. Baumgarte and S.L. Shapiro,  Phys. Rept. \bf 376 \normalfont 41 (2003), gr-qc/0211028.

\bibitem{thin sandwich1}      E.P. Belasco and H.C. Ohanian, J. Math. Phys. {\bf 10}, 1053 (1969).

\bibitem{BEAR}                G. Belot and J. Earman, in {\it Physics meets philosophy at the Planck scale},
                              eds. C. Callender and N. Huggett (Cambridge University Press, Cambridge, 2001).
 
\bibitem{SGBKL}               V.A. Belinskii, I.M. Khalatnikov and E.M. Lifshitz, Adv. in Phys. \bf 19 \normalfont 525 (1970).

\bibitem{SGBKLmod}            B.K. Berger, talk given at \it International Symposium on Frontiers of Fundamental Physics \normalfont 
                              Hyderabad, India, Dec 11-12, 1997 (1998); 
                              B.K. Berger, Living Rev. Relativity 2002-1, gr-qc/0201056. 

\bibitem{BishopBerkeley}      Bishop G. Berkeley, \it The Principles of Human Knowledge \normalfont (1710); 
                              \it Concerning Motion (De Motu) \normalfont (1721).

\bibitem{Bern}                Z. Bern et al, Nu. Phys. {\bf B530} 401 (1998). 

\bibitem{Cassini}             B. Bertotti, L. Iess and P. Tortora, Nature {\bf 425} 374 (2003).

\bibitem{Langlois1}           P. Binetruy, C. Deffayet and D. Langlois, Nucl. Phys. \bf B565 \normalfont 269 (2000), hep-th/9905012; 
                              P. Binetruy, C. Deffayet, U. Ellwanger and D. Langlois, Phys. Lett. \bf B477 \normalfont 285 (2000), hep-th/9910219.

\bibitem{BL22}                G.D. Birkhoff and R.E. Langer, {\it Relativity and Modern Physics}
                              (Harvard University Press, 1923).  

\bibitem{Bojowald}            See, for example, M. Bojowald and H.A. Morales-Tecotl, in proceedings of the Fifth Mexican School (DGFM): 
                              The Early Universe and Observational Cosmology, gr-qc/0306008, and references therein.  

\bibitem{bombelli}            See chapter III.4 written up by L. Bombelli in \cite{A88}.

\bibitem{Gregory}             F. Bonjour, C. Charmousis and R. Gregory, Phys. Rev. \bf D62 \normalfont 083504 (2000), gr-qc/0002063.

\bibitem{CGBD}                C. Brans and R. Dicke, Phys. Rev. \normalfont \bf 124 \normalfont 925 (1961).

\bibitem{BLdata}              D.R. Brill and R.W. Lindquist, Phys. Rev. {\bf 131} 471 (1963).

\bibitem{CGunique}            D.R. Brill and F. Flaherty Comm. Math. Phys. \normalfont \bf 50 \normalfont 157 (1976). 

\bibitem{CMClit}              D.R. Brill, J.M. Cavallo and J.A. Isenberg, J. Math. Phys. \bf 21 \normalfont 2789 (1980).  

\bibitem{exptinrin}           A. Brillet and J.L. Hall, Phys. Rev. Lett.  {\bf 42} 549 (1979).

\bibitem{B52}                 Y. Bruhat, Acta Mathematica \bf 88 \normalfont 141 (1952).  

\bibitem{B56}                 Y. Bruhat, J. Rat. Mech. Anal. \bf 5 \normalfont 951 (1956).

\bibitem{Cauchylit}           Y. Bruhat, in \it Gravitation: an introduction to current research\normalfont, ed. L. Witten (Wiley, New York, 1962).

\bibitem{MB}                  M. Bruni and P.K.S. Dunsby, Phys. Rev. \bf D66 \normalfont (2002) 101301 (2002), hep-th/0207189. 

\bibitem{Vhdl2}               I.L. Buchbinder and S.L. Lyahovich, Class. Quantum Grav \bf 4 \normalfont 1487 (1987).

\bibitem{buchdrag}            W. Buchm$\ddot{\textrm u}$ller and N. Dragon, Nu. Physics \textbf{B321} 207 (1989).

\bibitem{VTB}                 A. Burd and R. Tavakol, Phys. Rev. D \bf 47 \normalfont 5336 (1993).

\bibitem{VButterfield}        J.N. Butterfield, Brit. J. Phil. Sci. {\bf 53 } 289 (2002), gr-qc/0103055.

\bibitem{Campbell}            J. Campbell, \it A course of Differential Geometry \normalfont (Clarendon, Oxford 1926).  

\bibitem{Coley11}             A. Campos and C.F. Sopuerta, Phys. Rev. \bf D63 \normalfont 104012 (2001), hep-th/0101060.

\bibitem{Cantor}              M. Cantor, Commun. Math. Phys. \bf 57 \normalfont 83 (1977).

\bibitem{SGCapovilla}         R. Capovilla, Phys. Rev. \bf D46 \normalfont 1450 (1992), gr-qc/9207001.

\bibitem{Carlipbook}          S.J. Carlip, {\it Quantum Gravity in 2 + 1 Dimensions} (Cambridge University Press, Cambridge, 1998).

\bibitem{Carliprev}           S.J. Carlip, Rept. Prog. Phys. {\bf 64} 885 (2001)  , gr-qc/0108040.

\bibitem{ATTLG}               L. Carroll, \it Through the Looking Glass and What Alice Found There \normalfont 
                             (MacMillan, London, 1871).  

\bibitem{VCartan}             E. Cartan, J. Math. Pure Appl. \bf 1 \normalfont 141 (1922).

\bibitem{Cartan25}            E. Cartan, Mem Sci. Math. {\bf 8} (1925).

\bibitem{Cartansp}            E. Cartan, Ann. Ec. Norm. Sup. {\bf 40} 325 (1923); 
                              {\bf 41} 1 (1924).  

\bibitem{Castagnino}          M. Castagnino and R. Laura, Int J. Theor. Phys. {\bf 39} 1737 (2000), gr-qc/006012.

\bibitem{311lit}              A. Chamblin, H.S. Reall, H. Shinkai and T. Shiromizu, Phys. Rev. \bf D63 \normalfont 064015 (2001), hep-th/0008177.

\bibitem{Lovelockbranes}      See for example  C. Charmousis and J-F. Dufaux,  Class. Quantum Grav. \bf 19 \normalfont 4671 (2002), hep-th/0202107. 

\bibitem{CBY}                 Y. Choquet-Bruhat and J.W. York, in \it General Relativity and Gravitation \normalfont ed. A. Held, vol. 1 (Plenum Press, New York, 1980).

\bibitem{B76}                 Y. Choquet-Bruhat, in \it Differential Geometry and Relativity \normalfont, ed. M. Cahen and M. Flato (D.Reidel, Dordrecht 1976).  

\bibitem{CIY}                 Y. Choquet-Bruhat, J. Isenberg and J.W. York, Phys. Rev. \bf D61 \normalfont 084034 (2000), gr-qc/9906095.


\bibitem{Clarke}              C.J.S. Clarke, \it The Analysis of Space-Time Singularities \normalfont (Cambridge University Press, Cambridge 1993).

\bibitem{Coleyconj}           A.A. Coley, Class. Quantum Grav. \bf19 \normalfont L45 (2002), hep-th/0110117.


\bibitem{Cook}                G.B. Cook, Living Rev. Rel. \bf 3 \normalfont 5 (2000), gr-qc/0007085.                          

\bibitem{Cook2}               The Binary Black Hole Grand Challenge Alliance: G.B. Cook et al, Phys. Rev. Lett. \bf 80 \normalfont  2512 (1998), gr-qc/9711078.

\bibitem{otherpapers11}       E.J. Copeland, A.R. Liddle and J.E. Lidsey, Phys. Rev. \bf D64 \normalfont 023509 (2001), astro-ph/0006421. 

\bibitem{CH}                  R. Courant and D. Hilbert, \it Methods of Mathematical Physics \normalfont Vol. 2 (John Wiley and Sons, Chichester 1989).

\bibitem{CutThorne}           C. Cutler and K. Thorne, for Proceedings of GR16, gr-qc/0204090.  

\bibitem{BH}                  N. Dadhich, R. Maartens, P. Papadopoulos and V. Rezania, Phys. Lett. \bf B487 \normalfont 1 (2000), hep-th/0003061.

\bibitem{letter}              T. Damour and K. Nordtvedt, Phys. Rev. Lett. \bf 70 \normalfont 2217 (1993).

\bibitem{paper}               T. Damour and K. Nordtvedt, Phys. Rev. \bf D48 \normalfont 3437 (1993).

\bibitem{dampich}             T. Damour and B. Pichon,  Phys. Rev. \bf D59 \normalfont 123502 (1999), astro-ph/9807176.

\bibitem{Darmois23}           G. Darmois, Comptes Rendus {\bf 176} 646; 731 (1923); 
                              G. Darmois, Annales de Physique {\bf 1} 5 (1924).

\bibitem{Darmois27}           G. Darmois, Mem. Sc. Math. \bf 25 \normalfont (1927).  

\bibitem{Vsugy4}              P.D. D'Eath, Phys. Rev. \bf D29 \normalfont 2199 (1984).

\bibitem{Vsugy5}              P.D. D'Eath, \it Supersymmetric Quantum Cosmology \normalfont (Cambridge University Press, Cambridge, 1996).

\bibitem{2+2}                 R.A. d'Inverno and J. Stachel, J. Math. Phys. {\bf 19} 2447 (1978).

\bibitem{deDonder}            See T. de Donder, in Mem. Sci Math. {\bf 8} (1925) and references therein.   



\bibitem{Vsugy1}              S. Deser, J.H. Kay and K.S. Stelle, Phys. Rev. D \bf 16 \normalfont 2448 (1977).

\bibitem{DeWitt62}            B.S. DeWitt, in \it Gravitation: an introduction to current research \normalfont, ed. L. Witten (Wiley, New York 1962).

\bibitem{DeWitt}              B.S. DeWitt, Phys. Rev. \bf 160 \normalfont 1113 (1967).

\bibitem{DeWittcov}           B.S. DeWitt, Phys. Rev. \bf 160 \normalfont 1195 (1967).

\bibitem{VDeWitt70}           B.S. DeWitt, in \it Relativity \normalfont (Proceedings of the Relativity Conference in the Midwest, held at Cincinnati, Ohio June 2-6, 1969), 
                              ed. M. Carmeli, S.I. Fickler and L. Witten (Plenum, New York 1970).


\bibitem{Dirac51}             P.A.M. Dirac, Canadian J. Math. \bf 3 \normalfont 1 (1951).  

\bibitem{CGDiracPRC}          P.A.M. Dirac,  Proc. Roy. Soc. \textbf{A246}, 333 (1958).

\bibitem{VDiracrec}           P.A.M. Dirac, in \it Recent developments in General Relativity \normalfont  (Pergamon, Oxford, 1962).

\bibitem{Dirac}               P.A.M. Dirac, \it Lectures on Quantum Mechanics \normalfont (Yeshiva University, NY, 1964).

\bibitem{CGDirac73}           P.A.M. Dirac,   Proc. Roy. Soc. Lond. \bf A 333 \normalfont 403 (1973).

\bibitem{Drever}              R.W.P. Drever, Phil. Mag. {\bf 6} 683 (1961).  

\bibitem{KK}                  See, for example, M.J. Duff, Talk delivered at the Oskar Klein Centenary Nobel Symposium, hep-th/9410046.

\bibitem{SGVDTandAVDT1}       D. Eardley, E. Liang and R. Sachs, J. Math. Phys. \bf 13 \normalfont 99 (1972).

\bibitem{Eardley}             D.M. Eardley and L. Smarr, Phys. Rev. {\bf D19} 2239 (1979).  

\bibitem{Earman89}            J. Earman, \it World Enough and Space-Time: Absolute versus Relational Theories of Space and Time \normalfont (MIT Press, Cambridge MA, 1989).

\bibitem{Ehlers}              J. Ehlers, in {\it The Physicist's Conception of Nature}, ed. J. Mehra (Kluwer, Dordrecht, 1973).

\bibitem{OEOMB}               A. Einstein, Ann. Phys. (Germany) {\bf 17} 891 (1905).  The English translation is available in \it The Principle of Relativity \normalfont
                             (Dover, New York 1952, formerly published by Methuen, London 1923).

\bibitem{poe}                 A. Einstein, Ann. Phys. (Germany) {\bf 35} 898 (1911).  The English translation is available in \it The Principle of Relativity \normalfont
                             (Dover, New York 1952, formerly published by Methuen, London 1923).

\bibitem{EinGR}               A. Einstein finally gets his field equations in Preuss. Akad. Wiss. Berlin, Sitzber, 844 (1915); 
                              earlier work is quoted on p432-433 of \cite{MTW}.    
     
\bibitem{VEinstein}           A. Einstein reviews the outcome of the above work in Ann. Phys. (Germany) \bf 49 \normalfont 769 (1916).
                              The English translation is available in \it The Principle of Relativity \normalfont
                             (Dover, New York 1952, formerly published by Methuen, London 1923).
     
\bibitem{CGconstructive}      A. Einstein, The Times, November 28 (1919). 

\bibitem{CGEinstein1}         Einstein's objection was reported by Weyl at the end of \cite{CGWeyltheory}.

\bibitem{CGEinstein2}         A. Einstein, Sitzungsber. d. Preuss. Akad. d. Wissensch. 261 (1921). 

\bibitem{Ein34}               A. Einstein, {\it Mein Weltbild } (Querido Verlag, Amsterdam, 1933).  
                              The English translation is in {\it Essays in Science} (Philosophical Library, New York, 1934).

\bibitem{CGspacetime}         A. Einstein ``Autobiographical notes'' in: \textit{Albert Einstein: Philosopher--Scientist}, 
                              ed. P.A. Schilpp (Library of Living Scientists, Evanston 1949).

\bibitem{Ein50}               A. Einstein, {\it The Meaning of Relativity} (Princeton University Press, Princeton 1950).

\bibitem{EG13}                A. Einstein and M. Grossmann, Zeit. Math. Phys. {\bf 62 } 225 (1913), 
                              quoted in English in \cite{MTW}.  

\bibitem{Eisenhart26}         L.P. Eisenhart, {\it Riemannian Geometry} (Princeton University Press, Princeton 1926).  

\bibitem{Ellisthread}         G.F.R. Ellis, in \it General Realtivity and Cosmology\normalfont, ed. R.K. Sachs (Academic, New York, 1971).

\bibitem{sclass}              G.F.R Ellis and B.G. Schmidt, Gen. Rel. Grav. {\bf 8} 915 (1977).  

\bibitem{CGconscience}        G.F.R. Ellis, R. Maartens and S.D. Nel,   
                              Mon. Not. R. Astr. Soc. \bf 184 \normalfont 439 (1978).


\bibitem{VFP}                 M. Fierz and W. Pauli, Proc. R. Soc. Lond. A \bf 173 \normalfont (1939).

\bibitem{VFischer70}          A.E. Fischer, in \it Relativity \normalfont (Proceedings of the Relativity Conference in the Midwest, held at Cincinnati, Ohio June 2-6, 1969), 
                              ed. M. Carmeli, S.I. Fickler and L. Witten (Plenum, New York 1970).

\bibitem{Fischer86}           A.E. Fischer, J. Math. Phys {\bf 27} 718 (1986).  

\bibitem{CGFischMon}          See, for example, A.E. Fischer and V. Moncrief, Gen. Rel. Grav. \bf 28 \normalfont 221 (1996).

\bibitem{Fostercom}           B.Z. Foster, personal communication (Dec. 2003).  

\bibitem{FL}                  E.S Fradkin and V.Ya. Linetsky, Phys.  Lett. {\bf B261} 26 (1991).  

\bibitem{Vsugy2}              E.S. Fradkin and M.A. Vasiliev, Phys. Lett. \bf B72 \normalfont 70 (1977). 

\bibitem{Pilatilit3}          G. Francisco and M. Pilati, Phys. Rev. \bf D31 \normalfont 241 (1985).


\bibitem{sugy}                See, e.g, P.G.O. Freund, \it Introduction to Supersymmetry \normalfont (Cambridge University Press, Cambridge 1986).

\bibitem{FN}                  H. Friedrich and G. Nagy, Commun. Math. Phys. \bf 201 \normalfont 619 (1999).  

\bibitem{RF}                  H. Friedrich and A.D. Rendall, Lect. Notes Phys. \bf 540 \normalfont 127 (2000), gr-qc/0002074. 

\bibitem{dialogo}             G. Galilei, {\it Dialogo, Opere} {\bf 7} 299 (Edizione nazionale, Florence 1890-1909).

\bibitem{PG}                  R. Gambini and J. Pullin, {\it Loops, knots, gauge theories and quantum gravity} (Cambridge University Press, Cambridge 1996). 

\bibitem{TanGarr}             J. Garriga and T. Tanaka, Phys. Rev. Lett. \bf 84 \normalfont 2778 (2000), hep-th/9911055.
 
\bibitem{VGH}                 J. G\'{e}h\'{e}niau and M. Henneaux, Gen. Rel. Grav. \bf 8 \normalfont 611 (1977).

\bibitem{GMG}                 M. Gell-Mann and S.L. Glashow Ann. Phys. (N.Y.) \bf 15 \normalfont 437 (1961)

\bibitem{Gergely}             L.\'{A} Gergely, Class. Quantum Grav. {\bf 17} 1949 (2000), gr-qc/0003064;  
                              L.\'{A} Gergely and M. McKain, Class. Quantum Grav. {\bf 17} 1963, gr-qc/0003065.  

\bibitem{Gerlach}             U. Gerlach, Phys. Rev. {\bf 177} 1929 (1969).  

\bibitem{Geroch67}            R.P. Geroch, J. Math. Phys \bf 8\normalfont, 782 (1967).  


\bibitem{SGGibbons}           G.W. Gibbons, K. Hashimoto and P. Yi, hep-th/0209034.

\bibitem{sing2}               G.W. Gibbons, G.T. Horowitz and P.K. Townsend, Class. Quantum Grav. \bf 12 \normalfont 297 (1995), hep-th/9410073.
                     
\bibitem{SGGiulini}           D. Giulini, Phys. Rev. D \bf 51 \normalfont 5630 (1995), gr-qc/9311017.

\bibitem{thin sandwich3}      D. Giulini, J. Math. Phys. {\bf 40}, 1470 (1999).

\bibitem{giulini}             D. Giulini, personal communication (2001).

\bibitem{mathemb}             H.F. Goenner, in \it General Relativity and Gravitation \normalfont ed. A. Held (Plenum, New York 1980).
                     
\bibitem{stdbr4}              M. Gogberashvili, Europhys. Lett. {\bf 49} 396 (2000), hep-ph/9812365. 

\bibitem{cod2}                See, for example, M. Gogberashvili, J. Math. Phys. \bf 43 \normalfont 4886 (2002), gr-qc/0202061.


\bibitem{Gowdymodel}          R.H. Gowdy, Ann. Phys. (N.Y.) {\bf 83} 203 (1974). 

\bibitem{Gowdy}               See R.H. Gowdy, gr-qc/0107016 and references therein.  

\bibitem{VGSW}                M.B. Green, J.H. Schwartz and E. Witten, Superstring theory (Cambridge University Press, Cambridge 1987).

\bibitem{Greene}              R.E. Greene, Memoirs Amer. Math. Soc. {\bf 97} (1970).

\bibitem{SGstringyrefs12}     D.J. Gross, Phys. Rev. Lett \bf 60 \normalfont 1229 (1988).

\bibitem{GS}                  D.J. Gross and J.H. Sloan, Nu. Phys. \bf B291 \normalfont 41 (1987).

\bibitem{kuhaj}               P. Hajicek and K.V. Kucha\v{r}, Phys. Rev. {\bf D41 } 1091 (1990).  

\bibitem{Hallirev}            See, e.g, J.J Halliwell, in {\it Proceedings of the Seventh Jerusalem 
                              Winter School for Theoretical Physics: Quantum Cosmology and Baby Universes}, 
                              ed. S. Coleman, J.B. Hartle, T. Piran and S. Weinberg (World Scientific, Singapore, 1990). 

\bibitem{Hallimott}           J.J. Halliwell, Phys. Rev. {\bf D64} 044008 (2001), gr-qc/0008046; 
                              J.J. Halliwell and J. Thorwart, Phys. Rev. {\bf D65} 124018, gr-qc/0106095.  

\bibitem{HallHaw}             J.J. Halliwell and S.W. Hawking, Phys. Rev. {\bf 31} 1777 (1985).  

\bibitem{Hartlerev}           J.B. Hartle, in {\it Gravitation and Quantizations}, ed. 
                              B. Julia and J. Zinn-Justin (North Holland, Amsterdam, 1995), 
                              gr-qc/9304063. 

\bibitem{HH}                  J.B. Hartle and S.W. Hawking, Phys. Rev. {\bf D28} 2960 (1983).

\bibitem{Hawkingthread}       S.W. Hawking, Astrophys. J. \bf 145 \normalfont 544 (1966).

\bibitem{CGHawking}           For an introduction, see \cite{Wald}; \cite{HE}; there is also the original paper 
                              S.W. Hawking Proc. Roy. Soc. Lond. 
                              \bf A294 \normalfont 511 (1966).  

\bibitem{H84}                 S.W. Hawking, in {\it Relativity, Groups and Topology II}, eds. 
                              B.S. DeWitt and R. Stora (North Holland, Amsterdam, 1984).  

\bibitem{HE}                  S.W. Hawking and G.F.R. Ellis, \it The Large-Scale Structure of Space-Time \normalfont 
                             (Cambridge University Press, Cambridge 1973).

\bibitem{HP}                  S.W. Hawking and D.N. Page, Nu. Phys. {\bf B264} 185 (1986); 
                                                          Nu. Phys. {\bf B298} 789 (1988).  

\bibitem{Hen}                 M. Henneaux, Gen. Rel. Grav. {\bf 9} 1031 (1978).

\bibitem{SHenneaux}           M. Henneaux, Bull. Soc. Math. Belg. \bf 31 \normalfont 47 (1979).

\bibitem{AsADS}              For an introduction to the subtleties of asymptotics, see \cite{Wald}.
                             For the treatment of asymptotically AdS, see for example,  
                             M. Henneaux and C. Teitelboim Commun. Math. Phys. \bf 98 \normalfont 391 (1985).   

\bibitem{HTbook}             M. Henneaux and C. Teitelboim, {\it Quantization of Gauge systems} (Princeton University Press, Princeton, 1992).

\bibitem{Higgs}              P.W. Higgs, Phys. Rev. Lett. {\bf 13} 508 (1964).  

\bibitem{VKucharearly1}      S.A. Hojman and K.V. Kucha\v{r}, Bull. Amer. Phys. Soc. \bf 17 \normalfont 450 (1972).

\bibitem{VKucharearly3}      S.A. Hojman, K.V. Kucha\v{r} and C. Teitelboim, Nature (London) \bf 245 \normalfont 97 (1973).

\bibitem{HKT}                S.A. Hojman, K.V. Kucha\u{r} and C. Teitelboim, Annals of Physics \bf 96 \normalfont 88 (1976).

\bibitem{string1}             P. Ho\v{r}ava and E. Witten, Nucl. Phys. \bf B460 \normalfont 506 (1996), hep-th/9510209;
                              P. Ho\v{r}ava and E. Witten, Nucl. Phys. \bf B475 \normalfont  94 (1996), hep-th/9603142. 

\bibitem{Hooft}               G. t'Hooft, Phys Rev Lett \bf 37 \normalfont 8 (1976); G.t'Hooft, Phys. Rev.  \bf D14 \normalfont 3432 (1976).

\bibitem{Coley13}              R.J. van den Hoogen and J. Iba$\tilde{n}$ez, Phys. Rev. {\bf D67 } 083510 (2003), gr-qc/0212095;                        
                               R.J. van den Hoogen, A.A. Coley and Y. He,  Phys. Rev. {\bf D68 } 023502 (2003), gr-qc/0212094.  

\bibitem{Hughes}              V.W. Hughes, H.G. Robinson and V. Beltran-Lopez, Phys. Rev. Lett. {\bf 4} 342 (1960).

\bibitem{HKM}                 T.J.R. Hughes, T. Kato and J.E. Marsden, Arch. Rat. Mech. Anal. \bf 63 \normalfont 273 (1976).

\bibitem{Hughston}            L. Hughston and K. Jacobs Astrophys. J. {\bf 160} 147 (1970).  

\bibitem{SGHusain}            V. Husain, Class. Quantum Grav. \bf 5 \normalfont 575 (1988).

\bibitem{Isenberg81}          J.A. Isenberg, Phys. Rev. {\bf D24} 251 (1981).

\bibitem{Yam}                 J. Isenberg, GR16 talk, gr-qc/0203044. 

\bibitem{SGVDTandAVDT2}       J. Isenberg and V. Moncrief, Ann. Phys. \bf 199 \normalfont 84 (1990).

\bibitem{IN}                  J. Isenberg and J. Nester, Ann. Phys. (New York) {\bf 107} 56 (1977).  

\bibitem{IOY}                 J.A. Isenberg, N. \'{O} Murchadha and J.W. York \it Phys. Rev. D \normalfont \bf 13\normalfont 1532 (1976).  

\bibitem{WI79}                J. Isenberg and J.A. Wheeler, in \it Relativity, Quanta and Cosmology in the Development of the Scientific Thought of Albert Einstein \normalfont Vol. I, ed. M. Pantaleo and F. deFinis 
                              (Johnson Reprint Corporation, New York 1979)

\bibitem{SIsham}              C.J. Isham, Proc. R. Soc. Lond. A. \bf 351 \normalfont 209 (1976).

%

\bibitem{POTlit2}             C.J. Isham, in ``Integrable systems, quantum groups and quantum field theories" 
                              Eds. L.A. Ibort and M.A. Rodr\'{i}guez (Dordrecht: Kluwer 1993), preprint gr-qc/9210011.




\bibitem{qgravrevI00}         C.J. Isham and J. Butterfield, in {\it Physics meets philosophy at the Planck scale} 
                              Eds. C. Callender and N. Huggett (Cambridge University Press, Cambridge 2001).

\bibitem{KI85b}               C.J. Isham and K.V. Kucha\v{r}, Ann. Phys. (N.Y.) {\bf 164} 316 (1985).  

\bibitem{IL}                  C.J Isham and N. Linden, J. Math. Phys. {\bf 36} 5392 (1995), 
                              gr-qc/9503063.  

\bibitem{jns}                 W. Israel, Nuovo Cim. \bf 44B \normalfont (1966); errata \bf 48B \normalfont 463 (1966).  

\bibitem{Jacsugy}              T. Jacobson, Class. Quantum Grav. \bf 5 \normalfont 923 (1988).

\bibitem{Jacobbh}             T. Jacobson Phys. Rev. Lett. {\bf 75} 1260 (1995) gr-qc/9504004.  

\bibitem{SGstringyrefs2}      See for example, M. Kaku, \it Strings, Conformal Fields and Topology \normalfont (Springer--Verlag, New York 1991).

\bibitem{otherpapers13}       S. Kanno and J. Soda, hep-th/0303203.
\bibitem{scalarbulk}          See, e.g P. Kanti, S. Lee and K.A. Olive, Phys. Rev. \bf D67 \normalfont 024037 (2003), hep-ph/0209036.

\bibitem{Kelleher}            B. Kelleher, ``Gravity on Conformal Superspace" (PhD Thesis, University of Cork, 2003), gr-qc/0311034.    

\bibitem{Kellehertheory}      B. Kelleher, gr-qc/0310109.  

\bibitem{Turok}               J. Khoury, B.A. Ovrut, N. Seiberg, P.J. Steinhardt, N. Turok, Phys. Rev. \bf D65 \normalfont 086007 (2002),  hep-th/0108187; 
                              J. Khoury, B.A. Ovrut, P.J. Steinhardt, N. Turok, Phys. Rev. \bf D66 \normalfont 046005 (2002), hep-th/0109050.

\bibitem{VKiefer}             C. Kiefer, Phys. Lett. \bf B 225 \normalfont 227 (1989).  

\bibitem{KiefGiu}             C. Kiefer and D. Giulini, Phys. Lett. \bf A193 \normalfont 21 (1994),  gr-qc/9405040.

\bibitem{KM}                  C. Kiefer and E.A. Martinez, Class. Quantum Grav. \bf 10 \normalfont 2511 (1993), gr-qc/9306029.

\bibitem{Klainerman}          S. Klainerman and I. Rodnianski, preprint math.AP/0109173. 

\bibitem{Klauderlit1}         J.R. Klauder, Acta Physica Austriaca Suppl. \bf8 \normalfont 227 (1971). 

\bibitem{Klauderlit2}         J.R. Klauder, Commun. Math. Phys. \bf 18 \normalfont 307 (1970).

\bibitem{VKosmann}            Y. Kosmann, Ann. Mat. Pura Appl. (IV), \bf XCI \normalfont 317 (1972).

\bibitem{VKouletsis}          I. Kouletsis, Imperial College preprint TP/97-98/17, gr-qc/9801019; 
                              I. Kouletsis, ``Classical Histories in Hamiltonian Systems", (PhD Thesis, University of London, 2000), gr-qc/0108021.    

\bibitem{KoulKu}              I. Kouletsis and K.V. Kucha\v{r}, Phys. Rev. {\bf D65} 25026 (2002), gr-qc/0108022.

\bibitem{MC}                  H. Stephani, D. Kramer, M.A.H. MacCallum C. Hoenselaers and 
                              E. Herlt, \it Exact Solutions to {Einstein's} Field Equations\normalfont,
	                       (Cambridge University Press, Cambridge 2003).

\bibitem{CGKrasinski}         See Krasi\'{n}ski A 1997 ``Inhomogeneous Cosmological Models" (Cambridge: Cambridge University Press), esp. Appendix B and references therein.

\bibitem{Kucharcyl}           K.V. Kucha\v{r}, Phys. Rev. {\bf D4}, 955 (1971).

\bibitem{Kucharbubble}        K.V. Kucha\v{r}, J. Math. Phys. {\bf 13} 768 (1972).  

\bibitem {Kuchar74}           K.V. Kucha\v{r}, J. Math. Phys \bf 15 \normalfont 708  (1974).

\bibitem{VKucharearly2}       K.V. Kucha\v{r}, in \it Relativity, Astrophysics and Cosmology \normalfont, ed. W. Israel (Reidel, Dordrecht 1973).

\bibitem{KucharI}             K.V. Kucha\v{r}, J. Math. Phys. \bf 17 \normalfont 777 (1976).

\bibitem{KucharII}            K.V. Kucha\v{r}, J. Math. Phys. \bf 17 \normalfont 792 (1976).

\bibitem{KucharIII}           K.V. Kucha\v{r}, J. Math. Phys. \bf 17 \normalfont 801 (1976).

\bibitem{KucharIV}            K.V. Kucha\v{r}, J. Math. Phys. \bf 18 \normalfont 1589 (1977).

\bibitem{KuBacom}             K.V. Kucha\v{r}, personal communication to J.B. Barbour (1980), 
                              reported in \cite{BB82}.

\bibitem{Kuchar80}            K.V. Kucha\v{r}, in {\it Quantum Gravity 2: A Second Oxford Symposium}, eds. C.J. Isham, R. Penrose and D.W. Sciama (Clarendon, Oxford, 1981).

\bibitem{Ku81JMP}             K.V. Kucha\v{r}, J. Math. Phys. {\bf 22}  2640 (1981).  

\bibitem{POTlit1}             K.V. Kucha\v{r}, in \it Proceedings of the 4th Canadian Conference on 
                              General Relativity and Relativistic Astrophysics\normalfont,  
                              ed. G. Kunstatter, D. Vincent and J. Williams (World Scientific, Singapore, 1992).

\bibitem{CGK93}               K.V. Kucha\v{r}, in  ``General Relativity and Gravitation 1992"  
                             eds. R.J. Gleiser, C.N. Kozameh and O.M. Moreschi
                             (Institute of Physics, Bristol, 1993),  preprint gr-qc/9304012.

\bibitem{POTlit3}             K.V. Kucha\v{r}, in \it The Arguments of Time\normalfont, ed. J. Butterfield (Oxford University Press, Oxford 1999).

\bibitem{KucharRyan}          K.V. Kucha\v{r} and M.P. Ryan, Phys. Rev. {\bf D40} 3982 (1989).

\bibitem{KuTorre90}           K.V. Kucha\v{r} and C.G. Torre, Phys. Rev. {\bf D43} 419 (1991).  

\bibitem{KuTorre}             K.V. Kucha\v{r} and C.G. Torre, in {\it Conceptual Problems of Quantum 
                              Gravity}, eds. A. Ashtekar and J. Stachel (Birkh\"{a}user, Boston, 1991).  

\bibitem{Luminet}             See e.g M. Lachi\`{e}ze-Rey and J.P. Luminet, Phys. Rep. {\bf 254} 135 (1995).

\bibitem{LU}                  O.A. Ladyzhenskaya and N.N. Ural'tseva (Nauka Press, Moscow 1964). The English translation is \it Linear and Quasilinear Elliptic Equations \normalfont 
                             (Academic Press New York--London 1968). 

\bibitem{Lanczos22}           K. Lanczos, Ann. Phys. {\bf 13} 621 (1922). 

\bibitem{Lanczos23}           K. Lanczos, Phys Z. \bf 23 \normalfont 537 (1923). 

\bibitem{Lanczos}             C. Lanczos, \it The Variational Principles of Mechanics \normalfont (University of Toronto Press, Toronto 1949, Reprinted by Dover, New York 1986).

\bibitem{Lange}               L. Lange, Ber. Kgl. Gess. Wiss, Math-Phys. Kl. 333 (1885).  

\bibitem{SLMW}                D. Langlois, R. Maartens, M. Sasaki and D. Wands,  Phys. Rev. \bf D63 \normalfont 084009 (2001),  hep-th/0012044.        

\bibitem{LeibnizClarkecorr}   \it The Leibniz--Clark Correspondence\normalfont, ed. H.G. Alexander (Manchester 1956). 

\bibitem{otherpapers2}        B. Leong, A. Challinor, R. Maartens and A. Lasenby, Phys. Rev. \bf D66 \normalfont 104010 (2002), astro-ph/0208015; 

\bibitem{Leray}               J. Leray, \it Hyperbolic Differential Equations \normalfont (The Institute For Advanced Study, Princeton 1952).

\bibitem{Levin98}             J. Levin, E. Scannapieco, G. de Gasperis, J. Silk and J.D. Barrow, 
                              Phys. Rev. {\bf D58} 123006 (1998), astro-ph/9807206.  

\bibitem{CGLich}              A. Lichnerowicz, J. Math. Pures Appl.  \bf 23 \normalfont 37 (1944).

\bibitem{Lindquist63}         R.W. Lindquist, J. Math. Phys. {\bf 4} 938 (1963). 

\bibitem{Lindquist}           R.W. Lindquist and J.A. Wheeler, Rev. Mod. Phys. {\bf 29} 432 (1957).  

\bibitem{SGstringyrefs13}     U. Lindstrom, Invited lecture at INFN ELOISATRON Workshop ``From Superstrings To Supergravity", hep-th/9303173.  

\bibitem{Lorentz}             H.A. Lorentz, the English translation is available in \it The Principle of Relativity \normalfont
                             (Dover, New York 1952, formerly published by Methuen, London 1923).

\bibitem{VLovelock}           D. Lovelock, J. Math. Phys, \bf 12 \normalfont 498 (1971).

\bibitem{Lovelocktensor}      D. Lovelock, J. Math. Phys. \bf 13 \normalfont 874 (1972).

\bibitem{string2}             A. Lukas, B.A. Ovrut and D. Waldram, Phys. Rev. \bf D60 \normalfont 086001 (1999);
                              M. Braendle, A. Lukas, B.A. Ovrut, Phys. Rev. \bf D63 \normalfont 026003 (2001) hep-th/0003256. 

\bibitem{LB}                  D. Lynden-Bell, in \cite{buckets}.

\bibitem{Maartensdec}         R. Maartens, Phys. Rev. \bf D62 \normalfont 084023 (2000), hep-th/0004166;
                              R. Maartens, Reference Frames and Gravitomagnetism, eds. J. Pascual-S\'{a}nchez et al (World Scientific, Singapore, 2003), gr-qc/0101059.

\bibitem{Coley12}             R. Maartens, V. Sahni and T.D. Saini, Phys. Rev. \bf D63 \normalfont 063509 (2001), gr-qc/0011105.                        

\bibitem{MWBH}                R. Maartens, D. Wands, B. Bassett and I. Heard, Phys. Rev. \bf D62 \normalfont 041301 (2000), hep-th/9912464.

\bibitem{Mach}                E. Mach, {\it Die Mechanik in ihrer Entwickelung, Historisch-kritisch dargestellt} (J.A. Barth, Leipzig, 1883).  
                              The Enlish translation is \it The Science of Mechanics: A Critical and Historical Account of its Development \normalfont (Open Court, La Salle, Ill. 1960).    

\bibitem{Magaard}             L. Magaard, ``Zur einbettung riemannscher Raume in Einstein--Raume und konformeuclidsche Raume" (PhD Thesis, Kiel 1963).


\bibitem{MT}                  J.E. Marsden and F.J. Tipler, Phys. Rep. \bf 66 \normalfont 109 (1980).

\bibitem{CGconscience2}       D.R. Matravers, D.L. Vogel and M.S. Madsen, Class. Quantum Grav. \normalfont \bf 1 \normalfont 407 (1984).  

\bibitem{Mink}                H. Minkowski, the English translation is available in \it The Principle of Relativity \normalfont
                             (Dover, New York 1952, formerly published by Methuen, London 1923).
     
\bibitem{Misnerdata}          C.W. Misner, Ann. Phys. (N.Y.) {\bf 24} 102 (1963).

\bibitem{M69}                 C.W. Misner, Phys. Rev. \bf 186 \normalfont 1319 (1969).

\bibitem{VMisner}             C.W. Misner, in \it Magic without magic: John Archibald Wheeler\normalfont, ed. J.R. Klauder (Freeman, San Francisco, 1972).

\bibitem{MTW}                 C.W. Misner, K. Thorne and J.A Wheeler, Gravitation (Freedman, San Francisco, 1973).

\bibitem{MW57}                C.W. Misner and J.A. Wheeler, Ann. Phys. N.Y. \bf 2 \normalfont 525 (1957). 


\bibitem{TM72}                V. Moncrief and C. Teitelboim, Phys. Rev. {\bf D6} 966 (1972).

\bibitem{Mott}                N.F. Mott, Proc. Roy. Soc. {\bf A124} 375 (1929).

\bibitem{401N}                K. Nakao, K. Nakamura and T. Mishima, Phys. Lett. {\bf B564 }, 143 (2003), gr-qc/0112067;  
                              K. Nakamura, K. Nakao and T. Mishima, Progress of Theoretical Physics Supplement \bf 148 \normalfont \it Brane World: New Perspective in Cosmology \normalfont, gr-qc/0302058.

\bibitem{Vfermi}              J.E. Nelson and C. Teitelboim, Phys. Lett. \bf 69B \normalfont 81 (1977).
        
\bibitem{Principia13}         I. Newton, {\it Philosophiae Naturalis Principia Mathematica} (1686).  
                              For an English translation, see e.g I.B. Cohen and A. Whitman (University of California Press, Berkeley, 1999). 

\bibitem{Scholium}            I. Newton, Scholium on absolute motion in {\it Mathematical 
                              Principles of Natural Philosophy} (1729).  

\bibitem{cwb}                 N. \'{O} Murchadha and J.W. York, J. Math. Phys. \bf 11 \normalfont 1551 (1973).

\bibitem{SGNiall}             N. \'{O} Murchadha, Int. J. Mod. Phys. A \bf 20 \normalfont 2717 (2002).

\bibitem{Niall03}             N. \'{O} Murchadha, gr-qc/0305038.  

\bibitem{Wesson1}             J.M. Overduin and P.S. Wesson, Phys. Rept. \bf 283 \normalfont 303 (1997), gr-qc/9805018.

\bibitem{CGshapeshift2}       See, e.g, T. Padmanabhan, {\it Structure Formation in the Universe} (Cambridge University Press, Cambridge 1993).

\bibitem{PaWoo}               D.N. Page and W.K. Wooters, Phys. Rev. {\bf D27} 2885 (1983).  

\bibitem{Peccei}              R.D. Peccei, "the strong CP problem", CP Violation,
                              Advanced Series on Directions in High Energy Physics Vol. 3, ed C.Jarlskog (World Scientific, 1989)

\bibitem{Penrose60}           This work of Penrose's is mentioned in \cite{RMW2}.  

\bibitem{PenroseCCP69}        R. Penrose, Nuovo Cim. {\bf 1}, 252 (1969). 

\bibitem{Penrosespinnets}     R. Penrose, in {\it Advances in Twistor Theory}, ed. L.P. Hughston and R.S. Ward (Pitman, London, 1979).

\bibitem{Pen96}               R. Penrose Gen. Rel. Grav. {\bf 28} 581 (1996).  

\bibitem{PenRind}             R. Penrose and W. Rindler, \it Spinors and Space-time Vol. 1 Two-spinor calculus and relativistic fields \normalfont (Cambridge University Press,
                              Cambridge, 1984).

\bibitem{Perez}               See, e.g, A. Perez, Class. Quantum Grav. {\bf 20} R43 (2003), gr-qc/0301113. 

\bibitem{PS}                  M.E. Peskin and D.V. Schroeder, \it An Introduction to Quantum Field Theory \normalfont
                             (Addison-Wesley, and Perseus Books, 1995)


\bibitem{Vsugy3}              M. Pilati, Nucl. Phys. \bf B132 \normalfont 138 (1978).

\bibitem{SGPilatiID}          M. Pilati, in \it Quantum structure of space and time\normalfont, ed. M. Duff and C.J. Isham (Cambridge University Press, Cambridge 1982).

\bibitem{Pilatilit1}          M. Pilati,  Phys. Rev. D  \bf 26 \normalfont 2645 (1982).

\bibitem{Pilatilit2}          M. Pilati,  Phys. Rev. D  \bf 28 \normalfont 729 (1983).

\bibitem{Polchinski}          See, for example, J. Polchinski, \it String Theory \normalfont (Cambridge University Press, Cambridge, 1998).  

\bibitem{RRII}                O. Pooley, http://philsci-archive.pitt.edu/archive/00000221/index.html.

\bibitem{RRICM}               O. Pooley and H.R. Brown, Brit. J. Phil. Sci. \bf 53 \normalfont 183 (2002).

\bibitem{sing1}               J. Ponce de Leon, Gen. Rel. Grav. \bf 20 \normalfont 1115 (1988).

\bibitem{PDL01}               J. Ponce de Leon, Mod. Phys. Lett. \bf A16 \normalfont 2291 (2001), gr-qc/0111011.

\bibitem{ellbooks}            M.H. Protter and H.F. Weinberger, \it Maximal Principles in Differential Equations \normalfont (Prentice--Hall, Inc. Englewood Cliffs, N.J. 1967). 

\bibitem{Rainich}             G.Y. Rainich, Trans. Amer. Math. Soc. \bf 27 \normalfont 106 (1925).

\bibitem{stdw1}               L. Randall, R. Sundrum, Phys. Rev. Lett. \bf 83 \normalfont 3370 (1999), hep-ph/9905221.

\bibitem{stdw2}               L. Randall, R. Sundrum, Phys. Rev. Lett. \bf 83 \normalfont 4690 (1999), hep-th/9906064. 

\bibitem{CGshapeshift1}       M.J. Rees and D.W. Sciama, Nature \bf 217 \normalfont 511 (1968).  

\bibitem{RegWill}             T. Regge and R.M. Williams, J. Math. Phys. {\bf 41} (2000), gr-qc/0012035.  

\bibitem{Reissner}            H. Reissner (1914), English translation in \cite{buckets}.

\bibitem{Rendall02}           A.D. Rendall, Living Rev. Rel. \bf 5 \normalfont 6 (2002), gr-qc/0203012.


\bibitem{RTZ}                 C. Romero, R. Tavakol and R. Zalaletdinov, Gen. Rel. Grav. \bf 28 \normalfont 365 (1996).

\bibitem{SGRovelli}           C. Rovelli, Phys. Rev. D, \bf 35 \normalfont 2987 (1987).

\bibitem{RovAshStach}         C. Rovelli, page 126 in {\it Conceptual Problems of Quantum 
                              Gravity}, eds. A. Ashtekar and J. Stachel (Birkh\"{a}user, Boston, 1991).  

\bibitem{BBQ}                 C. Rovelli, page 292  in {\it Conceptual Problems of Quantum 
                              Gravity}, eds. A. Ashtekar and J. Stachel (Birkh\"{a}user, Boston, 1991).  


\bibitem{Rovelli97}           C. Rovelli, Living Rev. Rel. {\bf 1} 1 (1998), gr-qc/9710008.

\bibitem{RovSmo}              C. Rovelli and L. Smolin, Phys. Rev. Lett. {\bf 61} 1155 (1988); 
                              Nucl. Phys. {\bf B331} 80 (1990).  

\bibitem{RSspinnetworks}      C. Rovelli and L. Smolin, Nucl. Phys. {\bf B442} 593; 
                              erratum {\bf B456} 753 (1995), gr-qc/9411005; 
                              Phys. Rev {\bf D52} 5743 (1995), gr-qc/9505006.     

\bibitem{stdbr2}              V.A. Rubakov and M.E. Shaposhnikov, Phys. Lett. B \bf 125 \normalfont 136 (1985).  

\bibitem{Ryan}                M.P. Ryan, {\it Hamiltonian Cosmology} (Springer, New York, 1972).

\bibitem{otherpapers12}       V. Sahni, M. Sami and T. Souradeep, Phys. Rev. \bf D65 \normalfont 023518 (2002), gr-qc/0105121.

\bibitem{VSak}                The English translation is A.D. Sakharov, Sov. Phys. Doklady \bf 12 \normalfont 1040 (1968).

\bibitem{Savvidou}            K.N. Savvidou, J. Math. Phys. {\bf 43} 3053 (2002), gr-qc/0104053; 
                              K.N. Savvidou, Class. Quantum Grav. {\bf 18} 3611 (2001), gr-qc/0104081.  

\bibitem{SGstringyrefs11}     A. Schild, Phys. Rev. D, \bf 16 \normalfont 1722 (1977).

\bibitem{Schoen}              R.S. Schoen, J. Diff. Geom. \bf 20 \normalfont 45 (1984) and references therein .

\bibitem{anhHaj}              M. Sch\"{o}n and P. H\'{a}j\'{\i}\v{c}ek Class. Quantum Grav. {\bf 7} 861;  
                              P. H\'{a}j\'{\i}\v{c}ek, 871 (1990).

\bibitem{Schrodinger25}       E. Schr\"{o}dinger (1925), English translation in \cite{buckets}.

\bibitem{Schrodinger50}       E. Schr\"{o}dinger, {\it Space-Time Structure} Cambridge University Press, Cambridge 1950).

\bibitem{Wesson03}            S.S. Seahra and P.S. Wesson, Class. Quantum Grav. {\bf 20 } 1321 (2003), gr-qc/0302015.  


\bibitem{SGstringyrefs14}     N. Seiberg and E. Witten, Nu. Phys. B \bf 471 \normalfont 121 (1996), hep-th/9603003.  

\bibitem{Shibata}             M. Shibata, K. Taniguchi and K. Uryu, Phys. Rev. {\bf D68} 084020 (2003), gr-qc/0310030.  

\bibitem{SMS}                 T. Shiromizu, K. Maeda, M. Sasaki, Phys. Rev. \bf D62 \normalfont 024012 (2000), gr-qc/9910076.

\bibitem{401SS}               T. Shiromizu and M. Shibata, Phys. Rev. \bf D62 \normalfont 127502 (2000), hep-th/0007203.  

\bibitem{Smarrsim}            L. Smarr, in \it Sources of gravitational radiation\normalfont, ed. L. Smarr (Cambridge University Press, Cambridge 1979).

\bibitem{SYrad}               L. Smarr and J.W. York {\it Phys. Rev.} {\bf D17} 1945 (1978).   

\bibitem{CGSYkin}             L. Smarr and J.W. York  \it Phys. Rev. \normalfont \bf D17 \normalfont 2529 (1978).

\bibitem{Y79book}             See the theory chapters of \it Sources of gravitational radiation\normalfont, ed. L. Smarr (Cambridge University Press, Cambridge 1979).

\bibitem{VSmolin}             L. Smolin, in \it Time and the Instant\normalfont, Ed. R. Durie (Clinamen Press, Manchester 2000), gr-qc/0104097.

\bibitem{401lit2}             E. Sorkin and T. Piran, Phys. Rev. Lett. {\bf 90 } 171301 (2003), hep-th/0211210.  


\bibitem{Stachel}             J. Stachel, in {\it Stuies in the History of General Relativity}, 
                              ed. J. Eisenstaedt and A.J. Kox.

\bibitem{Stelle}              K.S. Stelle, Phys. Rev. {\bf D16} 953 (1977).  

\bibitem{Vhdl1}               K.S. Stelle, Gen. Rel. Grav. \bf 9 \normalfont 353 (1978).

\bibitem{Stellmacher38}       K. Stellmacher, Math. Ann. {\bf 115} 136 (1938). 

\bibitem{Stephani}            H. Stephani, {\it Allgemeine Relativit\"{a}tstheorie} (VEB, Berlin, G.D.R. 1977).  The English translation is {\it General relativity} (Cambridge University Press, Cambridge 1990).

\bibitem{Haus}                The result is in M.D. Stern (Princeton A.B. Senior Thesis, 1967); 
                              the result is reported in \cite{Wheeler} and in \cite{VFischer70}.  

\bibitem{Stewart}             J.M. Stewart, {\it Advanced General Relativity} (Cambridge University Press, Cambridge 1991).

\bibitem{VSydlowski}          M. Szydlowski and A. Lapeta, Phys. Lett. A \bf 148 \normalfont 239 (1990).

\bibitem{Teitelthesis}        C. Teitelboim, ``The Hamiltonian Structure of Spacetime" (PhD Thesis, Princeton 1973).  

\bibitem{T73}                 C. Teitelboim, Ann. Phys. N.Y. {\bf 79} 542 (1973).   

\bibitem{Vsqrtteitel}         C. Teitelboim, Phys. Rev. Lett. \bf 38 \normalfont 1106 (1977).

\bibitem{Teitelboim}          C. Teitelboim, in \it General Relativity and Gravitation \normalfont Vol 1, ed. A. Held (Plenum Press, NY, 1980).

\bibitem{STeitel}             C. Teitelboim, Phys. Rev. \bf D25 \normalfont 3159 (1982).

\bibitem{Lovelocksystem}      C. Teitelboim and J. Zanelli Class. Quantum Grav. \bf 4 \normalfont L125 (1987).  

\bibitem{Thiemann}            T. Thiemann, Living Rev. Rel (2001), gr-qc/0110034.

\bibitem{Thiemannpapers}      T. Thiemann, Class. Quant. Grav. {\bf 15} 839, gr-qc/9606089; 
                              875, gr-qc/9606090; 
                              1207, gr-qc/9705017 (1998).

\bibitem{Thiemannmatter}      T. Thiemann, Class. Quant. Grav. {\bf 15} 1281 (1998), gr-qc/9705019.  

\bibitem{Phoenix}             T. Thiemann, gr-qc/0305080.

\bibitem{MtheoryT}            P.K. Townsend,  Phys. Lett. B {\bf 350}, 184 (1995).

\bibitem{UW}                  W.G. Unruh and  R.M. Wald, Phys. Rev. {\bf D40} 2598 (1989).  

\bibitem{Utiyama}             R. Utiyama, Phys. Rev. {\bf 101} 1597 (1956). 

\bibitem{Vilenkin}            A. Vilenkin, Phys. Rev. {\bf D30} 509 (1984).

\bibitem{VW}                  M. Visser and D.L. Wiltshire, Phys. Rev. {\bf D67 } 104004 (2003), hep-th/0212333.

\bibitem{Wald}                R.M. Wald \it General Relativity \normalfont (University of Chicago Press, Chicago, 1984).

\bibitem{Weinberg}            S. Weinberg, \it Quantum Theory of Fields \normalfont volume II (Cambridge University Press, Cambridge, 1995).

\bibitem{VWeinberg3}          S. Weinberg, \it Quantum theory of Fields \normalfont Vol. III Cambridge University Press, Cambridge, 2000).


\bibitem{Wesson2}             P.S. Wesson, \it Space-Time-Matter \normalfont (World Scientific, Singapore 1999). 

\bibitem{CGWetterich}         C. Wetterich, Nuclear Physics \textbf{B302} 645; 668 (1988). 

\bibitem{CGWeyltheory}        H. Weyl, {Sitzungsber. d. Preuss. Akad. d. Wissensch.} 465 (1918)
                             (English translation: in L. O' Raifeartaigh  ``The Dawning of Gauge Theory" (Princeton: Princeton University Press 1997). 

\bibitem{VWeyltheory}         H. Weyl, \it Gravitation und Elektricitat\normalfont,
                              Sitzungsberichte der Preussichen Akad. d. Wissenschaften (1918),
                              available in English translation in same book as \cite{VEinstein}. 

\bibitem{VWeyl}               H. Weyl, \it Space-Time-Matter \normalfont 4th edition (English translation: Methuen, London 
                              1920), wherein note 8 of chapter 4 lists the original literature for the Einstein--Hilbert action, 
                              as regards the simplicity theorem, Weyl also credits the work of H. Vermeil, 
                              Nachr. Ges. Wiss. Gottingen (1917).

\bibitem{Weylkin}             H. Weyl, {\it Space-Time-Matter} (Dover, New York, 1950). 

\bibitem{CGWeylGauge}         H. Weyl, \textit{Zeitschrift f. Physik} \textbf{56} 330 (1929).

\bibitem{W59}                 J.A. Wheeler, Ann. Phys {\bf 2} 604 (1957).

\bibitem{RMW2}                J.A. Wheeler, \it Geometrodynamica \normalfont (Academic Press, New York 1962).

\bibitem{WheelerGRT}          J.A. Wheeler, in \it Groups, Relativity and Topology \normalfont, ed. B.S. DeWitt and C.M. DeWitt (Gordon and Breach, New York 1963); 

\bibitem{Wheeler}             J.A. Wheeler, in \it {Battelle rencontres: 1967 lectures in mathematics and physics} \normalfont 
                              ed. C. DeWitt and J.A. Wheeler (Benjamin, New York 1968).

\bibitem{W88}                 J.A. Wheeler, Int. J. Mod. Phys. A, \bf 3 \normalfont 2207 (1988).
 
\bibitem{Vostrograd}          See, for example, E.T. Whittaker, \it A Treatise on the Analytical Dynamics of Particles and Rigid Bodies
                              \normalfont (Cambridge University Press, Cambridge, 1904).

\bibitem{Will}                C.M. Will, Living Rev. Rel. \bf 4 \normalfont 4 (2001), gr-qc/0103036.

\bibitem{Winicour}            J. Winicour, Living Rev. Rel. (2001), gr-qc/0102085.

\bibitem{starapp}             T. Wiseman, Phys. Rev. \bf D65 \normalfont 124007 (2002), hep-th/0111057.                        

\bibitem{WittenKK}            E. Witten, Nu. Phys. \bf B186 \normalfont 412 (1981).  

\bibitem{SGstringyrefs15}     E. Witten, Nu. Phys.  \bf B471 \normalfont 195 (1996), hep-th/9602070.

\bibitem{MtheoryW}            E. Witten, Nu. Phys. {\bf B443},  85 (1995), hep-th/9503124.  

\bibitem{RMW1}                L. Witten, in \it Gravitation: an introduction to current research \normalfont, ed. L. Witten (Wiley, New York 1962).

\bibitem{CGYamabe}            H. Yamabe, Osaka Math. J. \bf 12 \normalfont 21 (1960).   

\bibitem{YM}                  C.N. Yang and R.L. Mills, Phys. Rev. \bf 96 \normalfont 191 (1954).

\bibitem{York71}              J.W. York, Phys. Rev. Lett. \bf 26 \normalfont 1656 (1971).

\bibitem{York72}              J.W. York, Phys. Rev. Lett. \bf 28 \normalfont 1082 (1972).

\bibitem{York73}              J.W. York, J. Math. Phys. \bf 14 \normalfont 456 (1973).

\bibitem{York74}              J.W. York, Ann. Inst. Henri Poincar\'{e} \bf 21 \normalfont 319 (1974).

\bibitem{York83}              J.W. York, in \it Gravitational Radiation\normalfont, ed. N. Deruelle and T. Piran (North-Holland, Amsterdam 1983). 

\bibitem{BY}                  This is nicely summarized in J.W. York, in \it Frontiers in Numerical Relativity \normalfont 
                              ed. C.R. Evans, L.S. Finn and D.W. Hobill (Cambridge University Press, Cambridge 1989).  

\bibitem{YCTS}                J.W. York, gr-qc/9810051.  

\bibitem{PfeifferYork}        J.W. York and H.P. Pfeiffer, Phys. Rev. {\bf D67} 044022 (2003), gr-qc/0207015.

\bibitem{Zan}                 H. Zanstra, Phys. Rev. {\bf 23} 528 (1923).  



\end{thebibliography}
\end{document}